\documentclass[iicol,sn-mathphys-num]{sn-jnl}

\usepackage[T1]{fontenc}
\usepackage{graphicx}%
\usepackage{multirow}%
\usepackage{amsmath,amssymb,amsfonts,mathtools}%
\usepackage{amsthm}%
\usepackage{commath}
\usepackage{mathrsfs}%
\usepackage[title]{appendix}%
\usepackage{xcolor}%
\usepackage{textcomp}%
\usepackage{manyfoot}%
\usepackage{booktabs}%
\usepackage{algorithm}%
\usepackage{algorithmicx}%
\usepackage{algpseudocode}%
\usepackage{listings}%
\usepackage{siunitx}
\usepackage{physics}
\usepackage{pifont}
\usepackage{bbm}
\usepackage{isotope}
\usepackage{empheq}
\usepackage{upgreek}
\usepackage{interval}
\usepackage{rotating}
\usepackage{lmodern} 
\usepackage{xspace}
\usepackage{cases}
\usepackage{arydshln}
\usepackage{url}
\usepackage{lmodern}
\usepackage{textcomp}%
\usepackage[utf8]{inputenc}
\usepackage{textgreek}
\usepackage{pdflscape}
\usepackage{rotating}
\usepackage{csquotes}

\newcommand{\xmark}{\ding{55}}
\newcommand{\e}{\mathrm{e}}  
\newcommand{\ii}{\mathrm{i}}  

\newcommand{\cev}[1]{\reflectbox{\ensuremath{\vec{\reflectbox{\ensuremath{#1}}}}}} 

\makeatletter
\renewcommand\paragraph{\@startsection{paragraph}{4}{\z@}%
            {-2.5ex\@plus -1ex \@minus -.25ex}%
            {1.25ex \@plus .25ex}%
            {\normalfont\normalsize\bfseries}}
\makeatother

\begin{document}

\title[Gogny force review]{Gogny interaction from beginnings to current challenges}


\author*[1,2]{\fnm{Nathalie} \sur{Pillet}}\email{nathalie.pillet@cea.fr}

\author[1,2]{\fnm{Geoffrey} \sur{Zietek}}

\author*[3,4]{\fnm{Luis} \sur{Robledo}}\email{luis.robledo@uam.es}

\author*[1,2]{\fnm{Marc} \sur{Dupuis}}\email{marc.dupuis@cea.fr}

\author*[4]{\fnm{Tomás R.} \sur{Rodríguez}}\email{trodrig@us.es}

\affil*[1]{ \orgname{CEA, DAM, DIF}, \orgaddress{\postcode{91297} \state{Arpajon}, \country{France}}}

\affil[2]{\orgdiv{Université Paris-Saclay}, \orgname{CEA, LMCE}, \orgaddress{\postcode{F-91680}, \state{Bruyères-le-Châtel}, \country{France}}}

\affil[3]{\orgdiv{Departamento Física Teórica and CIAFF}, \orgname{Universidad Aut\'onoma de Madrid}, \orgaddress{\postcode{28049}, \state{Madrid}, \country{Spain}}}


\affil[4]{\orgdiv{Departmento de F\'isica At\'omica, Molecular y Nuclear}, \orgname{Universidad de Sevilla}, \orgaddress{\postcode{E-41012}, \state{Sevilla}, \country{Spain}}}


\abstract{In the present review, various facets of the phenomenological effective Gogny interaction, which was originally proposed in the 70's, are  
gathered for the first time in the same place.
This involves both nuclear phenomena of interest that led to its creation and evolution as well as highly technical aspects that led the objectives to be achieved. With this in mind, 
a discussion structured around four key points is proposed. After a general introduction, the history and philosophy of the Gogny interaction is exposed. In particular, 
one highlights the way followed by D. Gogny to guide the determination of the parameters of its phenomenological interaction with results obtained from a realistic interaction 
using Hartree-Fock calculations and second order corrections and a G-matrix. One also shows how physical phenomena such as pairing or fission were essential to improve the 
initial parameterization. The evolution of the original analytical form over the years is then discussed. The second key point concern the emulator that was used for the generation 
of parameterizations. Its modifications, consistent with the evolution of its analytical form along the decades, are detailed. Other fitting procedures from other groups, 
more recent, are also evoked. 
The third key point is dedicated to the role of some nuclear matter properties in the fitting process and the acceptance of a parameterization. The objective of the last key 
point consists in highlighting some results obtained with the Gogny interaction in various domains such as nuclear structure, fission, reactions and nuclear data that have allowed 
to interpret many experimental data. Perpsectives are given explaining some paths that will be followed in the future concerning the functional itself as well as N-body approaches. 
}




\maketitle

\section{Introduction}\label{secintro}

The derivation of the nuclear interaction from first principles is one of the long-term research domain in theoretical nuclear physics. It is empirically known since the 1930s that the nuclear force is composed of an attractive part at medium and long ranges, and of a very strong repulsive part at short range, the so-called hard-core repulsion. Already at the two-body level, it is difficult to accurately separate the attractive from the repulsive regions, since it depends not only on the relative distance of the two implied nucleons $r \equiv |\vec{r}_1 - \vec{r}_2|$, but also on their velocities, spins and isospins. However, it is agreed that around $r \lesssim \SI{0.7}{\femto\metre}$, the hard-core repulsion dominates, while beyond it, the attractive character of the nuclear force prevails, with an exponential decline that practically vanishes at about $r \sim \SI{3}{\femto\metre}$.

As early as 1935, Yukawa \cite{Yukawa1935} attempted to describe this singular force by assuming that nucleons were exchanging particles, the mesons, that would mediate the nuclear interaction, in analogy with the electromagnetic force which was already understood as originating from photon exchanges. The Pion meson predicted by Yukawa was observed for the first time in cosmic rays in 1947. From there, many theories based on the exchange of various mesons were born and proved to be very fruitful.

In 1964, Gell-Mann \cite{GellMann1964} and Zweig \cite{Zweig1964} independently postulated that hadrons, of which nucleons and mesons are the representatives, were not elementary particles, but were composed of more fundamental objects, the quarks. In 1968, the SLAC's scattering experiments revealed their existence \cite{Riordan1992}. Since mesons were no longer seen as elementary particles, it was necessary to relegate meson theories to the level of models and to try to explain the origin of the nuclear force through the interaction between the quark themselves, the strong interaction. The quantum chromo-dynamics (QCD) appeared as the most fundamental theory during the 1970s \cite{Fritzsch1973,Fritzsch2012}. This strong force
revealed two extreme behaviors \cite{Bethke2007}. At large distances or equivalently low-energy transfers, the force is so strong that it prevents the existence of free quarks, the color confinement phenomenon \cite{Greensite2011}. At small distances or high-energy transfers, the force becomes weaker so that quarks behave like nearly free particles, the asymptotic freedom phenomenon \cite{Gross2005}. This last feature of the strong interaction was discovered by Gross and Wilczek \cite{Gross1973}, and independently by Politzer \cite{Politzer1973} in 1973.

Even if nucleons, which are made of three quarks, are color-
neutral particles, they interact at small distances 
($r \lesssim 2.5 \text{ fm}$). The nuclear interaction can be 
interpreted as the residual strong force, in the same way as, 
in the electromagnetism, two electrically neutral atoms still 
feel a residual electromagnetic force through the Van Der Waals 
interaction. During this last decades, a lot of progresses have 
been accomplished in order to derive an expression for the 
nuclear interaction from the QCD equations. As well-known, 
this research field faces two difficulties. Firstly, the 
interaction between two nucleons, which are composite 
particles, actually consists in a six quark problem. Secondly, 
the region in which the strong interaction is particularly 
intense corresponds to the low-energy regime of the nuclear 
physics, making QCD highly non-perturbative. Mainly two 
approaches have been developed to address these two issues. 
The first one tries to tackle the six-quark problem in which 
the system is embedded in a four-dimensional discretized 
lattice on which QCD equations are applied. This 
lattice quantum chromodynamics approach (LQCD) 
\cite{Wilson1974} provided a genuine breakthrough in 2007 by 
showing that the hard-core repulsion is also a consequence of 
the strong force \cite{Ishii2007}. Due to the enormous computer 
ressources needed, the LQCD developments cannot be considered 
for the time being as a standard technique for nuclear physics.
The second approach, which tries to tackle the non-perturbative 
character of QCD in the nuclear physics regime, uses 
effective field theories (EFTs) \cite{Hammer2020} to seek 
a relation between low-energy QCD and the nuclear interaction. 
The main idea here is to use the separation between the typical 
length scale or energy scale of the physical phenomena 
to describe and the typical length scale of the underlying dynamics, which is expressly ignored. 
The pioneering work of Weinberg \cite{Weinberg1979,Weinberg1991}
extended EFTs to low-energy QCD in the 1990s. This field is very vivid nowadays and nuclear EFTs provide a low-energy depiction of QCD in terms of hadrons instead of quarks. 
The first and surely the most widespread nuclear EFT is the chiral EFT ($\chi$EFT) \cite{Machleidt2011}.
The spontaneously breaking of the chiral symmetry brings 
out the pion as the Goldstone boson and makes it become the main mediator of the nuclear interaction. Its mass is chosen to be the typical energy scale $\Lambda_{\pi} \sim m_{\pi} \simeq \SI{140}{\MeV}$ of the phenomena and the chiral-symmetry breaking energy scale $\Lambda_{\chi} \sim \SI{1}{\GeV}$ is the limit beyond which the phenomena are not reachable by the theory. A chiral perturbation theory (ChPT) \cite{Leutwyler2012}, is then possible. It permits, by taking back perturbative methods to $\chi$EFT, to probe nuclear phenomena in a controlled manner, with the desired degree of accuracy. By accounting for the ever increasing orders of the perturbative expansion, the knowledge of the nuclear interaction is systematically refined.

Going back to the 50s, the problem of the hard-core repulsion was already on the minds of physicists. Even if, at normal density, the nucleons remain on average quite far from each other as a consequence of the Pauli principle, it cannot be excluded that sometimes two nucleons are momentarily at distances involving the hard-core repulsion. In that case, they scatter violently on each other, and feel the short range correlations. Several attempts appear in the 50s to take into account these correlations. One of the most widespread is that of Brueckner \cite{Brueckner1955,Day1967} who proposed a model consisting in re-normalizing the bare interaction into an effective interaction, the Brueckner G-matrix, by resuming the ladder diagrams. The effective interaction thus obtained has no hard-core and integrate in essence the effects of short range correlations. The effective interaction is more elaborated because it depends on the states of the nucleons and the density of the medium. Brueckner showed that such kind of effective interaction was compatible with Hartree-Fock (HF) theory. However, since the Brueckner-Hartree-Fock (BHF) theory is hard to set up in finite nuclei, especially for heavy and deformed nuclei, its application were mostly restricted to infinite nuclear matter (INM).
This led physicists to construct parameterizations of the nuclear effective interaction, depending on free parameters which were determined (i) from theoretical results obtained in Brueckner theory (ii) by adjusting them in order to reproduce key quantities of infinite nuclear matter (saturation point, incompressibility, etc.), and (iii) in a phenomenological way, by fitting them on experimental data of various types (binding energies, charge radii, etc.) to be able to describe as many nuclei as possible.
Most of them also retained the explicit density dependence of the medium which is essential in reproducing the standard properties of nuclei, but abandon the energy dependence which turns out to be less important, at least in the perspective of nuclear structure studies.

Although it is not yet possible to extract a general expression of the bare nuclear interaction from the more fundamental QCD, an analytical expression can be constrained by imposing some symmetries to be preserved \cite{Okubo1958,Ring2004,Berger2008}. Thus, for example, imposing the symmetry under the exchange of particles, the translational symmetry, the symmetry under Galilean transformation, the rotational symmetry, the parity symmetry, the time-reversal symmetry, the hermiticity and the charge independence, one obtains the following general form of a two-body interaction expressed in spin-isopsin channels $ST$:
\begin{equation} \label{v12form}
v_{12} = v^{00} + v^{10} (\vec{\sigma}_1 \cdot \vec{\sigma}_2) + v^{01} (\vec{\tau}_1 \cdot \vec{\tau}_2) + v^{11} (\vec{\sigma}_1 \cdot \vec{\sigma}_2)(\vec{\tau}_1 \cdot \vec{\tau}_2)
\end{equation}
where 
\begin{equation} \label{v12rpt}
v_{12}  \equiv v_{12}(\vec{r}, \vec{p}, \vec{L})
\end{equation}
and
\begin{equation} \label{componentsv12}
v^{ST} \equiv v^{ST}(\vec{r}, \vec{p}, \vec{L}) \equiv \sum_{k=1}^{5} f_k^{ST}(\vec{r}^{\, 2}, \vec{p}^{\, 2},\vec{L}^{2}) O_k
\end{equation}
for $S,T \in \{ 0 , 1 \}$. The functions $f_k^{ST}(\vec{r}^{\, 2}, \vec{p}^{\, 2},\vec{L}^{2})$ are the form factors and the operators $O_k$ take one of the 
following forms:
\begin{equation} \label{operatorsv12}
\begin{cases} 
  O_1 = \mathbbm{1}  \\
  O_2 = \vec{L} \cdot \vec{S}\\
  O_3 = S_{12}(\hat{r}) \equiv (\vec{\sigma}_1 \cdot \hat{r})(\vec{\sigma}_2 \cdot \hat{r}) - \frac{1}{3} \vec{\sigma}_1 \cdot \vec{\sigma}_2 \\
  O_4 = S_{12}(\hat{p}) \equiv (\vec{\sigma}_1 \cdot \hat{p})(\vec{\sigma}_2 \cdot \hat{p}) - \frac{1}{3} \vec{\sigma}_1 \cdot \vec{\sigma}_2\\
  O_5 = Q_{12} \equiv \frac{1}{2} \big[ (\vec{\sigma}_1 \cdot \vec{L})(\vec{\sigma}_2 \cdot \vec{L}) + (\vec{\sigma}_2 \cdot \vec{L})(\vec{\sigma}_1 \cdot \vec{L}) \big]
\end{cases}
\end{equation}
where one recognizes the so-called central $\mathbbm{1}$, spin-orbit $\vec{L} \cdot \vec{S}$, tensor $S_{12}(\hat{r})$, second tensor $S_{12}(\hat{p})$ and quadratic spin-orbit $Q_{12}$ components.
Here, several hypotheses have been postulated: the nucleons are the elementary particles of the theory; they are 
considered as point-like particles; they only interact through the nuclear interaction in the vacuum; the interaction 
propagates instantaneously as nucleons are assumed to obey a non-relativistic dynamics.

In the 60s, one saw the emergence of two specific ways of building the effective interaction. 
The former suggest that the interaction is reasonably re-normalized by the medium effects, so 
that the bare interaction can still serve as a guideline. The general expression of the effective 
interaction is then conducted by the available operators $O_k$ of the bare interactions, and its 
form factors $f_k^{ST}(\vec{r}^{\, 2}, \vec{p}^{\, 2},\vec{L}^{2})$ are chosen relatively simple 
to avoid too expensive calculation times. The latter claim that the various contributions are strongly 
mixed up when parameters are fitted on experimental data and a strong simplification of the analytical 
form is operated rendering the calculations in finite nuclei more systematics. As a consequence, the 
choice of the parameters is dictated by the desire to reproduce specific observables with the highest 
accuracy. This has led to two main phenomenological effective interactions, the Skyrme and Gogny interactions, 
that have been widely used and improved since then.
These last considerations guided D. Gogny in the construction of his phenomenological nucleon-nucleon 
interaction D1 \cite{Gogny1973,Gogny1975a,Decharge1980}.

The present review article has three main objectives.
The first one is to explain the original philosophy of 
the Gogny interaction and to take a stock of the 
different developments made since its beginning in the 
international community. The second objective is to 
explain the way this interaction was built and the 
connection to observables in its fitting procedure either 
deduced from the physics in finite nuclei, which was an 
essential point in the mind of D. Gogny, or from some 
nuclear matter properties. The third objective is to 
highlight some important applications in structure, 
fission and reactions, the fields of interest of D. Gogny 
for which he proposed his phenomenological effective 
force in the perspective of mean-field plus reasonable extensions to treat beyond-mean-field correlations. 
Their explicit treatment is supposed to bring an accurate description of observables. In that manner,
the fitting procedure of Gogny interaction is designed to leave room for them in such a way that, the mean-field
is not supposed to reproduce experimental data directly.
The application of a theory able to explain nuclear data was one of the fuel of D. Gogny research.
One hopes that this review will have the virtue of 
gathering for the first time many pieces concerning the 
Gogny interaction which are disseminated in many different articles or proceedings not always easily 
reachable or which have been never published,
preventing to build the global philosophy of this phenomenological effective interaction.

In section \ref{sec1}, the philosophy and the history
of the phenomenological effective Gogny interaction are
discussed, starting from the soft-core and local 
realistic interaction he proposed with his collaborators, 
then the first version of his phenomenological effective 
interaction known as D1,
and the other important parameterizations based on the D1 
analytical form that have been able to improve its 
predictive power, in particular in structure, fission and 
reactions. Then, the evolution of its analytical form 
will be discussed, which concerns both the inclusion of a 
tensor term or the fully finite-range character. 
These extensions will also be motivated by the 
improvement of the description of particular observables.
Sections \ref{sec2} and \ref{sec3} will be dedicated to the fitting procedure
of the Gogny interaction.
In section \ref{sec2}, the emulator that has been used
to deduce the values of the parameters is explained in
detail. In particular, the building of the associated 
meta-data to fix the mean-field is discussed. This 
includes several observables or pseudo-observables such as the binding energies, charge radii, pairing and asymmetry properties.
In section \ref{sec3}, the various filters 
coming from nuclear matter  
are explained and detailed expressions are given. Particular attention is paid to the equations of state, 
saturation density, 
incompressibility, symmetry energy and its slope, and Landau 
parameters. These last quantities are important as they test the particle-hole interaction, which is 
essential for RPA-type applications.
In section \ref{sec4}, highlights associated with  nuclear structure, fission, and reactions studies done
with the Gogny interaction are presented.
In section \ref{sec5}, conclusions and perspectives are proposed.

\section{Gogny interaction: History and Philosophy}\label{sec1}

In this section, one will go step by step to retrace as faithfully as one can the early stages that led to the creation of 
Gogny D1 phenomenological effective interaction. These considerations come directly from discussions with D. Gogny 
and his close collaborators, Jean-Francois Berger and M. Girod. Then, the more modern part will come either from the work of
some authors of the present review or from the work of numerous colleagues who published various studies in the scientific literature.

\subsection{From GPT to the D1 Gogny interaction}\label{GPTtoD1}

\begin{table*}[htb!]
    \centering
    \begin{tabular}{cc|cccccc|cc}
        \multicolumn{2}{c}{Channels} & \multicolumn{6}{c}{$v_{C}$} &  \multicolumn{2}{c}{$v_{LL}$} \\
        \hline
        S & T & {$v_{\mu}$} & {$\mu$} & {$v_{\nu}$} & {$\nu$} & {$v_{\rho}$} & {$\rho$} & {$v_{\mu}$} & {$\mu$} \\
        \hdashline
        0 & 1 & 560.0 & 0.8109 & -390.7 & 1.031 & -1.501 & 3.205 & -6.279 & 1.069 \\
        1 & 0 & 265.7 & 0.7482 & -113.9 & 1.418 & -1.000 & 2.412 & 50.82 & 0.8877 \\  
        0 & 0 & 240.0 & 1.200  & -0.9000& 2.000 & 0.1000 & 3.000 & -175.0 & 0.7000 \\
        1 & 1 & 9.335 & 1.184  & -1.37  & 2.099 & 0.1663 & 3.193 & 13.23 & 0.8097  \\
        \hline  \\
        \multicolumn{2}{c}{Channels} & \multicolumn{6}{c}{$v_{T}$} &  \multicolumn{2}{c}{$v_{LS}$} \\
        \hline
        S & T & {$v_{\mu}$} & {$\mu$} & {$v_{\nu}$} & {$\nu$} & {$v_{\rho}$} & {$\rho$} & {$v_{\mu}$} & {$\mu$} \\
        \hdashline
        1 & 0 & 82.49 & 0.5162 & -24.51 & 1.687 & -2.656 & 3.225 & 86.97 & 0.6718 \\  
        1 & 1 & 12.24 & 1.539  & -31.64 & 0.4039 & 0.8111 & 3.015 & -114.5 & 0.9296  \\
        \hline          
    \end{tabular}
    \caption{Parameter values of the GPT interaction.}
    \label{tab:GPT-ST}
\end{table*}

The first track of D. Gogny in the academic literature was through his proposition of a soft-core realistic nucleon-nucleon interaction, 
called GPT and published in 1970 with his collaborators P. Pir\`es and R. De Tourreil \cite{Gogny1970}. 
This work was part of the questioning at the time about whether or not it was possible "to fit the two-body data and at the same time to make, 
in finite nuclei, a HF plus second order perturbation treatment (as in the work of Kerman et al. \cite{Bassichis1967,Kerman1967}) successful". 
This GPT interaction was a fully finite range one. It was inspired by the potential of Hamada and Johnston \cite{Hamada1962}. In each spin-isospin channel (S,T), it writes {\it a priori} as:
\begin{equation} \label{v12GPT}
v_{12}^{GPT} = v_{C}(r) + v_{T}(r) S_{12} + v_{LS}(r) \vec{L}.\vec{S} + v_{LL}(r) L_{12}
\end{equation}
where 
\begin{equation} \label{ffGPT}
v_{i}(r) = \sum_{\alpha_{i}} v_{\alpha_{i}} e^{-r^2/\mu^2_{\alpha_{i}} }
\end{equation}
In Eq.(\ref{v12GPT}), $S_{12}$ was taken as three times the expression defined in Eq.(\ref{operatorsv12}) and 
$L_{12} = (\vec{\sigma}_{1}.\vec{\sigma}_{2}) \vec{L}^{2}-\frac{1}{2} \left[  
(\vec{\sigma}_{1}.\vec{L}) (\vec{\sigma}_{2}.\vec{L}) + (\vec{\sigma}_{2}.\vec{L}) (\vec{\sigma}_{1}.\vec{L})\right]$
was the adopted analytical form of the quadratic spin-orbit of in the Hamada-Johnston potential.
In Eq.(\ref{ffGPT}), the index "i" describes the central $v_{C}(r)$, tensor $v_{T}(r)$, spin-orbit 
$v_{LS}(r)$ and quadratic spin-orbit $v_{LL}(r)$ terms appearing in Eq.(\ref{v12GPT}). The number $\alpha_{i}$ 
represents the associated number of ranges by ST channel and $v_{\alpha_{i}}$ is the associated intensity of the nucleon-nucleon interaction. 
Gaussian form factors were chosen in order to allow simple HF calculations for spherical and deformed nuclei, thus strongly simplifying the original Hamada-Jonhston ones. 
Besides, the choice of Gaussians made this interaction similar to a separable interaction at the HF and HF-Bogoliubov (HFB) levels, as well as similar to a soft-core 
interaction with ultra-violet cutoff. At that time, other proposals were also made to transform the hard core Hamada-Johnston potential into a soft core nucleon-nucleon 
potential by introducing, for example, a soft-core of centrifugal type ($\rm \propto r^{-2}$) acting up to a separation radius \cite{Rochleder1968}.

The GPT interaction is a complex one as it contains all the key components of a realistic NN interaction.
Its fit was achieved on the following three criteria: (i) two-nucleon data (phase shifts up to $\SI{300}{\MeV}$, 
scattering lengths, effective ranges, deuteron properties) which were constrained by a $\chi^2$ minimization. 
At that time, there was no information about 3N and 4N systems from which behavior in light systems could be inferred 
(ii) the bulk properties of four spherical nuclei ($\isotope[16]{O}, \isotope[40]{Ca},\isotope[90]{Zr}$ and $\isotope[208]{Pb}$) 
with great accuracy for charge radii at the HF level assuming higher orders negligible while demanding an energy per particle 
of $\sim$3 MeV as important corrections were expected for higher orders with a rapidly convergent perturbation serie 
(iii) reasonable saturation properties in nuclear matter with a second order correction (SOC) of  $\sim 25 \%$ of the first 
order potential energy to make the perturbation serie convergence possible.
In Table \ref{tab:GPT-ST}, one has reported the parameters of the GPT interaction according to the spin-isospin channels (S,T). 
As previously mentionned, the values of ranges $\mu$, $\nu$ and $\rho$ depend on the (S,T) channels and describe short, medium and large distances.
In particular, the central and the tensor forces contain a range parameter which is large ($\sim$3 fm) to describe the long range associated with the 
one pion exchange potential (OPEP).

\begin{table}[htb!]
    \centering
    \begin{tabular}{ccccccc}
        {Nuclei} & E$^{(1)}$/A & E$^{(2)}$/A & E$^{(Exp.)}$/A  & r$^{(1)}_{c}$ & r$^{(Exp.)}_c$\\
        \hline
        $^{16}$O   & -2.57 & -7.96  & -7.98  & 2.85 &  2.73 \\
        $^{40}$Ca  & -2.78 & -10.00 & -8.55  & 3.48 &  3.50 \\
        $^{90}$Zr  & -2.76 & -10.78 & -8.71  & 4.11 &  4.30 \\
        $^{208}$Pb & -2.29 & -9.93  & -7.87  & 5.10 &  5.50 \\
        \hline 
        \end{tabular}
    \caption{Total binding energy per particle calculated at the HF (1) and HF plus SOC (2) levels. Charge radii have been evaluated at the HF approximation. A comparison with experiment (Exp.) is provided. Calculations were done for the $\isotope[16]{O}$, $\isotope[40]{Ca}$, $\isotope[90]{Zr}$ and $\isotope[208]{Pb}$ isotopes.}
    \label{tab:GPT-HF}
\end{table}

\begin{figure}[htb!]
    \centering
    \includegraphics[width=0.95\linewidth]{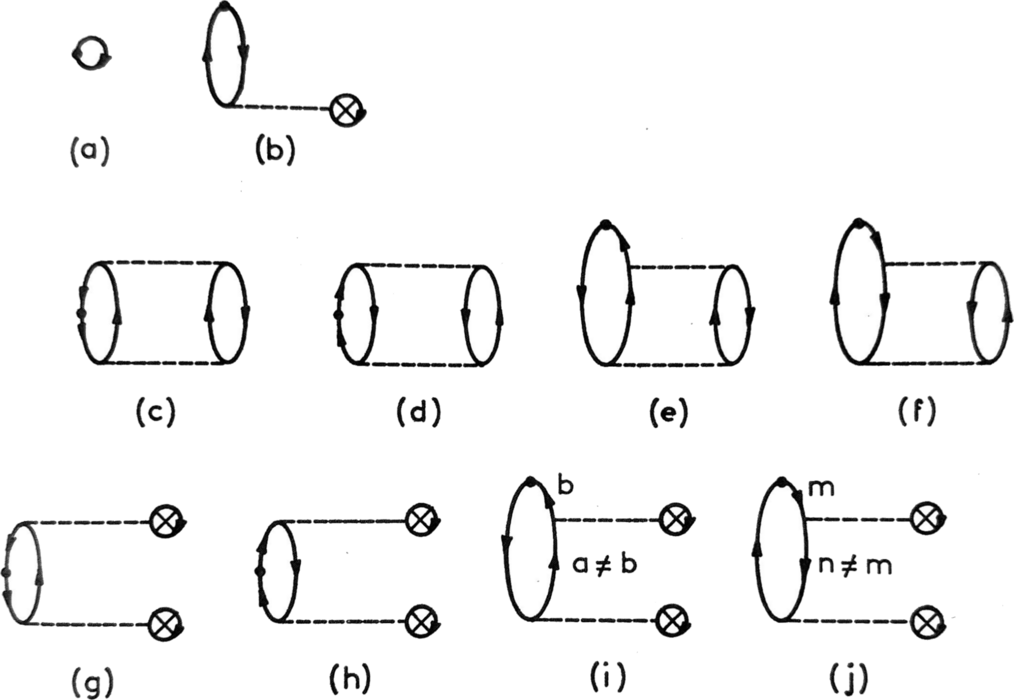}
    \caption{Goldstone diagrams included in the second order perturbation theory for the calculation of charge radii. The figure has been extracted from Ref. \cite{Maire1976}.}
    \label{fig:diagram}
\end{figure}

In Table \ref{tab:GPT-HF}, the binding energies calculated at both the HF, labeled by (1), and HF plus SOC,labeled by (2), levels are indicated 
for $\isotope[16]{O}$, $\isotope[40]{Ca}$, $\isotope[90]{Zr}$ and $\isotope[208]{Pb}$ isotopes (see PhD thesis of M. Maire \cite{Maire1976}). 
The charge radii r$_{c}^{(1)}$ evaluated at the HF level are also given. Associated experimental values, E$^{(Exp.)}$/A and r$_{c}^{(Exp.)}$ respectively, are shown. 
In these calculations, the SOC for binding energies were dominated by the tensor term in the ST=10 channel. It brought -2.44 MeV in $^{16}$O, to be compared to the 
central term with -0.87 MeV in the ST=10 channel and -0.79 MeV in the ST=01 one.
Concerning the charge radii, they were found relatively good at the HF approximation. It was obtained a global decrease of 7\% at second order in comparison 
with HF in the case of $^{16}$O providing a value of 2.64 fm. This decrease was essentially attributed to the second order diagram (e) of 
Fig. \ref{fig:diagram}, which is implicitly taken into account at the zero order of the BHF theory. It was obtained that the diagrams (c) and (d) canceled 
mutually and the diagrams (g,h,i) were found of one to two orders of magnitude smaller (diagram (j) being absent in $^{16}$O because of angular momentum conservation).

On the other hand, the nuclear matter properties of the GPT interaction were not as good as expected. In particular, the binding energy per particle was found too 
large as well as the value of the saturation density. Besides, at large distances, it was unable to reproduce the OPEP. P. Pir\`es showed that the saturation properties 
were greatly improved by replacing beyond 3 fm the gaussian analytical form by the OPEP one (noted POT2 in the following) \cite{Pires1973}.  
At this point, it is interesting to compare the INM with finite nuclei results in (S,T) channel. 
In Table \ref{tab:nm-HF}, the SOC in finite nuclei and nuclear matter calculated with both the GPT interaction and its version POT2 with Yukawa form factors are reported. They are expressed in percentage. The study was done for $\isotope[16]{O}$, $\isotope[40]{Ca}$, $\isotope[90]{Zr}$ and $\isotope[208]{Pb}$.
The striking feature is the similarity of the results obtained for $\isotope[208]{Pb}$ with GPT and for nuclear matter with POT2.
An important conclusion they drawn was that, while the OPEP is important in nuclear matter to recover basic properties, the asymptotic of the interaction is not determining in finite nuclei.

\begin{table}[htb!]
    \centering
    \begin{tabular}{cc|cccc|cc}
        \multicolumn{2}{c}{} & \multicolumn{4}{c}{Nuclei} &  \multicolumn{2}{c}{INM} \\
        \hline
        S & T & $\isotope[16]{O}$ & $\isotope[40]{Ca}$ & $\isotope[90]{Zr}$ & $\isotope[208]{Pb}$ & POT2 & GPT \\
        \hdashline
        0 & 0 & 16.0 & 22.0 & 26.0 & 30.0 & 32.0 & 38.0 \\
        0 & 1 & 14.5 & 12.5 & 12.0 & 11.5  & 10.0 & 10.0 \\ 
        1 & 0 & 61.5 & 55.0 & 49.0 & 43.0 & 41.0 & 37.0 \\ 
        1 & 1 & 8.0 & 11.0 & 13.0 & 15.0 & 16.0 & 15.0 \\
        \hline  \\
    \end{tabular}
    \caption{Percentages of SOC by (S,T) channels for various nuclei and INM.}
    \label{tab:nm-HF}
\end{table}

Finally, the study of the HF single-particle energies (SPEs) led to another significant conclusion that definitely persuaded D. Gogny to get involved in the newly appearing density-dependent interactions. In Table \ref{tab:spes}, the SPEs obtained for $\isotope[16]{O}$ with GPT including SOC are compared to those obtained with a Skyrme \cite{Vautherin1969a,Vautherin1967,Vautherin1968,Vautherin1969b}, Campi-Sprung (noted CS) \cite{Campi1972} and a mysterious Gogny \cite{Decharge1980} density-dependent interactions calculated at the HF level. Density-dependent interactions were able to retrieve SPE spectra at the HF level.

\begin{table}[htb!]
    \centering
    \begin{tabular}{c|cccc|c}
        \multicolumn{1}{c}{$\isotope[16]{O}$}  & \multicolumn{4}{c}{ GPT~~~  ~Skyrme~ ~~CS~~  ~~~Gogny  } &  \multicolumn{1}{c}{Exp.} \\
        \hline
        Neutrons    &        &        &         &      &  \\
        1s$_{1/2}$  & -38.69 & -35.20 & -36.36  & -39.65 & -47.00 \\
        1p$_{3/2}$  & -19.41 & -19.84 & -21.25  & -21.82 & -21.80 \\
        1p$_{1/2}$  & -15.44 & -15.14 & -15.81  & -16.02 & -15.80 \\
        \hdashline
        Protons     &        &        &         &        &  \\
        1s$_{1/2}$  & -35.39 & -31.21 & -33.16  & -36.35 & -40$\pm$8 \\
        1p$_{3/2}$  & -16.21 & -16.26 & -18.13  & -18.54 & -18.40 \\
        1p$_{1/2}$  & -12.33 & -11.64 & -12.78  & -12.83 & -12.10 \\
        \hline 
        \end{tabular}
    \caption{Comparison of SPEs in $\isotope[16]{O}$ obtained with GPT at the HF plus SOC level and with density-dependent effective interactions at the HF level. Experimental values are also indicated.}
    \label{tab:spes}
\end{table}

The original phenomenological Gogny interaction, known as D1, corresponds to a strong simplification of the GPT analytical form, conducted by the detailed analysis 
of the SOC by terms and channels. D. Gogny retained only the finite range central term in order to be able to treat in a consistent way mean-field and pairing field 
in the HFB approach. He introduced a density-dependent term, whose analytical expression was taken similar to the Skyrme interaction one. 
Its main role was to simulate the strong second order corrections generated by the GPT tensor term in the (S=1,T=0) channel. Then, the simple spin-orbit term of the Skyrme interaction was added. The abandonment of the finite range for the density-dependent and spin-orbit terms allowed to capture a large amount of the physics at a relatively 
low numerical cost, which was a strong argument at that time. Thus, the D1 phenomenological Gogny interaction writes as \cite{Gogny1973,Gogny1975a,Decharge1980}:
\begin{equation} \label{gognyD1}
\begin{split}
v_{12}^{\text{D1}} & = \sum_{i=1}^2 (W_i + B_i P_\sigma - H_i P_\tau - M_i P_\sigma P_\tau) \- ~e^{-\frac{(\vec{r}_1-\vec{r}_2)^2}{\mu_i^2}} \\
& \quad + t_0 (1+ x_0 P_\sigma) \delta(\vec{r}_1 - \vec{r}_2) \rho^{\alpha} \Big( \frac{\vec{r}_1 + \vec{r}_2}{2} \Big) \\
& \quad + i W_{ls} \big[ \vec{k}' \cross \delta(\vec{r}_1 - \vec{r}_2) \vec{k} \big] \cdot (\vec{\sigma}_1 + \vec{\sigma}_2).
\end{split}
\end{equation}
\noindent In Eq.(\ref{gognyD1}), one recognizes easily the three terms previously discussed. One remarks another simplification in comparison with GPT, assuming ranges independent of the (S,T) channel. 
Two ranges noted $\mu_i$ are introduced in the central term. The finite range nature of this term allow to introduce all the combinations of spin-isospin 
exchange operators $P_{\sigma}$ and $P_{\tau}$ introducing the sets of $\{W_i, B_i, H_i, M_i \}$ parameters. The spatial form factor of the density-dependent term $\delta(\vec{r}_1 - \vec{r}_2)$ is even under the exchange of particles. It contains solely two parameters, the global intensity $t_{0}$ and
$x_0$ which is chosen equal to 1 in order to cancel its contribution in proton and neutron pairing fields. Physically, the pairing interaction is known to be very close to the bare interaction and very few re-normalized by medium effects \cite{Sedrakian2003}. 
Besides, practically, this allows to avoid ultra-violet divergences introduced by a $\delta(\vec{r})$ function within the
HFB formalism. 
The power of the density $\alpha$ has been taken equal to 1/3. From the own words of D. Gogny who tested several values during the fitting process, this choice seemed optimal to him, with a single density-dependent term, ensuring the best saturation properties in INM.
As for the density-dependent term,
the spin-orbit force is represented by a contact term. It contains solely one
parameter, namely $W_{ls}$, which controls its intensity in the ST=11 channel. 
In Ref. \cite{Decharge1980}, the authors report a value of 115.00 MeV when the two-body center-of-mass contribution is excluded from the mean-field calculation. Otherwise, a value of 130.00 MeV is adopted, defining the D1' parameterization, which is identical to the D1 parameterization for all other terms.

In Table \ref{tab:STD1}, one presents with blue marks the contributions of D1 in the various spin-isospin channels ST for the three discussed terms. \\

\begin{table}[htb!]
    \centering
    \begin{tabular}{
        l
        S[table-format = 3]
        S[table-format = 3]
        S[table-format = 3]
        S[table-format = 3]
        }
        \multicolumn{1}{c}{} & 
        \multicolumn{4}{c}{Channels}\\
        \cmidrule{2-5}  
        {} & {ST=00} & {ST=01} & {ST=10} & {ST=11} \\
        \midrule
        C & ~~ \textcolor{blue}{\xmark} ~ \textcolor{teal}{\xmark} ~ \textcolor{red}{\xmark} & ~~ \textcolor{blue}{\xmark} ~ \textcolor{teal}{\xmark} ~ \textcolor{red}{\xmark} & ~~ \textcolor{blue}{\xmark} ~ \textcolor{teal}{\xmark} ~ \textcolor{red}{\xmark} & ~~ \textcolor{blue}{\xmark} ~ \textcolor{teal}{\xmark} ~ \textcolor{red}{\xmark} \\
        DD &  ~~ \textcolor{white}{\xmark} ~ \textcolor{teal}{\xmark} ~ \textcolor{red}{\xmark}  & ~~ \textcolor{white}{\xmark} ~ \textcolor{teal}{\xmark} ~ \textcolor{red}{\xmark}  & ~~ \textcolor{blue}{\xmark} ~ \textcolor{teal}{\xmark} ~ \textcolor{red}{\xmark} & ~~ \textcolor{white}{\xmark} ~ \textcolor{teal}{\xmark} ~ \textcolor{red}{\xmark} \\
        SO & ~~ \textcolor{white}{\xmark} ~ \textcolor{white}{\xmark} ~ \textcolor{white}{\xmark}  & ~~ \textcolor{white}{\xmark} ~ \textcolor{white}{\xmark} ~ \textcolor{white}{\xmark}  & ~~ \textcolor{white}{\xmark} ~ \textcolor{white}{\xmark} ~ \textcolor{red}{\xmark}  & ~~ \textcolor{blue}{\xmark} ~ \textcolor{teal}{\xmark} ~ \textcolor{red}{\xmark}  \\ 
        T & ~~ \textcolor{white}{\xmark} ~ \textcolor{white}{\xmark} ~ \textcolor{white}{\xmark}  & ~~ \textcolor{white}{\xmark} ~ \textcolor{white}{\xmark} ~ \textcolor{white}{\xmark}  & ~~ \textcolor{white}{\xmark} ~ \textcolor{white}{\xmark} ~ \textcolor{red}{\xmark} & ~~ \textcolor{white}{\xmark} ~ \textcolor{white}{\xmark} ~  \textcolor{red}{\xmark} \\         
        \bottomrule
    \end{tabular}
    \caption{Contributions of various analytical forms of the Gogny interaction to the $(S,T)$ channels. 
The D1-type (D2, DG) analytical form corresponds to the blue (green, red) marks. The central, density-dependent, spin-orbit and tensor terms are denoted by 
C, DD, SO and T, respectively.}
    \label{tab:STD1}
\end{table}

\begin{figure}
    \centering
    \includegraphics[width=0.95\linewidth]{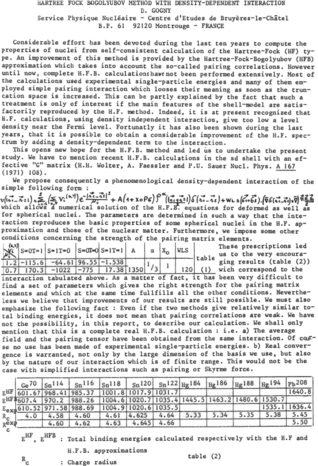}
    \caption{Proceeding publishing first calculations of binding energies and charge radii in a few spherical nuclei with a precursor of the D1 Gogny interaction (only the intensity of the spin-orbit term is different), International Conference on Nuclear Physics, Munich, August 27-September 1, 1973.}
    \label{fig:munich}
\end{figure}

\indent First finite nuclei results with a very close precursor of D1 Gogny interaction (only the intensity of the spin-orbit term is different with a value of 120.0 MeV)
were published furtively in a form of a one page proceeding
of the "International Conference on Nuclear Physics" that took place
in Munich in 1973 \cite{Gogny1973}. As can be seen from 
Fig. \ref{fig:munich}, binding
energies and charge radii are given for a few spherical nuclei and compared to experiment. 
The fitting procedure used to obtain this newly 
density-dependent nucleon-nucleon interaction was evoke in another 
proceeding of the international conference on nuclear self-consistent 
fields in 1975 \cite{Gogny1975a}. A more official article 
discussing about many observables obtained from the spherical HFB code was published in 1980 \cite{Decharge1980} with the D1 and D1' interactions. \\
\indent An elaborated scientific strategy was led by D. Gogny. 
The creation of D1 together with a spherical HFB code was only the first step \cite{Berger2017}.
In 1973, he started to create his research team at CEA located in 
Bruy\`eres-le-Ch\^atel, in France. Indeed, as mentioned previously, the 
philosophy of the Gogny interaction is to leave room for
an explicit treatment of nuclear long range correlations. The time had come to go beyond mean-field and to break symmetries. 
Thus, J. Decharg\'e and M. Girod joined firstly D. Gogny and J.F. Berger arrived in the mid-70s. 
In a few years, they built a RPA spherical code, an axial and triaxial 
HFB codes as well as a Five-Dimensional-Collective-Hamiltonian code dedicated to nuclear structure studies. For a microscopic description 
of fission process based on the TDGCM+GOA approach, a two-center HFB code was also elaborated. All these developments were done using the 
D1 Gogny force. Valuable collaborators such as B. Grammaticos, J.P. Blaizot, L. Sips and R. Padgen were part of the adventure. At the end 
of the 70s, the group of D. Gogny was able to analyze many nuclear phenomena, essential for nuclear applications
(see for example \cite{Decharge1975,Blaizot1976,Gogny1977,Girod1978a,Girod1978b,Decharge1982,Decharge1981,Berger1981,Berger1989,Berger1991,Haider1992,Chinn1992}. 
D. Gogny was passionate by applications and how a theory can be used in them, as his participation in the Harwell conference 
on "Neutron Physics and Nuclear Data"  in 1978 proves \cite{Gogny1978}. \\

\begin{figure*}[htb!]
    \centering
    \includegraphics[width=0.9\linewidth]{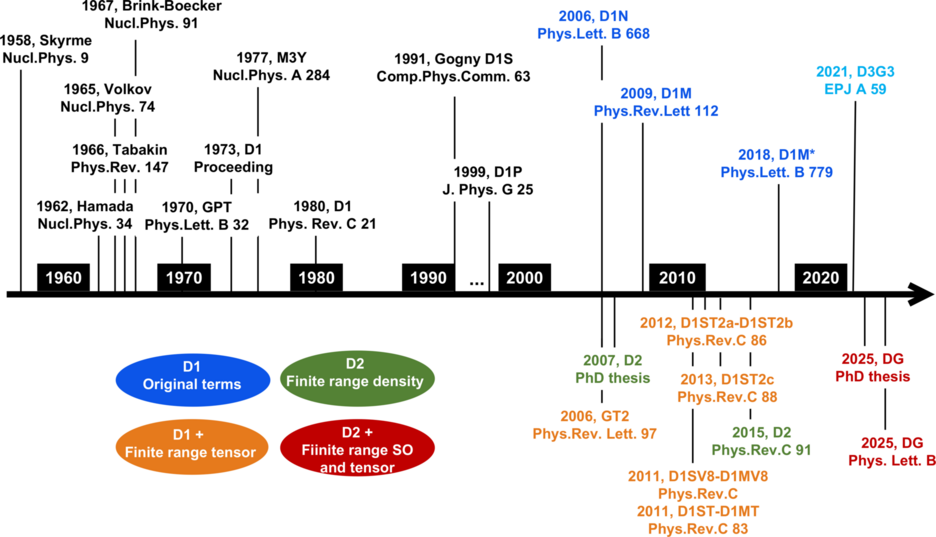}
    \caption{Timeline representing the context of the creation of the Gogny interaction and its evolution over the decades.}
    \label{fig:timeline}
\end{figure*}

Since the 60's, when the EDFs of Skyrme and Gogny type appeared, the desire to use them to describe numerous nuclear phenomena, coupled 
with the advent of numerous experimental results in nuclear structure, fission, reaction and astrophysics implying stable and exotic nuclei, 
has regularly led to the questioning of a parameterization or even to the extension of the analytical form. 
These two directions that have made the EDFs evolve in time are represented in Fig. \ref{fig:timeline} in the case of the Gogny interaction. 
One has also indicated the emergence of the Skyrme \cite{Skyrme1958} and M3Y \cite{Bertsch1977} interaction as well as other meaningful soft core interactions \cite{Volkov1965,Kerman1967,Bassichis1967,Brink1967,Hamada1962}.
Above the long black arrow that represents the increasing time
starting from the late 60's up to today, the various parameterizations based on the D1 analytical form, considered as important in the improvement of the functional for more predictive calculations, are indicated. Thus, beyond the D1 parameterization, one will discuss below the D1S, D1P, D1N, D1M, D1M* parameterizations. Below the black arrow, 
one has indicated extended analytical forms of the Gogny interaction. These extensions consist in either the replacement of zero-range term by a finite-range one or in the adding of a finite-range tensor term. In 
that respect, one will consider in our discussion D2 which is distinguished by a finite range density-dependent term, GT2 and D1ST-D1MT which contain a finite range tensor term, and the newly proposed DG which is fully finite range including a tensor term. As an example, one has reported in Table \ref{tab:STD1} with green and red marks the contributions of D2 and DG in the various spin-isospin channels ST. In
comparison with the D1 type parameterizations, the density-dependent term
acts now in all spin-isospin channel with D2 and DG, the spin-orbit 
has an additional contribution in the ST=10 channel with DG. This latter includes a tensor which plays a role in both the ST=10 and ST=11 channels, as expected. \\

In the following, one will discuss the physical motivations at the origin of the main steps in the evolution of the original D1 Gogny interaction. 
The details of the fitting procedure of the Gogny interaction, and its modifications since the original proposal, will be presented in parts 
\ref{sec2} and \ref{sec3}. One notes that each parameterization is a compromise of many properties and none of them is "perfect". However, the main
philosophy in fitting a new parameterization is to keep as much as possible the good properties obtained with the previous parameterizations and to 
improve other ones.

\subsection{Pairing and fission properties embodied by the D1S  parametrization}\label{paingfiss}

\begin{figure}
\centering
\includegraphics[width=0.95\linewidth]{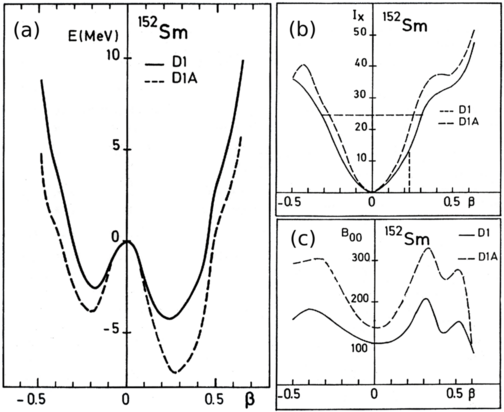}
\caption{(a) Potential energy surfaces of $^{152}$Sm according to the axial deformation $\beta$ calculated at the HFB approximation. (b) Moment of inertia I$_{x}$ of $^{152}$Sm according to $\beta$. The horizontal dashed line indicates the experimental value, the vertical line the dynamical deformation. 
(c) Axial mass parameters B$_{00}$ of $^{152}$Sm according to $\beta$. Calculations were done with the D1 and the intermediary D1A parametrizations (see text for explainations) \cite{Girod1983a}.} \label{fig:girodSm}
\end{figure}

The D1 Gogny interaction was tested in various approaches in order to estimate its ability to describe both static and dynamical properties of nuclei.
Refs. \cite{Gogny1975a,Decharge1980} display the results of many observables - binding energies, charge radii,
deformation, excitation spectra, densities, charge distribution, spectroscopic factors - obtained within the HFB approach, in spherical and deformed nuclei. 
The comparisons with experimental results were very encouraging, though there was still room for improvement.
A global reasonable agreement was obtained in comparison with experiment. 
The description of the spectroscopy using beyond HF and HFB methods was an even more stringent test. The dynamical properties were studied within both the RPA \cite{Gogny1977,Blaizot1977,Decharge1982,Decharge1983,Blaizot1977} and the 5DCH \cite{Girod1983a} approaches. 
The description of collective modes of small amplitude within the RPA approach was found 
in very good agreement with experiment, in particular for the states with high excitation energy in $^{208}$Pb \cite{Decharge1982,Decharge1983}.
Moreover, from a more formal point of view, in Ref. \cite{Blaizot1977}, it was shown that 
all eigenvalues were real, which means that the HF solution with the D1 interaction was stable against all particle-hole excitations. From Ref. \cite{Gogny1977}, it was shown that no instability was found close to the saturation density. The results of spectroscopy obtained in Ni, Ge and Sm 
isotopic chains within the 5DCH approach were less satisfactory \cite{Girod1983a}.
In general, the rotational bands were obtained too dilated indicating that the moments of inertia, calculated with the D1 interaction at the Cranking approximation, were too small at least in the neighborhood of the dynamical deformation. 
In addition, the 0$^{+}$ vibrational band heads were found with a too high excitation energy, which was linked to an underestimation of collective masses evaluated with the Cranking approximation. Independently of the approximations introduced at several levels in the present model, the systematic disagreements between theoretical and experimental results were attributed to the pairing content of the D1 parameterization. Therefore, a sensitivity 
study of the results to pairing correlations was performed. A parameterization, noted D1A in the following, owning exactly the same properties as the D1 interaction excepted a smaller pairing strength, was built. One notes that the results obtained with D1A were the same as the one obtained with D1 in magic nuclei such as $^{16}$O, $^{40-48}$Ca, $^{208}$Pb. In particular, the radii were unchanged. 
As an illustration, Fig. \ref{fig:girodSm} displays a comparison between the axial potential 
energy surfaces (panel (a)), the moments of inertia (panel (b)) and the axial mass parameters (panel (c)) 
in $^{152}$Sm obtained with the D1 and D1A parameterizations.
The decrease of the pairing intensity caused an appreciable increase of the collective masses, moments
of inertia and in some cases changes in the potential energy surfaces, as expected.
The fine adjustment of the pairing intensity was enough to modify the results in the expected direction. 

\begin{figure}
\centering
\includegraphics[width=0.8\linewidth]{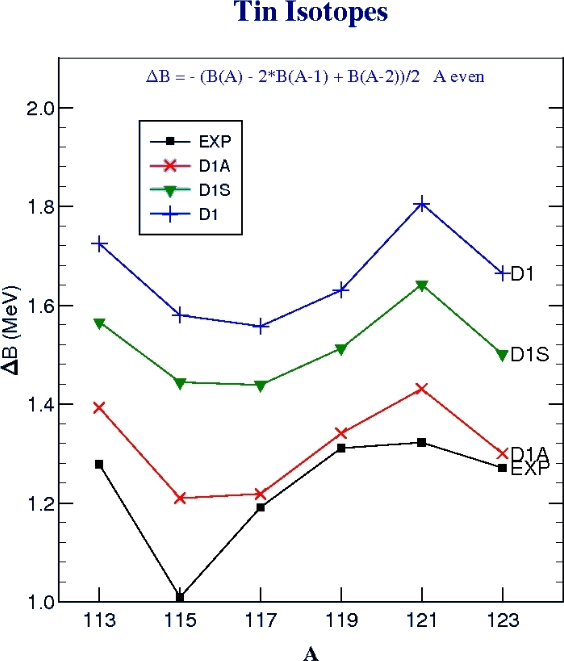}
\caption{Odd-even mass differences $\Delta$B in Sn isotopes calculated with the D1 and D1A parameterizations at the HFB approximation \cite{Girod1983a}.} \label{fig:oem}
\end{figure}

As discussed previously, the spectroscopy is an indirect test of the pairing strength. Actually, very few observables can
provide a quantitative indication on the intensity of pairing. The odd-even mass difference, noted $\Delta B$ in the following, is one of them as it 
measures the energy needed to break a pair. As shown in Fig. \ref{fig:oem} for the Sn isotopic chain \cite{Girod1983a}, this difference was evaluated within 
the HFB method, using the blocking approximation for odd nuclei. Its value was found larger than experiment with both D1 and to a lesser extent with D1A. 
A shifted value (higher) in comparison with experiment was desirable since it was expected that the particle-vibration residual interaction lowers the ground state energy of the odd nuclei. It was expected that this effect was of the order of $\sim$200-300 keV \cite{Decharge1980}. Thus, the value obtained 
with D1 overestimated it contrary to D1A that underestimated it.

\indent In addition to the need of a fine tuning of the pairing properties of the D1 Gogny interaction, another difficulty raised from
the problem of the second barrier heights in the context of fission studies \cite{Berger1989,Berger1991}. Several investigations using the 
Skyrme interactions put forward the link between the properties of the nucleon-nucleon effective interaction and the barrier heights
\cite{Flocard1974,Dutta1980,Bartel1982,Brack1985}. 
\begin{figure}
\begin{center}
\includegraphics[width=0.71\linewidth,angle=0]{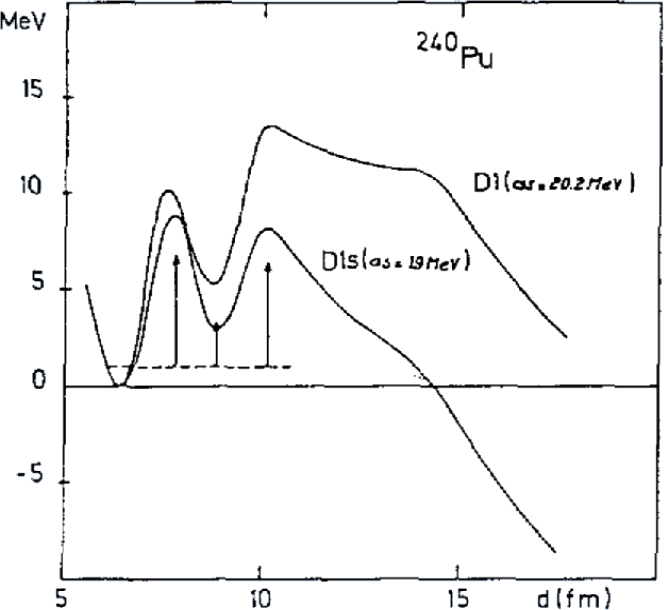} 
\includegraphics[width=0.7\linewidth,angle=0]{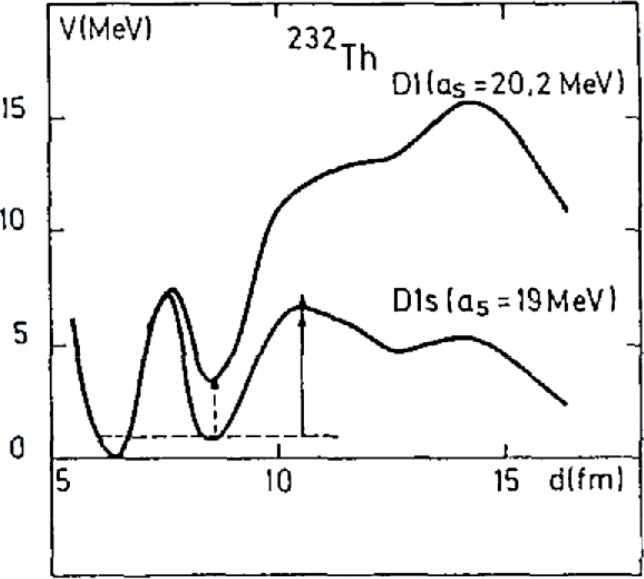} 
\end{center}
\caption{Asymmetric fission path drawn according to "d" obtained for the $^{240}$Pu (up) and $^{232}$Th (down) nuclei, calculated with the original D1 and the new D1S parameterizations. The surface coefficient is a$_{s}$ is also indicated for both parameterizations \cite{Berger1989,Berger1991}. 
The value of d represents the distance in fm between the centers of mass of the left and the right part of the nucleus defined with respect to the 
total center of mass. Axial-asymmetric deformations start to manifest in the vicinity of the top of the first barrier.}
\label{fiss1}
\end{figure} 
The heights appeared to scale like the values of the surface coefficient a$_{s}$ of the interactions. 
To be more precise, only the heights of the second barriers were significantly modified.
A similar study was done with the D1 interaction. As shown in Fig.(\ref{fiss1}) for $^{240}$Pu (up) and $^{232}$Th (down) nuclei, the shapes of
the second barrier were unrealistic with the D1 Gogny interaction. A decrease of the surface coefficient from 20.2 MeV to 19 MeV provided a better 
behavior of potential energy surfaces (these values include the contributions of the central and density dependent terms and were obtained using a semi-classical model to evaluate the binding energy of a spherical « nucleus » of very large radius $R$, thus extracting the $R^{2/3}$ contribution. This procedure provided coherent results with those obtained from the semi-infinite calculations of Ref. \cite{cote1978} in the case of the D1 parameterization).
The D1S Gogny interaction was created, incorporating a smaller surface coefficient a$_{s}$ and a pairing with a smaller intensity in comparison 
with the D1 Gogny interaction (larger than D1A as it can be seen from the odd-even mass difference displayed in Fig. \ref{fig:oem}), without spoiling 
as much as possible the other characteristics of the force \cite{Berger1989,Berger1991}. During more than 30 years, the D1S interaction has been used in many successful studies from different groups 
(Bruy\`eres-le-Ch\^atel, Madrid, Granada, Lecce, ...) employing different models at the mean-field and beyond levels, including the projection technics. Striking applications will be discussed in section \ref{sec4}.  \\

To end this section, an interesting point to raised is a first attempt to include more physical constraints on the Gogny interaction. 
The D1P parametrization \cite{Farine1999} was introduced at the end of the 90s as an improvement to D1S with respect to three points:
\begin{itemize}[label=$-$]
\item A depth of the optical potential in agreement with experimental data to energies beyond $\SI{200}{\MeV}$;
\item The first and second sum rules for the Landau parameters better fulfilled at and in a range 
around saturation density, and some unphysical instabilities cured in the high density 
region ($k_f\simeq 2 fm^{-1}$) due to the isovector breathing mode;
\item A realistic behavior of the neutron matter equation of state achieved at high densities.
\end{itemize}
Although this interaction is not often mentioned in the literature, it has the merit of introducing the need of a better reproduction of the neutron matter equation of state (high densities, high isospin), that will systematically  be discussed afterwards.

\subsection{Neutron matter and mass properties: from D1N to D1M parametrizations and beyond}\label{D1Nparam}

\begin{figure}
\centering
\includegraphics[width=0.99\linewidth]{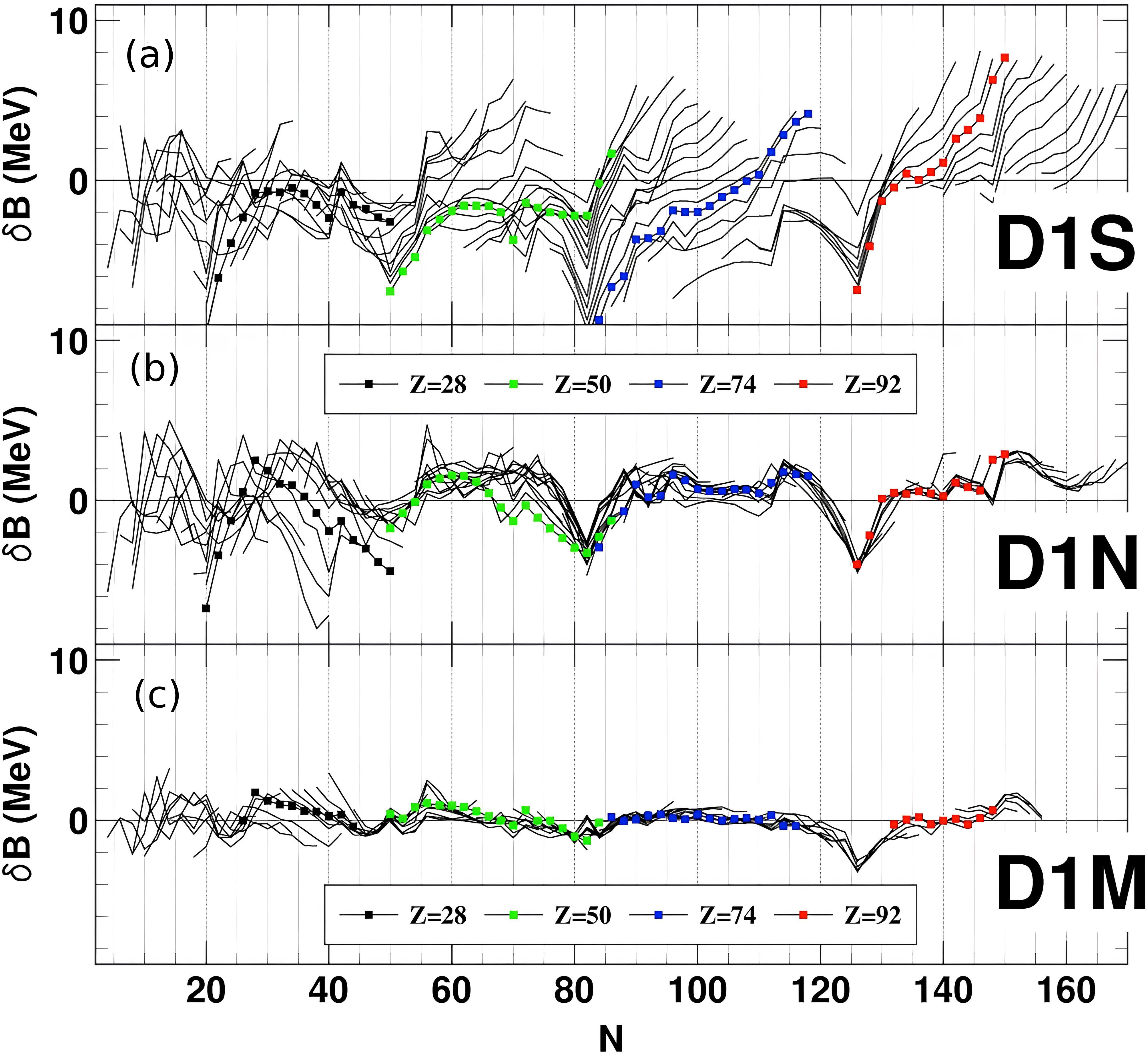}
\caption{Drift of masses along isotopic chains for D1S, D1N and D1M Gogny interactions \cite{Goriely2016a}. The energy difference between 5DCH and experimental masses, defined by $\delta B \equiv B_\text{5DCH} - B_{\text{exp}}$, is represented as a function of the neutron number $N$.} \label{fig:DriftMass}
\end{figure}

One interesting D1-type parametrization established in 2008 as part of F. Chappert's PhD thesis is the D1N interaction \cite{Chappert2007,Chappert2008}. In the 90's, technological breakthroughs led to the production of radioactive nuclear beams of exotic nuclei, opening the door to new regions of little known or unknown nuclei off the valley of stability \cite{Casten2000}. At the D1S time, no experimental data were available on such nuclei. As one will see in section \ref{sec2}, even if it has been introduced in the fitting procedure a constraint to control the proton-neutron asymmetry of the interaction, the D1S parametrization failed in reproducing adequately the binding energies of exotic neutron-rich nuclei. 
This is peculiarly striking when the energy difference between HFB predictions and experimental values is plotted for many isotopic chains as a function of the neutron number $N$. The same conclusion holds when 5DCH predictions were done. 
In Fig. \ref{fig:DriftMass}, panel (a), the energy difference $\delta B$ blows up for D1S, at the tails of isotopic chains, corresponding to exotic nuclei with 
slarge neutron numbers.

\begin{figure}
\centering
\includegraphics[width=0.99\linewidth]{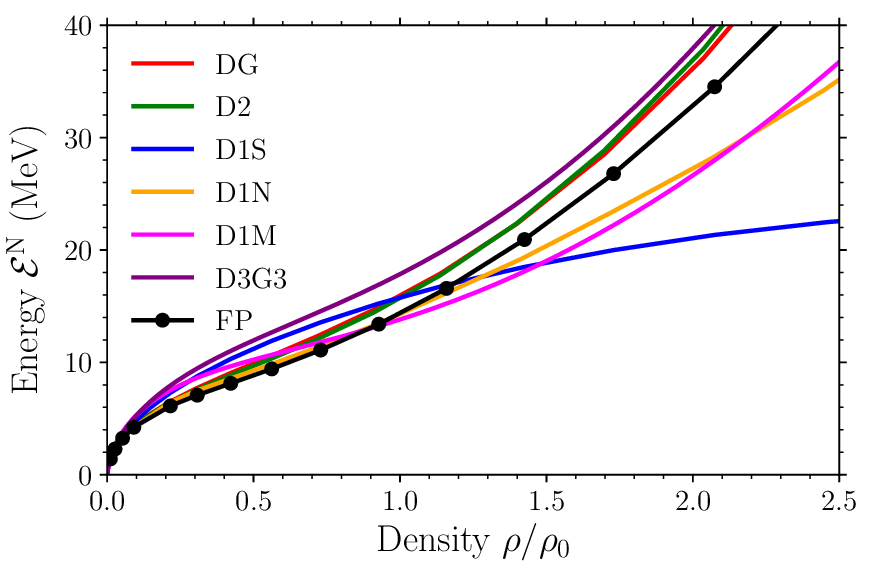}
\caption{Neutron equation of state (energy per nucleon) calculated for several Gogny interaction parameterizations, as a function of the total density $\rho$ normalized to the saturation density $\rho_0$. The figure have been extracted from \cite{Chappert2008}.} \label{fig:nEoS}
\end{figure}

To remedy this, a filter on the neutron matter equation of state was added to the D1S fitting procedure. Eight points of the equation of state obtained by realistic calculations of Friedman and Pandharipande (FP) \cite{Friedman1981} in neutron matter had to be reproduced, with chosen accuracies, by the interaction to pass the filter. Since the neutron density remains lower than the saturation density for atomic nuclei $\rho_0$, five points fixed to densities satisfying $\rho \leq \rho_0$ enabled a faithful description of nuclear structure data. Three points at densities such that $\rho > \rho_0$ were demanded, with a view to future astrophysical applications. The resulting parametrization was dubbed D1N and its benefits appears clearly in Fig. \ref{fig:DriftMass}, panel (b). 
The rms of $\delta B$ was of the order of 1 MeV.
This interaction is fundamental in the sense that all its successors similarly take up a filter/constraint on the neutron matter equation of state.
In Fig. \ref{fig:nEoS}, the neutron equation of states is shown for various Gogny interactions \cite{Chappert2008} according to $\rho/\rho_0$. One sees the improvement obtained with D1N
in the reproduction of the Friedman-Pandaripande data, in particular in the low density part below $\rho/\rho_0 \le 1$. One notes that D1S is unable
to follow the trend of the FP realistic calculations, leading to a collapse at high densities with negative values for $\rho/\rho_0 \ge 12.5$.

To go a step further with the problematic of masses, Goriely \textit{et al.}\ proposed in 2009 a new parameterization, called D1M \cite{Goriely2009a,Goriely2016a}. The main purpose of this interaction was to improve the overall reproduction of nuclear masses (and charge radii) at the 5DCH level, while further reducing the tiny energy drift of D1N (see Fig. \ref{fig:DriftMass}, panel (c)). 
The protocole to adjust the interaction was different from the one used for D1, D1S and D1N. In particular, a $\chi^{2}$ minimization on the mass difference between experiment and theory for $2149$ nuclei was used. Some standard physical quantities in infinite nuclear matter alike those of D1S were also included in the fit, as well as requisites to reproduce the neutron matter equation of state. With a better description of binding energies (at the 5DCH level, the rms of $\delta B$ is of the order of 0.8 MeV.), this interaction seems more suitable than previous D1-type parameterizations for astrophysical purposes.

\begin{figure}
\centering
\includegraphics[width=0.99\linewidth]{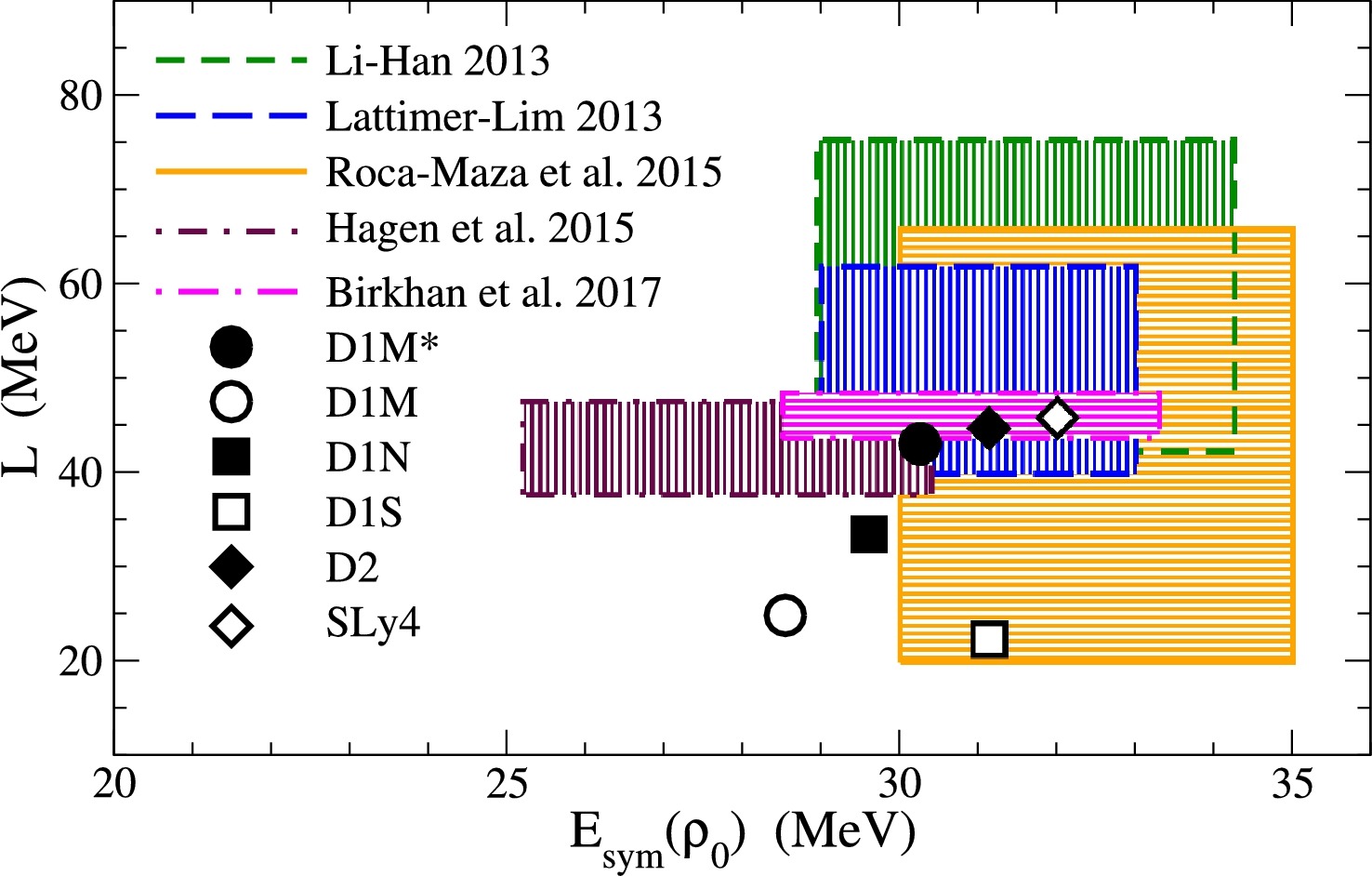}
\caption{Slope L of the symmetry energy for various Gogny interactions
as a function of the symmetry energy calculated at the saturation density $\rho_0$ \cite{GonzalezBoquera2018}. Experimental data coming from various sources are represented in colored rectangles.} \label{fig:sSE}
\end{figure}

In order to enlarge predictions of the Gogny interaction to astrophysical processes, a new parametrization, modeled on D1M, was investigated by Gonzalez-Boquera \textit{et al.} in 2018 \cite{GonzalezBoquera2018,Vinas2018,Mondal2020,GonzalezBoquera2017}. The aim of this interaction, known as D1M$^\star$, was to preserve the finite-nuclei properties of D1M while being competitive with modern Skyrme interactions in catching the physics of neutron stars. In particular, the slope of the symmetry energy was an important quantity to catch, as seen in Fig. \ref{fig:sSE} \cite{GonzalezBoquera2018}. One notes the result obtained with D2 (see section \ref{secD2}),
which is compatible with the experimental values.

To end the section, one would like to mention the recent D3G3M extension \cite{gorielyD3G3M} of the D1-type parameterizations that has been proposed recently following the physical context and philosophy of D1M. In this interaction, a third Gaussian central term is added following the work of Ref. \cite{Batail2023a,Davesne2017}. Two main objectives led to the development of D3G3M: 
\begin{itemize}[label=$-$]
\item A more faithful reproduction of physical quantities in infinite nuclear matter (symmetric and neutronic) than D1M for astrophysical applications 
\item A determination of Gaussian ranges in order to reproduce the same infinite nuclear matter properties as the ones obtained with Yukawa, 
in the spirit of the M3Y interaction for which the central term ranges (short, medium, long) are determined as if the force originated from 
the exchange of mesons ($\rho$, $\sigma$ and $\pi$ mesons, respectively).
\end{itemize}
The D3G3M interaction has been obtained using a $\chi^2$ minimization procedure similarly to D1M.

\subsection{Adding a finite range to the density-dependent term for exotic nuclei and matter: the D2 Gogny interaction family}\label{secD2}

In 2007, F. Chappert \textit{et al.} broke new ground by extending the expression of the original D1 Gogny interaction, rather than looking for another D1-type parametrization \cite{Chappert2007,Chappert2015,Pillet2017a}. In that context, the density-dependent term acquired a finite range
and this extended form of the Gogny interaction was named D2.

There were essentially two reasons that have led to this new extension of the Gogny force, which corresponds to a first step
towards a fully finite-range interaction. 
The first one is based on the observation that an interaction of finite range has fundamentally different properties from those of a contact force, even at the mean-field approximation. In particular, a zero-range force leads to a local mean-field, whereas a finite-range force produces a non-local component, the exchange field, which significantly affects the structure of the one-particle states. 
It should also be pointed out that the finite range of the effective nuclear force also plays a role in the strength and structure of the long-range correlations that govern the deformation properties of nuclei \cite{Decharge1980}.
The second reason lies in the prospect of using the effective Gogny force in approaches beyond the mean-field. In the spirit of a fully microscopic approach to the structure of nuclei, it seems logical, although there is no fundamental justification for this, to seek to construct the residual interaction responsible for correlations beyond the mean-field by employing the same nuclear force that used for the mean-field. In view of what has been said above, this is only possible if the two-body matrix elements of the nuclear interaction decrease as a function of the transferred momentum, i.e. if the effective interaction is not a contact one. 


    In the context of symmetry restoration with projection techniques it is of great importance to avoid the inconsistencies that appear when the generalized Wick theorem is used in situations where the overlap in the denominator of the overlap density and pairing tensor matrices vanishes, rendering them divergent quantities \cite{Sheikh2021}. In the evaluation of Hamiltonian overlaps, the divergence cancels out only if the three contributions, namely direct, exchange and pairing, from a two body potential are fully considered \cite{Anguiano2002a}. This is a real problem as very often some of the aforementioned terms are neglected as happens with the pairing contribution from the Coulomb interaction. It is also common practice to use a pairing interaction different from the central potential determining the long range contribution to the Hartree-Fock field. Additionally, it is also frequent to replace Coulomb exchange by the Slater approximation. For the Gogny interaction, the pairing field comes from the long range Gaussian potential and therefore the cancellation of the divergence is guaranteed.  However, taking care of the problem forces the additional consideration of full Coulomb exchange and antipairing as well as the pairing field form the spin-orbit potential \cite{Anguiano2001a}. 

The new analytical form proposed by the D2 Gogny force writes as:

\begin{multline}\label{gognyD2}
v_{12}^{\text{D2}} = \sum_{i=1,2} (W_i + B_i P_{\sigma} - H_i  P_{\tau} - M_i P_{\sigma} P_{\tau}) V_i(r_{12}) \\
+ (W_3 + B_3 P_{\sigma} - H_3  P_{\tau} - M_3 P_{\sigma} P_{\tau}) \times \\  \frac{V_3(r_{12})}{(\mu_3 \sqrt{\pi})^3} 
\frac{\rho^{\alpha}(\vec{r}_1) + \rho^{\alpha}(\vec{r}_2)}{2} \\
+ i W_{ls} \big[ \vec{k}' \cross \delta(\vec{r}_1 - \vec{r}_2) \vec{k} \big] \cdot (\vec{\sigma}_1 + \vec{\sigma}_2)
\end{multline}

The form factor associated with the finite-range density-dependent term is, like the central terms, of Gaussian shape, i.e.\
\begin{equation}
V_j(r_{12}) \equiv e^{-(\vec{r}_1 - \vec{r}_2)^2/\mu_j^2}
\end{equation}
for $j \in \{1,2,3 \}$.
One notes that the denominator $(\mu_3 \sqrt{\pi})^3$ in Eq.(\ref{gognyD2}) ensures that the exact zero-range density-dependent term of the original D1 Gogny interaction is recovered by means of the formula:
\begin{equation}
\lim_{\mu \rightarrow 0} \frac{e^{-(\vec{r}_1 - \vec{r}_2)^2/\mu^2}}{(\mu \sqrt{\pi})^3} = \delta(\vec{r}_1 - \vec{r}_2).
\end{equation}

Why was the density term first extended to a finite-range? The traditional Gogny force (see Eq.(\ref{gognyD1})) contains two zero-range terms: the spin-orbit component
and the density-dependent component. There are indications that it would be desirable to introduce, already at the mean field level, a finite-range in the spin-orbit component. In particular, this extension would make it possible to improve the
dependence of the spin-orbit field on the isospin, by allowing it to act in both the T=0 and T=1 sub-spaces of two-nucleon states.
Concerning the contribution of this
component to the residual interaction, it appears that, to a first approximation, it can be neglected. In fact, it plays a small role in the nucleus pairing field
and its influence on the particle-hole matrix elements of the RPA and QRPA approaches is small in relative value.
 One postpones this discussion to section \ref{DGsection}.
The density-dependent component of the effective force behaves very differently with respect to correlations beyond the mean-field. On the one hand, concerning the pairing correlations, theoretical arguments indicate that, at first order, the part of the residual interaction that generates them is not renormalized by medium effects. It is expected to be practically independent of density. Consequently, the density-dependent component of the effective force is expected to contribute very little to the pairing correlations. The D2 interaction has been fitted in order to take into account this physical constraint. On the other hand, this component is known to be essential in the description of RPA and QRPA type correlations \cite{Decharge1980}. It is also expected to play a very important role in the more general particle-vibration correlations that renormalize the single-particle states of the mean-field, as well as in odd nuclei.
In the traditional D1-type parametrizations of the Gogny interaction, the action of the density-dependent component was restricted to the even-triplet (S=1,T=0) subspace of two-nucleon states so that only the finite-range components of the force are involved in the pairing interaction between nucleons of the same type. 
This limitation was necessary in order to be able to apply the HFB method, reduced to particle-like pairing, without having to artificially truncate the space of one-particle states likely to pair. However, such a restriction appears excessive when it comes to describing correlations other than pairing. In this case, it is desirable for the density-dependent component to be able to act in other subspaces of the two-nucleon states than just the triplet-even space. In these conditions, as one has pointed out, it is
 necessary for this component to be of finite range so that the matrix elements of the interaction possess non-pathological behavior with the transferred momenta.
Given the dominant role played by the density-dependent component in approaches beyond the mean-field, the work
was concentrated on the finite-range extension of the density-dependent term alone in the D2 Gogny force family, leaving for the future a similar generalization of the spin-orbit component. In Table \ref{tab:STD1}, one sees with green marks the contribution of D2 in the various spin-isospin channels, that have to be compared to D1-type ones indicated with blue marks. 
\begin{figure*}[htb!]
\centering
\includegraphics[width=0.55\linewidth]{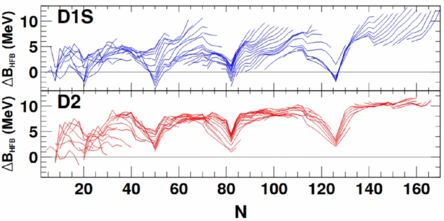}
\caption {Difference $\rm \Delta B_{HFB}$ between experimental and HFB binding energies calculated for D1S and D2 for many the isotopic chains \cite{Chappert2007,Pillet2017a}. Energies are expressed in MeV.} \label{fig:bindingD2}
\end{figure*}
As seen from Fig. \ref{fig:nEoS}, the neutron equation of state associated with D2 (green line) follows in a relatively satisfactory way the FP predictions at low as well as higher densities, in an even better way than D1N, D1M and D3G3. The drift in the binding energy has been also relatively well-controlled following the improvement brought by D1N, as it can been seen in Fig. \ref{fig:bindingD2} where the difference in binding energies between experimental and HFB ones has been drawn \cite{Chappert2007,Pillet2017a}.

\begin{figure*}
\centering
\includegraphics[width=0.29\linewidth]{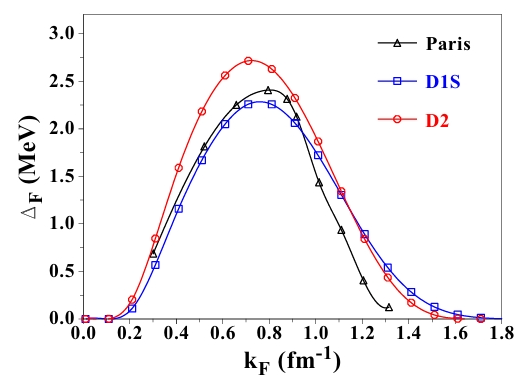}
\includegraphics[width=0.33\linewidth]{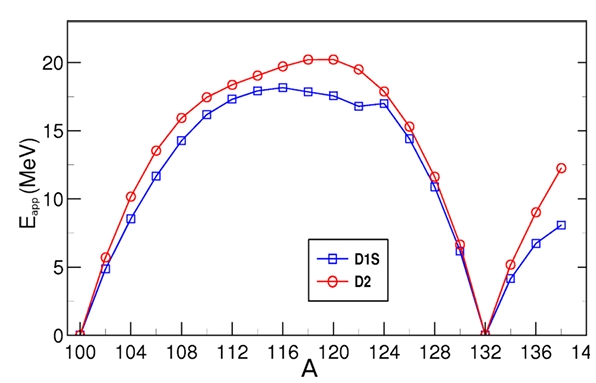}
\includegraphics[width=0.34\linewidth]{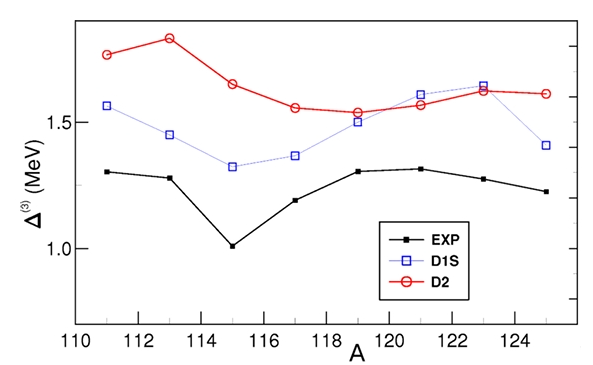}
\caption{Gap equation for symmetric nuclear matter (left) according to the Fermi momentum k$_F$, HFB pairing energy E$_{app}$ in Sn isotopes (right) and odd-even mass difference $\Delta^{(3)}$ in Sn isotopes deduced from the HFB approach, calculated with D1S and D2. Energies are expressed in MeV. A comparison to the gap equation calculated with the Paris force (in black) is provided (left) as well as experimental values (in black) for the odd-even mass difference \cite{Chappert2015}.} \label{fig:pairing1}
\end{figure*}

\begin{figure*}
\centering
\includegraphics[width=0.45\linewidth]{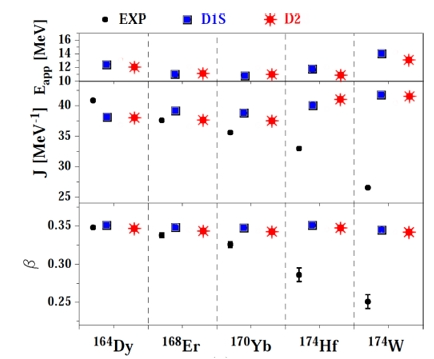}
\includegraphics[width=0.45\linewidth]{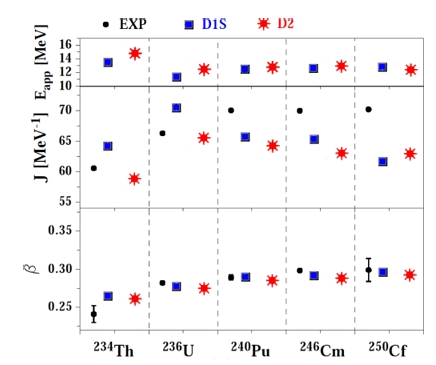}
\caption{HFB pairing energies (top panel), the moments of inertia calculated at the Cranking approximation (middle panel) and the axial quadrupole deformations (bottom panel) of well-deformed nuclei (left figure) and actinides (right figure) in their ground states, calculated with D1S and D2 \cite{Chappert2015}. }\label{fig:pairingdeformed}
\end{figure*}

The pairing properties of D2 have been fitted in order to be close to the D1S ones.
These can be seen by calculating, for example, the gap equation in symmetric nuclear matter and the HFB pairing energy in the Sn isotopic chain.
Results are shown in Fig. \ref{fig:pairing1}, respectively on the left and middle panels, for D1S and D2. Both quantities demonstrates their similarities in this respect. 
The odd-even mass difference in the Sn isotopes is displayed in the right panel. The difference to experimental data in black, which is desired according to original arguments associated to particle-vibration effects in odd nuclei, is of the same order of magnitude with D1S and D2.
The HFB pairing energies (top panel), the moments of inertia calculated at the Cranking approximation (middle panel) and the axial quadrupole deformations (bottom panel) of well-deformed nuclei (left figure) and actinides (right figure) in their ground states are presented in 
Fig. \ref{fig:pairingdeformed}. A comparison to experimental data is provided in black. These results confirm the proximity between D1S and D2 concerning the pairing properties (and quadrupole deformation).

\begin{figure*}
\centering
\includegraphics[width=0.8\linewidth]{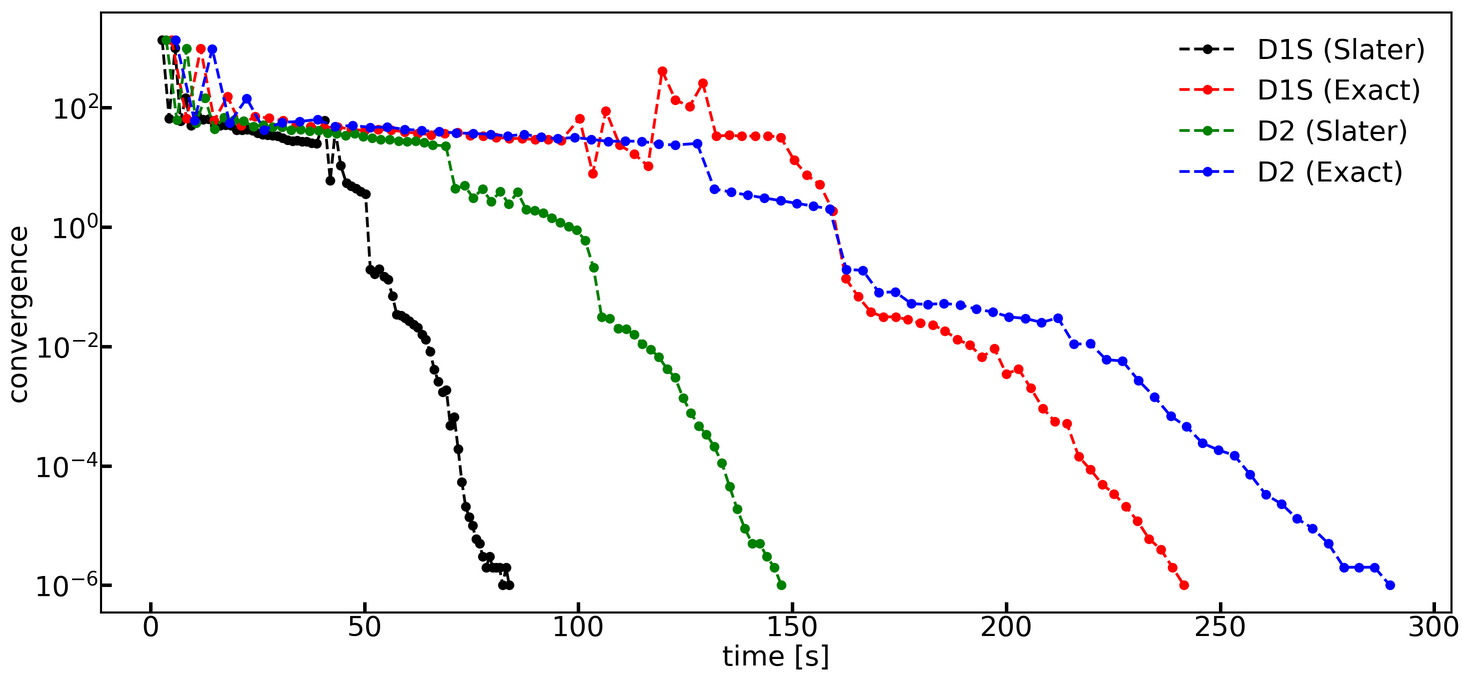}
\caption{HFB convergence study performed for $^{240}$Pu using the HFB3 solver \cite{Carpentier2024a}.} \label{fig:D2HFB3}
\end{figure*}

The introduction of a finite-range density dependent term in
HFB or beyond mean-field methods induces a more challenging
numerical implementation. As discussed in \cite{Chappert2007},
a careful and optimized treatment was performed, rendering the D2 Gogny force exploitable for systematic calculations. 
The original implementation of D2 in the axially deformed HFB AMEDEE solver 
using a one-center HO basis provided with a factor $\simeq$5 longer for the full convergence in comparison with a D1-type parameterization. In Fig. \ref{fig:D2HFB3}, a comparison between D1S and D2 is made with the newly developed HFB3 solver \cite{Dubray2025a,Carpentier2024a}, which is a two-center axial HFB solver breaking the parity. Calculations have been performed with an HO basis of 2$\times$11 shells in the case $^{240}$Pu. 
The evolution of the convergence based on the maximum variation of the elements of the one-body density matrix between two successive iterations is shown as a function of time expressed in second. The results are shown for both the Slater approximation and the exact treatment of the Coulomb fields.
The performances, obtained on a simple computer without any internal parallelization of the HFB3 solver, are very encouraging for a more extensive utilization of the D2 interaction.

\subsection{Spectroscopy in odd nuclei and fission barrier properties with finite range spin-orbit and tensor terms: the DG Gogny interaction family}\label{DGsection}

Although the adding of a tensor term to the Gogny interaction appeared in private documents from J. Decharg\'e in the 80's, the first published paper including a finite-range tensor term to the D1-type Gogny interaction appeared at the same period as 
the work of F. Chappert and collaborators on D1N and D2,
as seen in the timeline represented in Fig. \ref{fig:timeline} \cite{Chappert2007,Chappert2008,Chappert2015}.
After this first work by Otsuka \textit{et al.} \cite{Otsuka2006},
who performed a full refitting of the force, several
other works of Co', Anguiano \textit{et al.} \cite{Co2011,Anguiano2011,Anguiano2012}
who pursued a perturbative adding of a tensor term on top
of the D1S and D1M parameterizations appeared in the literature.
Recently, a fully finite-range Gogny interaction, named DG, including a finite-range spin-orbit and tensor terms, has been proposed by Zietek \textit{et al.} \cite{Zietek2023,Zietek2025}. In line with the D2 interaction and its philosophy, the authors of DG 
have applied a full refitting of the parameters.

\subsubsection{Pioneer works with tensor-dependent interactions based on the D1 analytical form}\label{pioneer}

Otsuka \textit{et al.}\ \cite{Otsuka2006} were the first to introduce in 2006 a tensor term in the Gogny interaction. Basically, they took the D1 analytical expression \eqref{gognyD1} and supplemented it by an isospin-dependent tensor force with a Gaussian shape of the form: \footnote{Actually the notations used in the article are a bit different but equivalent to those which are more convenient to be compared with the other tensor-dependent interactions.}
\begin{equation}
v_{12}^{\text{T}} = W_{\text{T}} (\vec{\tau}_1 \cdot \vec{\tau}_2) \widetilde{S}_{12} ~ e^{-(\vec{r}_1 - \vec{r}_2)^2/\mu_{\text{T}}^2},
\end{equation}
where their tensor operator is defined as
\begin{equation} \label{S12first}
\widetilde{S}_{12} \equiv 3 (\vec{\sigma}_1 \cdot \hat{r}_{12}) (\vec{\sigma}_2 \cdot \hat{r}_{12}) - \vec{\sigma}_1 \cdot \vec{\sigma}_2.
\end{equation}
All the parameters (including the spin--orbit one) were refitted in a fully consistent way, based on five requirements:
\begin{itemize}[label=$-$]
\item the reproduction of the saturation point of D1S in symmetric nuclear matter
\item an incompressibility within empirical values in symmetric nuclear matter
\item binding energies of $\isotope[16]{O}$, $\isotope[40]{Ca}$, $\isotope[48]{Ca}$, $\isotope[56]{Ni}$, $\isotope[132]{Sn}$ and $\isotope[132]{Sn}$ reproduced with an accuracy of about $1\%$ at the HF level
\item the part of the central terms proportional to $(\vec{\sigma}_1 \cdot \vec{\sigma}_2)(\vec{\tau}_1 \cdot \vec{\tau}_2)$ showing a positive overall strength
\item the strength $W_{\text{T}}$ such that the volume integral of the Gaussian form factor is equal to that of the AV8' potential \cite{Pudliner1997}.
\end{itemize}
Additionally, the range of the tensor force was chosen to be equal to the longest range of the central part of the D1S interaction, namely $\mu_{\text{T}} = \SI{1.2}{\femto\metre}$. The corresponding parametrization, named GT2, was only tested regarding its predictions on the single-particle energies at the HF approximation and their evolution along isotopic chains. 
To our knowledge, there have been no subsequent studies to probe its pairing or beyond mean-field properties. \\

Five years later, Co' \textit{et al.}\ \cite{Co2011} also added an isospin-dependent tensor force but perturbatively on top of the D1 analytical expression, whose expression is:
\begin{equation}\label{tensorco}
v_{12}^{\text{T}} = v^{\tau}_{\text{AV8'}}(r) \big( 1 - e^{-b r^2} \big).
\end{equation}
In Eq.(\ref{tensorco}), the tensor-isospin dependence is fully contained in the tensor-isospin term of the Argonne V8' potential, denoted as $v^{\tau}_{\text{AV8'}}(r)$ \cite{Pudliner1997}. It is multiplied by a function which incorporates the effects of the short-range correlations 
through the parameter $b$. Indeed, small values of $b$ increase the active range of the correlations, reducing the strength of the force \cite{AriasdeSaavedra2007}.
The free parameter $b$ was adjusted in order to reproduce the $0^-$ state in $\isotope[16]{O}$ using an iterative procedure.
The authors first made HF calculations starting from D1S and D1M parameterizations with no tensor forces to get starting wave functions and SPEs. Then, they switch on the tensor forces in RPA calculations and selected the value of $b$ which reproduced the experimental energy of the $0^-$ state. Finally, they evaluated new wave functions and SPEs with the new interactions and tuned the spin-orbit parameter to find back the neutron $(1p_{3/2}$ - $1p_{1/2})$ splitting in $\isotope[16]{O}$. New RPA calculations followed and the procedure was repeated until convergence. The out-coming parameterizations were labeled D1SV8 and D1MV8, respectively.
This fitting process was rapidly taken up step by step by Anguiano \textit{et al.}\ \cite{Anguiano2011} with the isospin-dependent Argonne V18 potential \cite{Wiringa1995} in place of the V8' version, and generated the D1ST and D1MT parametrizations from the tensor-free D1S and D1M 
interactions.

This fitting procedure is fundamentally different from that used to elaborate GT2 as the D1S and D1M parameterizations were kept like they were and the tensor forces (as well as the spin-orbit force) were adjusted aside, on separate observables. 
In a sense, this is exactly what was already done for the spin-orbit term in the original Skyrme and Gogny interactions. This approach offers the advantage of isolating the effects produced by the term fitted perturbatively, but it is incomplete since each term composing the interaction participates in reproducing the observables and will be re-normalized by the contributions of all the other terms.\\

In 2012, Anguiano \textit{et al.}\ \cite{Anguiano2012} generalized the expression of the tensor force to describe both proton-neutron and particle-like tensor effects. In particular, they studied the experimental increase of the $N=28$ neutron gap between the $2p_{3/2}$ and $1f_{7/2}$ shells
when going from $\isotope[40]{Ca}$ to $\isotope[48]{Ca}$, which is often attributed to the tensor force \cite{Sorlin2008}.  
Indeed, according to Otsuka's picture, the tensor has almost no effect on the $\isotope[40]{Ca}$ SPEs as this nucleus is both proton and neutron spin-saturated. The situation is different for $\isotope[48]{Ca}$ which is spin-unsaturated on the neutron side (the particle-like tensor component is then active). 
Whereas the SLy5 Skyrme interaction predicts an energy decrease for the $N=28$ gap, the same interaction supplemented by a tensor force managed to recover the experimental trend \cite{Colo2007}. With D1S, the neutron $2p_{3/2}$ - $1f_{7/2}$ gap also moves in the wrong direction and the D1ST interaction worsens the predictions by diminishing even more its value. Modifying the sign of the tensor term would cured the problem, but it would altered the $0^-$ excitation energy of $\isotope[16]{O}$ on which it was adjusted. In order to cope with this issue, the analytical expression of the tensor force was extended in a formulation identical to that introduced by Onishi and Negele \cite{Onishi1978}:
\begin{align}
v_{12}^{\text{T}} & = (V_{\text{T1}} + V_{\text{T2}} P_\tau) ~e^{-(\vec{r}_1 - \vec{r}_2)^2/\mu_\text{T}^2} \widetilde{S}_{12}
\end{align}
where the isospin-exchange operator $P_\tau = (1 + \vec{\tau}_1 \cdot \vec{\tau}_2) /2$, introduces a pure 
and isospin-dependent components. 
In even-even nuclei, the direct component of the HF tensor field is 
zero and only the exchange tensor field contributes. It involves two-body 
matrix elements whose isospin part, labeled by $t$, reads \cite{Bernard2016a}:
\begin{multline} \label{pnpl}
\mel{t_a t_b}{(V_{\text{T1}} + V_{\text{T2}} P_\tau)}{t_b t_a} = 
(V_{\text{T1}} + V_{\text{T2}}) \delta_{t_a t_b} \\ + V_{\text{T2}} 
\delta_{t_a,- t_b}
\end{multline}
The strength of the tensor interaction acting between particle-like 
pairs is given by the sum $V_{\text{T1}} + V_{\text{T2}}$, and by 
$V_{\text{T2}}$ between protons and neutrons, for even-even nuclei. 
The combination $V_{\text{T1}} + V_{\text{T2}}$ was thus fitted on 
the experimental energy difference between the neutron $1f_{5/2}$ and 
$1f_{7/2}$ shells in $\isotope[48]{Ca}$, using the HF 
approximation \cite{Cottle2008}. The parameter $V_{\text{T2}}$, 
directly related to the isospin-dependent component of the tensor 
force, was chosen so as to reproduce the $0^-$ excitation energy of 
$\isotope[16]{O}$. The adjustment was performed according 
to the iterative procedure of Co' \textit{et al.}\ detailed for the 
previous parameterizations. The parameter of the spin-orbit parameter 
remained that of the D1S parametrization, and it was the particle-like 
part of the tensor term, that was fitted on SPEs. This parametrization was dubbed D1ST2a. 
Another parametrization, where the particle-like part rather had to reproduce the $N=28$ neutron gap 
increase was also set up, and referred to as D1ST2b. The tensor range 
was chosen to be equal to the longest range of the central part of 
D1S, as for GT2, namely $\mu_{\text{T}} = \SI{1.2}{\femto\metre}$. 

Finally, M. Anguiano and M. Grasso \cite{Grasso2013} built up another 
parametrization for which the spin-orbit and tensor parameters were 
jointly adjusted so as to reproduce the neutron single-particle 
energies of three doubly-magic nuclei: $\isotope[40]{Ca}, \isotope[48]
{Ca}$ and $\isotope[56]{Ni}$. This procedure, already employed for 
other interactions \cite{Zalewski2008,Zalewski2009}, is interesting as the 
spin-orbit and tensor interactions produce fine effects on SPEs, that 
either tend to add to or cancel each other out. However,
it is important to keep in mind that, depending on the nucleus, this procedure can have dramatic effects on nuclear masses, with variations ranging from a few MeV to a few tens of MeV. A global refit of the interaction parameters is always the best approach, even though it is more challenging.
In the procedure of Ref.\cite{Grasso2013}, the splitting of $\isotope[40]{Ca}$ is first reproduced by tuning the 
spin-orbit parameter $W_{ls}$ alone, as this nucleus is spin-saturated. 
Then, keeping the new value for the spin-orbit parameter, the 
splitting of $\isotope[48]{Ca}$ is analyzed. Since $W_{ls}$ is fixed, 
only the particle-like tensor strength $V_{\text{T1}} + V_{\text{T2}}$ 
is adjusted to recover the experimental value. Finally, the remaining 
proton-neutron tensor intensity is modulated on the experimental 
splitting of $\isotope[56]{Ni}$, a spin-unsaturated nucleus, with 
$W_{ls}$ 
and $V_{\text{T1}} + V_{\text{T2}}$ fixed above. It is worth noting 
that a maximum action of the tensor force is guaranteed in 
$\isotope[48]{Ca}$ for the particle-like contribution as the 
$(1f_{7/2})_\nu$ shell 
is fully filled and the $(1f_{5/2})_\nu$ shell is fully empty, and in $
\isotope[56]{Ni}$ for the proton-neutron contribution as the 
$(1f_{7/2})_\pi$ shell is fully filled and the $(1f_{5/2})_\pi$ is 
fully empty. The tensor range 
of this new parameterization labeled D1ST2c was fixed to the same 
value as the one of the other D1ST2 versions. Structure studies
were performed, in particular the interplay between the tensor term 
and deformation in the ground state of light and medium mass 
nuclei using the HFB formalism and the D1ST2a parameterization
\cite{Bernard2016a}.

\begin{figure*}
\centering
\includegraphics[width=0.24\linewidth]{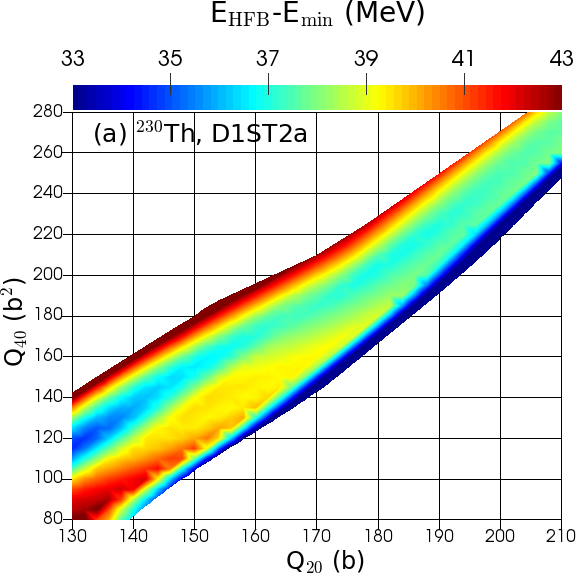}
\includegraphics[width=0.24\linewidth]{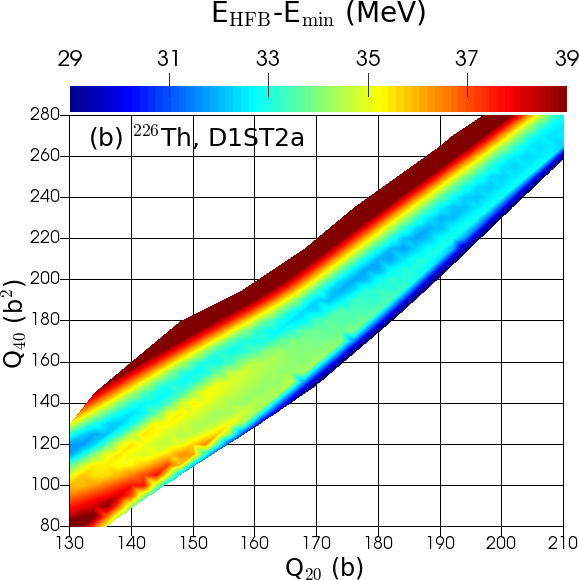}
\includegraphics[width=0.24\linewidth]{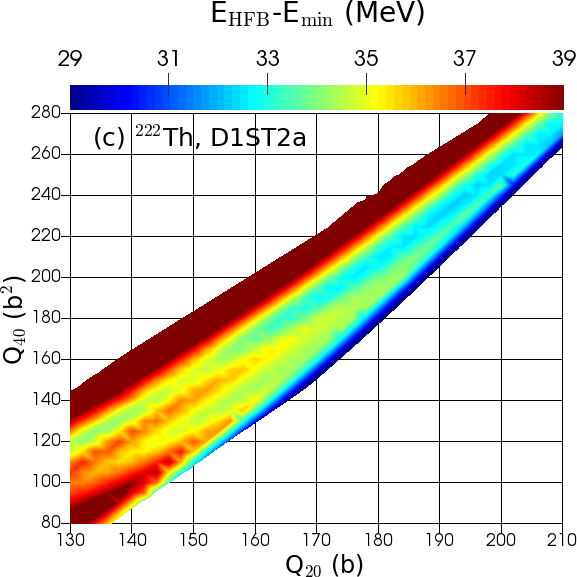}
\includegraphics[width=0.24\linewidth]{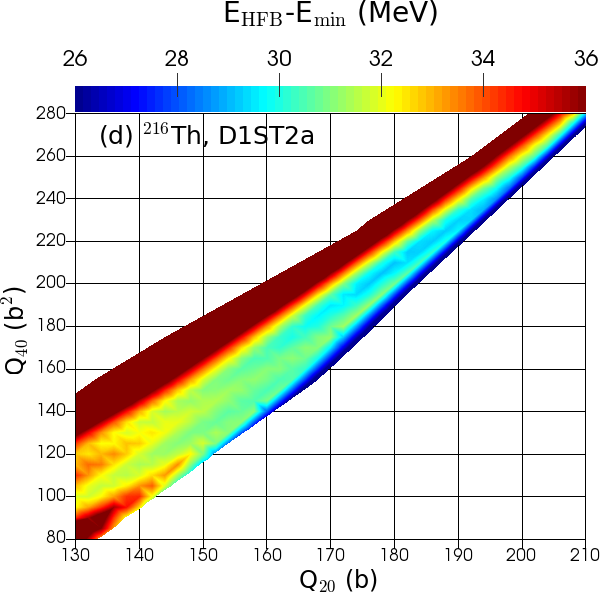}
\includegraphics[width=0.24\linewidth]{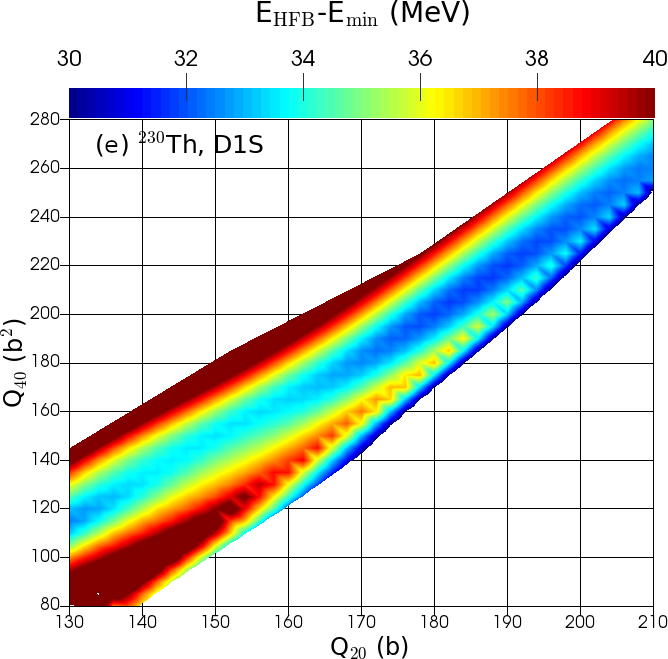}
\includegraphics[width=0.24\linewidth]{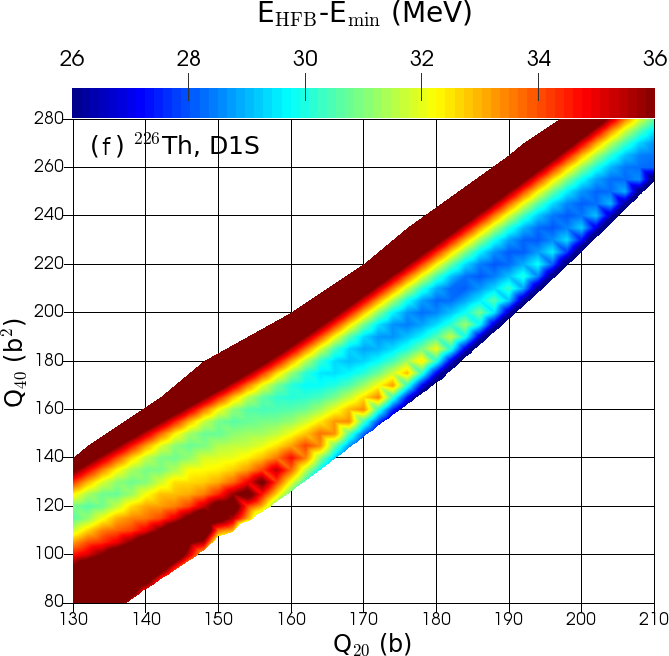}
\includegraphics[width=0.24\linewidth]{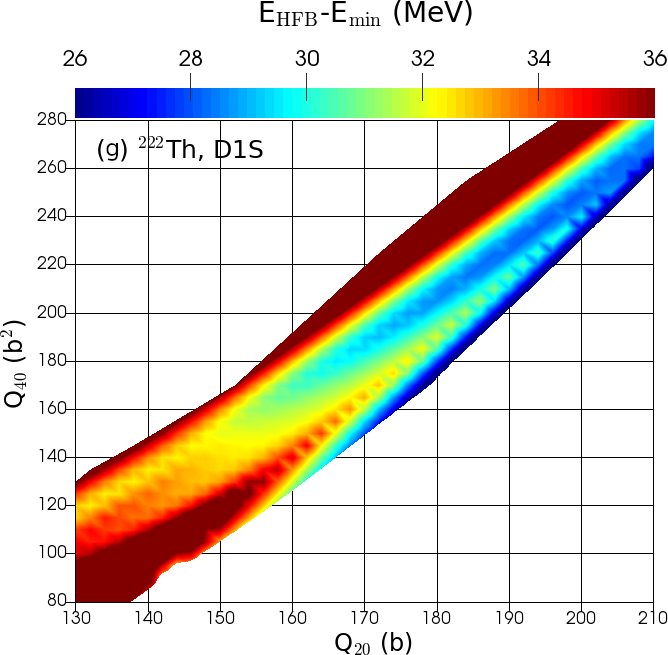}
\includegraphics[width=0.24\linewidth]{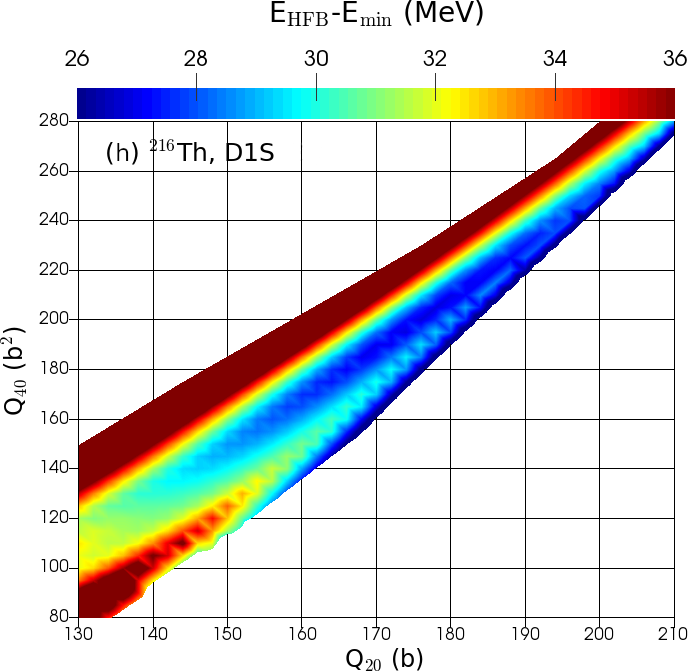}
\caption{Evolution of the HFB total binding energy, normalized to the ground state HFB energy, according to Q$_{20}$ and 
Q$_{40}$ collective coordinates in $^{230}$Th up to $^{216}$Th 
( Q$_{30}$=0) calculated
with the D1ST2a (panels (a) to (d)) and the D1S (panels (e) to (h)) parameterizations \cite{Bernard2020}. } \label{fig:thori2}
\end{figure*}
More recently, it was shown that the tensor term may have some impacts 
at very large quadrupole deformations, characteristic 
of fissioning systems \cite{Bernard2020}. This theoretical analysis
which focuses on Thorium isotopes, was based on Hartree-Fock-
Bogoliubov calculations under constraints and the D1ST2a 
parameterization. 
This study of the neutron-deficient thorium
isotopes fission was motivated by the experimental data obtained 
during the 2012 SOFIA campaign at GSI Darmstadt 
\cite{Chatillon2019,Chatillon2020}. The experimental results confirmed the already observed asymmetric to symmetric transition in the 
mass distribution of the fragments  \cite{Schmidt2000}. In addition, they suggested the existence of a new bimodal symmetric 
fission in this region, composed of the standard super-long mode and a new compact one. This last one was experimentally
characterized by the strong decrease of the average neutron multiplicity along the isotopic chain when decreasing neutron
number. The interplay between the tensor force and the deformation was proposed as a possible microscopic 
explanation of the appearance of this new fission compact mode. Indeed, the tensor force of the D1ST2a parameterization 
tends firstly to decrease the energy difference between the symmetric and the asymmetric second barrier in comparison to D1S, 
the effect being more and more pronounced when going to more exotic Thorium isotopes, leading to similar second barrier heights in $^{216}$Th. Secondly, for the symmetric path, a second valley corresponding to lower Q$_{40}$ values starts 
to develop with D1ST2a, which is predicted as the lowest one in energy in the $^{216}$Th isotope, as seen from Fig. \ref{fig:thori2}. This last one is
compatible with the new experimental symmetric compact mode.

\subsubsection{Fully finite-range Gogny force including a tensor term: the DG Gogny interaction family}\label{DGinter}

Recently, G. Zietek \textit{et al.} built an extended Gogny 
interaction, starting from the analytical form of the D2 interaction 
\eqref{gognyD2}, by promoting the spin-orbit term to a finite range and 
by adding a finite-range tensor term \cite{Zietek2023,Zietek2025}. This 
generalized Gogny interaction, labeled DG, has the following analytical 
expression:
\begin{multline} \label{gognyDG}
v_{12}^{\text{DG}} = \sum_{i=1,2} (W_i + B_i P_{\sigma} - H_i  
P_{\tau} - M_i P_{\sigma} P_{\tau}) V_i(r_{12}) \\
 \quad + (W_3 + B_3 P_{\sigma} - H_3  P_{\tau} - M_3 P_{\sigma} 
P_{\tau}) \\
\times \frac{V_3(r_{12})}{(\mu_3 \sqrt{\pi})^3} \frac{\rho^{\alpha}(\vec{r}_1) + \rho^{\alpha}(\vec{r}_2)}{2} \\
 + (W_4 - H_4  P_{\tau}) ~B(\mu_4)~ V_4(r_{12}) \,  \vec{L} 
\cdot \vec{S} \\
+ (W_5 - H_5  P_{\tau}) V_5(r_{12}) \,  S_{12}
\end{multline}
where the potentials $V_j(r_{12})$ are kept as Gaussian form factors:
\begin{equation} \label{gaussianpot}
V_j(r_{12}) = e^{-(\vec{r}_1 - \vec{r}_2)^2/\mu_j^2}
\end{equation} 
with $j \in \{1,2,3,4,5 \}$. In Eq.(\ref{gaussianpot}), $\mu_j$ is the range of the associated term, 
with the relative distance between 
the two nucleons given by $\vec{r}_{12} \equiv \vec{r}_1 - \vec{r}_2$. 

The finite-range spin-orbit force is identified by the spin-orbit operator $\vec{L} \cdot \vec{S}$ 
composed of a total intrinsic spin $\vec{S} \equiv (\vec{\sigma}_1 + 
\vec{\sigma}_2)/2$ and a total relative 
orbital momentum
$\vec{L} \equiv \vec{r}_{12} \times \vec{p}_{12}$,
where the relative momentum of the two-nucleon system is
defined by $\vec{p}_{12} = (\vec{p}_1-\vec{p}_2)/2$
with the momentum of the particle $j$ equal to $\vec{p}_j \equiv - i \hbar
\vec{\nabla}_j$. The coefficient $B(\mu_4)$ chosen such that:
\begin{equation} \label{coeffB}
B(\mu_4) = - \frac{4}{\mu_4^2} \frac{1}{(\mu_4 \sqrt{\pi})^3}
\end{equation}
ensures that one recovers the exact zero-range spin-orbit term of the 
original Gogny interaction at the zero-range limit (i.e.\ when $\mu_4 \rightarrow 0$).

The finite-range tensor force, characterized by the operator $S_{12}$, 
has been chosen in line with the formulation introduced by 
Onishi and Negele and pursued by Anguiano \textit{et al.}. Here, the 
expression of 
the tensor operator has been conventionally taken as:
\begin{equation}\label{tensoroperatormain}
S_{12} = (\vec{\sigma}_1 \cdot \hat{r}_{12}) (\vec{\sigma}_2 \cdot 
\hat{r}_{12}) - \frac{1}{3} \vec{\sigma}_1 \cdot \vec{\sigma}_2
\end{equation}
where $\hat{r}_{12}$ corresponds to the unit vector
$\hat{r}_{12} \equiv \vec{r}_{12} / \abs{\vec{r}_{12}}$. Unlike the 
density-dependent and the spin-orbit terms, no global multiplicative 
factor has been added as no zero-range tensor term had to be recovered in the context of the Gogny 
interaction. From Table \ref{tab:STD1}, one sees that the DG interaction acts in all possible allowed
spin-isospin channels, contrary to the D1 and D2 families. An interesting point is the possibility to 
recover very easily the D2, D1, D1ST analytical forms from DG, which is a strength either for 
the fitting procedure or its use inside a given solver. The numerical matchings are rapidly and stably 
obtained, already for values of $\mu$ of the order of 10$^{-2}$-10$^{-3}$.
\\

\begin{figure}
\centering
\includegraphics[width=0.7\linewidth,angle=-90]{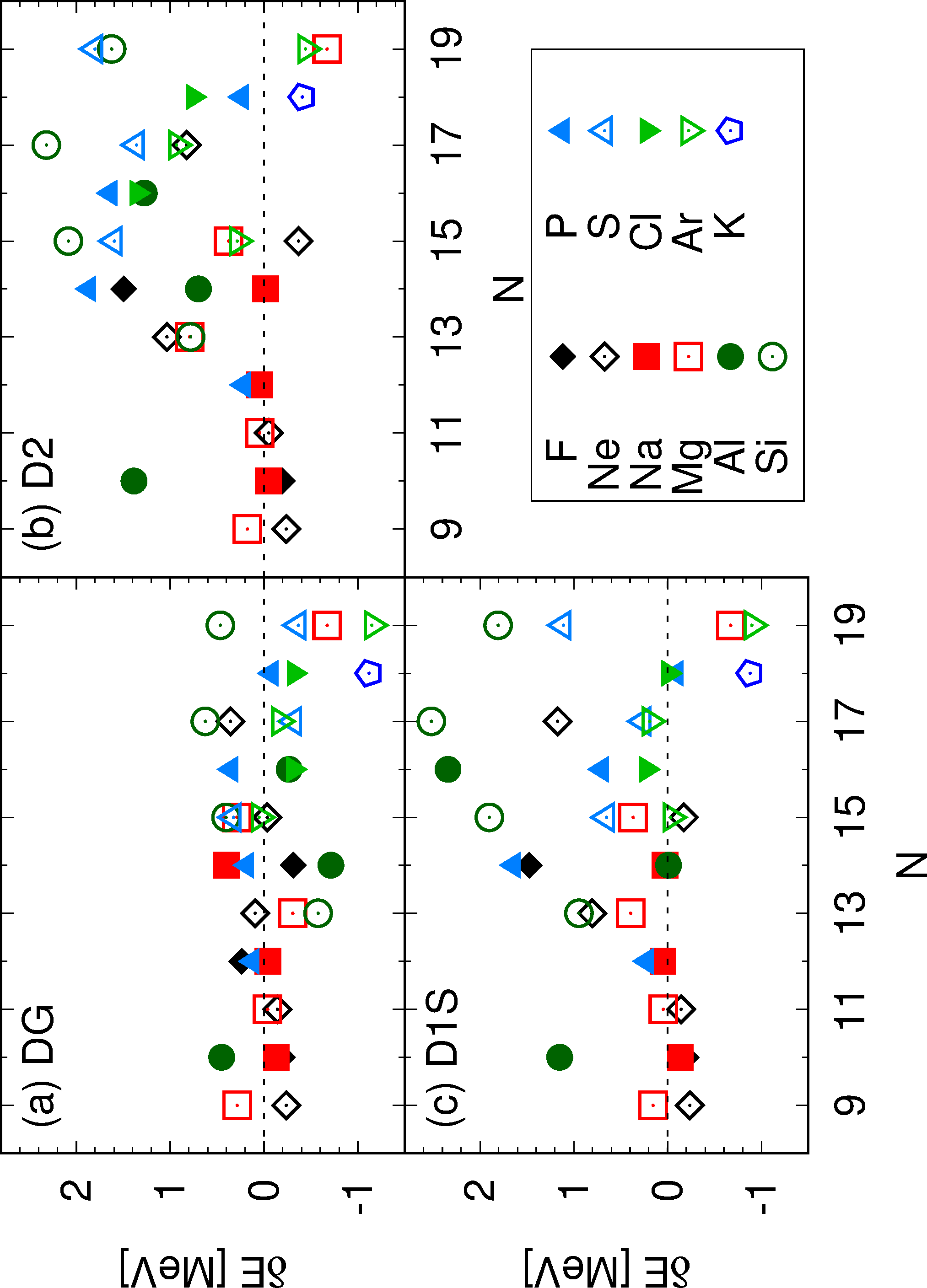}
\caption{Excitation energy $\delta$E of the first excited state of $sd$-shell odd nuclei calculated with (a) DG, (b) D2 
and (c) D1S parameterizations. Figure extracted from \cite{Zietek2025}.} \label{fig:mpmh}
\end{figure}

\begin{figure}
\centering
\includegraphics[width=0.55\linewidth,angle=-90]{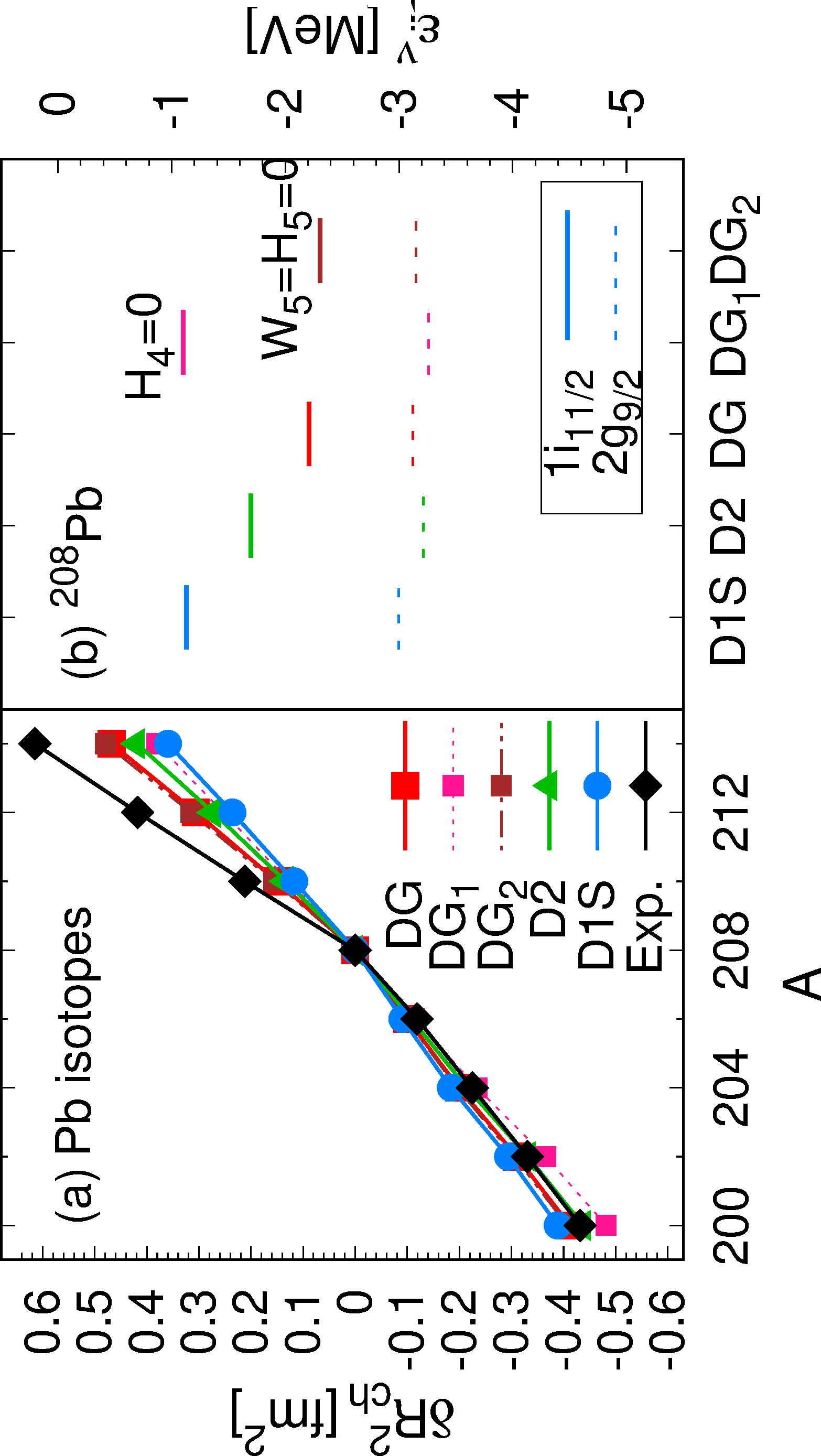}
\caption{Panel (a): Kink $\delta {\rm R_{ch}^{2} (A)= R_{ch}^{2}(A)- R_{ch}^{2}(208)}$ in Pb isotopes. Experimental
data have been extracted from \cite{Angeli2013}. 
Panel (b): The 1i$_{11/2}$-2g$_{9/2}$ neutron gap in $^{208}$Pb. A comparison is done for 
D1S, D2 and DG Gogny interactions. Figure extracted from \cite{Zietek2025}.} \label{fig:kink}
\end{figure}

Why introducing fully finite-range spin-orbit and tensor terms? 
On the one hand, for the finite-range nature, the general 
arguments are shared with those exposed in the section \ref{secD2} 
dedicated to D2. Indeed, this ensures that no divergence occurs at the mean-field level and beyond. Moreover, 
as the tensor term has a spin-orbit effect, it is important to 
consider them together. 
On the other hand, these two terms have 
specific roles on single-particle states and thus fine structure 
effects. Few of them have already been discussed in the previous 
sub-section \ref{pioneer} in the case of the particle-like and tensor isospin 
components of the tensor term. 
In the case of the DG parameterization, the spin-orbit and tensor 
term intensities have not been fitted in order to reproduce the 
experimental gap values at the HF approximation, eventhough these quantities have been checked to have reasonable values after. 
In order to fine-tune the intensity of the spin-orbit term which has a global effect on spin-orbit partner gaps 
and the intensity of the tensor term which is expected to have a local effect whose magnitude depends on the occupancy of 
the spin-orbit partners, it has been preferred to use beyond mean-field calculations 
\cite{Pillet2008,Robin2014a,Robin2016,Robin2017,LeBloas2014} to calculate first excited states in $sd$-shell even-even and odd 
nuclei in order. In Fig. \ref{fig:mpmh}, one sees the improvement in the excitation energy 
$\delta$E of the first excited states of $sd$-shell odd nuclei brought by DG (panel (a)) in comparison with D2 (panel (b)) and 
D1S (panel (c))  \cite{Zietek2025}.
In addition, the kink in the even-even Pb isotopes charge radii has also been considered in the selection of the DG parameterization 
as the $H P_{\tau}$ component of the spin-orbit term is evoked as partly responsible of the effect \cite{Sharma1995}. The 
comparison between DG, D2 and D1S, shown in Fig. \ref{fig:kink}, panel (a), indicates possible improvement in presence of 
finite-range spin-orbit and tensor \cite{Zietek2025}. With DG, one observes that the effect is strongly linked
to the parameter H$_4$ which governs the neutron gap between the 
1i$_{11/2}$ and 2g$_{9/2}$ shells (see panel (b)). Indeed, an increase of the gap is obtained with DG$_1$ 
whose parameter values are the same as the DG ones excepted H$_4$ which has been set to zero
(see panel (b)). The kink is deteriorated. In contrast, setting $W_5$ and $H_5$ to zero has essentially no impact on the gap (DG$_2$ case) and thus on the kink
in comparison with DG. 

\begin{figure}
\centering
\includegraphics[width=0.70\linewidth,angle=-90]{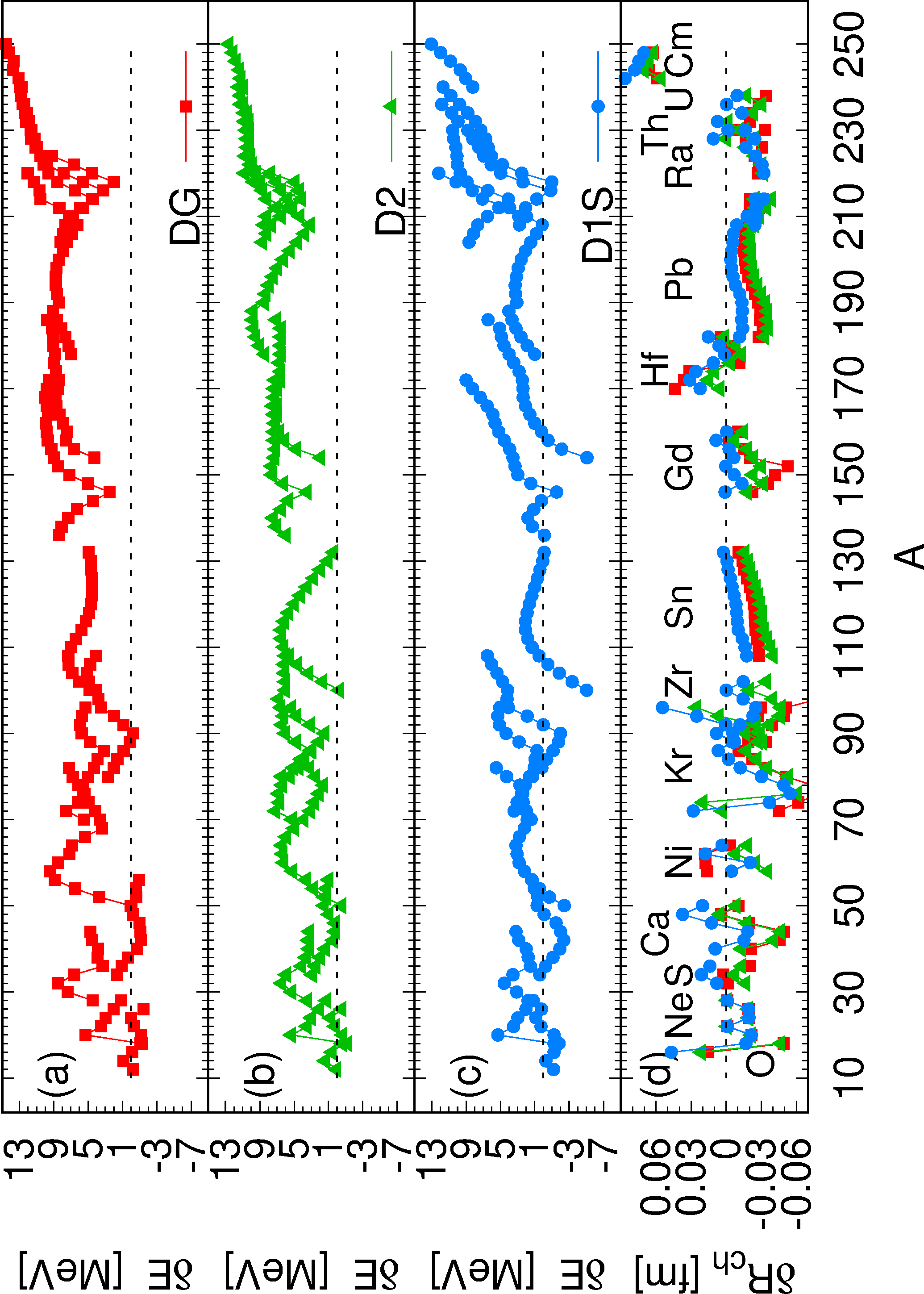}
\caption{Difference $\delta B$ in MeV between experimental and HFB binding energies for DG (panel (a)), D2 (panel (b)) and D1S (panel (c)).
In panel (d), the difference $\delta \rm R_{ch}$ between experimental and HFB charge radii (expressed in fm) is shown for the three interactions. Figure extracted from \cite{Zietek2025}.} \label{fig:BE-R2}
\end{figure}

Global calculations of binding energies in even-even nuclei of numerous isotopic chains reveals that the drift in binding energies 
has been fully controlled for DG, as displayed in Fig.\ref{fig:BE-R2}, panel (a) for the difference $\delta B$ between experimental and HFB binding 
energies. This was to main objective. The underbinding of nuclei was intentional, in anticipation of beyond-mean-field treatments that would introduce additional correlation energy. The same quantity has been plotted for D2 and D1S in panels (b) and (c), respectively, showing a similar behavior for D2 and a drift for D1S. 
Concerning the difference $\delta \rm R_{ch}$ between experimental and HFB charge radii (panel (d)), one observes that, for spherical nuclei and many 
deformed ones, it is found negative, which is desirable to include beyond mean-field effects. Many of the $\delta \rm R_{ch}$ cases correspond to soft nuclei
for which dynamical quadrupole correlations may have a non negligible impact. The light Hf isotopes and Cm anomaly is not explained but one indicates the existence of large experimental error bars in comparison with the other isotopic chains. New measures would be welcome. Similarly, the globally smaller values obtained for DG (and D2) were intentional, as the charge radii calculated with D1S using the 5DCH beyond-mean-field approach were found too large \cite{Delaroche2010}. Indeed, one recalls that the philosophy of the Gogny interaction is to leave room for an explicit treatment of nuclear long range correlations.

Finally, in Table \ref{tab:energies}, one presents both HFB and dynamical (from GCM+GOA), first and second, fission barrier heights in selected actinides
\cite{Zietek2025}. One observes that DG tends to decrease the height of the first symmetric barriers in comparison with D2 and D1S.
\begin{table}[htb!]
  \begin{tabular}{ll|cc|cc}
\hline
    Nucleus & Inter & $\rm B^{HFB}_I$  & $\rm B^{GCM}_I$  & $\rm B^{HFB}_{II}$  & $\rm B^{GCM}_{II}$ \\
    \hline
    $^{236}$U & D1S & 11.0   & 7.7      & 8.0   & 6.5  \\
              & D2  &  9.1   & 7.1      & 7.6   & 6.3   \\
              & DG  &  9.0   & 7.3      & 8.0   & 6.9  \\
\hline
    $^{240}$Pu& D1S & 11.1   & 9.3      & 7.9   & 6.7  \\
              & D2  & 10.5   & 8.6      & 7.6   & 6.4   \\
              & DG  & 10.1   & 8.2      & 8.5   & 7.5  \\
\hline
    $^{252}$Cf& D1S & 11.2   & 9.6      & 3.7   & 2.3   \\
              & D2  & 10.6   & 9.0      & 3.5   & 2.1   \\
              & DG  &  9.3   & 7.6      & 2.8   & 1.9  \\
\hline
  \end{tabular}
  \caption{First and second barrier heights (in MeV) obtained from the TDGCM+GOA approach and compared to a pure HFB estimation.}
  \label{tab:energies}
\end{table}
  
The actual numerical implementation of the finite-range spin-orbit and tensor terms in the HFB3 solver does not contain particular optimization.
For the example given in Fig. \ref{fig:D2HFB3}, a factor of the order of $\simeq$5 is obtained in comparison to D2. One third of the additional time is due to
the spin-orbit mean- and pairing- fields and the rest comes from the building of the tensor fields. 
These performances, which will be improved in the future, allow to envisage already systematic calculations for structure and fission studies.

\section{The restricted Hartree-Fock emulator for the generation of parameterizations}\label{sec2}

In section \ref{sec1}, the evolutions
of the Gogny force leading either to other parameterizations or to extended
analytical forms have been discussed. The physical motivations behind them have been highlighted. 
In the original fitting procedure of the D1-type Gogny interaction, several 
successive steps allow the extraction of a parameterization 
\cite{Gogny1973,Gogny1975a,Chappert2007,Zietek2023}. Firstly, some initial 
data associated with finite nuclei are collected to allocate 
quantitative values to constraints involving the parameters 
to be determined. Through the inversion of three systems gathering these constraints, a set of parameters is deduced. Secondly, this set faces 
filters associated with nuclear matter properties. If 
it passes through all the filters, it becomes a potential parameterization. This procedure
and its extension have led to D1, D1S, D1N, D2 and DG.
In the present section, one will describe the first step of the Gogny interaction fitting procedure that concerns the finite nuclei constraints. The second step, related to 
the nuclear matter filters, will be detailed further in section \ref{sec3}.
The procedures associated with other parameterizations 
, as D1M or D1ST2a for example, have already 
been slightly evoked in part \ref{sec1}, and one advises the readers to 
refer to the associated articles.

Once the analytical form of the interaction is 
fixed, the question to be solved is
how to determine the values of the parameters 
that best describe the 
nuclear properties within the framework of mean-
field theories
and beyond, knowing that the philosophy of the Gogny interaction is to leave room for an explicit treatment 
of the nuclear long range correlations. 
A systematic use of the codes associated with these 
theories can prove very or even too costly 
numerically for the fitting procedure, especially 
when the number of parameters is large. 
In the late 60's, it was even truer.
The idea of an emulator of the theories, i.e. a 
simplified model, was and
is still of great help. This is what has been 
imagined by D. Gogny in the late 60's. 
In the context of the Gogny's interaction 
fitting code, the restricted Hartree-Fock 
formalism plays the role of an emulator which 
is constrained by experimental data.
Initially formulated by D. Gogny for D1-type 
parameterizations, it is explained in details in the following, as well as its generalizations to D2 
and DG which were formulated later.

\subsection{Restricted Hartree-Fock model}\label{s1}

The determination of phenomenological effective interactions is closely linked to the knowledge of experimental data that can be used to assign values to the parameters on which they depend. 
Binding energies and charge radii are key observables in this respect.
An ingenious procedure would be to constrain the interaction parameters directly from the experimental data. 
In this proposal, experimental data are the starting point. The delicate point is to match them with theoretical results. Indeed, to get back to the interaction parameters, the calculation method needs to be inverted. This inversion is
only possible if the relation between the
theoretical estimates and the interaction parameters is analytical. In other words, to achieve this inversion, it is necessary to be able to write theoretical results as a linear analytical function of the interaction parameters.
This is what is proposed by the Restricted Hartree-Fock approximation (HFR), the principle of which is explained below. 

One assumes that the nucleus, composed of $A$ interacting nucleons, is described by an Hamiltonian $\hat{H}$ built from a two-body density-dependent interaction:

\begin{equation} \label{HHFR}
\hat{H} = \sum_{ab} \! \mel*{a}{t_{{K}}}{b} \! c^{\dagger}_a c_b + \frac{1}{4} \sum_{abcd} \! \mel*{ac}{v_{12}^{\text{({a})}}}{bd} \! c^{\dagger}_a c^{\dagger}_c c_d c_b
\end{equation}
where the operators $c^{\dagger}_a$ and $c_a$ are the creation and annihilation operators of nucleons, the one-body kinetic energy operator $t_{K}$ is written as a function of the momentum operator $\hat{p} \equiv \hbar \hat{k}$ such that $t_{\text{K}} \equiv p^2/{2M}$, and the two-body Gogny interaction $v_{12}$ has been anti-symmetrized:
\begin{equation}
v_{12}^{{(a)}} \equiv v_{12}(1 - P_r P_{\sigma} P_{\tau}).
\end{equation}

\noindent If $\vert \Psi \rangle $ represents the wave function of the $A$ nucleon system and $\mathcal{E}$ the corresponding total binding energy, 
the Schrödinger equation used to determine them, $\hat{H} \vert \Psi \rangle = \mathcal{E} \vert \Psi \rangle$,
is equivalent to the variational equation $\delta \mathcal{E}[\Psi] =0$ where $\mathcal{E}[\Psi]= \langle \Psi \vert \hat{H} \vert \Psi \rangle /\langle \Psi \vert \Psi \rangle$.
To solve this equation, the Hartree-Fock (HF) method proposes approximating $\vert \Psi \rangle$\ by a Slater determinant, denoted $\vert \Phi_{\rm HF}\rangle$ in the
following. It is written as the antisymmetric product of $A$ one-body wave functions 
$\vert \varphi^{\text{HF}}_i\rangle$ which are the unknowns of the problem.
This approximation corresponds to an independent particle picture. It assumes that nucleons evolve independently within an average potential created by the nucleon themselves. The energy minimization procedure is used to determine the individual wave functions $\vert \varphi^{\text{HF}}_i\rangle$. In practice, they are developed on a basis of known functions. In the original Gogny interaction fitting code using the HFR approach, 
spherical harmonic oscillator (HO) basis is used. 
Thus, the expansion of $\vert \varphi^{\rm HF}_{i} \rangle$ in this basis is written as:
\begin{equation}\label{ttt1}
\ket*{\varphi^{\text{HF}}_i} = \sum_{a=1}^{n} U_{ia}(\hbar \omega) \! \ket*{\phi_a (\hbar \omega)}
\end{equation}
for  $i \in \{1, \ldots, A \}$. In Eq.\eqref{ttt1}, the function $\vert \phi_{(a)} (\hbar \omega) \rangle$ represents a spherical HO wave function (see Appendix \ref{anexa}) characterized by the spherical quantum numbers $ (a) \equiv (n_a,l_a)$ and the HO frequency $\hbar \omega$.
The maximum number $n$ of states in the basis, is determined by its size.
Determining the individual functions amounts to determining the coefficients $U_{ia}(\hbar \omega)$ of the expansion. \\

The total binding energy $\mathcal{E}_{\text{HF}}$ calculated in the HF approximation is simply written as a function of the one-body density matrix $\rho = \langle{\Phi_{\text{HF}}} | \hat{\rho} | {\Phi_{\text{HF}}}\rangle$:
\begin{equation} \label{HFenergymain}
\mathcal{E}_{\text{HF}} = \sum_{ab} \! \mel*{a}{t_{{K}}}{b} \! \rho_{ba} + \frac{1}{2} \sum_{abcd} \! \mel*{ac}{v_{12}^{\text{({a})}}}{bd} \! \rho_{ba} \rho_{db}
\end{equation}
where
\begin{equation}
\rho_{ab} = \sum_{i=1}^A U^*_{ia}(\hbar \omega, v_{12}) U_{ib}(\hbar \omega, v_{12}).
\end{equation}
The non-linearity of the HF equations leads to an iterative resolution that makes it impossible to extract a linear relation between
the coefficients $U_{ia}(\hbar \omega, v_{12})$ and the parameters of the Gogny interaction. In other words, it is not possible to express the energy directly as a function of the interaction parameters.
The restricted HF approximation (HFR) removes this complication by assuming that the unitary transformation $U$ is diagonal, i.e. 
$U_{ia}= \delta_{ia}$. In that case, the individual wave functions $\ket*{\varphi^{\text{HF}}_i}$ reduce to pure HO ones:
\begin{equation}
\ket*{\varphi^{\text{HFR}}_i } = \ket*{\varphi^{\text{HFR}}_i (\hbar \omega)} = \ket*{\phi_a (\hbar \omega) } \delta_{ia}
\end{equation}
for $ i \in \{1, \ldots, A \}$ and the one-body density matrix acquires a diagonal form:
\begin{equation}
\rho_{ab} = \delta_{ab} \rho_{aa} = 
\begin{cases} 
  1 \quad & \text{if} \ a \le A,  \\
  0 \quad & \text{otherwise}. 
\end{cases}
\end{equation}
The total binding energy $\mathcal{E}^{\text{HFR}}$, which contains a kinetic $\mathcal{E}^{\text{HFR}}_K$ and 
potential $\mathcal{E}^{\text{HFR}}_P$ energy terms, simplifies:
\begin{equation}\label{EHFR} 
\begin{split}
\mathcal{E}^{\text{HFR}} & = \mathcal{E}^{\text{HFR}}_K + \mathcal{E}^{\text{HFR}}_P \\
& = \sum_{a} K_{aa}^{\text{HFR}} \rho_{aa} +  \sum_{a} \Gamma_{aa}^{\text{HFR}} \rho_{aa} \\
& =\sum_{a} \! \mel*{a}{t_{{K}}}{a} \! \rho_{aa} + \frac{1}{2} \sum_{ab} \! \mel*{ab}{v_{12}^{\text{({a})}}}{ab} \! \rho_{aa} \rho_{bb} 
\end{split}
\end{equation}
The one-body HFR Hamiltonian $h^{\text{HFR}}$, to which each nucleon of the nucleus is submitted, satisfies the relation $h_{aa}^{\text{HFR}} \equiv \partial\mathcal{E}^{\text{HFR}}/\partial \rho_{aa}$, so that:
\begin{equation} \label{hHFR}
h_{aa}^{\text{HFR}} = K_{aa}^{\text{HFR}} + \Gamma_{aa}^{\text{HFR}} + \partial \Gamma_{aa}^{\text{HFR}}
\end{equation}
\noindent The rearrangement field $\partial \Gamma_{aa}$ appearing in Eq.(\ref{hHFR}), which comes from the density dependence of the interaction, is expressed as:
\begin{equation} \label{rfHFB}
\partial \Gamma_{ab} \equiv \frac{1}{2} \sum_{a'b'} \! \mel*{a'b'}{\pdv{v_{12}^{{(a)}}}{\rho_{aa}}}{a'b'} \! \rho_{a' a'} \rho_{b' b'}.
\end{equation}
Since each field contributing to $h_{aa}^{\text{HFR}}$ of \eqref{hHFR} is diagonal, $h_{aa}^{\text{HFR}}$ is itself diagonal. Therefore, each diagonal term corresponds to an eigenvalue $\varepsilon_a^{\text{HFR}}$ identified as a one-body energy, i.e.:
\begin{equation} \label{epsHFR}
\varepsilon_a^{\text{HFR}} = h_{aa}^{\text{HFR}} = K_{aa}^{\text{HFR}} + \Gamma_{aa}^{\text{HFR}} + \partial \Gamma_{aa}^{\text{HFR}}.
\end{equation}
General expressions of $h_{aa}^{\text{HFR}}$ and $\mathcal{E}^{\text{HFR}}$ are derived in Appendix \ref{appenb} in the case of the DG interaction, from which the expressions for D2 and D1 are easily deducible.

\begin{figure}
\centering
\includegraphics[angle=-0,width=0.75\linewidth]{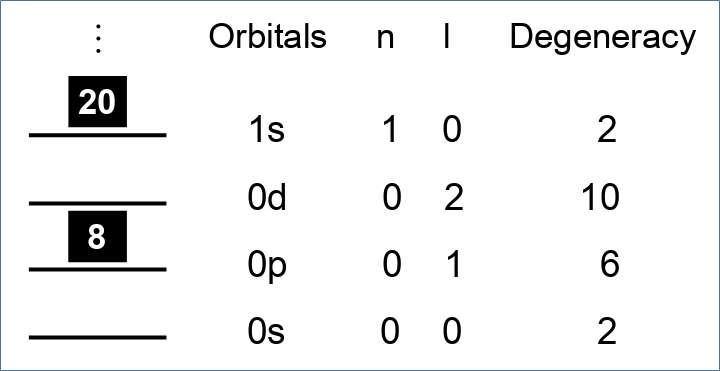}
\caption{Representation of the one-body spherical HO basis used in the HFR emulator.}
\label{fig:shells}
\end{figure}
\noindent Another important approximation done in the context of the HFR approximation is that the spin-orbit coupling is neglected in such a way that the spherical shells are labeled only by their principal quantum numbers $n$ and their angular orbital momenta $l$. A representation is displayed in Fig.\ref{fig:shells}. \\

\begin{figure*}
\centering
\includegraphics[angle=-0,width=0.95\linewidth]{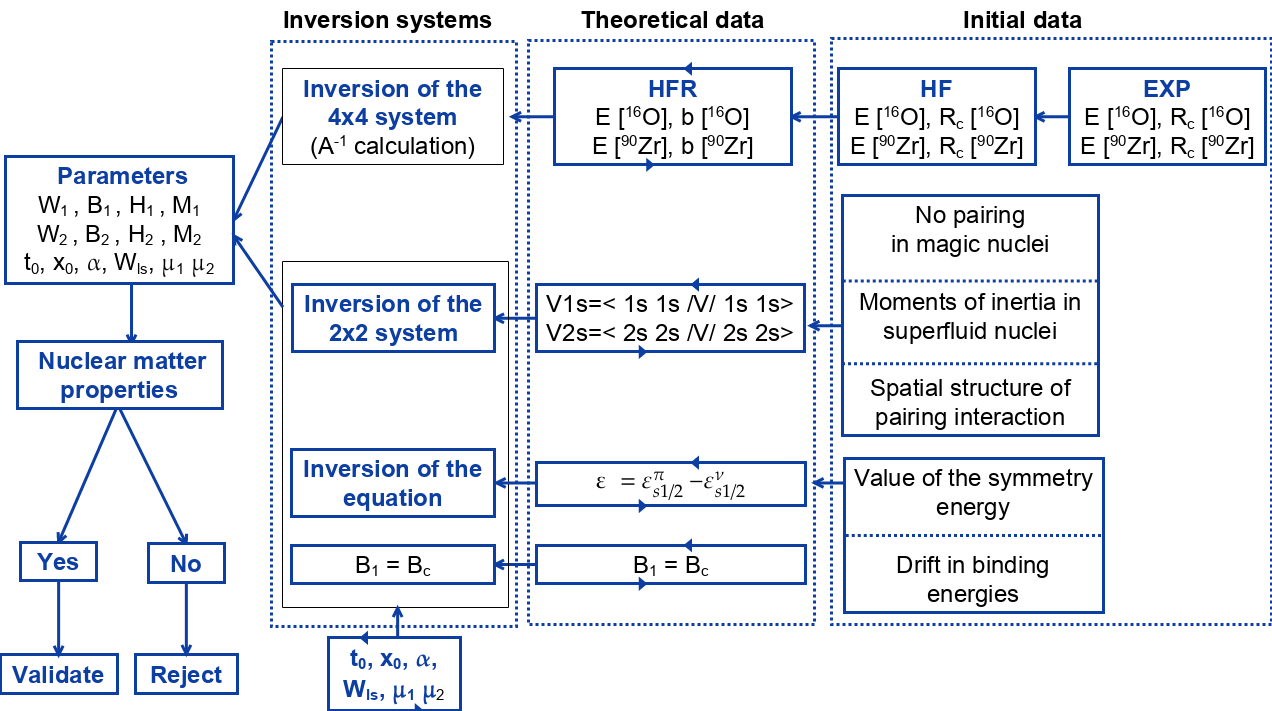}
\caption{Global adjustment scheme for a D1-type Gogny interaction \cite{Chappert2007}. The arrows indicate the quantities which are varied in the fitting code.}
\label{fig:schema_fit}
\end{figure*}

From Eq.(\ref{EHFR}), the total binding energy $\mathcal{E}^{\text{HFR}}$ can be expressed directly as a function of the parameters of the 
interaction, which suggests the possibility of inverting the calculation method to determine the interaction parameters from experimental data.
At this level, it is important to discuss the link between the HF and HFR theoretical results to establish the meta-modeling, and bring credit to the HFR approximation
which is nothing less than an emulator. 
It is not appropriate to determine HFR data directly from experimental ones because the model is not accurate enough. However, a linear behavior can be established between the HF and HFR models, for both total binding energies and charge radii. Therefore, the HF data will be compared to the experimental binding energies $ \rm \mathcal{E}^{HF}=\mathcal{E}^{exp.}$ and charge radii $\rm R_{c}^{HF}=R_{c}^{exp.}$ while the HFR ones will be deduced by the linear relations.
The determination of pairing correlation intensity and symmetry energy are also used to determine the parameters.
A schematic representation of the global fitting procedure of D1-type parameterizations (\ref{gognyD1}) is displayed in Fig.\ref{fig:schema_fit}. A similar approach has been adopted for D2 (\ref{gognyD2}) and DG (\ref{gognyDG}) Gogny interactions with specific changes that will be further mentioned . 
The three mentioned inversion systems thus introduced enable to impose the physical properties associated with basic experimental data. Doing so, a reduction of the
number of free parameters is also performed. For D1-type parameterizations, only four are left free (not defined by the inversion systems), 
including the two ranges, the spin-orbit and density-dependent term intensities. Two parameters are fixed, namely $x_0=1$ and $\alpha =1/3$. 
On the left-hand side of the Fig.\ref{fig:schema_fit}, the properties of infinite nuclear matter, which will be commented in section \ref{sec3}, appear as filters. \\

The explicit equations forming the HFR emulator are detailed below. They are expressed firstly for D1-type parameterizations. Modifications related to the D2 and DG analytical forms are indicated in a second time.

\subsubsection{Adjustment of global volume properties}\label{ss1}

As shown in Eq.(\ref{EHFR}), the HFR total binding energy $\rm \mathcal{E}^{\text{HFR}}$ is composed of the kinetic $ \mathcal{E}^{\text{HFR}}_{K}$ and the potential $ \mathcal{E}^{\text{HFR}}_{P}$ terms. Their explicit expressions are given below in the case of the two spherical nuclei $^{16}$O and $^{90}$Zr
used in the fitting process. 

The kinetic energy contribution $\mathcal{E}^{\text{HFR}}_{K}$ is given in the spherical HO representation by:
\begin{equation} \label{kineticfieldHFR}
\mathcal{E}^{\text{HFR}}_{K}= \sum_{n_a l_a m_{l_{a}}s_a t_a} \frac{\hbar \omega}{2} \Big( 2 n_a + l_a + \frac{3}{2} \Big) \rho_{aa}
\end{equation}
\noindent Using the Virial theorem, the application to $^{16}$O and $^{90}$Zr leads to:
\begin{equation}\label{cin}
\begin{split}
\mathcal{E}^{\text{HFR}}_{K} [^{16}O]= \frac{9}{8} \frac{\hbar^2}{Mb^{2}_{\rm [^{16}O]}} \\
\mathcal{E}^{\text{HFR}}_{K} [^{90}Zr]= \frac{71}{36} \frac{\hbar^2}{Mb^{2}_{[\rm ^{90}Zr]}} 
\end{split}
\end{equation}
where the HO length $b$ is related to the HO frequency $\omega$ by $b=\sqrt{\hbar/M\omega}$. \\

The potential energy $\mathcal{E}^{\text{HFR}}_{P}$ contains three contributions associated with the two finite-range terms, 
$\mathcal{E}^{\text{HFR}}_{P \mu_{1}}$ and $\mathcal{E}^{\text{HFR}}_{P \mu_{2}}$,
and the density-dependent one, $\mathcal{E}^{\text{HFR}}_{P 0}$. For a given nucleus X:
\begin{equation}\label{pote}
\mathcal{E}^{\text{HFR}}_{P} [X] = \mathcal{E}^{\text{HFR}}_{P \mu_{1}} [X] 
 +\mathcal{E}^{\text{HFR}}_{P \mu_{2}} [X] + \mathcal{E}^{\text{HFR}}_{P 0} [X]
\end{equation}
where
\begin{equation}\label{noyasym}
\begin{array}{lcl}
\mathcal{E}^{\text{HFR}}_{P \mu_{i}}[X] &=  & F_i^{{D}}[X] (4W_i + 2B_i - 2H_i - M_i) \\ 
&+& F_i^{{E}}[X] (4M_i + 2H_i - 2B_i - W_i)
\end{array}
\end{equation}
\noindent and
\begin{equation}
\begin{array}{lcl}
 \displaystyle \mathcal{E}^{\text{HFR}}_{P 0}[X] &= & G(t_0, x_0) [X] \\ 
 &= &\frac{1}{2} t_{0} \int d^3r  \rho^{\alpha} (\vec{r})  \times \displaystyle \big( (1+ \frac{x_0}{2}) \rho^{2}(\vec{r}) \\
&-& (x_0 + \frac{1}{2}) \sum_{t} [\rho_t(\vec{r})]^2 \big)
\end{array}
\end{equation} 
\noindent The direct $F_i^{{D}}$ and exchange $F_i^{{E}}$ functions are given by:
\begin{equation} \label{Fdirect}
F_i^{{D}}[X] \equiv 2 \sum_{(a)_{\pi}} \sum_{(b)_{\nu}} \widetilde{S}_{(a)(b)}^{\text{C}} \vert_{\text{D}}
\end{equation}
\begin{equation} \label{Fexchange}
F_i^{{E}}[X] \equiv 2 \sum_{(a)_{\pi}} \sum_{(b)_{\nu}} \widetilde{S}_{(a)(b)}^{\text{C}} \vert_{\text{E}}
\end{equation}
with
\begin{equation} \label{sss1}
\begin{array}{lcl}
\widetilde{S}^{\text{C}}_{ab} \vert_{\text{D}} &\equiv& \displaystyle \frac{(2l_a+1)(2l_b+1) }{4 \pi} \sum_{n_\mu n_\nu} T_{(a) (a)}^{(n_\mu 0)} \times \\
 & &T_{(b) (b)}^{(n_\nu 0)} I_{n_\mu n_\nu 0}
 \end{array}
\end{equation}
and
\begin{equation} \label{sss2}
\begin{array}{lcl}
\displaystyle \widetilde{S}^{\text{C}}_{ab} \vert_{\text{E}} &\equiv& \displaystyle \frac{(2l_a+1)(2l_b+1) }{4 \pi}
 \sum_{n_\mu  n_\nu l_\mu} (2l_\mu+1) \times \\
 & &\displaystyle \begin{pmatrix} l_a & l_b & l_\mu \\ 0 & 0 & 0 \end{pmatrix}^2 T_{(a) (b)}^{(n_\mu l_\mu)} T_{(b) (a)}^{(n_\nu l_\mu)} I_{n_\mu n_\nu l_\mu}
 \end{array}
\end{equation}
In the above equations, the indices $(a)$ and $(b)$ represent the quantum numbers $(n_a l_a)$ and $(n_b l_b)$, respectively.
In Eqs. \eqref{sss1} and \eqref{sss2}, the quantities T represents the Talman coefficients defined in Eq.\eqref{talman}. Finally, the quantity $I_{n n' l}$ is defined by:
\begin{equation} \label{Jsph}
\begin{array}{lcl}
I_{n n' l} &\equiv& \displaystyle \frac{1}{2^{n+n'+l}} \times \\
 & & \displaystyle \frac{(n+n'+l)! \, \Gamma(n+n'+l+3/2)}{[n! n'! \, \Gamma(n+l+3/2) \, \Gamma(n'+l+3/2)]^{1/2}} \\
 & & \displaystyle \times \sum_{i=0}^{n+n'+l} \, \frac{(-)^i G^{-i-3/2}}{i! (n+n'+l-i)!}.
 \end{array}
\end{equation}
\noindent The Eq.(\ref{noyasym}) is exact for nuclei with equal numbers of protons and neutrons. For nuclei with an asymmetric number of protons and neutrons, an additional contribution appears, displaying two new parameter combinations, namely $\rm (2W+B-2H-M)$ and $\rm (2M+H-2B-W)$. However, in weakly asymmetrical nuclei, such as $^{90}$Zr, its contribution is negligible compared with those of Eq.(\ref{noyasym}). Consequently, the potential energy Eq.(\ref{noyasym}) gives a very satisfactory estimate of the exact result and is adopted in the fitting code.\\

Furthermore, the value of the HO length $b$ must correspond to the energy minimum, which imposes the condition $d \mathcal{E}^{\rm HFR}/db =\mathcal{E}^{\prime}_{\rm HFR} = 0$. Thus:
\begin{equation}\label{rayonc}
\begin{array}{lll}
\displaystyle  \mathcal{E}^{\prime}_{\text{HFR}}[X]& \equiv& \mathcal{E}_{\text{K}}^{\prime \text{HFR}}[X]
+ \mathcal{E}_{{P}}^{\prime \text{HFR}}[X] \\
 & =& \displaystyle \mathcal{E}_{{K}}^{\prime \, \text{HFR}}[X] + G^{\prime}[X](t_0, x_0) \\
 &+ &\displaystyle  \sum_{i=1,2} F_i^{\prime {D}}[X] (4W_i + 2B_i - 2H_i - M_i) \\ 
 & + & \displaystyle \sum_{i=1,2} F_i^{\prime {E}}[X] (4M_i + 2H_i - 2B_i - W_i)  \\
 &=& 0
\end{array}
\end{equation}
\noindent From the expressions of the kinetic energies, one can easily deduce:
\begin{equation}
\begin{split}
\frac{\mathcal{E}_{{K}}^{\prime \, \text{HFR}} [\isotope[16]{O} ]}{16} = -\frac{9}{4} \frac{\hbar^2}{Mb^3_{\rm[^{16}O]}} \\
\frac{\mathcal{E}_{{K}}^{\prime \, \text{HFR}} [\isotope[90]{Zr} ]}{90} = -\frac{71}{18} \frac{\hbar^2}{Mb^3_{\rm [^{90}Zr]}}
\end{split}
\end{equation}

\noindent A system of four equations is obtained from Eqs. (\ref{pote}) and (\ref{rayonc}):
\begin{equation} \label{constraint1}
BY=F
\end{equation}
where $B$ is a 4x4 matrix, $Y$ and $F$ are two column vectors such that:
\begin{subequations}
\begin{gather}
A \equiv
\begin{bmatrix}
F_1^{\text{D}}[\isotope[16]{O}] & F_2^{\text{D}}[\isotope[16]{O}] & F_1^{\text{E}}[\isotope[16]{O}]  & F_2^{\text{E}}[\isotope[16]{O}]  \\
F_1^{\prime \text{D}}[\isotope[16]{O}] & F_2^{\prime \text{D}}[\isotope[16]{O}] & F_1^{\prime \text{E}}[\isotope[16]{O}] & F_2^{\prime \text{E}}[\isotope[16]{O}] \\
F_1^{\text{D}}[\isotope[90]{Zr}]  & F_2^{\text{D}}[\isotope[90]{Zr}]  & F_1^{\text{E}}[\isotope[90]{Zr}]  & F_2^{\text{E}}[\isotope[90]{Zr}]  \\
F_1^{\prime \text{D}}[\isotope[90]{Zr}]  & F_2^{\prime \text{D}}[\isotope[90]{Zr}]  & F_1^{\prime \text{E}}[\isotope[90]{Zr}]  & F_2^{\prime \text{E}}[\isotope[90]{Zr}]
\end{bmatrix} \\
Y \equiv
\begin{bmatrix}
4W_1 + 2B_1 - 2H_1 - M_1 \\
4W_2 + 2B_2 - 2H_2 - M_2 \\
4M_1 + 2H_1 - 2B_1 - W_1 \\
4M_2 + 2H_2 - 2B_2 - W_2 
\end{bmatrix}  \\
F \equiv
\begin{bmatrix}
\mathcal{E}^{\text{HFR}}[\isotope[16]{O}] - \mathcal{E}_{\text{K}}^{\text{HFR}}[\isotope[16]{O}] - G(t_0, x_0)[\isotope[16]{O}] \\
\mathcal{E}_{\text{K}}^{\prime \text{HFR}}[\isotope[16]{O}] - G^{\prime}(t_0, x_0)[\isotope[16]{O}] \\
\mathcal{E}^{\text{HFR}}[\isotope[90]{Zr}] - \mathcal{E}_{{K}}^{\text{HFR}}[\isotope[90]{Zr}] - G(t_0, x_0)[\isotope[90]{Zr}] \\
\mathcal{E}_{{K}}^{\prime \text{HFR}}[\isotope[90]{Zr}] - G^{\prime}(t_0, x_0)[\isotope[90]{Zr}] \\
\end{bmatrix}
\end{gather}
\end{subequations}

The inversion of the 4x4 system allows to constrain the values of four linear combinations of the parameters. However, it gives no information on the
other parameters $t_0$, $\mu_{1}$ and $\mu_{2}$. Therefore, they must be known elsewhere, since they are involved in the construction of B and F. The adopted solution consists in to giving initial values to $ t_0$, $\mu_{1}$ and $\mu_{2}$ which are not fixed but describe an interval.

The Virial theorem is once again of great use, as in the HFR approximation it allows to relate the charge radius of a nucleus to the HO parameter $b$:
\begin{equation}\label{virial}
\begin{split}
R_{{ch}}^{\text{HFR}} \big[\isotope[16]{O} \big] = \frac{\sqrt{15}}{2} 
b_{[\isotope[16]{O}]} \\
R_{{ch}}^{\text{HFR}} \big[\isotope[90]{Zr} \big] = \frac{\sqrt{25}}{2} b_{[\isotope[90]{Zr}]}
\end{split}
\end{equation}
If the value of the charge radius $R_{{ch}}^{\text{HFR}}$ evaluated at the HFR approximation is known, the value of $b$ can be immediately deduced. \\
\begin{figure}
\centering
\includegraphics[angle=-0,width=0.8\linewidth]{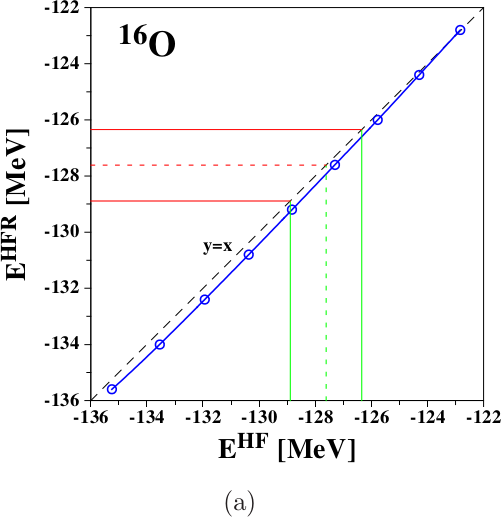}
\includegraphics[angle=-0,width=0.8\linewidth]{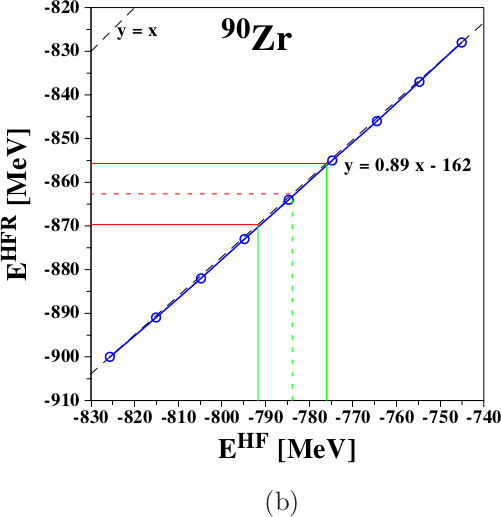}
\caption{Evolution of the HFR binding energies as a function of the HF binding energy for (a) $^{16}$O and (b) $^{90}$Zr. Energies are expressed in MeV. The experimental value is indicated by the green vertical dotted lines \cite{Chappert2007}.}
\label{fig:EHFR}
\end{figure}
A specific procedure as been applied to find in which acceptable ranges he HFR binding energies and charge radii vary, starting from experimental data. 
Concerning the binding energies of $^{16}$O and $^{90}$Zr, the first step is to generate several parameterizations $\{ v_{12}^{D1} \} = \{ W_{1}$, $B_{1}$, $H_{1}$, $M_{1}$, $\mu_{1}$, $W_{2}$, $B_{2}$, $H_{2}$,
$M_{2}$, $\mu_{2}$, $t_0$, $x_0$, $\alpha$, $W_{ls} \}$, nine in the present example.
For these nine parameterizations, HF and HFR calculations are performed. Fig. \ref{fig:EHFR} shows that the $^{16}O$ (top) and $^{90}Zr$ (bottom) binding energies obtained from HF and HFR can be deduced from each other by an almost linear transformation: 
\begin{equation}
\rm \mathcal{E}^{HFR}[^{16}O] \simeq \mathcal{E}_{HF}[^{16}O]
\end{equation}
\begin{equation}
\rm \mathcal{E}^{HFR}[^{90}Zr] \simeq 0.89 \mathcal{E}_{HF}[^{90}Zr] -162 ~MeV
\end{equation}

\noindent Using the experimental binding energies as a reference, the HF binding energies are chosen so that they reproduce the formers to within 1\%. Variation intervals, represented by the solid green lines, are thus deduced. The experimental value is indicated by the green vertical dotted lines:
$$\rm \mathcal{E}^{HF}[^{16}O] \in [-128.892, -126.340]~MeV$$ 
$$\rm \mathcal{E}^{HF}[^{90}Zr] \in [-791.730, -776.052] ~MeV$$
The corresponding variation intervals for the HFR data can then be extracted: 
$$\rm \mathcal{E}^{HFR}[^{16}O] \in [-128.892, -126.340]~MeV$$ 
$$\rm \mathcal{E}^{HFR}[^{90}Zr] \in [-868.509, -854.511] ~MeV$$
\noindent While it is identical for $^{16}$O, it is quite different for $^{90}$Zr. \\

\begin{figure}
\centering
\includegraphics[angle=-0,width=0.85\linewidth]{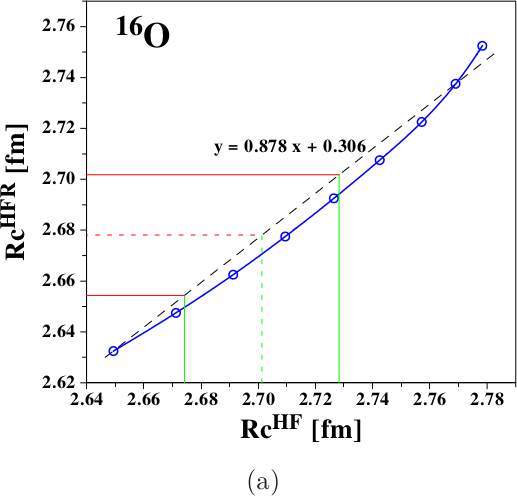}
\includegraphics[angle=-0,width=0.86\linewidth]{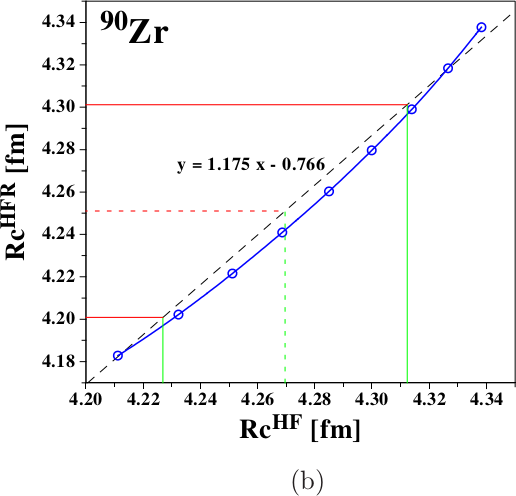}
\caption{Evolution of the HFR charge radii as a function of the HF ones for (a) $^{16}$O and (b) $^{90}$Zr. Radii are expressed in fm. The experimental value is indicated by the vertical green dotted lines \cite{Chappert2007}.}
\label{fig:RcHFR}
\end{figure}

The same analysis has been carried out for the charge 
radii $R_{c}$ of $^{16}$O and $^{90}$Zr. To a good approximation, a linear relation has also been found, as shown 
in Fig. \ref{fig:RcHFR} for $^{16}$O (top) and $^{90}$Zr (bottom). One deduces:
$$R_{c}^{\rm HFR} {\rm [^{16}O]} \simeq 0.878 R_{c}^{\rm HF}\rm [^{16}O] + 0.306 \rm ~fm$$ 
$$R_{c}^{\rm HFR} {\rm [^{90}Zr]} \simeq 1.175 R_{c}^{\rm HF}\rm [^{90}Zr] -0.766 \rm ~fm$$

\noindent If the HF theoretical charge radii are required to reproduce the experimental values within 1\%, the following charge radius variation intervals are obtained :
$$R_{c}^{\rm HF}\rm [^{16}O] \in [2.6743, 2.7283]  \rm fm$$ 
$$ R_{c}^{\rm HF}\rm[^{90}Zr] \in [4.2269, 4.3123] ~ \rm fm $$
And for the HFR one:
$$R_{c}^{\rm HFR}\rm[^{16}O] \in [2.6543, 2.7018] \rm fm$$
$$R_{c}^{\rm HFR}\rm [^{90}Zr] \in [4.2008, 4.3012]  \rm fm$$
\noindent The HFR approach tends to reduce their values.

All the results discussed so far depend on the value of the spherical HO parameter $b$  used for $^{16}$O and $^{90}$Zr. The limits on the possible values
are obtained by inverting the Eqs.(\ref{virial}), which directly relate the HFR charge radius and the HO parameter $b$:
$$ b_{\rm O} \in [1.7695,1.8012]  \rm fm$$ 
$$b_{\rm Zr} \in [2.1693,2.2211]  \rm fm$$


\subsubsection{Adjustment of pairing properties}\label{ss2}

The HFB calculations in finite nuclei routinely take into account the pairing correlations between protons and between neutrons which are essential. Since the proton-neutron pairing is much rarer in nuclei, very few HFB codes take it into account. In that context, the pairing properties of the Gogny interaction were originally controlled only in the  $(S=0,T=1)$ channel. The pairing associated with the channel $(S=1,T=1)$ has a much lower intensity that disadvantages it in nuclei.

Pairing properties are not directly linked to a specific observable. The most exhaustive way of characterizing the nuclear interaction
in this channel is to explore the set of two-body matrix elements coupled to S=0 and T=1. Originally, 
D. Gogny chose to select two two-body matrix elements
(TBME) implying the 1$s$ and 2$s$ states in the $^{32}$S nucleus,
providing with a new system composed of two new equations. In this channel, these TBME involve only the finite-range central part of the Gogny's D1-type interaction. 
Indeed, the parameter $x_0$ has been chosen equal to one in order to exclude the density-dependent term to participate 
to the pairing $(S=0,T=1)$ channel which is known to be very few re-normalized by medium effects. The pairing force is considered as behaving as the bare interaction
\cite{Sedrakian2003}. Thus,
the two new equations write as:
\begin{equation} \label{constraint2}
\begin{array}{lcl}
V_{1s} &\equiv& \displaystyle \! \mel*{1s 1s}{v^{\text{D1}}_{12}}{1s  1s}^{(S=0, T=1)} \\ 
 &=&  \displaystyle \sum_{i=1,2} f_{1s}^i (W_i-B_i-H_i+M_i)
\end{array}
\end{equation}
\begin{equation} \label{constraint2bis}
\begin{array}{lcl}
V_{2s} &\equiv& \displaystyle \! \mel*{2s 2s}{v^{\text{D1}}_{12}}{2s  2s}^{(S=0, T=1)} \\ 
& =& \displaystyle \sum_{i=1,2} f_{2s}^i (W_i-B_i-H_i+M_i).
\end{array}
\end{equation}
where the functions $f_{1s}^i$ et $f_{2s}^i$ have the following analytical expressions:
\begin{equation} \label{f1s2s}
\begin{array}{lcl}
\displaystyle f_{1s} &  =& G^{-3/2}
\end{array}
\end{equation}
\begin{equation} \label{f1s2sbis}
\begin{array}{lcl}
\displaystyle f_{2s} & = &G^{-3/2} ( \frac{41}{64} - \frac{79}{48}G^{-1} + \frac{385}{96} G^{-2} -  \frac{175}{48} G^{-3} \\ 
& + & \frac{105}{64} G^{-4} )
\end{array}
\end{equation}
with $G_{i}= 1+ 2b^{2}/\mu_{i}^{2}$. \\

\noindent The two TBMEs \eqref{constraint2}-\eqref{constraint2bis} are expressed as a linear new combination of the parameters. This new system of equations can be inverted to determine the values of the two linear combinations. 
One remarks that the TBMEs depend on the ranges $\mu_1$ and $\mu_2$ which are varied in the procedure.\\

TBMEs are not experimentally measurable quantities. In order their values on rely on experimental data or nuclei properties, four criteria were investigated: the absence of pairing correlation in magic nuclei, the moment of inertia in well-deformed nuclei, the spatial shape of the interaction in the singlet-even channel and the odd-even staggering in Sn isotopes.

\begin{figure}
\centering
\includegraphics[angle=-0,width=0.8\linewidth]{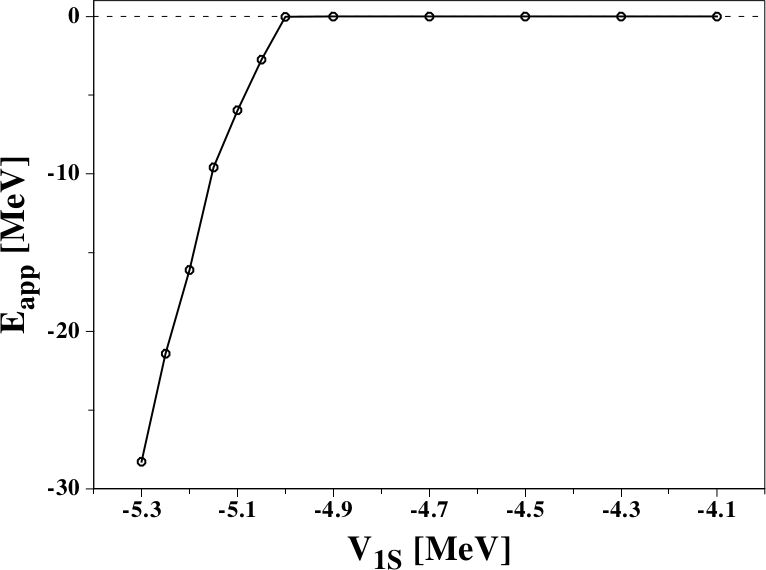}
\caption{Relation between the HFB pairing energy $\rm E_{app}$ and the value $V_{1s}$ of the interaction matrix element.}
\label{fig:Epair_O16}
\end{figure}

The absence of pairing correlations in magic nuclei is 
due to a large energy gap between the last occupied single-particle level and the first empty one, rendering nucleon pair scattering inoperative. Taking up the idea of the previous section, twelve parameterizations corresponding to different values of the pairing TBME have been produced, such that $\rm V_{1s,i=1,..,12}^{S=0,T=1} \in [-5.3,-4.1]$ MeV. For each of these parameterizations, an HFB calculation has been performed.
The values of the HFB pairing energy in $^{16}$O are shown in Fig.\ref{fig:Epair_O16} as a function of parameterization. The interactions with too negative values of $ V_{1s}$, typically $\rm V_{1s} \le -5.0~MeV$, generate pairing correlations in $^{16}$O with a transition between superfluid and normal phases very abrupt. The lower limit of $\rm V_{1s}$ has thus been estimated to be $\simeq -5.0 \rm~MeV$. 

\begin{figure}
\centering
\includegraphics[angle=-0,width=0.8\linewidth]{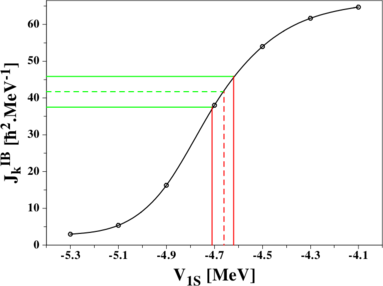}
\caption{Relation between the moment of inertia $\mathcal{J}_{exp}$ in the $^{158}$Sm and the value $\rm V_{1s}$ of the matrix element.}
\label{fig:inertie_Sm}
\end{figure}

In order to refine an interval for acceptable values of $V_{1s}$, a study dedicated to the dependence of the value of the moment of inertia on the pairing intensity has been achieved, using the twelve parameterizations generated previously. HFB calculations have been performed for the well-deformed $^{158}$Sm nucleus, with the moment of inertia calculated at the Inglis-Belyaev approximation. Moments of inertia
$\mathcal{J}_{exp}$ are not directly measurable quantities, but they are extracted from the rotational spectra of well-deformed nuclei:
$$E^{*}_{exp} \simeq \frac{J(J+1)}{2 \mathcal{J}_{exp}}$$
The energy of the first 2$^+$ excited state is equal to $E^{*}_{exp}= 72~ \rm keV$ from which $\mathcal{J}_{exp} \rm (^{158}Sm) \simeq 41.7~ \hbar^2.MeV^{-1}$. As shown in Fig. \ref{fig:inertie_Sm}, this value (horizontal green dashed line) corresponds to $V_{1s} \simeq -4.66 \rm  MeV$ (vertical red dashed line). To reproduce
the experimental value to within 10\% (horizontal green solid lines), an acceptable interval is such that $ V_{1s} \in [-4.71,-4.62] \rm~MeV$.

\begin{figure}
\centering
\includegraphics[angle=-0,width=0.8\linewidth]{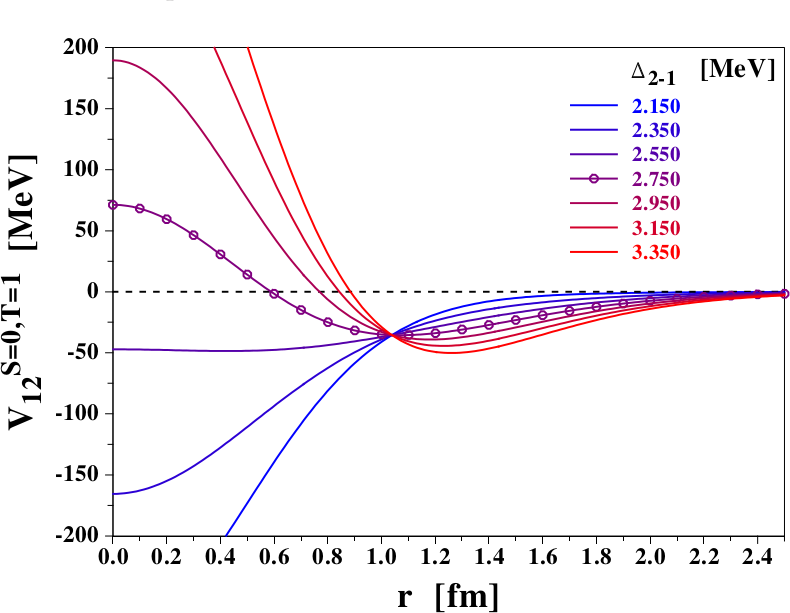}
\caption{Spatial shape of the interaction in the singlet-even channel (S=0,T=1) for different values of $\rm \Delta_{2-1}$.}
\label{fig:spatial}
\end{figure}

The value of the second matrix element $\rm V_{2s}$ has been deduced from the difference: $$\rm \Delta_{2-1}= V_{2s} - V_{1s}$$
To fix the value of $\Delta_{2-1}$, the spatial form of the pairing potential was studied.
Setting the value of $V_{1s}$ to -4.65 MeV,
several values of $\rm \Delta_{2-1}$ were tried between 2.150 and 3.350 MeV with a step of 200 keV, thus generating
seven different solutions $ (W_{1}-B_{1}-H_{1}+M_{1})_{i=1,...,7}$ and $ (W_{2}-B_{2}-H_{2}+M_{2})_{i=1,...,7}$ by solving the system of equations (\ref{constraint2})-(\ref{constraint2bis}). The seven corresponding interactions are fully defined in the 
singlet-even channel $(S=0,T=1)$ by:
\begin{equation}
\begin{array}{lcl}
v_{12,i}^{S=0,T=1}(r) &=& (W_{1}-B_{1}-H_{1}+M_{1})_{i}~ e^{-r^2/\mu_{1}^{2}} \\
 & + & (W_{2}-B_{2}-H_{2}+M_{2})_{i}~ e^{-r^2/\mu_{2}^{2}}
\end{array}
\end{equation}
where $r$ represents the inter-nucleon distance. The values of the ranges were imposed to $\rm \mu_{1} = 0.7~fm$ and $\rm \mu_{2} = 1.2~fm$.
The corresponding spatial shape of the nuclear potential is shown in Fig. \ref{fig:spatial} for the different values of $\Delta_{2-1}$.
Thus, the value of $\Delta_{2-1}$ has been selected on two criteria, the spatial shape of the potential (Van-Der-Waals type) and its value at zero distance. Indeed,
to have a good description of the pairing properties in nuclei, the potential must be repulsive (attractive) at short (long) distance. A value around $\rm \Delta_{2-1} \simeq 2.750$ MeV allows such a description. Thus, the following values were adopted for the two pairing matrix elements:
\begin{equation}
V_{1s} \simeq -4.65 \rm MeV
\, \, \text{and} \, \, 
V_{2s} \simeq -1.90 \rm MeV
\end{equation}
To finally validate the pairing properties, the odd-mass staggering has been calculated in the Sn isotopic chain, as described in section \ref{paingfiss}.

\subsubsection{Adjustment of the symmetry energy}\label{ss3} 

To fully specify the parameterization, two additional equations are needed. 
The symmetry energy properties are exploited to constrain the Gogny interaction to correctly describe their isospin-dependent observables. 
Among possible observables, the total binding energy is one of them. Its dependence on
isospin $T_{z} = \frac{N-Z}{2}$ is explicit in the Bethe and Weizsäker's semi-empirical mass formula:
\begin{equation}\label{beth}
\mathcal{E}(N,Z) = a_{v} A + a_{s} A^{2/3} + a_{c} \frac{Z^2}{A^{1/3}} + a_{\tau} \frac{4 T^{2}_{z}}{A} + ...
\end{equation}
In particular, Eq.\eqref{beth} shows that the good reproduction of the experimental binding energies requires the accurate determination of the symmetry coefficient $\rm a_{\tau}$. The best mass formula fits are obtained 
for $\rm a_{\tau}$ values between 
28 and 32 MeV. 
By analogy with the empirical formulas, the symmetry energy $\mathcal{E}_{\text{sym}}$ can be defined in infinite nuclear matter as:
\begin{equation}
\mathcal{E}_{\text{sym}} \equiv \frac{1}{2} \pdv[2]{\mathcal{E}/A}{\beta}\Big\vert_{\beta=0}.
\end{equation}
where  $\beta= (\rho_{\nu}-\rho_{\pi})/\rho$ is the asymmetry parameter of the medium, equal to zero for symmetric matter and one for neutron one.

\noindent However, D. Gogny didn't work directly 
with the symmetry energy to adjust the isospin 
properties of its force and kept this quantity as a nuclear matter filter (see section \ref{sec3}). He preferred to introduce the difference in energies $\Delta \varepsilon$ between the neutron $\varepsilon^{\nu}_{2s1/2}$ and 
proton $\varepsilon^{\pi}_{2s1/2}$ states in $^{48}$Ca.
In the HFR approximation, this difference is written as $\Delta \varepsilon$:
\begin{equation} \label{constraint3}
\begin{array}{lcl}
\displaystyle \Delta \varepsilon &=& \displaystyle \sum_{i=1,2} \big[ f_{{D}}(\mu_i) ~(2H_i + M_i) \\
 & +& \displaystyle  f_{{E}}(\mu_i) ~(W_i + 2B_i) \big] + g(t_0, x_0,\alpha)
\end{array}
\end{equation}
where the quantities $f_{{D}} (\mu_i)$ and $ f_{{E}} (\mu_i)$ depend on the ranges and have for expressions:
\begin{subequations} \label{quantitiesDeltaeps}
\begin{align}
f_{{D}} \equiv - \frac{4}{7} \widetilde{S}_{10 \, 03}^{{C}} \vert_{{D}} \\
f_{{E}} \equiv + \frac{4}{7} \widetilde{S}_{10 \, 03}^{{C}} \vert_{{E}}
\end{align}
\end{subequations}
The term $g(t_0, x_0,\alpha)$ in Eq.(\ref{constraint3}) corresponds to the contribution of the zero-range density-dependent term:
\begin{equation}
\begin{array}{lcl}
g(t_0, x_0,\alpha) &=& t_0 \left( x_0 + \frac{1}{2} \right) \int d^3r |\phi_{100} (\vec{r})|^2 \times \\
& & \rho^{\alpha}(\vec{r}) \left[ \rho_{\pi} (\vec{r}) - \rho_{\nu} (\vec{r}) \right]
\end{array}
\end{equation}
where the HO wave function $\phi_{100} (\vec{r})$ is defined by Eq.\eqref{swavef}.
A correction due to the coulomb term has also been taken into account for proton single particle energies. 

\begin{figure}
\centering
\includegraphics[angle=-0,width=0.8 \linewidth]{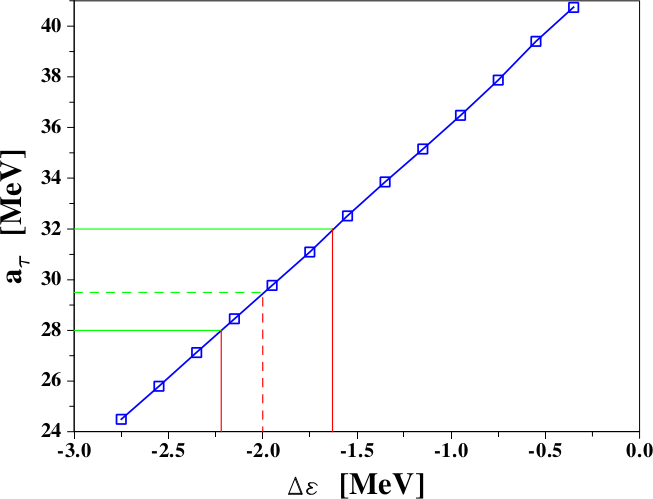}
\caption{Relation between the symmetry energy a$_{\tau}$ and the difference $\Delta \varepsilon$ defined between the $\rm 2s_{1/2}$ neutron and proton states in $^{48}$Ca \cite{Chappert2007}.}
\label{fig:a_tau}
\end{figure}

\noindent To specify the interval of variation of $\Delta \varepsilon$, the method is similar to that developed in subsections \ref{ss1} and \ref{ss2}.
As an example, thirteen parameterizations have been generated giving rise to thirteen parameter sets. 
The linearity relation obtained between the symmetry energy coefficient $\rm a_{\tau}$ and the difference $\Delta \varepsilon$ is shown in Fig. \ref{fig:a_tau}.
It makes it easy to establish an interval for values of $\Delta \varepsilon$ compatible with experimental data:
$$\Delta \varepsilon \in [-2.2, -1.6] \rm ~MeV$$

\begin{figure*}
\centering
\includegraphics[angle=-0,width=0.45\linewidth]{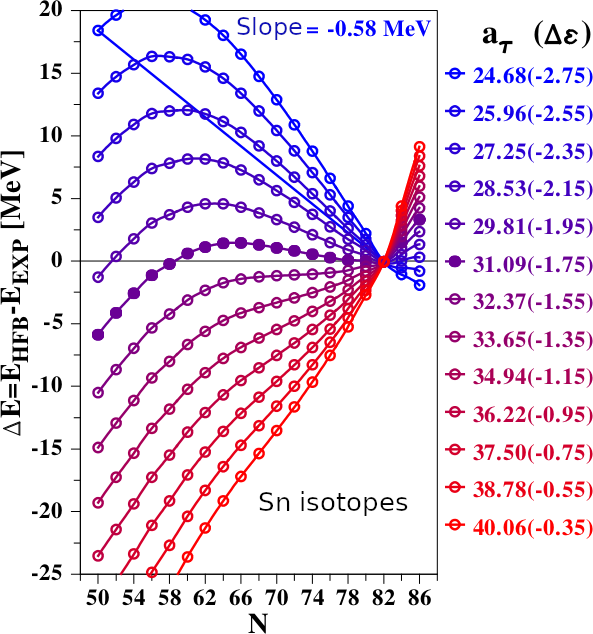} \hspace{1cm}
\includegraphics[angle=-0,width=0.322\linewidth]{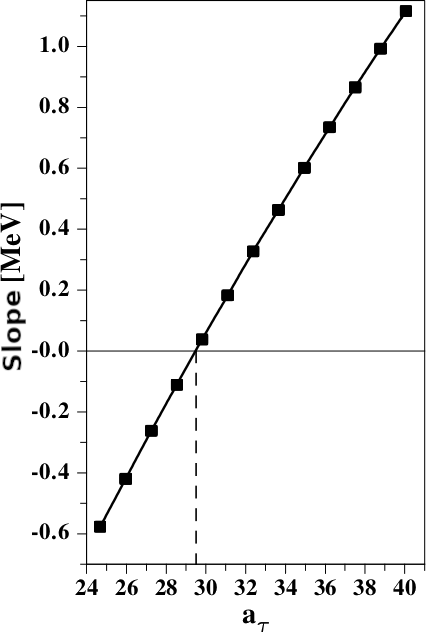}
\caption{Left: Evolution of the binding energies drift $\Delta E$ in the Sn isotopic chain for various values of a$_{\tau}$. Right: Slope between $^{100}$Sn and $^{132}$Sn as a function of a$_{\tau}$ \cite{Chappert2007}.}
\label{fig:a_tauSn}
\end{figure*}

\noindent In Fig. \ref{fig:a_tauSn}, left panel, the impact of the value of a$_{\tau}$ on the binding energies drift $\Delta E$ is shown in the Sn isotopic chain. The results (in MeV) have been obtained by calculating the difference $\Delta E = E_{HFB}-E_{exp.}$ between HFB and experimental energies (in MeV).
The right panel shows the slope between $^{100}$Sn and $^{132}$Sn as a function of a$_{\tau}$. For the fifth interaction, the slope is found to be very close to zero.
It corresponds to a symmetry energy $\rm a_{\tau} \simeq 29.81 ~MeV$. This study in Sn isotopes clarifies the choice of the value of $\rm \Delta \epsilon$ which must be around $\rm \simeq -2.0 ~MeV$.
A seventh equation can be established to determine a new combination of parameters based on the other three combinations. For example: 
\begin{equation}
\begin{array}{lcl}
\displaystyle 2 H_1 +M_1 &=& \displaystyle \frac{1}{f_{{D}} (\mu_1)} [ \Delta \varepsilon -  f_{{D}} (\mu_2) (2 H_2 +M_2) \\
& &\displaystyle  - g(t_0, x_0,\alpha) 
  - f_{{E}} (\mu_1) (2 B_1 +W_1)  \\
&  & \displaystyle  - f_{{E}} (\mu_2) (2 B_2 +W_2) ]
\end{array}
\end{equation}

\noindent Finally, the eighth equation is chosen so as to assign an arbitrary 
value to $B_1$:
\begin{equation}
B_1 = B_c
\end{equation}
For a greater flexibility, $B_c$ is varied in the fitting process.

\subsubsection{Extension of the adjustment procedure for the D2 Gogny force}\label{D2}

When going from D1 to D2 Gogny interactions (see Eq.(\ref{gognyD2})), 
the parameters $\{t_{0},x_{0} \}$ of the zero range density-dependent term are 
replaced by the five parameters $\{W_{3},B_{3},H_{3},M_{3},\mu_{3} \}$ associated with the finite range one. The extension of the fitting procedure in the HFR approach therefore consists of:
\begin{itemize}[label=$-$]
\item the extension of the procedure itself for determining the new parameters
\item the generalization of the expressions presented in subsections \ref{ss1}, \ref{ss2}, \ref{ss3} to a finite range density-dependent term.
\end{itemize}

\paragraph{The new parameter determination procedure for D2}

With D2, the five new parameters $\rm \{ W_3, B_3, H_3, M_3, \mu_3 \}$ subsume the 
two parameters $\rm \{t_0, x_0\}$ of the density-dependent term of D1-type interactions.
The parameter $\alpha$ corresponding to the power of the density is retained, with a value of 1/3.
In the D2 adjustment procedure \cite{Chappert2007}, the range $\mu_3$  is left free. As with the other ranges of the central term, a sufficiently large interval of values is allowed. 
To determine the other four parameters 
$\rm \{ W_3, B_3, H_3, M_3\}$, a first extension of the fitting code was attempted by adding four additional constraints to fully specify D2. 
F. Chappert et al. initially chose to reproduce the binding energy and charge radius of another double magic nucleus, $\isotope[100]{Sn}$, as well as two 
points of the neutron equation of state at low energy. 
However, the inversion system shown instabilities for certain values of $\mu_3$. Thus, the abandonment of this protocol was decided. 
Instead, the four parameters $\rm \{ W_3, B_3, H_3, M_3 \}$ were left free and variation intervals were defined for each parameter, in the manner of
$\mu_3$. The neutron matter equation was relegated to the role of filter, as it will be discussed in section \ref{nmes}. 
No additional constraints was added to the D2 adjustment procedure.

\paragraph{HFR equations with D2}

With D2, new expressions are obtained for the binding energies, the charge radii, the pairing matrix elements and the symmetry energy. 
Their derivations can be found in both Appendix \ref{appenb} and Refs. \cite{Chappert2007,Zietek2023}. \\

\noindent \textit{Binding energies}\label{it1} \\

\noindent The binding energies contain four contributions: the 
kinetic energy, the two potential energy contributions due to the two ranges of the central term, and the potential energy contribution due to the finite range density-dependent term. They are given by the following formula:
\begin{equation}\label{beD2}
\mathcal{E}^{\text{HFR}} [X] =  \mathcal{E}^{\text{HFR}}_{K} [X]+ \mathcal{E}^{\text{HFR}}_{P \mu_{1}} [X] + \mathcal{E}^{\text{HFR}}_{P \mu_{2}} [X] + \mathcal{E}^{\text{HFR}}_{\mu_{3}} [X]
\end{equation}
where
\begin{equation}\label{noyasymm}
\begin{array}{lcl}
\mathcal{E}^{\text{HFR}}_{P \mu_{i}}[X] &=&  F_{i}^{{D}}[X] (4W_{i} + 2B_{i} - 2H_{i} - M_{i}) \\
& +& F_{i}^{{E}}[X] (4M_{i} + 2H_{i} - 2B_{i} - W_{i})
\end{array}
\end{equation}
\begin{equation}
\begin{array}{lcl}
\mathcal{E}^{\text{HFR}}_{P \mu_{3}}[X] &=&  G^{{D}}[X] (4W_3 + 2B_3 - 2H_3 - M_3) \\
&+& G^{{E}}[X] (4M_3 + 2H_3 - 2B_3 - W_3) 
\end{array}
\end{equation}
\noindent where $i=1,2$. The functions $F_{i}^{{D}}[X]$ and $F_{i}^{{E}}[X]$ are defined by Eqs.(\ref{Fdirect}) and (\ref{Fexchange}), and  
the functions $G^{{D}}[X]$ and $G^{{E}}[X]$ related to the finite range density-dependent term have for expression:
\begin{equation}\label{noyasym1}
\rm G^{{D}}[X] \equiv 2  \sum_{(a)_{\pi}} \sum_{(b)_{\nu}} \widetilde{S}_{(a) (b)}^{{DD}} \vert_{{D}}
\end{equation}
\begin{equation}\label{noyasym2}
G^{{E}}[X] \equiv 2  \sum_{(a)_{\pi}} \sum_{(b)_{\nu}} \widetilde{S}_{(a) (b)}^{{DD}} \vert_{{E}}
\end{equation}
with
\begin{equation}\label{genD22}
\begin{array}{lcl}
\displaystyle \widetilde{S}^{{DD}}_{(a) (b)} \vert_{{D}} &\equiv& \displaystyle \frac{(2l_a+1)(2l_b+1)}{4 \pi}  \sum_{n_\mu n_\nu} T_{(a) (a)}^{(n_\mu 0)} \\
&\times&  T_{(b) (b)}^{(n_\nu 0)} \frac{1}{2} \big[ J_{n_\mu n_\nu 0} + J_{n_\nu n_\mu 0} \big]
\end{array}
\end{equation}
\begin{equation}\label{genD23}
\begin{array}{lcl}
\displaystyle \widetilde{S}^{{DD}}_{(a) (b)} \vert_{{E}} & \equiv& \displaystyle \frac{(2l_a+1)(2l_b+1)}{4 \pi} \sum_{n_\mu  n_\nu l_\mu} (2l_\mu+1) \\
& \times & \displaystyle  \begin{pmatrix} l_a & l_b & l_\mu \\ 0 & 0 & 0 \end{pmatrix}^2 T_{(a) (b)}^{(n_\mu l_\mu)} T_{(b) (a)}^{(n_\nu l_\mu)} \\
& \times & \displaystyle  \frac{1}{2} \big[ J_{n_\mu n_\nu l_\mu} + J_{n_\nu n_\nu l_\mu} \big]
\end{array}
\end{equation}
and
\begin{equation} \label{intcentralJ}
\begin{array}{lcl}
\displaystyle J_{n n' l} &=& K \lambda_{(n' l)} \int_0^{\infty} \dd{r} r^2 \phi_{(00)}(r) \phi_{(nl)}(r)  \rho^{\alpha}(r)  \\
&\times& \displaystyle \phi_{(00)}(r, b\sqrt{g}) \phi_{(n'l)}(r, b\sqrt{g})
\end{array}
\end{equation}
The quantities $K$, $g$ and $\lambda$ are defined by: 
$$\displaystyle  K \equiv \pi [b \sqrt{g-1}]^{3}/2$$ 
$$ \displaystyle  g \equiv 1+\mu^2/b^2$$ 
$$ \displaystyle \lambda_{(r_\nu)} \equiv g^{-n_\nu - l_\nu/2}$$.

\noindent The kinetic energy $\mathcal{E}^{\text{HFR}}_{K}[X]$ appearing in Eq.
(\ref{beD2}) is defined for $^{16}$O and $^{90}$Zr by Eq.(\ref{cin}). \\

\noindent \textit{Charge radii}\label{it2} \\

\noindent Determined so as to minimize the energy with respect to the HO parameter $b$, the constraints on charge radii lead to the following new expressions, which differ from Eq.(\ref{rayonc}) by the contribution of the density-dependent term:
\begin{equation}\label{radD2}
\begin{array}{lll}
\displaystyle \mathcal{E}^{\prime}_{\text{HFR}}[X]& \equiv& \displaystyle \mathcal{E}_{{K}}^{\prime \text{HFR}}[X] + \mathcal{E}_{{P}}^{\prime \text{HFR}}[X] \\
 & =& \displaystyle \mathcal{E}_{{K}}^{\prime \, \text{HFR}}[X] \\
 &+& \displaystyle \sum_{i=1,2} F_i^{\prime {D}}[X] (4W_i + 2B_i - 2H_i - M_i) \\
 &+& \displaystyle \sum_{i=1,2} F_i^{\prime {E}}[X] (4M_i + 2H_i - 2B_i - W_i) \\ 
 &+&  G^{\prime {D}}[X] (4W_3 + 2B_3 - 2H_3 - M_3) \\
 &+& G^{\prime {E}}[X] (4M_3 + 2H_3 - 2B_3 - W_3) \\
 &=& 0 
\end{array}
\end{equation}
where $G^{\prime {D}}[X]$ and $G^{\prime {E}}[X]$ represent the 
derivatives of the functions $ G^{{D}}[X]$ and $G^{{E}}[X]$ 
defined by Eqs. (\ref{noyasym1}) and (\ref{noyasym2}), respectively. \\

\noindent \textit{Pairing matrix elements}\label{it3}\\

\noindent The analytical expressions of the two pairing matrix elements 
coupled to $(S=0, T=1)$ contain an additional term associated with the finite-range density term in comparison with Eqs.(\ref{constraint2})-(\ref{constraint2bis}):
\begin{equation}\label{constraint2D2}
\begin{array}{lcl}
\displaystyle V_{1s} 
 &=& \displaystyle \sum_{i=1,2} f_{1s}^i (W_i-B_i-H_i+M_i) \\
 &+&  g_{1s} (\mu_3,\alpha) (W_3-B_3-H_3+M_3)
\end{array}
\end{equation}
\begin{equation}
\begin{array}{lcl}
V_{2s} 
&=& \displaystyle  \sum_{i=1,2} f_{2s}^i (W_i-B_i-H_i+M_i) \\
&+&  g_{2s} (\mu_3,\alpha) (W_3-B_3-H_3+M_3)
\end{array}
\end{equation}

\noindent The functions $f_{1s}^i$ and $f_{2s}^i$ are defined by Eqs.(\ref{f1s2s}) and (\ref{f1s2sbis}). The functions $g_{1s} (\mu_3,\alpha)$ and 
$g_{2s} (\mu_3,\alpha)$ have for expressions:
\begin{equation}
\begin{array}{lcl}
g_{1s} (\mu_3,\alpha) &\equiv& \displaystyle \langle 1s ~ 1s\vert e^{-(\vec{r}_1-\vec{r}_2)^2/\mu^2_3} \times \\
 & & \displaystyle \frac{1}{2} \big(\rho^{\alpha}(\vec{r_1}) - 
\rho^{\alpha}(\vec{r_2}) \big) \vert 1s ~ 1s\rangle
\end{array}
\end{equation}
\begin{equation}
\begin{array}{lcl}
g_{2s} (\mu_3,\alpha) &\equiv& \displaystyle \langle 2s ~ 2s\vert e^{-(\vec{r}_1-\vec{r}_2)^2/\mu^2_3} \times \\ 
 & & \displaystyle \frac{1}{2} \big(\rho^{\alpha}(\vec{r_1}) - 
\rho^{\alpha}(\vec{r_2}) \big) \vert 2s ~ 2s\rangle
\end{array}
\end{equation}

\noindent where the matrix elements are evaluated for particular values of the spherical HO wave functions. Noting 
${G^{(3)} =} e^{-(\vec{r}_1-\vec{r}_2)^2/\mu^2} ~ \frac{1}{2} \big(\rho^{\alpha}(\vec{r_1}) - 
\rho^{\alpha}(\vec{r_2}) \big)$ and $r_a \equiv \{ n_a l_a m_a \}$, these matrix elements can be written as:
\begin{multline}
\displaystyle \langle r_a ~ r_b\vert G^{(3)} \vert r_c ~ r_d \rangle = \displaystyle \sum_{r_{\mu},r_{\nu}} T^{(r_{\mu})}_{(r_a)(r_c)} T^{(r_{\nu})}_{(r_b)(r_d)} \times \\
  \displaystyle \langle l_a m_a\vert Y^{m_{\mu}*}_{l_{\mu}} \vert l_c m_c \rangle  \langle l_b m_b\vert Y^{m_{\nu}*}_{l_{\nu}} \vert l_d m_d \rangle \langle 00 \vert G^{(3)} \vert r_{\mu} ~ r_{\nu} \rangle
\end{multline}
with
\begin{equation}
\displaystyle \langle 00 \vert G^{(3)} \vert r_{\mu} ~ r_{\nu} \rangle = J_{r_{\mu} r_{\nu}} + J_{r_{\nu}  r_{\mu}}
\end{equation}
where
\begin{equation}
\begin{array}{lcl}
\displaystyle J_{r_{\mu} r_{\nu}} &=& \frac{1}{2} K^{3/2} \lambda_{(r_{\nu})} \int d^3 r_1 \phi_{0} (\vec{r_1})  \phi_{r_{\mu}} (\vec{r_1}) 
\rho^{\alpha} (\vec{r_1}) \\
& \times& \displaystyle e^{- \frac{1}{2} \frac{r_1^2}{b^2 g}} \phi_{r_{\nu}} (\vec{r_1}, b \sqrt{g}) 
\end{array}
\end{equation}
The quantities $g$, $K$ and $\lambda$ are defined by $g= 1 + \frac{\mu_3^2}{b^2}$, $K= \sqrt{\pi} b \frac{g-1}{\sqrt{g}}$ et $\lambda_{(r_{\nu})} = g^{-n_{\nu}-l_{\nu}/2}$. The integrals $J_{r_{\mu} r_{\nu}}$ are evaluated numerically. \\

\noindent \textit{Symmetry energy}\label{it4} \\

\noindent As discussed in section \ref{ss3}, 
the symmetry energy is not used directly in the 
fitting code but the difference $\rm \Delta \epsilon$ (see Eq.\eqref{constraint3})
between the energy of the 2s$_{1/2}$ proton and 
neutron levels levels in $^{48}$Ca. 
With the D2 interaction, the analytical expression (\ref{constraint3}) of 
$\rm \Delta \epsilon$ is modified 
in the following way due to the finite-range density-dependent term:
\begin{equation}\label{constraint3D2}
\begin{array}{lcl}
\Delta \varepsilon &=& \displaystyle \sum_{i=1,2} f_{{D}}(\mu_i) (2H_i + M_i) + f_{{E}} (\mu_i) (W_i + 2B_i) \\
&+& \displaystyle g_{{D}}(\mu_3) (2H_3 + M_3) + g_{{E}} (\mu_3) (W_3 + 2B_3)
\end{array}
\end{equation}
where the functions $f_{{D}}(\mu_i)$ and $f_{{E}}(\mu_i)$ are given by Eqs.(\ref{quantitiesDeltaeps}).
The quantities $g_{\text{D}}(\mu_3)$ and $\rm g_{\text{E}}(\mu_3)$ have for expressions:
\begin{equation} \label{quantitiesDeltaepsDD}
g_{{D}} \equiv - \frac{4}{7} \widetilde{S}_{10 \, 03}^{{DD}} \vert_{{D}} {\rm ~~and~~}
g_{{E}} \equiv + \frac{4}{7} \widetilde{S}_{10 \, 03}^{{DD}} \vert_{{E}}
\end{equation}
where $\widetilde{S}_{10 \, 03}^{{DD}} \vert_{{D}}$ and 
$\widetilde{S}_{10 \, 03}^{{DD}} \vert_{{E}}$ are defined by Eqs. (\ref{genD22}), (\ref{genD23}) and (\ref{intcentralJ}). 

\subsubsection{Extension of the adjustment procedure for DG Gogny force}\label{DG}

Going back to the analytical form (\ref{gognyDG}) 
of DG, the parameter $\{ W_{ls}\}$ 
characterizing the spin-orbit term 
appearing in D2 has been replaced by the three new parameters 
$\{ W_{4},H_{4},\mu_{4} \}$ due to the finite range character. 
The $-4/ \mu^5_4 \sqrt{\pi}^3$ coefficient 
which depends on the range $\mu_{4}$ allows to recover the zero-range spin-orbit term of D1 and D2 in the limit $\mu_{4} \rightarrow 0$.
Similarly, the tensor term is characterized by three parameters $\{ W_{5},H_{5},\mu_{5} \}$.
The extension of the fitting procedure to DG in the HFR approach 
therefore consists of:
\begin{itemize}[label=$-$]
\item the extension of the procedure itself to determine the new parameters
\item the generalization of the expressions set out in subsections \ref{it1} to finite range spin-orbit and tensor terms.
\end{itemize}

\paragraph{The new parameter determination procedure}

The starting point is the D2 adjustment procedure.
Several changes associated with the finite range 
spin-orbit and tensor terms have been included in 
order to extract parameterizations having a DG analytical form.
The expressions associated with binding energies and charge radii 
are unchanged in the HFR approximation. 
The same remark applies for the expressions of the two pairing matrix elements
coupled to $(S=0,T=1)$ implying the 0s$_{1/2}$ and 1s$_{1/2}$ states in the 
$^{32}$S nucleus. Indeed, the spin-orbit and tensor terms act only between 
two nucleons coupled to S=1. As for the difference $\Delta \epsilon$, 
the tensor contribution is zero. With regard to the spin-orbit term, its contribution 
has been neglected, as it was already the case for the D1 and D2 
interactions. The numerical value of $\Delta \epsilon$ has been modified to avoid 
any drift in the binding energies of Sn isotopes. \\

\begin{figure}\label{USDGog}
\centering
\includegraphics[width=0.95\linewidth]{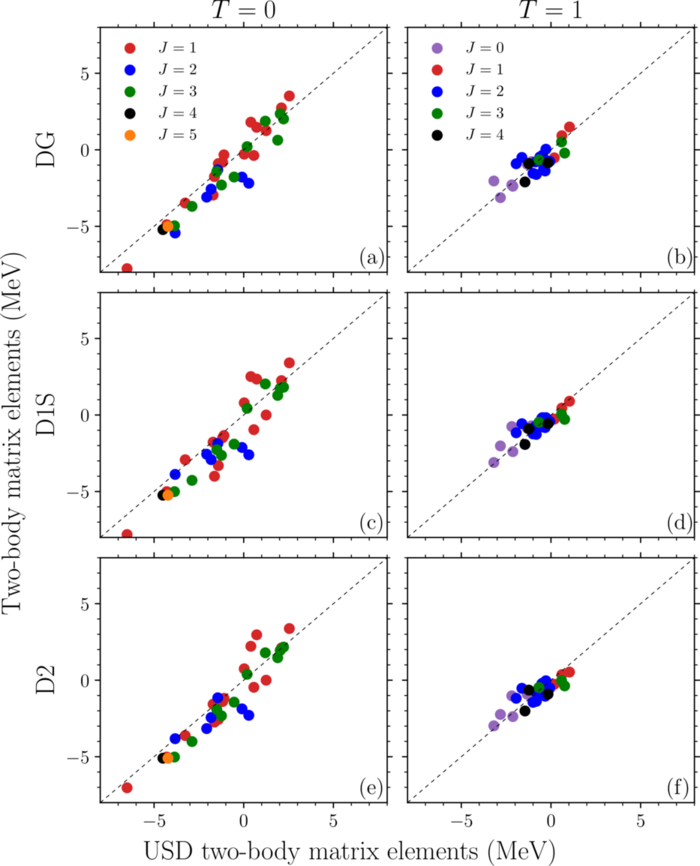}
\caption{Comparison of T=0 and T=1 TBMEs between the USD interaction and DG, D1S and D2 Gogny ones, for the various values of J.
Calculations have been done for $^{18}$O.}
\end{figure}

Apart from the two ranges, the new spin-orbit and 
tensor terms bring four new parameters, $\{W_4,H_4,W_5,H_5 \}$.
In order to determine them, the idea of constraining more general matrix elements than the
pairing ones has been adopted, in order to characterize the residual part of the interaction
involving contributions from the spin-orbit and tensor terms. 
Among these new matrix elements, some associated with proton-neutron pairing coupled to 
$(J=1, T=0)$ have been chosen. As a guide to control the residual part of the 
interaction, one has used the values provided by the shell model USD and GPF interactions, 
allowing a variation of 10\% of their original values. This idea was based on the 
empirical remark associated with the similarity between T=1 TBMEs of the USD and D1S Gogny interactions calculated with spherical HO wave functions. 
This is exemplified in Fig.\ref{USDGog} in the case of $^{18}$O for DG (top, panels (a) and (b)), D1S (middle, panels (c) and (d)) and D2 (bottom, panels (e) and (f)). One sees that the dispersion is found larger for the T=0 channel (left) in comparison with the T=1 one (right).

In addition to the constraints discussed above, three additional filters associated 
with the spin-orbit and tensor terms have been added in order to select relevant parameterizations from the tens of thousands allowed by the standard filters of 
the D2 procedure. The first one concerns the intensity of the 
spin-orbit term. One has filtered the interactions so that, in the T=1 channel,
the combination $\rm W_4-H_4$ takes values similar to $\rm W_{ls}$ in the limit $\mu_4 \rightarrow 0$. Indeed, the following correspondence can be established:
\begin{equation}
\lim_{\mu \rightarrow 0} B(\mu) G(r_{12}, \mu) = \delta(\vec{r}_1 - \vec{r}_2)
\end{equation}
with $G(r_{12}, \mu) = -\frac{\mu^2}{4} e^{-(\vec{r}_1 - \vec{r}_2)^2/\mu^2}$ a Gaussian potential. Thus: 
\begin{equation}
B(\mu) = - \frac{4}{\mu^2} \frac{1}{(\mu \sqrt{\pi})^3}
\end{equation}
\noindent Quantitatively, $(W_4-H_4) \in [110,150]~ \rm MeV~fm^5$ has been authorized.

The second filter concerns the value of $H_4$. Previous studies in the context of extended
Skyrme interactions have shown that it drives the value of the isotopic charge radius shift in 
Pb isotopes \cite{Sharma1995}. These studies have made it possible to define an order of magnitude 
for the isospin-dependent part of the spin-orbit force, without giving any indication of its sign. Given the difference in the analytical form of the Skyrme and Gogny interactions, 
$H_4$ has been allowed to vary in the range $H_4 \in [-50,50]~ \rm MeV~fm^5$.

The third filter has focused on the intensity of the tensor term. Here, a 
previous study which added a perturbative tensor to D1S in order to reproduce the 
spin-orbit splittings in Ca and Ni \cite{Grasso2013} has set $(W_5-H_5) = 225~ \rm MeV$ with $H_5=-180 \rm MeV$. In the DG case, the interaction has been completely readjusted and 
ranges of values that was not too restricted was adopted: 
$(W_5-H_5) = \in [-275,-175] ~ \rm MeV$ and $H_5 = \in [-230,-130] ~ \rm MeV$. 
In particular, as previously discussed for the Gogny interaction, room is left for an 
explicit treatment of the nuclear long range correlations. In that context, the experimental spin-orbit splitting is not desired to be reproduced precisely at the HF or HFB approximation in contrast with Ref.\cite{Grasso2013}. 
Indeed, the particle-vibration coupling is known to shrink the energy gap. To check the relevance of the gap values, first excited states have been calculated in relevant nuclei. \\

In order to validate the parametrization, a final 
step has consisted in performing HFB and MPMH \cite{Pillet2008,LeBloas2014,Robin2016,Robin2017}
calculations which enabled to calculate specific 
observables in finite nuclei, as already evoked
in section \ref{DGinter}:
\begin{itemize}[label=$-$]
\item the binding energies of $\isotope[16]{O}, \isotope[90]{Zr}$ and $\isotope[208]{Pb}$
\item the pairing energy in double magic nuclei
\item the single-particle states in $\isotope[40]{Ca}, \isotope[48]{Ca}$ and $\isotope[56]{Ni}$ isotopes;
\item the kink in Pb isotopes charge radii
\item the absence of energy drift in neutron-rich nuclei (Sn isotopic chain)
\item the energies of the first excited states in the $sd$-shell even-even and odd nuclei.
\end{itemize}

\paragraph{HFR equations with DG}

The new expressions for binding energies, charge radii, 
pairing matrix elements and symmetry energy are presented without demonstration. \\

\noindent \textit{Binding energies, charge radii 
and pairing matrix elements}\label{it5}

\noindent No change is made to the expressions defined by Eqs.
(\ref{beD2}), (\ref{radD2}) and 
(\ref{constraint2D2}) developed for D2, 
concerning total binding energies, charge radii and pairing matrix elements, respectively. \\

\noindent \textit{More general two-body matrix elements}\label{it8} \\

\noindent In the following, 
$\expval*{V^{{SM}}}^{(J, T)}$ designates 
a shell model TBME coupled
to a total angular momentum J and a total isospin T.
Besides, the contribution of each
term of DG to the total TBME will be noted by $\! \expval*{V_a^{{C}}}^{(J, T)}$, $\! \expval*{V_b^{{C}}}^{(J, T)}$, $\! \expval*{V_a^{{DD}}}^{(J, T)}$,  $\!\expval*{V_b^{{DD}}}^{(J, T)}$, $\!\expval*{V^{{SO}}}^{(J, T)} \!$ and $\! \expval*{V^{{T}}}^{(J, T)}$, respectively, for the central, density-dependent, spin-orbit and tensor terms. 
Their expressions can be found in Appendix \ref{anexa}. Here,
the indices $a$ and $b$ are associated to two different combinations of parameters.
The new set of equations is created in the following way by defining the quantity $E$ which
correspond to the sum of the spin-orbit and tensor contributions:

\begin{multline}\label{Eequation}
E =  \! \expval*{V^{{SM}}}^{(J, T)} \\  - \sum_{i=1,2} \Big[ \big(W_i - B_i + (-)^T H_i - (-)^T M_i \big) \times \\ \expval*{V_{a,i}^{{C}}}^{(J, T)}\! + \big(B_i  - (-)^T M_i \big) \! \expval*{V_{b,i}^{{C}}}^{(J, T)}\! \Big] \\
- \Big[ \big(W_3 - B_3 + (-)^T H_3 - (-)^T M_3 \big) \expval*{V_{a}^{{DD}}}^{(J, T)}
\\ + \big(B_3  + (-)^T M_3 \big) \! \expval*{V_{b}^{{DD}}}^{(J, T)}\! \Big] \\
= \big(W_5 + (-)^T H_5 \big)  \expval*{V^{{SO}}}^{(J, T)} \\ + \big(W_7 + (-)^T H_7\big) \! \expval*{V^{{T}}}^{(J, T)}
\end{multline}

\noindent Since the parameters associated with the central terms are extracted from the previous systems and the ones of the density-dependent interaction are left free, four equations are needed, two written for the component $T=0$ and two for the other component $T=1$, to unequivocally determine the parameters of the tensor and spin-orbit interactions. The parameters have been obtained from constraints on both $T=0$ proton-neutron pairing and $T=1$ particle-like TBMEs. The four equations, written $E_1, E_2, E_3$ and $E_4$ in the following, that shape the third system of the fitting code of the DG Gogny interaction, read:
\begin{eqnarray} \label{3rdsystem}
E_1 &= \big(W_5 + H_5 \big) \! \expval*{V^{\text{SO}}}^{(J_1=1, T=0)}_{sd}\! \\ \nonumber
    &+ \big(W_7 + H_7\big) \! \expval*{V^{\text{T}}}^{(J_1=1, T=0)}_{sd} \\
E_2 &= \big(W_5 - H_5 \big) \! \expval*{V^{\text{SO}}}^{(J_2=2, T=1)}_{sd}\! \\ \nonumber
    &+ \big(W_7 - H_7\big) \! \expval*{V^{\text{T}}}^{(J_2=2, T=1)}_{sd} \\
E_3 &= \big(W_5 + H_5 \big) \! \expval*{V^{\text{SO}}}^{(J_3=1, T=0)}_{pf}\! \\ \nonumber
    &+ \big(W_7 + H_7\big) \! \expval*{V^{\text{T}}}^{(J_3=1, T=0)}_{pf} \\
E_4 &= \big(W_5 - H_5 \big) \! \expval*{V^{\text{SO}}}^{(J_4=2, T=1)}_{pf}\! \\ \nonumber
    &+ \big(W_7 - H_7\big) \! \expval*{V^{\text{T}}}^{(J_4=2, T=1)}_{pf} 
\end{eqnarray}
\noindent where the quantities $E_i$, with $i \in \{1,2,3,4\}$, can be deduced from Eq.(\ref{Eequation}). In practice, a set of four shell-model TBMEs (two in the $sd$-shell and two in the $pf$-shell) satisfying the above-mentioned requirements is taken and all the set of parameters reproducing them in the range $\!\expval*{V^{{SM}}}^{(J, T)}\! \pm 0.1 \!\expval*{V^{{SM}}}^{(J, T)}\!$ are kept. \\

\noindent \textit{Symmetry energy}\label{it9} \\

\noindent As discussed previously, the symmetry energy is not used as 
a constraint in the fitting code. It has been preferred to use the 
difference $\Delta \epsilon$ between proton and neutron 2s$_{1/2}$ 
single-particle states in $^{48}$Ca.
It can be shown that the tensor interaction does not contribute to
$\Delta \epsilon$ at the HFR approximation.
For the spin-orbit term, the situation is different as the direct 
contribution to the mean-field is not zero. However, as already 
mentioned for the D1-type and D2-type interactions, this contribution
is neglected in the fitting procedure. Indeed, no spin-orbit term
is introduced in the HFR fields.
So, the expression of $\Delta \epsilon$ that has been used in the fitting procedure of DG 
is exactly the same as the one obtained for D2 
(see Eq.(\ref{constraint3D2})).

\subsection{Parameter values of various Gogny parameterizations}

To end this section, we summarize the parameter values of most Gogny parameterizations discussed previously, namely D1, D1S, D1P, D1N, D1M, D3G3M, D2, D1ST2a and DG. 
Regarding the value of the power $\alpha$ appearing in the density-dependent term, we have reported it as it appears in publications, where 
no mention is made of how it has been numerically implemented in the various codes.

Concerning the Bruyères-le-Châtel group codes (HFB, MPMH and fitting codes) which have been partially or fully used in the fitting process of 
the previous cited parameterizations (which excludes the D1ST2a parameterization) 
the adopted values of the main constants are:
\begin{itemize}
\item the proton mass: $M_{\pi}= 938.27 ~MeV\/c^2$,
\item the neutron mass: $M_{\nu}=939.57 ~MeV\/c^2$,
\item the nucleon mass: $M_{n}=(M_{\pi}+M_{\nu})/2$,
\item $\hbar c = 1.9732698045930246e+02~ MeV.fm$, 
\item the electric constant: $e = 1.602176634e-19 ~ A.s $,
\item the fine structure constant: 7.2973525664e-03, 
\item the power of the density: $\alpha=0.333333$.
\end{itemize}
Moreover, the charge radii $R_{ch}$ are calculated by means of the formula:
\begin{eqnarray}
R_{ch}^2 = R_{\pi}^2 + \frac{3}{2} \left( P^2 - B^2 \right) + \frac{N}{Z} \langle r^2_{\nu}\rangle
\end{eqnarray}
where the first term involves the proton radius $R_{\pi}$, the second and third terms correspond to corrections accounting for the spatial 
extension of the proton and the charge distribution within the neutron, respectively. The calculation of $P$ is done through the formula
$P^2 = 2/3 \langle r^2_{\pi} \rangle$ where $\langle r^2_{\pi} \rangle $ is the mean square radius of the proton and 
$B^2 = \hbar/ M \omega A$ with $\hbar^2/M \simeq 41.47$. The frequency $\omega$ is linked to the number of particles A through the 
Bethe's formula $\hbar \omega = 1.85+35.5 A^{-1/3}$.
The values of $\langle r^2_{\pi} \rangle $ and $\langle r^2_{\nu} \rangle$ used in the old codes can be found in Ref.\cite{Delaroche2010}.
In the new HFB3 solver \cite{Dubray2025a}, we have adopted more recent evaluations \cite{tsiesinga,atac}: $\sqrt{\langle r^2_{\pi} \rangle} \simeq  0.8414 ~fm$ and 
$\langle r^2_{\nu} \rangle \simeq -0.110 ~fm^2 $.

Besides, for the third step of the fitting procedure (the validation of a parametrization with HFB and beyond mean-field calculations), all parameterizations have been validating using the Slater approximation for the evaluation of the exchange term in the mean-field, without contribution to the pairing channel. Only the DG Gogny interaction has benefited from an exact treatment of the Coulomb force for the mean-field, including also its contribution to the pairing field, thanks to the HFB3 solver which take them into account \cite{Dubray2025a}.
The two-body contribution of the center of mass have been taken into account in the mean-field for all the parameterizations excepted D1. Besides, DG was validated with its additional contribution to the pairing channel.

Finally, the use of binding energies as input data in the fitting procedures has been implemented in various ways, depending on the parameterization. For the D1M and D3G3M mass models, an infinite basis correction was applied.
For the older parameterizations (D1, D1S, and D1P), no such correction was added. Instead, the masses were manually verified in the third part of the protocol for selected spherical nuclei using a sufficiently large harmonic oscillator (HO) basis size.
For D1N, as previously discussed, a filter on the neutron matter equation of state was introduced, but no infinite basis correction was included. The earlier manual verification process was retained, including the absence of drift.
For D2 and DG, the same procedure as for D1N was maintained, as the primary objective of these two parameterizations was to extend the analytical form while preserving global standard properties. From this perspectives, there is room for improvement in this last two parameterizations.

\begin{onecolumn}
\begin{sidewaystable}
    \centering
    \begin{tabular}{l|lllll|lll|l}
        \multicolumn{1}{c}{} & \multicolumn{5}{c}{$Central$} &  \multicolumn{3}{c}{$Density$} &  \multicolumn{1}{c}{$Spin-orbit$}\\
        \hline
         & & & & & & & &  \\
        D1 & {$W_{1}$=-402.40} & {$B_{1}$=-100.00} & {$H_{1}$=-496.20} & {$M_{1}$=-23.56} & {$\mu_{1}$=0.7} & {$t_{0}$=+1350.00} & {$x_{0}$=1.0} & {$\alpha = 1/3$} & {$W_{ls}$=115.00} \\
           & {$W_{2}$=-21.30} & {$B_{2}$=-11.77} & {$H_{2}$=+37.27} & {$M_{2}$=-68.81} & {$\mu_{2}$=1.2} & & &\\
           & & & & & & & &\\
        \hdashline
         & & & & & & & & \\
        D1S & {$W_{1}$=-1720.30} & {$B_{1}$=+1300.00} & {$H_{1}$=-1813.53} & {$M_{1}$=+1397.60} & {$\mu_{1}$=0.7} & {$t_{0}$=+1390.60} &  {$x_{0}$=1.0} & {$\alpha = 1/3$} &{$W_{ls}$=130.00} \\
           & {$W_{2}$=+103.64} & {$B_{2}$=-163.48} & {$H_{2}$=+162.81} & {$M_{2}$=-223.93} & {$\mu_{2}$=1.2} & & & \\
            & & & & & & & & \\
        \hdashline
         & & & & & & & & \\
        D1P & {$W_{1}$=-372.89} & {$B_{1}$=+62.69} & {$H_{1}$=-464.51} & {$M_{1}$=-31.49} & {$\mu_{1}$=0.9} & {$t_{0}$=+1025.90} & {$t_{0} x_{0}$=+1190.04} & {$\alpha = 0.33$} & {$W_{ls}$=130.00} \\
           & {$W_{2}$=+34.62} & {$B_{2}$=-14.08} & {$H_{2}$=+70.95} & {$M_{2}$=-20.96} & {$\mu_{2}$=1.44} & {$t_{0}$=+256.02} & {$t_{0} x_{0}$=-513.84} & {$\alpha =$0.92} & \\
            & & & & & & & & \\
        \hdashline
         & & & & & & & &\\
         D1N & {$W_{1}$=-2047.61} & {$B_{1}$=+1700.00} & {$H_{1}$=-2414.93} & {$M_{1}$=+1519.35} & {$\mu_{1}$=0.8} & {$t_{0}$=+1609.46} & {$x_{0}$=1.0} & {$\alpha = 1/3$}& {$W_{ls}$=115.00} \\
           & {$W_{2}$=+293.02} & {$B_{2}$=-300.78} & {$H_{2}$=+414.59} & {$M_{2}$=-316.84} & {$\mu_{2}$=1.2} & & &\\
            & & & & & & & & \\
        \hdashline
         & & & & & & & &\\
        D1M & {$W_{1}$=-12797.57} & {$B_{1}$=+14048.85} & {$H_{1}$=-15144.43} & {$M_{1}$=+11963.89} & {$\mu_{1}$=0.5} & {$t_{0}$=+1562.22} & {$x_{0}$=1.0} & {$\alpha = 1/3$} & {$W_{ls}$=115.36} \\
           & {$W_{2}$=+490.95} & {$B_{2}$=-752.27} & {$H_{2}$=+675.12} & {$M_{2}$=-693.57} & {$\mu_{2}$=1.0}& & & \\
            & & & & & & & &\\
        \hdashline
         & & & & & & & & \\
        D3G3M & {$W_{1}$=-8552.859} & {$B_{1}$=+15470.967} & {$H_{1}$=-17650.205} & {$M_{1}$=+6910.033} & {$\mu_{1}$=0.0.475} & {$t_{0}$=+1602.030} & {$x_{0}$=1.0} & {$\alpha = 1/3$} & {$W_{ls}$=119.804} \\
        & {$W_{2}$=+878.129} & {$B_{2}$=-2370.137} & {$H_{2}$=+2475.527} & {$M_{2}$=-1037.0530} & {$\mu_{2}$=0.716} & & & \\
        & {$W_{3}$=+5.000} & {$B_{3}$=-10.000} & {$H_{3}$=+10.000} & {$M_{3}$=-20.000} & {$\mu_{3}$=1.780} & & & \\
         & & & & & & & & \\
        \hline         
    \end{tabular}
    \caption{Parameter values of the D1-type Gogny interactions. They are expressed in MeV, excepted the ranges $\mu_i$ which are in fm.}
    \label{tab:GognyparamD1}
\end{sidewaystable}

\begin{sidewaystable}
    \centering
    \begin{tabular}{l|lllll|lllll}
        \multicolumn{1}{c}{} & \multicolumn{5}{c}{$Central$} &  \multicolumn{5}{c}{$Density$} \\
        \hline
        & & & & & & & & & & \\
        D2 & {$W_{1}$=-1176.440} & {$B_{1}$=+800.000} & {$H_{1}$=-927.366} & {$M_{1}$=+1115.573} & {$\mu_{1}$=0.8} 
        & {$W_{3}$=+1800.000} & {$B_{3}$=+600.000} & {$H_{3}$=+400.000} & {$M_{3}$=-600.000} & {$\mu_{3}$=0.6}\\
        & {$W_{2}$=+93.741} & {$B_{2}$=-162.161} & {$H_{2}$=+122.414} & {$M_{2}$=-223.859}
            & {$\mu_{2}$=1.3} & &  & & & \\
           & & & & & & & & & &  \\
        \hline       
    \end{tabular}
    \caption{Same as Table \ref{tab:GognyparamD1} but for the D2 Gogny interaction. The spin-orbit parameter $W_{ls}$ of the D2 interaction is equal to +130.000 MeV.}
    \label{tab:GognyparamD2}

    \hspace{2cm}
    
    \begin{tabular}{l|lllll|lll}
        \multicolumn{1}{c}{} & \multicolumn{5}{c}{$Central$} &  \multicolumn{2}{c}{$Density$}  \\
        \hline
        & & & & & & &   \\
        D1ST2a & {$W_{1}$=-1720.30} & {$B_{1}$=+1300.00} & {$H_{1}$=-1813.53} & {$M_{1}$=+1397.60} & {$\mu_{1}$=0.7} & {$t_{0}$=+1390.60} &  {$x_{0}$=1.0}  \\
        & {$W_{2}$=+103.64} & {$B_{2}$=-163.48} & {$H_{2}$=+162.81} & {$M_{2}$=-223.93} & {$\mu_{2}$=1.2} \\
         & & & & & & &   \\
        \hline
    \end{tabular}
    \begin{tabular}{l|l|lll}
        \multicolumn{1}{c}{} & \multicolumn{1}{c}{$Spin-orbit$} & \multicolumn{3}{c}{$Tensor$} \\
        \hline
         & & & & \\
        D1ST2a  &{$W_{ls}$=130.00}  & {$V_{\text{T1}}$=-135.00} & {$V_{\text{T2}}$=+115.00} & {$\mu_5=1.2$} \\
         & & & & \\
        \hline         
    \end{tabular}
    \hspace{1cm}
    \begin{tabular}{l|lllll|lllll}
        \multicolumn{1}{c}{} & \multicolumn{5}{c}{$Central$} &  \multicolumn{5}{c}{$Density$} \\
        \hline
        & & & & & & & & & & \\
         DG & {$W_{1}$=-1190.016} & {$B_{1}$=+800.000} & {$H_{1}$=-877.422} & {$M_{1}$=+1198.923} & {$\mu_1=0.8$} &{$W_{3}$=+1836.200} & {$B_{3}$=+581.600} & {$H_{3}$=+377.600} & {$M_{3}$=-633.220} &{$\mu_3=0.6$} \\
            & {$W_{2}$=+109.179} & {$B_{2}$=-191.226} & {$H_{2}$=+133.441} & {$M_{2}$=-277.509} &{$\mu_2=1.24$} & & & & & \\
        \hline
    \end{tabular}
    \begin{tabular}{l|lll|lll}
        \multicolumn{1}{l}{} & \multicolumn{3}{c}{$Spin-orbit$} &  \multicolumn{3}{c}{$Tensor$} \\
        \hline
        & & & & & & \\
         DG & {$W_{4}$=+145.483} & {$H_{4}$=+29.634} & {$\mu_4=0.2$} &{$W_{5}$=-392.544} & {$H_{5}$=-196.481} &{$\mu_5=1.10$} \\
         & & & & & & \\
        \hline
    \end{tabular}
    \caption{Same as Table \ref{tab:GognyparamD1} but for Gogny interactions including a tensor term. The power of the density dependent term is equal to 1/3.}
    \label{tab:GognyparamDG}
\end{sidewaystable}

\end{onecolumn}

\twocolumn

\section{Nuclear matter properties in the parameterization selection 
process}\label{sec3}

Once a set of parameters is proposed by the HFR emulator (see section \ref{sec2}),
the new parameterization is tested against a well defined set of nuclear matter properties. They include  symmetric infinite nuclear 
matter properties, neutron equation of state,  evolution of proton and neutron effective masses as well as  
Landau parameters
(see Fig. \ref{fig:schema_fit}). Other quantities like the slope of the symmetry energy, important 
for astrophysical applications, have been considered in addition in the case of the D2 and DG fitting procedure.
This section is divided in three parts:
in the first part, the expression of the nuclear matter equation of state is derived. In the second part,
the expressions for other physical quantities such as the incompressibility, the symmetry energy,
the effective masses and the Landau parameters are discussed. In the third part, the values for those
quantities obtained for various Gogny interactions are obtained and discussed.
In order to provide the most general formulas, all the derivations are given for the DG interaction. 
The expressions for D2 can easily be obtained from the DG ones. The specific zero-range density-
dependent term encountered with the D1-type parameterization leads to simpler expressions  
that are also  explicitly given.

\subsection{Infinite nuclear matter}


In the context of INM, the single-particle basis consists in plane waves. 
Then, an HF state is built as an antisymmetrized product of such states, each being characterized by their momentum $\hbar \vec{k}$ as well 
as the spin and isospin quantum numbers:
\begin{equation} \label{PWstate}
\ket*{\vec{k} s t} \!,
\end{equation}
where $\vec{k}$ is the wave number vector describing the spatial degrees of 
freedom, and $s = \pm 1/2$ and $t = \pm 1/2$ are the projections of 
the spin and isospin degrees of freedom along the quantization axis, chosen to be $Oz$. 

\noindent The plane wave functions then take the form
\begin{equation} \label{planew}
\varphi_{\vec{k} s t}(\vec{r}, \sigma, \tau) = \braket*{\vec{r} \sigma \tau}{\vec{k} s t} = \phi_{\vec{k}}(\vec{r}) \chi_s(\sigma) \zeta_t(\tau),
\end{equation}
where $\xi_{s}(\sigma)$ and $\zeta_{t}(\tau)$ are the normalized spin 
and isospin wave functions, associated with $\sigma$ and $\tau$, the 
Pauli matrices describing the spin and isospin degrees of freedom, 
respectively, and where the spatial plane wave functions 
are expressed as
\begin{equation}
\phi_{\vec{k}}(\vec{r}) = \frac{1}{\sqrt{\mathcal{V}}} \, 
e^{i \vec{k} \cdot \vec{r}}
\end{equation}
with the normalization coefficient $1/\sqrt{\mathcal{V}}$ involving an arbitrary large volume $\mathcal{V}$. 
If the nucleons are moving in a finite volume, the situation is 
equivalent to particles trapped in a three-dimensional box in such a 
way that the boundary conditions impose quantized momenta $\vec{k}$. 
If, instead, the volume is infinite, one has no such conditions and the 
momenta are considered continuous. To go from a finite to an infinite 
volume, one must therefore perform the following transformation: 
\begin{equation} \label{transfocontinu}
\sum_{\vec{k}} \rightarrow \frac{\mathcal{V}}{(2 \pi)^3}\int 
\dd[3]{k} 
\end{equation}
One can show that the plane wave functions satisfy an orthogonality relation, 
for continuous momenta, such that:
\begin{multline} \label{orthoINM}
\sum_{\sigma \tau} \int \dd[3]{r} \varphi^*_{\vec{k} s t}(\vec{r}, \sigma, \tau) \varphi_{\vec{k}' s' t'}(\vec{r}, \sigma, \tau) = \\ \delta(\vec{k} - \vec{k}') \delta_{ss'} \delta_{tt'}
\end{multline}

In order to evaluate the energy in INM at the HF 
approximation, one evaluates first the local nuclear density 
whose expression 
is given by:
\begin{multline}\label{densityorigin}
\rho(\vec{r}) = \sum_{\sigma \tau} \mel*{\vec{r} 
\sigma \tau}{\rho}{\vec{r} \sigma \tau} \\ 
 = \sum_{\sigma \tau} \sum_{ab} \varphi^*_a(\vec{r}, \sigma,
 \tau) \varphi_b(\vec{r}, \sigma, \tau) \rho_{ba}
\end{multline}
To obtain Eq.(\ref{densityorigin}), two completeness 
relations associated with the sets of quantum numbers 
$a \equiv \{ \vec{k}_a,s_a,t_a \}$ and 
$b \equiv \{ \vec{k}_b,s_b,t_b \}$ 
characterizing basis states $\varphi$ have been used. 
The quantities $\rho_{ba} \equiv \! \mel*{a}{\rho}{b}$ are the matrix 
elements of the one-body density matrix expressed in this basis. \\
The density of a given isospin $t$ reads
\begin{multline}\label{eeqq1}
\displaystyle \rho_t (\vec{r}) = \displaystyle \sum_{\sigma \tau} \sum_{a_1} \sum_{a_2} \! \varphi^*_{a_1}(\vec{r}, \sigma, \tau) \varphi_{a_2}(\vec{r}, \sigma, \tau) \times \\ \rho_{a_1, a_2} \delta_{t_1,t_2}\\
 = \displaystyle \frac{\mathcal{V}}{(2\pi)^3} \sum_s \int \dd[3]{k} \phi^*_{a}(\vec{r}, \sigma, \tau) \phi_{a}(\vec{r}, \sigma, \tau) \\
= \displaystyle \frac{2 \mathcal{V}}{(2 \pi)^2} \sum_s \int_0^{k_{{F}}^t} \dd{k} k^2 \phi^*_{a}(\vec{r}, \sigma, \tau) \phi_{a}(\vec{r}, \sigma, \tau) \\
 = \displaystyle \frac{(k_{{F}}^t)^3}{3 \pi^2}
\end{multline}
To obtain Eq.\eqref{eeqq1}, the particular form (diagonal) of the density matrix in 
the HF basis as been exploited, as well as the plane wave functions
\eqref{planew} and the orthogonality relation 
\eqref{orthoINM}. 
The quantity $k_{{F}}^t$ represents the Fermi momentum of 
the isospin-$t$. 
The local density is constant in INM which induces
its invariance by translation. \\
The total nuclear density is thus equal to the sum of the 
neutron $ \rho_\nu$ and proton $\rho_\pi$ density:
\begin{equation} \label{densityINM}
\rho = \rho_\nu+ \rho_\pi = \frac{2 k_{{F}}^3}{3 \pi^2}
\end{equation}
where the auxiliary quantity $k_{{F}}$ is expressed as:
\begin{equation}
k_{{F}}^3 = \frac{(k_{{F}}^\nu)^3 + 
(k_{{F}}^\pi)^3}{2}.
\end{equation}

\noindent In order to specify the proportion of neutrons and protons in 
INM, the asymmetry parameter $\beta$ is introduced such that:
\begin{equation} \label{asymparameter}
\beta \equiv \frac{\rho_\nu - \rho_\pi}{\rho}.
\end{equation}
Symmetric INM is characterized by $\beta=0$  whereas
$\beta=1$ corresponds to pure neutron matter. For intermediate values of $\beta$ one talks about asymmetric INM.

In the HF approximation the energy is the sum of kinetic and potential contributions
\begin{align} \label{eeqq2}
\mathcal{E} & = \mathcal{E}_{{K}} + 
\mathcal{E}_{{P}} 
= \sum_{a} \! \mel*{a}{t_{{K}}}{a} \! \rho_{a, a} \\ \nonumber
& + \frac{1}{2} \sum_{a_1} \sum_{a_2} \! \! \! \mel*{a_1 \, a_2}{v_{12}^{{(a)}}}{a_1 \, a_2 } \! 
\rho_{a_1, a_1} \! \rho_{a_2, a_2}.
\end{align}
where $a$ stands for the set of quantum numbers $\{\vec{k} s t\}$.
In the following, one will evaluate separately the kinetic and potential contributions in asymmetric, symmetric and neutron infinite nuclear matter, as it is useful to compute 
the physical quantities of interest.

\paragraph{Kinetic energy}

\noindent The total kinetic energy $\mathcal{E}_{{K}}$ can be re-expressed such that:
\begin{align}
\mathcal{E}_{{K}} & = \sum_{\vec{k} s t} \frac{\hbar^2 k^2}{2m}
= \frac{\mathcal{V}}{\pi^2} \sum_{t} \int_0^{k_{{F}}^t} \dd{k} \frac{\hbar^2 k^4}{2m} \\ \nonumber
& = \sum_t \frac{3}{5} \frac{\hbar^2 (k_{{F}}^t)^2}{2m} \rho_t \mathcal{V}
\end{align}
Thus, the total kinetic energy per nucleon is equal to:
\begin{equation} \label{kineticenergy}
\frac{\mathcal{E}_{{K}}}{A} = \sum_t \frac{3}{5} \frac{\hbar^2 (k_{{F}}^t)^2}{2m} \frac{\rho_t}{\rho}. 
\end{equation}
This expression is well suited for computing the different particular cases considered in the following.

\paragraph{Potential energy}

The potential energy is given by
\begin{equation}\label{potentialenergyINM}
\mathcal{E}_{{P}} = \frac{1}{2} \sum_{a_1} \sum_{a_2}  \mel*{a_1 \, a_2}{v_{12}^{{(a)}}}{a_1 \ a_2} 
\rho_{a_1, a_1} \rho_{a_2, a_2}
\end{equation}
where $a_1$ and $a_2$ is a short hand notation for the quantum numbers $\vec{k},s,t$. The potential energy contains contributions from the central, density-dependent, spin-orbit and tensor terms that will be detailed in the following.
The expressions below are given for the most general DG interaction, which allow to recover the expressions associated with D2 and D1S by performing the 
appropriate limits. For detailed derivations, please check Refs   \cite{Zietek2023,Chappert2007}. \\

\noindent \textit{Contributions of the central and density-dependent terms}\\

\noindent The antisymmetrized central and density-dependent (CDD) interactions can be encompassed in the following expression:
\begin{multline}\label{CDDpot}
v_{12}^{\text{CDD,(a)}}  = v_{12}^{\text{CDD}} (1-P_{r} P_{\sigma} P_{\tau}) \\
 = \mathcal{P}_{{D}} V(r_{12}) D[\rho] + \mathcal{P}_{{E}} V(r_{12}) D[\rho] P_r
\end{multline}
with the Gaussian potential 
\begin{equation} \label{gaussianpotential}
V(r_{12}) \equiv e^{-(\vec{r}_1 - \vec{r}_2)^2/ \mu^2}
\end{equation}
and the functional of the density,
\begin{equation} \label{FrhoINM}
D[\rho] \equiv \frac{\rho^{\alpha}(\vec{r}_1) + \rho^{\alpha}(\vec{r}_2)}{2}.
\end{equation}
The central contributions to the potential energy are obtained by:
\begin{itemize}[label=$-$]
\item setting $\alpha = 0$ 
\item summing over the central ranges $\mu_1$ and $\mu_2$. 
\end{itemize}
To get the finite-range density-dependent term from Eq.(\ref{CDDpot}), it suffices to:
\begin{itemize}[label=$-$]
\item consider $\alpha \ne 0$ 
\item add a global multiplicative factor $1/(\mu_3 \sqrt{\pi})^3$ not indicated for simplicity.
\end{itemize}
The spin-isospin components of the direct and exchange components of the CDD interactions are, respectively:
\begin{multline} \label{WBHMCDDINM}
\mathcal{P}_{{D}} = W + B P_{\sigma} - H P_{\tau} - M P_{\sigma} P_{\tau},  \\
\mathcal{P}_{{E}} = M + H P_{\sigma} - B P_{\tau} - W P_{\sigma} P_{\tau}.
\end{multline}

\noindent Separating the spatial and spin-isospin part of the TBME as well as the direct and exchange components of the CDD interaction, the potential energy becomes:
\begin{multline} \label{VpotCDD}
\mathcal{E}_{{P}\mu}^{CDD} = \\ \frac{1}{2} \sum_{\vec{k}_1 s_1 t_1} \sum_{\vec{k}_2 s_2 t_2} \big[  \mel*{\vec{k}_1 \vec{k}_2}{V(r_{12}) D[\rho]}{\vec{k}_1 \vec{k}_2} \mathcal{P}_{{D}_{12} } \\
\quad + \mel*{\vec{k}_1 \vec{k}_2}{V(r_{12}) D[\rho]}{\vec{k}_2 \vec{k}_1} \mathcal{P}_{{E}_{12}}  \big]
\end{multline}
where $ (\mathcal{P}_{{D}})_{12} = \mel*{s_1 t_1 s_2 t_2}{\mathcal{P}_{{D}}}{s_1 t_1 s_2 t_2}$ and $(\mathcal{P}_{{E}})_{12} = \mel*{s_1 t_1 s_2 t_2}{\mathcal{P}_{E}}{s_1 t_1 s_2 t_2}$.
The spatial part of the direct component reads:
\begin{equation} \label{directterm}
\mel*{\vec{k}_1 \vec{k}_2}{V(r_{12}) D[\rho]}{\vec{k}_1 \vec{k}_2}
= \frac{\rho^{\alpha}}{\mathcal{V}} (\mu \sqrt{\pi})^3
\end{equation}
where the translational invariance of INM as been used, $\rho(\vec{r}_1) = \rho(\vec{r}_2) = \rho$.
Since the above matrix element does not explicitly depend on the momenta $\vec{k}_1$ and $\vec{k}_2$, the contribution of the direct spatial term to the potential energy is straightforward. At the continuous limit \eqref{transfocontinu}, reads:
\begin{multline} \label{spacedirect}
\displaystyle \sum_{\vec{k}_1 \vec{k}_2} \mel*{\vec{k}_1 \vec{k}_2}{V(r_{12}) D[\rho]}{\vec{k}_1 \vec{k}_2} \\ = \displaystyle \frac{\mathcal{V} \rho^{\alpha}}{(2 \pi)^6} (\mu \sqrt{\pi})^3 \frac{4 \pi (k_{F}^{t_1})^3}{3} \frac{4 \pi (k_{F}^{t_2})^3}{3} 
\end{multline}
where $k_{F}^{t_1}$ and $k_{F}^{t_2}$ are the Fermi momenta associated with $\vec{k}_1$ and $\vec{k}_2$, respectively. 

Concerning the spatial part of the exchange component, one finds:
\begin{multline} \label{elmatPW}
 \displaystyle \mel*{\vec{k}_1 \vec{k}_2}{V(r_{12}) D[\rho]}{\vec{k}_2 \vec{k}_1} \\
= \displaystyle \frac{\rho^{\alpha}}{\mathcal{V}} (\mu \sqrt{\pi})^3 
  \displaystyle e^{-(\vec{k}_1 - \vec{k}_2)^2 \mu^2/4}
\end{multline}
The spatial contribution of the exchange term to the potential energy depends explicitly on the momenta $\vec{k}_1$ and $\vec{k}_2$. 
After some manipulations and in the continuous limit \eqref{transfocontinu}, one obtains:
\begin{multline} \label{spaceexchange}
\displaystyle \sum_{\vec{k}_1 \vec{k}_2} \! \mel*{\vec{k}_1 \vec{k}_2}{V(r_{12}) D[\rho]}{\vec{k}_2 \vec{k}_1} \\  = \displaystyle\frac{\mathcal{V} \rho^{\alpha}}{(2 \pi)^6} (\mu \sqrt{\pi})^3  I(X_{t_1}, X_{t_2})
\end{multline}
where the dimensionless quantity $X_t$ associated with a given isospin $t$ is such that $X_t \equiv \mu k_{{F}}^t$, and the quantity $I(X_{t_1}, X_{t_2})$ is given by:
\begin{equation} 
I(X_{t_1}, X_{t_2}) = \bigg( \frac{2}{\mu} \bigg)^6 \frac{\pi^2}{6} Q(X_{t_1}, X_{t_2}),
\end{equation}
with the function $Q(X_{t_1}, X_{t_2})$ defined by:
\begin{multline}
Q(X_{t_1}, X_{t_2}) \\ = e^{-\left(\frac{X_{t_1}+X_{t_2}}{2} \right)^2} \big( X_{t_1}^2 + X_{t_2}^2 - X_{t_1}X_{t_2} -2 \big) \\ 
 - e^{-\left(\frac{X_{t_1}-X_{t_2}}{2} \right)^2} \big( X_{t_1}^2 + X_{t_2}^2 + X_{t_1}X_{t_2} -2 \big) \\
+ \frac{\sqrt{\pi}}{2} \erf \bigg(\frac{X_{t_1}+X_{t_2}}{2} \bigg) \big(X_{t_1}^3 + X_{t_2}^3\big) \\
- \frac{\sqrt{\pi}}{2} \erf \bigg(\frac{X_{t_1}-X_{t_2}}{2} \bigg) \big(X_{t_1}^3 - X_{t_2}^3\big)
\end{multline}

\noindent Concerning the direct spin-isospin part of the potential energy \eqref{VpotCDD}, one gets, going from the uncoupled two-particle representation to the coupled one:
\begin{multline}
\! \mel*{s_1 t_1 s_2 t_2}{W + B P_\sigma - H P_\tau - M P_\sigma P_\tau}{s_1 t_1 s_2 t_2} \\
\displaystyle = \sum_{S M_S} \sum_{T M_T} \braket*{1/2 s_1 1/2 s_2}{S M_S}^2 \\
\times \braket*{1/2 t_1 1/2 t_2}{T M_T}^2 \\ 
\displaystyle \times (W - (-)^S B + (-)^T H - (-)^{S+T} M)
\end{multline}
where the quantities $\braket*{1/2 s_1 1/2 s_2}{S M_S}$ and
$\braket*{1/2 t_1 1/2 t_2}{T M_T}$ denote Clebsch-Gordan coefficients. Using the same process for the exchange spin-isospin part, 
the potential energy \eqref{VpotCDD} becomes:
\begin{multline}
\displaystyle \mathcal{E}_{{P}\mu}^{CDD} = 8 \sqrt{\pi} \sum_{t_1 t_2} \sum_{ST} \mathcal{L}^{T}_{t_1 t_2} \times \\
\displaystyle \sum_{\vec{k}_1 \vec{k}_2} \big[ \mathcal{A}^{ST} \! \mel*{\vec{k}_1 \vec{k}_2}{V(r_{12}) D[\rho]}{\vec{k}_1 \vec{k}_2} \\
\displaystyle  - \mathcal{B}^{ST} \! \mel*{\vec{k}_1 \vec{k}_2}{V(r_{12}) D[\rho]}{\vec{k}_2 \vec{k}_1} \! \big]
\end{multline}
where the coefficients $\mathcal{L}^{T}_{t_1 t_2}$, $\mathcal{A}^{ST}$ and $\mathcal{B}^{ST}$ have for expressions:
\begin{multline}
\displaystyle \mathcal{L}^{T}_{t_1 t_2} = \frac{1}{2T+1} \sum_{M_T} \! \braket*{1/2 t_1 1/2 t_2}{T M_T}^2 \label{firsteq} \\ 
\displaystyle \mathcal{A}^{ST} = \frac{(2S+1) (2T+1)}{16 \sqrt{\pi}} \times \\
\big(W - (-)^S B + (-)^T H - (-)^{S+T} M \big) \\
\displaystyle \mathcal{B}^{ST}  = -\frac{(2S+1) (2T+1)}{16 \sqrt{\pi}} \times \\
\big(M - (-)^S H + (-)^T B - (-)^{S+T} W \big)
\end{multline}
Noticing that the volume can be written $\mathcal{V} = A/\rho = A \times 3 \pi^2 \mu^3/2 X^3$ 
and plugging the expressions of the spatial parts \eqref{spacedirect} and \eqref{spaceexchange}, the potential energy per nucleon is then:
\begin{multline} \label{vpotintermediate}
\displaystyle \frac{\mathcal{E}_{{P}\mu}^{CDD}}{A} = 2 \frac{\rho^{\alpha}}{X^3}  \sum_{ST} \sum_{t_1 t_2} \mathcal{L}^{T}_{t_1 t_2}  \times \\
\displaystyle \bigg[ \mathcal{A}^{ST} \frac{(X_{t_1} X_{t_2})^3}{6} - \mathcal{B}^{ST} Q(X_{t_1}, X_{t_2}) \bigg]
\end{multline}

\noindent Performing the summations on (S,T) and ($t_1,~t_2$), the contributions from the central and density-dependent terms to the potential energy per nucleon 
in asymmetric INM reads:
\begin{multline} \label{potentialenergyINM2}
\displaystyle \frac{\mathcal{E}_{{P}\mu}^{CDD}}{A} = \frac{\rho^{\alpha}}{X^3} \bigg[ \frac{\mathcal{A}}{6} \big( X_\pi^6 + X_\nu^6 \big) - \frac{\mathcal{A}'}{6} \big( X_\pi^3 - X_\nu^3 \big)^2 \\
\displaystyle + (\mathcal{B}'-\mathcal{B}) \big( Q(X_\pi, X_\pi) + Q(X_\nu, X_\nu) \big) \\
\displaystyle - 2 \mathcal{B}' Q(X_\pi, X_\nu) \bigg]
\end{multline}
with the summed quantities $\mathcal{A}$, $\mathcal{B}$,
$\mathcal{A}'$ and $\mathcal{B}'$:
\begin{multline}\label{ABAPBP}
\displaystyle \mathcal{A} = \sum_{ST} \mathcal{A}^{ST} = \frac{1}{2 \sqrt{\pi}} \Big( 2W + B - H  - \frac{M}{2}  \Big) \\
\displaystyle \mathcal{B}= \sum_{ST} \mathcal{B}^{ST} = \frac{1}{2 \sqrt{\pi}} \Big( \frac{W}{2} + B - H  - 2M \Big) \\
\displaystyle \mathcal{A'} = \sum_{ST} \frac{\mathcal{A}^{ST}}{2T+1} = \frac{1}{2 \sqrt{\pi}} \Big( W + \frac{B}{2} \Big) \\
\displaystyle \mathcal{B'} = \sum_{ST} \frac{\mathcal{B}^{ST}}{2T+1} = -\frac{1}{2 \sqrt{\pi}} \Big( M + \frac{H}{2} \Big)
\end{multline}

In the particular case of the zero-range density-dependent term
of the D1-type parameterizations, one finds:
\begin{equation}
\begin{array}{lcl}
\displaystyle \frac{\mathcal{E}_{{P}0}^{DD}}{A} & =& \displaystyle \frac{1}{24 \pi^2}
\frac{\rho^{\alpha}}{k^3_{F}} t_0 [ (1-x_0) \left( (k_F^{\pi})^6 + (k_F^{\nu})^6 \right) \\ 
& +& \displaystyle \left( 4+2 x_0 \right) (k_F^{\pi} k_F^{\nu})^3 ]
\end{array}
\end{equation}

\noindent \textit{Contribution of the spin-orbit term}\label{sopotzero} \\

\noindent To evaluate the spin-orbit contribution, one considers first an equivalent definition. Indeed, one starts from the standard expression of the spin-orbit potential: 
\begin{equation} \label{SOannex}
\begin{array}{lcl}
v_{12}^{\text{SO}} & =& B(\mu_4) (W-H P_\tau) V(r_{12}, \mu_4) \vec{L} \cdot \vec{S} 
\end{array}
\end{equation}
Replacing $\vec{L}$ and $\vec{S}$ by their definitions, one obtains:
\begin{equation}
\begin{array}{lcl}
v_{12}^{\text{SO}} & =& \displaystyle - \frac{i}{4} B(\mu_4) (W-H P_\tau) V(r_{12}, \mu_4) \\ & & \qquad \times \big[ \vec{r}_{12} \cross \vec{\nabla}_{12} \big] \cdot (\vec{\sigma}_1 + \vec{\sigma}_2)
\end{array}
\end{equation}
Then, one considers a potential $V(r, \mu)$ from which one can find a function $G(r, \mu)$ satisfying:
\begin{equation}
\vec{\nabla} G(r, \mu) = V(r, \mu) \vec{r}.
\end{equation}
In the particular case of the Gaussian potential $V(r_{12}, \mu_4)$, one obtains:
\begin{equation}
G(r_{12}, \mu_4) = - \frac{\mu_4^2}{4} e^{-(\vec{r}_1 - \vec{r}_2)^2/\mu_4^2}
\end{equation}
Thus, the equivalent expression of $v_{12}^{\text{SO}}$ can be deduced:
\begin{eqnarray} \label{SOLS} 
v_{12}^{\text{SO}} = - \frac{i}{4} B(\mu_4) (W-H P_\tau) \qquad \qquad \nonumber \\ \big[ \cev{\nabla}_{12} G(r_{12}, \mu_4) \cross \vec{\nabla}_{12} \big] \cdot (\vec{\sigma}_1 + \vec{\sigma}_2)
\end{eqnarray}
with $B(\mu_4) = -4/ \big(\mu_4^2 (\mu \sqrt \pi )^3\big)$.
To simplify the notation, we have omitted the $\mu_4$ dependence of the function $G(r_{12}, \mu_4) \equiv G(r_{12})$ in the following. 
\noindent The anti-symmetrized spin-orbit interaction can then be written as:
\begin{multline} \label{SOpot}
\displaystyle v_{12}^{\text{SO,{(a)}}} = v_{12}^{\text{SO}}(1-P_{r}P_{\sigma}P_{\tau}) =\\
\displaystyle  \frac{i B(\mu)}{4} \left( \mathcal{P}_{{D}} \big[ \cev{\nabla}_{12} G(r_{12}) \cross \vec{\nabla}_{12} \big] \cdot (\vec{\sigma}_1 + \vec{\sigma}_2) \right.\\
\displaystyle \left. \quad + \mathcal{P}_{{E}} \big[ \cev{\nabla}_{12} G(r_{12}) \cross \vec{\nabla}_{12} \big] \cdot (\vec{\sigma}_1 + \vec{\sigma}_2) P_r \right)
\end{multline}
The factor $i B(\mu)/4$ will be omitted in the following for simplicity. Besides, the presence of the operator $\vec{\sigma}_1 + \vec{\sigma}_2 = 2 \vec{S}$, acting symmetrically on the spin variables, 
makes the operator $P_{\sigma}$ the identity. The isospin components of the direct and exchange spin-orbit interaction are, respectively:
\begin{subequations}
\begin{align}
\mathcal{P}_{{D}} \equiv W - H P_{\tau}, \\
\mathcal{P}_{{E}} \equiv H - W P_{\tau}.
\end{align}
\end{subequations}

\noindent Separating the spatial and spin-isospin part of the TBMEs as well as the direct and exchange components of the spin-orbit interaction, 
the potential energy becomes:
\begin{multline}
\displaystyle \mathcal{E}_{{P}}  = \frac{1}{2} \sum_{\vec{k}_1 s_1 t_1} \! \sum_{\vec{k}_2 s_2 t_2} \Big[ \! \mel*{\vec{k}_1 \vec{k}_2}{ \big[ \cev{\nabla}_{12} G(r_{12}) \cross \vec{\nabla}_{12} \big] }{\vec{k}_1 \vec{k}_2}  \\ 
\displaystyle \mel*{s_1 t_1 s_2 t_2}{\mathcal{P}_{{D}} (\vec{\sigma}_1 + \vec{\sigma}_2)}{s_1 t_1 s_2 t_2} \\
\displaystyle + \! \mel*{\vec{k}_1 \vec{k}_2}{\big[ \cev{\nabla}_{12} G(r_{12}) \cross \vec{\nabla}_{12} \big]}{\vec{k}_2 \vec{k}_1} \\ 
\displaystyle  \mel*{s_1 t_1 s_2 t_2}{\mathcal{P}_{{E}} (\vec{\sigma}_1 + \vec{\sigma}_2)}{s_1 t_1 s_2 t_2} \! \Big]
\end{multline}
One  considers firstly the direct spatial part of the TBMEs:
\begin{multline}\label{directSOelmat}
\displaystyle \mel*{\vec{k}_1 \vec{k}_2}{\big[ \cev{\nabla}_{12} G(r_{12}) \cross \vec{\nabla}_{12} \big]}{\vec{k}_1 \vec{k}_2} =\\
\displaystyle \frac{1}{\mathcal{V}^2} \int \dd[3]{r_1} \int \dd[3]{r_2}  \big( e^{- i \vec{k}_1 \cdot \vec{r}_1} e^{- i \vec{k}_2 \cdot \vec{r}_2} \cev{\nabla}_{12} \big) G(r_{12}) \\
\displaystyle \cross \big( \vec{\nabla}_{12} e^{i \vec{k}_1 \cdot \vec{r}_1} e^{i \vec{k}_2 \cdot \vec{r}_2} \big) \\
\displaystyle =  \frac{1}{\mathcal{V}^2} \int \dd[3]{r_1} \int \dd[3]{r_2} G(r_{12}) \big[ (\vec{k}_1 - \vec{k}_2) \cross (\vec{k}_1 - \vec{k}_2) \big] \\
\displaystyle = 0
\end{multline}
The direct spatial component of the spin-orbit interaction does not contribute to the potential energy. 
Concerning the the direct spin-isospin part of the TBMEs, it splits up in such a way that:
\begin{multline}
\displaystyle \mel*{s_1 t_1 s_2 t_2}{\mathcal{P}_{{D}} [\vec{\sigma}_1 + \vec{\sigma}_2]^{(1)}_l}{s_1 t_1 s_2 t_2} \\ 
\displaystyle  = \! \mel*{t_1 t_2}{\mathcal{P}_{{D}}}{t_1 t_2} \! \! \mel*{s_1 s_2}{[\vec{\sigma}_1 + \vec{\sigma}_2]^{(1)}_l}{s_1 s_2}
\end{multline}
where 
\begin{equation}
\mel*{s_1 s_2}{[\vec{\sigma}_1 + \vec{\sigma}_2]^{(1)}_l}{s_1 s_2} = 2 (s_1+s_2) \delta_{l,0}
\end{equation} 
and
\begin{equation}
\sum_{s_1 s_2} \! \mel*{s_1 s_2}{[\vec{\sigma}_1 + \vec{\sigma}_2]^{(1)}_l}{s_1 s_2} = 0.
\end{equation}
The direct component of the spin-orbit interaction does not contribute to the potential energy. 

One can shows that the same conclusions hold for the exchange part. Indeed,
\begin{multline} \label{exchangeSOelmat}
\displaystyle \mel*{\vec{k}_1 \vec{k}_2}{\big[ \cev{\nabla}_{12} G(r_{12}) \cross \vec{\nabla}_{12} \big]}{\vec{k}_2 \vec{k}_1} =\\
\displaystyle \frac{1}{\mathcal{V}^2} \int \dd[3]{r_1} \int \dd[3]{r_2}  \big( \e^{-\ii \vec{k}_1 \cdot \vec{r}_1} \e^{-\ii \vec{k}_2 \cdot \vec{r}_2} \cev{\nabla}_{12} \big) G(r_{12}) \\
\displaystyle \cross \big( \vec{\nabla}_{12} \e^{\ii \vec{k}_2 \cdot \vec{r}_1} \e^{\ii \vec{k}_1 \cdot \vec{r}_2} \big) \notag \\
\displaystyle = -\frac{1}{\mathcal{V}^2} \int \dd[3]{r_1} \int \dd[3]{r_2} G(r_{12}) \big[ (\vec{k}_1 - \vec{k}_2) \cross (\vec{k}_1 - \vec{k}_2) \big] \\
\displaystyle = 0
\end{multline}
In INM, at the HF approximation, the exchange component of the spin-orbit interaction does not contribute to the potential energy either. As for the exchange spin-isospin TBMEs, they also split up according to:
\begin{multline}
\displaystyle  \mel*{s_1 t_1 s_2 t_2}{\mathcal{P}_{{E}} [\vec{\sigma}_1 + \vec{\sigma}_2]^{(1)}_l}{s_1 t_1 s_2 t_2} \\
 = \displaystyle \mel*{t_1 t_2}{\mathcal{P}_{{E}}}{t_2 t_1} \mel*{s_1 s_2}{[\vec{\sigma}_1 + \vec{\sigma}_2]^{(1)}_l}{s_2 s_1}
\end{multline}
Then, the spin TBMEs can be evaluated by means of Eq.\eqref{spinSO}. One obtains:
\begin{equation}
\mel*{s_1 s_2}{[\vec{\sigma}_1 + \vec{\sigma}_2]^{(1)}_l}{s_2 s_1} = 2 (s_1+s_2) \delta_{s_1 s_2} \delta_{l,0}
\end{equation}
so that the spin part of the potential energy associated with the direct spin-orbit interaction vanishes, i.e.:
\begin{equation}
\sum_{s_1 s_2} \! \mel*{s_1 s_2}{[\vec{\sigma}_1 + \vec{\sigma}_2]^{(1)}_l}{s_1 s_2} = 0.
\end{equation}

\noindent One concludes that both the spatial and the spin parts of the spin-orbit interaction have a null contribution to the potential energy. \\

\noindent \textit{Contribution of the tensor term}\label{tensorpotzero}  \\

\noindent The anti-symmetrized tensor interaction can be expressed as:
\begin{multline} \label{Tpot}
\displaystyle v_{12}^{\text{T,({a})}} = v_{12}^{\text{T}}(1-P_{r}P_{\sigma}P_{\tau}) \\
= \displaystyle (W - H P_{\tau}) V(r_{12}) S_{12} (1 - P_r P_{\tau}) \\
= \displaystyle \mathcal{P}_{{D}} V(r_{12}) S_{12}  + \mathcal{P}_{{E}} V(r_{12}) S_{12} P_r
\end{multline}
where the Gaussian potential is expressed in Eq.\eqref{gaussianpotential}. The tensor operator $S_{12}$ removes the operator $P_{\sigma}$ since it acts symmetrically on the spin variables. As for the isospin components of the direct and exchange tensor interaction, they are defined by, respectively:
\begin{subequations} \label{WHtensorINM}
\begin{align}
\mathcal{P}_{{D}} \equiv W - H P_{\tau} \\
\mathcal{P}_{{E}} \equiv H - W P_{\tau}.
\end{align}
\end{subequations}

\noindent Using an equivalent form of the tensor operator, $S_{12} = [\hat{r}_{12} \otimes \hat{r}_{12}]^{(2)} \cdot [\vec{\sigma}_1 \otimes \vec{\sigma}_2]^{(2)}$, and separating the spatial and spin-isospin parts of the TBMEs as well as the direct 
and exchange components, the potential energy becomes:
\begin{multline}
\mathcal{E}_{{P}}  = \displaystyle \frac{1}{2} \sum_l (-)^l \sum_{\vec{k}_1 s_1 t_1} \sum_{\vec{k}_2 s_2 q_2} \\
\displaystyle \times \big[ \mel*{\vec{k}_1 \vec{k}_2}{V(r_{12}) [\hat{r}_{12} \otimes \hat{r}_{12}]^{(2)}_{-l}}{\vec{k}_1 \vec{k}_2} \times\\ 
 \displaystyle  \mel*{s_1 t_1 s_2 t_2}{\mathcal{P}_{{D}} [\vec{\sigma}_1 \otimes \vec{\sigma}_2]^{(2)}_l}{s_1 t_1 s_2 t_2} \\
\displaystyle + \mel*{\vec{k}_1 \vec{k}_2}{V(r_{12}) [\hat{r}_{12} \otimes \hat{r}_{12}]^{(2)}_{-l}}{\vec{k}_2 \vec{k}_1} \times \\
\displaystyle  \mel*{s_1 t_1 s_2 t_2}{\mathcal{P}_{{E}} [\vec{\sigma}_1 \otimes \vec{\sigma}_2]^{(2)}_l}{s_1 t_1 s_2 t_2}  \big]
\end{multline}
The direct spin-isospin part of the TBMEs splits up in such a way that:
\begin{multline}
\displaystyle \mel*{s_1 t_1 s_2 t_2}{\mathcal{P}_{{D}} [\vec{\sigma}_1 \otimes \vec{\sigma}_2]^{(2)}_l}{s_1 t_1 s_2 t_2} \\
\displaystyle = \! \mel*{t_1 t_2}{\mathcal{P}_{{D}}}{t_1 t_2} \! \! \mel*{s_1 s_2}{[\vec{\sigma}_1 \otimes \vec{\sigma}_2]^{(2)}_l}{s_1 s_2}
\end{multline}
where 
\begin{equation}
\mel*{s_1 s_2}{[\vec{\sigma}_1 \otimes \vec{\sigma}_2]^{(2)}_l}{s_1 s_2} = 4 \sqrt{\frac{2}{3}} s_1 s_2 \delta_{l,0}.
\end{equation}
This implies that
the direct component of the tensor interaction does not contribute to the potential energy as it can be shown that 
$$\displaystyle \sum_{s_1 s_2} \! \mel*{s_1 s_2}{[\vec{\sigma}_1 \otimes \vec{\sigma}_2]^{(2)}_l}{s_1 s_2} = 0.$$

\noindent For the exchange spin-isospin part of the TBMEs, it splits up according to:
\begin{multline} 
\displaystyle \mel*{s_1 t_1 s_2 t_2}{\mathcal{P}_{{E}} [\vec{\sigma}_1 \otimes \vec{\sigma}_2]^{(2)}_l}{s_2 t_2 s_1 t_1} \\
\displaystyle = \mel*{t_1 t_2}{\mathcal{P}_{{E}}}{t_2 t_1} \! \! \mel*{s_1 s_2}{[\vec{\sigma}_1 \otimes \vec{\sigma}_2]^{(2)}_l}{s_2 s_1}
\end{multline}
where 
\begin{multline}
\displaystyle \mel*{s_1 s_2}{[\vec{\sigma}_1 \otimes \vec{\sigma}_2]^{(2)}_l}{s_2 s_1} \\
\displaystyle = 4 \sqrt{\frac{2}{3}} s_1 s_2 \delta_{l,0} (\delta_{s_1 s_2} + \delta_{s_1,-s_2})
\end{multline}
The spin part of the potential energy associated with the exchange component of the tensor interaction vanishes as:
\begin{equation}
\displaystyle \sum_{s_1 s_2} \! \mel*{s_1 s_2}{[\vec{\sigma}_1 \otimes \vec{\sigma}_2]^{(2)}_l}{s_2 s_1} = 0.
\end{equation}
The exchange component of the tensor interaction does not contribute to the potential energy either from spin-saturation properties of the system. \\

\paragraph{Energy in symmetric nuclear matter}

In the case of the symmetric INM characterized by the asymmetry parameter $\beta = 0$, one has $k_{{F}}^{\nu} = k_{{F}}^{\pi} = k_{{F}}$, and then $X_{\nu} = X_{\pi} = X$.  
The total kinetic energy per nucleon \eqref{kineticenergy} acquires the following expression:
\begin{equation} \label{Ekinsym}
\frac{\mathcal{E}_{{K}}^{{S}}}{A} = \frac{3}{5} \frac{\hbar^2 k_{{F}}^2}{2m}.
\end{equation}
As for the potential energy per nucleon defined in Eq.\eqref{potentialenergyINM}, it simplifies for each finite-range  central ($\alpha=0$) or density-dependent ($\alpha \ne 0$ and global factor $1/(\mu_3 \sqrt{\pi})^3$) terms as:
\begin{equation} \label{Epotsym}
\frac{\mathcal{E}^{{S}}_{{P}{\mu}}}{A} = 2 \rho^{\alpha} \sum_{ST} \bigg[ \mathcal{A}^{ST} \frac{X^3}{6}
- \mathcal{B}^{ST} f(X) \bigg].
\end{equation}
where 
\begin{equation}\label{Epotsyma}
    f(X)=  e^{-X^2} \Big( \frac{1}{X} - \frac{2}{X^3} \Big) - \frac{3}{X} + \frac{2}{X^3} + \sqrt{\pi} \erf(X) 
\end{equation}
In the above equations the $S$ and $T$ quantum numbers have been made explicit again to be able to evaluate the contribution of the energy in symmetric matter in each spin-isospin $(S,T)$ channel of the interaction, which will be useful in the following.
In the case of the zero-range density-dependent term of D1-type parameterizations the expression simplifies to 
\begin{equation}\label{eeqq3}
\frac{\mathcal{E}^{{S}}_{{P}{0}}}{A} = \frac{3}{8} t_0 \rho^{\alpha+1}
\end{equation}
for the contribution in the (S=1, T=0) channel.

\paragraph{Energy in neutron nuclear matter}

In the case of the neutron INM characterized by the asymmetry parameter $\beta = 1$,  one has $k_{{F}}^{\pi} = 0$ and $k_{{F}}^{\nu} = 2^{1/3} k_{{F}}$.
Thus, $X_{\pi} = 0$ and $X_{\nu} = 2^{1/3} X$.  The total kinetic energy per nucleon \eqref{kineticenergy} becomes:
\begin{equation}
\frac{\mathcal{E}_{K}^{{N}}}{A} = \frac{3}{5} \frac{\hbar^2 (k_{{F}}^{\nu})^2}{2m}.
\end{equation}
As for the potential energy per nucleon \eqref{potentialenergyINM}, it simplifies for each finite-range  central ($\alpha=0$) or density-dependent ($\alpha \ne 0$ and global factor $1/(\mu_3 \sqrt{\pi})^3$) terms as:
\begin{equation} 
\frac{V^\nu_{{P}\mu}}{A} = 4 \rho^{\alpha} \sum_{ST} \frac{T}{2T+1} \bigg[ \mathcal{A}^{ST} \frac{X^3}{6} - \mathcal{B}^{ST} \frac{Q(X_{\nu},X_{\nu})}{X_{\nu}^3} \bigg]
\end{equation}
where the dependence in $S$ and $T$ has been left explicit.

In the case of the zero-range density-dependent term of D1-type parameterizations, one has:
\begin{equation}\label{eeqq3bis}
\frac{\mathcal{E}^{\text{S}}_{{P}{0}}}{A} = \frac{1}{4} t_0 (1-x_0) \rho^{\alpha+1}
\end{equation}

\subsection{Physical quantities}\label{physicalquantities}

In this sub-section, one will derive the expressions of properties that
can be extracted from INM. One will focus on the incompressibility, the 
symmetry energy, its slope as well as the Landau parameters that are 
used in the selection process of the parameterization.
Details of the derivation are available in \cite{Zietek2023,Chappert2007} as well as in Appendix \ref{Landauparam}.

\subsubsection{Incompressibility $K_{\infty}$}

The incompressibility corresponds to the amount of energy needed to modify the density of the medium around the equilibrium density $\rho_0$. 
It is defined by:
\begin{equation} \label{Kinfdef}
K_{\infty} \equiv 9 \rho^2 \pdv[2]{\mathcal{E}/A}{\rho} \bigg\vert_{\rho_0}
\end{equation}
This formula is only valid at the energy minimum corresponding to the saturation density $\rho_0$. 
Given the expression of the total kinetic \eqref{kineticenergy} and potential \eqref{potentialenergyINM} energies, the derivation will be done according to 
the Fermi momentum $k_{{F}}$ rather than the density. The relation between the density and the Fermi momentum given in Eq.\eqref{densityINM} allows to write:
\begin{equation}
K_{\infty} \equiv k_{{F}}^2 \pdv[2]{\mathcal{E}/A}{k_{{F}}} \bigg\vert_{k_{{F}}^0}
\end{equation}
where the Fermi momentum $k_{{F}}^0$ is such that $k_{{F}}^0 = (3 \pi^2 \rho_0 / 2)^{1/3}$.

\paragraph{Kinetic energy contribution}
Combining Eq. \eqref{kineticenergy} and the Fermi momenta as functions of the asymmetry parameter $\beta$, $k_{{F}}^{\nu} = k_{{F}} (1+ \beta)^{1/3}$ and $k_{{F}}^{\pi} = k_{{F}} (1-\beta)^{1/3}$, one gets 
\begin{equation} \label{Kinfkin}
k_{{F}}^2 \pdv[2]{\mathcal{E}_{{K}}/A}{k_{{F}}} \bigg\vert_{k_{{F}}^0} = 2 \frac{\mathcal{E}_{{K}}}{A}\bigg\vert_{k_{{F}}^0}.
\end{equation}

\paragraph{Potential energy contribution}
From the potential energy \eqref{potentialenergyINM},  the calculation is quite long. 
The result obtained in INM with some asymmetry parameter $\beta$ leads to the following expression for each finite-range  central ($\alpha=0$) or density-dependent ($\alpha \ne 0$ and global factor $1/(\mu_3 \sqrt{\pi})^3$) terms:
\begin{multline} \label{KinfVpot}
\displaystyle k_{{F}}^2 \pdv[2]{\mathcal{E}_{{P}\mu}/A}{k_{{F}}} = \\
\displaystyle \frac{3(\alpha-1)(3\alpha-4) \rho^\alpha}{X^3} \bigg[ \frac{\mathcal{A}}{6} \big( X_\pi^6 + X_\nu^6 \big) \\
\displaystyle - \frac{\mathcal{A}'}{6} \big( X_\pi^3 - X_\nu^3 \big)^2 \\
\displaystyle + (\mathcal{B} - \mathcal{B}') \big( Q(X_\pi, X_\pi) + Q(X_\nu, X_\nu) \big) \\
\displaystyle - 2 \mathcal{B}' Q(X_\pi, X_\nu) \bigg] \\
\displaystyle - \frac{6(\alpha-1)\rho^\alpha}{X^2} \bigg[ \mathcal{A} \big( (1-\beta)^{1/3} X_\pi^5 + (1+\beta)^{1/3} X_\nu^5 \big) \\
\displaystyle - \mathcal{A}' \big( (1-\beta)^{1/3} X_\pi^5 - (1+\beta)^{1/3} X_\nu^5 \big) \big( X_\pi^3 - X_\nu^3 \big) \\
\displaystyle + (\mathcal{B} - \mathcal{B}') \bigg( \pdv{X} Q(X_\pi, X_\pi) + \pdv{X} Q(X_\nu, X_\nu) \bigg) \\
\displaystyle - 2 \mathcal{B}' \pdv{X} Q(X_\pi, X_\nu) \bigg] \\
 + \frac{\rho^\alpha}{X} \bigg[ 5 \mathcal{A} \big( (1-\beta)^{2/3} X_\pi^4 + (1+\beta)^{2/3} X_\nu^4 \big) \\
\displaystyle - \mathcal{A}' \Big( 2 \big( (1-\beta)^{2/3} X_\pi^4 - (1+\beta)^{2/3} X_\nu^4 \big) \big( X_\pi^3 - X_\nu^3 \big) \\
\displaystyle + 3 \big( (1-\beta)^{1/3} X_\pi^2 - (1+\beta)^{1/3} X_\nu^2 \big)^2 \Big) \\
 + (\mathcal{B} - \mathcal{B}') \bigg( \pdv[2]{X} Q(X_\pi, X_\pi) + \pdv[2]{X} Q(X_\nu, X_\nu) \bigg) \\
\displaystyle -2\mathcal{B}' \pdv[2]{X} Q(X_\pi, X_\nu)  \bigg]
\end{multline}
that has to be evaluated at $k_{{F}} = k_{{F}}^0$.  The derivatives of the function $Q$ have been performed with respect to 
$X = \mu k_{{F}}$ instead of $k_{{F}}$ to get rid of overall $\mu$ factors. 
The first derivative reads:
\begin{multline}
\displaystyle \pdv{X} Q(X_\pi, X_\nu) =\\ 
\displaystyle \e^{-\big( \frac{X_\pi + X_\nu}{2} \big)^2} \bigg[ (1-\beta)^{1/3} (2 X_\pi - X_\nu) \\
\displaystyle~~~~~~~~+ (1+\beta)^{1/3} (2 X_\nu - X_\pi) \\
\displaystyle~~~~~~~~ - \big( (1-\beta)^{1/3} + (1+\beta)^{1/3} \big) \\
\displaystyle\times \bigg( \Big( \frac{X_\pi+X_\nu}{2} \Big) \big( X_\pi^2 + X_\nu^2 - X_\pi X_\nu - 2 \big) \\
\displaystyle - \Big( \frac{X_\pi^3 + X_\nu^3}{2} \Big) \bigg) \bigg] \\
\displaystyle - \e^{-\big( \frac{X_\pi - X_\nu}{2} \big)^2} \bigg[ (1-\beta)^{1/3} (2 X_\pi + X_\nu) \\
\displaystyle~~~~~~~~+ (1+\beta)^{1/3} (2 X_\nu + X_\pi) \\
\displaystyle~~~~~~~~ - \big( (1-\beta)^{1/3} - (1+\beta)^{1/3} \big) \\
\displaystyle \times \bigg( \Big( \frac{X_\pi-X_\nu}{2} \Big) \big( X_\pi^2 + X_\nu^2 + X_\pi X_\nu - 2 \big) \\
\displaystyle - \Big( \frac{X_\pi^3 - X_\nu^3}{2} \Big) \bigg) \bigg] \\
\displaystyle + \frac{3 \sqrt{\pi}}{2} \bigg[ \big( (1-\beta)^{1/3} X_\pi^2 + (1+\beta)^{1/3} X_\nu^2 \big) \\
\times \displaystyle \erf\Big( \frac{X_\pi+X_\nu}{2} \Big) \\
\displaystyle - \big( (1-\beta)^{1/3} X_\pi^2 - (1+\beta)^{1/3} X_\nu^2 \big) \\
\displaystyle \erf\Big( \frac{X_\pi-X_\nu}{2} \Big) \bigg]
\end{multline}
For the second derivative, one obtains:
\begingroup
\allowdisplaybreaks
\begin{multline}
\displaystyle \pdv[2]{X} Q(X_\pi, X_\nu) = \\
\displaystyle \e^{-\big( \frac{X_\pi + X_\nu}{2} \big)^2} \bigg\{ 2 \big( (1-\beta)^{2/3} + (1+\beta)^{2/3} \\
\displaystyle - (1-\beta)^{1/3} (1+\beta)^{1/3} \big) 
\displaystyle - \big( (1-\beta)^{1/3} + (1+\beta)^{1/3} \big) \\
\displaystyle \times \bigg[ \Big( \frac{(1-\beta)^{1/3} + (1+\beta)^{1/3}}{2} \Big) \\
\displaystyle \bigg( \big( X_\pi^2 + X_\nu^2 - X_\pi X_\nu - 2 \big) \\
\displaystyle + (X_\pi + X_\nu) \Big( \frac{X_\pi^3 + X_\nu^3}{2} \Big) \bigg) \\
\displaystyle - 3 \big( (1-\beta)^{1/3} X_\pi^2 + (1+\beta)^{1/3} X_\nu^2 \big) \\
\displaystyle + \Big( \frac{X_\pi + X_\nu}{2} \Big) \Big( 2 \big( (1-\beta)^{1/3} (2 X_\pi - X_\nu) \notag \\
\displaystyle + (1+\beta)^{1/3} (2 X_\nu - X_\pi) \big) - \big( (1-\beta)^{1/3} + (1+\beta)^{1/3} \big) \\
\displaystyle \times \Big( \frac{X_\pi + X_\nu}{2} \Big) \big( X_\pi^2 + X_\nu^2 - X_\pi X_\nu - 2 \big) \Big) \bigg] \bigg\} \\
\displaystyle - \e^{-\big( \frac{X_\pi - X_\nu}{2} \big)^2} \bigg\{ 2 \big( (1-\beta)^{2/3} + (1+\beta)^{2/3}  \\
\displaystyle + (1-\beta)^{1/3} (1+\beta)^{1/3} \big) - \big( (1-\beta)^{1/3} - (1+\beta)^{1/3} \big)  \\
\displaystyle \times \bigg[ \Big( \frac{(1-\beta)^{1/3} - (1+\beta)^{1/3}}{2} \Big) \\
\displaystyle \times \bigg( \big( X_\pi^2 + X_\nu^2 + X_\pi X_\nu - 2 \big) \\
\displaystyle + (X_\pi - X_\nu) \Big( \frac{X_\pi^3 - X_\nu^3}{2} \Big) \bigg) \\
\displaystyle - 3 \big( (1-\beta)^{1/3} X_\pi^2 - (1+\beta)^{1/3} X_\nu^2 \big)  \\
\displaystyle + \Big( \frac{X_\pi - X_\nu}{2} \Big) \Big( 2 \big( (1-\beta)^{1/3} (2 X_\pi + X_\nu)  \\
\displaystyle + (1+\beta)^{1/3} (2 X_\nu + X_\pi) \big) \\
\displaystyle - \big( (1-\beta)^{1/3} - (1+\beta)^{1/3} \big)  \\
\times \Big( \frac{X_\pi - X_\nu}{2} \Big) \big( X_\pi^2 + X_\nu^2 + X_\pi X_\nu - 2 \big) \Big) \bigg] \bigg\}  \\
\displaystyle + 3 \sqrt{\pi} \bigg[ \big( (1-\beta)^{2/3} X_\pi + (1+\beta)^{2/3} X_\nu \big) \\
\displaystyle \erf\Big( \frac{X_\pi+X_\nu}{2} \Big)  \\
\displaystyle - \big( (1-\beta)^{2/3} X_\pi - (1+\beta)^{2/3} X_\nu \big) \\
\displaystyle \erf\Big( \frac{X_\pi-X_\nu}{2} \Big) \bigg]
\end{multline}
\endgroup
The derivatives of $Q(X_\pi, X_\pi)$ and $Q(X_\nu, X_\nu)$ associated with either protons or neutrons can easily be deduced from the above ones by making the substitutions $X_\nu \rightarrow X_\pi$ and $1+\beta \rightarrow 1-\beta$, and $X_\pi \rightarrow X_\nu$ and $1-\beta \rightarrow 1+\beta$, respectively.

\paragraph{Incompressibility in symmetric nuclear matter}

By definition of the symmetric INM ($\beta = 0$), $k_{{F}}^{\nu} = k_{{F}}^{\pi} = k_{{F}}$, and then $X_{\nu} = X_{\pi} = X$.  Thus, the contribution of the potential energy to the incompressibility \eqref{KinfVpot} is greatly simplified, and becomes:
\begin{equation} \label{Kinfsym}
\begin{split}
k_{{F}}^2 \pdv[2]{\mathcal{E}_{{P}}^{\text{S}}/A}{k_{{F}}} & = \frac{3(\alpha-1)(3 \alpha-4) \rho^\alpha}{X^3} \bigg[ \frac{\mathcal{A}}{3} X^6 - 2 \mathcal{B} Q(X,X) \bigg] \\
& + \frac{12 (\alpha-1) \rho^\alpha}{X^2} \bigg[ \mathcal{A} X^5 - 2 \mathcal{B} \pdv{X} Q(X,X) \bigg] \\
& + \frac{2\rho^\alpha}{X} \bigg[ 5 \mathcal{A} X^4 - \mathcal{B} \pdv[2]{X} Q(X,X) \bigg]
\end{split}
\end{equation}
that has to be evaluated at $k_{{F}} = k_{{F}}^0$. 
The function $Q$ is simply:
\begin{equation} 
Q(X,X) = \e^{-X^2}(X^2-2) - 3X^2 + 2 + \sqrt{\pi} X^3 \erf(X)
\end{equation}
while the derivatives are given by:
\begin{equation}
\frac{1}{3} \pdv{X} Q(X,X) = 2X\big(\e^{-X^2}-1\big) + \sqrt{\pi} X^2 \erf(X)
\end{equation}
and
\begin{equation}
\frac{1}{6} \pdv[2]{X} Q(X,X) = \e^{-X^2}(1-X^2) - 1 + \sqrt{\pi} X \erf(X).
\end{equation}

In the case of the zero-range density-dependent term of D1-type parameterizations, one obtains:
\begin{equation}
k_{{F}}^2 \pdv[2]{\mathcal{E}^S_{{P}0}/A}{k_{{F}}} =
3\left( \alpha +1 \right) \left( 3 \alpha + 2\right)
\frac{\mathcal{E}^S_{P 0}}{A} \bigg\vert_{k_{{F}}^0}
\end{equation}



\subsubsection{Symmetry energy $a_{\tau}$ and its slope $L$}

The symmetry energy quantifies the variation of the total energy as a function of the asymmetry parameter $\beta$ of the medium. 
The symmetry energy $\mathcal{E}_{{sym}}$ is defined by:
\begin{equation}
\mathcal{E}_{{sym}} \equiv \frac{1}{2} \pdv[2]{\mathcal{E}/A}{\beta}\Big\vert_{\beta=0}.
\end{equation}
The derivative is evaluated at zero asymmetry ($\beta = 0$) as the symmetric matter is the reference point around which the energy fluctuations 
due to the asymmetry are evaluated.

From Eq.\eqref{kineticenergy}, one obtains the kinetic energy contribution:
\begin{equation}\label{kinesym}
\frac{1}{2} \pdv[2]{\mathcal{E}_{{K}}/A}{\beta}\Big\vert_{\beta=0} = \frac{1}{3} \frac{\hbar^2 (k_{{F}}^t)^2}{2M} = 	\frac{5}{9} \frac{\mathcal{E}_{K}^{\text{S}}}{A}
\end{equation}
which is expressed in terms of the total kinetic energy per nucleon of symmetric matter \eqref{Ekinsym}. 

Concerning the potential energy contribution,
the calculation is quite long and tedious, from the potential energy expression \eqref{potentialenergyINM}. 
The evaluation at $\beta = 0$ simplifies the result so that
for each finite-range central ($\alpha=0$) or density-dependent ($\alpha \ne 0$ and global factor $1/(\mu_3 \sqrt{\pi})^3$) terms:
\begin{multline}\label{potsym}
\displaystyle\frac{1}{2} \pdv[2]{\mathcal{E}_{{P}\mu}/A}{\beta}\Big\vert_{\beta=0} = \frac{\rho^\alpha}{3} \bigg[ (\mathcal{A} - 2 \mathcal{A}') X^3 \\
\displaystyle + 2 \mathcal{B} \bigg( \Big( \frac{1}{X} + X \Big) \e^{-X^2} - \frac{1}{X} \bigg) \\
\displaystyle  + 2 \mathcal{B}' X \big( 1-\e^{-X^2} \big)  \bigg]
\end{multline}
where the coefficient $\mathcal{A}$, $\mathcal{A}'$, $\mathcal{B}$, $\mathcal{B}'$s are defined by Eqs.\eqref{ABAPBP}.
In the case of the zero-range density-dependent term of D1-type parameterizations, one obtains:
\begin{equation}\label{symzero}
\frac{1}{2} \pdv[2]{\mathcal{E}_{{P}0}/A}{\beta}\Big\vert_{\beta=0} = -\frac{\rho^{\alpha+1}}{8} t_0 (1+2 x_0)
\end{equation}

The slope $L$ of the symmetry energy is by definition:
\begin{equation}\label{defl}
L = 3 \rho \frac{\partial \mathcal{E}_{{sym}}}{\partial \rho}
\Big\vert_{\rho_0} =L_K + L_{P}
\end{equation}
where $L_K$ and $L_{P}$ are the kinetic and potential
contributions.

Applying the previous formula to the kinetic contribution
expressed in Eq.\eqref{kinesym}, one obtains the kinetic
contribution $L_K$ to the slope:
\begin{equation}
L_K = \frac{2 \hbar^2}{3} (\frac{3 \pi^2}{2} \rho)^{2/3}\Big\vert_{\rho_0}
\end{equation}

Starting from the definition \eqref{defl} to each 
central and density-dependent term $L_{P \mu}$ of the DG interaction leads to, using Eq.\eqref{potsym}:
\begin{eqnarray} \begin{cases}
L_{P \mu} = 0 ~~~~ \text{for $\mu \equiv \mu_1$ and $\mu_2$} \\
\displaystyle L_{P \mu_3} = \frac{\displaystyle \rho^{1/3}}{\displaystyle \left(\mu_3 \sqrt{\pi}\right)^3} \bigg[ (\mathcal{A} - 2 \mathcal{A}') X^3 \\
\displaystyle + 2 \mathcal{B} \bigg( \Big( \frac{1}{X} + X \Big) \e^{-X^2} - \frac{1}{X} \bigg) \\
\displaystyle  + 2 \mathcal{B}' X \big( 1-\e^{-X^2} \big)  \bigg] \Big\vert_{\rho_0}
\end{cases}
\end{eqnarray}
The expression of the contribution of the zero-range density-dependent term of the D1-type parameterization is 
straightforward from Eq.\eqref{symzero}.

\subsubsection{Effective masses}

The effective mass $m^*$ corresponds to the renormalized mass the nucleons would have if they were considered as moving freely, rather than submitted to the nuclear potential with their actual mass $m$.  By definition, the effective mass of an isospin-$t$ particle is defined by:
\begin{equation}
\frac{1}{m_t^*} \equiv \frac{1}{\hbar^2 k} \dv{\varepsilon_t(\vec{k})}{k}\bigg\vert_{k=k_{{F}}^t}
\end{equation}
where the single-particle energy of an isospin-$t$ particle, is given in INM by:
\begin{equation}
\varepsilon_t(\vec{k}) = \frac{\hbar^2 k^2}{2m} + \Gamma_t(\vec{k}) + \partial \Gamma.
\end{equation}
As the rearrangement field $\partial \Gamma$ is the same for protons and neutrons and more importantly it does not depend on the momentum 
$\vec{k}$, it will not contribute to the effective mass defined. Thus,
\begin{equation} \label{effectivemass}
\frac{1}{m_t^*} = \frac{1}{m} + \frac{1}{\hbar^2 k} \dv{\Gamma_t(\vec{k})}{k}\bigg\vert_{k=k_{{F}}^t}.
\end{equation}
The calculation of the effective mass implies to derive the mean field $\Gamma_t(\vec{k})$ for each term of the generalized Gogny interaction which are
linked to the previous expressions of the potential energy by $\mathcal{E}_P= \frac{1}{2}  \sum_a \Gamma_{aa} \rho_{aa}$.   

\paragraph{Central and density-dependent term contributions}

The central and density-dependent interactions are encompassed in the expression \eqref{CDDpot}. Following the exact same steps that led to the contributions of those interactions to the potential energy per nucleon given by Eq.\eqref{vpotintermediate}, one finds for the expression of the mean-field $\Gamma_t(\vec{k})$ associated with each finite-range  central ($\alpha=0$) or density-dependent ($\alpha \ne 0$ and global factor $1/(\mu_3 \sqrt{\pi})^3$) terms:
\begin{multline*}
\displaystyle \Gamma_t(\vec{k}) = 4 \rho^\alpha \sum_{ST} \sum_{t'} \frac{1}{2T+1} \big(1+(2T-1) \delta_{tt'} \big) \times \\
\displaystyle \bigg[ \frac{\mathcal{A}^{ST}}{6} X_{q'}^3 + \times \\
\mathcal{B}^{ST} \Big( g_{\mu k} \Big( \mu \frac{k - k_{{F}}^{t'}}{2} \Big) - g_{\mu k} \Big( \mu \frac{k + k_{{F}}^{t'}}{2} \Big)  \Big) \bigg]
\end{multline*}
where the function $g_{\mu k}$ is defined by:
\begin{equation} \label{fonctiong}
g_{\mu k}(x) \equiv \frac{\sqrt{\pi}}{2} \erf(x) + \frac{1}{{\mu k}} \, \e^{-x^2}.
\end{equation}
Taking the derivative of $\Gamma_t(\vec{k})$ and evaluating it at $k = k_{{F}}^t$, one obtains:
\begin{multline*}
\dv{\Gamma_t(\vec{k})}{k}\bigg\vert_{k=k_{{F}}^t} = -4 \rho^\alpha \sum_{ST} \sum_{t'} \frac{1}{2T+1} \times \\
\big(1+(2T-1) \delta_{tt'} \big) \times \\
\mathcal{B}^{ST} \bigg[ \e^{-\big( \frac{X_t - X_{t'}}{2} \big)^2} \Big( \frac{1}{X_t k_{{F}}^t} + \frac{X_t - X_{t'}}{2k_{{F}}^t} - \frac{\mu}{2} \Big) \\
 - \e^{-\big( \frac{X_t + X_{t'}}{2} \big)^2} \Big( \frac{1}{X_t k_{{F}}^t} + \frac{X_t + X_{t'}}{2k_{{F}}^t} - \frac{\mu}{2} \Big) \bigg]
\end{multline*}
where the following formula has been used for the derivatives:
\begin{multline*}
\dv{k} g_{\mu k} \Big( \mu \frac{k \pm k_{{F}}^{t'}}{2} \Big) = - \e^{\big( \mu \frac{k \pm k_{{F}}^{t'}}{2} \big)^2} \times \\
\Big( \frac{1}{\mu k^2} + \mu \frac{k \pm k_{{F}}^{t'}}{2k} - \frac{\mu}{2}  \Big)
\end{multline*}
Finally, the contribution of the central and density-dependent terms to the effective mass can be written as:
\begin{equation} \label{effmassformula}
- \frac{4 \mu^2 \rho^\alpha}{\hbar^2} \sum_{ST} \frac{1}{2T+1} \frac{\mathcal{B}^{ST}}{X_t^3} \big( 2T f(t,t) + f(t,-t) \big)
\end{equation}
where the isospin-dependent function $f(t,t')$ is defined by:
\begin{multline}
f(t,t') \equiv \bigg( 1 - \frac{X_t X_{t'}}{2} \bigg) \, \e^{ -\big( \frac{X_t - X_{t'}}{2} \big)^2} \\
- \bigg( 1 + \frac{X_t X_{t'}}{2} \bigg) \, \e^{ -\big( \frac{X_t + X_{t'}}{2} \big)^2}
\end{multline}
\noindent The zero-range density-dependent term of D1-type parameterizations
does not participate to the effective mass.

\paragraph{Spin-orbit contribution}

The mean field associated with the spin-orbit interaction \eqref{SOpot} does not contribute to the effective mass \eqref{effectivemass}. 
Indeed, the spatial matrix elements are zero in the plane wave function representation, as shown in subsection \ref{sopotzero}.

\paragraph{Tensor contribution}

In a similar manner that the potential energy associated with the tensor interaction is zero (see subsection \ref{tensorpotzero}), 
one shows
that the corresponding mean-field vanishes because of its spin part. Then, the tensor interaction does not contribute to the effective mass defined in Eq.\eqref{effectivemass}.

\subsubsection{Landau parameters}\label{lpder}

In this section, one presents the main steps leading to the definition of the Landau parameters, the stability criteria 
and the sum rules associated with a DG-type analytical form of
the Gogny interaction. 
Some of them are used in the fitting process as filters.
They are derived within the theory of normal Fermi liquids in symmetric nuclear matter. Those quantities will serve as good 
indicators of the consistency of the parameterizations. 
In view of that, one will remind firstly the general context which allows to introduce the notion of the effective quasiparticle interaction, that will be linked to the Landau parameters.

The effective quasiparticle interaction is defined in the context of the time-dependent Hartree-Fock (TDHF) theory in the limit of small amplitudes.
One starts from the TDHF equations and assumes a weak external field (only small amplitudes oscillations of the stationary density are allowed, which is itself solution of the stationary HF equations). Working in the particle-hole representation within the HF basis, one shows \cite{Ring2004,Gogny2004,Gogny1977} that the Random Phase Approximation (RPA) equations are recovered, with a residual interaction given by:
\begin{equation} \label{Vpheffective}
\mel*{ph'}{V_{{ph}}}{hp'} = \pdv[2]{\mathcal{E}_{\text{HF}}}{\rho_{hp}}{\rho_{p'h'}},
\end{equation}
where $\mathcal{E}_{\text{HF}}$ is the HF total binding energy defined as:
\begin{equation}
\mathcal{E}_{\text{HF}} = \sum_{ab} \! \mel*{a}{t_{{K}}}{b} \! \rho_{ba} + \frac{1}{2} \sum_{abcd} \! \mel*{ac}{v_{12}^{{({a})}}}{bd} \! \rho_{ba} \rho_{db}.
\end{equation}
The residual interaction at the RPA level traduces the oscillations of small amplitudes of 
the system and is referred to as the quasiparticle (or particle-hole) interaction, denoted $V_{{ph}}$ in the following. 
For a density-dependent interaction like the DG Gogny interaction defined in Eq.\eqref{gognyDG}, the second derivative of the HF energy with respect to the density matrices provides:

\begin{multline} \label{MEph}
\mel*{ph'}{V_{{ph}}}{hp'} = \mel*{ph'}{v_{12}^{{(a)}}}{hp'} + \\ 
\sum_i \mel*{h'i}{\pdv{v_{12}^{{(a)}}}{\rho_{hp}}}{p'i} + \sum_i \! \mel*{pi}{\pdv{v_{12}^{{(a)}}}{\rho_{p'h'}}}{hi}  \\
+ \frac{1}{2}  \sum_{ij} \mel*{ij}{\pdv[2]{v_{12}^{{(a)}}}{\rho_{hp}}{\rho_{p'h'}}}{ij}
\end{multline}

In Eq.\eqref{MEph} the summations are performed on the occupied states, i.e.\ the hole states in the particle-hole representation. 
The last three quantities of the right-hand side are specific to density-dependent interactions as they involve derivatives of the density. They 
corresponds to rearrangement terms which play a crucial role in the RPA theory. Since the particle-hole interaction $\mel*{ph'}{V_{{ph}}}{hp'}$ is here expressed in terms of the effective interaction $v_{12}$, it defines the effective quasiparticle (or particle-hole) interaction.

\paragraph{Theory of normal Fermi liquids}

The general theory of normal Fermi liquid gives a general frame to the residual interation in the context of RPA theory.
References on the subject can be found in \cite{Landau1957a,Landau1957b,Landau1959,Nozieres1964,Backman1968,Brown1971}. In the following, one presents the general context in order
to make the link with the effective quasiparticle (or particle-hole) interaction.

To build this theory, an ideal system made up of a non-interacting homogeneous gas of fermions at zero temperature is considered. 
The system presents translational invariance and the one-body wave functions are the plane waves defined in Eq.\eqref{planew}. 
Accordingly, the wave function of the whole system corresponds to an antisymmetric linear combination of products of those plane waves.
An eigen-state of the system is defined by listing which are the occupied plane waves. This information is provided by the distribution function $\rho(\vec{k})$, where $\vec{k}$ is the momentum of the plane wave. 
Under the above hypotheses, the distribution function of the system in the ground state corresponds to an isotropic distribution of the type:
\begin{equation} \label{distrib}
\rho(\vec{k}) =  
\begin{cases} 
  1 \quad & \text{for} \ \ |\vec{k}| \le k_{{F}},  \\
  0 \quad & \text{for} \ \ |\vec{k}| > k_{{F}}.
\end{cases}
\end{equation}
This implies that the Fermi surface of a normal Fermi liquid in momentum space is spherical. \footnote{In the case of a spin-polarized medium,  corresponding to $N_\uparrow + Z_\uparrow \ne N_\downarrow + Z_\downarrow$, two distinct Fermi surfaces, one for spin up and another for spin down nucleons show up. Besides, the problem becomes more difficult since the spin excess introduces an anisotropy in the system so that the Fermi surfaces are not spherical in general. In particular, the Fermi surfaces deviate from spherical shapes because of the coupling between momentum and spin implied by the \textit{tensor term} of the interaction. These observations remain true in a spin-isospin polarized medium, corresponding to $N_\uparrow + Z_\downarrow \ne N_\downarrow + Z_\uparrow$, but with four Fermi surfaces (one for each spin and isospin projection) \cite{Dabrowski1976}.}
The fundamental assumption of the Landau's theory of Fermi liquids is that when one gradually turns on the interaction, the ideal system adiabatically develops into the real system in such a way that, in the interacting system, the one-body wave functions remain the plane waves and the distribution function in the ground state is still given by the definition above. 
Note, however, that the distribution function now describes the occupation of quasiparticle states. Indeed, if one adds a particle with momentum $|\vec{k}| > k_{{F}}$ (or, equivalently, a hole with momentum $|\vec{k}| \le k_{{F}}$) to the ideal ground state and turns on the interaction, the system becomes a real ground state plus a quasiparticle of momentum $\vec{k}$.
If a system obeys Landau's assumption, it is said to be a normal Fermi liquid. For quasiparticles to be a meaningful concept, they must have reasonably long lifetimes. 
In a normal Fermi liquid, the characteristic length of a quasiparticle with momentum $\vec{k}$ at zero temperature is proportional to $(k-k_{{F}})^2$, because of the Pauli principle. Since the typical lifetime of a quasiparticle is inversely proportional to its characteristic length, assuming the quasiparticles to be close enough to 
the Fermi surface guarantees them sufficiently long lifetimes.

The energy $\mathcal{E}$ of the real system depends on the distribution function of the quasiparticles. Landau investigated how this energy is modified when one changes the distribution function $\rho(\vec{k})$ by an amount $\updelta \rho(\vec{k})$ \cite{Landau1957a,Landau1957b,Landau1959}. Geometrically, this equates to translate the Fermi sphere by a small momentum (giving a uniform velocity to the system) without modifying its size or its shape. The resulting change in energy at the first order in $\updelta \rho(\vec{k})$ is
\begin{equation}
\updelta \mathcal{E} = \sum_{\vec{k}} \varepsilon(\vec{k}) \updelta \rho (\vec{k}),
\end{equation}
where the summation should also run over spin and isospin projections. To simplify the treatment, they have been omitted temporary. The quantity $\varepsilon(\vec{k})$ is interpreted as the energy of a quasiparticle with momentum $\vec{k}$. The quasiparticle energy is obtained by varying the energy with respect to the quasiparticle distribution, i.e.\ 
\begin{equation} \label{dvvareps}
\varepsilon(\vec{k}) \equiv \frac{\updelta \mathcal{E}[\rho]}{\updelta \rho(\vec{k})}.
\end{equation}
Actually, Landau found out that result demanding the conservation of the quasiparticle momentum \cite{Brown1971}.  
Moreover, the energy of a quasiparticle is modified if the distribution of the other quasiparticles changes. 
Thus, Landau also took into account the second order variations of the distribution function by writing the total change in energy as:
\begin{equation} \label{deltaE}
\updelta \mathcal{E} = \sum_{\vec{k}} \varepsilon_0(\vec{k}) \updelta \rho (\vec{k}) + \frac{1}{2} \sum_{\vec{k} \vec{k}'} \mathcal{F}(\vec{k}, \vec{k}') \updelta \rho (\vec{k}) \updelta \rho (\vec{k}').
\end{equation}
It appears that $\varepsilon_0(\vec{k})$ corresponds to the energy of a non-interacting quasiparticle.  By means of relation \eqref{dvvareps}, a variation of the distribution function of the other quasiparticles changes $\varepsilon_0(\vec{k})$ into:
\begin{equation}
\varepsilon(\vec{k}) = \varepsilon_0(\vec{k}) + \sum_{\vec{k}'} \mathcal{F}(\vec{k}, \vec{k}') \updelta \rho(\vec{k}'),
\end{equation}
so that the function $\mathcal{F}(\vec{k}, \vec{k}')$ describes the interaction between the quasiparticles. Formally, it can be written as the second functional derivative of the energy with respect to the distribution function:
\begin{equation}\label{quasiinter}
\mathcal{F}(\vec{k}, \vec{k}') \equiv \frac{\updelta \varepsilon [\rho (\vec{k})]}{\updelta \rho(\vec{k}')} \equiv \frac{\updelta^2 \mathcal{E} [\rho]}{\updelta \rho(\vec{k}) \updelta \rho(\vec{k}')}
\end{equation}
where $\vec{k}$ and $\vec{k}'$ denote the momenta of the two particles. 
This quantity $\mathcal{F}(\vec{k}, \vec{k}')$ is called the quasiparticle (or particle--hole) interaction in the light of its physical interpretation. 
It is only defined at the Landau limit, i.e.\ near the Fermi surface where $|\vec{k}| \simeq |\vec{k}'| \simeq k_{{F}}$, because of the reasonably long lifetime requirement of quasiparticles discussed earlier. 
The Landau limit is sometimes called the limit of equal momenta or, accordingly, the long-wavelength limit. Thus, at this limit, the particle-hole interaction only 
depends on the angle between the momenta $\vec{k}$ and $\vec{k}'$, sometimes referred to as the Landau angle $\theta$.  

This quasiparticle interaction \eqref{quasiinter} is defined in the same way as the effective quasiparticle interaction calculated at the TDHF approximation in the limit of small amplitude oscillations \eqref{Vpheffective}. 
Those quantities coincide at the Landau limit, that is to say: 
\begin{multline}
\mathcal{F}(\vec{k}_{{F}}, \vec{\sigma}_1, \vec{\tau}_1;\vec{k}_{{F}}', \vec{\sigma}_2, \vec{\tau}_2) = V^{{L}}_{{ph}} \\
=\mel*{\vec{k}_{F} \vec{\sigma}_1 \vec{\tau}_1 ; \vec{k}_{F}' \vec{\sigma}_2 \vec{\tau}_2}{V_{{ph}}}{\vec{k}_{F} \vec{\sigma}_1 \vec{\tau}_1 ; \vec{k}_{F}' \vec{\sigma}_2 \vec{\tau}_2}
\end{multline}
where one has restored the spin and isospin dependences of the quasiparticle interaction. Midgal \cite{Migdal1967} was the first to establish this link between the theory of normal Fermi liquids and heavy atomic nuclei, in the framework of INM. This equivalence was in particular showed by Gogny and Padjen for the D1 interaction in Ref.\cite{Gogny1977}.
Therefore, in homogeneous and symmetric INM, the particle-hole interaction can be parametrized as:
\begin{multline} \label{QPinteraction}
V^{{L}}_{{ph}} = N_0^{-1} \Big[ f(\theta) + g(\theta) (\vec{\sigma}_1 \cdot \vec{\sigma}_2) + f'(\theta) (\vec{\tau}_1 \cdot \vec{\tau}_2) \\
+ g'(\theta) (\vec{\sigma}_1 \cdot \vec{\sigma}_2) (\vec{\tau}_1 \cdot \vec{\tau}_2) \\
+ h(\theta) \frac{q_{{F}}^2}{k_{{F}}^2} S_{12}(\hat{q}_{12}) \\
+ h'(\theta) \frac{q_{{F}}^2}{k_{{F}}^2} S_{12}(\hat{q}_{12}) (\vec{\tau}_1 \cdot \vec{\tau}_2) \Big]
\end{multline}

where the density of quasiparticle states $N_0$ at the Fermi surface is defined by $N_0 \equiv 2 m^{*} \mathcal{V} k_{{F}}/{\pi^2 \hbar^2}$.
This factor ensures that the parameters $f,f',g,g',h$ and $h'$ are dimensionless quantities. 
These parameters only depend on the angle $\theta$ between the momenta $\vec{k}_{{F}}$ and $\vec{k}_{{F}}'$ of the quasiparticle pair. 
As for $S_{12}(\hat{q}_{12})$, it is the tensor operator in momentum space defined, by analogy with \eqref{tensoroperatormain} in coordinate space, as
\begin{equation} \label{S12q}
S_{12}(\hat{q}_{12}) = (\vec{\sigma}_1 \cdot \hat{q}_{12}) (\vec{\sigma}_2 \cdot \hat{q}_{12}) - \frac{1}{3} \vec{\sigma}_1 \cdot \vec{\sigma}_2
\end{equation}
where $\hat{q}_{12} \equiv \vec{q}_{12}/|\vec{q}_{12}|$ is the direction pointed by the relative momentum. 

The expression \eqref{QPinteraction} is the most general form of the quasiparticle interaction considering the symmetry constraints to be satisfied. 
The first four terms are called the central terms as they are related to the central part of the interaction while the two last ones are the tensor terms which are related to the non-central part of the interaction. Accordingly, the central parameters $f,g,f'$ and $g'$ respectively contribute to the $(S=0,T=0)$,  $(S=1,T=0)$, $(S=0,T=1)$ and $(S=1,T=1)$ channels of the interaction while the tensor parameters $h$ and $h'$ respectively contribute to the $(S=1,T=0)$ and $(S=1,T=1)$ channels of the interaction. 
In the following, these parameters will be casted into the quantities $f^{ST} (\theta)$, with $f^{00}(\theta) \equiv f(\theta)$, $f^{10}(\theta) \equiv g(\theta)$, $f^{01}(\theta) \equiv f'(\theta)$ and $f^{11}(\theta) \equiv g'(\theta)$, and $h^{1T}(\theta)$, with $h^{10}(\theta) = h(\theta)$ and $h^{11}(\theta) = h'(\theta)$.

Since the particle-hole interaction only depends on the Landau angle $\theta$ between the quasiparticles, these quantities can be expanded in series of Legendre polynomials with argument $\hat{k} \cdot \hat{k}' = \cos \theta$, according to:
\begin{subequations} \label{legendreexpansion}
\begin{align}
f^{ST}(\theta) & = \sum_l f_l^{ST} P_l(\cos \theta)\label{expansionf} \\
h^{1T}(\theta) & = \sum_l h_l^{1T} P_l(\cos \theta) \label{expansionh}
\end{align}
\end{subequations}
where the coefficients $f_l^{ST}$ and $h_l^{1T}$ of these expansions are the so-called Landau parameters that one needs to evaluate, in principle for all values of $l$, to fully characterize the particle--hole interaction.  In fact, the serie converges quite rapidly in the sense that the absolute value of the coefficients decreases sharply as $l$ increases. Thus, the calculation of the coefficients for the first values of $l$ is important in practice. 
Finally, contrary to the central terms, the tensor terms exhibit a factor $q_{{F}}^2/k_{{F}}^2$, where the relative momentum at the Fermi surface is defined by $\vec{q}_{{F}} \equiv \vec{k}_{{F}} - \vec{k}'_{{F}}$. Its magnitude reads $q_{{F}} = k_{{F}} \sqrt{2 (1 - \cos \theta)}$.
This factor is conventional but often chosen as it is well-suited to calculate the response function.  
The response function will not be calculated here, but this convention has been kept to simplify the contribution of the tensor force to the sum rules 
(see further).
Some authors have defined the tensor terms without this factor, claiming that it leads to a faster convergence of the corresponding Landau parameters \cite{Olsson2004}. In any case, there exists a recurrence relation linking the two conventions \cite{Olsson2004}.

In the following, the analytical expressions for the Landau parameters $f_l^{ST}(\theta)$ and $h_l^{1T}(\theta)$ of the DG Gogny interaction are given. The contributions of the various terms are
detailed. Besides, the contribution of the zero-range density-dependent term of the D1-type parameterization will be indicated.
One notes that the spin-orbit interaction does not
contribute to the Landau parameters in INM.
Full derivations of calculations can be found in Appendix \ref{Landauparam}. \\

\noindent \textit{Central and density-dependent term contributions} \\

\noindent The contributions of each central ($\alpha=0$) and 
density-dependent ($\alpha \ne 0$ and global factor $1/(\mu_3 \sqrt{\pi})^3$) interactions to the Landau parameters, 
defined in Eq.\eqref{legendreexpansion}, can be 
evaluated for $l \in \mathbb{N}$ by means of the 
following relations:
\begin{multline}
f_l^{00}  =  \widetilde{G}_\mu^\alpha(0) \bigg[ \Big( W + \frac{B}{2} - \frac{H}{2} - \frac{M}{4} \Big) \delta_{l,0} \\ + \Big( M + \frac{H}{2} - \frac{B}{2} - \frac{W}{4} \Big) G_l \bigg] \\
f_l^{10}  = \widetilde{G}_\mu^\alpha(0) \bigg[ \Big( \frac{B}{2} - \frac{M}{4} \Big) \delta_{l,0} + \Big( \frac{H}{2} - \frac{W}{4} \Big) G_l \bigg] \\
f_l^{01}  = - \widetilde{G}_\mu^\alpha(0) \bigg[ \Big( \frac{H}{2} + \frac{M}{4} \Big) \delta_{l,0} + \Big( \frac{B}{2} + \frac{W}{4} \Big) G_l \bigg] \\
f_l^{11}  = - \widetilde{G}_\mu^\alpha(0) \bigg[ \frac{M}{4} \delta_{l,0} + \frac{W}{4} G_l \bigg]
\end{multline}

where 
\begin{equation}
\widetilde{G}_\mu^\alpha(\vec{q}_{{F}}) = N_0 \frac{\rho^{\alpha}}{\mathcal{V}} (\mu \sqrt{\pi})^3 \, \e^{-\mu^2 k_{{F}}^2 (1- \cos \theta)/2}
\end{equation}
\begin{equation}\label{eeqq4}
\widetilde{G}_\mu^\alpha(\vec{q}_{{F}}) = \widetilde{G}_\mu^\alpha(0) \sum_l G_l P_l (\cos \theta)
\end{equation}
\begin{equation}
\widetilde{G}_\mu^\alpha(0) \equiv N_0 G_\mu^\alpha(0) = N_0 \frac{\rho^\alpha}{\mathcal{V}}
\end{equation}
and the coefficients $G_l$ of the expansion in Legendre polynomials in Eq.\eqref{eeqq4} are defined by:
\begin{equation}
G_l \equiv \frac{\sqrt{\pi}}{2} (2l+1) \, \e^{-\mu^2 k_{{F}}^2/2} \sum_m \frac{\big(\mu^2 k_{{F}}^2/4\big)^{2m+l}}{m! \- \Gamma(m+l+3/2)}.
\end{equation}

In the case of the zero-range density-dependent interaction of the D1-type Gogny parameterization, one obtains a contribution only in the $l=0$ partial wave:
\begin{subequations}
\begin{empheq}{align}
f_0^{00} & = \frac{3}{4} t_0 \rho^{\alpha}  \\
f_0^{10} & =  \left( \frac{x_0}{2} - \frac{1}{4}\right)  t_0 \rho^{\alpha}  \\
f_0^{01} & =   \left( -\frac{x_0}{2} - \frac{1}{4}\right)  t_0 \rho^{\alpha}\\
f_0^{11} & =  - \frac{1}{4} t_0 \rho^{\alpha}
\end{empheq}
\end{subequations}
These results are consistent with those of Ref.\cite{Gogny1977}.\\ 

\noindent \textit{Rearrangement contributions} \\

\noindent According to Eq.\eqref{MEph}, there are three rearrangement term contributions to compute. 
Their calculations is tedious and is explained in Appendix \ref{Landauparam}. 
It is shown that the rearrangement terms acts 
only in the $(S=0, T=0)$ channel of the quasiparticle 
interaction and that only the component $l=0$ participates. 
Their total contribution to the Landau parameters reduces to:
\begin{multline}
f^{00}_l = \widetilde{G}_\mu^\alpha(0) \delta_{l,0} \bigg[ \frac{(3+\alpha)\alpha}{2} \Big(W + \frac{B}{2} - \frac{H}{2} - \frac{M}{4} \Big) \\
+ \frac{3 \alpha}{X^3} \big( 4 K_1(X) + \frac{\alpha-1}{X^3} K_2(X) \big) \Big(M + \frac{H}{2} - \frac{B}{2} - \frac{W}{4} \Big) \bigg]
\end{multline}
where the functions $K_1$ and $K_2$ are defined by
\begin{multline}
K_1(X) = \frac{1}{X} \big( e^{-X^2} - 1 \big) + \frac{\sqrt{\pi}}{2} \erf(X) \\
K_2(X) = e^{-X^2}(X^2-2) - (3X^2-2) \\ + \sqrt{\pi} X^3 \erf(X)
\end{multline}
and the function $\widetilde{G}_\mu^\alpha(0)$ has for expression 
$\widetilde{G}_\mu^\alpha(0) \equiv N_0 G_\mu^\alpha(0) = N_0 \frac{\rho^\alpha}{\mathcal{V}}$. \\

In the case of the zero-range density-dependent interaction of the D1-type Gogny parameterization, one obtains for its contribution in the $l=0$ partial wave:
\begin{equation}
\begin{split}
f^{00}_0 & = \frac{3}{8} \alpha \left( \alpha + 3 \right) t_0 \rho^{\alpha}
\end{split}
\end{equation}

\noindent \textit{Tensor term contribution} \\

\noindent The Landau parameters associated with the tensor interaction are given by: 
\begin{subequations} \label{Landautensor}
\begin{align}
h_l^{10} & = \Big( H - \frac{W}{2} \Big) H_l(q_{{F}}) \\
h_l^{11} & = - \frac{W}{2} H_l(q_{{F}})
\end{align}
\end{subequations}
where, after the change of variable $u \equiv \cos \theta$, the function $H_l(k)$ is defined by:
\begin{multline} \label{Hl}
H_l(k) \equiv - \frac{2 \pi}{\mathcal{V}} (2l+1) \int_{-1}^1 \dd{u} \frac{P_l(u)}{2(1-u)} \times \\ \int_0^{\infty} \dd{r} r^2 V(r) j_2(kr)
\end{multline}
After some manipulations, one finds:
\begin{multline}
H_l(q_{{F}}) = - \frac{2 \pi}{\mathcal{V}} (2l+1) \sum_{i=0}^{l} \frac{(-)^i (l+i)!}{(i!)^2 (l-i)!} \times \\
\int_0^{\infty} \dd{r} r^{2(1-i)} \, \e^{-r^2/\mu^2} I_{2(i-2)}(r)
\end{multline}
where the integral $I_{2(i-2)}(r)$ has for expression,
for $n=i-2 \ge -2$:
\begin{multline}
I_{2n}(r) = \int_0^{2 x_{{F}}} \dd{x} \big( 3 x^{2n} \sin x - x^{2n+2} \sin x \\ - 3x^{2n+1} \cos x \big)
\end{multline}
This final expression is similar to the one obtained in Ref.\cite{Pastore2015a} for the linear response function in homogeneous INM.
For $n \ge 0$,
\begin{multline}
I_{2n}(r) = (2n+2)(2n+4)J_{2n}(r) \\ - (2n+5)(2 x_{{F}})^{2n+1} \sin 2 x_{{F}} \\ + (2 x_{{F}})^{2n+2} \cos 2 x_{{F}}
\end{multline}
with 
\begin{multline}
J_{2n} = (2n)!  \Bigg[ (-)^n (1 - \cos 2 x_{{F}}) \\ + \sum_{t=0}^{n-1} (-)^{t} \frac{(2x_{{F}})^{2n-2t}}{(2n-2t)!} \times \\
\bigg( (2n-2t) \frac{\sin 2x_{{F}}}{2x_{{F}}} - \cos 2x_{{F}} \bigg) \Bigg]
\end{multline}
For $n=-1$,
\begin{equation}
I_{-2}(r) = \cos 2 x_{{F}} - \frac{3 \sin 2 x_{{F}}}{2 x_{{F}}} + 2.
\end{equation}
For $n=-2$, 
\begin{multline}
I_{-4}(r) = \bigg[ \frac{x \cos x - \sin x}{x^3} \bigg]^{2 x_{{F}}}_0 \\
= \frac{2 x_{{F}} \cos 2 x_{{F}} - \sin 2 x_{{F}}}{(2 x_{{F}})^3} + \frac{1}{3}
\end{multline}
Setting $X \equiv \mu k_{{F}}$, the explicit expressions of the function $H_l(q_{{F}})$ for the first seven values of $l$ are equal to:
\begin{multline}
H_0 = -\frac{\pi^{3/2}}{4 \mathcal{V} k_{{F}}^3} \bigg[ \frac{2}{3} X^3 + X \e^{-X^2} - \frac{\sqrt{\pi}}{2} \erf(X) \bigg] \\
H_1 = -\frac{3 \pi^{3/2}}{4 \mathcal{V} k_{{F}}^3} \bigg[ \frac{2}{3} X^3 - X \big( \e^{-X^2} + 4 \big) + \frac{5}{2} \sqrt{\pi} \erf(X) \bigg] \\
H_2 = -\frac{5 \pi^{3/2}}{4 \mathcal{V} k_{{F}}^3} \bigg[ \frac{2}{3} X^3 + X \big( \e^{-X^2} - 12 \big) + \frac{24}{X} \big( \e^{-X^2} - 1 \big) \\ + \frac{35}{2} \sqrt{\pi} \erf(X) \bigg] \\
H_3 = -\frac{7 \pi^{3/2}}{4 \mathcal{V} k_{{F}}^3} \bigg[ \frac{2}{3} X^3 - X \big( \e^{-X^2} + 24 \big) + \frac{40}{X} \big( \e^{-X^2} - 3 \big) \\ 
- \frac{80}{X^3} \big( \e^{-X^2} - 1 \big)
+ \frac{105}{2} \sqrt{\pi} \erf(X) \bigg] \\
H_4 = -\frac{9 \pi^{3/2}}{4 \mathcal{V} k_{{F}}^3} \bigg[ \frac{2}{3} X^3 + X \big( \e^{-X^2} - 40 \big) \\
+ \frac{8}{X} \big(17 \- \e^{-X^2} - 45 \big) + \frac{112}{X^3}\big( \e^{-X^2} + 5 \big) \\
+ \frac{672}{X^5} \big( \e^{-X^2} - 1 \big) + \frac{231}{2} \sqrt{\pi} \erf(X) \bigg]  \\
H_5 = -\frac{11 \pi^{3/2}}{4 \mathcal{V} k_{{F}}^3} \bigg[ \frac{2}{3} X^3 - X \big( \e^{-X^2} + 60 \big) \\+ \frac{8}{X} \big(23 \- \e^{-X^2} - 105 \big) 
- \frac{64}{X^3}\big( 8 \- \e^{-X^2} - 35 \big) \\ - \frac{864}{X^5} \big( 3 \- \e^{-X^2} + 7 \big) - \frac{8 \, 640}{X^7} \big( \e^{-X^2} + 1 \big) \\
+ \frac{429}{2} \sqrt{\pi} \erf(X) \bigg]  \\
H_6  = -\frac{13 \pi^{3/2}}{4 \mathcal{V} k_{{F}}^3} \bigg[ \frac{2}{3} X^3 + X \big( \e^{-X^2} - 84 \big) \\ + \frac{80}{X} \big(5 \- \e^{-X^2} - 21 \big) 
- \frac{320}{X^3}\big( 2 \- \e^{-X^2} + 21 \big) \\ + \frac{480}{X^5} \big( 19 \- \e^{-X^2} - 63 \big) + \frac{10 \, 560}{X^7} \big(5 \- \e^{-X^2} + 9 \big) \\
 + \frac{147 \, 840}{X^9} \big( \e^{-X^2} - 1 \big) + \frac{715}{2} \sqrt{\pi} \erf(X) \bigg] \\
H_7  = -\frac{15 \pi^{3/2}}{4 \mathcal{V} k_{{F}}^3} \bigg[ \frac{2}{3} X^3 - X \big( \e^{-X^2} + 112 \big) \\ + \frac{16}{X} \big(31 \- \e^{-X^2} - 189 \big) 
 \quad - \frac{160}{X^3}\big( 11 \- \e^{-X^2} - 105 \big) \\ - \frac{480}{X^5} \big( 47 \- \e^{-X^2} + 231 \big) - \frac{1 \, 920}{X^7} \big(11 \- \e^{-X^2} - 297 \big) \\
 \qquad \quad - \frac{174 \, 720}{X^9} \big(7 \- \e^{-X^2} + 11 \big) \\ - \frac{3 \, 144 \, 960}{X^{11}} \big( \e^{-X^2} - 1 \big) + \frac{1 \, 105}{2} \sqrt{\pi} \erf(X) \bigg] 
\end{multline}

\paragraph{Stability criteria}

In this section, one introduces and derives 
stability criteria associated with the Landau 
parameters in INM, without and 
with the tensor force, in the context of the Gogny
interactions. \\

\noindent \textit{Without tensor forces} \\

\noindent One starts by establishing stability criteria \cite{Nozieres1964,Backman1985,Brown1971,Chappert2007} considering the quasiparticle interaction defined in Eq.\eqref{QPinteraction} without tensor terms, i.e.\ with $h(\theta) = h'(\theta) = 0$. 
In the present case, the quasiparticle ground state defines a spherical Fermi surface in momentum space. This state corresponds effectively to a ground state if it is stable against small deformations, i.e.\ if it is associated with a minimum of the free energy $F$:
\begin{equation}\label{freeenergy}
F \equiv \mathcal{E} - \lambda_{{c}} A
\end{equation}
where $\lambda_{{c}}$ is the chemical potential and $A$ the number of nucleons. In order to slightly distort the Fermi surface, one defines an infinitesimal displacement $u(\theta, \varphi)$ in momentum space in terms of spherical harmonics according to:
\begin{equation} \label{udev}
u(\theta, \varphi) = \sum_{lm} u_{lm} Y_l^m(\theta,\varphi)
\end{equation}
for which the reality condition $u^*_{lm} = (-)^m u_{l-m}$ is assumed. The change in the distribution function given in Eq.\eqref{distrib} becomes:
\begin{equation}
\updelta \rho(\vec{k}) =  
\begin{cases} 
  1 \quad & \text{if} \ \ k_{{F}} \le |\vec{k}| \le k_{{F}} + u(\theta, \varphi),  \\
  0 \quad & \text{otherwise}.
\end{cases}
\end{equation}
Considering the continuous limit \eqref{transfocontinu} to transform the summations of Eq.\eqref{deltaE}, while considering the above distribution function, the variation in free energy is given by:
\begin{multline}
\updelta F = \frac{\mathcal{V}}{(2 \pi)^3} \sum_{sq} \int \dd{\hat{r}} \int_{k_{{F}}}^{k_{{F}} + u(\theta, \varphi)} \dd{k} k^2 (\varepsilon_0(k) - \mu_{{c}}) \\
+ \frac{1}{2} \frac{\mathcal{V}^2}{(2 \pi)^6} \sum_{sq} \sum_{s'q'} \int \dd{\hat{r}} \int \dd{\hat{r}^{\- \prime}} \times \\ \int_{k_{{F}}}^{k_{{F}} + u(\theta, \varphi)} \dd{k} k^2 \int_{k_{{F}}}^{k_{{F}} + u(\theta', \varphi')} \dd{k}' k'^2 \mathcal{F}(\vec{k}, \vec{k}')
\end{multline}
Since small displacements of the Fermi surface are considered, only the lowest orders in $u(\theta, \varphi)$ are kept. 
For the first right-hand side term, the lowest order is $2$ as $\varepsilon_0(k_{{F}}) = \lambda_c$, so that the terms of order $1$ vanish. 
For the second right-hand side term, it is $1$, for each integral.  Performing the integrations over momenta, one gets:
\begin{multline} \label{deltaAdernier}
\updelta F = \frac{\mathcal{V}}{(2 \pi)^3} \frac{\hbar^2 k_{{F}}^3}{2 m^*} \sum_{sq} \int \dd{\hat{r}} u^2(\theta,\varphi) + \frac{\mathcal{V}^2 k_{{F}}^4}{(2 \pi)^6} \times  \\\sum_{sq} \sum_{s'q'} \int \dd{\hat{r}} u(\theta, \varphi) \int \dd{\hat{r}^{\- \prime}} u(\theta', \varphi') \mathcal{F}(\xi)
\end{multline}
where $\xi$ is the angle between the directions $(\theta, \varphi)$ and $(\theta', \varphi')$. As the quasiparticle interaction only depends on this angle, one can expand it in Legendre polynomials:
\begin{equation} \label{Fdev}
\mathcal{F}(\xi) = \sum_l \mathcal{F}_l P_l(\cos \xi)
\end{equation}
Here, one has assumed the displacement \eqref{udev} and the quasiparticle interaction \eqref{Fdev} to be independent on the spin and isospin variables for simplicity. Further, one will see how the result generalizes when these dependencies are restored. Inserting Eq.\eqref{udev} and Eq.\eqref{Fdev} in \eqref{deltaAdernier}, and using properties of the spherical harmonics as well as the Legendre expansion, one obtains:
\begin{equation}
\updelta F = \frac{\mathcal{V}}{(2 \pi)^3} \frac{\hbar^2 k_{{F}}^3}{2 m^*} \sum_{lm} |u_{lm}|^2 \Big( 1 + N_0 \frac{\mathcal{F}_l}{2l+1} \Big)
\end{equation}
where $N_0$ is the density of quasiparticle states at the Fermi surface. Thus, if the free energy is minimal for a spherical Fermi surface corresponding to the ground state, any variation of the free energy must be positive, i.e.\ $\updelta F > 0$. It implies the stability criterion
\begin{equation}
f^{00}_l > -(2l+1), \quad \text{for } l \in \mathbb{N},
\end{equation}
where the notations leading to Eq.\eqref{legendreexpansion} have been used.  Since Eqs.\eqref{udev} and \eqref{Fdev} have been assumed to be independent on the spin and isospin variables, this result only holds in the $(S=0, T=0)$ channel. However, it can be extended to the other channels by repeating the calculation while displacing the Fermi surface in the directions $\vec{u}(\theta, \varphi) \cdot \vec{\sigma}$, $u(\theta, \varphi) \tau_z$ and $\vec{u}(\theta, \varphi) \cdot \vec{\sigma} \tau_z$, where $Oz$ is chosen as quantization axis. It brings the same stability criteria for the channels $(S=1, T=0)$, $(S=0, T=1)$ and $(S=1, T=0)$, respectively, i.e.\
\begin{equation} \label{stabnotensor}
f^{ST}_l > -(2l+1), \quad \text{for } l \in \mathbb{N}
\end{equation}
For the ground state to be stable when the quasiparticle interaction is purely composed of central terms, the above stability criteria have to be fulfilled at any time.\\

\noindent \textit{With tensor forces} \\

\noindent One focuses on the stability criteria considering the full quasiparticle interaction definde in Eq.\eqref{QPinteraction} with tensor terms, i.e.\ with $h(\theta), h'(\theta) \ne 0$.  One will give the new stability criteria and only sketch their derivation \cite{Backman1979}, this latter being pretty similar (although more intricate) to the one without tensor terms of the previous section.

As the tensor terms only act in the $S=1$ channel of the quasiparticle interaction, one can immediately state that the stability criteria in the $S=0$ channel remain the same as in Eq.\eqref{stabnotensor}, i.e.\
\begin{equation}
f^{0T}_l > -(2l+1), \quad \text{for } l \in \mathbb{N}
\end{equation}
Nevertheless, they are modified in the $S=1$ channel. To find this modification, one considers as previously a variation of the free energy engendered by some displacement of the Fermi surface. The displacements in the $(S=1, T=0)$ and $(S=1, T=1)$ channels are respectively $\vec{u}(\theta, \varphi) \cdot \vec{\sigma}$ and $\vec{u}(\theta, \varphi) \cdot \vec{\sigma} \tau_z$, in agreement with the fact that the tensor interaction explicitly involves the spin variables. 

In the $(S=1, T=0)$ channel, one shows that the change in free energy is eventually given by its matrix elements coupled to the particle-hole quantum numbers $J$ and $J'$ as \cite{Backman1979}:
\begin{multline} \label{meA}
\mel*{l' J'}{F}{l J} = \delta_{ll'} \delta_{JJ'} \bigg( 1 + \frac{1}{2l+1} f_l^{10} \\ + \frac{2l}{(2l-1)(2l+1)}h_{l-1}^{10} + \frac{2(l+1)}{(2l+1)(2l+3)} h_{l+1}^{10} \bigg) \\
  + \delta_{JJ'} \Bigg( 15 (-)^J \begin{pmatrix} 1 & 1 & 2 \\ 0 & 0 & 0 \end{pmatrix} \begin{pmatrix} l & l' & 2 \\ 0 & 0 & 0 \end{pmatrix} \begin{Bmatrix} 1 & l' & J \\ l & 1 & 2 \end{Bmatrix} \times \\ \bigg[ \sqrt{\frac{2l+1}{2l'+1}} h_{l'}^{10} + \sqrt{\frac{2l'+1}{2l+1}} h_{l}^{10} \bigg] \\
 - \frac{3 \sqrt{(2l+1)(2l'+1)}}{2J+1} \begin{pmatrix} l & 1 & J \\ 0 & 0 & 0 \end{pmatrix}  \begin{pmatrix} l' & 1 & J \\ 0 & 0 & 0 \end{pmatrix} h_J^{10} \\
 - 3 \sqrt{(2l+1)(2l'+1)} \sum_{l''} \begin{pmatrix} l & 1 & l'' \\ 0 & 0 & 0 \end{pmatrix}  \begin{pmatrix} l' & 1 & l'' \\ 0 & 0 & 0 \end{pmatrix} \times \\ \begin{Bmatrix} 1 & l' & J \\ 1 & l & l'' \end{Bmatrix}  h_{l''}^{10} \Bigg)
\end{multline}
where the parentheses and the brackets denote the Wigner-$3j$ and $6j$ symbols, respectively. 
\noindent For the $(S=1,T=1)$ channel, one finds out the exact same relation after substituting $f_l^{10}$ by $f_l^{11}$ and $h_l^{10}$ by $h_l^{11}$, for all $l$:
\begin{multline} \label{meA2}
\mel*{l' J'}{F}{l J} = \delta_{ll'} \delta_{JJ'} \bigg( 1 + \frac{1}{2l+1} f_l^{11} \\
+ \frac{2l}{(2l-1)(2l+1)}h_{l-1}^{11}  + \frac{2(l+1)}{(2l+1)(2l+3)} h_{l+1}^{11} \bigg) \\
+ \delta_{JJ'} \Bigg( 15 (-)^J \begin{pmatrix} 1 & 1 & 2 \\ 0 & 0 & 0 \end{pmatrix} \begin{pmatrix} l & l' & 2 \\ 0 & 0 & 0 \end{pmatrix} \begin{Bmatrix} 1 & l' & J \\ l & 1 & 2 \end{Bmatrix} \times \\ \bigg[ \sqrt{\frac{2l+1}{2l'+1}} h_{l'}^{11} + \sqrt{\frac{2l'+1}{2l+1}} h_{l}^{11} \bigg] \\
 - \frac{3 \sqrt{(2l+1)(2l'+1)}}{2J+1} \begin{pmatrix} l & 1 & J \\ 0 & 0 & 0 \end{pmatrix}  \begin{pmatrix} l' & 1 & J \\ 0 & 0 & 0 \end{pmatrix} h_J^{11} \\
 - 3 \sqrt{(2l+1)(2l'+1)} \sum_{l''} \begin{pmatrix} l & 1 & l'' \\ 0 & 0 & 0 \end{pmatrix}  \begin{pmatrix} l' & 1 & l'' \\ 0 & 0 & 0 \end{pmatrix} \times \\ \begin{Bmatrix} 1 & l' & J \\ 1 & l & l'' \end{Bmatrix}  h_{l''}^{11} \Bigg)
\end{multline}

Thus, the condition for stability of the ground state becomes that the matrix $F$ has only positive eigenvalues.  Note in passing that switching off the tensor interaction, i.e.\ demanding $h_l^{10} = 0$ in Eq.\eqref{meA} and $h_l^{11} = 0$ in Eq.\ref{meA2}, for all $l$, consistently brings back the stability criteria without tensor terms given in Eq.\eqref{stabnotensor} in the associated channels. Now, one analyzes how to extract the eigenvalues of the matrix $F$ from the above expression. First, one notes that $J$ and $J'$ are the total momenta that couple the angular and intrinsic momenta of two quasiparticle pairs, in such a way that $|l-1| \le J \le l+1$ and  $|l'-1| \le J' \le l'+1$ (as can be inferred in particular from the Wigner-$6j$ symbols).  By the way, those momenta have to be equal, $J=J'$, otherwise the matrix elements are zero. Besides, the properties of the Wigner-$3j$ symbols impose $l$ and $l'$ to be of the same parity. Therefore, for a given $J$ value, there is only one $2 \times 2$ matrix to diagonalize, the case $J=l=l'$ being a diagonal sub-block. The case $J=0$, $l=l'=1$ is also diagonal since the sub-block $J=l=l'=1$ is, as stated above. One then gives the stability criteria for the first values of $J, l$ and $l'$. From the diagonal matrix elements, one obtains:
\begin{equation}\label{stabdiag}
\begin{array}{l}
\displaystyle \text{for } J  = 0, l=l'=1, \qquad \\ \displaystyle 1 + \frac{1}{3}f_1^{10} - \frac{10}{3}h_0^{10} + \frac{4}{3} h_1^{10} - \frac{2}{15} h_2^{10} > 0 \\ \\
\displaystyle \text{for } J  = 1, l=l'=1, \qquad \\ \displaystyle 1 + \frac{1}{3}f_1^{10} + \frac{5}{3}h_0^{10} - \frac{2}{3} h_1^{10} + \frac{1}{15} h_2^{10} > 0   \\ \\
\displaystyle \text{for } J  = 2, l=l'=2, \qquad \\ \displaystyle 1 + \frac{1}{5}f_2^{10} + \frac{7}{15}h_1^{10} - \frac{2}{5} h_2^{10} + \frac{3}{35} h_3^{10} > 0 \\ \\ 
\displaystyle \text{for } J = 3, l=l'=3, \qquad \\ \displaystyle 1 + \frac{1}{7}f_3^{10} + \frac{9}{35}h_2^{10} - \frac{2}{7} h_3^{10} + \frac{5}{63} h_4^{10} > 0 \\ \\
\displaystyle \text{for } J  = 4, l=l'=4, \qquad \\ \displaystyle 1 + \frac{1}{9}f_4^{10} + \frac{11}{63}h_3^{10} - \frac{2}{9} h_4^{10} + \frac{7}{99} h_5^{10} > 0 \\ \\
\displaystyle \text{for } J  = 5, l=l'=5, \qquad \\ \displaystyle 1 + \frac{1}{11}f_5^{10} + \frac{13}{99}h_4^{10} - \frac{2}{11} h_5^{10} + \frac{9}{143} h_6^{10} > 0 
\end{array}
\end{equation}
\noindent The first four conditions are identical to those of Ref.\cite{Backman1979}. The two last ones have been derived additionaly.

For a given $J$,  the sub-matrix $2 \times 2$ of $F$ to diagonalize is defined by
\begin{equation} \label{submatrix}
F^J \equiv \begin{pmatrix} \mel*{J-1 \, J}{F}{J-1 \,  J} & \mel*{J-1 \,  J}{F}{J+1 \,  J} \\ \mel*{J+1 \,  J}{F}{J-1 \,  J} & \mel*{J+1 \,  J}{F}{J+1 \,  J} \end{pmatrix}
\end{equation}
The corresponding eigenvalues $\lambda_{\pm}^J $ of Eq.\eqref{submatrix} define the stability criteria simply by the condition:
\begin{equation} \label{submatrixstab}
\lambda_{\pm}^J \equiv \frac{-b_J \pm \sqrt{\Delta_J}}{2} > 0,
\end{equation}
with
\begin{multline}
b_J \equiv - \big( \! \mel*{J-1 \, J}{F}{J-1 \,  J} + \! \mel*{J+1 \, J}{F}{J+1 \,  J} \! \big) \\
c_J \equiv - \mel*{J-1 \, J}{F}{J+1 \,  J} \! \! \mel*{J+1 \, J}{F}{J-1 \,  J}  \\
\Delta_J \equiv b_J^2 - 4 c_J
\end{multline}
The matrix elements involved in the sub-matrices of the form \eqref{submatrix}, from which the stability criteria are obtained thanks to Eq.\eqref{submatrixstab},  are:

\begin{equation} \label{stabnondiag}
\begin{array}{l}
\text{for } J =1, \\
\begin{cases} \mel*{01}{F}{01} = 1 + f_0^{10} \\ \mel*{01}{F}{21} = \mel*{21}{F}{01} = - \sqrt{2} \big( h_0^{10} - \frac{2}{3} h_1^{10} + \frac{1}{5} h_2^{10} \big)  \\ \mel*{21}{F}{21} = 1 + \frac{1}{5} f_2^{10} - \frac{7}{15} h_1^{10} + \frac{2}{5} h_2^{10} - \frac{3}{35} h_3^{10}\end{cases} \\ \\
\text{for } J  =2, \\
\begin{cases} \mel*{12}{F}{12} = 1 + \frac{1}{3} f_1^{10} - \frac{1}{3} h_0^{10} + \frac{2}{15} h_1^{10} - \frac{1}{75} h_2^{10} \\ \mel*{12}{F}{32} = \! \mel*{32}{F}{12} = - \sqrt{6} \big( \frac{1}{5} h_1^{10} - \frac{6}{25} h_2^{10} + \frac{3}{35} h_3^{10} \big)  \\ \mel*{32}{F}{32} = 1 + \frac{1}{7} f_3^{10} - \frac{36}{175} h_2^{10} + \frac{8}{35} h_2^{10} - \frac{4}{63} h_4^{10} \end{cases} \\ \\
\noindent \text{for } J  =3, \\
\begin{cases} \mel*{23}{F}{23} = 1 + \frac{1}{5} f_2^{10} - \frac{2}{15} h_1^{10} + \frac{12}{105} h_2^{10} - \frac{6}{245} h_3^{10} \\ \mel*{23}{F}{43} = \! \mel*{43}{F}{23} = - \sqrt{3} \big( \frac{6}{35} h_2^{10} - \frac{12}{49} h_3^{10} + \frac{10}{105} h_4^{10} \big)  \\ \mel*{43}{F}{43} = 1 + \frac{1}{9} f_4^{10} - \frac{55}{441} h_3^{10} + \frac{10}{63} h_4^{10} - \frac{5}{99} h_5^{10}\end{cases} \\ \\
\text{for } J  =4, \\
\begin{cases} \mel*{34}{F}{34} = 1 + \frac{1}{7} f_3^{10} - \frac{3}{35} h_2^{10} + \frac{2}{9} h_3^{10} - \frac{29}{189} h_4^{10} \\ \mel*{34}{F}{54} = \! \mel*{54}{F}{34} = - \sqrt{5} \big( \frac{2}{21} h_3^{10} - \frac{4}{27} h_4^{10} + \frac{2}{33} h_5^{10} \big)  \\ \mel*{54}{F}{54} = 1 + \frac{1}{11} f_5^{10} - \frac{26}{297} h_4^{10} + \frac{4}{33} h_5^{10} - \frac{18}{143} h_6^{10} \end{cases}
\end{array}
\end{equation}
\noindent The two first sets of matrix elements, extracted from \cite{Backman1979}, have been checked. The two last ones have been evaluated. One gives two additional off-diagonal matrix elements needed in section \ref{Landauparameters}, namely:
\begin{equation} \label{stabadd}
\begin{array}{l}
\displaystyle \text{for } J = 5,  \mel*{45}{F}{45} = 1 + \frac{1}{9}f_4^{10} - \frac{4}{63}h_3^{10} \\ + \frac{8}{99} h_4^{10} - \frac{46}{363} h_5^{10} \\ \\
\displaystyle \text{for } J = 6, \mel*{56}{F}{56} = 1 + \frac{1}{11}f_5^{10} - \frac{5}{99}h_4^{10} \\ + \frac{10}{143} h_5^{10} - \frac{45}{1 \, 859} h_6^{10}
\end{array}
\end{equation}
\noindent Those matrix elements given for the $(S=1,T=0)$ channel can immediately be deduced for the $(S=1,T=1)$ channel, applying the changes $f_l^{10} \rightarrow f_l^{11}$ and $h_l^{10} \rightarrow h_l^{11}$, for all $l$.

\paragraph{Sum rules}

One ends the Landau parameter sub-section by introducing and
deriving the sum rules associated with the Landau parameters, still in homogeneous and symmetric INM, without and with tensor terms. One starts with the general framework that applies to both cases.

Within the theory of normal Fermi liquids, one can express the forward scattering amplitude $\mathcal{S}$ of a quasiparticle pair on the Fermi surface \cite{Backman1985,Friman1979} in terms of parameters in a similar way as for the quasiparticle interaction defined in Eq.\eqref{QPinteraction}, i.e.\
\begin{multline} 
\mathcal{S}  = N_0^{-1} \Big[ b(\theta) + c(\theta) (\vec{\sigma}_1 \cdot \vec{\sigma}_2) + b'(\theta) (\vec{\tau}_1 \cdot \vec{\tau}_2) \\
+ c'(\theta) (\vec{\sigma}_1 \cdot \vec{\sigma}_2) (\vec{\tau}_1 \cdot \vec{\tau}_2)
+ d(\theta) \frac{q_{{F}}^2}{k_{{F}}^2} S_{12}(\hat{q}_{12}) \\ + d'(\theta) \frac{q_{{F}}^2}{k_{{F}}^2} S_{12}(\hat{q}_{12}) (\vec{\tau}_1 \cdot \vec{\tau}_2) \Big]
\end{multline}
where $\mathcal{S}$ depends on $k$, $\sigma$ and $\tau$ in such a way that $\mathcal{S} \equiv \mathcal{S}(\vec{k}_{{F}}, \vec{\sigma}_1, \vec{\tau}_1;\vec{k}_{{F}}', \vec{\sigma}_2, \vec{\tau}_2)$ and $N_0$ the density of quasiparticle states at the Fermi surface defined previously, ensuring that $b$, $b'$, $c$, $c'$, $d$, $d'$ are dimensionless quantities. 
These parameters to be determined only depend on the momenta $\vec{k}_{{F}}$ and $\vec{k}_{{F}}'$ of the quasiparticle pair. 
Like the Landau parameters, one will cast these parameters into common notations $b^{ST}(\theta)$ and $d^{1T}(\theta)$ reflecting the $(S,T)$ channels in which they contribute, namely $b^{00}(\theta) \equiv b(\theta)$, $b^{10}(\theta) \equiv c(\theta)$, $b^{01}(\theta) \equiv b'(\theta)$, $b^{11}(\theta) \equiv c'(\theta)$, and \mbox{$d^{10}(\theta) \equiv d(\theta)$}, $d^{11}(\theta) \equiv d'(\theta)$.  As done for the parameters of the quasiparticle interaction in Eq.\eqref{legendreexpansion}, these parameters are expanded in Legendre polynomials, as they only depend on the Landau angle $\theta$ between the two quasiparticles:
\begin{subequations}
\begin{align}
b^{ST}(\theta) & = \sum_l b_l^{ST} P_l(\cos \theta) \label{bexpand} \\
d^{1T}(\theta) & = \sum_l d_l^{1T} P_l(\cos \theta).
\end{align}
\end{subequations}

\noindent The Pauli exclusion principle imposes the forward scattering amplitude to be antisymmetric under the exchange of the two ingoing or outgoing quasiparticles, thus
leading to the expression:
\begin{equation} \label{antisymcond}
P \mathcal{S} \equiv P_r P_\sigma P_\tau \mathcal{S} = -\mathcal{S}
\end{equation}
In Eq.\eqref{antisymcond}, the operators $P_r$, $P_\sigma$ and $P_\tau$ are the exchange momenta, spins and isospins of the two quasiparticles, respectively. 
One evaluates this action when the momenta of the quasiparticles are equal, i.e.\ when $\vec{k}_{{F}} = \vec{k}'_{{F}}$, and then $P_r = \mathbbm{1}$. Since $\vec{q}_{12} = 0$, the tensors terms do not contribute. 
By means of the definitions of the exchange operators as well as the properties $(\vec{\sigma}_1 \cdot \vec{\sigma}_2)^2 = 3  - 2 \vec{\sigma}_1 \cdot \vec{\sigma}_2$ and $(\vec{\tau}_1 \cdot \vec{\tau}_2)^2 = 3 - 2 \vec{\tau}_1 \cdot \vec{\tau}_2$, one obtains:
\begin{multline}
4 P \mathcal{S}(\vec{q}_{12}=0) = (b+3b'+3c+9c') \\ + (b+3b'-c-3c') (\vec{\sigma}_1 \cdot \vec{\sigma}_2) \\
 + (b-b'+3c-3c') (\vec{\tau}_1 \cdot \vec{\tau}_2) \\ + (b-b'-c+c') (\vec{\sigma}_1 \cdot \vec{\sigma}_2) (\vec{\tau}_1 \cdot \vec{\tau}_2)
\end{multline}
Considering the antisymmetrization condition of the forward scattering amplitude together with the above relation in each of the four $(S,T)$ channels, one gets,
the four following equations:
\begin{equation}\label{eeqq5}
\begin{cases}
5b+3b'+3c+9c' = 0 \\
b+3b'+3c-3c' = 0 \\
b+3b'+3c-3c' = 0 \\
b-b'-c+5c' = 0.
\end{cases}
\end{equation}
One notices that the second and third equations of the 
system \eqref{eeqq5} are the same. Moreover, multiplying the second line by two, the fourth line by three and adding the result, 
one falls back on the first equation. Thus, only two of the three remaining equations are linearly independent. 
One decides to keep the first and second ones. 
Using the notations introduced to bring some physics insight, one writes down the two linearly independent equations as:
\begin{equation} \label{lineareq}
\begin{cases}
5b^{00}+3b^{10}+3b^{01}+9b^{11} = 0 \\
b^{00}+3b^{10}+3b^{01}-3b^{11} = 0.
\end{cases}
\end{equation}
Summing the equations and plugging the expansions \eqref{bexpand} in the result (keeping in mind that $\theta = 0$ as $\vec{k}_{{F}} = \vec{k}'_{{F}}$), one gets the first sum rule: 
\begin{equation} \label{sr1}
\sum_l \big(b_l^{00} + b_l^{10} + b_l^{01} + b_l^{11} \big) = 0.
\end{equation}
Dividing the first equation by two and subtracting it three half of the second one, one obtains the second sum rule,
\begin{equation} \label{sr2}
\sum_l \big(b_l^{00} -3 b_l^{10} -3 b_l^{01} +9 b_l^{11} \big) = 0.
\end{equation}
One notes that other combinations of the two linearly independent equations \eqref{lineareq} could have been chosen, leading to different sum rules.
Nevertheless, those ones are convenient since their physical interpretation is obvious. 
The first and second sum rules are the relations obtained when considering the antisymmetrization condition \eqref{antisymcond} in the triplet-odd $(S=1, T=1)$ and singlet--odd $(S=0, T=0)$ channels, respectively. 
As such, they project onto the triplet-odd and singlet-odd states respectively, ensuring that the forward scattering amplitude of two quasiparticles with same momenta vanishes in odd partial waves. 
There are no such sum rules in even-parity states since the forward scattering amplitude is not expected to vanish for quasiparticles with same momenta in even partial waves. This is indeed what it is found out when considering the antisymmetrization condition \eqref{antisymcond} in the triplet-even $(S=1, T=0)$ and singlet-even $(S=0, T=1)$ channels. 
In conclusion, the Pauli exclusion principle is satisfied when the forward scattering amplitude is only composed of central terms if and only if the above 
sum rules defined in Eq.\eqref{sr1} and Eq.\eqref{sr2} hold.\\

\noindent \textit{Without tensor forces} \\

\noindent One writes down the sum rules in terms of Landau parameters when there is no tensor force \cite{Backman1985,Friman1979,Chappert2007}.
In that case, the link between the forward scattering amplitude and the quasiparticle interaction is quite simple so that their respective parameters 
can be expressed in relation to each other according to:
\begin{align} \label{bTS}
b_l^{ST} & = \frac{f_l^{ST}}{1+f_l^{ST}/(2l+1)}, \quad \text{for } l \in \mathbb{N}.
\end{align}
\noindent Then, the sum rules \eqref{sr1} and \eqref{sr2} merely become:
\begin{multline} \label{notensor1}
\sum_l  \bigg[ \frac{f_l^{00}}{1+f_l^{00}/(2l+1)} + \frac{f_l^{10}}{1+f_l^{10}/(2l+1)} \\+ \frac{f_l^{01}}{1+f_l^{01}/(2l+1)} + \frac{f_l^{11}}{1+f_l^{11}/(2l+1)} \bigg] = 0
\end{multline}
\noindent as well as
\begin{multline} \label{notensor2}
\sum_l  \bigg[ \frac{f_l^{00}}{1+f_l^{00}/(2l+1)} - \frac{3 f_l^{10}}{1+f_l^{10}/(2l+1)}  \\ - \frac{3 f_l^{01}}{1+f_l^{01}/(2l+1)} + \frac{9 f_l^{11}}{1+f_l^{11}/(2l+1)} \bigg] = 0
\end{multline}
\noindent In the absence of tensor forces, the two equations above must be satisfied for the Pauli exclusion principle to hold.\\

\noindent \textit{With tensor forces} \\

\noindent In this part, one write down the sum rules in terms of Landau parameters in the presence of tensor forces \cite{Friman1979}.
As seen in the framework outlined earlier, the tensor parameters $d(\theta)$ and $d'(\theta)$ do not take part in the sum rules since they are derived in the limit of equal momenta for which $\vec{q}_{12} = 0$. However, the spin-dependent nature of the tensor forces complicates the equation connecting the forward scattering amplitude to the quasiparticle interaction. To solve it, one needs to couple the angular and intrinsic momenta of the quasiparticle pair to the total momentum $J$. Then, one shows that the parameters in the $S=1$ channel can be written in terms of the forward scattering amplitude coupled to $J$ as:
\begin{equation}
b_l^{1T} = \frac{1}{3} \sum_J \frac{2J+1}{2l+1} \mathcal{S}^{JT}_{ll}
\end{equation}
where $\mathcal{S}^{JT}_{ll}$ is the forward scattering amplitude composed of a diagonal part (for $J = l = l'$ and $J=0, l=l'=1$) and a non-diagonal part (for $l = J \pm 1$ and $l' = J \mp 1$), such that:
\begin{equation} \label{defmathcalA}
\mathcal{S}^{JT}_{ll} = 
\begin{cases} 
  \text{For the diagonal part:}  \\
\displaystyle ~~~~~~~~~~ \frac{\mathcal{F}^{JT}_{ll}}{1+\mathcal{F}^{JT}_{ll}/(2l+1)} \\
\text{For the non diagonal part:} \\
\displaystyle \mathcal{D}^{-1} \bigg[ \mathcal{F}^{JT}_{ll} \Big( 1 + \frac{\mathcal{F}^{JT}_{l'l'}}{2l'+1} \Big) - \frac{(\mathcal{F}^{JT}_{ll'})^2}{2l'+1} \bigg]
\end{cases}
\end{equation}
with the quantity
\begin{equation} \label{defD}
\mathcal{D} \equiv \Bigg( 1 + \frac{\mathcal{F}^{JT}_{ll}}{2l+1} \Bigg) \Bigg( 1 + \frac{\mathcal{F}^{JT}_{l'l'}}{2l'+1} \Bigg) - \frac{(\mathcal{F}^{JT}_{ll'})^2}{(2l+1)(2l'+1)}
\end{equation}
The quasiparticle interaction coupled to $J$ is related to the matrix elements of the free energy $F$ expressed in the formalism defined before, according to:
\begin{equation} \label{FJT}
\mathcal{F}^{JT}_{ll'} = \sqrt{(2l+1)(2l'+1)} \big(\! \mel*{l' J}{F}{l J} - \delta_{ll'} \big)
\end{equation}
where the matrix elements are given by Eq.\eqref{meA} for $T=0$ and Eq.\eqref{meA2} for $T=1$.\\

As for the parameters $b_l^{0T}$ in the $S=0$ channel, they are still given by Eq.\eqref{bTS} since the tensor forces do not act in this channel, i.e.\
\begin{align}
b_l^{0T} & = \frac{f_l^{0T}}{1+f_l^{0T}/(2l+1)}, \quad \text{for } l \in \mathbb{N}.
\end{align}

\noindent Then, the sum rules defined in Eqs.\eqref{sr1} and \eqref{sr2} become:
\begin{multline} \label{tensorSR1}
\sum_l \bigg[ \frac{f_l^{00}}{1+f_l^{00}/(2l+1)} + \frac{f_l^{01}}{1+f_l^{01}/(2l+1)} \\ + \frac{1}{3} \sum_J \frac{2J+1}{2l+1} \big( \mathcal{S}^{J0}_{ll} +\mathcal{S}^{J1}_{ll} \big) \bigg] = 0
\end{multline}
and
\begin{multline} \label{tensorSR2}
\sum_l \bigg[ \frac{f_l^{00}}{1+f_l^{00}/(2l+1)} - \frac{3 f_l^{01}}{1+f_l^{01}/(2l+1)} \\ + \sum_J \frac{2J+1}{2l+1} \big( 3 \mathcal{S}^{J1}_{ll} - \mathcal{S}^{J0}_{ll} \big) \bigg] = 0
\end{multline}
where the first values of $\mathcal{S}^{JT}_{ll}$ can easily be deduced from the first values of the diagonal and coupled  matrix elements defined in Eqs. \eqref{stabdiag} and \eqref{stabnondiag}, together with the relations \eqref{defmathcalA} and \eqref{FJT}\footnote{In Ref.\cite{Friman1979}, the first and second sum rules when tensor forces are taken into account read, respectively:
\begin{equation}
\sum_l \bigg[ \frac{f_l^{00}}{1+f_l^{00}/(2l+1)} + \sum_J \frac{2J+1}{2l+1} \mathcal{S}^{J1}_{ll} \bigg] = 0
\end{equation}
and
\begin{multline}
\sum_l \bigg[ \frac{2}{3} \frac{f_l^{00}}{1+f_l^{00}/(2l+1)} + \frac{f_l^{01}}{1+f_l^{01}/(2l+1)} \\ + \frac{1}{3} \sum_J \frac{2J+1}{2l+1} \mathcal{S}^{J0}_{ll} \bigg] = 0
\end{multline}
These are equivalent expressions of  Eq.\eqref{tensorSR1} and Eq.\eqref{tensorSR2} since the first relation can be obtained by multiplying the first sum rule
\eqref{tensorSR1} by three and adding it the second one \eqref{tensorSR2}, and the second relation by multiplying the first sum rule \eqref{tensorSR1} by nine and subtracting it the second one \eqref{tensorSR2}. In practice, these sum rules are violated (see section \ref{SRcorps}), so that the two writings are no longer equivalent. As a generalization of the sum rules \eqref{notensor1} and \eqref{notensor2} is proposed in the presence of tensor forces, the forms \eqref{tensorSR1} and \eqref{tensorSR2} that restore these when the tensor forces are set to zero have been implemented.}. 

\noindent By switching off the tensor interaction in Eq.\eqref{meA} and Eq.\eqref{meA2}, i.e.\ by demanding $h_l^{10} = 0$ and $h_l^{11} = 0$ for all $l$, one finds out:
\begin{equation}
\mel*{l' J}{F}{l J} = \delta_{ll'} \bigg( 1 + \frac{f_l^{1T}}{2l+1} \bigg),
\end{equation}
Thus, Eq.\eqref{defmathcalA} reduces, for both the diagonal and non diagonal parts, to:
\begin{equation} 
\mathcal{S}^{JT}_{ll} = \frac{f^{1T}_{l}}{1+f^{1T}_{l}/(2l+1)}
\end{equation}
which allows to recover Eq.\eqref{bTS} for all $S$ and $T$ values. One consistently falls back on the sum rules without tensor forces, given by Eq.\eqref{notensor1} and Eq.\eqref{notensor2}.\\

\subsection{Nuclear matter filters}

In the context of the original fitting procedure of the Gogny interaction, some nuclear matter properties are used as filters. Indeed, for each parameterization proposed by the HFR emulator, one
calculates at least the saturation density $\rho_0$, the total binding energy per particle $\mathcal{E}_0/A$, the incompressibility, the effective masses $m^*$, the symmetry energy $a_{\tau}$ and its slope $L$.
If their values are acceptable, considering the
experimental or empirical values, then the parameterization is kept for the next step which implies finite nuclei calculations. 
The aim of this section is to propose an overview
of the results associated with various Gogny 
interactions. In the fitting process, the priority is given to the quantities that have an experimental counterpart or empirical evaluation. However, the other INM properties are considered as the
sum rules for example.
In the following, one will discuss firstly the equation of state which allows to extract $\rho_0$, $\mathcal{E}_0/A$, $K_{\infty}$, $m^*$, $a_{\tau}$ and $L$. Secondly, one will show the (S,T) decomposition
and compare to G-matrix results since the G-matrix serves as a guide for the Gogny interaction in order to fix the sign of each (S,T) component to the equation of state according to the density and avoid at maximum bad compensation between channels. Then a deeper inspection has been recently proposed to see the role of the various terms of the interaction in a partial wave decomposition. One will continue the discussion with the effective masses and Landau parameters including stability criteria and sum rules.
Depending on the quantity, calculations include results from various
Gogny interactions, namely DG, D1S, D1M, D2, D3G3, D1ST2a and D1ST2c.

\subsubsection{Equation of state in infinite symmetric nuclear matter}

In this sub-section, one is interested by the properties of the equation of state of the symmetric nuclear matter (SNM).
Numerical values of several common physical quantities (see section \ref{physicalquantities}) evaluated in SNM 
are given in Table \ref{tab:physicalquantities} for D1S, D1M, D3G3, D2 and DG Gogny interactions. The saturation density $\rho_0$, the binding energy per nucleon $\mathcal{E}_0/A \equiv \mathcal{E}^{{S}}(\rho_0)/A$ (see Eq.(\ref{Ekinsym}) and Eq.(\ref{Epotsym})), the incompressibility $K_\infty$ \footnote{The standard value found in the literature for D1S is 210 MeV which is a misprint. The right value is 203 MeV.} (see Eq.\eqref{Kinfkin} and Eq.\eqref{Kinfsym}), the effective mass $m^{*}/m$ (see Eq.\eqref{effmassformula}), the symmetry energy $a_\tau \equiv \mathcal{E}_{{sym}} (\rho_0)$ (see Eq.\eqref{kinesym} and \eqref{potsym}) and the slope $L$ of the symmetry energy, all evaluated at saturation density $\rho_0$, are tabulated. 
For comparison, some empirical values extracted from experiment
are also indicated. 

\begin{table}[htb!]
    \centering
    \begin{tabular}{
        l
        S[table-format = 1.3]
        S[table-format = -2.2]
        S[table-format = 3]
        S[table-format = 0.2]
        S[table-format = 2.1]
        S[table-format = 2.1]
        }
        \toprule
        \multicolumn{1}{c}{} & 
        \multicolumn{6}{c}{Physical quantities in SNM}\\
        \cmidrule(lr){2-7}        
        & {$\rho_0$} & {$\mathcal{E}_0/A$} & {$K_{\infty}$} & {$m^{*}/m$} & {$a_\tau$} & L\\
        Inter. & {(\si{\femto \metre^{-3}})} & {(\si{\MeV})} & {(\si{\MeV})} & {} & {(\si{\MeV})} & {(\si{\MeV})} \\
        \midrule
        DG &  0.163 & -16.01 & 210 & 0.74 & 31.1   & 43.02 \\     
        D1S &  0.163 & -16.02 & 203 & 0.70 & 31.1  & 22.43 \\
        D1M &  0.165 & -16.03 & 225 & 0.75 & 28.6  & 24.83 \\
        D3G3 &  0.165 & -16.05 & 227 & 0.68 & 32.6 & 36.70 \\
        D2 &  0.163 & -16.00 & 209 & 0.74 & 31.1   & 44.85 \\   
        Emp. &  0.17(2)  & -16(1)  &  215(15) &  0.70(5) & 30(2) & 50(10)  \\
        \bottomrule
    \end{tabular}
    \caption{Various physical quantities evaluated in symmetric nuclear matter (SNM) for different Gogny interactions. Empirical values are also indicated.}
    \label{tab:physicalquantities}
\end{table}

The saturation density $\rho_0$ has been obtained from the charge distribution of heavy nuclei taking into account the corrections induced by the Coulomb repulsion and the surface tension. The value commonly adopted is $\rho_0 = \SI{0.17(2)}{\femto \metre^{-3}}$ \cite{Frois1987}. Besides, the energy per nucleon $\mathcal{E}_0/A$ and symmetry energy $a_{\tau}$ appear explicitly in the semi-empirical mass formula. Their values have been then deduced from the successive adjustments of this formula on experimental binding energies \cite{Myers1966,Myers1969,Myers1970,Moller1990,Moller1995}, leading to $\mathcal{E}_0/A = \SI{16(1)}{\MeV}$ and $a_\tau = \SI{30(2)}{\MeV}$. The incompressibility has been determined using various phenomenological effective interactions to reproduce experimental data of heavy nuclei and the energy of the breathing mode in $\isotope[208]{Pb}$, in such a way that $K_{\infty} = \SI{215(15)}{\MeV}$ \cite{Blaizot1995}. 
Finally,  an experimental nucleon-nucleus scattering analyzed in the framework 
of the optical potential furnished values for the effective mass, $m^{*}/m = \SI{0.70(5)}{}$ \cite{Mahaux1991}. From astrophysical measurements, the value 
of the slope $L$ of the symmetry energy has been estimated to be $L=50(10)$ \cite{GonzalezBoquera2018,Vinas2018,Mondal2020,GonzalezBoquera2017}.

Unsurprisingly, the DG interaction, like its counterparts, fall within the ranges of empirical values. 
Indeed, in general, these empirical values behave like filters in the fitting code (or constraint in the protocols that have produced D1M and D3G3). 
The DG values are pretty close to those of D2, which corroborates the initial will not to move too far from D2, underpinned by the similarity of the central and density-dependent part of these interactions (see Refs. \cite{Chappert2015,Zietek2025,Zietek2023}). Finally, one notes that the properties of these two interactions are not very different from the others, except for the incompressibility, which is lower, and the effective masses of D1S and D3G3, which are the smallest.
Only D2 and DG satisfy the experimental requirement of the slope L of the symmetry energy.


\begin{figure}
\centering
\includegraphics[width=0.95\linewidth]{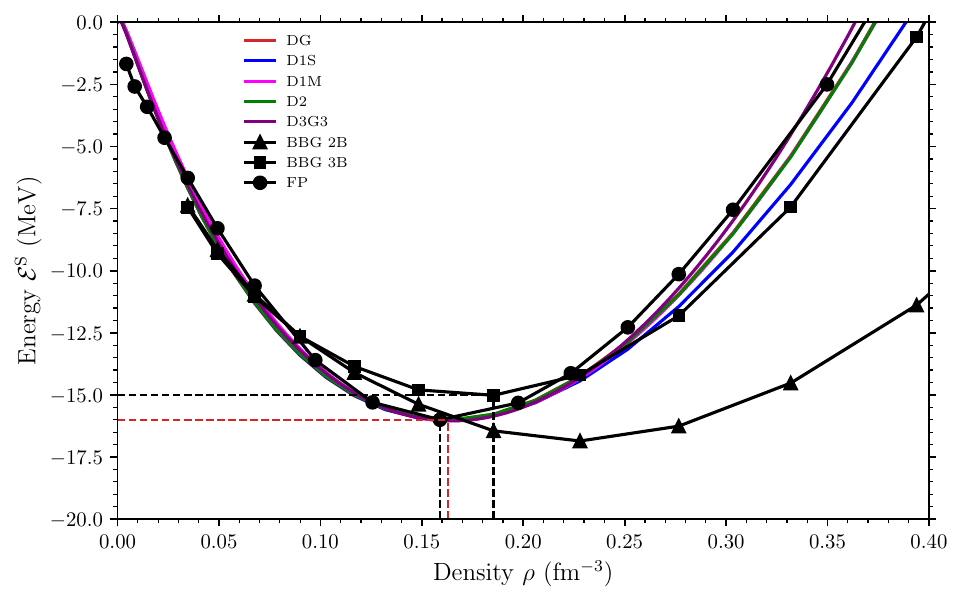}
\caption{Equations of state in symmetric nuclear matter (SNM) for D1S, D1M, D3G3, D2 and DG interactions. Realistic calculations based on Friedman--Pandharipande (FP) and Bethe--Brueckner--Goldstone approaches, with (BBG 3B) and without three-body corrections (BBG 2B), are also represented.}
\label{fig:Esym}
\end{figure}

In Fig. \ref{fig:Esym} are drawn the SNM equations of state representing $\mathcal{E}_0/A $ as a function of the medium density $\rho$, for several Gogny interactions listed in Table \ref{tab:physicalquantities}, which are compared to two type of realistic calculations.
The first one utilizes a variational method applied by Friedman and Pandharipande (FP) \cite{Friedman1981} to the realistic interaction UV14 \cite{Lagaris1981}. The Urbana potential possesses two- and three-body contributions. It reproduces the scattering data in $S$-, $P$-, $D$- and $F$-waves and fits the ground-state energy, density and compressibility in symmetric nuclear matter. 
The second one takes advantage of the Bethe-Brueckner-Goldstone (BBG) method, a generalization of the $G$-matrix appearing in the Goldstone development, applied by Baldo \textit{et al.} \cite{Baldo2008} to the realistic interaction AV14 \cite{Wiringa1984}. The Argonne potential presents a good fit to deuteron properties and neutron--proton scattering below $\SI{330}{\MeV}$. Unfortunately the BBG method does not natively retain the three-body contributions, so that only two-body interactions are taken into account within this approach. Nonetheless, by averaging the position of a third particle by means of Lejeune \textit{et al.}\ technique \cite{Lejeune1986}, a density dependence simulating three-body effects can be adjoined to the BBG method, noted “BBG-3B” here.

All the Gogny interactions, including DG, correctly reproduce the parabolic behavior observed in realistic FP calculations in this low-density regime. It is worth noticing the limitation of realistic two-body calculations noted BBG-2B, whose parabola is wider, with a minimum located at a lower energy for a higher density. The coordinates of the saturation points $(\rho_0, \mathcal{E}_0/A)$, corresponding to the minima of the curves, are spotted by black dashed lines for three-body realistic calculations, and by a red dashed line for DG. One checks that, as well as falling within the intervals of empirical values, the DG saturation point is very close to that of the FP calculation $(0.159, -16.00)$, with the saturation density of DG a little bigger than the FP one. This is a strong result for an interaction to provide reliable results in finite nuclei. The other Gogny interaction provide with very similar results. The saturation density predicted by BBG-3B calculations, $\rho \simeq \SI{0.185}{\femto\metre^{-3}}$, then appears to be too high. 
Since incompressibility is proportional to the second derivative of the energy evaluated at the saturation density (see Eq.\eqref{Kinfdef}), its magnitude tells us about the curvature of the equation of state around that point. The greater the incompressibility, the steeper the rise of the curve. This is indeed the case, as the D1M and D3G3 curves are rising more sharply than D1S, D2 and DG around the saturation point, in line with the inequalities deduced from Table \ref{tab:physicalquantities}, $K_\infty^{\text{D1M}} \simeq K_\infty^{\text{D3G3}} > K_\infty^{\text{D1S}} \simeq K_\infty^{\text{D2}} \simeq K_\infty^{\text{DG}}$. Note that this trend is more pronounced at high densities, due to the factor $9 \rho^2$ in the definition of $K_\infty$. These curves are therefore closer to the FP results in the range $\interval{\rho_0}{0.30}$ than are D1S, D2 and DG, even though D2 and DG present a steeper behavior than D1S.

\subsubsection{Energy in (S,T) channels in symmetric INM}

\begin{figure*}
\centering
\includegraphics[width=0.95\linewidth]{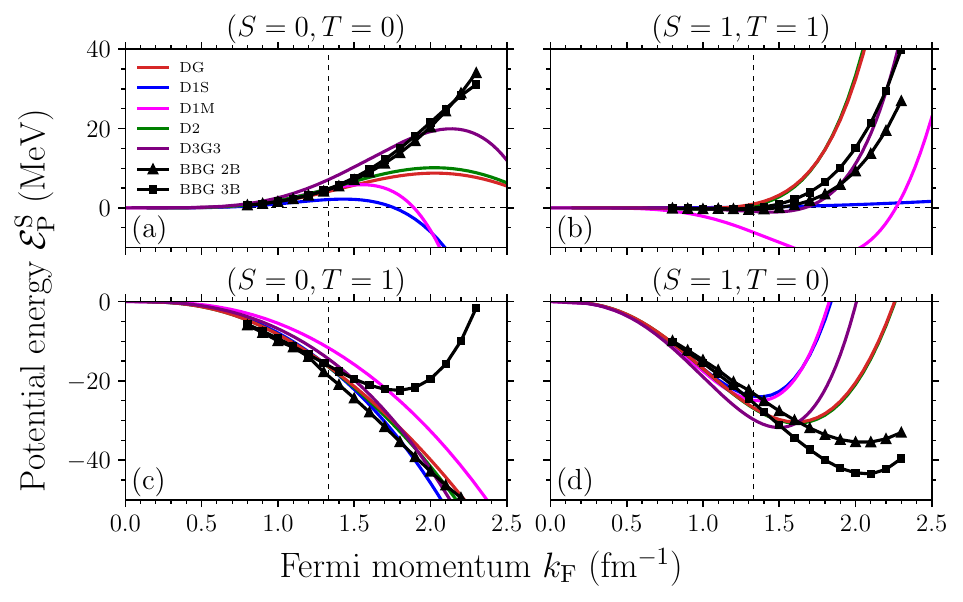}
\caption{Potential energy as a function of the Fermi momentum in symmetric nuclear matter for D1S, D1M, D2, D3G3 and DG interactions. The energy is decomposed in the $(S=0,T=0), (S=1,T=1), (S=0,T=1)$ and $(S=1, T=0)$ channels in panels (a), (b), (c) and (d), respectively. The results are compared to the two-body (BBG 2B) and three-body (BBG 3B) Bethe-Brueckner-Goldstone realistic calculations. The $\mathcal{E}_{P}^{{S}}/A = 0$ equation is represented by a dashed line in panels (a) and (b) for clarity.}
\label{fig:STchannels}
\end{figure*}

In addition to the SNM equation of state, an analysis by (S,T) channels is always performed, with a comparison to G-matrix results that serve as a 
guide to ensure coherent attractivity and repulsivity with the phenomenological effective Gogny interaction. Indeed, it is desired to avoid cross-
channel compensation wherever possible. A concrete example is the pairing which is linked to the (S=0, T=1) channel. At low density, 
this channel has to be attractive in order to provide good pairing properties in coherence with the physics of nuclei. The total binding energy per nucleon $\mathcal{E}^{{S}}/A$ can be divided into kinetic and potential parts as already discussed previously:
\begin{equation}
\frac{\mathcal{E}^{{S}}}{A} = \frac{\mathcal{E}_{K}^{\text{S}}}{A} + \frac{\mathcal{E}_{P}^{{S}}}{A},
\end{equation}
As well-known, the potential contribution \eqref{Epotsym} can itself be expressed as a sum over the coupled spin $S \in \{ 0, 1 \}$ and isospin $T \in \{ 0, 1 \}$ of the two-nucleon system, i.e.:
\begin{equation}
\frac{\mathcal{E}_{P}^{{S}}}{A} = \sum_{ST} \left. \frac{\mathcal{E}_{P}^{{S}}}{A} \right\vert_{ST},
\end{equation}
where $\mathcal{E}_{{P}}^{{S}}/A \vert_{ST}$ corresponds to the potential energy per nucleon in the $(S, T)$ channel. According to this decomposition, the nuclear potential can be viewed as four independent interactions, each acting in its own subspace $(S, T)$. 
The four possible configurations therefore give rise to four different contributions to the interaction with different meaning implying different mechanisms. The evolution of the potential energy in the $(S, T)$ channels according to the Fermi momentum $k_{{F}}$ are plotted in Fig. \ref{fig:STchannels}, and compared to the former BBG realistic calculations.

In the singlet-odd channel $(S=0, T=0)$, realistic calculations are very similar, with or without three-body effects taken into account. They predict positive potential energy and a repulsion between nucleons ($k_{F}$ is homogeneous to the inverse of a length) that increases with the Fermi momentum. The tested Gogny interactions are also repulsive and fit rather well realistic curves for small values of $k_{{F}}$, before collapsing and then becoming attractive. For D1S, the collapse starts at $k_{{F}} \simeq \SI{1.5}{\femto \metre^{-1}}$ and makes the energy negative-valued from $k_{{F}} \simeq \SI{1.8}{\femto \metre^{-1}}$. The behaviour is very similar in the case of D1M, with a steeper descent. For D2 and DG, it is carried forward to $k_{{F}} \simeq \SI{2.1}{\femto \metre^{-1}}$ and $k_{{F}} \simeq \SI{2.0}{\femto \metre^{-1}}$, with negative-valued energies from $k_{{F}} \simeq \SI{2.8}{\femto \metre^{-1}}$ and $k_{{F}} \simeq \SI{2.7}{\femto \metre^{-1}}$, respectively. Equivalently, D1S, D2 and DG become attractive from around $\rho \simeq 1.4 \rho_0$, $\rho \simeq 3.8 \rho_0$ and $\rho \simeq 3.3 \rho_0$, respectively. This is less pathological for D2 and DG in comparison to D1S, in particular in view of astrophysical applications. For nuclei, the situation is more favorable. 
The apparent improvement obtained with D3G3 is polluted by the presence of a collapse with negative values appearing around $k_{{F}} \simeq \SI{2.7}{\femto \metre^{-1}}$
and a steeper slope than D2 and DG.

In the triplet-odd channel $(S=1, T=1)$, the two types of realistic calculations are nearly identical until $k_{{F}} \simeq \SI{1.5}{\femto \metre^{-1}}$; beyond, the inclusion of the three-body effects implies a steeper rise. Their contributions are globally repulsive, except around $k_{{F}} \simeq \SI{1.3}{\femto \metre^{-1}}$, where they exhibit a tenuous attraction that does not exceed $\SI{500}{\keV}$. The D1S interaction is repulsive at all densities, but fails to reproduce the rise of realistic calculations, contrary to D2 and DG. The ascent is a little too strong with D2 and DG, beginning at $k_{{F}} \simeq \SI{1.2}{\femto \metre^{-1}}$ compared to $k_{{F}} \simeq \SI{1.5}{\femto \metre^{-1}}$ for realistic calculations, although very slightly closer to that of the BBG 3B curve in the case of DG. As for the attraction, D2 and DG succeeded in reproducing it even though it takes place a bit earlier for D2 and DG, at $k_{{F}} \simeq \SI{1}{\femto \metre^{-1}}$, and reaches a maximum value of $\SI{350}{\keV}$ and $\SI{200}{\keV}$, respectively.
The D1M parametrization develops a too strong attractivity.
Here again, the D3G3 seems to behave more properly than the
other interactions. However, to compensate its behavior
at large $k_{{F}}$ in the (S=0, T=0) channel, the slope
present a to high rigidity, which is also the case for D2 and DG.

In the singlet-even $(S=0, T=1)$ channel, two- and three-body realistic calculations are attractive and blend into each other up to $k_{{F}} \simeq \SI{1.2}{\femto \metre^{-1}}$. Afterwards, while BBG curve continues decreasing, the descent of BBG 3B stops at $k_{{F}} \simeq \SI{1.7}{\femto \metre^{-1}}$ and evolves into a highly repulsive behavior. This is clearly the channel in which the three-body effects are the most visible. The BBG results are properly reproduced by D1S, without the physics of the three-body interactions.
This is an important point as this channel correspond to the physics of superfluidity in nuclei on the low Fermi momentum side (surface effects). 
By tuning the intensity of the finite-range density-dependent term in this subspace, Chappert \textit{et al.} managed to create a D2-type interactions following the BBG-3B curve \cite{Chappert2007,Chappert2015}. Regrettably, these interactions manifested non-zero pairing in magic nuclei because of their excessive attractivity at low Fermi momenta. It is known that pairing interaction should be very close to the bare interaction, i.e. it is not very renormalized by medium effect. Thus, it should not depend very much on the density of the medium.
The conclusion was that an interaction should not be too far from D1S in this channel so as not to show pairing in magic nuclei, hence the D2 curve. 
Maybe, a second finite-range density-dependent term acting at larger values of $k_{{F}}$ should be investigating.
One notes that all the Gogny interactions manifest the same trend in 
this channel, with only a difference with D1M that exhibits less
attractive for medium and large values of $k_{{F}}$.

In the triplet-even $(S=1, T=0)$ channel, realistic calculations reveal attraction with decreasing energy until $k_{{F}} \simeq \SI{2}{\femto \metre^{-1}}$, which is a bit more pronounced with the three-body correction. A reduction of the potential energy per particle of about $\SI{8}{\MeV}$ with respect to pure BBG previsions is induced, at the minimum, by this correction. Indeed, it is about $\SI{-35.5}{\MeV}$ without, compared to $\SI{-43.5}{\MeV}$ with three-body effects taken into consideration. Then the curves go up, as a signature of repulsion. This tendency is observed by D1S, even if the minimum of $\SI{-24}{\MeV}$ appears sooner, at $k_{{F}} \simeq \SI{1.4}{\femto \metre^{-1}}$ (D1M displays similar results). Predictions are amended with D2 and DG. Both are attractive and stick to the BBG curve before reaching their minima at the higher Fermi momentum $k_{{F}} \simeq \SI{1.7}{\femto \metre^{-1}}$, with respective energies $\SI{-30.5}{\MeV}$ and $\SI{-29.5}{\MeV}$. These very close points mark the start of the final ascent predicted by realistic calculations.

One would like to end this section with a final remark. As commented in the section \ref{GPTtoD1}, the density-dependent term may include some of the short-range effects of the tensor interaction in the (S=1,T=0)
channel. Albeit the tensor force does not contribute to the energy in nuclear matter equation of state, its presence renormalizes the coefficients of the density-dependent term and should, ultimately, affect the behavior of the DG interaction in $S=1$ channels. 
However, the D2 and DG curves are very similar in these channels. This observation can be attributed to the relatively long range associated with the tensor term and its intensity, which is much less than that of the density-dependent interaction. 

From this (S,T) channel analysis, one can infer that there is
still a  work to be done to desintricate the (S,T) channels in the case of the Gogny interaction,
in particular the odd ones, in order to avoid at maximum 
compensations between themselves. 

\subsubsection{Partial wave decomposition in symmetric INM} \label{partialwaves}

\begin{figure}
\centering
\includegraphics[width=0.95\linewidth]{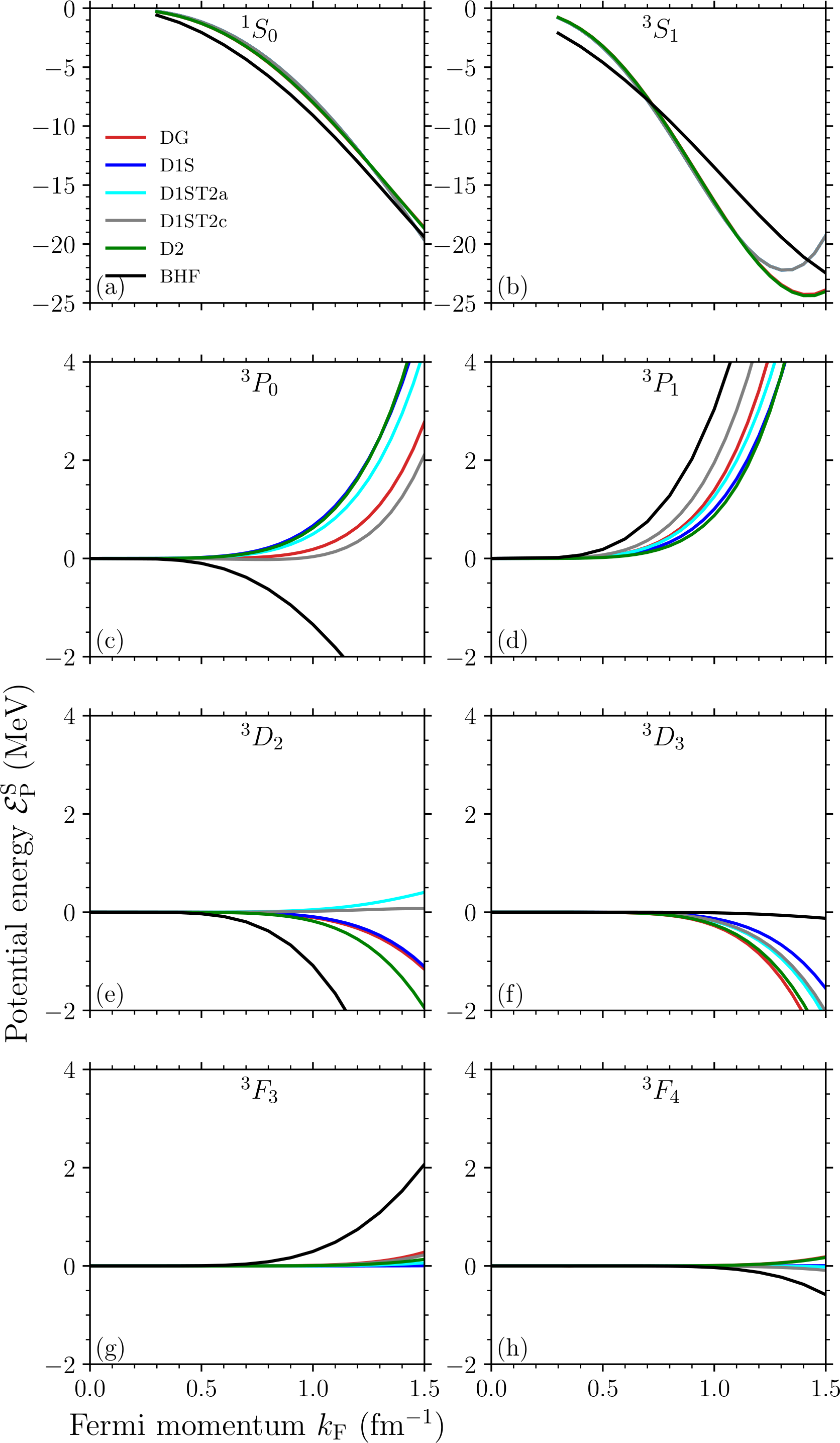}
\caption{Potential energy per particle of the partial waves $^1S_0$, $^3S_1$, $^3P_0, ^3P_1, ^3D_2, ^3D_3, ^3F_3$ and $^3F_4$ versus the Fermi momentum, evaluated in symmetric nuclear matter for DG, D1S, D1ST2a, D1ST2c and D2 interactions. The predictions are compared with realistic calculations using Brueckner--Hartree--Fock approximation \cite{Arellano2015,Arellano2016}.}
\label{fig:Indivpartialwaves}
\end{figure}

As already explained, neither the spin-orbit nor the tensor 
terms of the Gogny interactions contribute to INM equation of state. 
Thus, the previous quantities studied so far do not allow to 
figure out from nuclear matter whether these two terms have 
been fitted in the right way. 
Actually, one can get some clues by performing a partial wave decomposition in symmetric INM. This is achieved by projecting the potential energy per particle onto a basis characterized 
by the orbital $\vec{L}$, intrinsic $\vec{S}$ and total $\vec{J} \equiv \vec{L} + \vec{S}$ angular momenta, according to \cite{Rotival2008}:
\begin{equation} \label{partialwavesexp}
\frac{\mathcal{E}_{P}^{\text{S}}}{A} = \sum_{LSJ} \frac{\mathcal{E}_{P}^{\text{S}}}{A} \smash{\big[} ^{2S+1}L_J \big],
\end{equation}
where $\mathcal{E}_{P}^{\text{S}}/A \smash{[} ^{2S+1}L_J]$ denotes the contribution  to the potential energy of the corresponding partial wave $^{2S+1}L_J$.
Finite-range interactions participate to all partial waves
contrary to zero-range ones.
An analysis of BHF calculations, as the one done in Refs.\cite{Arellano2015,Arellano2016} with the realistic interaction AV18 \cite{Wiringa1995}, reveals that the contributions of partial waves decrease as $L$ grows. For $L>3$, they become negligible, so that the expansion \eqref{partialwavesexp} can be truncated to order $L=3$. Thus, in the following a focus will be propose on the first partial waves $S$, $P$, $D$ and $F$, which allow to reveal
effects coming from the spin-orbit and tensor terms.
Results are shown in Fig. \ref{fig:Indivpartialwaves} for DG, D2, D1S, D1ST2a and D1ST2c.

The $^1S_0$ and $^3S_1$ partial waves are found close to the BHF
predictions for all Gogny interaction. The constraint of the pairing properties in the fitting procedure is probably at the origin of this result.
One sees that the reproduction of $P$ waves is enhanced by the tensor-dependent interactions DG, D1ST2a and D1ST2c, with respect to D1S and D2, although the monotony of $^3P_0$ is not recovered. 
Only in the cases of D1ST2c and DG, the partial wave starts to develop a small attractivity that, unfortunately, disappears 
at $k_{{F}} \simeq 0.75$ fm$^{-1}$ and $\simeq 0.4$ fm$^{-1}$, respectively. The $^3P_1$ partial wave is also improved with 
tensor-dependent interactions.
In the $D$ waves, the descent of the BHF curve is less pronounced but observed with Gogny interactions, excepted for D1ST2-type ones in $^3D_2$. In $^3D_3$, the predictions of the Gogny interactions have the good sign but manifest a too strong attractivity in comparison with the BHF predictions. 
Finally, the results obtained with all the Gogny interactions in $F$ waves are relatively flat and close to zero.  
However, this is not pathological since the main contribution to the partial wave decomposition is due to the $P$ waves, and to a lesser extent, to the $D$ waves. 
Indeed, for instance, DG and D2 predictions differ from about $\SI{2.5}{\MeV}, \SI{0.8}{\MeV}$ and $\SI{150}{\keV}$ in $^3P_0, ^3D_2$ and $^3F_3$ waves, respectively, at $k_{{F}} = \SI{1.5}{\femto\metre^{-1}}$. Consequently, even if D2 is closer to the BHF calculations in $^3D_2$, the important improvement is obtained with DG in $^3P_0$.

\begin{figure}
\centering
\includegraphics[width=0.95\linewidth]{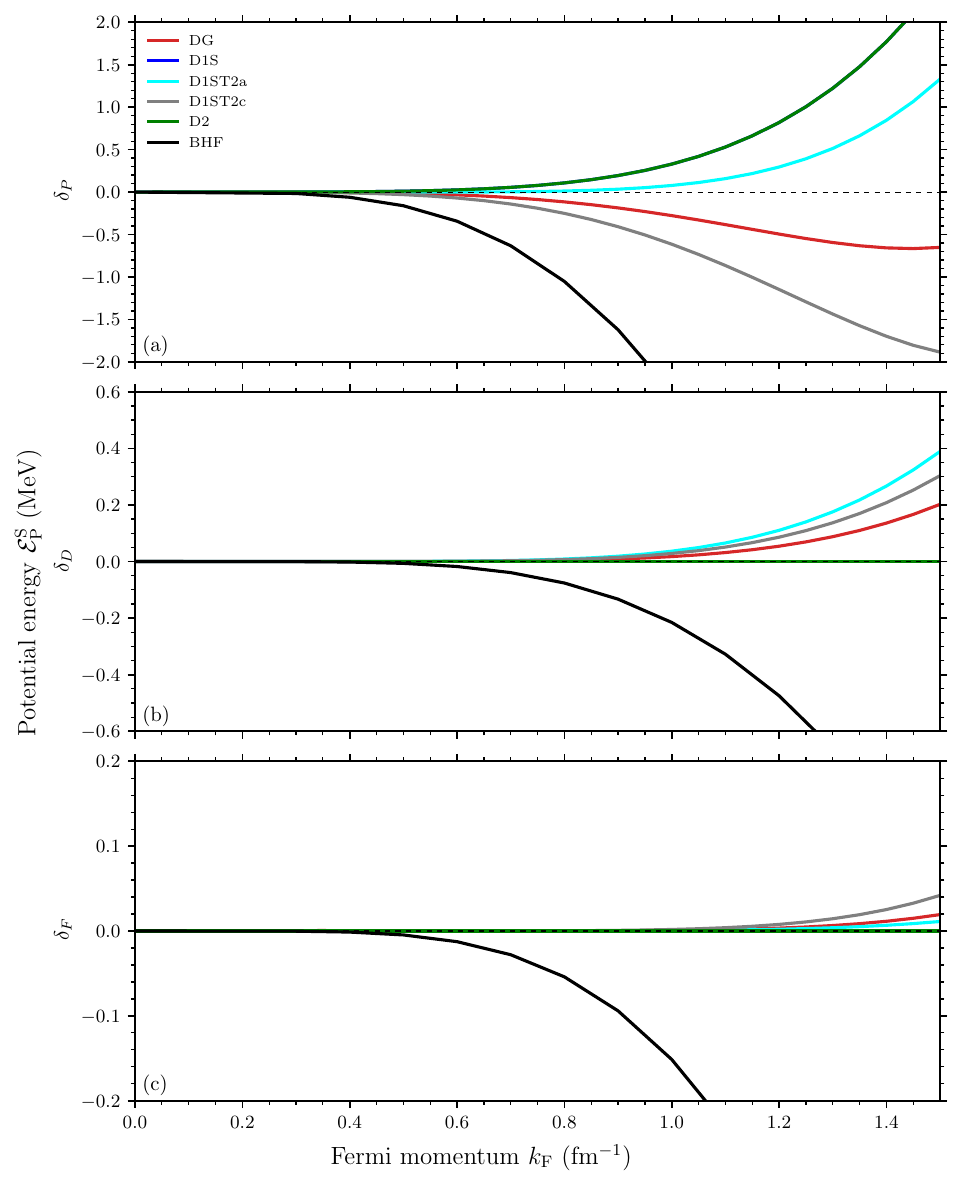}
\caption{Potential energy per particle of the differences in partial waves $\delta_P, \delta_D$ and $\delta_D$ given in \eqref{deltapartial} versus the Fermi momentum, evaluated in symmetric nuclear matter for DG, D1S, D1ST2a, D1ST2c and D2 interactions. The predictions are compared with realistic calculations using Brueckner-Hartree-Fock approximation \cite{Arellano2015,Arellano2016}. Zero differences are displayed by dashed lines for clarity.}
\label{fig:Partialwaves}
\end{figure}

Even more convincing that the previous analysis, specific differences
between partial waves allow to cancel the contributions of the central
and density-dependent terms. 
Indeed, it has been shown \cite{Rotival2008} that for given values of $L$ and $S$, the contributions of central and density-dependent terms to $\mathcal{E}_{P}^{\text{S}}/A \smash{[} ^{2S+1}L_J]$ are the same for all values of $J$, up to a global factor $(2J+1)$. On the other hand, the spin-orbit and tensor terms contribute only to the $S=1$ channel.
It can be shown that the differences $\delta_{P}$, $\delta_{D}$ and $\delta_{F}$ defined below depend only on spin-orbit and tensor contributions: 
\begin{subequations}
\begin{align} \label{deltapartial}
\delta_P & \equiv \frac{\mathcal{E}_{P}^{\text{S}}}{A} \smash{\big[} ^{3}P_0 \big] - \frac{1}{3} \frac{\mathcal{E}_{P}^{\text{S}}}{A} \smash{\big[} ^{3}P_1 \big], \\
\delta_D & \equiv \frac{1}{5} \frac{\mathcal{E}_{P}^{\text{S}}}{A} \smash{\big[} ^{3}D_2 \big] - \frac{1}{7} \frac{\mathcal{E}_{P}^{\text{S}}}{A} \smash{\big[} ^{3}D_3 \big], \\
\delta_F & \equiv \frac{1}{7} \frac{\mathcal{E}_{P}^{\text{S}}}{A} \smash{\big[} ^{3}F_3 \big] - \frac{1}{9} \frac{\mathcal{E}_{P}^{\text{S}}}{A} \smash{\big[} ^{3}F_4 \big], 
\end{align}
\end{subequations}
The first analysis of this type can be found in 
Ref.\cite{Davesne2015a} with the Skyrme N3LO pseudo-potential and in Ref.\cite{Davesne2016a} for D1-type Gogny interactions. 
Here, this work has been extended to DG, D1ST2a, D1ST2c and D2 \cite{Guillaume}. The results are displayed in Fig. \ref{fig:Partialwaves}, including a comparison with BHF results obtained with AV18 interaction \cite{Arellano2015,Arellano2016}.

Firstly, one notices that the magnitudes of the differences in partial waves evaluated with the various Gogny interactions decreases sharply as $L$ increases, until becoming very small in $F$ waves, in accordance with BHF predictions. 
As expected, the contributions of D1S and D2 interactions are the same since they are tensor-independent and their spin-orbit terms 
are identical. They additionally provide no contributions in $D$ and $F$ waves as the zero-range spin-orbit force is zero in $L>1$ states, as shown in Ref.\cite{Rotival2008}. 
The predictions of $\delta_P$ are improved by all tensor-dependent interactions, but a little deteriorated for the differences $\delta_D$ and 
$\delta_F$, with respect to those of D1S and D2. 
This is rather encouraging since, among the differences under study, the main contribution to the partial wave decomposition is related to $P$ waves. In particular, a striking change of monotony is observed with DG and D1ST2c curves to get closer to the realistic calculations. It is worth noticing that such behavior is not shared with D1ST2a, whose difference with D1S and D2 is only due to the tensor force, the strength of the spin-orbit term being the same ($W_{ls} = \SI{130.0}{\MeV\femto\metre^5}$). Although modifying the outputs in the right direction, the tensor term alone here seems unable to reverse the monotony obtained with D1S and D2 interactions. 
This seems to indicate that the strength of the spin-orbit force may be reduced as well, as for DG ($W_5 - H_5 = \SI{115.849}{\MeV\femto\metre^5}$) and D1ST2c ($W_5 = \SI{103}{\MeV\femto\metre^5}$) interactions. 
Among the tensor-dependent interactions, DG has the least negative impact on $\delta_D$ while it is slightly worse than D1ST2a in the reproduction of $\delta_F$, but the difference is minuscule, reaching at most $\SI{10}{\keV}$ at $k_{{F}} = \SI{1.5}{\femto\metre^{-1}}$. 
Gathering the previous remarks, even though not perfect, DG appears as the best comprise, its spin-orbit and tensor part globally enhancing the description in terms of partial waves. 
An interesting proposal would be to add a second ranges to this two terms to see the impact on the partial waves and the differences $\delta_{P}$, $\delta_{D}$ and $\delta_{F}$.
Future extensions of the DG interaction may be inspired by the results
obtained with the M3Y interaction which possesses several ranges in each terms \cite{Davesne2015a,Davesne2016a}.

\subsubsection{Neutron matter equation of state}\label{nmes}

As already discussed in the subsections \ref{D1Nparam}, \ref{secD2} and \ref{DGsection}, a reduction of the energy drift along isotopic chains was achieved by the D1N, D2, D1M, D3G3 and DG interactions. To do so, requirements on the neutron matter ($\beta = 1$) equation of state was put in the fitting protocols of those interactions. In the specific cases of D1S and D2, eight points of the equation of state obtained by Friedman and Pandharipande (FP) \cite{Friedman1981} in neutron matter had to be reproduced with chosen accuracies. Contrary to D1M and D3G3, the energy drift was also controlled through a constraint on the energy difference $\Delta \varepsilon$ which can contribute to shrink that drift.

\begin{figure}
\centering
\includegraphics[width=0.95\linewidth]{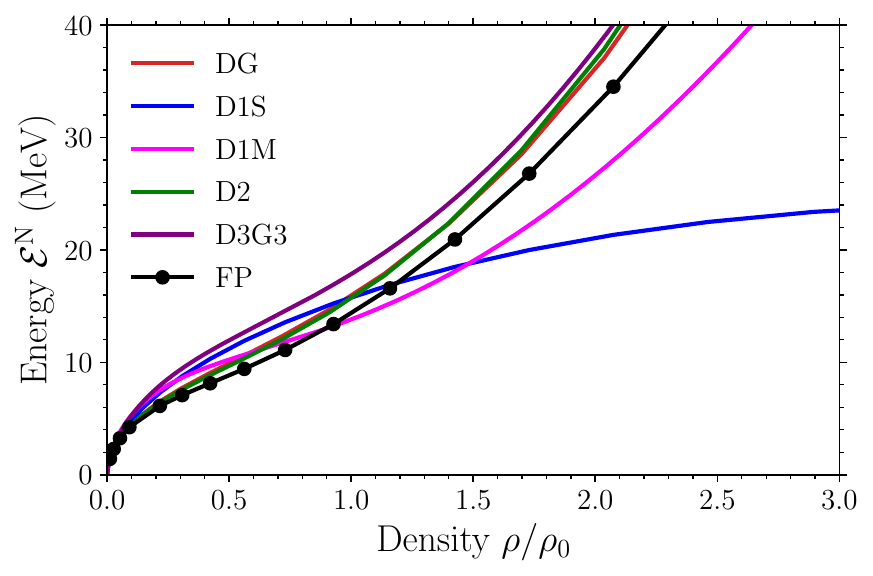}
\caption{Equations of state in neutron matter for D1S, D1M, D3G3, D2 and DG interactions. Realistic calculations based on the work of Friedman and Pandharipande (FP) are also represented. Densities are normalized by the saturation density $\rho_0$. The results are presented at all densities in the left-hand panel, and at a restricted range of densities in the right-hand panel.}
\label{fig:EqState}
\end{figure}

In Fig. \ref{fig:EqState}, the neutron matter equations of state for several Gogny interactions are shown according to the neutron density, and compared to FP predictions. 
It can be seen in panel (a) that the realistic calculations describe a strictly increasing and positive energy-valued equation of state at all densities, which ensures the stability of neutron matter. 
This is not the case of the D1S interaction as it starts decreasing at $\rho \simeq 3.5 \rho_0$ until vanishing identically at $\rho = 12 \rho_0$. 
Albeit this is not a good news for astrophysical perspectives, it is not fully satisfactory for exotic nuclei. 
However, even if in the range $\rho \leq \rho_0$ the D1S equation of states overestimates the realistic energies per particle by $\SI{3}{\MeV}$ at worst, this is actually problematic in neutron-rich nuclei for which the difference grows fast. 
A real improvement has been obtained in the parameterizations that has followed D1S.

In the $\rho \leq \rho_0$ regime, relevant for atomic nuclei, the D2 and DG interactions are no more than $\SI{1.5}{\MeV}$ above the realistic curve. This was expected since the fitting code tolerates a maximum deviation of $10\%$ to the FP values in this regime, and that they reach a maximum energy of $\SI{16}{\MeV}$ at saturation density.

In the $\rho > \rho_0$ regime, potentially interesting for astrophysical purposes, the energy difference to FP predictions enlarges a bit for the D2 and DG interactions, becoming about $\SI{2}{\MeV}$ at $\rho = 1.5 \rho_0$, but remaining under control. The D1S interaction, in contrast, goes completely off course and announces its collapse. Beyond $2 \rho_0$, the difference with FP growes but the most striking feature is that the curves do not break up. This is indeed the case for D2 and DG, with a somewhat better fit for DG at high densities.

\subsubsection{Splitting of the effective masses}

\begin{figure}
\centering
\includegraphics[width=0.95\linewidth]{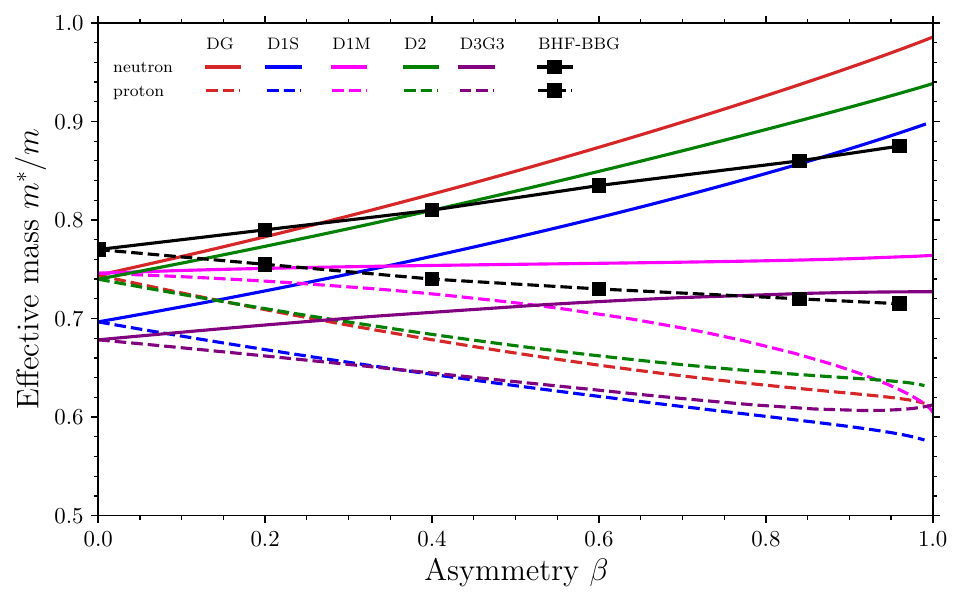}
\caption{Evolution of the proton (full lines) and neutron (dashed lines) effective masses as a function of the asymmetry parameter $\beta$ for D1S, D2, D1M, D3G3 and DG interactions. Microscopic calculations, founded on the Brueckner--Hartree--Fock approximation of the Bethe--Brueckner--Goldstone (BHF-BBG) theory are also displayed.}
\label{fig:Meff}
\end{figure}

The values of the equal neutron and proton effective masses in symmetric nuclear matter have been given in Table \ref{tab:physicalquantities} for different Gogny interactions. 
This quantity is controlled in the fitting procedure in such a way that its value is lower than one in order leave
room for beyond mean-field correlations on the single-particle spectrum (for example particle-vibration).
It is also interesting to control its evolution according to the asymmetry $\beta$ of the medium.
In Fig. \ref{fig:Meff}, the evolution of the effective
mass according to $\beta$ is shown for various Gogny interactions.
The results are compared to microscopic calculations based on the so-called Brueckner-Hartree-Fock (BHF) approximation of the Bethe-Brueckner-Goldstone (BBG) theory \cite{Bombaci1991}, which serve as a guide. 
Within this approach, the effective interaction to be used is evaluated by the self-consistent treatment of the $G$-matrix. They are denoted BHF-BBG from now on. 
One first notices that, for zero asymmetry, the effective mass of the microscopic calculations $m^{*}/m = 0.77$ slightly overshoots the empirical values used, $m^{*}/m = \SI{0.70(5)}{}$. This is, however, not the case of the other interactions that fit the interval. The D1M, D2 and DG interactions in particular are the closest to the microscopic results, being inside the empirical uncertainty range. 
As the asymmetry grows, the neutron and proton effective masses of the BHF-BBG scheme split up, with the neutron effective mass always greater than the proton effective mass, reaching a maximum difference of $0.16$ for pure neutron matter ($\beta = 1$). 
It turns out that this behavior is described by both non-relativistic (as considered here) and relativistic (using Dirac-Brueckner-Hartree-Fock approximation) \cite{vanDalen2005} microscopic calculations for neutron-rich nuclear matter. This salient phenomenon must be respected in order to obtain a good reproduction of the neutron single-particle energies. It is predicted by all the Gogny interactions displayed. 
The maximum difference in neutron and proton effective masses is about $0.32$, $0.30$ and $0.36$, reached at $\beta = 1$, for D1S, D2 and DG, respectively. The D2 interaction is the closest to the microscopic value and DG the furthest away. 
One has expected the filter on the neutron matter equation of state to substantially improve the difference in neutron and proton effective masses in neutron matter ($\beta = 1)$ for DG, as it did for D2. An interesting study would be to reduce
the difference between neutron and proton effective masses,
while preserving a neutron value close to the BHF prediction at $\beta \simeq 0.2$ which is characteristic of $^{208}$Pb.
Indeed, one ingredient of the improvement in the
reproduction of the kink in Pb isotopes is the increase
of the neutron effective mass with D2 (and DG) in comparison
with D1S.

\subsubsection{Landau parameters}\label{Landauparameters}

In this part, one wants to illustrate the role of the 
Landau parameters in
the selection process of a parameterization. One will present 
results for D1S, D2 and DG. 
The Landau parameters are fast quantities to calculate which already
allow a first discrimination of parameterizations as
\begin{itemize}[label=$-$]
\item several physical quantities are related to some central Landau parameters as the effective mass, the incompressibility and the symmetry energy \cite{Migdal1967}
\item they test the part of the residual interaction used 
in RPA calculations, the previously defined particle-hole interaction
\item they participate in the calculation of criteria ensuring the stability of nuclear matter and some rules  derived from the Pauli exclusion principle.
\end{itemize} 
As an example, one discusses in the following results obtained for 
the parameterizations D1S, D1St2a, D1ST2c, D2 and DG.

\paragraph{Landau parameters}

One recalls that only the central, density-dependent and tensor terms
participate to the particle-hole interaction defined in 
Eq.\eqref{MEph}. 
Table \ref{tab:LandauParametersDG} gives the values of the first six central and non-central Landau parameters, evaluated at saturation density $\rho_0$, for the DG parametrization. The values of the Landau parameters associated with D1S and D2 interactions are also displayed in Tables \ref{tab:LandauParametersD1S} and \ref{tab:LandauParametersD2}, for which non-central components do not exist.
\begin{table}[h]
    \centering
    \begin{tabular}{
        S[table-format = 1]
        S[table-format = -1.3]
        S[table-format = 1.3]
        S[table-format = 1.3]
        S[table-format = -1.3]
        S[table-format = 1.3]
        S[table-format = -1.3]
        }
        \toprule   
        \multicolumn{7}{c}{DG}\\
        \cmidrule(lr){1-7}      
        {$l$} & {$f_l^{00}$} & {$f_l^{10}$} & {$f_l^{01}$} & {$f_l^{11}$} & {$h_l^{10}$} & {$h_l^{11}$} \\
        \midrule
        0 &  -0.300 & 0.226 & 0.869 & 0.922 & 0.001 & -0.800 \\
        1 &  -0.775 & 0.027 & 0.528 & 0.501 & 0.001 & -0.559  \\
        2 &  -0.845 & 0.229 & 0.503 & -0.027 & {} & -0.148  \\
        3 &  -0.299 & 0.084 & 0.159 & -0.041 & {} & -0.025  \\
        4 &  -0.064 & 0.018 & 0.032 & -0.011 & {} & -0.003 \\        
        5 &  -0.010 & 0.003 & 0.005 & -0.002 & {} & {} \\
        6 &  -0.001 & {} & 0.001 & {} & {} & {} \\
        \bottomrule
    \end{tabular}
    \caption{Numerical values of the first Landau parameters of the interaction DG evaluated at saturation density. Empty boxes correspond to Landau parameters with a magnitude of less than $10^{-3}$.}
    \label{tab:LandauParametersDG}
\end{table}
\begin{table}[!h]
    \begin{minipage}{.5\linewidth}
      \centering
    \begin{tabular}{
        S[table-format = 1]
        S[table-format = -1.3]
        S[table-format = -1.3]
        S[table-format = 1.3]
        S[table-format = -1.3]
        }
        \toprule   
        \multicolumn{5}{c}{D1S}\\
        \cmidrule(lr){1-5}      
        {$l$} & {$f_l^{00}$} & {$f_l^{10}$} & {$f_l^{01}$} & {$f_l^{11}$} \\
        \midrule
        0 &  -0.376 & 0.463 & 0.746 & 0.633 \\
        1 &  -0.908 & -0.182 & 0.470 & 0.607  \\
        2 &  -0.555 & 0.244 & 0.341 & -0.038 \\
        3 &  -0.156 & 0.090 & 0.099 & -0.035 \\
        4 &  -0.029 & 0.018 & 0.019 & -0.008  \\        
        5 &  -0.004 & 0.003 & 0.003 & -0.001 \\
        6 &  {} & {} & {} & {} \\
        \bottomrule
    \end{tabular}
    \end{minipage}%
    \caption{Same as in Table \ref{tab:LandauParametersDG}, but for the D1S interaction. \protect\footnotemark}
    \label{tab:LandauParametersD1S}
\end{table}
\begin{table}[!h]
    \begin{minipage}{.5\linewidth}
      \centering
    \begin{tabular}{
        S[table-format = 1]
        S[table-format = -1.3]
        S[table-format = -1.3]
        S[table-format = 1.3]
        S[table-format = -1.3]
        }
        \toprule   
        \multicolumn{5}{c}{D2}\\
        \cmidrule(lr){1-5}      
        {$l$} & {$f_l^{00}$} & {$f_l^{10}$} & {$f_l^{01}$} & {$f_l^{11}$} \\
        \midrule
        0 &  -0.312 & 0.197 & 0.850 & 0.962 \\
        1 &  -0.785 & -0.008 & 0.462 & 0.498 \\
        2 &  -0.828 & 0.277 & 0.532 & -0.047  \\
        3 &  -0.310 & 0.111 & 0.184 & -0.053 \\
        4 &  -0.072 & 0.026 & 0.041 & -0.015 \\        
        5 &  -0.012 & 0.004 & 0.007 & -0.003 \\
        6 &  -0.002 & 0.001 & 0.001 & {} \\
        \bottomrule
    \end{tabular}
    \end{minipage} 
    \caption{Same as in Table \ref{tab:LandauParametersDG}, but for the D2 interactions. \protect\footnotemark}
    \label{tab:LandauParametersD2}
\end{table}

Empirical values for the first Landau parameters can be deduced from experimental data associated with collective states in $\isotope[208]{Pb}$ (excitation energies, transition probabilities, etc.). For a zero-range particle-hole interaction with no tensor forces, Ring \textit{et al.}\ \cite{Ring1973,Ring1974,Speth1974} obtained quantities that can be directly related to Landau parameters using Migdal theory \cite{Migdal1967}, assuming the $l=0$ component in Eq.\eqref{legendreexpansion} to be the main contribution in the expansions to a large extent. 
The empirical values read $f_0^{00} \simeq 0.137, f_0^{10} \simeq 1.150, f_0^{01} \simeq 0.663$ and $f_0^{11} \simeq 1.450$. One sees that $f_0^{00}$ and $f_0^{11}$ are closer to the empirical values with DG than with D1S and D2 interactions. As for $f_0^{01}$, the agreement remains pretty good with all three interactions, but less with $f_0^{10}$. Obviously, a perfect match with these empirical values is
not expected as our interaction is of finite range and the tensor force may renormalize the central Landau parameters. On top of that, one sees explicitly that the main contribution to the Landau parameter $f_l^{00}$ is not contained in the $l=0$ component. 
A comparison with empirical values taking into account one or more of the above hypotheses would certainly be more relevant.

As previously announced, several physical quantities, namely effective mass $\frac{m^{*}}{m}$, incompressibility $K_\infty$ and symmetry energy $a_\tau$, already calculated in SNM in section \ref{physicalquantities}, can be related to some central Landau parameters as \cite{Migdal1967}:

\begin{equation} \label{listquantitiesLandau}
\begin{array}{lcl}
\displaystyle \frac{m^{*}}{m} & =& \displaystyle 1 + \frac{1}{3} f_1^{00} \\
\displaystyle K_\infty & =& \displaystyle \frac{3 \hbar^2 k_{{F}}^2}{m^{*}} \big(1 + f_0^{00} \big) \\
\displaystyle a_\tau & =& \displaystyle \frac{\hbar^2 k_{{F}}^2}{6 m^{*}} \big(1 + f_0^{01} \big)
\end{array}
\end{equation}

\noindent These three quantities are not impacted by the Landau parameters associated with the tensor force, so they are solely judges of the appropriateness of the central Landau parameters. 

\paragraph{Stability criteria}

One recalls that, in the presence of tensor forces, the stability criteria in the $S = 0$ channels simply read:
\begin{equation} \label{f0T}
f^{0T}_l + (2l+1) > 0, \quad \text{for } l \in \mathbb{N}.
\end{equation}
This expression remains the same without tensor forces since they do not contribute in this channel. The above quantities evaluated for the first six values of $l$ in the case of the interaction DG are displayed in Table \ref{tab:StabilityS=0}, and in Tables \ref{tab:StabilityS=0D1S} and \ref{tab:StabilityS=0D2} for the interactions D1S and D2, respectively. One sees that all the 
interaction respect the stability criteria in the S=0 channels. 
Moreover, with increasing $l$, the
quantities related to the stability criteria tend to $(2l+1)$ in bothe T=0 and T=1 channels, which manifests
the fast decreasing of the Landau parameters.

\begin{table}[h]
    \centering
    \begin{tabular}{
        S[table-format = 1]
        S[table-format = 2.3]
        S[table-format = 2.3]
        }
        \toprule 
        \multicolumn{3}{c}{DG}\\
        \cmidrule(lr){1-3}    
        {$l$} & {$f_l^{00} + (2l + 1)$} & {$f_l^{01} + (2l + 1)$}  \\
        \midrule
        0 &  0.700 & 1.869 \\
        1 &  2.225 & 3.528  \\
        2 &  4.155 & 5.503  \\
        3 &  6.701 & 7.159  \\
        4 &  8.936 & 9.032 \\        
        5 &  10.990 & 11.005 \\
        6 & 12.999 & 13.001 \\
        \bottomrule
    \end{tabular}
    \caption{Numerical values of the first quantities related to the stability criteria \eqref{f0T} of the interaction DG in the $S=0$ channel. Columns two and three give the results in the $T=0$ and $T=1$ channels, respectively.}
    \label{tab:StabilityS=0}
\end{table}

\begin{table}[!h]
    \begin{minipage}{.5\linewidth}
      \centering
    \begin{tabular}{
        S[table-format = 1]
        S[table-format = 2.3]
        S[table-format = 2.3]
        }
        \toprule   
        \multicolumn{3}{c}{D1S}\\
        \cmidrule(lr){1-3}  
        {$l$} & {$f_l^{00} + (2l + 1)$} & {$f_l^{01} + (2l + 1)$}  \\
        \midrule
        0 &  0.624 & 1.633 \\
        1 &  2.092 & 3.607  \\
        2 &  4.445 & 4.962  \\
        3 &  6.844 & 6.965  \\
        4 &  8.971 & 8.992 \\        
        5 &  10.996 & 10.999 \\
        6 & 13.000 & 13.000 \\
        \bottomrule
    \end{tabular}
    \end{minipage}%
    \caption{Same as in Table \ref{tab:StabilityS=0}, but for the D1S interaction.}
    \label{tab:StabilityS=0D1S}
\end{table}

\begin{table}[!h]
    \begin{minipage}{.5\linewidth}
      \centering
    \begin{tabular}{
        S[table-format = 1]
        S[table-format = 2.3]
        S[table-format = 2.3]
        }
        \toprule   
        \multicolumn{3}{c}{D2}\\
        \cmidrule(lr){1-3}  
        {$l$} & {$f_l^{00} + (2l + 1)$} & {$f_l^{01} + (2l + 1)$}  \\
        \midrule
        0 &  0.688 & 1.197 \\
        1 &  2.215 & 2.992  \\
        2 &  4.172 & 5.277  \\
        3 &  6.690 & 7.111  \\
        4 &  8.928 & 9.026 \\        
        5 &  10.988 & 11.004 \\
        6 & 12.998 & 13.001 \\
        \bottomrule
    \end{tabular}
    \end{minipage} 
    \caption{Same as in Table \ref{tab:StabilityS=0}, but for the D2 interaction.}
    \label{tab:StabilityS=0D2}
\end{table}

The stability criteria in the $S = 1$ channel are effectively complicated when the tensor force is taken into account. The condition becomes that the eigenvalues of the matrix $F$ defined by Eq.\eqref{submatrix} must all be positive as discussed in the previous section, Eq.\eqref{submatrixstab} \cite{Backman1979}. One also has shown that,
for some value of the total two-nucleon angular momentum $J$, the associated $3 \times 3$ matrix is divided into a diagonal $1 \times 1$ sub-block corresponding to $J = l = l'$ (the sub-block $J=0$ being diagonal as well), and a non-diagonal $2 \times 2$ sub-matrix. One then distinguishes between two types of eigenvalues. On the one side, the eigenvalues denoted $\lambda_0^{(J, T)}$ are directly given by the TBMEs belonging to the diagonal sub-blocks.
On the other side, one has 
two eigenvalues denoted $\lambda_{\pm}^{(J, T)}$
which are extracted from
the non-diagonal sub-matrices by means of the procedure described in Eq.\eqref{submatrix}. 
Their values are provided in Table \ref{tab:StabilityS=1}. One observes that all
the eigenvalues are found positive.
One sees that, in the S=1 channel, 
the eigenvalues $\lambda_0^{(J, 0)}$ tend
towards 1 and $\lambda_{\pm}^{(J, T)}$
towrds 2 (up to J=4), as expected.
One can not compare those values to tensor-free interactions like D1S and D2 since the related stability criteria obey different relations.
In conclusion, one deduces that the stability criteria are fulfilled in both $S=0$ and $S=1$ channels, explicitly up to $l_{{max}}$ and $J_{{max}}$ - as all quantities in Tables \ref{tab:StabilityS=0} and \ref{tab:StabilityS=1} are positive -, and implicitly for all values $l \in \mathbb{N}$ and $\abs{l-1} \le J \le l + 1$. It should be noted that these criteria are fairly restrictive and not easy to fulfill.

\begin{table}[h]
    \centering
    \begin{tabular}{
        S[table-format = 1]
        S[table-format = 1.3]
        S[table-format = 1.3]
        S[table-format = 1.3]
        S[table-format = 1.3]
        S[table-format = 1.3]
        S[table-format = 1.3]
        }
        \toprule   
        {$J$} & {$\lambda_0^{(J, 0)}$} & {$\lambda_+^{(J, 0)}$} & {$\lambda_-^{(J, 0)}$} & {$\lambda_0^{(J, 1)}$} & {$\lambda_+^{(J, 1)}$} & {$\lambda_-^{(J, 1)}$} \\
        \midrule
        0 &  1.007 & {} & {} & 3.106 & {} & {} \\
        1 &  1.010 & 2.452 & 2.091 & 0.197 & 4.600 & 1.641 \\
        2 &  1.046 & 2.024 & 2.017 & 0.791 & 2.895 & 1.865  \\
        3 &  1.012 & 2.092 & 2.004 & 0.963 & 2.139 & 1.969 \\
        4 &  1.002 & 2.024 & 2.001 & 0.995 & 2.011 & 1.993 \\        
        5 &  1.000 & {} & 1.002 & 0.999 & {} & 1.000 \\
        6 & {} & {} & 1.000 & {} & {} & 1.000 \\
        \bottomrule
    \end{tabular}
    \caption{Numerical values of the first eigenvalues of the free energy \eqref{freeenergy} for the interaction DG. The stability criteria in the $S=1$ channel require these values to be positive. Columns two to four and five to seven give the results in the $T=0$ and $T=1$ channels, respectively.}
    \label{tab:StabilityS=1}
\end{table}

Furthermore, finite-size instabilities require the 
calculation of the full RPA response which is a step 
beyond the Landau parameter, 
stability criteria  
and sum rules calculation but numerically very 
time-consuming. This factual situation prevents to include it 
in a fitting process which is designed to be fast. 
A post-determination of the full RPA response is,
however, desired to exclude parameterizations presenting 
finite-size 
instabilities. Already several studies have been done
including various type of effective interactions including
the Gogny ones \cite{DePace2016,Pastore2012a,Pastore2012b,Martini2019,Davesne2021,Hellemans2013}.

\begin{figure}
    \centering
    \includegraphics[width=0.95\linewidth]{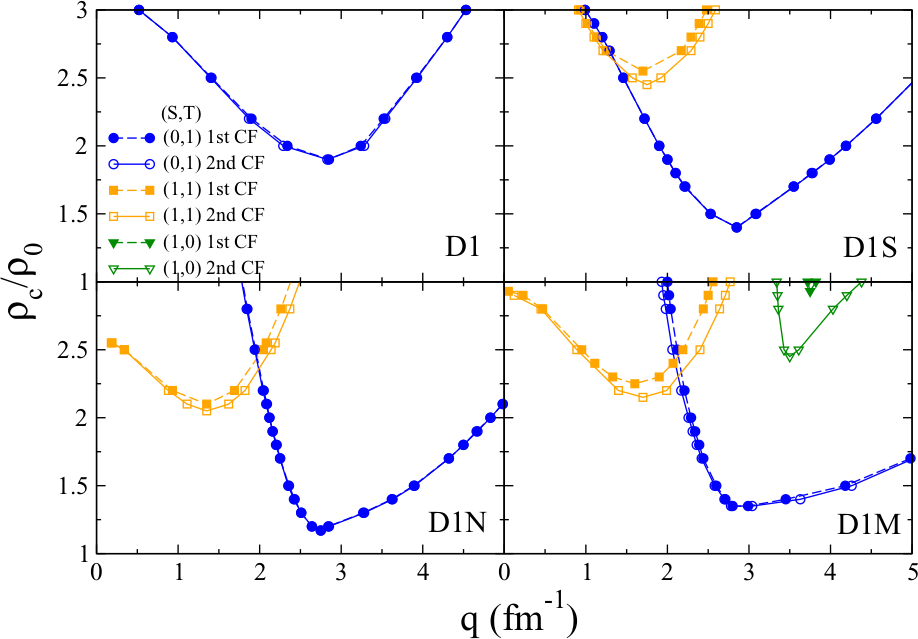}
    \caption{Critical densities $\rho_c$ divided by a constant value of the
saturation density ρ 0 = 0.16 fm$^{-3}$ as a function of the transferred
momentum q (in fm$^{-1}$ ) for the most commonly used parameterizations
of the Gogny force. The calculations of RPA response R(S,T) (q,$\omega=0$), through
which the critical densities are deduced, are performed at first and second order in the continued fraction expansion \cite{DePace2016}.}
\label{fig:martini1}
\end{figure}

\begin{figure*}
    \centering
    \includegraphics[width=0.95\linewidth]{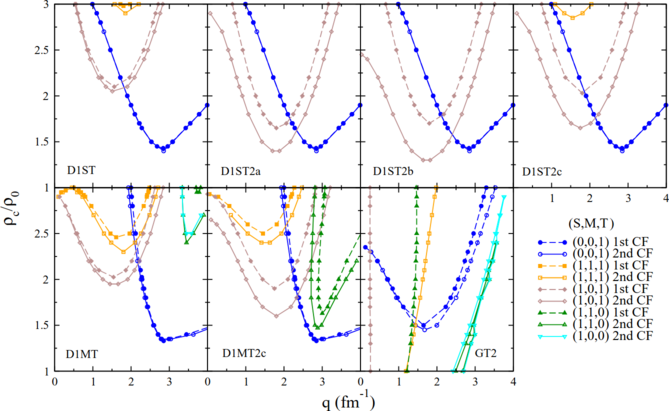}
    \caption{Same as Fig.\ref{fig:martini1} but for Gogny forces
    including a tensor term \cite{DePace2016}.}
    \label{fig:martini2}
\end{figure*}
As an example, one shows in Fig.\ref{fig:martini1} the
evolution of the critical density $\rho_c$ according to
the transfer moment $q$ for standard D1-type interactions without
tensor component. One sees that the D1 interaction presents the best results \cite{DePace2016}.
The same analysis was done for the D1ST-type and D1MT-type as
well as GT2 interaction which presents a tensor component.
The results are shown in Fig.\ref{fig:martini2}.
The GT2 interaction displays huge instabilities unlike
the D1ST-type and D1MT-type interactions which behaves in a similar way as D1S and D1M, respectively.
In the future, it may be interesting to enlarge those studies 
to more general expression of the Gogny interaction as D2 and DG.

\paragraph{Sum rules}\label{SRcorps}

One has demonstrated in the sub-section \ref{lpder} that the Pauli principle holds if and only if the so-called sum rules expressed in Eqs. \eqref{tensorSR1} and \eqref{tensorSR2} are verified, here in the presence of tensor forces. In the absence of a tensor term, these expressions of sum rules simplify. 
In practice, one truncates these two summations at the finite order $l_{{max}} = 5$ as one has done for the stability criteria, so that they here become \cite{Friman1979}:

\begin{multline}
S_1  \equiv \sum_{l=0}^{l_{{max}}} \bigg[ \frac{f_l^{00}}{1+f_l^{00}/(2l+1)} + \frac{f_l^{01}}{1+f_l^{01}/(2l+1)} \\ + \frac{1}{3} \sum_J \frac{2J+1}{2l+1} \big( \mathcal{S}^{J0}_{ll} +\mathcal{S}^{J1}_{ll} \big) \bigg] = \sum_i T_1^i = 0  \label{SR1truncated}
\end{multline}
\begin{multline}
S_2 \equiv \sum_{l=0}^{l_{{max}}} \bigg[ \frac{f_l^{00}}{1+f_l^{00}/(2l+1)} - \frac{3 f_l^{01}}{1+f_l^{01}/(2l+1)} \\ + \sum_J \frac{2J+1}{2l+1} \big( 3 \mathcal{S}^{J1}_{ll} - \mathcal{S}^{J0}_{ll} \big) \bigg] = \sum_i T_2^i = 0  \label{SR2truncated}
\end{multline}

The sum rules are almost always violated by effective interactions \cite{Gogny1977} and DG is not an exception. As sum rules are not observables, it is difficult to say whether a definite non-zero value accounts for a significant violation, or not. Then, one usually compares the predictions of the interactions with each other, as it is exemplified in Table \ref{tab:SumRulesComparison} for the interactions containing tensor components, namely DG, D1ST2a and D1ST2c, and tensor-free interactions as D1S and D2. 
One clearly sees that the interaction including a tensor term
have $S_1$ and $S_2$ values which are strongly different from zero. As an example, DG and D2 which have very similar central
and density-dependent terms differs strongly in the values of
$S_1$ and $S_2$. If one remove the tensor term to DG, one recovers
$S_1$ and $S_2$ values close to the ones of D2 and even more
importantly close to zero. One observes that the situation is even worse in the case of the D1ST2a and D1ST2c interactions.
\begin{table}[h]
    \centering
    \begin{tabular}{
        l
        S[table-format = -2.3]
        S[table-format = -3.3]
        }
        \toprule
        \multicolumn{1}{c}{} & 
        \multicolumn{2}{c}{Sum rules}\\
        \cmidrule(lr){2-3}        
        & {$S_1$} & {$S_2$} \\
        Interaction & {} & {} \\
        \midrule
        DG &  -4.297 & -40.826  \\
        DG (no tensor) &  0.048 & -1.718  \\    
        D1S &  -0.175 & -0.650 \\
        D1ST2a & 28.113 & 254.948 \\
        D1ST2c & 28.131 & 254.897 \\
        D2 &  0.007 & -2.104 \\  
        \bottomrule
    \end{tabular}
    \caption{Numerical values of the sum rules for the D1S, D2, D1ST2a, D1ST2c and DG Gogny interactions. The values obtained by switching off the tensor interaction in DG are also tabulated as “DG (no tensor)”.}
    \label{tab:SumRulesComparison}
\end{table}

In order to evaluate their relative contributions, the two sum 
rules have been decomposed into nine components $T_i$, as 
indicated in Eq.\eqref{SR2truncated}. 
The two first contributions correspond to the sums over $l$ of the terms involving the Landau parameters $f_l^{00}$ and $f_l^{01}$. 
The remaining ones correspond to the sums over $l$ of the last term for some given value of $J$, implying the forward scattering amplitude $\mathcal{S}^{JT}_{ll}$ coupled to (J,T). There are $J_{{max}} + 1 = 7$ such quantities. These contributions together with the full sum rules $S_1$ and $S_2$ for the interaction DG are listed in Table \ref{tab:SumRules}.
\begin{table}[h]
    \centering
    \begin{tabular}{
        S[table-format = 1]
        S[table-format = 1]
        S[table-format = -1.3]
        S[table-format = -2.3]
        }
        \toprule   
        {$i$} & {$J$} & {$T_1^i$} & {$T_2^i$}  \\
        \midrule
        1 & {} &  -2.877 & -2.877 \\
        2 & {} &  1.564 & -4.691  \\
        3 & 0 &  0.228 & 2.027  \\
        4 & 1 & -3.482 & -34.195  \\
        5 & 2 & 0.061 & -0.739 \\        
        6 & 3 & 0.161 & -0.168 \\
        7 & 4 &0.041 & -0.139 \\
        8 & 5 &0.007 & -0.038 \\        
        9 & 6 & 0.001 & -0.005 \\  
        \cdashline{3-4}
        {} & {} & -4.297 & -40.826 \\    
        \bottomrule
    \end{tabular}
    \caption{Relative contributions of the various terms $T_1^i$ and $T_2^i$ appearing in the sum rules \eqref{SR1truncated} and \eqref{SR2truncated}. The numerical values of the sum rules $S_1$ and $S_2$ are displayed in the last row, under the dashed line.}
    \label{tab:SumRules}
\end{table}
This decomposition unveils that the dominant contributions to these sums correspond to $i=4$, that is the fourth terms of $S_1$ and $S_2$ obtained for $J=1$. In the following, one focuses 
on the contribution $J=1$ of $S_2$ as it is the largest. The fitting code reveals that 
the component $l=1$ is mainly responsible for it, which corresponds to the combination $3 \mathcal{S}_{11}^{J1} - \mathcal{S}_{11}^{J0}$. According to the expression obtained in Eq.\eqref{defmathcalA} of the coupled forward scattering amplitude, one has, in the channel $T$,
\begin{equation}
\mathcal{S}_{11}^{1T} = 3 \frac{\mel{11}{F}{11}^T - 1}{\mel{11}{F}{11}^T},
\end{equation}
where the associated TBMEs are given by Eq.\eqref{stabdiag}, i.e.\
\begin{equation}
\mel{11}{F}{11}^T = 1 + \frac{1}{3}f_1^{1T} + \frac{5}{3}h_0^{1T} - \frac{2}{3} h_1^{1T} + \frac{1}{15} h_2^{1T}.
\end{equation}
Plugging the values of the Landau parameters indicated in Table \ref{tab:LandauParametersDG}, one obtains $\! \mel{11}{F}{11}^0 \simeq \SI{1.010}{}$, hence $\mathcal{S}_{11}^{10} \simeq \SI{0.030}{}$, and $\! \mel{11}{F}{11}^1 \simeq \SI{0.197}{}$, hence $\mathcal{S}_{11}^{11} \simeq \SI{-12.228}{}$. It is therefore the high value of $\mathcal{S}_{11}^{11}$, resulting from the low value of $\! \mel{11}{F}{11}^1 \!$, that causes the explosion of the magnitude of $S_2$. Actually, the low value of $\! \mel{11}{F}{11}^1 \!$ comes from the prevailing contribution of the tensor Landau parameter $h_0^{11}$, since $5h_0^{11}/3 \simeq \SI{-1.333}{}$. A similar study can be carried out to explain the high value of the component $i=4$ of $S_1$. As a consequence, an important part of the violation of the DG sum rules can be attributed to the tensor interaction. 
To get to the bottom of it, one looks at the sum rules for the DG interaction when the tensor interaction is set to zero (see Table \ref{tab:SumRulesComparison}). Accordingly, one finds out that the Pauli principle is less violated, the values being of the order of magnitude of those of D2, as expected; $S_2$ is even closer to zero. Although the value of $S_2$ is still a little bigger than that of D1S, $S_1$ is better.

As already noted, one has compared the previous 
values with the sum rules of the D1ST2a and D1St2c effective interactions. Alas, no such study exists in the literature. Only an article in which sum rules are calculated for a zero-range tensor-dependent Skyrme interaction has been found \cite{Pastore2012a}. However, these are of a different kind.
The perturbative nature of the D1ST2a and D1ST2c interactions allows to isolate the tensor effects on the sum rules, while their D1S part is little violated. Table \ref{tab:SumRulesComparison} indicates that the D1ST2a and D1ST2c sum rules have similar values and are by far the most broken of all tested interactions. They are the only ones to be overestimated to that extent. Yet, the D1ST2-type interactions manifest some valuable characteristics, from the point of view of their nuclear structure \cite{Anguiano2012,Anguiano2016a}, deformation \cite{Bernard2016a,Co2021} and fission \cite{Bernard2020} properties. Comparatively, the DG tensor-dependent interaction produces reasonable values. In order to settle whether the sum rules are problematic or not, one would definitely need to test it in finite-nuclei RPA calculations, which are beyond the scope of this review article. 
It would also be enlightening to know if there exists other tensor-dependent effective interactions capable of reducing the sum rules to values comparable to those of DG, or even better. On the other hand, it was shown that the rearrangement terms associated with the density-dependent interaction reduced strongly the violation of the sum rules \cite{Chappert2015,Gogny1977}. It is then legitimate to think that adding a density dependence to the tensor term might induce new rearrangement terms that may produce the same effect.

\section{Some highlights on applications}\label{sec4}

\subsection{Nuclear Structure}\label{subsec2}

Numerous nuclear structure applications have used the Gogny interaction, investigating various phenomena. This section briefly discusses only recent ones, mainly using the D1-type Gogny interaction, published from 2019 on. For earlier studies, the reader is referred to the reviews 
in Refs. \cite{Peru2014a,Egido2016a,Robledo2019}.

\subsubsection{Shape evolution, shape coexistence and/or shape mixing}\label{app:gcm}

One of the most widely used applications of the Gogny interaction is in the study of nuclear shape evolution and/or the emergence of shape coexistence across various regions of the nuclear chart. This is because the concept of intrinsic nuclear shape is intimately connected to self-consistent mean-field approaches, in which Gogny interactions are deeply rooted. In fact, the first step in the vast majority of shape studies consists of computing the potential energy surfaces (PES) at the HFB level and/or with particle-number projection \footnote{These surfaces are often referred to as potential energy surfaces (PES), though this designation may be misleading, as they include kinetic as well as potential contributions.}. These are obtained by solving the HFB or PN-VAP equations under constraints on the expectation values of multipole deformation operators. Depending on the topography of the PES, nuclei can be identified as spherical, deformed, soft, and/or exhibiting shape co-existence if the surfaces display a minimum at spherical or deformed configurations, a soft landscape with quasi-degenerate configurations, or two or more distinct minima, respectively. The next step in obtaining excitation energies and electromagnetic properties depends on the specific beyond-mean-field (BMF) method employed. 

\begin{figure}
    \centering
    \includegraphics[width=1.0\linewidth]{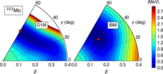}
    \caption{ Contour plot of the PES surface in the
($\beta$, $\gamma$) plane for $^{102}Mo$ computed with the constrained HFB method by using the Gogny functional D1M (left) and with the mapped IBM
(right). The red dot indicates the minimum of the energy surface plots
and the difference between two neighboring contours is 100 keV. Taken from \cite{Esmaylzadeh2021}.}
    \label{fig:IBM}
\end{figure}

In recent years, the interacting boson model-mapping method, originally designed to study spectra involving triaxial quadrupole shapes, has been extensively applied to investigate axial quadrupole-octupole couplings \cite{Nomura2020a,Nomura2021a,Nomura2021b,Nomura2021c,RodriguezGuzman2021,RodriguezGuzman2022,RodriguezGuzman2023} and axial quadrupole-hexadecapole couplings \cite{Lotina2025,RodriguezGuzman2025} in Xe and Ba isotopes, as well as in the lanthanide and actinide regions. 
In this model, the derivation of the Hamiltonians is obtained through the mapping of PES deduced from constraint HFB calculations, using in particular the Gogny interaction \cite{Nomura2011a}.
The IBM parameters for each nucleus are determined by mapping the fermionic PES onto the expectation value of the bosonic Hamiltonian evaluated in the boson-coherent state.
Once the parameters are fixed, the IBM Hamiltonian is diagonalized to compute the excitation spectra of the corresponding nuclei.
With respect to triaxial quadrupole degrees of freedom, the IBM-mapping approach has recently been extended to odd-mass nuclei through the inclusion of the Interacting Boson-Fermion Model (IBFM), allowing the study of shape co-existence in neutron-rich nuclei around $N\approx60$ \cite{Esmaylzadeh2019,Esmaylzadeh2021,Gerst2022}, and shape evolution in Rh and Pd nuclei \cite{Nomura2022}. An example of PES is given in Fig.\ref{fig:IBM}.

\begin{figure*}
    \centering
    \includegraphics[width=1.0\linewidth]{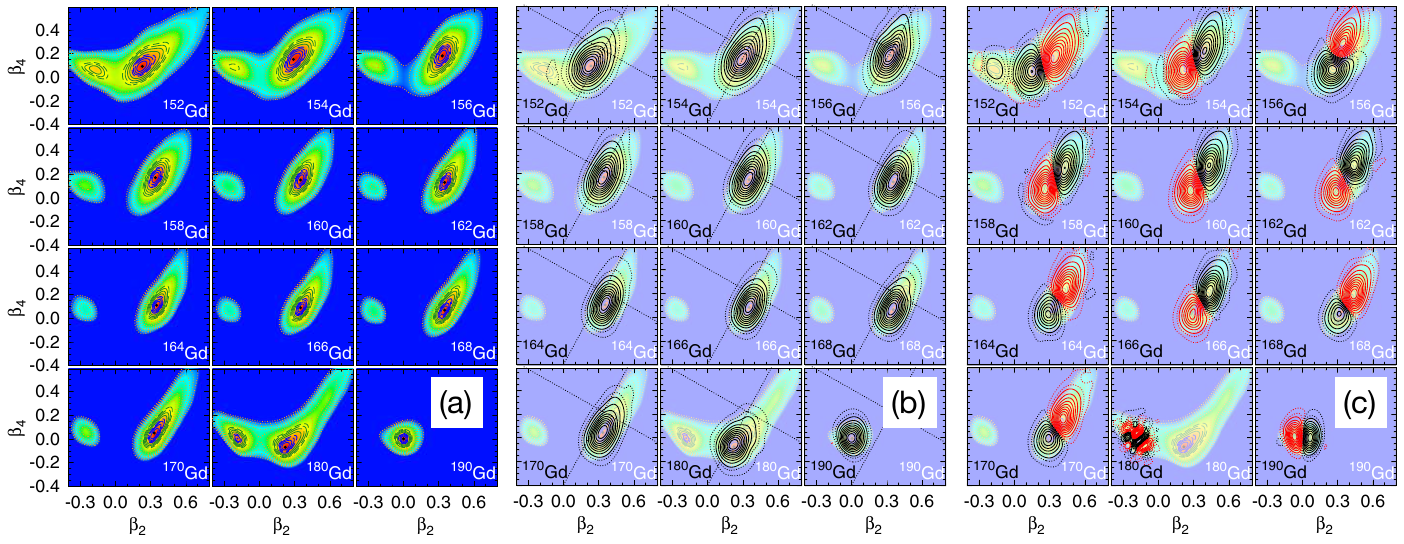}
    \caption{(a) HFB energy surfaces and (b)-(c) collective wave functions for the two lowest GCM energies as a function of the quadrupole and hexadecapole deformation for Gd isotopes. Gogny D1S is used (adapted from Ref. \cite{Kumar2023}).}
    \label{Gd_q2_q4}
\end{figure*}

The symmetry conserving configuration mixing approach (SCCM), which displays a strong computational complexity, has been applied to selected nuclei and regions of the nuclear chart to explore shape evolution, as well as potential configuration mixing and shape co-existence. 
This approach is based on the Generator Coordinate Method (GCM) which is a general approach to solving the nuclear many-body problem. In that 
context, nuclear wave functions, $|\Psi^{\sigma}\rangle$, are defined as linear combinations
of simpler many-body wave functions, $|\Phi(\vec{q})\rangle$, built along a set of collective generator coordinates, $\vec{q}$:
\begin{equation}
    |\Psi^{\sigma}\rangle = \int f^{\sigma}(\vec{q}) |\Phi(\vec{q})\rangle \, d\vec{q}.
\end{equation}
This general form of the wave function underlies many of the methods developed using the Gogny interaction. First, if the generator coordinates correspond to symmetry group parameters and the coefficients are fixed by symmetry considerations, the resulting GCM wave function becomes a wave function projected onto the quantum numbers associated with those symmetries. 
In such cases, the multidimensional integral corresponds to a representation of projection operators onto good quantum numbers. 
Projections onto good parity, proton and neutron number, angular momentum, and linear momentum have been implemented with the Gogny interaction (specifically the D1 family).
However, the GCM coefficients are typically treated as variational parameters, determined by solving the Hill-Wheeler-Griffin (HWG) equation:
\begin{equation}
    \int \left[ \mathcal{H}(\vec{q}, \vec{q}') - E^{\sigma} \mathcal{N}(\vec{q}, \vec{q}') \right] f^{\sigma}(\vec{q}') \, d\vec{q}' = 0,
\end{equation}
where the norm and Hamiltonian overlap kernels are defined as:
\begin{align}
    \mathcal{N}(\vec{q}, \vec{q}') &= \langle \Phi(\vec{q}) | \Phi(\vec{q}') \rangle, \\
    \mathcal{H}(\vec{q}, \vec{q}') &= \langle \Phi(\vec{q}) | \hat{H} | \Phi(\vec{q}') \rangle.
\end{align}
Depending on the type of wave functions used to evaluate these overlaps, different GCM implementations have been developed. 
Normally, the states $|\Phi(\vec{q})\rangle$ are projected onto good quantum numbers $\{ \xi_i \}$ using the corresponding projection operators $\{ \hat{P}^{\xi_i} \}$, such that:
\[
|\Phi(\vec{q})\rangle = \prod_i \hat{P}^{\xi_i} |\phi(\vec{q})\rangle.
\]
Two main types of projected GCM implementations can be distinguished depending on the intrinsic wave functions $|\phi(\vec{q})\rangle$. 
In the Antisymmetrized Molecular Dynamics with GCM (AMD+GCM), the wave functions are Slater determinants built from Gaussian single-particle wave packets, 
where the centroids and spin orientations of the nucleons are variational parameters \cite{Kimura2004}. The collective coordinates in this case are typically the distances between the AMD wave functions representing nuclear clusters.

On the other hand, the class of methods known as Symmetry Conserving Configuration Mixing (SCCM) is based on projection and mixing of Hartree-Fock-Bogoliubov (HFB)-type wave functions \cite{Egido2016a,Robledo2019}. 
The generator coordinates are usually related to multipole deformations, primarily quadrupole but also octupole and, more recently, hexadecapole deformations (or a combination of them). 
Additionally, particle-number fluctuations and intrinsic rotations (cranking) have been considered.

The SCCM method typically involves three main steps: (i) generating a set of HFB-like states by solving HFB or variation-after-projection (VAP) equations with constraints on the collective variables that define $\vec{q}$; (ii) projecting these HFB states onto good quantum numbers (parity, particle number, angular momentum); and (iii) mixing the projected configurations by solving a discretized version of the HWG equations for each set of angular momentum and parity quantum numbers (if those symmetry restorations are performed).

Multiple quadrupole shape coexistence has been studied in light Pb and Hg nuclei using axial wave functions \cite{Siciliano2020,Stryjczyk2023,MontesPlaza2025}. Axially symmetric wave functions have also been employed to explore axial quadrupole-octupole coupling using SCCM methods including angular momentum, particle number, and parity projection in light Xe and Te nuclei \cite{Illana2024,Testov2021}, and in the doubly magic nucleus $^{208}$Pb \cite{Henderson2025}, as well as axial quadrupole-hexadecapole coupling using HFB wave-function mixing only in the rare-earth region \cite{Kumar2023}. Taking the latter reference as an example of a recent study on the evolution of nuclear shapes using the Gogny interaction, Fig. \ref{Gd_q2_q4}(a) shows the HFB energy surfaces for the isotopes $^{152-170,180,190}$Gd as a function of the axial quadrupole and hexadecapole deformations, $(\beta_{2},\beta_{4})$. In all cases, prolate-deformed minima are observed (except for the spherical semi-magic nucleus $^{190}$Gd), and the deformation involves not only a quadrupole component but also a significant hexadecapole contribution. Furthermore, Figs. \ref{Gd_q2_q4}(b)–(c) display the collective wave functions for the two lowest-energy states obtained with the generator coordinate method (without projection). The ground-state wave function corresponds to the prolate energy minimum, while the first excited state, within the axial approximation, may be interpreted as a vibration exhibiting strong $\beta_{2}-\beta_{4}$ coupling.

Recent applications have incorporated triaxial quadrupole deformations in various regions of the nuclear chart, providing theoretical interpretations of experimental data. In particular, the emergence of triaxiality and/or shape coexistence has been investigated in medium-mass nuclei such as Zn \cite{Rocchini2021,Hellgartner2023}, Ge \cite{Elekes2024}, Se \cite{Lizarazo2020,Marchini2023,Elekes2023}, Kr \cite{Wimmer2020,Algora2025}, Zr \cite{Pasqualato2023a,Moon2024}, Mo \cite{Rodriguez2025}, Ru \cite{Garrett2020a}, Cd \cite{Garrett2019,Garrett2020b,Siciliano2021}, and Ce \cite{Alexa2022}, as well as in superheavy nuclei (Fl, Lv, Og, Cn) \cite{Egido2020,Egido2021,SamarkRoth2021,SamarkRoth2023,Cox2023}.
\begin{figure*}
    \centering
    \includegraphics[width=0.85\linewidth]{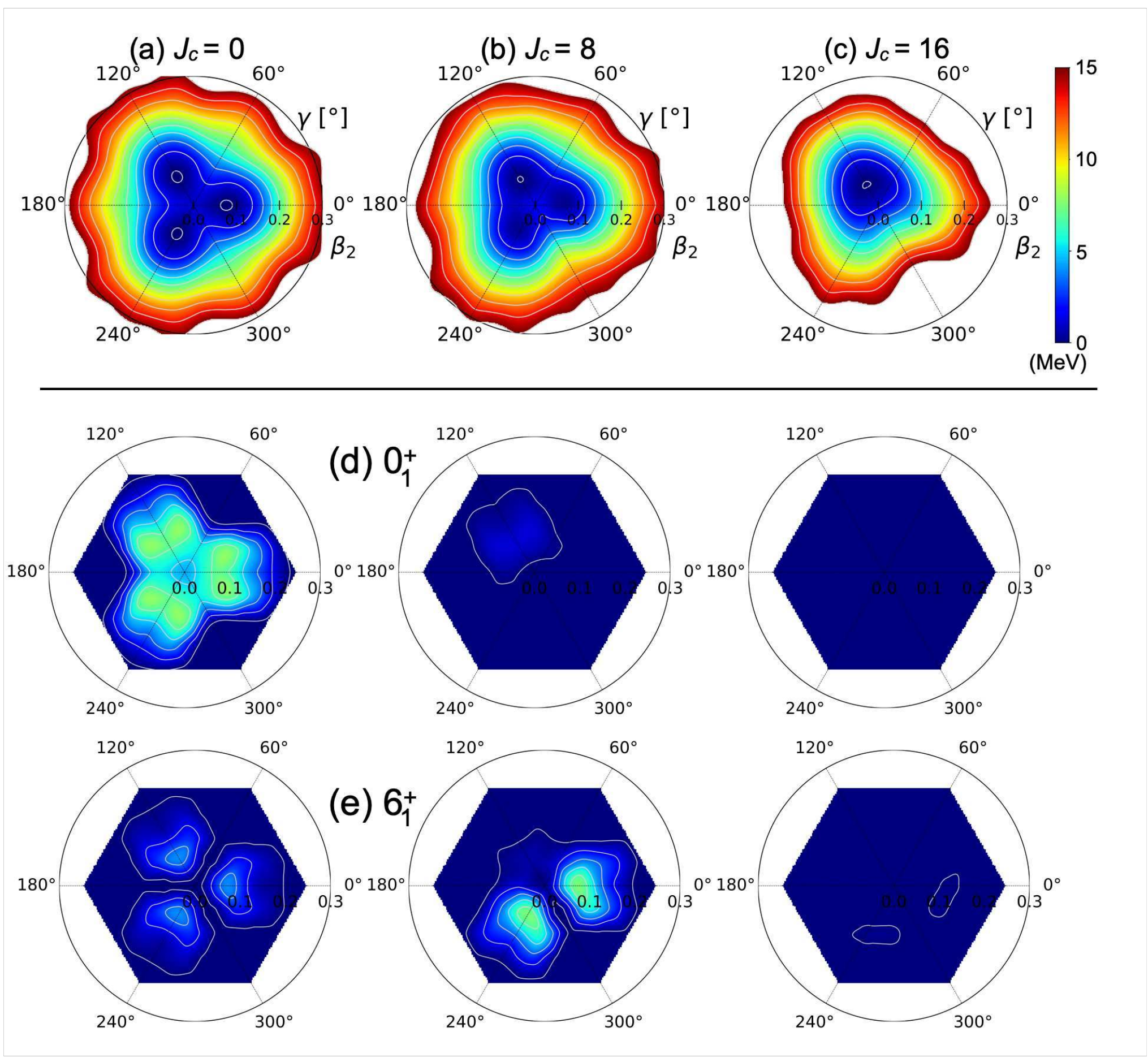}
    \caption{Top panel: Particle-number energy surfaces for different values of the cranking angular momentum, $J_c$, in the $(\beta_{2},\gamma)$-plane calculated with Gogny D1S energy density functional for $^{128}$Cd (energies are normalized to the minimum of each surface). Bottom panel: Components of the (d) $0^{+}_{1}$ and (e) $6^{+}_{1}$ collective wave functions in the $(\beta_{2},\gamma)$-plane and $J_{c}$.}
    \label{TES_128Cd}
\end{figure*}
\begin{figure}
    \centering
    \includegraphics[width=0.95\linewidth]{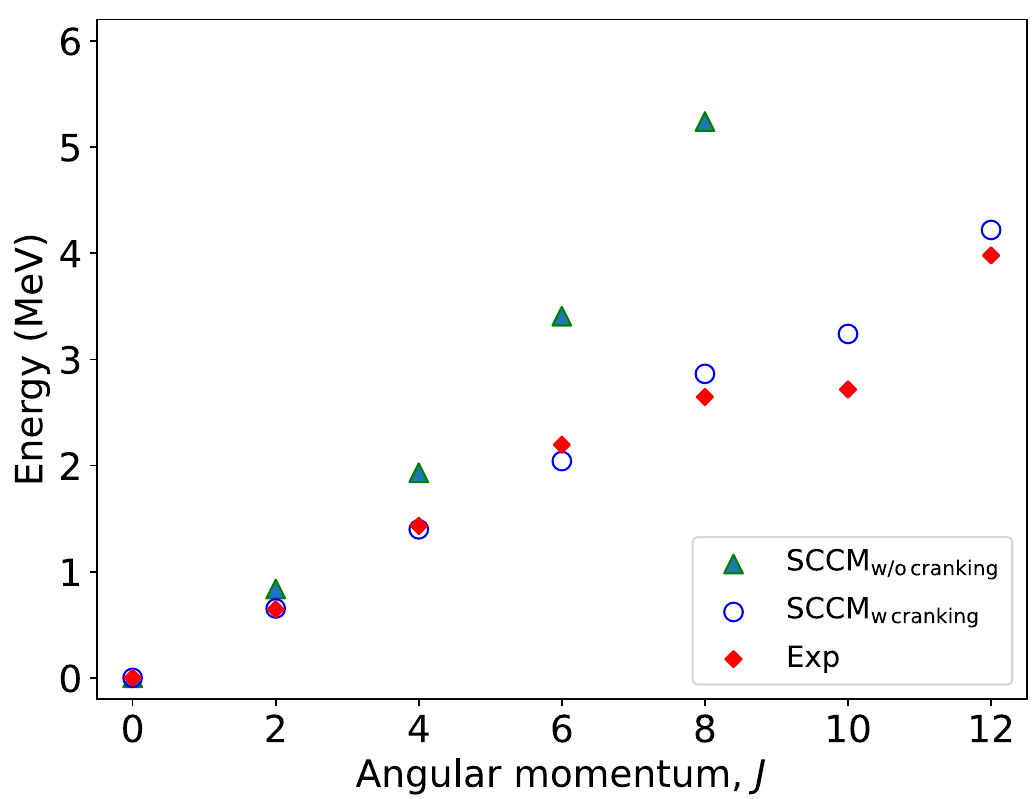}
    \caption{Excitation energies of the yrast positive parity states calculated with the SCCM method including: static shapes ($J_{c} = 0$, without cranking, green triangles); static and cranking states ($J_{c}$ = 0, 8, and 16, blue circles). Red diamonds represent the experimental data (taken from Ref. \cite{Jungclaus2017}).}
    \label{SPECT_128Cd}
\end{figure}
One of the aspects that has been improved within the SCCM framework in recent years is the ability to reproduce not only qualitatively but also quantitatively the experimental excitation energies. This is achieved by enlarging the variational space for the excited states, allowing for the mixing of cranking wave functions. As an example, SCCM calculations with the Gogny D1S interaction for the nucleus $^{128}$Cd are shown.
Fig. \ref{TES_128Cd} displays the particle-number–projected (VAP-PN) energy surfaces in the $(\beta_{2},\gamma)$ plane for three values of the cranking angular momentum, $J_{c}=0$ (static shapes), 8, and 16 (intrinsic rotations around the $x$ axis). The standard triaxial SCCM calculation includes only states with $J_{c}=0$ in one sextant, since the $D_{2h}$ symmetry is preserved and all six sextants in the plane are equivalent \cite{Borrajo2015a}. In this case, the collective wave functions for the yrast band are built on a slightly prolate/triaxial deformation close to the minimum found in the total energy surface. The energies of this band exhibit a rather rotational character, and the stretching of the spectrum with respect to the experimental result is quite evident (see Fig. \ref{SPECT_128Cd}).
Once the cranking degree of freedom is introduced, the $D_{2h}$ symmetry is broken \cite{Borrajo2015a}, and two equivalent half-planes are obtained (Figs. \ref{TES_128Cd}(b)–(c)). In addition, the geometry of the surface evolves to display an absolute minimum at small deformations and $\gamma=120^{\circ}$. When all wave functions are mixed (after projection onto good particle number and angular momentum), the effect on the excitation energies of the ground-state band becomes very clear, resulting in a compression that brings the spectrum closer to the experimental data (see Fig. \ref{SPECT_128Cd}).
The collective wave function of the ground state contains only negligible components from configurations with $J_{c}\neq0$, but as the angular momentum increases, the contributions from the intrinsic cranking wave functions become increasingly relevant, becoming dominant for $J\geq6$ (see bottom panel of Fig. \ref{TES_128Cd}). Although such calculations are very demanding from a computational point of view, they open the way to a quantitative study of excitation energies and to the description of isomeric states that arise in cranked configurations and are absent in more standard calculations.

\begin{figure*}
    \centering
    \includegraphics[width=0.37\linewidth]{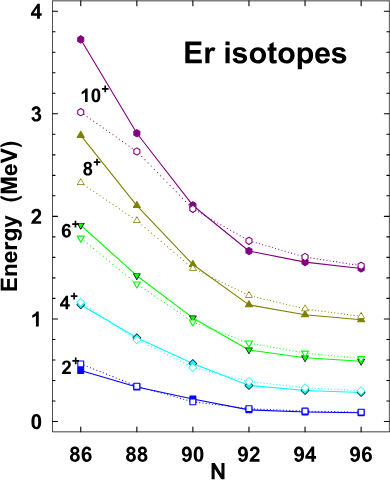}
    \includegraphics[width=0.34\linewidth]{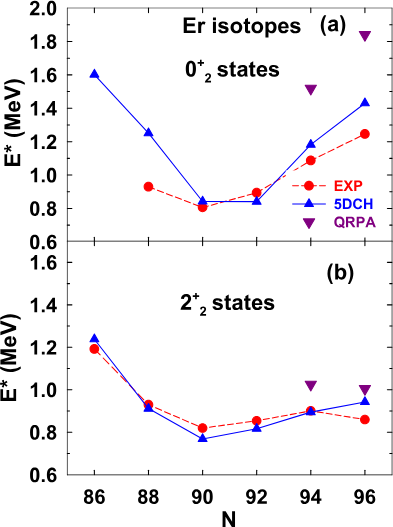}
    \caption{Left: Ground-state band levels of Erbium isotopes with neu-
tron numbers N = 86, 88, 90, 92, 94, and 96. Energies are in MeV.
Comparisons are between experimental values (open symbols) and
5DCH calculations (solid dots). Right: Panels (a) and (b) show excitation energies (MeV) of the $0^+_2$ and $2^+_2$ bandhead levels, respectively. Red dots are for experimental data. Purple and blue symbols are 
for QRPA and 5DCH values, respectively. Taken from \cite{Turturica2025}.}
    \label{erbium}
\end{figure*}

Historically, the GCM framework has served as the starting point for the derivation of Collective Hamiltonians based on microscopic input, under the assumption of Gaussian overlap approximations to the kernels in the HWG equation. In the case of the D1-type Gogny interaction, the collective quadrupole variables $(\beta_2, \gamma)$ have been used as generator coordinates to extract a collective Hamiltonian known as the five-dimensional collective Hamiltonian (5DCH) \cite{Delaroche2010}.The five dynamical variables of this Hamiltonian are the three Euler angles and the quadrupole deformations $(\beta_2, \gamma)$.  Due to the computational efficiency, systematics studies on the whole nuclear chart have been possible with such approach, for even-even nuclei. 
The 5DCH approach has recently been used to provide theoretical interpretations of experimental data in neutron-deficient Nd isotopes \cite{Turturica2021}, neutron-rich nuclei near $N\approx60$ \cite{Gerst2022,Tocabens2025}, and in the neutron-deficient region of Er isotopes \cite{Turturica2025}, as can be seen from Fig.\ref{erbium} with an additional comparison with QRPA.
Moreover, the 5DCH framework has been used to conduct a systematic study of $E0$ transitions over a wide region of the nuclear chart, showing reasonable agreement with experiment in cases where the transitions occur between collective states \cite{Delaroche2024}.

\subsubsection{Shell evolution}\label{app:mpmh}

In addition to the implications of deformation on the shell structure and vice versa, Gogny functionals have been used to study shell closure.
Early studies can be found for example in Refs. \cite{Peru2000,Gaudefroy2009} with the D1S interaction.
More recently, it has been employed to study other shell closures. For example, the evolution of single-particle energies in Ni isotopes around $N=50$ has been analyzed within the mean-field framework by adding a tensor term to the functional \cite{Routray2021}; $B(E2)$ transition probabilities near the doubly magic nucleus $^{100}$Sn have been investigated using QRPA and MPMH configuration mixing \cite{Pillet2008,Robin2014a,Robin2016,Robin2017,LeBloas2014}methods \cite{Pasqualato2023b}; the onset of collectivity for argon isotopes close to $N=32$ with HFB+QRPA approaches \cite{Linh2024}; the spectrum of $^{62}$Ti (neighbor of the magic nucleus $^{60}$Ca) has been studied using SCCM methods with triaxial projection \cite{Cortes2020}; and the possible emergence of shell closures at $N=32$ and $N=34$ in Ti isotopes has been explored with SCCM methods including triaxial angular momentum projection and cranked wave functions \cite{Rodriguez2020}. Furthermore, the experimental spectrum of the doubly magic nucleus $^{78}$Ni has been compared with the results obtained using the Gogny D1M functional within the QRPA framework \cite{Taniuchi2019}. Other effects related to quasiparticle excitations, such as the occurrence of $K$-isomers, have been addressed using HFB calculations with axial blocked wave functions in various nuclei ($^{254}$No, $^{178}$Hf, $^{270}$Ds, W isotopes) \cite{Robledo2024}.



\subsubsection{Other applications}\label{app:qrpa}
One of the most prolific applications of the Gogny interaction is in the study of nuclear response to electromagnetic fields. 
The HFB+QRPA approach is particularly effective for calculating $\gamma$-strength functions \cite{Goriely2019,Krticka2019,Utsunomiya2019,Utsunomiya2019,Brits2019}, which are of interest in nuclear astrophysics, especially for determining radiative neutron capture cross sections relevant to the $s$-process nucleosynthesis. 
Indeed, this approach is widely used to calculate both collective and single-particle excited states, as well as to study giant resonances \cite{Ring2004,Peru2008}. In this framework, an excited nuclear state, $|n\rangle$, is obtained by creating a phonon on top of the QRPA correlated ground state, $|\mathrm{QRPA}\rangle$:
\begin{equation}
|n\rangle = \hat{Q}^{\dagger}_{n}|\mathrm{QRPA}\rangle, \qquad \hat{Q}_{n}|\mathrm{QRPA}\rangle = 0,
\end{equation}
where the quasiboson operators are defined as two-quasiparticle excitations:
\begin{equation}
\hat{Q}^{\dagger}_{n} = \frac{1}{2}\sum_{ij} \left( X^{ij}_{n}\beta^{\dagger}_{i}\beta^{\dagger}_{j} - Y^{ij}_{n}\beta_{j}\beta_{i} \right),
\end{equation}
with $\beta^{\dagger}_{i}$ being a quasiparticle creation operator defined through the HFB transformation. The coefficients $X_{n}$ and $Y_{n}$ represent the forward and backward two-quasiparticle amplitudes, respectively, and are determined from the QRPA eigenvalue equations:
\begin{eqnarray}
\left(\begin{array}{cc}
A & B \\ B^{*} & A^{*} \end{array}\right)
\left(\begin{array}{c} X_{n} \\ Y_{n} \end{array}\right)
= \omega_{n}
\left(\begin{array}{c} X_{n} \\ -Y_{n} \end{array}\right),
\end{eqnarray}
with the matrices
\begin{eqnarray}
A_{kk'll'} &=& \left(E_{k}+E_{k'}\right)\delta_{kl}\delta_{k'l'} + H^{22}_{kk'll'}, \nonumber\\
B_{kk'll'} &=& 4! \, H^{40}_{kk'll'},
\end{eqnarray}
where $H^{22}$ and $H^{40}$ denote the two-body and four-body quasiparticle components of the Hamiltonian, respectively \cite{Ring2004}. Due to the explicit density dependence of the Gogny interaction, these matrices also contain rearrangement terms that must be included (see, e.g.,  \cite{Peru2014a}).
State-of-the-art QRPA implementations with Gogny interactions incorporate two-quasiparticle excitations built on axially deformed configurations, thereby enabling the study of deformed nuclei (see Refs. \cite{Peru2008,Chimanski2025} and references therein).
Using the same theoretical framework, electric dipole (E1), magnetic dipole (M1) strength distributions, as well as magnetic dipole and electric quadrupole moments, have been evaluated across different regions of the nuclear chart \cite{Peru2019,Goriely2020,Peru2021,Zamora2021,Barzakh2021,Weber2023,Chimanski2025}. This includes the study of pygmy dipole resonances (in the nucleus $^{140}$Ce) \cite{Vandebrouck2024} and scissor modes (in $^{254}$No) \cite{BelloGarrote2022}. Nuclear level densities have also been computed within the HFB+QRPA framework, extending traditional combinatorial approaches using the recently proposed QRPA plus boson expansion method \cite{Hilaire2023,Goriely2024}. These applications will be further discussed in sections \ref{ssec:ld} and \ref{gsf} dedicated to reaction applications.

The Gogny interaction has also been used in recent years to predict nuclear charge radii and compare them with new experimental data from laser spectroscopy in the isotopes Bi \cite{Barzakh2021}, Au \cite{Cubiss2023}, Cf \cite{Weber2023}, and Nb and Fm \cite{Warbinek2024}, using HFB with the equal filling approximation for the treatment of odd-mass nuclei. Neutron skin thicknesses for $^{208}$Pb, $^{40,48}$Ca, and several isotopes of O, N, C, Ni, and Sn have been investigated using the antisymmetrized molecular dynamics with generator coordinate method \cite{Matsuzaki2021,Tagami2021,Wakasa2023,Tagami2023a,Tagami2023b,Tagami2024}.

Finally, other recent applications of the Gogny interaction to more exotic aspects of nuclear structure 
are worth mentioning, such as the study of $\alpha$-particle and cluster correlations in $^{48}$Ti 
\cite{Taniguchi2021} and $^{24}$Mg \cite{Chiba2021} using the AMD+GCM approach. One can also mention 
the investigation of various excited states in hypernuclei  throughout the nuclear chart 
\cite{Kumar2024}. In this study, the $\Lambda-N$ interaction used  is a simple 
phenomenological potential sum of two Gaussian  with only Wigner and 
Barlett exchange terms. The six parameters entering the potential are 
fitted to reproduce the experimental 1s $\Lambda$ single particle state 
in various (eight) hypernuclei where experimental data exists.


\subsection{Fission}\label{subsec3}

Fission is one of the most characteristic and intriguing phenomena in 
the physics of the atomic nucleus. It is  unique to heavy mass nuclei 
due to the peculiarities of the nuclear interaction and it is a direct 
consequence of the competition between Coulomb repulsion and the 
surface tension energy emerging from the saturation property of the 
nuclear force \cite{Bohr1939,Brack1972,Bjrnholm1980}. In its way from 
the initial configuration \footnote{The initial configuration can be 
the ground state (spontaneous fission), any excited state built on top 
of it or any other excited state like fission isomeric states (induced 
fission)} to scission, the nucleus goes through various regions where 
the potential energy can be higher than that of the initial one (the one of 
the ground state, in the spontaneous fission case). In those cases, the 
magic of quantum mechanics, in the form of tunneling through a barrier, 
is required to explain fission \cite{Schunck2016}. Following the 
seminal ideas introduced by Bohr and Wheeler to explain fission dynamics 
\cite{Bohr1939}, it is essential to define a relevant coordinate 
characterizing the sequence of configurations populated in the 
transition from ground state to scission. Typically, the preferred 
coordinate is nothing but a measure of the deformation of the nucleus 
(like, for instance, its quadrupole moment) although more sophisticated 
models require more degrees of freedom (higher order multipoles like 
octupole, hexadecapole, neck, mass asymmetry, etc) to fully 
characterize the process. Once a fission driving coordinate is 
introduced, one can get an initial understanding of  the evolution 
towards scission by looking at the energy landscape of the system as a 
function of that coordinate, what is usually refereed to as the 
potential energy surface (PES) of fission. In this initial approach, 
concepts like fission barrier height and shape, fission isomeric 
states, second fission barriers, etc emerge in a natural way. However, 
the knowledge  of the PES alone is not enough to characterize fission, 
as tunneling probability depends upon the "action" associated to the 
fission coordinate and this quantity depends on an "inertia" to be 
associated to that coordinate (the momentum conjugate of the operator 
defining the coordinate). The inertia measures how fast the system 
transitions from one configuration to the next and therefore depends on 
the level density of the target configuration. As level densities 
around Fermi level depend on pairing properties of the configurations 
considered, the "inertia" strongly depends on the amount of pairing 
correlations present in those configurations that is often 
characterized in terms of, for instance, the pairing gap. This leads to 
the standard assumption that the inertia depends on the inverse of the 
square of the pairing gap \cite{Brack1972,Moretto1974}.

\begin{figure}
\centering
\includegraphics[width=0.71\linewidth,angle=0]{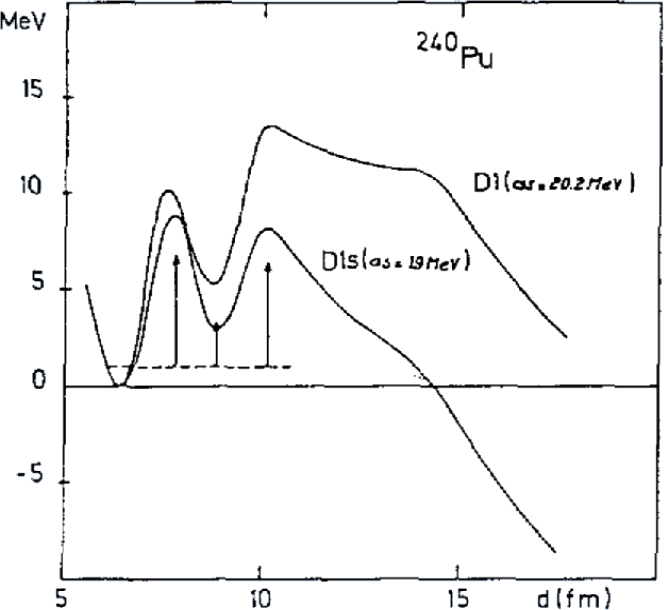}
\caption{Fission barrier heights computed with different Gogny parameterizations in
the case of $^{240}Pu$. The figure shows that D1S is the only one of the parameterizations considered that satisfy the expectations. Taken from \cite{Berger1989}}
\label{fig:Berger1989}
\end{figure}

A reasonable interaction willing to successfully describe fission 
phenomena must satisfy, at least, the requirement of having good 
surface properties as well as providing a consistent treatment of 
pairing. As discussed in the introduction, the first parameterization 
of Gogny D1 has a too strong surface energy coefficient and 
therefore fission barriers turn out to be too high and too wide as to 
reproduce fission phenomenology. This deficiency  was the main 
motivation to refit the force parameters leading to the new D1S 
parametrization (S stands for "surface") with a reduced surface energy 
coefficient $a_S\simeq19$ (evaluated without the spin-orbit contribution) and providing a much improved fission barrier in 
$^{240}$Pu and $^{232}$Th (see Fig.\ref{fig:Berger1989}). Since the inception of Gogny D1S, it has been thoroughly 
used to describe fission phenomena following the methodology discussed 
below. In this respect, it has to be mentioned that more modern 
parametrizations of the force preserve the surface properties of D1S 
and therefore they are also suited for fission studies at least in the
actinide and superheavy region.

Regarding pairing properties it has been long argued that the finite
range of the Gogny force allows for a consistent treatment of the pairing
channel along with the long range part of the Gaussian central interaction.
In this respect, one can consider that the different incarnations of the
Gogny interaction starting from D1S with its good surface properties are well suited for fission studies, as it will
be shown in the remaining part of the section.

In the following, we will discuss our present understanding of fission but
from the perspective of the Gogny force. There are many interesting studies
in the literature using other interactions or functionals, both relativistic and non relativistic, but we are not
going to discuss them here. We refer the reader to recent review papers
on the subject \cite{Schunck2016} for further information on results with
other interactions.

We discuss here the main ingredients entering the 
description of fission dynamics as well as some particularities to be 
considered in the treatment of odd mass nuclei and induced fission 
described in the framework of finite temperature. Recent developments 
including the impact of tensor interaction as well as an alternative
characterization of fission paths will be considered at the end of the 
section. Except for some specific examples, the bulk of fission 
treatment is carried out at the mean field level including pairing, 
i.e. using the HFB approximation and eventually its 
extensions. In the following, we will use the HFB method to 
characterize the different ingredients of a microscopic description of 
fission. See also \cite{Younes2017} for a historical perspective of
fission in connection with Daniel Gogny. 

The study of fission has been recently supplemented 
by including beyond mean field effects with the Gogny force \cite{Bernard2019} in the form of 
calculations including particle number projection (PNP) 
\cite{Robledo2007,Robledo2010a}. Subsequent studies 
\cite{RodrguezGuzman2022} in the Fermium and Nobelium isotopes using
the Gogny D1M* interaction have shown that the expected impact of 
PNP (an increase of pairing correlations due to symmetry restoration leads to a  decrease of 
the collective inertias with the subsequent reduction of $t_{SF}$) is 
somehow canceled out by the fact that PNP requires exact Coulomb 
exchange and  anti-pairing contributions to be fully taken into account 
to get rid of the self-energy and self-pairing effects that may appear 
in restoring symmetries \cite{Anguiano2001a,Anguiano2002a,Robledo2007,Sheikh2021}. 
Coulomb anti-pairing significantly reduces proton pairing correlations
compensating the increase of pairing correlations due to symmetry restoration.

The above result raises the question whether Coulomb anti-pairing should be
considered or not already at the mean field level.
The introduction of Coulomb anti-pairing increases substantially the 
computational cost of any mean field calculation and therefore its 
impact is often not considered in realistic applications. However, the 
independence of charge of the nuclear interaction and the fact that the 
pairing channel of the Gogny force is fitted to reproduce pairing gaps 
in the tin isotopes, where proton pairing is zero, leads to the 
conclusion that the Coulomb anti-pairing should be always considered 
(see \cite{RodrguezGuzman2023} for a recent calculation on the impact at the mean field level of
Coulomb anti-pairing in fission).
From the previous discussion, one can conclude that indeed Coulomb anti-pairing should be included at the mean field level in fission studies. However, its inclusion increases by a factor of around 5 the computational cost of calculations and, to avoid it, one can argue from a pragmatic point of view, that the increase of pairing correlations associated to PNP is going to compensate the impact of Coulomb anti-pairing making unnecessary its consideration at the mean field level. Clearly, the compensation is not exact and it depends on the nuclei considered. Finally, the combination of the two effects surely affect other quantities in different ways. Therefore, from a fundamental point of view, both Coulomb anti-pairing and PNP should be taken into account for a consistent microscopic description of fission.


Including 
exact Coulomb is a project that is already in the agenda of many 
practitioners of the Gogny force and the latest published results are increasingly 
taking this effect into account. 

\subsubsection{Potential energy surfaces}

In a first approach to fission, the potential energy surface (PES) as a function
of the relevant shape parameters  is 
computed by using the constrained HFB method using several constraints 
on quantities like multipole moments of the mass distribution or the
amount of matter in the neck that develops in the nucleus in its way to
scission - see Ref \cite{Han2021} for a recent account. It is also possible to consider other 
constraints aimed at changing the amount of pairing correlation in the 
system \cite{Giuliani2014,Zdeb2025a}. Typical calculations are restricted to axial symmetry, but 
reflection symmetry breaking (relevant to describe asymmetric fission) 
is allowed. Breaking of axial symmetry is allowed for configuration at 
the top of the first \cite{Delaroche2006a,Warda2002} (and also second) fission barriers, but its high 
computational cost prevents its systematic consideration throughout the fission 
landscape. In addition, triaxial paths seem not to be favored if least action paths are considered \cite{Delaroche2006a}, see below.  

The HFB calculations are carried out by expanding the quasiparticle operators as linear combinations of harmonic 
oscillator basis states to take advantage of the many analytical results that 
can be used in the evaluation of matrix elements, especially for the 
Gaussian terms present in the central part of the interaction. In the first studies of fission with the Gogny D1S parametrization, Berger and coworkers in Bruy\`eres-le-Ch\^atel used a two-centered HO basis and used the separation between the centers as an 
additional parameter to optimize the energy of the different PES configurations \cite{Berger1984,Berger1989,Berger2000,Goutte2005}. 
Using a two-center basis formalism allows the efficient and precise calculation of very elongated nuclear states. 
With such a rich basis, the description of the fragments spatial separation becomes possible, enabling the exploration of the Coulomb valley. Of course, such a code can be also used 
with a one center HO basis. 
Later \cite{Egido1997}, it has been shown that a single-center harmonic oscillator basis including a sufficiently large number of HO shells is able to study the fission process but requires of a careful evaluation of 
matrix elements \footnote{The evaluation of matrix elements of the 
interaction involves sums with very large terms that alternate in sign
leading to strong cancellations  which are prone to large numerical 
errors coming from the incomplete representation of floating point 
numbers in standard computer arithmetic}. 
For a detailed description of
how to evaluate the matrix elements of the Gogny force in a HO basis 
see, for instance, Refs \cite{Girod1983b,Egido1997,Younes2019}. 

To our knowledge, until very recently, all the computer codes, exception made of the Bruyères-le-Châtel ones, 
use one center basis \cite{Younes2009a,Robledo2011a,Dobaczewski2021,Marevic2022}.
A recent and modern re-writting of the Berger code with 2-center HO basis has been done by the Bruy\`eres-le-Ch\^atel group and released into
the public domain \cite{Dubray2025a}. 
This new solver, referred as HFB3, contains D1, D2 and DG-type
Gogny interactions and includes an exact treatment of the Coulomb term in both
mean and pairing fields. The public version has firstly been released without DG.
One notes that, recently, a version of the HFODD code has introduced two-center HO basis with Skyrme interaction \cite{HFODD2025}.

As mentioned above, the different mean field wave functions required to build the PES are often 
restricted to preserve axial symmetry in order to reduce the 
computational cost of the calculation \cite{Berger1984,Younes2009a}. On the other hand, reflection 
symmetry must be allowed to be broken in order to obtain configuration with 
mass asymmetry that might eventually lead to a mass asymmetric 
splitting of the nucleus required to explain asymmetric fragment mass 
distributions. The breaking of reflection symmetry imposes a constraint 
on the center of mass coordinate, to fix the nucleus at the origin of 
coordinates and prevent the presence of spurious configurations. 
It has been known since a long time \cite{Brack1972,Bjrnholm1980} that allowing 
triaxial shapes in fission might lead to a substantial reduction of the first 
fission barrier height, and to a lesser extent the one of the second. 
For the second barrier, reflection asymmetry has already set in and 
therefore, triaxial octupole shapes are to be considered. The reduction 
of the first barrier height can be of the order of a few MeV, not 
exceeding 5 MeV and it is often claimed that this reduction can be 
instrumental in bringing predictions for spontaneous fission half lives 
closer to experimental data. There is, however, a caveat in this line 
of reasoning: the reduction of barrier heights is due to a decreasing 
of level density at the top of the barrier that leads to an increase in 
the collective inertia that might eventually compensate the reduction 
in the barrier height \cite{Delaroche2006a}. It is therefore quite common 
that fission practitioners restrict their calculations to axial symmetry.
\begin{figure}
	\includegraphics[width=\columnwidth]{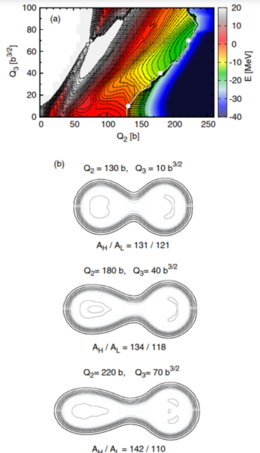}
	\caption{In the upper panel the potential energy surface of $^{252}$Cf is displayed. The scission line is
depicted in white. In the lower panels,  density profiles corresponding to the
configurations marked by dots in the upper panel. Taken from \cite{Zdeb2017}}
     \label{fig:zdebQ2Q3}
\end{figure}
Nowadays, it is customary to carry out PES calculations as a function of
two variables, typically multipole moments of various kinds 
including the axial quadrupole moment along with the octupole or the 
hexadecapole moment. One dimensional calculations are reserved to systematic
studies like the one of Ref \cite{RodriguezGuzman2024} covering a very wide margin of
isotopes in the superheavy region. On the other hand, a nice example of two dimensional calculations can be found
in Refs  \cite{Bernard2023,Bernard2024} where the potential energy surface of $^{180}$Hg is
analyzed to explain the unusual asymmetric fragment mass distribution experimentally 
observed in this nucleus. Another typical example for the PES as a function of the quadrupole
and octupole moments is presented in Fig.\ref{fig:zdebQ2Q3} for the nucleus
$^{252}$Cf. In the same figure, the nuclear matter density distribution for some 
relevant configurations before scission are also portrayed. Similar calculations
have been discussed in Ref \cite{Lemaitre2018} for the $^{236}$U and $^{240}$Pu nuclei using
the Gogny D1M interaction. One can also mention the work in Ref \cite{Zdeb2021} dealing with
multidimensional potential-energy surfaces obtained for $^{252}$Cf and $^{258}$No.
Quadrupole triaxial degrees of freedom are included in some calculations,
but this is not very common. One of the reasons is that in the calculation of  \cite{Delaroche2006a} it
was shown that the triaxial path in the first fission barrier is often
energetically favored but this is not the case for the corresponding
action and therefore, fission seems to avoid this path. An example
of calculation including triaxial degrees of freedom is shown in Fig.\ref{fig:zdebTri}.
\begin{figure*}
	\includegraphics[width=\textwidth]{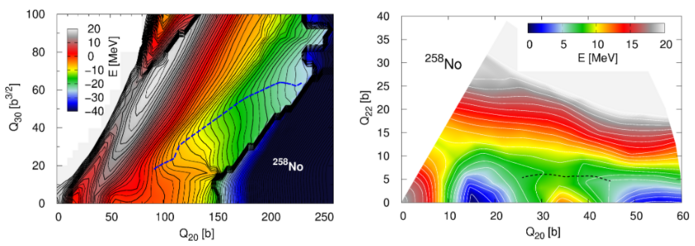}
	\caption{Left: PES of $^{258}$No as a function
of the axial quadrupole and octupole moments are shown as contour and color plots. 
Constant energy contour lines are plotted every 1 MeV. The asymmetric fission path is
represented by a blue dashed line. Right: PES in the $Q_{20}-Q_{22}$ plane. The dashed line shows the region
of the fission path around the first barrier with nonzero triaxial
deformation. Taken from \cite{Zdeb2021}}
    \label{fig:zdebTri}
\end{figure*}
Another recent example including triaxial configurations around the first
fission barrier is given in Ref \cite{Lemaitre2018}. 
Calculations including three degrees of freedom are scarce, and we can
mention the results discussed in \cite{Zdeb2021} where $Q_{3}-Q_{4}$ PES
for different values of $Q_{2}$ are used in the analysis of fission
in $^{252}$Cf and $^{258}$No nuclei. These calculations are computationally
very demanding explaining why there are so few examples of these three
dimensional calculations in the market.

\begin{figure}
	\includegraphics[width=\columnwidth]{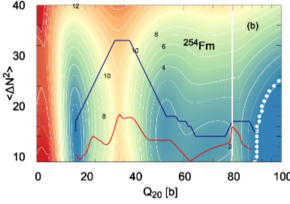}
	\caption{PES (in MeV) of $^{254}$Fm in the $Q_{20}-\langle \Delta N^{2}\rangle$ coordinates.
	 Red (Blue) lines correspond to the least-energy (least-action) paths. The white dots show the
     exit points towards scission. Taken from \cite{Zdeb2025a}}
     \label{fig:zdebdelta}
\end{figure}

Finally, pairing has been also included as dynamical variable just to 
analyse the impact of this degree of freedom in the action entering
the spontaneous fission life-time. As seen in Fig.\ref{fig:zdebdelta}
increasing the pairing gap $\Delta$ increases the energy, but the shape
of the PES along the quadrupole degree of freedom does not change substantially.
The most relevant impact of pairing is on the collective inertia, as it
will be discussed below.
%

\subsubsection{Inertias and pairing}
In our vision of fission, where the potential energy of the collective 
variable is the most important ingredient driving the dynamics, one 
also needs to define an inertia (or collective mass) for the 
corresponding dynamical degrees of freedom. In the beginning, those 
inertias were obtained by using hydrodynamical models of the nucleus 
\cite{Brack1972}, but it is clear that a microscopic description 
demands a more fundamental approach.

Traditionally, there are two ways to introduced microscopic collective 
inertias. The first is connected with the generator coordinate method: 
in the Gaussian overlap approximation one can approximate the 
Hill-Wheeler equation of the GCM by a collective Schr\"odinger equation 
on the collective variable \cite{Schunck2016}. The collective potential 
is defined by the HFB energy given as a function of the collective 
variable and supplemented by a zero point energy correction (rotational 
and/or vibrational). The inertia is  given in terms of second 
derivatives of the Hamiltonian overlap kernel with respect to the 
collective variable. As the derivatives with respect to the collective 
variable can be assimilated to a momentum-like operator in 
quasiparticle representation one is led with an expression for the 
inertia involving the mean value of the Hamiltonian times two 
collective momentum operators (see P. Ring and P. Schuck book \cite{Ring2004} for 
details). On the other hand, it is also possible to define a classical 
Hamiltonian for the collective variables by starting from the time 
dependent HFB method and using its adiabatic limit (ATDHFB) to obtain a 
quadratic expansion on the collective velocities in order to obtain a 
kinetic energy that defines de collective inertia. In this approach, 
the collective inertia also depends on the collective momentum operator 
expressed in the quasiparticle representation.

The two approaches are similar in what concerns the quantities entering
the expressions of the inertia. In both cases those expressions involve
the inverse of the stability matrix of the HFB theory which complicates
enormously the numerical evaluation of the collective inertia due the
huge dimensionality of the aforementioned matrix. In order to reduce the
computational cost two approximations are often used leading to the 
so-called "perturbative inertia". Expressions of the "perturbative inertia"
in terms of momenta of the collective variables can be found, for instance,
in \cite{Schunck2016}. When compared with exact GCM inertias \cite{Giuliani2018a}, it turns
out that the perturbative ones are larger than the exact inertias meaning 
that the spontaneous fission lifetimes (see below Eqs. (\ref{eq:tsf}) and (\ref{eq:action}) are overestimated by
the perturbative approach. An alternative is the use of the so-called
"non-perturbative" inertias where the momentum operator is computed 
using numerical derivatives. A comparison with the exact result can be
obtained in \cite{Giuliani2018a}.


\subsubsection{Spontaneous fission lifetime}
%

The most important observables related to fission are the fission half-life, the multiplicities of
emitted neutrons and gamma rays, 
and the fragment's properties like their mass distribution, kinetic energies,
excitation energy, angular momentum, etc . The fission half-life is not 
a uniquely defined quantity as it depends on the excitation energy of the parent nucleus that can
be acquired by different mechanisms (neutron absortion, beta decay to an
excited state, etc). This is the main reason why most of the studies tend
to focus on spontaneous fission half-life, which is associated to the time scale of the spontaneous 
decay of the ground state of the parent nucleus. In the traditional view
of fission, with a potential energy and an inertia for a given collective
coordinate, the evaluation of spontaneous fission half-life amounts to compute
the transmission coefficients through the fission barrier(s). Although the
task can be performed exactly (provided there are not too many collective variables), it is customary to use a semi-classical formula
for the calculation, the Wentzel–Kramers–Brillouin (WKB) formula, as it reveals
an important aspect of the transmission coefficient, namely its exponential
dependence with the classical action computed in the classically forbidden
region \cite{Schunck2016}
\begin{equation}\label{eq:action}
	S=\int_{\textup{in}}^{\textup{out}}\sqrt{2B(Q)(V(Q)-E_0)}dQ
\end{equation}
In the above expression, $Q$ is the collective coordinate (usually the
quadrupole moment), $V(Q)$ is the HFB energy supplemented by zero point
energy corrections and $B(Q)$ is the collective inertia discussed in the
previous section. Once the action is computed the spontaneous fission
lifetime can be evaluated by means of the formula
\begin{equation}\label{eq:tsf}
t_{sf}=2.86\times 10^{-21}(1 + \exp(2S))
\end{equation}
where $ t_{sf}$ is given in seconds. It is clear from the above formula
that the preferred path to scission is the one minimizing the action, i.e.
the nucleus, in its way to scission, is going to follow the path in the
multidimensional collective space that minimizes the action $S$ in Eq. (\ref{eq:action}). This principle
requires the exploration of the action in the multidimensional space of HFB states and therefore it is
difficult to implement unless there are dominant (and relevant) degrees of freedom. An 
approximation to it is to consider that the collective inertia is roughly 
constant in the path to scission as in this situation the minimization of the action is equivalent to the minimization
of the energy in $V(Q)$. This is a simpler task than the minimization of the action due to the variational 
origin of the HFB method that provides directly the least energy path. Most of the fission calculations carried out
with the Gogny force and mentioned in this review rely on the least 
energy (LE) principle. Only a few use the least action (LA) principle: in
Ref  \cite{Lemaitre2018} the LA principle is used with the Gogny D1M
interaction and considering the quadrupole and octupole degrees of freedom
as the relevant ones to explore the action. The results look very similar
to the ones obtained with the LE approach, casting serious doubts about the convenience
of using the LA principle. However, other studies with the Gogny force 
\cite{Giuliani2014,RodriguezGuzman2018,RodrguezGuzman2023,Zdeb2025b} including the
pairing degree of freedom as relevant coordinate have shown that the 
strong dependence of the collective inertia with the inverse of the square
of the pairing gap move the path to fission towards regions of larger
pairing correlation \cite{Moretto1974} reducing by several orders of magnitude the predicted
$t_{SF}$ in the LE scheme, bringing the theoretical predictions for $t_{SF}$ in much better agreement with experimental data than in the LE case.

\begin{figure*}
	\includegraphics[width=\textwidth]{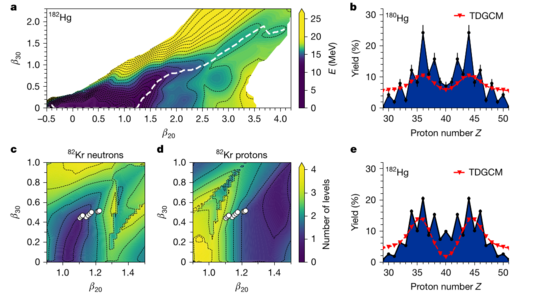}
	\caption{ Panel a: Two-dimensional PES of $^{182}$Hg as a function of its quadrupole and octupole
moments. Panels b and e: Experimental charge yields with statistical uncertainties as
error bars, compared to those obtained from the TDGCM calculations for
$^{180}$Hg (b) and $^{182}$Hg (e). Panels c and d: Smoothed number of levels around the neutron (c)
and proton (d) Fermi levels for the favoured light pre-fragment $^{82}$Kr emitted in the fission of $^{182}$Hg.
Taken from \cite{Morfouace2025}}
    \label{morfouace}
\end{figure*}

Systematic calculations of $t_{SF}$ show that even in the LE case, the theoretical predictions
follow the isotopic trend with neutron number and in some cases, and after a careful adjustment of
parameters, they reproduce reasonably well experimental data over a wide range of isotopes in the actinide and
superheavy region \cite{Warda2002,Warda2012,RodriguezGuzman2024}.

\subsubsection{Properties of fragments}
One of the most characteristic set of observables in fission is the fragment's mass 
distribution, deformation of the fragments and the kinetic energy that the fragments
carry after scission. In order to compute those quantities one has first
to define when the nucleus transitions from a configuration as a whole to a 
two fragment (or three fragment in the case of ternary fission) solution. One can use some ad-hoc criteria based on the number
of particles in the neck separating the nascent fragments \cite{Dubray2008a,Han2021} to identify the scission line \cite{Younes2009b} or one
can follow the time dependent evolution of the nucleus after crossing the
fission barrier. In the later respect, there
are two main approaches: the Time Dependent HFB (TDHFB) method and the Time Dependent Generator Coordinate method (TDGCM).
In both cases one studies the time evolution of the system but from a different perspective. In the TDHFB one follows
the dynamic evolution of an initial HFB state with the condition for the wave function
to remain a HFB state along the whole evolution. In the TDGCM one starts 
with a GCM ansatz for the initial wave function written as a linear 
combination of HFB states labeled by the collective coordinates used in
the description of fission. The dynamic evolution corresponds to the time
evolution of the amplitudes in the GCM ansatz. The two approaches are 
computationally demanding being the TDHFB the worst scenario. This is the
reason why the TDHFB method has only been implemented
with contact forces of the Skyrme type and strong assumptions about the 
non-local pairing field. Both assumptions aim to make the equations governing the time dependent evolution local in coordinate space in
order to save computational resources. As there are no implementations with
the Gogny force we will not discuss this approach here. On the other hand,
the TDGCM \cite{Verriere2020} has successfully been implemented with the Gogny 
force, using the Gaussian overlap approximation (GOA) for the evaluation of
the norm and Hamiltonian kernels \cite{Goutte2005,GOUTTE2006,DUBRAY2008,Dubray2008a}. In the GOA it is
assumed that the overlap between the different HFB configurations entering
the collective space of the GCM decreases very rapidly in such a way it can be considered
as a decaying Gaussian with the appropriate choice of collective variables.
In this way, the intrinsic non-locality of the GCM is approximated by a 
local treatment that only requires the knowledge of at most second derivatives
of the general overlaps. This represents a great simplification because the
calculation of the general overlaps is far from straightforward in fission: given
the very large set of deformations involved in fission, HO basis with 
tailored oscillator lengths have to be used for every configuration. As HO
basis with different oscillator lengths span non-equivalent  (under unitary
transformations) subspaces of the total Hilbert space, a special treatment
is required \cite{Robledo1994,Robledo2022a,Robledo2022b}. So far, a full 
TDGCM calculation with the Gogny force has not been performed.

Properties of fragments obtained in the induced fission of mercury 
isotopes have been analyzed with the Gaussian overlap approximation as 
implemented in \cite{Regnier2018} and using a two dimensional PES 
($\beta_{2}-\beta_{3}$) obtained after mean field HFB calculations
with Gogny D1S \cite{Morfouace2025} (see Fig.\ref{morfouace}).

Other results obtained in the same framework with the Gogny force include the 
work of Ref \cite{Regnier2018} used to describe the transition from 
asymmetric to symmetric fragment mass distribution in the fermium 
isotopes. In \cite{Goutte2005,GOUTTE2006} the kinetic energy and 
mass distribution of fragments in the fission of $^{238}$U is studied. A similar
analysis is performed in \cite{DUBRAY2008,Dubray2008a} but this time for
the fission of $^{226}$Th and $^{256-260}$Fm.

\begin{figure}
	\includegraphics[width=\columnwidth]{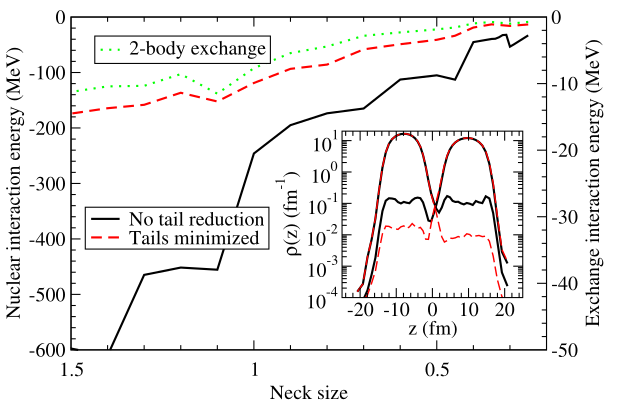}
	\caption{ Interaction energies plotted as a function
of neck size (QN ). The solid black and red dashed curves are, respectively, the
nuclear interaction energies before and after localization. The dotted green
line is the exchange part of the 2-body component of the
interaction energy (energy scale on the right y axis). The inset
shows the effect of the localization on the densities of the
fragments at scission. Taken from \cite{Younes2011}}
    \label{younesg}
\end{figure}

When the two fission fragments start to separate, it is not 
possible at the beginning to assign separated and well defined wave functions to any of 
them. In the standard quantum mechanics language, the fragments are said to be entangled. Obviously, in the 
asymptotic regime, when the two fragments are  well apart the 
problem disappears. But this region of separation is scarcely calculated with
standard HFB technics. 
One way that can be interesting is to find ways 
to "disentangle" as soon as possible  
the wave functions of the fragments to characterize their properties. 
This can be done, for instance, by using the existing freedom in  the 
Bogoliubov transformation associated to a unitary transformation among 
quasiparticles. The energy variational principle is not sensitive to 
that transformation and therefore it is possible to use it to minimize 
the overlap between the wave functions of the fragments without 
altering their properties  \cite{Younes2011}. In this way it is 
possible to assign a well defined entity to each of the fragments what 
allows a much better estimate of the Coulomb repulsion energy and, as a 
consequence, of the total kinetic energy of the fragments (see Fig.\ref{younesg}). See also
the discussion in \cite{Qiang2025} in this respect.

Another alternative is to use projection techniques on a spatial domain 
(see, for instance, \cite{Simenel2010,Verriere2019} ) to 
obtain wave functions restricted to a spatial domain with the required 
quantum numbers (mostly, particle number, to facilitate the calculation 
of fission mass distributions). The projection on a domain for HFB wave 
functions is more involved than the one for pure HF states 
\cite{Simenel2010}, and requires the use of the pfaffian formalism of Ref 
\cite{Bertsch2012} (see \cite{Scamps2013} for an implementation in the fission
context) and also requires to consider some technical issues 
discussed in \cite{Robledo2025,Robledo1994}.

\subsubsection{Odd mass systems}
%

%
%
%
%

The mean field treatment of nuclei with pairing correlations and an odd mass number
(either protons or neutrons) require an extension of the HFB method to 
consider the so-called "blocked states" which are nothing but a one-quasiparticle
excitation built on an even-even core
$$
|\phi_{\mu}\rangle = \beta_{\mu}^{+}|\phi_\textrm{even}\rangle{}
$$
This type of excitation breaks time reversal invariance and often axial symmetry. Therefore,
its treatment adds an extra layer of complexity to the calculations. Another 
difficulty is associated with the new degree of freedom introduced, namely, the 
characteristics of the blocked quasiparticle. 
The variational principle does not guarantee that the blocked state obtained by 
starting the self-consistent procedure with the
blocked configuration corresponding to the lowest one-quasiparticle energy will lead to the
lowest energy blocked configuration. As a consequence several quasiparticle
states with small quasiparticle energies have to be considered as
targets for the blocked quasiparticle. It has been
found \cite{DeLaIglesia2009} that typically 3-5 initial configurations have to be considered to reach with 
confidence the ground state, increasing 
the already costly computational time of fission by a factor corresponding to the number of
initial blocked configurations considered. In addition,
the breaking of time-reversal invariance represents an additional complication
as  time-odd fields now contribute to the Hartree-Fock potential and pairing
tensor (see Ref \cite{Robledo2012a,Robledo2014a} in the context of the Gogny force).
In the Gogny force case, the time-odd fields come directly from the 
folding of the time-odd density with the matrix elements of the 
interaction and therefore there is no room to readjust the behavior of 
the time-odd channel. The impact of time odd-fields is manifest in the 
behavior of rotational bands, where many calculations have shown that 
the Gogny force compares well with experimental data. Time-odd fields 
are also important in the description of odd mass nuclei and 
multiquasiparticle excitations. In those cases, the Gogny force is also 
doing well as the observed discrepancies with respect to experimental data are 
partly due to many body effects not included in mean field treatment of 
the problem.

\begin{figure}
	\includegraphics[width=\columnwidth]{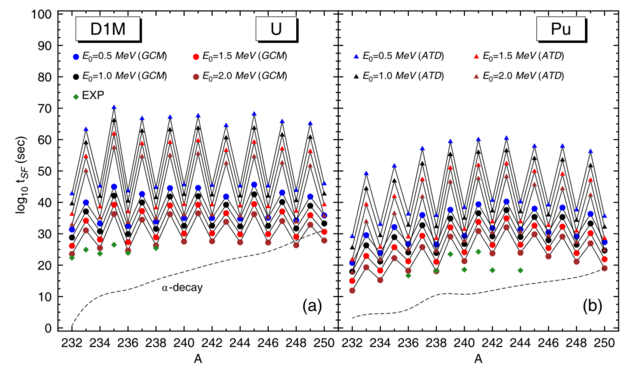}
	\caption{ Spontaneous fission lifetimes computed for several
	even and odd mass isotopes of uranium and plutonium showing 
	the characteristic odd-even staggering. Experimental data
	is also given. From the figure one concludes that irrespective
	of the value of $E_{0}$ taken in the calculation of the
	action, the staggering obtained in the calculation is far too
	large as compared to experimental data. In addition, $\alpha$-decay half-lives are plotted with short dashed lines. Taken from \cite{RodriguezGuzman2017}.}
    \label{fig:Odd-U-Pu}
\end{figure}

The outcome
of a "blocking" calculation is not only the ground state but a set of low
lying states that are usually characterized (in an axially symmetric calculation) by the $K$ quantum-number.
In the spirit of the strong coupling assumption of the particle plus rotor model \cite{Ring2004}, the $K$ quantum number 
is assimilated with the spin of the physical state in the laboratory system. 
The relative position of those states changes with deformation and therefore
change along the path to scission. It is usually assumed that along the
path to fission, the total angular momentum of the state $J$ or its $K$ 
quantum number is conserved. Another characteristic of blocking is the
severe quenching of pairing correlations with the associated increase
of the collective inertia. Therefore it is to be expected that odd-A nuclei
are going to have a longer spontaneous fission half-lives than the neighboring
even-even ones. The odd-even staggering in $t_{sf}$ has been observed experimentally
and typically represents two or three orders of magnitude. There is an
additional problem in the application of the standard methods of fission
to odd-A nuclei: the expressions of the perturbative inertias are not well 
defined in this situation. In order to circumvent some of these problems it
is common practice to use the Equal Filling Approximation (EFA) that
intuitively distributes the unpaired nucleon in a given orbital and
its time-reversed companion with probability 1/2. From a more formal
perspective \cite{PerezMartin2008}, the EFA treats the nucleus as a statistical collection
of quantum states in which the probability of finding a system with
a given orbital blocked is the same as the probability of finding
the system in the time-reversed orbital, and both equal 1/2. As a 
consequence, the EFA do not break time-reversal invariance and do not
require the evaluation of time-odd fields. In extensive studies with
both the Skyrme interaction \cite{Schunck2010} and the Gogny one \cite{Giuliani2024} it has been proven
that the results obtained with full blocking and the EFA are qualitatively
the same as they only differ in minor quantitative aspects. It seems then 
justified  to use the EFA as an alternative to full blocking in fission
calculations. For the evaluation of the collective inertias in the EFA, one
uses the inertias given by the finite temperature HFB (which is also 
a quantum statistical approach). Fission calculations with the Gogny force
and the EFA for odd-A systems were first carried out in \cite{Perez2005,PerezMartin2009a,DeLaIglesia2009} for
the nucleus $^{235}$U. Subsequent studies  have
analyzed fission properties of several odd-A systems like uranium, plutonium and nobelium isotopes
\cite{RodriguezGuzman2016,RodriguezGuzman2017}. The  odd-even staggering of $t_{SF}$ obtained in the calculations is huge and
can amount to up to twenty orders of magnitude (see Fig.\ref{fig:Odd-U-Pu}). 
It has been argued that the quenching of pairing correlations 
associated to the EFA (quantitatively and qualitatively similar to the 
one of full blocking) leads to a too large increase in the inertias for 
odd-A systems explaining thereby the huge staggering obtained in the 
calculations. It has also been argued that in a least action framework, 
the minimization of the action will lead to paths with stronger pairing 
correlations and therefore to smaller inertias, reducing thereby the 
observed staggering. In order to simulate this effect, calculations 
increasing by 5\% the pairing strength of Gogny D1S have been carried 
out \cite{RodriguezGuzman2016,RodriguezGuzman2017}, leading to a much 
sorter half-lives of the odd-A isotopes and a much reduced odd-even 
staggering.
  
\subsubsection{Dissipation and two quasiparticle excitations}\label{diabatic}

Part of the energy available in the initial configuration can be dissipated 
during fission by transferring it to internal degrees of freedom of the fissioning
nucleus (or the fragments). Among the many excitation modes available in the atomic nucleus
single particle (two quasiparticle) excitations are the predominant ones and
therefore they have to be accounted for in a microscopic description of 
dissipation. The most direct approach to include two quasiparticle (and/or
higher quasiparticle excitations) into the microscopic description is
to allow for general wave functions in the spirit of the GCM but including
also 2qp (and/or higher order) excitations built on top of the configurations visited in the fission
process. What one would call a multiquasiparticle TDGCM (mqp-TDGCM). The
approach is very appealing but its implementation requires overcoming 
many technical issues mostly dealing with the calculation of the overlap
of the Hamiltonian between multi-quasiparticle excitations. Also, the Hill-Wheeler
equation of the GCM (and TDGCM) require an orthogonalization of the norm overlap that can be
ill behaved if the norm overlap is not smooth enough. When including 2qp excitations
the norm overlap is no longer a smooth function of the collective coordinates hampering the
solution of the HW equation of the GCM. As a consequence,
it is not surprising that such microscopic approach including dissipation have rarely  been implemented
in the literature, and when implemented it has been in the framework of a 
Schr\"odinger Collective-Intrinsic Method \cite{Bernard2011}. The formalism presented
in \cite{Bernard2011} make use of the symmetric moment expansion approach
but the complexity of the expressions involved restricts its application
to just the calculation of a few relevant overlaps. From the results obtained
it is clear that the traditional ideas like the Gaussian overlap approximation
often used to simplify fission dynamics are no longer valid and alternatives have to 
be considered.

Studies of dissipation into collective degrees of freedom using the GCM framework
with the GOA and the Gogny force can be found in Ref \cite{Younes2012}. See also
\cite{Berger1984} for an early study of collective dissipation. 

\subsubsection{Induced fission and finite temperature}

\begin{figure}
	\includegraphics[width=\columnwidth]{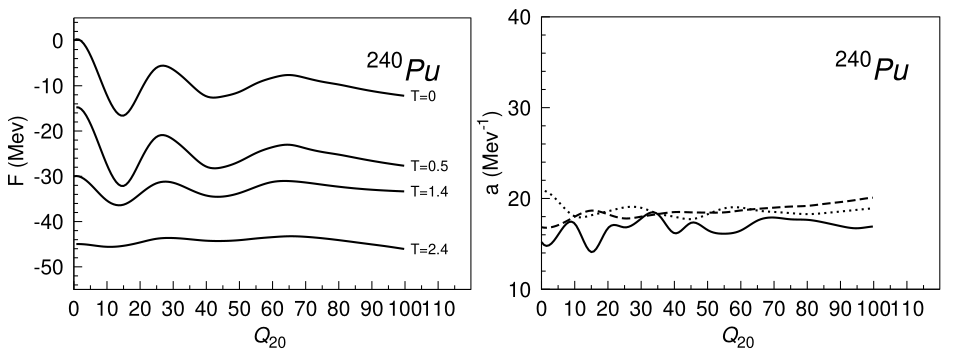}
		\caption{ In the left hand side panel, the free energy (in MeV) as a function of the quadrupole moment
Q20 (in b) for the nucleus $^{240}$Pu and for different temperatures. For representation purposes, the
absolute position of the curves has been modified to make them closer, but the ordering has been
preserved. On the right hand side panel, the level density parameter $a$  is
plotted as a function of the quadrupole moment. Taken from \cite{MARTIN2009}}
 \label{barrierT}
\end{figure}

In the case of induced fission, instead of starting from the ground state the nucleus starts from
an excited state like in $\beta$-delayed fission, or neutron induced fission. The excited state can
be collective, for instance the member of a rotational band with angular momentum
$J$. Under the assumption that the $J$ value does not change along the 
fission process one has to consider PES at finite angular momentum computed
within the cranking HFB approach that breaks not only axial symmetry but
also time-reversal invariance. This calculation was carried out in Ref
\cite{Egido2000a} for the nucleus $^{254}$No with the Gogny D1S interaction. A
transition from a two humped to one humped fission barrier regime was observed. Also a gradual
decrease of the fission barrier height as a function of angular momentum
was observed. It was concluded that at angular momentum $J=40\hbar$ the
fission barrier was so low that fission happens instantaneously, in 
good agreement with experimental data. If the excited state is two-quasiparticle
in character, the preferred approach is to use a finite temperature mean
field approach because the level density of two quasiparticle excitations
is so high that it is virtually impossible to follow the evolution of the
initial state in its way to scission. Finite temperature mean field is 
formulated in the canonical or grand canonical ensemble and the concept
of temperature is introduced to replace the average value of the energy. Obviously,
talking about canonical and grand canonical ensembles in a finite, closed
system like the atomic nucleus is not the most appropriate approach. However,
finite temperature mean field represents a good starting point to study
fission from excited states. The finite temperature HFB (FTHFB) method
was formulated long ago \cite{Goodman1981} and applied to studies with
the Gogny D1S force in \cite{Egido2000b,Martin2003a,DeLaIglesia2009}. The evolution of the
fission barrier height (see Fig.\ref{barrierT}), collective inertias and spontaneous fission 
half life in $^{240}$Pu and computed with the Gogny D1S force is analyzed in Ref \cite{MARTIN2009}.
Other studies with the Gogny D1S force are analyzed in \cite{DeLaIglesia2009}.
The results of the FTHFB theory are also used to describe key elements 
of fission of odd-A systems within the EFA given the quantum statistical
origin of this approximation  \cite{PerezMartin2008}.

\subsubsection{Tensor interaction}

\begin{figure}
	\includegraphics[width=\columnwidth]{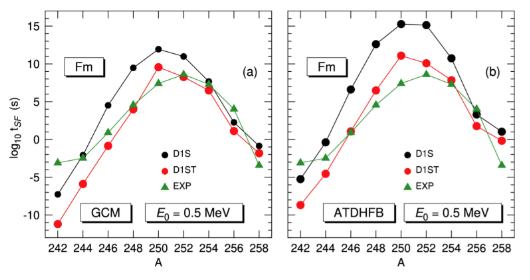}
		\caption{ Evolution with neutron number of the 
		spontaneous fission half-life $t_{SF}$ in several
		isotopes of fermium along with a comparison with
		experimental data. The plot shows how the inclusion
		of the perturbative tensor term in the interaction helps
		to improve substantially the agreement with experimental
		data. Taken from \cite{RodriguezGuzman2024}}
        \label{fig:fistensorbis}
\end{figure}

The tensor force is an important part of the nuclear interaction but it 
is often disregarded in mean field calculations with effective forces 
because its consequences are assumed to be partially taken into account by the 
density dependent part of the interaction. The impact of the tensor 
force in the deformation properties of nuclei has not been extensively 
studied, not to mention in fission. As the tensor interaction rearranges 
the position of certain single particle orbitals it means that it might 
affect the quantities relevant to fission like fission 
barrier heights and widths, collective inertias, etc. In Ref \cite{Bernard2020} 
the possibility that the tensor interaction could be the 
missing ingredient required to explain a symmetric bimodal fission mode 
recently found in some neutron-deficient thorium isotopes was explored. In  this 
study the perturbative D1ST2a Gogny+tensor  interaction \cite{Anguiano2012,Grasso2013} was 
compared to standard D1S results in several isotopes of thorium using 
$(Q_{20},Q_{30})$ as the main collective space and considering also 
$Q_{40}$ in several instances. The perturbative tensor interaction improves the description of fission properties of neutron deficient thorium isotopes. A tensor term should be integrated into the long range part of future  effective interaction (with a full refit of all the parameters, as done for DG) as it may provide a better description of fission, in particular in exotic pre-actinides.
In a subsequent publication \cite{RodriguezGuzman2024} the impact of the tensor force in the 
fission properties of the heavy
and superheavy nuclei in the isotopic chains of curium, californium and fermium was analyzed, confirming the results of Ref \cite{Bernard2020}. 
The inclusion of the tensor term tends to reduce the height of the first fission 
barrier improving the agreement with experimental data for the $t_\textrm{SF}$,
as can be seen in Fig.\ref{fig:fistensorbis}. 
The non-negligible impact of the tensor term in fission properties calls a
thoroughly exploration of fission properties of the new Gogny force DG
discussed above.

\subsubsection{Alternative determination of the fission path}

\begin{figure}
	\includegraphics[width=\columnwidth]{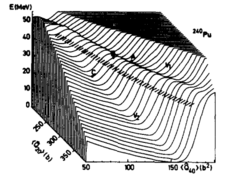}	
	\includegraphics[width=\columnwidth]{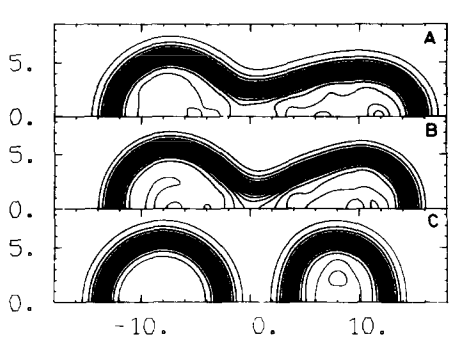}
	\caption{ In the upper panel, potential energy surface of $^{240}$Pu as a function
	of the axial quadrupole and hexadecapole moments showing the
	two relevant valleys in fission: The valley V1 starts in the
	ground state and is followed by the nucleus in its way to scission
	until it reaches a region where it is favourable energetically
	to descend to valley V2 that leads to two fragment solutions. In the
	lower panel, the nucleus' matter density at points A, B and C displayed
	in the PES are displayed. Taken from \cite{Berger1984}}
    \label{fig:fisBerger1984}
\end{figure}

In the traditional description of fission, the collective variable driving
fission is chosen based on heuristic arguments having to do with the way one parametrizes
nuclear shapes in terms of multipole moments of the mass distribution. This is so 
as in the fission process one has to consider how the nucleus deform in its way from the initial configuration to scission. Typically, the
quadrupole moment of the nuclear density is chosen as such driving coordinate. Already in the 
eighties of the past century it was argued that describing the transition
from compact to very elongated shapes implies going from a valley in the
energy landscape to a neighboring one characterized by a larger hexadecapole
deformation \cite{Berger1984,Berger1989}. In Fig.\ref{fig:fisBerger1984}, in the upper plot, the
potential energy surface of $^{240}$~Pu is shown as a function of the quadrupole
and hexadecapole degrees of freedom. Two valleys are clearly identified, the
one with larger hexadecapole moments corresponds to compact shapes (first panel in the lower plot)
while the one with lower hexadecapole correspond to a two fragment solution where
the neck connecting fragments is already broken (third panel in the lower plot),
the intermediate situation corresponding to the ridge connecting the two valleys with density matter distribution depicted in the middle panel B.
From the plot it is also clear that there are many possibilities to switch
between valleys and how to climb the ridge from the compact to the elongated valley
is still subject of debate and only a genuine time dependent description
can tell the answer. 
On the other hand, in the calculation of PES, it is
common \cite{Dubray2012,Regnier2019,Zdeb2021} to find discontinuities in the PES as a function of the driving
coordinate (or coordinates in the 2D and beyond case) that are associated
to the abrupt transition from one valley to another in the multidimensional
parameter space. The discontinuity can be characterized by a sudden change in
some observables characterizing the nuclear configuration, typically multipole
moments of the mass distribution or the amount of pairing correlations. How to soften the transition is a delicate problem often requiring
heuristic arguments. Typically, one argues that the discontinuity calls for the
introduction of the observable that changes suddenly across it as a new collective
variable. 
A first attempt in that direction was proposed by Ref.\cite{Lau2022} in which the
discontinuities are softened by increasing locally the dimension of the PES by one collective variable, for the adiabatic path. The algorithm was developed starting from one- and two- dimensional
PES. This solution is numerically challenging for PES of larger dimensionality. The discontinuities represent a problem in the 
description of the dynamical evolution in the framework of the TDGCM as both 
the HFB wave function and Hamiltonian overlaps change abruptly in the region
around the discontinuity altering in a substantial way both the smoothness of
the collective wave functions and the time dependent evolution of the system. 

It has been proposed recently \cite{Carpentier2024b} a novel method to solve
this problem of discontinuities in quantum states, without increasing the number of collective variables. This approach allows to generate continuous paths for both adiabatic and excited states. Using three different protocols (Link, Drop and Continuous Deflation) based on HFB calculations under overlap constraints ensuring continuity or/and orthogonality, it is possible to determine the configurations linking an initial state (located in the ground state well) with the scission
point and beyond (the two-center HO basis allows to inspect the Coulomb valley without any difficulty). 
\begin{figure*}
	\includegraphics[width=\textwidth]{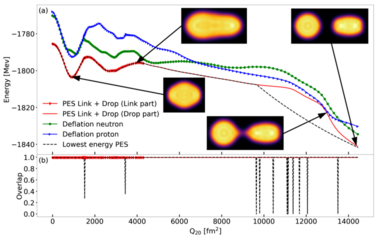}
	\caption{ (a) Some PESs (red, green, and blue curves) obtained with the different variational protocols based on HFB energy minimization
	with constraints. The PESs in red
have been calculated for $^{240}$Pu using the link and drop method using $d = 0.990$ and are displayed as a function of the mass quadrupole
moment. For comparison, the black dashed line corresponds to the solutions obtained with conventional minimization techniques
(adiabatic limit). All PESs present a continuous evolution of the total binding energy. More precisely, solutions corresponding to red diamonds are based on the link method, while the solutions forming the
continuous red curve result from the drop method. The different insets show spatial and local one-body total nuclear density slices for some
striking states. The deflation method has been used to produce a neutron (proton) excited PES in green (blue), imposing orthogonality
with the corresponding red HFB solution while ensuring the same overlap value. Note that ten excited states have been calculated with the
deflation method and have obtained very stable results (not shown here). (b) Overlap between each state and its adjacent state for the red
and black curves. We can see some isolated single low values of the overlap for the black curve, corresponding to discontinuities as
discussed in the text. The link and drop methods allow a constant high value of the overlap shown here for the red curve. Taken from \cite{Carpentier2024b}}
\label{fig:fisCarpentier2024}
\end{figure*}
Exploratory calculations with the D1S Gogny force for the 1D asymmetric path 
in $^{240}$Pu and using the two-center HO basis code HFB3 \cite{Dubray2025a} shows the suitability of the method as can be observed in Fig.\ref{fig:fisCarpentier2024}. 
The required overlaps are computed paying attention to the use of different HO
basis parameters for each of the configurations \cite{Robledo1994,Robledo2022a,Robledo2022b}, but extending the previous formulas to a two-center HO basis \cite{Carpentier2024a,CarpentierArxiv1}. This has allowed for the first time the implementation in one dimension of the  Schr\"odinger Collective-Intrinsic Model mentioned in section \ref{diabatic}, with more general excited states produced by the Continuous Deflation than two quasiparticle states initially proposed which display discontinuities \cite{Carpentier2024a,CarpentierArxiv1,CarpentierArxiv2,CarpentierArxiv3}. An energy balanced at scission was calculated including the intrinsic excitation energy of the fissioning system.
An extension of the approach to two dimensions is underway.


\subsection{Gogny interaction and nuclear reactions}
\label{sec:reaction}
\subsubsection{Introduction}
\label{ssec:reac-intro}
The development of the Gogny effective interaction in the early 1970s, driven by needs in nuclear structure, immediately opened the door to its application in reaction theory.  Its strength lay in a fully consistent mean-field framework (HFB) that already provided a satisfactory account, given its broad application range in terms of the number of nuclei described, of ground-state masses and matter densities, augmented by beyond–mean-field methods (RPA, QRPA, GCM - add refs. to previous sections) to describe a rich variety of excited modes: giant resonances, low-energy surface vibrations, non natural-parity excitations, and collective GCM states.  This unified description supplies all microscopic ingredients, single-particle spectra, matter densities, pairing properties, excitation spectra, and more, required for reaction calculations.

From the outset, collaborations between the Bruyères-le-Châtel structure group, reaction-theory experts such as Ch. Lagrange, and experimental teams led by G. Haouat, implemented Gogny D1 structure predictions directly in scattering and reaction models.  Microscopic calculations were systematically compared with fresh experimental measurements from the 1980s to the present, particularly for elastic and direct inelastic scattering,  while parallel experimental efforts improved knowledge of nuclear structure. These applications, along with the other topics introduced below, will be illustrated in the following subsections.  In recent years, Gogny-based calculations have started to be applied in nuclear data fields, and recent advances and new production promise deeper and more systematic integration in nuclear data evaluation in the near future.

Nuclear reactions induced by nucleons or light ions at energies ranging from a few keV up to several hundred MeV per nucleon generally proceed via three main mechanisms: direct, pre-equilibrium, and compound reactions.  
A concise overview of these mechanisms and the essential theoretical formalisms that underpin them is given in Appendix \ref{app:reac}.  This foundation will prove useful when discussing the specific Gogny-based applications in optical model potentials (OMPs), direct inelastic scattering, pre-equilibrium emission, level densities, and gamma-strength functions presented in the following subsections.

\subsubsection{Optical Model Potentials}
\label{ssec:reac-omp}
The optical potential is a central ingredient for modeling a wide range of nuclear reaction mechanisms. Using the Gogny interaction, several studies have been carried out to predict nucleon–nucleus optical potentials, either within nuclear-matter-based folding approaches or using nuclear-structure inputs embedded in Green’s-function formalisms. In the following, we describe the main applications of these Gogny-based optical potentials.

\paragraph{Nuclear-matter approaches}
\label{sssec:omp-nm}
One powerful strategy for constructing microscopic OMPs is to start from an effective, density-dependent nucleon–nucleon interaction in infinite nuclear matter.  A local density approximation (LDA) can then bridge back to the finite nucleus by assuming the nuclear matter $G$-matrix is valid at each local density of the target. Two applications presented below leverage this idea, implementing it in different ways.

\subparagraph{\textbf{JLM and JLMB Semi-Microscopic Optical Potentials}}
\label{par:omp-jlm}

The Jeukenne-Lejeune-Mahaux (JLM) optical model, developed in the mid-1970s, was a pioneering microscopic approach to nucleon-nucleus interactions. Its theoretical foundation lies in the Brueckner $G$-matrix (Brueckner-Hartree-Fock theory) for infinite nuclear matter, using a realistic nucleon-nucleon interaction (Reid’s hard-core potential) \cite{Jeukenne1977a}. The authors calculated the complex nucleon self-energy in symmetric nuclear matter as a function of density and energy, yielding isoscalar and isovector components of the optical potential (plus a Coulomb term) self-consistently \cite{Jeukenne1977b}. 

This folding of microscopic nuclear matter results gives a real and an imaginary potential that are generated from the same underlying $G$-matrix (unlike phenomenological models that fit them separately), making the model predictive and globally applicable rather than fit individually to each nucleus. 
Importantly, the original JLM model included only a single free parameter, an effective interaction range to simulate finite-range exchange effects omitted by the strict LDA.

Jeukenne–Lejeune–Mahaux (JLM) derived analytic expressions for an energy‑ and density‑dependent nucleon–nucleus potential that can be applied globally, without refitting nucleus by nucleus. To improve agreement with elastic‑scattering data, later work introduced smooth, energy‑dependent renormalization coefficients that remain close to unity. Because the original JLM framework lacks a spin–orbit term, a microscopic but ad hoc complex spin–orbit potential was added. The resulting approach has two facets: (i) microscopic/predictive, supplying the density profiles, the dominant energy dependence, and the real and imaginary central parts from many‑body calculations and theoretical matter densities; and (ii) phenomenological, encompassing the overall renormalization factors, the spin–orbit term, and an effective range. The phenomenological part has been treated in two ways: first, local adjustments tailored to specific studies; later, global adjustments that determine smooth, purely energy‑dependent renormalization factors valid over broad energy ranges, constrained by scattering data across many nuclei.

Soon after the JLM model’s introduction, researchers began testing its predictions against experimental data for heavy nuclei. In the late 1970s and early 1980s, C.~Lagrange and collaborators at CEA Bruyères-le-Châtel carried out some of the first applications of the JLM folding model to real elastic and inelastic scattering observable. One representative study by Lagrange and Brient  \cite{Lagrange1983a} considered nucleon scattering on the doubly-magic lead nucleus $^{208}$Pb. They applied the JLM approach to construct microscopic folded potentials for $^{208}$Pb
. Using local nuclear densities for HF and RPA calculations performed with the Gogny D1 interaction, they successfully analyzed both elastic and inelastic cross sections of nucleons on $^{208}$Pb over the 8.5–61 MeV energy range. 
In particular, they examined the first excited state of $^{208}$Pb and emphasized the energy-dependence of its excitation probability. 

\begin{figure*}
  \centering
  \includegraphics[width=\linewidth]{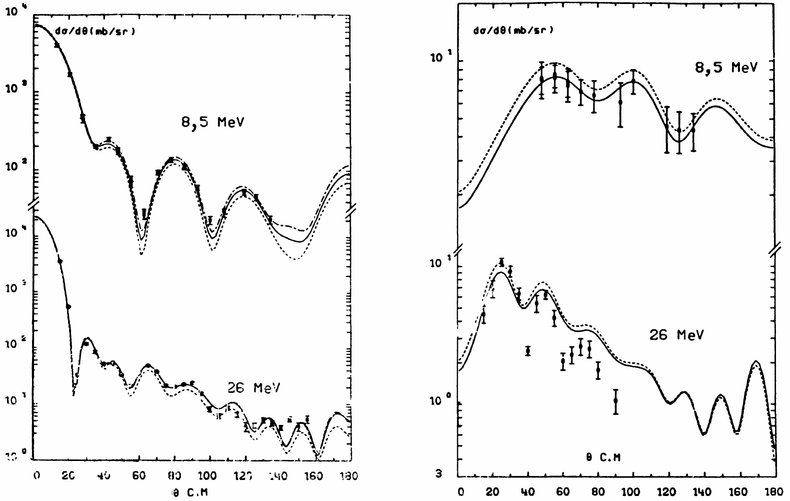}
  \caption{Figures from \cite{Lagrange1983a}. Theory–experiment comparison for neutron scattering from $^{208}\mathrm{Pb}$: elastic (1a) and inelastic (1b) angular distributions. Experimental data: Refs.~[11] ($E_n=8.5$~MeV) and [12] ($E_n=26$~MeV). See text for details of the calculated curves.}
  \label{fig:pb208-n-ela-inela}
\end{figure*}

This was an important insight: the JLM model naturally predicts how inelastic transition strengths (form factors) vary with incident energy, thanks to the built-in energy dependence of the microscopic $G$-matrix interaction. 
Fig. \ref{fig:pb208-n-ela-inela} compares the calculated angular distributions, obtained using the ground-state density from HF/D1 and the transition density from HF+RPA/D1, with the experimental data for neutron elastic and inelastic scattering from $^{208}$Pb.
The early studies showed that the semi-microscopic folding approach could describe angular distributions and cross sections with reasonable accuracy using a global parametrization rather than adjusting optical potentials specifically for lead. 

\begin{figure*}
\centering
\includegraphics[width=\linewidth]{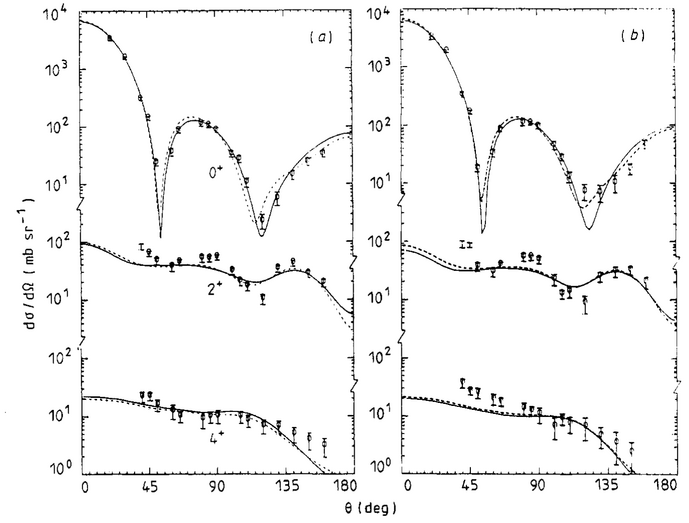}
\caption{Neutron scattering cross sections at $E_n=3.4$~MeV for the $0^+$ ground state, $2^+$, and $4^+$ levels of $^{238}$U (a) and $^{232}$Th (b).
Solid lines: semi-microscopic coupled-channel calculations using HFB deformed densities (Gogny D1) and the JLM interaction.
Dashed lines: phenomenological optical model results (adapted from Lagrange and Girod, \textit{J.~Phys.~G:~Nucl.~Phys.} \textbf{9}, L97 (1983)).}
\label{fig:UTh_scattering}
\end{figure*}

Building on the success for spherical targets, the JLM folding model was extended in the 1980s to heavy, deformed nuclei using microscopic structure inputs based on the Gogny force. In this semi-microscopic scheme, deformed optical potentials are obtained by folding the JLM interaction with deformed HFB ground-state densities (Gogny D1). The multipole expansion of these Gogny-HFB densities provides the coupling form factors for coupled-channels (CC) calculations, which then describe elastic scattering and inelastic excitation within the ground-state rotational band \cite{Lagrange1983b,Lagrange1986}. Fig. \ref{fig:UTh_scattering} shows the quality of such CC results for $^{238}$U and $^{232}$Th at 3.4~MeV when Gogny-HFB densities are used in the folding.

A key development was the consistent use of Gogny-based transition densities. Lagrange, Madland, and Girod applied the JLM folding model to $n+^{239}$Pu from $\sim$10~keV to 10~MeV, coupling states within the rotational band via microscopically derived transition densities \cite{Lagrange1986}. More broadly, CC studies that folded JLM with HFB ground-state densities  densities reproduced elastic and inelastic data from about 8.5 to 61~MeV without ad-hoc form factors, with only modest global depth renormalizations. In parallel, a generalized CC framework embedded a microscopic Bohr Hamiltonian—where the collective potential and six inertial functions come from microscopic dynamics—so that $(\beta,\gamma)$ softness and triaxial couplings could be treated on the same footing as spherical limits \cite{Kumar1985}. This provided a practical bridge from spherical to transitional and well-deformed targets.

Using Gogny-HFB inputs became standard in later applications: for $n+^{232}$Th and $n+^{238}$U at 14.1~MeV, and for $n+^{239}$Pu over a broad energy range, semi-microscopic CC analyses with deformed Gogny-HFB densities reproduced elastic and inelastic angular distributions (and, for $^{239}$Pu, total cross sections) at the level of state-of-the-art phenomenology, while preserving  a direct structure–reaction link \cite{Hansen1986,Lagrange1986}. These studies tested the sensitivity of the optical potential to the deformation and collectivity predicted by Gogny-HFB/D1, and they highlighted the improved predictive power from microscopic densities.
: for example, the calculated $^{239}$Pu neutron total cross section as a function of energy agreed well with experiment without local refits that are often required by purely phenomenological OMPs \cite{Hansen1986}. 

In the late 1990s, the work of Bauge, Delaroche, and Girod \cite{Bauge1998} marked a major step in connecting microscopic nuclear structure and reaction modeling. Using ground-state densities from Hartree–Fock–Bogoliubov (HFB) calculations with the Gogny D1S force, available across the nuclear chart, they constructed a global nucleon–nucleus optical potential within the JLM framework. This development (known as JLMB for JLM-Bruyères) demonstrated that structure inputs derived from a self-consistent Gogny interaction could quantitatively reproduce a wide range of nucleon scattering observables, from a few keV up to 200 MeV and for nearly all nuclei with $A>40$. The approach, later refined to ensure Lane consistency between proton and neutron channels \cite{Bauge2001}, provided the first fully semimicroscopic, structure-based global potential—highlighting the predictive power of Gogny-HFB densities in reaction modeling.

\begin{figure}[!htb]
    \centering
    \includegraphics[width=0.75\linewidth]{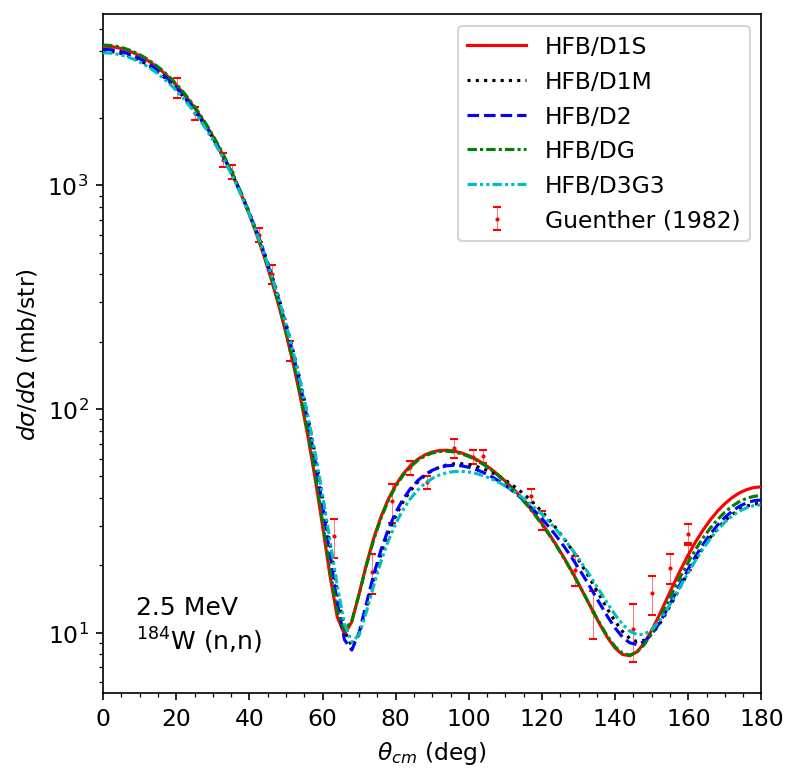}\\
    \includegraphics[width=0.75\linewidth]{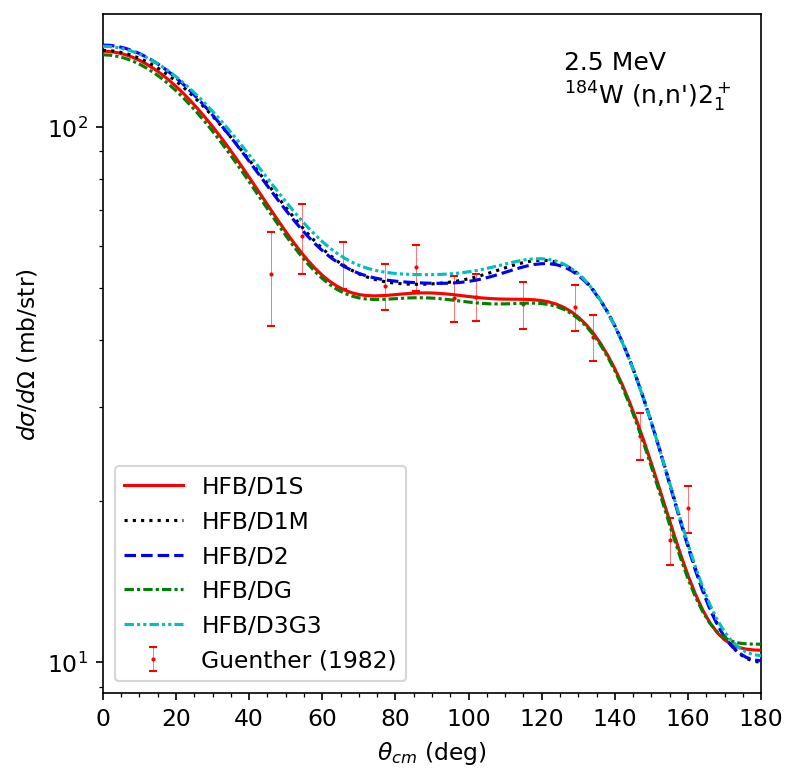}
    \includegraphics[width=0.75\linewidth]{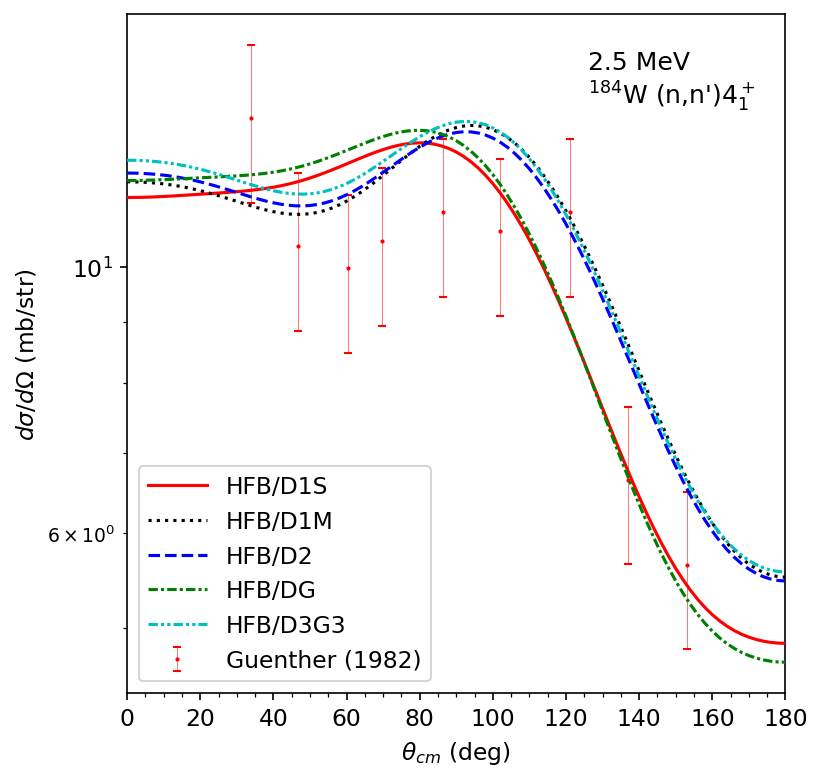}
    \caption{Elastic (top panel), and inelastic to the first 2$^+$ (middle panel) and 4$^+$ levels (bottom panel), differential cross sections for 2.5~MeV neutrons impinging on a $^{184}$W target. Experimental data (symbols) are compared to JLMB/Gogny-HFB calculations with various Gogny forces (D1S, D1M, D3G3, D2, DG).}
    \label{fig:WcompGogny}
\end{figure}

Building of the previous successes of Ch. Lagrange and collaborators for deformed nuclei, Bauge et al.
applied the JLMB model to
 the rare-earth region. Neutron scattering on $^{155,156,157,158,160}$Gd at 2.5 and 4.1~MeV was described using the deformed HFB densities with Gogny-D1S interaction within the coupled-channels framework  \cite{Bauge2000}. Later, even integral observables were shown to be sensitive to the same structure inputs: the characteristic oscillations of neutron total-cross-section differences among $^{182,184,186}$W between 5 and 100~MeV were reproduced up to about 200~MeV using Gogny-HFB neutron and proton densities within a Lane-consistent JLMB framework (spherical and coupled channels), underlining the role of realistic surface isovector densities \cite{Dietrich2003}. This approach has been applied to the impact of newly developed D3G3, D2 et DG Gogny interactions on HFB deformed densities through neutron scattering cross section calculations. Fig. \ref{fig:WcompGogny} compares calculations using densities with various Gogny interactions for elastic and inelastic to levels of the ground state rotational band and experimental data.

\subparagraph{\textbf{OMP from G-matrices}}
\label{sssec:omp-gmatrix}
Besides JLM folding model, optical potential where also built considering the solution of the Brueckner-Bethe-Goldstone equations : the $G$-matrix.
Two well-established implementations of the $G$-matrix  are often referred to by the names of the groups that developed them: the Melbourne and Santiago $G$-matrix approaches. Both yield complex, nonlocal, and energy-dependent optical potentials with no free adjustable parameters (aside from choices in the nuclear structure inputs and the underlying bar NN force). They have been successfully applied to nucleon elastic and inelastic scattering at intermediate energies (typically tens to a few hundreds of MeV). We summarize each in turn.

The Melbourne  group employed a $G$-matrix originally computed by the Melbourne group for nucleon–nucleon interactions in nuclear matter \cite{Amos2000}. This effective interaction is provided in coordinate space and has a finite range, being expressed as a sum of Yukawa form factors. It includes central, spin–orbit, and tensor components for all relevant spin–isospin channels. The $G$-matrix is density-dependent  and energy-dependent, and it is complex to account for the loss of flux into non-elastic channels. The full folding procedure produces a nonlocal optical potential $U(\mathbf{r},\mathbf{r'})$ for the nucleon–nucleus system as described in \cite{Amos2000}.

\begin{figure}[!htb]
  \centering
  \includegraphics[width=\linewidth]{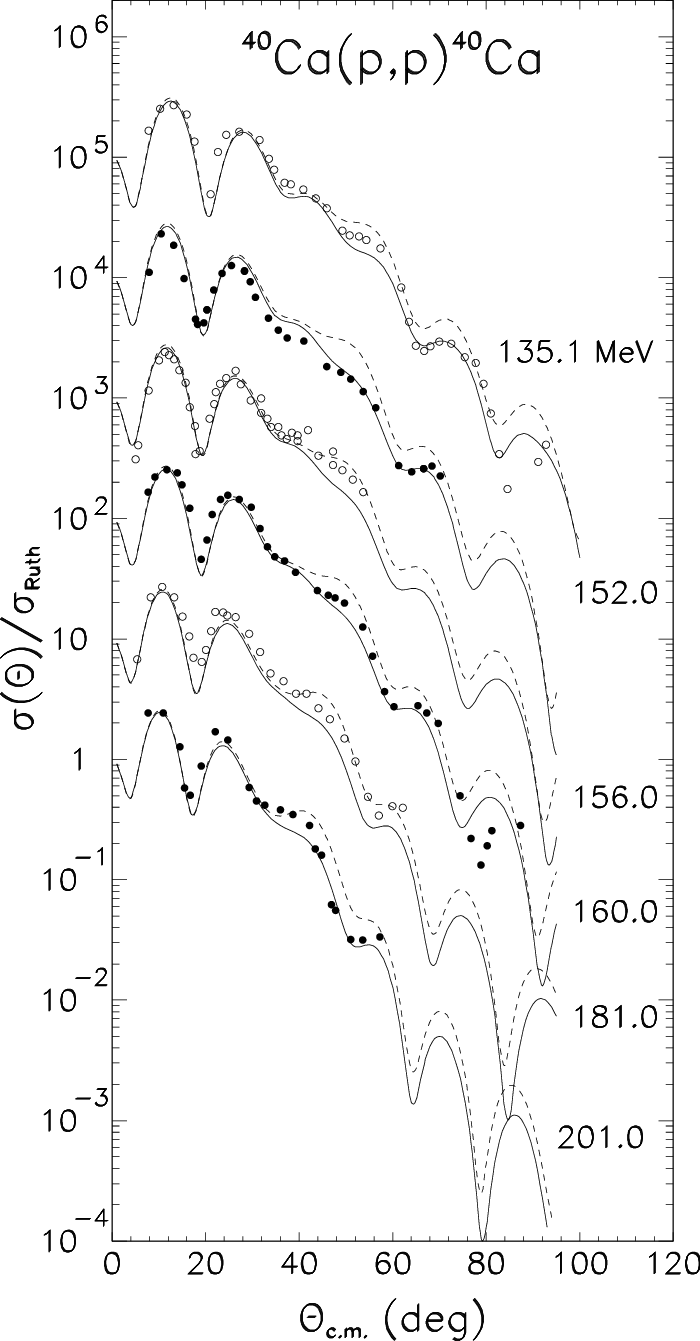}
  \caption{Differential \(p,p^{\prime}\) cross sections on \(^{40}\mathrm{Ca}\),
  normalized to the Rutherford cross section, for incident proton energies between
  80 and 200~MeV. The curves are predictions obtained with optical potentials constructed
  by folding the Melbourne \(g\)-matrix with either Hartree-Fock (HF) densities
  (dashed) or random-phase-approximation (RPA) densities (solid). Experimental
  measurements (symbols) are shown for comparison.}
  \label{fig:ca40_pp}
\end{figure}

The Melbourne $G$-matrix folding model has demonstrated very good predictive power at medium energies. Dupuis \textit{et al.} \cite{Dupuis2006a,Dupuis2006b} applied this approach to proton scattering on doubly magic nuclei ($^{16}$O, $^{40,48}$Ca, $^{208}$Pb) using Gogny D1S HF+RPA one-body density matrices, finding that the elastic scattering observables (angular distributions, analyzing powers) were well reproduced for incident energies from about $E\approx 60$~MeV up to $E\sim 200$~MeV without any parameter tweaking. Angular distributions were shown to be sensitive to long range RPA correlations. 
This behavior, illustrated for $^{208}\rm{Pb}(p,p^{\prime})$ in Ref. \cite{Dupuis2006a},
is further demonstrated for $^{40}\rm{Ca}$ in Fig. \ref{fig:ca40_pp}:  RPA correlations reduce cross section more and more when transfered momumentum increases, reflecting the change in the density profile.

At lower energies ($E \lesssim 50$~MeV), the same calculations showed increasing discrepancies with experiment. This indicates that additional reaction mechanisms (in particular, explicit couplings to collective excitations of the target) become important as the energy decreases.

A closely related line of development has been pursued by the Santiago group. The Santiago approach also begins with a nuclear matter $G$-matrix folding philosophy, but it introduced from the outset a more exact treatment of the folding integration \cite{Arellano2011a} and a careful separation of various components of the interaction \cite{Arellano2007a}. 

Nuclear‑matter G‑matrix approaches—such as the Melbourne and Santiago models—perform very well for medium‑energy nucleon scattering. At lower energies, however, they require additional corrections; a common choice is the semi‑microscopic JLM folding model. For a fully predictive low‑energy optical model potential, without any ad‑hoc renormalization, one must go beyond the nuclear‑matter framework. An alternative is to derive the optical potential directly in finite nuclei within the so called nuclear structure approach based on Green function formalism, as discussed in the next paragraph.

\subsubsection{Gogny-based Green's-function RPA optical potential 
}
\label{ssec:omp-ns}

At low incident energies ($E\!\lesssim\!30$--50\,MeV), explicit couplings between the projectile and the target's collective excitations must be included in the optical model. In the Green's-function RPA approach, the nucleon optical potential is built from the microscopic self-energy, with a static Hartree-Fock part and a dynamic polarization term $\Delta U(E)$ accounting for couplings to particle-hole (phonon) excitations. This program originates from Vinh~Mau and collaborators, who established the role of RPA correlations and second-order contributions to the self-energy in optical potentials \cite{Mau1970,Bouyssy1981}.

In its modern implementation, the finite-range, density-dependent Gogny D1S force is used consistently for the ground state, the RPA spectrum and transition densities, and the particle-hole interaction entering the self-energy \cite{Blanchon2015,Blanchon2015}. The resulting nonlocal, energy-dependent potential is used directly in the scattering equation (no ad hoc localization), with intermediate single-particle resonances in the continuum treated explicitly, a subtraction scheme to avoid double counting of the second-order $ph$ term, and damping widths assigned to RPA states to average escape and compound-nucleus spreading \cite{Blanchon2015}.

Within this framework, the Nuclear Structure Method, initially proposed by N. Vinh Mau and A. Bouyssy, reproduces neutron and proton elastic scattering on $^{40}$Ca up to about $30$\,MeV, including differential cross sections, analyzing powers, and reaction/total cross sections (see Figs.~1-3 in \cite{Blanchon2015}). Its extension to $^{48}$Ca confirms the robustness of the approach: neutron scattering is well described and proton angular distributions are reasonable, while the remaining lack of absorption for $p\!+\!^{48}$Ca clearly signals missing two-fold charge-exchange/Gamow-Teller couplings that can be approximately recovered with a Lane-consistent correction to the imaginary part \cite{Blanchon2017}.

\subsubsection{Other Gogny-based optical-potential studies}
\label{ssec:omp-other}
Beyond the work described above, other attempts to use Gogny forces in microscopic optical potentials have been made:
\begin{itemize}[label=$-$]
  \item  Using a self-consistent HF\,+\,RPA scheme with \emph{Gogny D1S}, low-energy ($\sim$5-10\,MeV) nucleon elastic scattering on doubly-magic targets (e.g.\ $^{16}$O, $^{208}$Pb) was described without phenomenological renormalization, validating Gogny-based surface absorption from RPA couplings \cite{Hao2015}; see also the broader neutron–nucleus systematics in \cite{Hao2018}.
  \item A parameter-free semi-microscopic OMP is obtained by computing the on-shell mass operator in symmetric nuclear matter with the Gogny force in Brueckner-Hartree--Fock (real part from first order, imaginary from second order), then mapping to finite nuclei via the local density approximation. The resulting Gogny-based OMP provides a good account of elastic scattering data across a broad mass/energy range \cite{Morana2021}; it has also been used as input for light-ion elastic scattering ($d,t,^{3}$He,$\alpha$) in a folding (Watanabe) framework \cite{Morana2023}.
\end{itemize}

\subsubsection{Direct inelastic scattering towards intrinsic excitations}
\label{ssec:inel}
This section reviews microscopic descriptions of inelastic nucleon--nucleus scattering that fold Gogny-based structure inputs into reaction models, namely transition densities from RPA/QRPA (see Secs. \ref{app:mpmh}, \ref{app:qrpa} and references therein), from collective form factors from the five-dimensional collective Hamiltonian (5DCH, see Sec. \ref{app:gcm} and references therein), and from the multiconfiguration mixing method (MPMH, see Sec. \ref{app:mpmh} and references therein). These frameworks describe intrinsic excitations in both spherical and deformed targets. As in the optical-potential section, we follow two complementary nuclear-matter routes: (i) a semi-microscopic folding with the Jeukenne-Lejeune-Mahaux (JLM) effective interaction, and (ii) a full-folding approach based on the Melbourne $G$-matrix. 

\paragraph{JLM folded with Gogny and RPA/QRPA densities}
\label{sssec:inel-jlm}
Historically, the first Gogny-driven inelastic applications were the $(n,n')$ and $(p,p')$ studies on $^{208}$Pb and excited levels of rational bands for axial nuclei within the JLM folding model using RPA-D1 transition densities by Lagrange et al., as noted earlier. Further work concerning excitation beyond rotational bands were performed within the 2000's and later on first considering 5DCH then newly implemented QRPA Gogny-force based nuclear structure methods.

Using Gogny-D1S constrained HFB + 5DCH to generate microscopic ground-state and $2^+_1$ transition densities, which are then folded (via JLMB) purely as a bridge to $(p,p')$ observables, inverse-kinematics data around 30–33 MeV/u across medium-mass chains were described. Along $^{32\text{–}40}$S, the Gogny-based collective densities reproduce absolute $2^+_1$ angular distributions with a single, energy-independent renormalization close to unity, and the surface-peaked neutron transitions indicate a developing neutron skin in $^{38,40}$S \cite{Marechal1999}. Transported unchanged to $^{36\text{–}44}$Ar at 33 MeV/u, the same Gogny-D1S 5DCH framework accounts for elastic and $2^+_1$ $(p,p')$ data with the same near-unity renormalization, yielding neutron-skin thicknesses of about 0.11–0.15 fm in $^{42,44}$Ar and quadrupole deformations consistent with Coulomb excitation \cite{Scheit2000}.

In the 2010's, at higher energy and away from stability, the same logic remains efficient when QRPA(Gogny) can be used directly for transition densities. For semi-magic tin at 150~MeV/u, JLM folding with local $n/p$ transition densities from HFB+QRPA (Gogny D1M) reproduced the shapes and absolute magnitudes of the $2^+_1$ and $3^-_1$ angular distributions for $^{112}$Sn and, without normalization, the first $(p,p')$ result on the neutron-deficient $^{104}$Sn. 
The analysis traces the trend to a strong decrease of the neutron matrix element \(M_n\) within a dominantly isoscalar quadrupole mode (\(M_n/M_p > 1\)) \cite{Corsi2015}. 
Here \(M_{n}\) (\(M_{p}\)) denotes the neutron (proton) matrix element of the multipole transition operator between the ground state and the excited state, \(M_{n(p)}=\langle J_f \\
Vert \sum_{i\in n(p)} r_i^{\lambda}Y_{\lambda}(\hat{r}_i)\Vert 0^+_{\mathrm{gs}}\rangle\) with \(\lambda=2\) in the present case, so that the ratio \(M_n/M_p\) measures the relative neutron and proton contributions to the excitation and reduces to \(N/Z\) for a purely isoscalar (homogeneous) transition.

\begin{figure}[!htb]
  \centering
  \includegraphics[width=.35\textwidth]{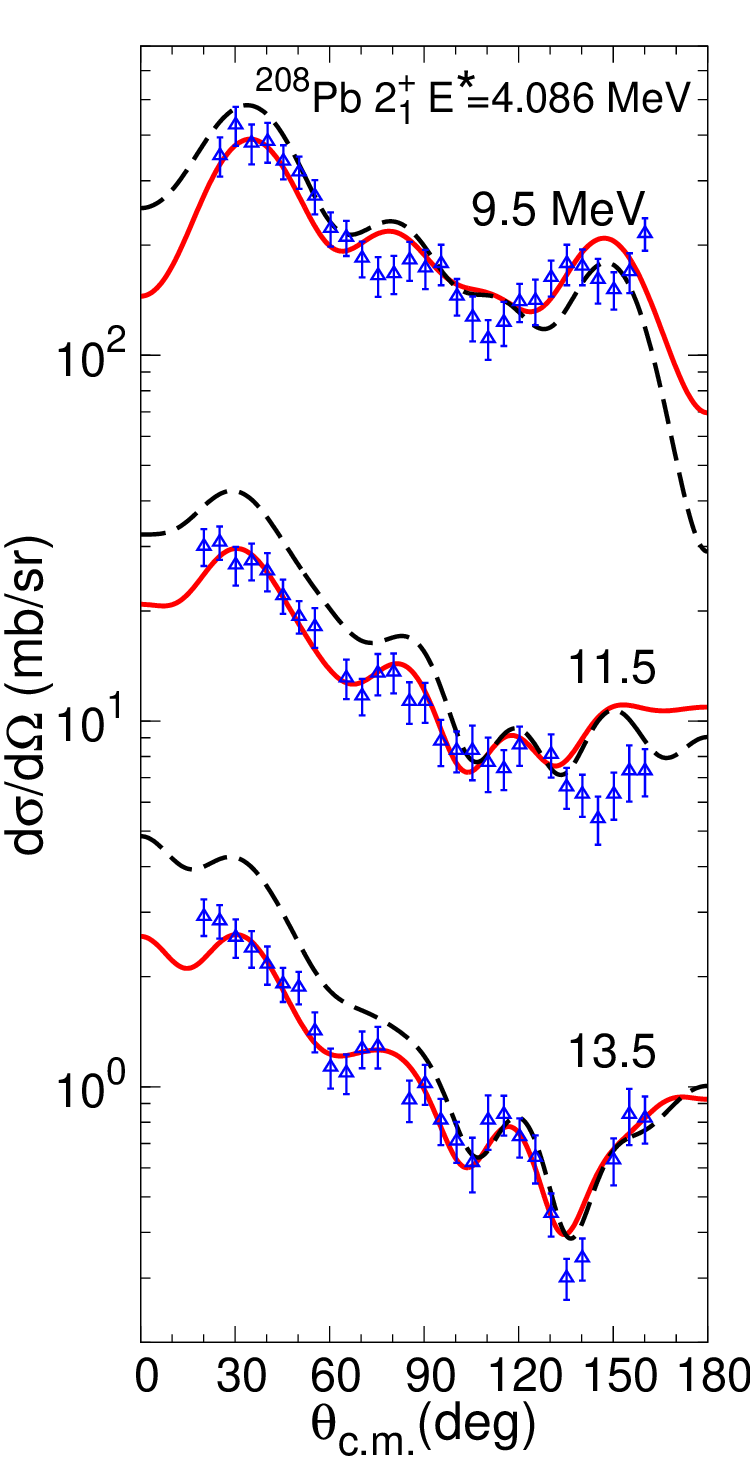}
  \caption{Differential cross sections for neutron inelastic scattering off 
  $^{208}$Pb to the yrast $2^+$ states. Incident energies are indicated 
  above each curve. Cross sections are offset by successive factors of 10.
  Symbols denote experimental values. JLM calculations with HFB/QRPA matter and transition densities 
  are plotted as red solid curves, and calculations performed within the collective 
  model as black dashed curves. Figure taken from 
  Ref. \cite{Dupuis2019}.}
  \label{fig:nnPb2082+1}
\end{figure}

An extensive analysis of neutron elastic and inelastic scattering data, including new measurements between 1 and 30~MeV, was performed within the JLMB+QRPA(D1S) framework for $^{206,207,208}$Pb and $^{209}$Bi \cite{Dupuis2019}. In this work, elastic scattering is described with the JLMB optical potential, while direct inelastic scattering to discrete states is treated within a DWBA framework using QRPA transition densities based on the Gogny D1S force. Complementary $(p,p')$ calculations for the same set of excitations further demonstrate the predictive power of this approach, and reveal possible inconsistencies in the experimental data for the $2^+_1$ excitation in $^{206}$Pb.  Overall, elastic and direct inelastic channels to discrete states are described consistently using a single HFB/QRPA structure input combined with the same JLMB interaction, yielding an accuracy comparable to global phenomenological optical potentials. An example is shown in Fig. \ref{fig:nnPb2082+1} for the $(n,n')$ excitation of the $2^+_1$ state in $^{208}$Pb.

Near the $N\!=\!50$ shell closure: for $(p,p')$ on $^{72,74}$Ni and $^{76,80}$Zn at \(\sim 235\)~MeV/u, an extension of JLM beyond 200~MeV folded with QRPA(Gogny D1M) densities indicates neutron-dominant quadrupole strength in Ni (consistent with a preserved \(Z\!=\!28\) shell) and proton-dominant character in Zn (consistent with a robust \(N\!=\!50\) gap toward $^{78}$Ni) \cite{Cortes2018}. Finally, a recent program with \((n,n'\gamma)\) at SPIRAL2/NFS adopts QRPA(Gogny D1M) \(1^-\) transition densities around \(9\)~MeV combined with JLM in DWBA to provide a common, parameter-light baseline for comparing neutron and proton probes of pygmy E1 surface strength, illustrated on \(^{140}\)Ce \cite{Vandebrouck2024}.

\begin{figure}[!htb]
    \centering
    \includegraphics[width=0.8\linewidth]{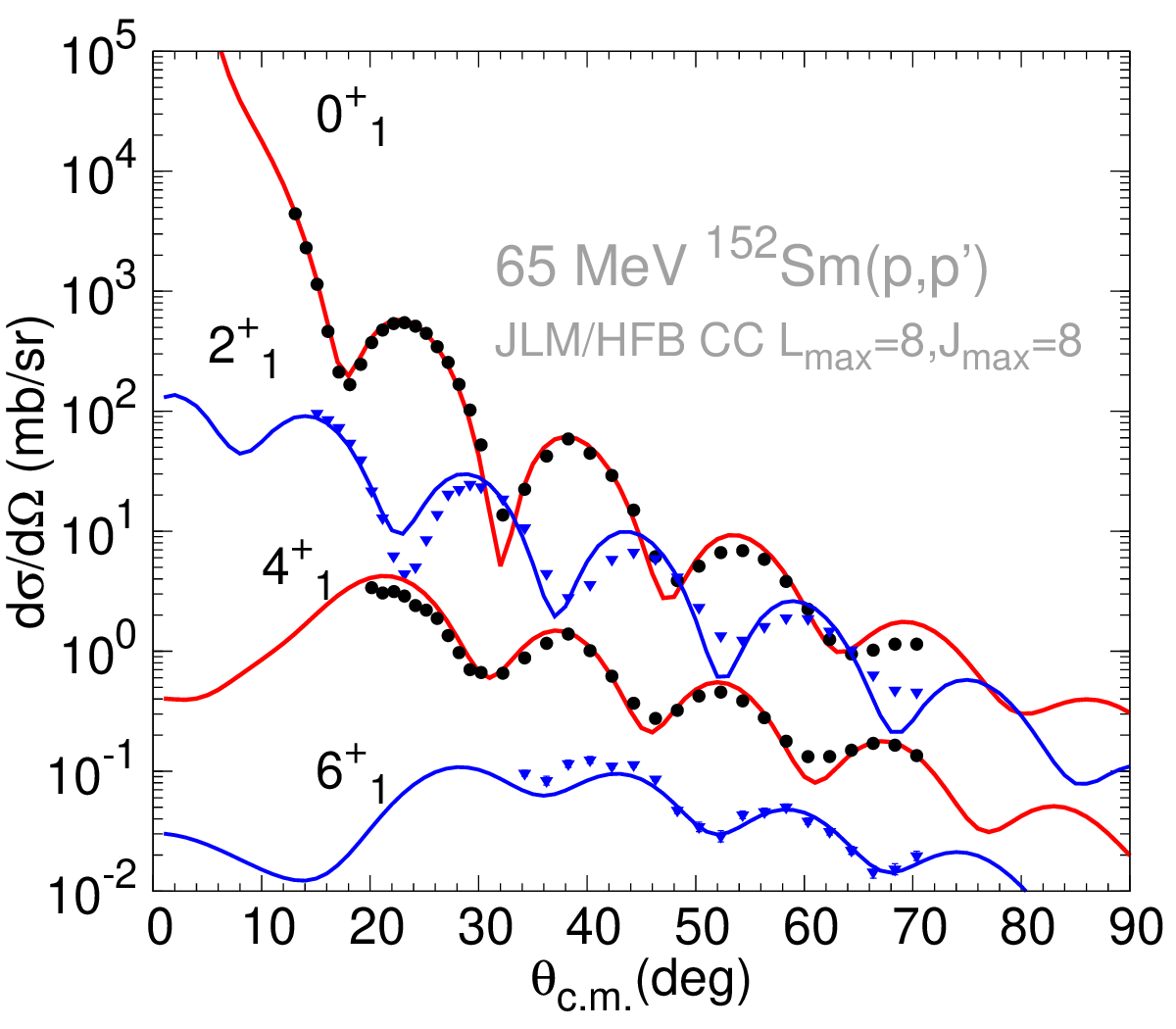}
    \includegraphics[width=0.8\linewidth]{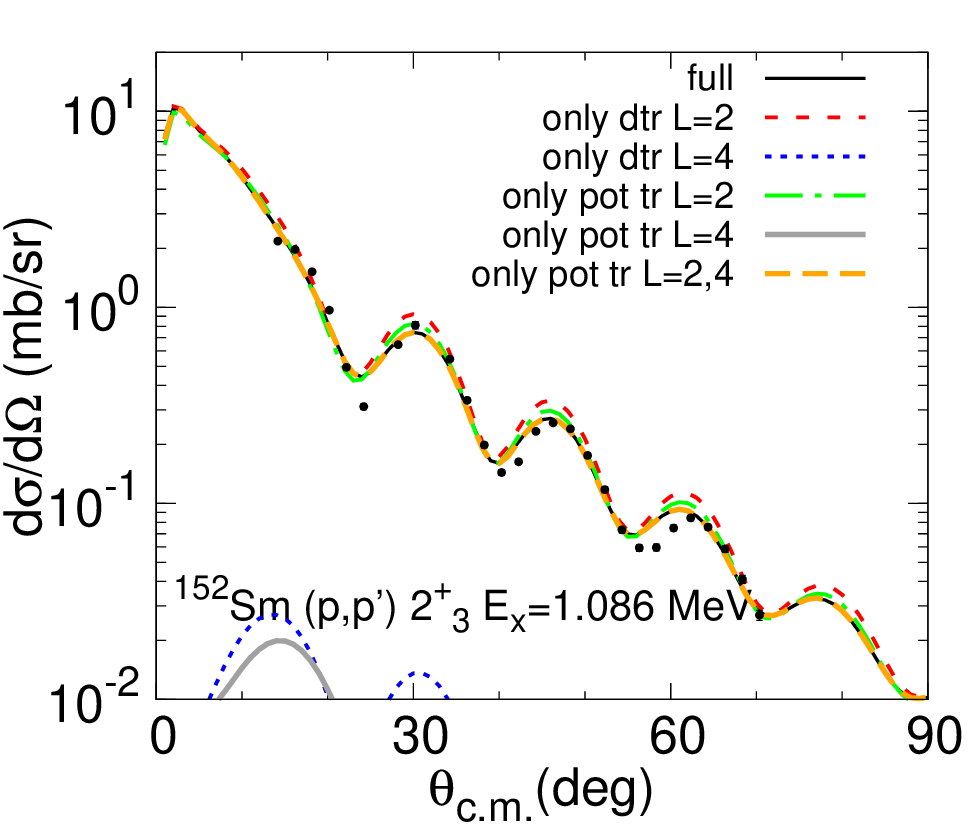}
    \includegraphics[width=0.8\linewidth]{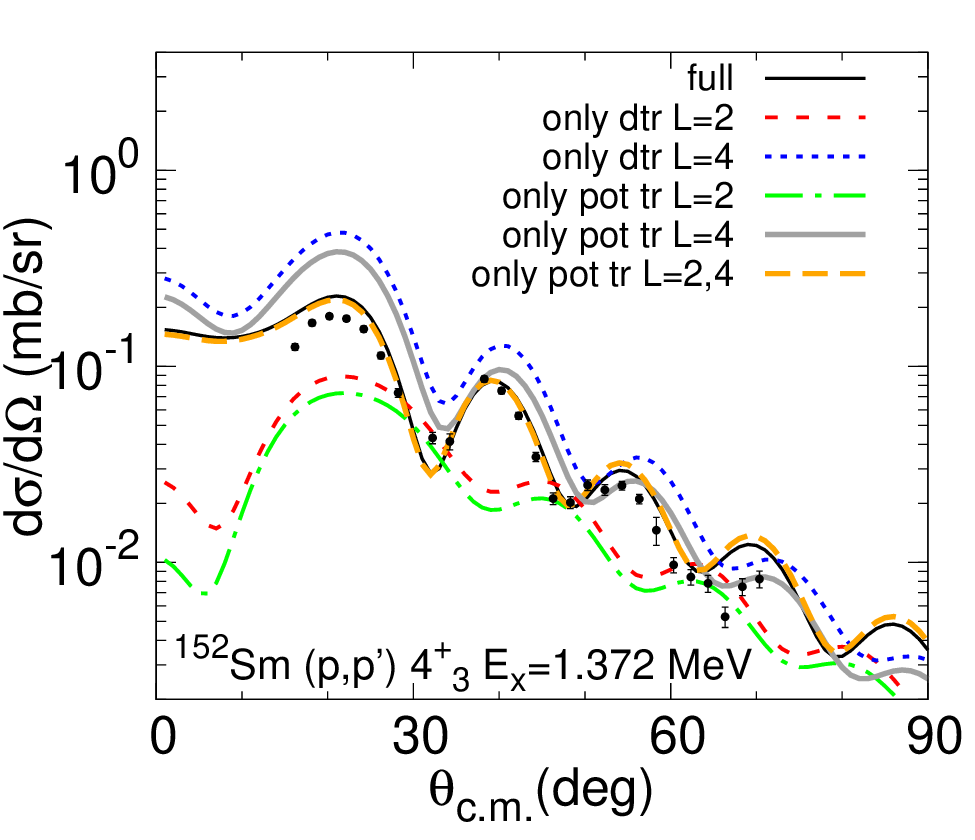}
    \caption{Cross sections for 65 MeV proton elastic (indicated as level $0^+_1$) and inelastic scattering to discrete levels belonging to the GS (top panel) and the $2^+$ (middle panel) and $4^+$ (bottom panel) of the first gamma-band in $^{152}$Sm.}
    \label{fig:152Smpinl}
\end{figure}

For axially deformed targets, the coupled-channels description has been extended to include
\textit{interband} couplings between the ground-state rotational band and bands built on one-phonon
QRPA intrinsic excitations computed with Gogny D1S. In practice, JLM optical and transition
potentials are obtained by folding the Lane-consistent JLM interaction with (i) HFB/D1S
ground-state densities for intraband couplings and (ii) QRPA/D1S transition densities for
interband couplings; laboratory states are generated by angular-momentum projection and
the CC equations are solved with a rotational approximation for the couplings.
A benchmark on $^{152}$Sm$(p,p')$ at 65~MeV shows that elastic and inelastic angular
distributions to the $2^+_1,4^+_1,6^+_1,8^+_1$ levels of the ground band and to the
$\gamma$-band members $2^+_3$ and $4^+_3$ are reproduced in absolute scale \cite{Dupuis2024}, as illustrated on Fig. \ref{fig:152Smpinl} for (p,p') on $^{152}$Sm to levels belonging to the gamma-excited band. A detailed
analysis demonstrates that both $L{=}2$ and $L{=}4$ components of the QRPA transition
densities \emph{and} of the resulting coupling potentials must be retained to match the data,
reflecting constructive/destructive interferences specific to interband CC. The underlying
structure inputs are validated by the proximity between QRPA/D1S and measured reduced
transition strengths (e.g.\ $B(E2\uparrow)$ and $B(E4\uparrow)$) for the $\gamma$ band.
The same framework, applied to neutron scattering on actinides, allows one to treat, within
a single microscopic scheme, low-energy discrete $(n,n')$ channels (via CC among rotational
bands) and the first-step direct pre-equilibrium emission as a coherent sum over QRPA
one-phonon excitations, with quantified implications for inclusive $(n,xn)$ spectra, as discussed in the forthcoming
pre-equilibirum section.

Overall, the JLM route shows three robust qualities: (i) a \emph{parameter-light} absolute scale across isotopic chains once densities are fixed by Gogny-based structure; (ii) genuine sensitivity to the surface isovector profile (neutron skins) and to the neutron–proton partition \(M_n/M_p\) of low-lying multipoles; and (iii) and straigthforward extension to coupled channels treatments based on deformed HFB, deformed QRPA, and collective dynamics (5DCH/Bohr), which is essential for deformed nuclei.

\paragraph{Melbourne \texorpdfstring{$G$}{G}-matrix folded transition densities (50-200~MeV)}
\label{sssec:inel-melbourne}
In the energy window \(\sim 50\text{–}200\)~MeV, the Melbourne $G$-matrix offers a parameter-light, spin-sensitive description of inelastic scattering.

When folded with Gogny-RPA transition densities and solved in DWBA for $^{208}$Pb(p,p'), the same set-up describes in a unified way the absolute cross sections for low-lying \(3^-\), \(5^-\), \(2^+\), \(4^+\) and selected high-spin states \cite{Dupuis2008a} of both natural and unnatural parity, while also reproducing analyzing powers that probe the spin-orbit interaction. 
\begin{figure}[!htb]
  \centering
  \includegraphics[width=0.55\textwidth]{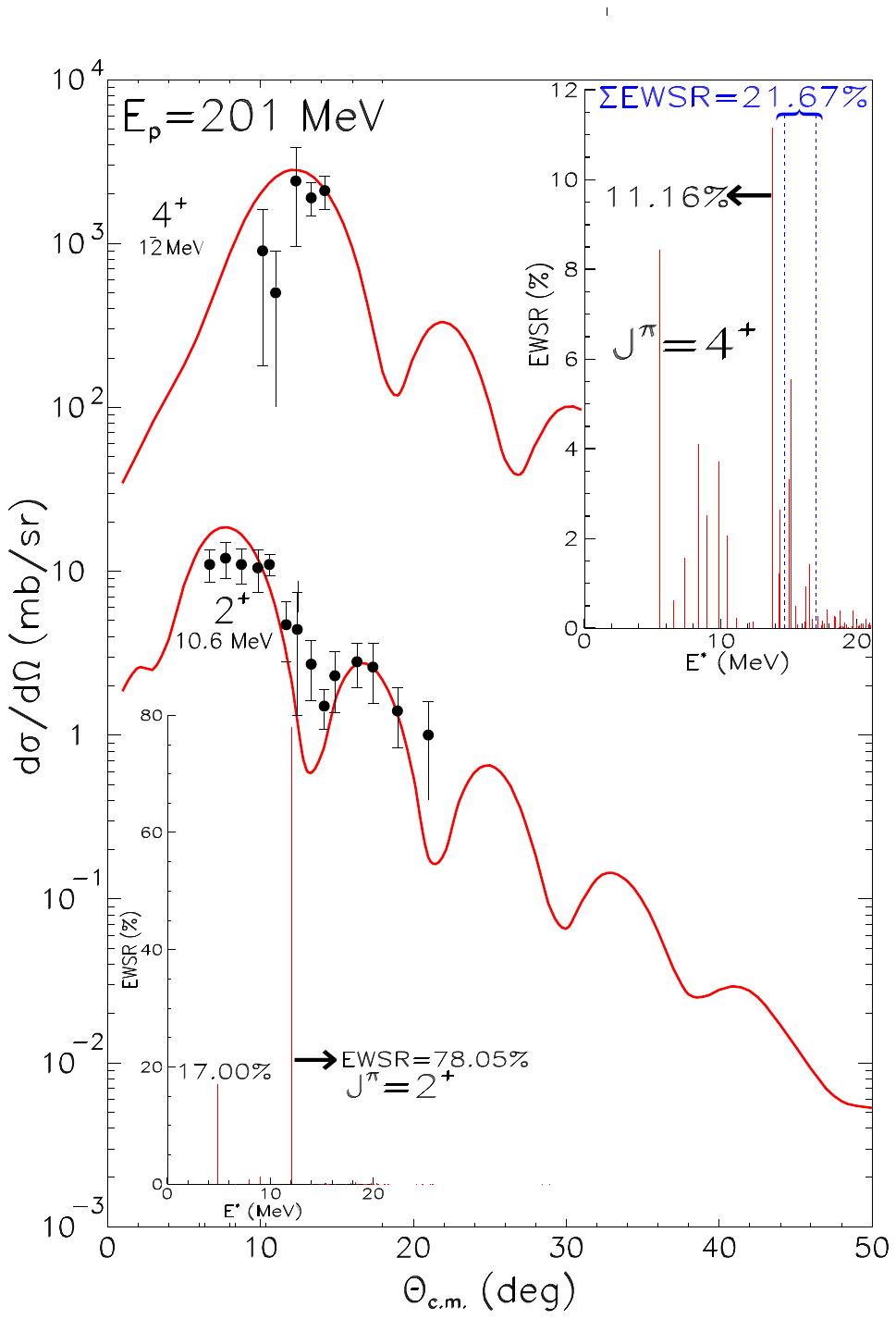}
  \caption{Proton inelastic scattering on $^{208}$Pb at 201~MeV leading to the collective $2^+$ and $4^+$ giant–resonance excitations. The panels inside the figure display the fraction of the EWSR associated with each mode, as used in the reaction analysis. The scattering cross sections were computed in DWBA with the Melbourne $G$-matrix, while the transition strengths originate from RPA calculations based on the finite-range Gogny D1S interaction. Figure adapted from Refs. \cite{Dupuis2006b,Dupuis2024}.}
  \label{fig:Pb208_GR}
\end{figure}
The framework extends to giant resonances with cross sections correlating with the fraction of the Energy-Weighted Sum Rule (EWSR) carried by the RPA states as illustrated on Fig. \ref{fig:Pb208_GR}.
It also applies to light doubly-magic nuclei where the expected limits for very low-lying \(0^+/2^+\) appear, and to the first inverse-kinematics applications such as \(p+^{56}\)Ni at 101~MeV and predictions for \(^{24}\)O \cite{Dupuis2006b}. 

We have also studied inelastic proton scattering within the multiparticle-multihole (MPMH) configuration mixing framework, folding the Melbourne interacdtion with Gogny D1S transition densities for sd-shell nuclei. The $0^+_1 \rightarrow 2^+_1$ excitations of $^{24}$Mg, $^{28}$Si, and $^{32}$S \cite{Robin2017,Dupuis2017}. were calculated and compared to $(p,p')$ data. The shape of the calculated angular distributions agrees well with experiment, but the cross sections remain too low in magnitude. This behaviour is also observed in $(e,e')$ Coulomb form factors, where the diffraction pattern is correctly reproduced but the absolute strength is clearly underestimated. These results indicate that the MPMH wave functions capture the correct spatial structure of the transition densities, but miss collective components. The origin of this deficiency is known; it lies in the use of a restricted configuration space, limited to 0$\hbar\omega$ sd-shell excitations. To overcome this,  the model space will be extended by including 2$\hbar\omega$ configurations with carefully chosen truncations, possibly based on excitation energy or pairing content \cite{LeBloas2014,Pillet2017a}. Such improvements are expected to increase the transition strength and enhance the agreement with experimental observables.

\subsubsection{Microscopic pre-equilibrium with Gogny inputs: from Melbourne to JLMB}
\label{sec:preeq}

Here, we discuss microscopic modeling of one-step and two-step direct processes  based on the nuclear structure ingredients stemming from the (Q)RPA approach for both spherical and deformed targets in various energy regimes.

Pre-equilibrium emission lies between direct and compound mechanisms and becomes important in calculating particle emission for incident energies beyond a few MeV (depending of the target).
Usually, pre-equilibrium model are tested comparing measurements to predicted doubly differential cross-sections (in emission angles and in outgoing energies) of emitted particles such as inclusive (n,$x$n) (inclusive means that it accounts for any number of particle emission: (n,n')+(n,2n)+(n,3n) ...), (p,x$p$), (p,x$n$) ...
We will discuss its important role in predictive partial cross sections for
(n,xn$\gamma$), focusing on (n,n'$\gamma$).

\paragraph{Strength of Gogny structure inputs.}
A key point of this program is the quality of the structure calculations. With the finite-range Gogny force, the (Q)RPA method gives a unified description of nuclear
excitations: from the lowest-lying collective states up to giant resonances, for all multipolarities, both isoscalar and isovector. This wide coverage ensures that the same interaction provides the doorways to the continuum that drive the pre-equilibrium stage, without the need for extra phenomenology.

\begin{figure}[!htb]
  \centering
   \includegraphics[width=0.45\textwidth]{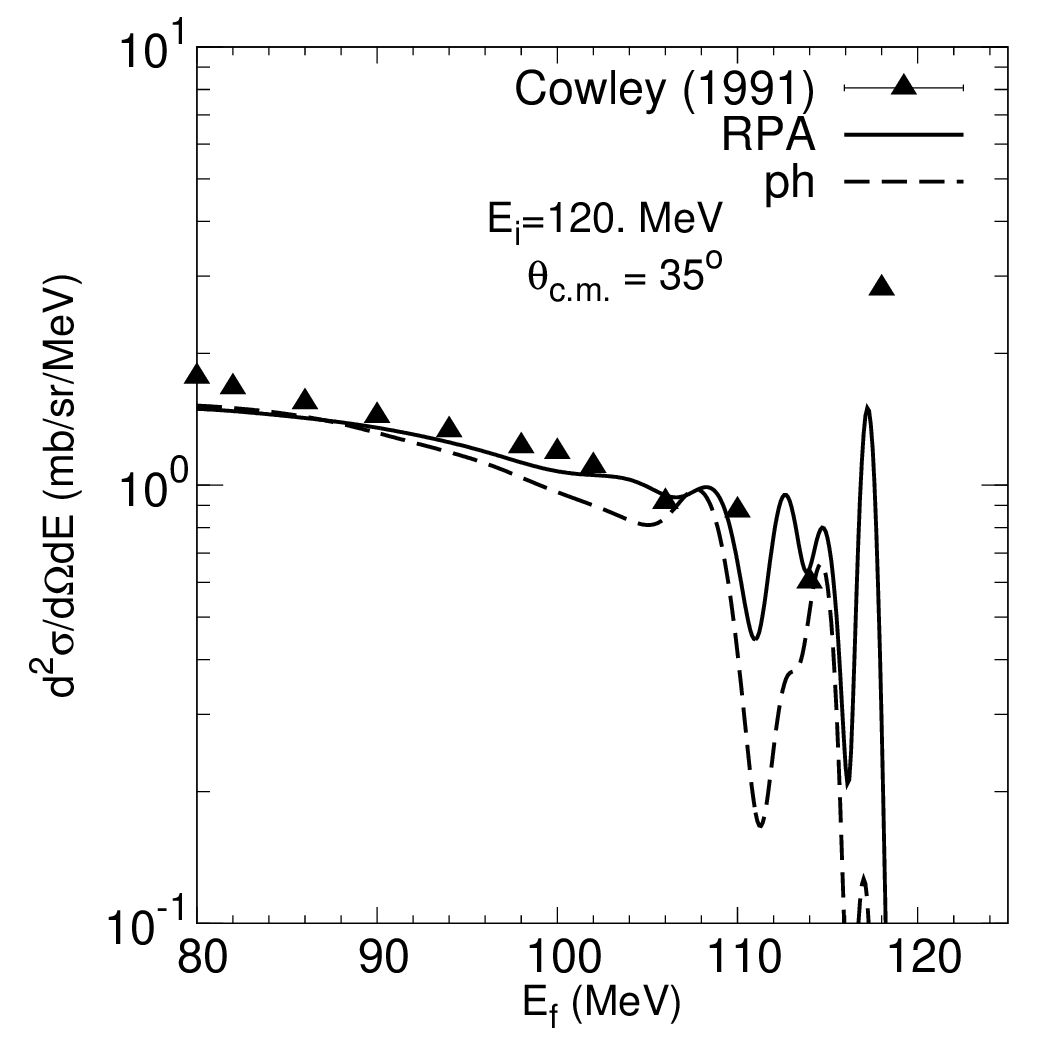}
  \includegraphics[width=0.45\textwidth]{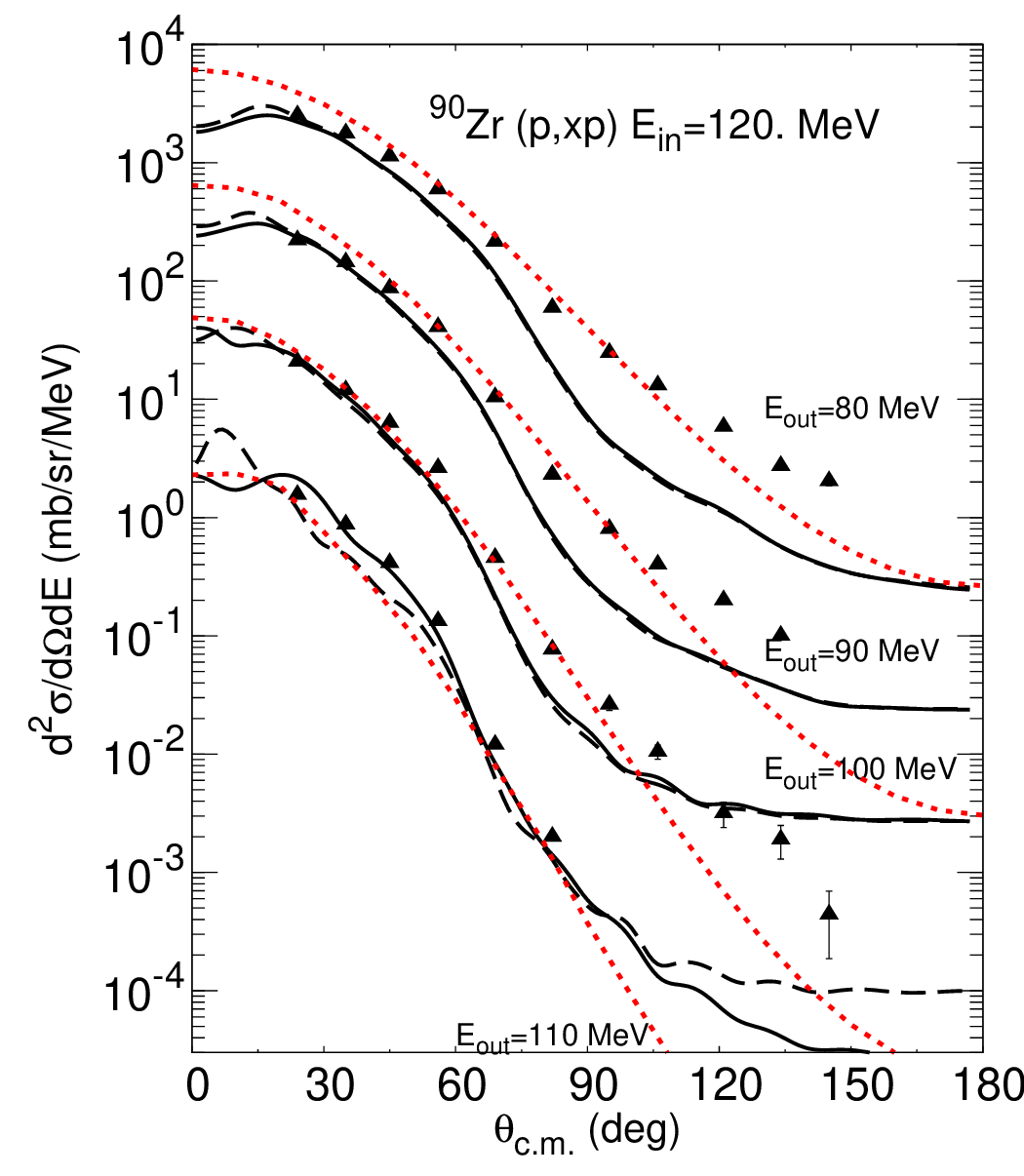}
  \caption{Proton inelastic scattering on $^{90}$Zr at 120~MeV. 
  The upper panel shows the double differential cross sections as a function 
  of the outgoing proton energy at $\theta_{\mathrm{c.m.}} = 35^\circ$. 
  The lower panel presents angular distributions for several outgoing energies 
  (indicated in each subpanel); successive curves are shifted by factors of~10 
  for clarity. Calculations including RPA excitations are plotted as solid lines, 
  while particle--hole contributions are shown as dashed lines. The red dashed 
  curves correspond to the sum of exciton components together with direct 
  contributions to low-lying states and giant resonances obtained with the 
  \textsc{TALYS} code. Experimental data are plotted as symbols. 
  Figure adapted from Ref. \cite{Dupuis2017}.}
  \label{fig:Zr90_120MeV}
\end{figure}

\paragraph{First microscopic pre-equilibrium with Gogny (Melbourne folding).}
The first microscopic description of direct pre-equilibrium emission based on Gogny inputs relied on RPA(D1S) transition densities folded with the Melbourne $G$-matrix in a full-folding DWBA framework, as already discussed for direct inelastic scattering in Sec. \ref{ssec:inel}. This approach was extended in \cite{Dupuis2006b,Dupuis2017} to describe the first-order term of the multistep direct (MSD) mechanism for (p,$x$p) emission on $^{90}$Zr and $^{208}$Pb at incident energies between 60 and 200~MeV. As shown in Fig. \ref{fig:Zr90_120MeV}, this first-order MSD contribution accounts well for the measured strength up to an energy transfer of about 20~MeV and for forward emission angles, where direct processes dominate. The increasing deviations from experiment at higher excitation energies and more backward angles reflect the missing higher-order MSD components, which were not included in this calculation.

In Ref.\cite{Dupuis2011}, the \emph{first} (one-step) (n,$x$n)
emission for $^{90}$Zr and $^{208}$Pb was calculated at incident energies below
30~MeV. In that case, the Gogny-RPA transition densities were folded with a
density-dependent M3Y interaction to generate transition potentials, and the optical potential was derived from the Koning-Delaroche parametrization, which is well-suited for this energy range. Even if the method needed further approximation (a different model for optical than for transition potential), this showed that the method could be used in
the energy range relevant for applications, bridging the gap with the higher-energy Melbourne calculations.

\begin{figure}[!htb]
    \centering
    \includegraphics[width=\linewidth]{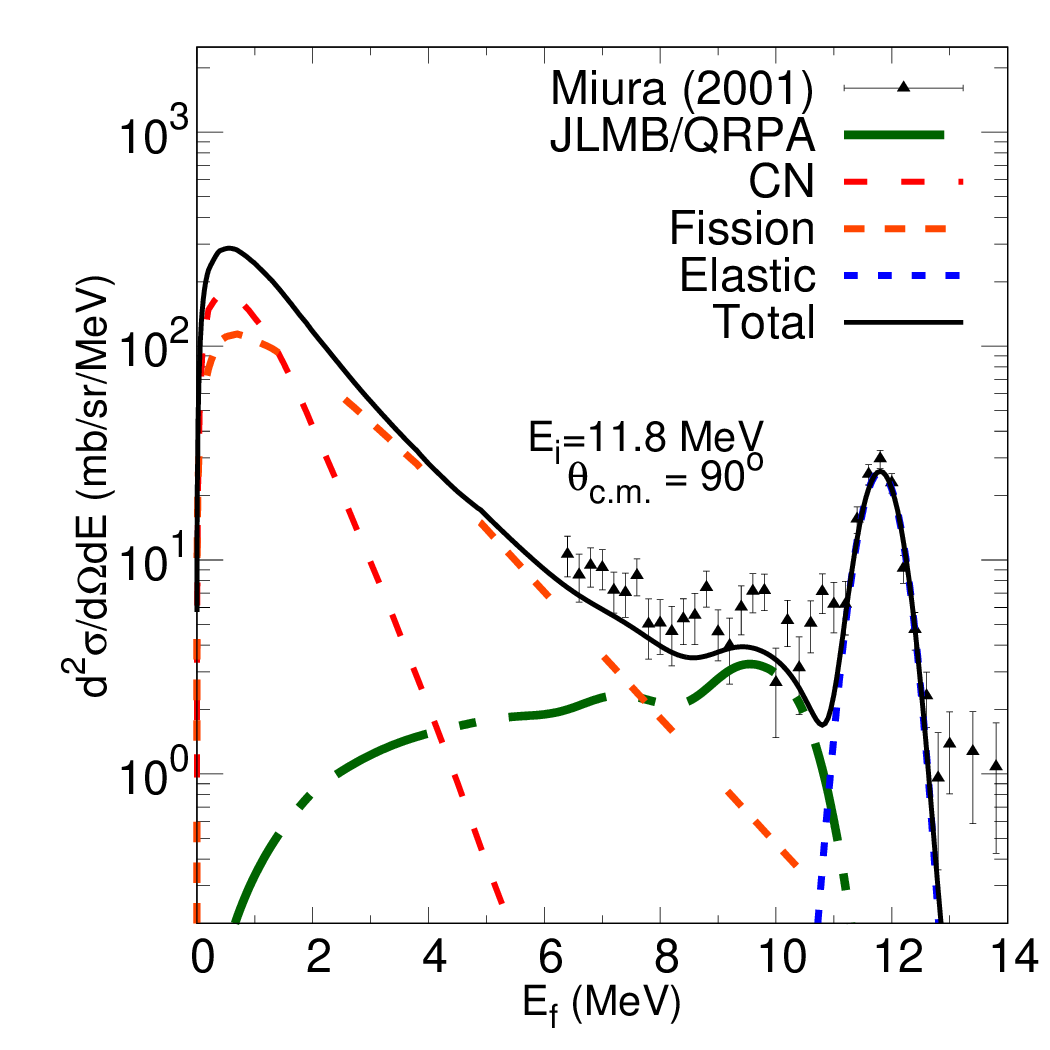}
    \caption{Inclusive neutron emission for $^{238}$U$(n,xn)$ at $E_n=11.8$~MeV and 
    $\theta_{\rm lab}=90^\circ$. Contributions from the different emission mechanisms 
    are compared with the measured spectrum. The JLM/QRPA component includes the full 
    set of intrinsic one-phonon excitations predicted by QRPA and energetically 
    accessible at this incident energy.}
    \label{fig:nxnU238}
\end{figure}

\paragraph{JLM folding model below $\sim$30~MeV}
For incident energies below $\sim$~30~MeV, the range most relevant to energy-production applications, one can rely on the local JLMB optical potentials folded with HF/HFB ground-state densities (and their deformed extensions). The domain of validity of these potentials extends from 1~keV to about 200~MeV (see Sec.~\ref{par:omp-jlm}) and thus fully covers this low-energy region.
This made systematic studies of
neutrons on actinides possible, where strong deformation and low-lying collective states play a major role. Using deformed HFB densities to generate diagonal and transition JLMB potentials, and adding QRPA(D1S) one-phonon states to extend the
basis (see Sec. \ref{sssec:inel-jlm}), JLMB+CC calculations reproduced the magnitude and shape of (n,xn) spectra
in $^{238}$U around 9\,MeV \cite{Dupuis2017}. 
By providing microscopic transition strengths, the approach made it unnecessary to introduce the ad-hoc “pseudo-states’’ that phenomenological studies had used to mimic missing collective excitations \cite{Young2007,Capote2018,Romain2016}. Those studies typically combine a collective model for direct reactions with an exciton model for pre-equilibrium emission, and they inevitably miss the direct excitation of low-lying collective states (2–6 MeV). Such states are absent from both ingredients: they are not generated in the exciton model, and—when not experimentally established—they are also not included in the collective model used for the direct-reaction component. 
An illustration of the prediction of this microscopic model for neutron emission is depicted in Fig. \ref{fig:nxnU238} for the 11.8~MeV(n,$x$n) on $^{238}$U reaction. The contribution from inelastic scattering to QRPA intrinsic one-phonon excitations is compared to the measured inclusive neutron emission, along with contributions from other emission mechanisms (elastic, compound nucleus evaporation, and evaporation from fragments after fission).

\paragraph{From (n,xn) to (n,xn$\gamma$): spin-parity distributions and surrogate.}
The same microscopic ingredients (QRPA level schemes and transition densities
folded into direct and pre-equilibrium amplitudes) also provide the spin–parity
distributions of the residual nucleus. These distributions are essential to
describe $\gamma$ cascades, for example, in (n,xn$\gamma$) reactions and for the
surrogate-reaction method. On the $\gamma$ side, precise (n,n$'\gamma$) data on
actinides (e.g.\ from GRAPhEME at GELINA) have been compared with models that
combine pre-equilibrium and discrete structure, giving constraints on both the
reaction mechanism and the nuclear input \cite{Kerveno2021}. 

\begin{figure}[!htb]
    \centering
    \includegraphics[width=\linewidth]{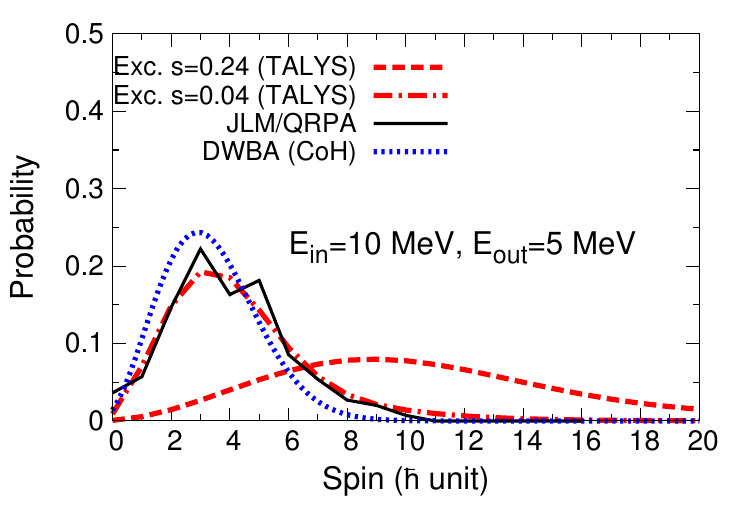}
    \includegraphics[width=\linewidth]{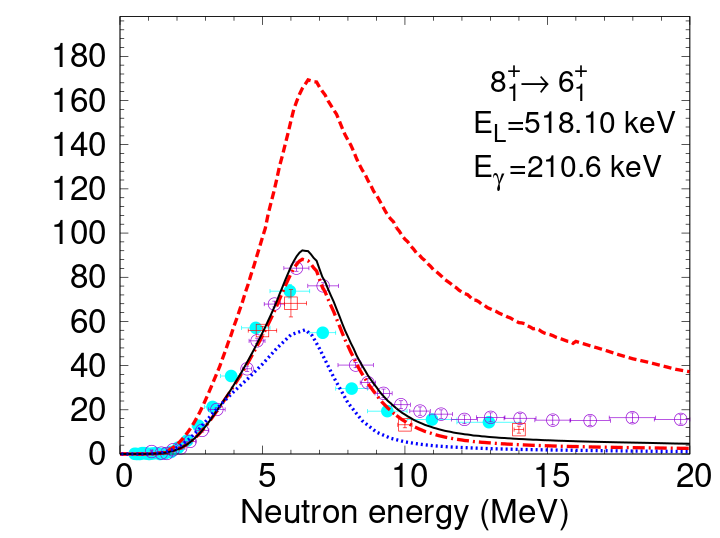}

    \caption{Impact of the microscopic JLMB/QRPA approach on spin distributions 
    and $\gamma$-decay observables in $^{238}$U. Top panel: spin distribution at 
    $E_x = 5$~MeV for incident neutrons of $E_n = 10$~MeV, and average spin as a 
    function of the excitation energy for two incident energies. The microscopic 
    JLMB/QRPA framework modifies the spin distribution and reduces the population 
    of high-spin states compared with standard exciton prescriptions. 
    Bottom panel: $^{238}$U$(n,n'\gamma)$ cross section for the 
    $8^+\!\rightarrow 6^+$ transition within the ground-state rotational band. 
    Calculations based on four pre-equilibrium models are compared: JLMB/QRPA 
    (full black line), the exciton model with the standard spin cut-off parameter 
    $s=0.24$ (dashed red line), the exciton model with $s=0.04$, a value guided 
    by the JLMB/QRPA calculation (dotted--dashed red line), and DWBA based on 
    particle--hole excitations (short dashed blue line). Figure adapted from 
    Kerveno (2021), where further details can be found.}
    \label{fig:U238_spin_nng}
\end{figure}

Fig. \ref{fig:U238_spin_nng} illustrates how a microscopic treatment modifies both the spin distribution and the associated $\gamma$-decay observables.
The top panel shows that the JLMB/QRPA framework produces a markedly different spin distribution at $E_n=10$ MeV and reshapes the evolution of the average spin with excitation energy. This microscopic constraint reduces the population of high-spin states relative to standard exciton prescriptions.
The bottom panel demonstrates the consequence on measurable quantities: the calculated $^{238}$U$(n,n'\gamma)$ cross section for the $8^+\rightarrow 6^+$ transition is better reproduced when using the spin distribution guided by JLMB/QRPA, in agreement with $\gamma$-decay data from such states. 

On the surrogate side, the first consistent extraction of neutron-induced fission and radiative-capture cross sections from the same surrogate measurement was achieved by combining measured decay probabilities with calculated $J^\pi$ distributions \cite{PerezSanchez2020}.

More recently, the first surrogate experiment in a storage ring employing inverse kinematics was devoted to determining the radiative capture cross section of $^{208}\mathrm{Pb}(n,\gamma)$ \cite{Sguazzin2025a,Sguazzin2025b} via a $^{208}\mathrm{Pb}(p,p')$ measurement, using microscopically calculated spin distributions to infer the capture cross section.
 
\begin{figure}
  \centering
  \includegraphics[width=0.43\textwidth]{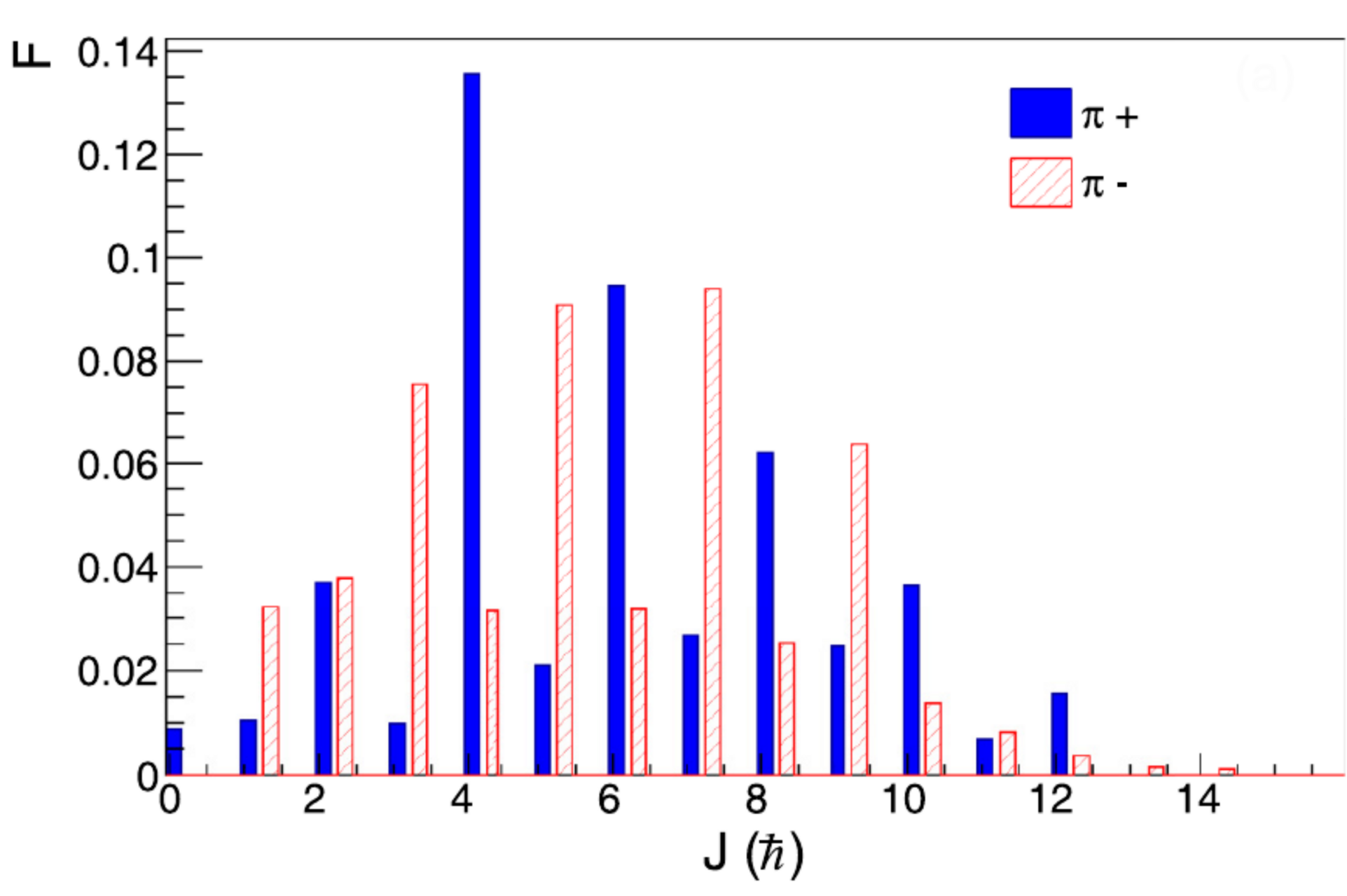}
  \includegraphics[width=0.45\textwidth]{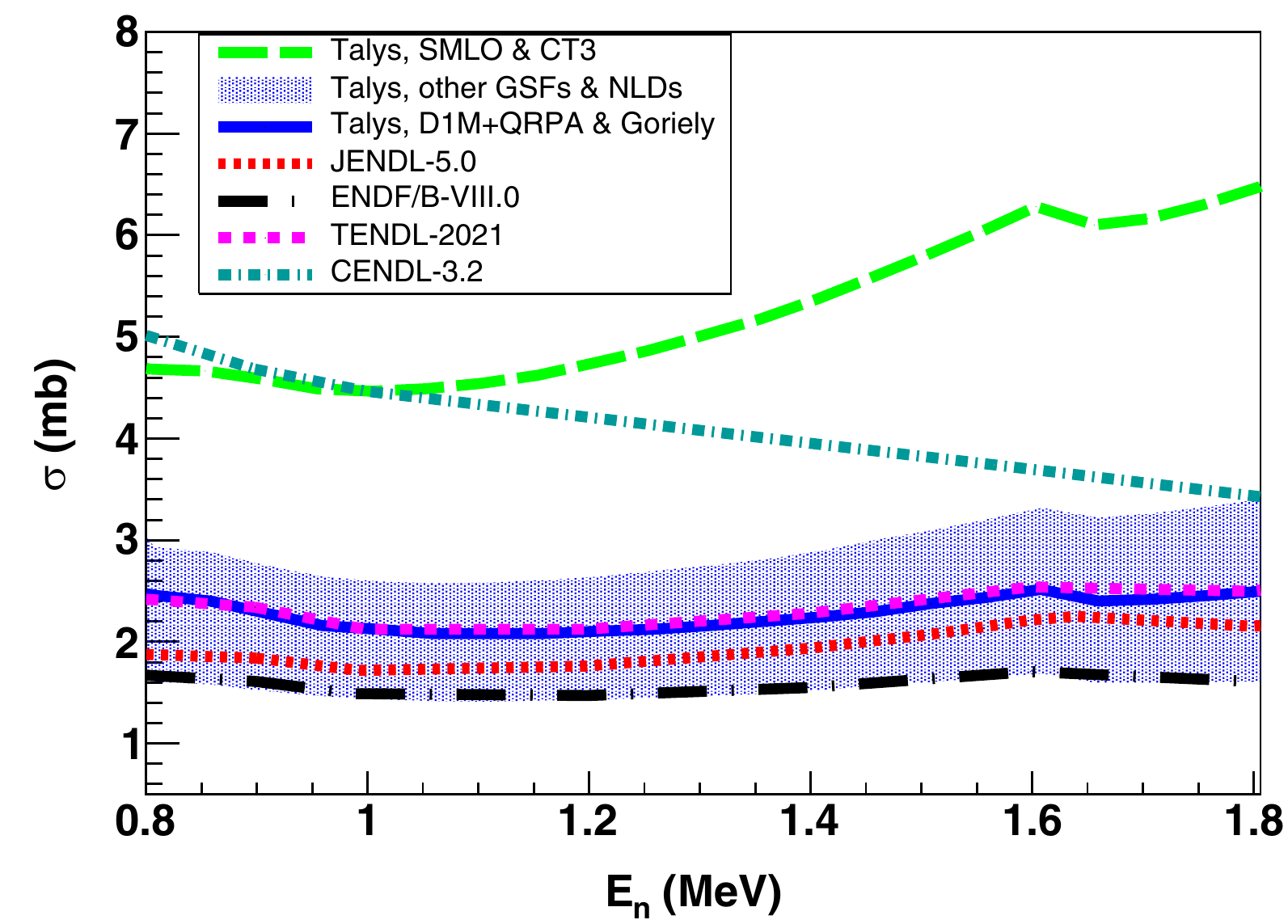}
  \caption{Top: Calculated spin-parity distribution of states populated in the 
  $^{208}$Pb$(p,p')$ reaction at $E^{*}=8$ MeV. 
  Bottom: Calculated $^{207}$Pb$(n,\gamma)$ cross section as a function of the 
  neutron energy. Both panels are taken from 
  Refs. \cite{Sguazzin2025b,Sguazzin2025a}.}
  \label{fig:Pb208surrogate}
\end{figure}

In Fig. \ref{fig:Pb208surrogate} we illustrate how Gogny-based ingredients enter this surrogate-reaction analysis:
the upper panel shows the compound nucleus spin–parity distribution populated in the $^{208}$Pb$(p,p')$ reaction at $E^*=8$~MeV, calculated with the JLMB/RPA and Melbourne/RPA models with the Gogny force. The lower panel displays the corresponding $^{207}$Pb$(n,\gamma)$ cross section obtained from Hauser–Feshbach calculations constrained by the measured neutron-emission probability, the microscopic spin distributions and QRPA $\gamma$-ray strength functions.

A recent surrogate study by Thapa \textit{et al.} \cite{Thapa2025} leveraged the $\mathrm{^{90}Zr}(p,p'\gamma)$ reaction to indirectly determine the neutron capture cross section of the short-lived $\mathrm{^{89}Zr}$ nucleus. The JLMB/QRPA approach provide a realistic compound-nucleus spin-parity population distribution completed by a the population stemming for a two-step process with intermediate deuteron (p$\rightarrow$d$\rightarrow$p'). The predicted population was subsequently used in a Hauser-Feshbach decay analysis where a Bayesian procedure adjusted the nuclear level density and $\gamma$-ray strength function parameters to fit the observed (p,p'$\gamma$) coincidence data. This approach enabled a successful extraction of the $^{89}$Zr(n,$\gamma$) cross section from the surrogate measurement in close agreement with evaluated data libraries (JEFF-3.3, JENDL-5.0).

\paragraph{Two-step direct process}

\begin{figure}[!htb]
  \centering
  \includegraphics[width=0.45\textwidth]{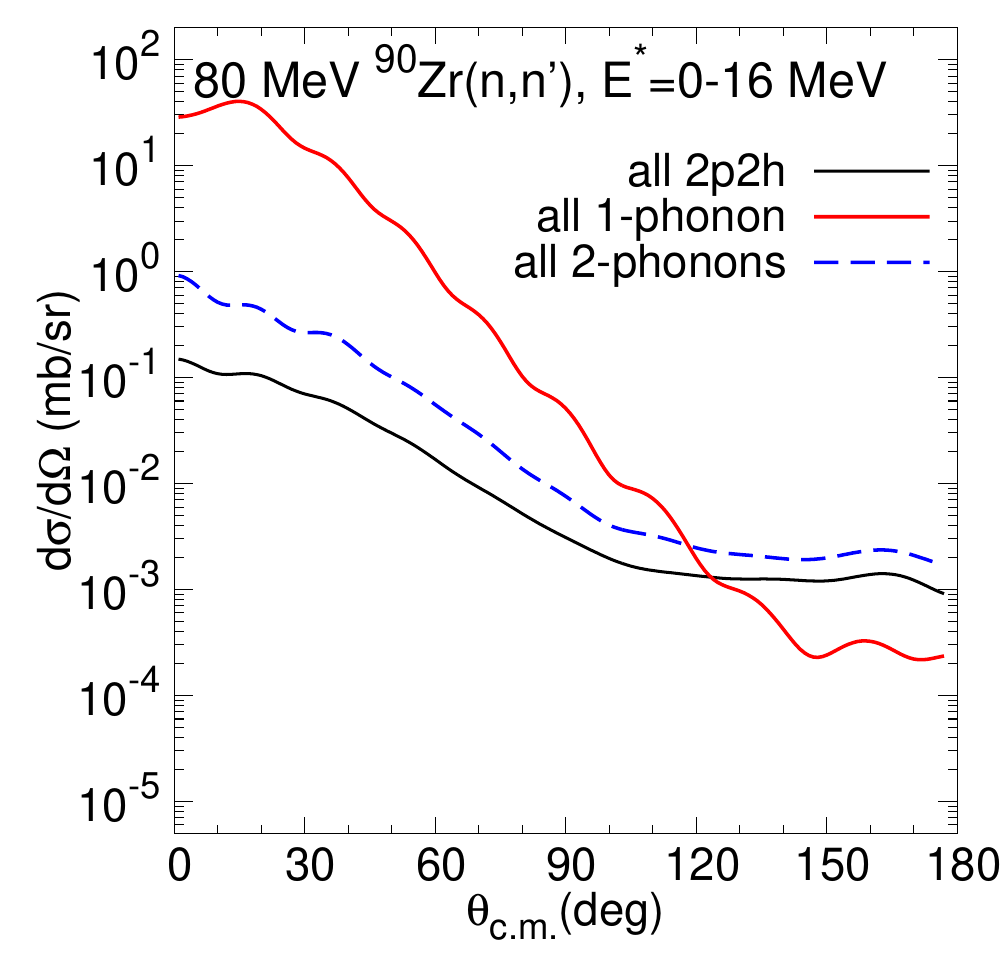}
  \caption{Comparison of the angular distributions for the $^{90}$Zr(n,n') cross section at 80 MeV obtained using RPA calculations for one-phonon (full red curve) and two-phonon (blue dashed curve) states, and using an uncorrelated 2p-2h (full black curve) state. The excitation energies considered range from 0 to 16 MeV. Figure adapted from \cite{Nasri2019}.}
  \label{fig:two-step}
\end{figure}

The extension from one-step to two-step direct mechanisms represents a natural evolution in microscopic pre-equilibrium modeling. Early 
exploratory work by M. Dupuis \cite{Dupuis2006b} already demonstrated the feasibility of second-order amplitudes, work revisited recentrly in 
the PhD thesis of A. Nasri \cite{Nasri2018}. In these studies, all building blocks entering the two-step formalism are obtained from Gogny-
based structure calculations: single-particle and particle-hole configurations stem from HF solutions with the D1S interaction, while 
collective excitations entering the intermediate channels are taken from Gogny-RPA, including both one-phonon and two-phonon states. 
The resulting two-step amplitudes incorporate coherent sums of various paths that lead to the same 2p-2h final state: each of these paths is 
characterized by an intermediate 1p-1h excitation. The comparison between cross sections calculated from the two-step amplitudes stemming 
from  a coherent or an incoherent sum allows to measure the interference effect. It was shown that collective Gogny-RPA 2-phonon excitation 
could play a dominant role in the second step, enhancing the amplitude and modifying angular distributions as illustrated in Fig. 
\ref{fig:two-step}.
One point that still needs clarification is the distinction between the genuine enhancement arising from the collective content of the RPA 
phonon excitations, and the spurious enhancement generated by Pauli-violation effects inherent to the quasi-boson approximation, which may 
become significant when several phonons are combined.

\subsubsection{Microscopic nuclear level densities}
\label{ssec:ld}
Nuclear level densities (NLDs) are a key quantity underpinning statistical models of nuclear reactions (Sec. \ref{sec:intro:cn}). Historically, NLDs were described by empirical formulas (e.g., Fermi gas models), but modern approaches seek a microscopic description that natively includes shell structure, pairing and collective effects. Starting in the 1970s, self-consistent mean-field theory enabled NLD calculations to naturally include shell effects. In recent decades, the Gogny effective interaction has become a prominent choice for such calculations, providing a reliable mean-field with pairing correlations well-suited for microscopic NLD studies. 

In the late 1990s, efforts began to develop fully microscopic NLD models. In 1997, S.~Hilaire introduced a combinatorial approach based on Hartree-Fock-Bogoliubov (HFB) theory with the D1S Gogny interaction \cite{Hilaire1997}. This method explicitly counts many-body excitations (particle-hole or quasiparticle configurations) using single-particle level schemes, pairing correlations, and deformations provided by HFB calculations. 

In 2001, a Gogny-based combinatorial nuclear level density (NLD) model for 65 even-even nuclei \cite{Hilaire2001} achieved good agreement with measured level counts and neutron-resonance spacings. The NLDs were built by explicitly treating spin, parity and pairing, and incorporating collective rotational effects (moment of inertia from the Inglis–Belyaev HFB approximation with an ad-hoc renormalisation), and vibrational enhancements (relying on known low-lying collective level energies). In doing so, it naturally produced realistic spin and parity distributions of levels, unlike the Gaussian spin distribution and equal-parity assumptions of conventional phenomenological models. At the same time, Goriely et al. developed a similar combinatorial NLD approach with alternative mean-field inputs for astrophysical applications \cite{Goriely1996,Demetriou2001}, these studies demonstrated that microscopic models can match the predictive power of global formulas while improving the reliability of extrapolations.

By 2006, the HFB+combinatorial method, applied in that case with a Skyrme interaction, had been extended within a deformed framework to a global model covering $\sim$8500 nuclei \cite{Hilaire2006}.

\begin{figure*}[!htb]
  \centering
  \includegraphics[width=\linewidth]{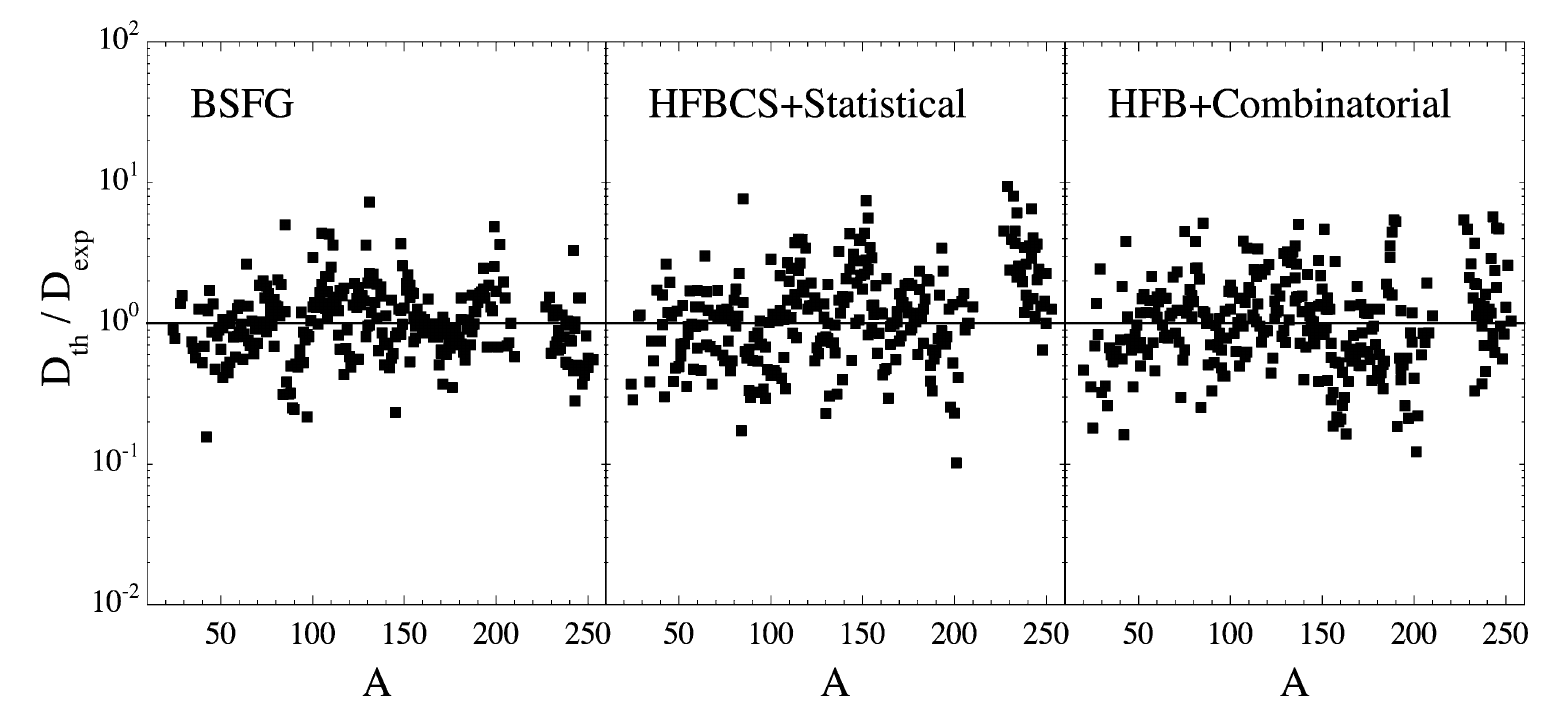}
  \caption{
  Ratio of theoretical to experimental $s$-wave neutron resonance spacings,
  $D_{\mathrm{th}}/D_{\mathrm{exp}}$, as a function of mass number $A$ for the
  295 nuclei of the RIPL-2 compilation. The three panels show, from left to
  right, the results obtained with (i) a phenomenological back-shifted
  Fermi-gas (BSFG) systematics, (ii) a microscopic HFBCS plus statistical
  model, and (iii) the fully microscopic HFB plus combinatorial level-density
  prescription \cite{Hilaire2006}. 
  Figure adapted from ref. \cite{Hilaire2006}.
  }
  \label{fig:ld_hfb_combi}
\end{figure*}

The resulting NLD library reproduced $s$-wave neutron resonance spacings at the neutron separation energy with accuracy comparable to the best fitted models, confirming the competitiveness of the microscopic approach (see Fig. \ref{fig:ld_hfb_combi}). Vibrational collective effects were still added phenomenologically in this model, while rotational enhancements came directly from HFB moments of inertia. In 2008, an improved microscopic model refined the treatment of vibrational and rotational contributions \cite{Goriely2008a}. This yielded energy-, spin-, and parity-dependent NLDs that better matched experimental discrete level data and resonance spacings across the mass table. 

\begin{figure}[!htb]
    \centering
    \includegraphics[width=0.45\textwidth]{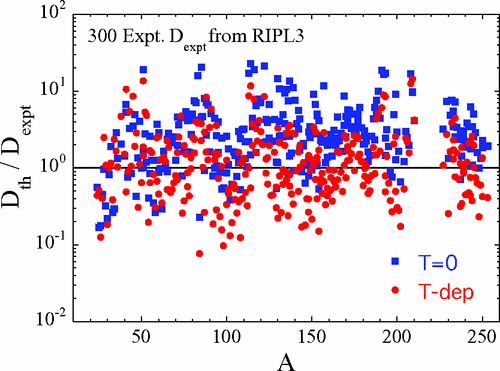}
    \caption{Ratio of HFB plus combinatorial ($D_\mathrm{th}$) to experimental ($D_\mathrm{expt}$) $s$-wave neutron resonance spacings \cite{Capote2009}. The temperature-dependent treatment is shown with circles and the $T=0$ case with squares. Figure adapted from \cite{Hilaire2012}.}
    \label{fig:ld-tdep-ratio}
\end{figure}

A major advancement came in 2009 with the Gogny D1M interaction \cite{Goriely2009a}. Building on D1M, a next-generation HFB+combinatorial model was developed to more rigorously incorporate collective excitations and the evolution of nuclear structure with excitation energy \cite{Hilaire2012}. It used vibrational frequencies and moments of inertia from HFB calculations directly, making vibrational collective enhancements self-consistent rather than phenomenological. For deformed nuclei, temperature-dependent HFB calculations at increasing excitation energies accounted for the gradual shape change from deformation to sphericity, capturing the damping of shell and deformation effects at high energies. As illustrated in Fig. \ref{fig:ld-tdep-ratio}, it significantly improved agreement with experimental $s$-wave neutron resonance spacings compared to a $T=0$ calculation.

In the past decade, Gogny-based microscopic NLDs have been widely applied in both basic and applied nuclear science. For example, in nuclear astrophysics they provide NLD inputs for $r$- and $s$-process reaction network calculations, improving reaction rate predictions for thousands of unstable nuclei. In nuclear data evaluation, these models have been integrated into reference libraries (e.g., IAEA’s RIPL-3 \cite{Capote2009}) and reaction codes like TALYS \cite{Koning2017}, enhancing cross-section predictions especially for nuclei lacking experimental data. 

\begin{figure}[!htb]
    \centering
    \includegraphics[width=0.45\textwidth]{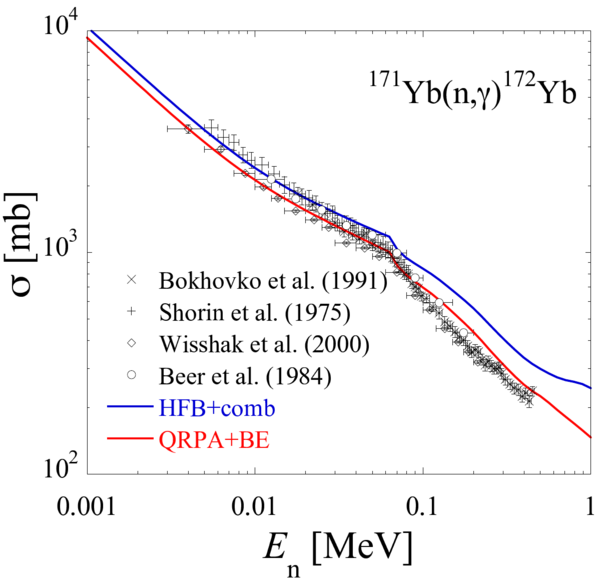}
    \caption{Comparison between experimental (black symbols) and theoretical $^{171}$Yb(n,$\gamma$) cross sections using either the QRPA+BE (red curve) or HFB+comb (black curve) NLDs. Figure adapted from \cite{Hilaire2023}}
    \label{fig:QRPA-BE}
\end{figure}

Meanwhile, research on Gogny-based NLD models has continued. A 2022 study \cite{Goriely2022} compared several global NLD models with experimental observables, finding that the Gogny-HFB combinatorial approach generally performs well, though some deficiencies remain in certain regions, motivating further improvements. One such innovation is the QRPA plus boson expansion (QRPA+BE) method proposed in 2023 \cite{Hilaire2023} that goes beyond the independent quasiparticle picture. Here, collective vibrational excitations from QRPA on the HFB ground state are treated as bosons and added to the combinatorial counting, including multi-phonon states. This is especially important for near-spherical nuclei with strong vibrational collectivity. This approach predicts a shift of the spin distribution to lower values, in line with recent (n,n'$\gamma$) observations \cite{Kerveno2021} (see Sec. \ref{sec:preeq}). As illustrated in Fig. \ref{fig:QRPA-BE}, the QRPA+BE approach improves agreement with experimental cross sections, as shown for the $^{171}$Yb(n,$\gamma$)$^{172}$Yb reaction.

These ongoing developments, together with the foundation built since 1997, ensure that Gogny-based microscopic NLD models remain at the forefront of nuclear data research.

\subsubsection{Gamma strength functions}\label{gsf}

In the Hauser–Feshbach statistical model (Sec. \ref{sec:intro:cn}), the $\gamma$-ray strength function is a key ingredient controlling the competition between $\gamma$ emission and particle emission during compound nucleus decay, thereby determining the average radiative width $\langle\Gamma_\gamma\rangle$ and $(n,\gamma)$ cross section. While E1 transitions typically dominate, M1 contributions—including the scissors mode in deformed systems and the low-energy “upbend”—can become decisive near or below $S_n$ when the level density and transmission coefficients favor low-$E_\gamma$ transitions. A microscopic, globally applicable strength function is thus essential to reduce model dependence and improve predictive power for nuclear data and astrophysical reaction rates.

\paragraph{Electric Dipole (E1) Strength}
\label{sssec:E1}

Fully self-consistent Gogny-HFB+QRPA calculations of nuclear $\gamma$-ray strength functions have been performed using the finite-range Gogny D1M interaction in a large axially-deformed basis (typically up to $N_{\rm sh}\sim13$ shells for heavy nuclei) that includes both $K^\pi=0^-$ and $1^-$ components of the dipole response. Using up to $N_{\rm sh}\approx 13$–17 shells ensures the GDR centroid is essentially converged; any remaining $\sim$2~MeV overestimation (e.g.\ in $^{238}$U) reflects genuine QRPA limitations (missing complex configurations) rather than basis truncation.

\begin{figure}[!htb]
  \centering
  \includegraphics[width=0.48\textwidth]{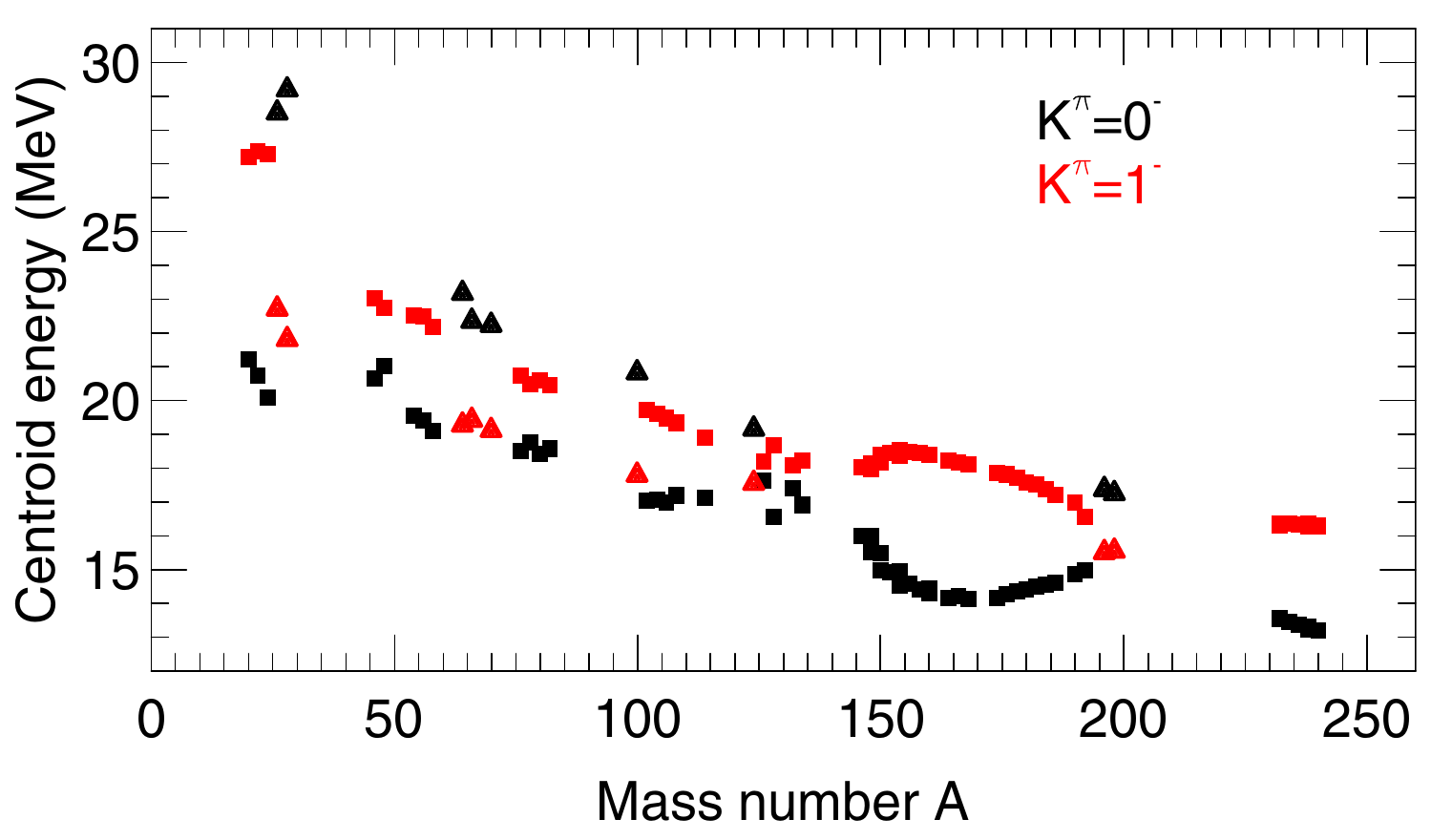}
  \caption{Centroid energies of the $K^\pi = 0^-$ and $1^-$ components
    for  deformed nuclei, obtained from
    QRPA calculations based on the Gogny D1M interaction. Squares
    correspond to prolate nuclei and triangles to oblate ones.
    Figure taken from Ref. \cite{Martini2016a}.}
  \label{fig:e1-split}
\end{figure}

The Gogny-D1M force enables a systematic, accurate description of dipole strength across the nuclear chart \citep{Goriely2009a}, providing a microscopic foundation to predict E1 strength functions far from stability and naturally including deformation effects (e.g., the split GDR peaks in deformed nuclei) that earlier spherical QRPA studies had to treat phenomenologically. However, standard QRPA (limited to 1p–1h excitations) neglects couplings to more complex configurations (2p–2h, vibrational phonons) that fragment the GDR and shift its centroid downward. Consequently, early Gogny-QRPA calculations \cite{Peru2011} produced GDR peaks about 2~MeV too high and with too narrow a width. To better reproduce the data, phenomenological corrections were applied: (i) a downward energy shift of $\sim$2~MeV to simulate the missing coupling effects, and (ii) a smearing of each QRPA spike with an energy-dependent width to reproduce the observed GDR damping \citep{Hilaire2017}. In deformed nuclei, the calculated $K=0$ and $K=1$ components already yield a double-humped GDR shape, so no additional splitting is required (see Fig. \ref{fig:e1-split}). By 2016, global Gogny-QRPA E1 strength calculations with these corrections were completed for even–even nuclei \citep{Martini2016a,Hilaire2017}. They reproduced the main GDR systematics: e.g., the mass-dependent trend of GDR centroid energies from light to heavy nuclei (including characteristically lower centroids in the rare-earth region) is well captured, and after a $\sim$2~MeV shift the GDR peak energies agree with experiment to within 0.5–1~MeV. Moreover, the Gogny-HFB+QRPA E1 strength functions, after these calibrations, can reproduce measured photoabsorption cross sections. For example, the shifted Gogny-QRPA GDR centroids align well with evaluated RIPL-3 data \citep{Hilaire2017,Capote2009}, illustrating the improvement over phenomenological models (RIPL-3, compiled in 2009, relied on generalized Lorentzian fits). A complete tabulation of Gogny-QRPA E1 strength functions for $\sim$2000 even–even nuclei has been made available for reaction codes \citep{Goriely2018a}. These calculations also predict a pygmy dipole resonance (PDR) around 5–9~MeV in neutron-rich nuclei.

\paragraph{Magnetic Dipole (M1) Strength: Spin-Flip and Scissors Modes}
\label{sssec:M1}

The nuclear M1 response has two main components: a high-energy spin-flip resonance (around 7–9~MeV) and, in well-deformed nuclei, a low-energy orbital scissors mode (2–4~MeV) \citep{Peru2021}. The spin-flip mode involves isovector spin excitations, while the scissors mode corresponds to a coherent oscillation of proton and neutron shapes against each other. In spherical nuclei, most M1 strength is concentrated in the $\sim$8~MeV spin-flip peak, whereas deformed nuclei also exhibit a scissors peak at lower energy. Gogny-QRPA calculations naturally include both types of M1 excitations.

Using the same D1M Gogny force and HFB mean fields as for E1, systematic M1 strength functions have been computed for even–even nuclei from $Z\sim20$ up to the actinides. The raw QRPA spin-flip M1 strength was found to be too weak and at too high an energy compared to systematics. As with E1, global adjustments were made: a downward shift of about 2~MeV (bringing the spin-flip peak to $\sim$7–8~MeV) and a strength scaling of $\sim$2 to reproduce known M1 cross sections, effectively compensating for missing 2p–2h (quenching) effects \cite{Kopecky2017}. Meanwhile, the scissors mode arises naturally in deformed nuclei, and the calculated $K=0$ and $K=1$ components recover the expected deformation splitting. Experiments support these predictions: NRF and Oslo measurements confirm the calculated scissors strength and M1/E1 balance near $S_n$, and analyses of average resonance capture data show that the Gogny strengths reproduce both the total dipole strength and the $M1/E1$ ratio within uncertainties \citep{Renstrm2018,Sieja2018,Kopecky2017}. (For odd-$A$ and odd–odd nuclei, an even–even approximation is currently used pending fully blocked QRPA calculations.)

\paragraph{Low-Energy Enhancement (\enquote{Upbend})}
\label{sssec:upbend}

\begin{figure}[!htb]
  \centering
  \includegraphics[width=\linewidth]{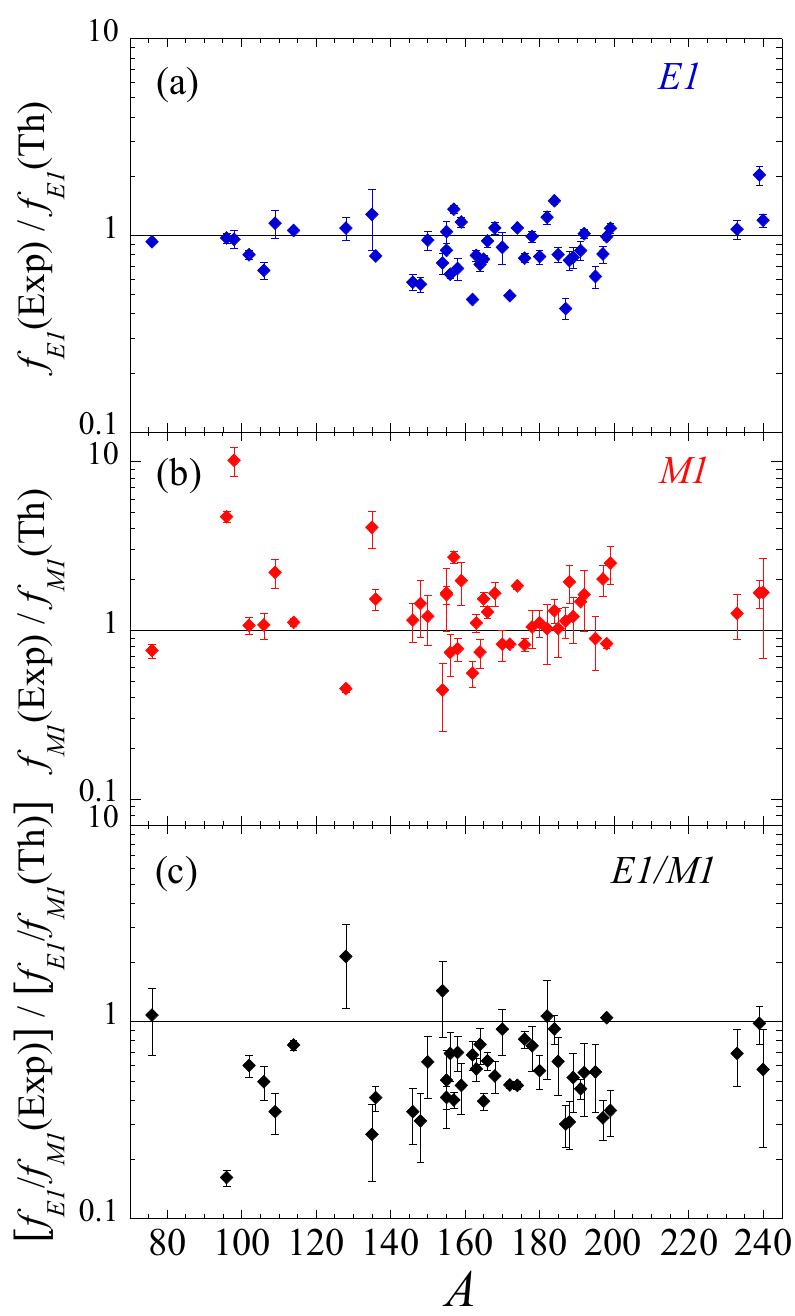}
  \caption{Comparison of the electric (E1) and magnetic (M1) dipole strength functions deduced from average resonance capture (ARC) $\gamma$-ray data , from neutron-capture experiments on isolated resonances, with the D1M+QRPA+0lim predictions for the 47 nuclei included in the recent reanalysis. The panels show (a) ratios of E1 strengths, (b) ratios of M1 strengths, and (c) the corresponding E1-to-M1 strength ratios. Figure adapted from \cite{Goriely2018a}}
  \label{fig:E1M1}
\end{figure}

\begin{figure}
  \centering
  \includegraphics[width=\linewidth]{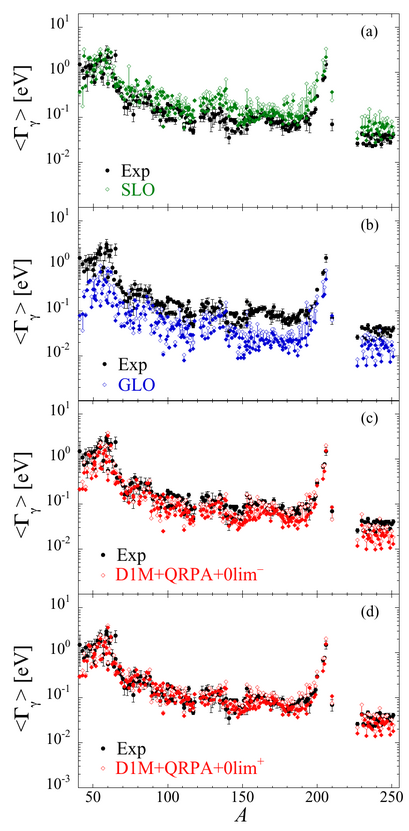}
  \caption{%
    Comparison of experimental (black circles) and theoretical
    (colored diamonds) average radiative neutron widths
    $\langle\Gamma_\gamma\rangle$ as a function of the mass number
    $A$. The theoretical values are obtained using four prescriptions
    for the dipole $\gamma$-ray strength function:
    (a) the standard Lorentzian (SLO) model for both the E1 and M1
    components,
    (b) the generalized Lorentzian (GLO) model,
    (c) the D1M+QRPA+0lim$^{-}$ model, and
    (d) the D1M+QRPA+0lim$^{+}$ model.
    Error bars on the theoretical points represent the spread caused by using two different nuclear level–density models: open diamonds
    correspond to the microscopic combinatorial (HFB+Combinatorial)
    model, and filled diamonds to the phenomenological constant–
    temperature model. Figure adapted from Ref. \cite{Goriely2018a}}
  \label{fig:width}
\end{figure}

Multiple experiments (NRF, Oslo, and two-step $\gamma$-cascade measurements following neutron capture) have revealed an unexpected enhancement of dipole strength at very low $\gamma$ energies ($E_\gamma \lesssim 3$~MeV), often termed a low-energy “upbend.” Theoretically, large-scale shell-model calculations indicate that many near-degenerate $M1$ transitions between excited states can produce a nonzero $M1$ strength as $E_\gamma \to 0$, whereas the $E1$ contribution in that region remains weak \citep{Sieja2018}. In well-deformed rare-earth nuclei, NRF measurements of the scissors mode (ground-state $M1$ transitions) yield a summed $B(M1)$ significantly smaller than the total $M1$ strength extracted from Oslo data (which includes transitions between excited states). This implies that a substantial fraction of $M1$ strength resides in transitions among excited levels — consistent with an $M1$ upbend.

For reaction modeling, the downward (de-excitation) $\gamma$-strength function is constructed by adding a low-energy tail to the Gogny-QRPA E1+M1 absorption strengths so that $f(E_\gamma)$ does not vanish as $E_\gamma \to 0$ (denoted as the D1M+QRPA+0lim model). This hybrid approach, calibrated to shell-model insights and experimental constraints (Oslo and NRF), respects detailed balance and reproduces the observed upbend \citep{Goriely2018a,Renstrm2018} (see Fig. \ref{fig:E1M1}). As illustrated in Fig. \ref{fig:width}, including this low-$E_\gamma$ enhancement can significantly increase average radiative widths $\langle\Gamma_\gamma\rangle$, especially for neutron-rich nuclei, and thereby affect $(n,\gamma)$ rates (in extreme neutron-rich cases, $\langle\Gamma_\gamma\rangle$ may increase by orders of magnitude).

\begin{figure}
  \centering
  \includegraphics[width=0.9\linewidth]{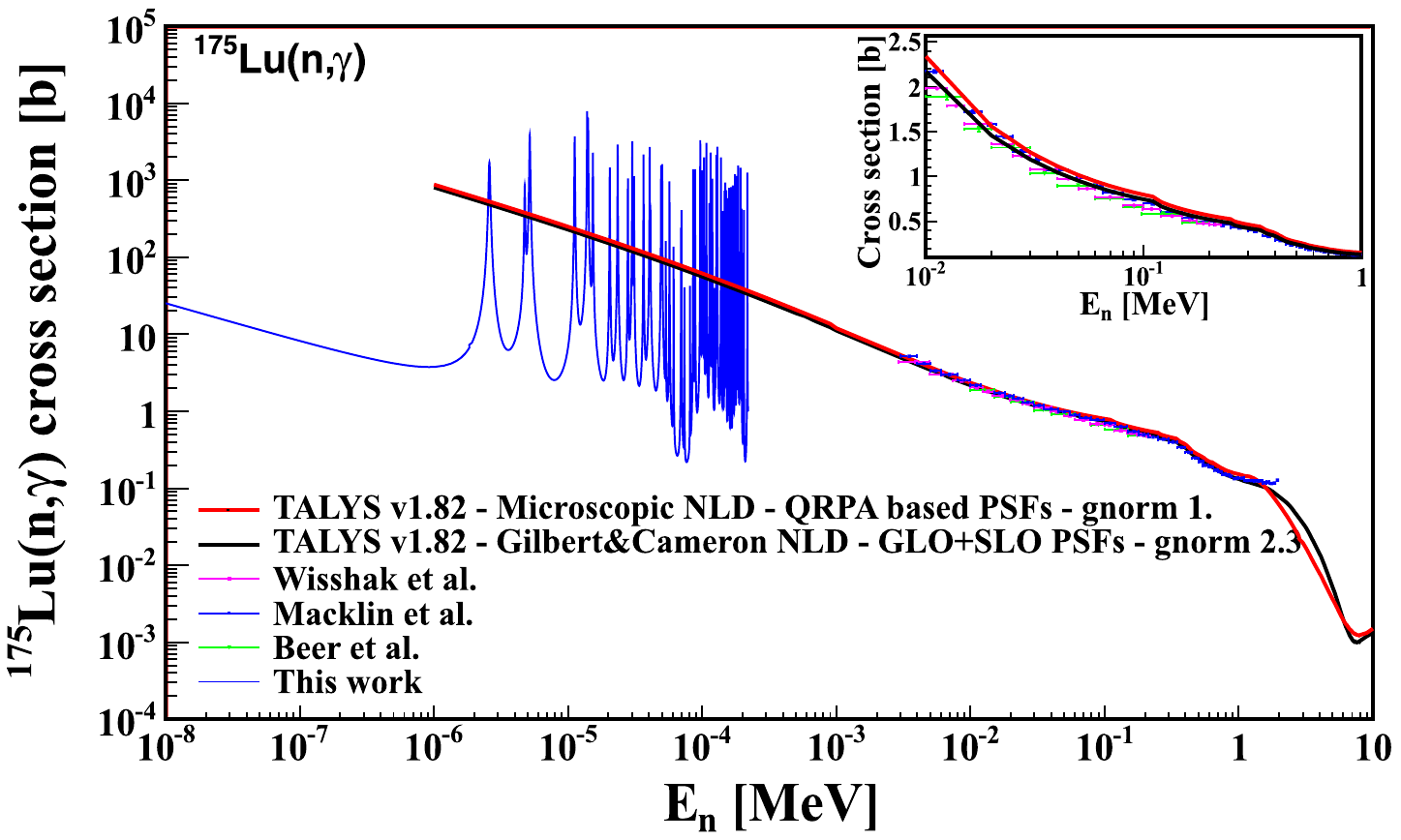}
  \caption{%
    Radiative neutron–capture cross section for the $^{175}$Lu(n,$\gamma$) reaction. 
    Experimental results are shown as symbols. 
    The standard TALYS 1.82 calculation 
    is displayed in black, while the calculation using E1 and M1 photon–strength 
    functions derived from QRPA models is shown in red. 
    Figure adapted from Ref. \citep{Ebran2019}.}
  \label{fig:Lu175ng}
\end{figure}

\paragraph{Applications in Reaction Modeling and Astrophysics}
\label{sssec:gsf-reac-astro}

Incorporating Gogny-QRPA strength functions into Hauser–Feshbach calculations (e.g., using TALYS) has improved the predictive power for $(n,\gamma)$ cross sections and radiative widths \cite{Koning2017}. For instance, the $^{173}$Lu$(n,\gamma)$ cross section was reproduced using Gogny E1 strengths \cite{Ebran2019}. At the LANL DANCE facility, Gogny-based E1/M1 strengths were used to analyze precision measurements on $^{176}$Lu, enabling the disentanglement of ground-state vs.\ isomeric $\gamma$-decay feeding \cite{Roig2016,DenisPetit2016}. Likewise, NewSUBARU photoneutron data on Tl isotopes, analyzed with Gogny E1 strengths plus the low-$E_\gamma$ upbend, led to improved constraints on the $^{204}$Tl$(n,\gamma)$ rate and $^{205}$Pb–$^{205}$Tl cosmochronometry \cite{Utsunomiya2019}. Looking ahead, these microscopic photon strength functions are strong candidates for future evaluated libraries (RIPL-4, JEFF, ENDF), promising to reduce model-dependent uncertainties and provide physics-based inputs \cite{Capote2009}.

\section{Summary and perspectives}\label{sec5}

In this review article, the main and detailed
information concerning the origin of the Gogny interaction and its main fitting protocol have been gathered for the first time in the same place, as well as its 
evolution along five decades.
After an historical introduction motivating the philosophy and the physical
context in which this interaction was designed, its evolution has been
discussed on two aspects: a) the generation of new parameterizations incorporating the description for new observables, as for example the 
astrophysical applications, preserving as much as possible its previous good
properties b) the extension of its analytical form to fully finite-range terms including a tensor term, preserving its standard properties and
improving detailed structure description with beyond-mean field approaches.
In a second step, the principle of the emulator based on the restricted Hartree-Fock approximation, which allows to propose parameterization candidates, has been discussed 
in details for the three incarnations of the Gogny interaction, namely D1, D2 and DG-type analytical forms. In particular, the construction of meta-data related to total binding 
energies, charge radii, pairing and isospin-asymmetry properties 
used in the emulator and the choice of experimental data is explained in details, highlighting a linear behavior between the restricted HF and HF models. Then, a full section 
has been dedicated to the nuclear matter properties 
(saturation density, total energy per nucleon, incompressibility, asymmetry energy and its slope, symmetric and neutron matter equation of state as well as the Landau 
parameters) which serve to filter the parameterization candidates proposed by the emulator. A final section is devoted to some highlights on structure,
fission, reaction and nuclear data applications showing the variety of phenomenons that have been analyzed with the help of the Gogny interaction coupled to mean-field and 
various beyond mean-field approaches.
For nuclear structure applications, its reliability concerning the description of shape evolution, shape coexistence, shape mixing, shell evolution, shell effects, low-lying 
spectroscopy, giant resonnaces, gamma-strength functions, using HFB, GCM, QRPA, MPMH approaches, is no longer in doubt. Its ability to describe the fission phenomenon for which 
pairing properties are particularly essential in the descent from the saddle point towards scission and using HFB and Time-dependent GCM approaches is stricking. The reaction 
applications are not left behind. Indeed, Gogny-HFB densities provide reliable inputs to JLM/JLMB optical models, deformed nuclei and rotational bands are well reproduced with 
microscopic CC calculations, Gogny-QRPA transition densities yield absolute inelastic cross sections without tuning, Gogny-QRPA doorway states enable microscopic MSD pr-equilibirum 
modeling, Gogny-based spin distributions improve $(n,xn\gamma)$ and surrogate analyses, HFB+combinatorial NLDs reproduce level spacings with good global accuracy, Gogny-D1M inputs 
ensure realistic spin/parity distributions and robust extrapolation, Gogny-QRPA provides unified $E1$ and $M1$ strengths (GDR, scissors), small empirical corrections yield accurate 
photo-absorption and capture data, Gogny-based structure inputs allow a coherent, predictive reaction framework linking elastic, inelastic, pre-equilibrium, and statistical decay.

The Gogny force in any of its forms has proven a strong potential in many facets of nuclear structure, fission, reactions and nuclear data. 
Its perpectives for the next decades are numerous.
At the level of the force itself, a lot of developments are envisaged. In particular, one can cite its extension to at least two ranges by type of term in order to allow, as in the 
case of the two standard central gaussians, attractive and repulsive contributions for the various fields. This will help in increasing the accuracy of the predictions, provided good 
constraint are used in the fitting protocol. Still missing is a clear understanding of how to use the phenomenological
density dependent term in the evaluation of energy kernels as required by the Generator Coordinate method and symmetry restorations. The introduction of three body finite-range force 
is one of the way we want to follow in the future. 
A modern version of the fitting protocol code is under development, going from the emulator up to calibration, passing through an optimization procedure according 
to several observables. The calibration of the force which evaluates the sensitivity of the observables to the parameters is underway and will allow uncertainty quantification. 
In parallel, an effort will be pursued to introduce extended forms of the Gogny force in mean-field and beyond approache codes.

We anticipate that the Gogny force will continue to be used to interpret experimental data and help experimentalists to propose new experiments. We believe the force has 
a brilliant future ahead in this respect. From this point of view, the exploration of new degrees of freedom as triaxial-octupole shapes, proton-neutron pairing correlations, 
GCM with conjugate coordinates, GCM with multi-quasiparticle excitations constitutes one of the perspectives.
There are several areas of nuclear structure where the Gogny force has been used only testimonially. One can mention the physics of the nucleus at finite temperature as well as 
all the phenomena related with the breaking of time-reversal like study of odd mass nuclei, high spin physics, multiquasiparticle excitations \dots{}. It will be interesting to
make some progress in this directions in the future, which will benefit both for the functional itself and the interpretation of experiments.
The advent of powerful computers and more powerfull ones will allow more complex calculations, rendering the breaking of more symmetries, the use of larger bases possible 
and customary, extending the realm of the Gogny force to exotic phenomena and extreme regions of the nuclear chart near the neutron drip line as well as to  
complex many-body methods.

All the previous perspectives will benefit to reaction applications and thus to nuclear data, allowing in particular the study of observables that depends on the tensor term, 
as for example the non-natural parity states and beta decays.

New applications to fission with the Gogny interaction in all its analytical form will be actively pushed in order to extract fragments properties at scission, including diabatic 
effects in the low \cite{CarpentierArxiv1,CarpentierArxiv2,CarpentierArxiv3}, intermediate and high energy regimes.

\bibliography{main-refcorr}

\begin{appendices}

\section{Spherical one-center harmonic oscillator representation and the Gogny separable expansion}\label{anex0}

\subsection{Spherical harmonic oscillator representation} 

In the fitting code, the one-center spherical HO representation is used. The states $\ket{a}$ of this basis are characterized by the set of quantum numbers 
\begin{equation} \label{HOstateS}
\ket{a} = \ket{r_a u_a} = \ket{n_a l_a m_{la} s_a t_a} \!,
\end{equation}
where the triplet $r_a = (n_a, l_a, m_{la})$ specifies the orbitals of the spherical HO, while the doublet $u_a = (s_a = \pm 1/2, t_a = \pm 1/2)$ corresponds to the projections of the spin and isospin along the quantization axis, chosen to be $Oz$, respectively. 

The spherically symmetric HO wave functions then take the form
\begin{equation} \label{wavefunctionsph}
\Phi_{a}(\vec{r}, \sigma, \tau) \equiv \braket*{\vec{r} \tau \sigma}{a} = \phi_{r_a}(\vec{r}) \xi_{s_a}(\sigma)\zeta_{t_a}(\tau),
\end{equation}
where $\xi_{s_a}(\sigma)$ and $\zeta_{t_a}(\tau)$ are the (normalized) spin and isospin wave functions,  associated with $\sigma$ and $\tau$, the Pauli matrices describing the spin and isospin degrees of freedom, respectively, and where the spatial spherical wave functions are expressed, in spherical coordinates $\vec{r} = (r, \hat{r}) = (r, \theta, \varphi)$, as
\begin{multline} \label{swavef}
\phi_{r_a} (\vec{r}) = \left[\frac{2}{b^3} \frac{n_a!}{\Gamma(n_a + l_a +3/2)}\right]^{1/2} \times \\ \e^{-r^2/2b^2} \Big( \frac{r}{b} \Big)^{l_a} L_{n_a}^{l_a+1/2} \bigg( \frac{r^2}{b^2} \bigg) Y_{l_a}^{m_{la}}(\hat{r})
\end{multline}
where $b = \sqrt{\hbar/ M \omega}$ is the oscillator length,  with $\hbar \omega$ the oscillator frequency and $M$ the mass of the nucleus considered. Note that $L_{n_a}^{l_a+1/2}(r^2/b^2)$ refers to the generalized Laguerre polynomials and $Y_{l_a}^{m_{la}}(\hat{r})$ the spherical harmonics. The spherical HO wave functions satisfy the orthogonality relation
\begin{equation} \label{orthogonalitysph}
\sum_{\sigma \tau} \int \dd[3]{r} \Phi_{a}^*(\vec{r}, \sigma, \tau) \Phi_{b}(\vec{r}, \sigma, \tau) = \delta_{r_a r_b} \delta_{s_a s_b} \delta_{t_a t_b}
\end{equation}
In the following, the spherical HO wave functions \eqref{swavef} with their quantum numbers equal to zero will be often meet. For convenience, one writes them down once and for all,
\begin{equation} \label{phi0spherique}
\phi_0(\vec{r}) = \frac{1}{\sqrt{4 \pi}} \bigg[ \frac{4}{\sqrt{\pi} \-  b^3} \bigg]^{1/2} \e^{-r^2/2b^2}
\end{equation}
with the associated spherical harmonics given by $Y_0^0(\hat{r}) = 1/\sqrt{4 \pi}$.
The angular dependence of the spherical wave functions \eqref{swavef} being entirely contained in the spherical harmonics, one can simply define their radial part as 
\begin{equation} \label{phisansSH}
\phi_{(r_a)} (r) \equiv \phi_{r_a} (\vec{r})/Y_{l_a}^{m_{la}}(\hat{r})
\end{equation}
where one has introduced the notation $(r_a) = (n_a, l_a)$.  Note that one has also the particular case $\phi_{(0)}(r) = \sqrt{4 \pi} \phi_{0}(\vec{r})$. Finally, one can define the spherical HO wave functions deprived of their exponential as
\begin{equation} \label{phihatsph}
\hat{\phi}_{(r_a)}(r) \equiv \e^{r^2/2 b^2} \phi_{(r_a)}(r)
\end{equation}
where one has additionally removed the angular dependence according to \eqref{phisansSH}.

\subsection{Formulas useful for the spherical symmetry}

In this sub-section, one gives the main formulas used to carry out the above calculations in the framework of the spherical harmonic oscillator representation. One will not attempt to justify the developments leading to those and refer the reader in particular to \cite{Gogny1975b,Berger1985} for more detailed presentations. 

\subsubsection{Spherical harmonics}

The spherical harmonics considered in the spherical HO wave functions \eqref{swavef} are conventionally defined as:
\begin{multline} \label{sphericalharmonics}
Y_l^{m_l}(\theta, \varphi) \equiv (-)^{m_l} \sqrt{\frac{(2l+1)}{4 \pi} \frac{(l-m_l)!}{(l+m_l)!}} \times \\ P_l^{m_l}(\cos \theta) \- \e^{\ii m_l \varphi}
\end{multline}
where $P_l^{m_l}(\cos \theta)$ are the associated Legendre polynomials. By definition, the spherical harmonics satisfy the orthogonality relation:
\begin{equation} \label{orthogonalitySH}
\int \dd[2]{\hat{r}} Y_l^{m_l}(\hat{r}) Y_{l'}^{m'_l *}(\hat{r}) = \delta_{l l'} \delta_{m_l m_l'}, 
\end{equation}
as well as the so-called addition theorem applied for identical angles:
\begin{equation} \label{addition}
\sum_{m_l} Y_l^{m_l *}(\hat{r}) Y_l^{m_l}(\hat{r}) = \frac{2l+1}{4 \pi},
\end{equation}
where one recalls the notation for the angular variables, $\hat{r} = (\theta, \varphi)$, and where one has used the reality condition:
\begin{equation} \label{etoileSH}
Y_l^{m_l *}(\hat{r}) = (-)^{m_l} Y_l^{-m_l}(\hat{r}).
\end{equation}

In the calculations, one often meets the spherical harmonics \eqref{sphericalharmonics} with their quantum numbers equal to zero. They simply read:
\begin{equation} \label{Y00}
Y_0^0(\hat{r}) = \frac{1}{\sqrt{4 \pi}}.
\end{equation}

One shows that the spherical harmonics can be written as a product of irreducible tensors of the form \cite{Khersonskii1988}
\begin{multline}
Y_l^{m_l}(\hat{r}) = \frac{1}{r^l} \sqrt{\frac{1}{4 \pi} \frac{(2l+1)!}{2^l (l!)^2}} \times \\ \big[ \ldots [[\vec{r} \otimes \vec{r}]^{(2)} \otimes \vec{r}]^{(3)} \ldots \otimes \vec{r} \big]^l_{m_l}
\end{multline}
In the particular case $l=1$, useful for spin-orbit calculations, this relation brings:
\begin{equation} \label{harmoSO}
Y_1^{m_l} (\hat{r}) = \frac{1}{r} \sqrt{\frac{3}{4 \pi}} r^{m_l}
\end{equation}
and in the particular case $l=2$, useful for tensor calculations,
\begin{equation} \label{harmoTS}
Y_2^{m_l} (\hat{r}) = \frac{1}{r^2} \sqrt{\frac{15}{8 \pi}} [\vec{r} \otimes \vec{r}]^{(2)}_{m_l}
\end{equation}
Finally, the integral of the product of three spherical harmonics reads:
\begin{multline} \label{product3SH}
\int \dd[2]{\hat{r}} Y_{l_a}^{m_{la}*}(\hat{r}) Y_{l_\mu}^{m_{l\mu}*}(\hat{r}) Y_{l_c}^{m_{lc}}(\hat{r})  = \\ (-)^{m_{lc}} \sqrt{\frac{(2l_a+1)(2l_\mu+1)(2l_c+1)}{4 \pi}} \\
 \times \begin{pmatrix} l_a & l_\mu & l_c \\ 0 & 0 & 0 \end{pmatrix} \begin{pmatrix} l_a & l_\mu & l_c \\ m_{la} & m_{l\mu} & -m_{lc} \end{pmatrix}
\end{multline}
which is non-zero if and only if the conditions:
\begin{subequations} \label{condY}
\begin{gather} 
m_{li} \le \abs{l_i} \ \text{for } i \in \{ a, \mu, c \} \\
\abs{l_a - l_c} \le l_\mu \le l_a + l_c \\
m_{l\mu} = -(m_{la} + m_{lc})
\end{gather}
\end{subequations}
imposed by the above Wigner-$3j$ symbols, denoted by the parentheses, are satisfied. Besides, the parity condition imposes that $l_a+l_c+l_\mu$ must be an even integer.

\subsubsection{Action of the gradient operator in spherical symmetry}

One can decompose the action of the gradient operator on the spherical harmonic oscillator wave functions according to its components in the spherical basis. Then, each spherical 
component $\beta = 0, \pm 1$ of the gradient operator acts in the following way on a wave function $\phi_{r}(\vec{r}) = \phi_{(n, l)} (r) Y_{l}^{m_l}(\hat{r})$ \cite{Gogny1975b},
\begin{multline} \label{grad}
\nabla_{\beta} \phi_{r}(\vec{r}) \equiv \sum_{j=\pm 1} (-)^{l + m_l + \beta + 1} \begin{pmatrix} l & 1 & l + j \\ m_l & \beta & -m_l - \beta \end{pmatrix} \\ g_j^{l}(r) \phi_{(n,  l)} (r) Y_{l + j}^{m_l + \beta}(\hat{r})
\end{multline}
where the operator $g_j^{l}(r)$ is defined by
\begin{equation}
g_j^{l}(r) \equiv j \sqrt{l + \frac{1}{2} \- (j+1)} \, \bigg[\dv{r} - j \- \frac{l \- (j-1)}{r} \bigg]
\end{equation}
Using the definition of the spherical HO wave functions \eqref{swavef}, it is not difficult to prove the recurrence 
relation:
\begin{multline} \label{recurrence}
g_j^{l}(r) \phi_{(n, l)}(r) = \frac{1}{b} \big[ a_j^{(n,l)} \phi_{(n - j, l + j)}(r) \\ + c_j^{(n,l)} \phi_{(n,  l + j)}(r) \big]
\end{multline}
with the coefficients
\begin{multline} \label{coefs}
a_j^{(n,l)} = -\sqrt{\Big(n-\frac{1}{2}(j-1)\Big)\Big(l +\frac{1}{2}(j+1)\Big)},  \\
c_j^{(n,l)} = -\sqrt{\Big(n + l + \frac{1}{2} (j+1)\Big)\Big(l +\frac{1}{2} \- (j+1)\Big)}
\end{multline}
Combining equations \eqref{grad} and \eqref{recurrence} leads to the final form of the gradient formula, i.e.\ for $\beta = 0, \pm 1$:
\begin{multline} \label{grafinal}
\nabla_{\beta} \phi_{r}(\vec{r}) = \frac{1}{b} \sum_{j=\pm 1} (-)^{l + m_l + \beta + 1} \times \\ \begin{pmatrix} l & 1 & l + j \\ m_l & \beta & -m_l - \beta \end{pmatrix} \times \\
\big[ a_j^{(n,l)} \phi_{n - j,  l + j,  m_l + \beta}(\vec{r}) + c_j^{(n,l)} \phi_{n,  l + j, m_l + \beta} (\vec{r}) \big]
\end{multline}

\subsubsection{Talman coefficients in spherical symmetry}

The Gogny separable expansion \cite{Gogny1975b} (see below section \ref{Gognyseparable} for a quick presentation) applied on spherical HO wave functions provides the relation:
\begin{multline} \label{Gognysepspherique}
\phi_{r_a}^*(\vec{r}) \phi_{r_b}(\vec{r}) = \sum_{r_\mu} T_{(r_a)(r_b)}^{(r_\mu)}  \mel*{l_a m_{la}}{Y_{l_\mu}^{m_{l \mu}*}}{l_b m_{lb}} \times \\ \phi_0(\vec{r}) \phi_{r_\mu}(\vec{r}),
\end{multline}
where, by definition:
\begin{multline}
\mel*{l_a m_{la}}{Y_{l_\mu}^{m_{l \mu}*}}{l_b m_{lb}} \equiv \int \dd[2]{\hat{r}} Y_{l_a}^{m_{la}*}(\hat{r}) \times \\ Y_{l_\mu}^{m_{l\mu}*}(\hat{r}) Y_{l_b}^{m_{lb}}(\hat{r}) 
\end{multline}
with the integral over three spherical harmonics given in Eq.\eqref{product3SH}, and where it appears a coefficient called the spherical Talman coefficient, that reads:
\begin{multline}\label{talman}
T_{(r_a)(r_b)}^{(r_\mu)} = (-)^{n_\mu} N_{(r_a)(r_b)}^{(r_\mu)} \times \\\sum_{i j} \frac{(-)^{i+j}}{i!(n_a-i)! j!(n_b-j)! } \\
\times \frac{\Gamma\big[\frac{1}{2}(l_a + l_b + l_\mu) +i + j +3/2 \big] }{\Gamma[l_a + i +3/2] \, \Gamma[l_b + j +3/2]} \\
\times \frac{ \Gamma\big[\frac{1}{2}(l_a + l_b - l_\mu) +i + j +1 \big]}{\Gamma\big[\frac{1}{2}(l_a + l_b - l_\mu) - n_\mu +i + j +1\big]} 
\end{multline}
with
\begin{multline} \label{Ncoeff}
N_{(r_a)(r_b)}^{(r_\mu)} = \bigg[ 2\pi^{3/2} \frac{n_a ! \, \Gamma[n_a + l_a +3/2] \, n_b ! }{n_\mu ! \, \Gamma[n_\mu + l_\mu +3/2]} \\ \times \Gamma[n_b + l_b +3/2]\bigg]^{1/2}
\end{multline}
The spherical Talman coefficient is non-zero if and only if the following inequalities are satisfied, 
\begin{equation} \label{condTalmansph}
\max\bigg( 0, \frac{|N_a-N_b|-l_\mu}{2} \bigg) \le n_\mu \le \frac{N_a + N_b - l_\mu}{2}
\end{equation}
with $X_{a} \equiv 2 n_a + l_a$.

\subsubsection{Moshinsky coefficients in spherical symmetry}

The Moshinsky transformation allows to move from the nucleon coordinates $(\vec{r}_1, \vec{r}_2)$ to the relative and center-of-mass coordinates $(\vec{r},  \vec{R})$, defined as\footnote{This convention is important; another definition of the relative and center-of-mass coordinates $(\vec{r}, \vec{R})$ would bring a different expression of the Moshinsky coefficient.}
\begin{equation} \label{rRspherique}
\vec{r} \equiv \frac{\vec{r}_1-\vec{r}_2}{\sqrt{2}}, \quad \text{and} \quad \vec{R} \equiv \frac{\vec{r}_1+\vec{r}_2}{\sqrt{2}}
\end{equation}

The Moshinky transformation applied on spherical HO wave functions furnishes the relation:
\begin{equation} \label{Moshinskytransfospherique}
\phi_{r_\mu}(\vec{r}_1) \phi_{r_\nu}(\vec{r}_2) = \sum_{r_\lambda r_\sigma} M_{r_\mu r_\nu}^{r_{\lambda} r_{\sigma}} \phi_{r_{\lambda}}(\vec{R}) \phi_{r_{\sigma}}(\vec{r})
\end{equation}
where it appears a coefficient called the spherical Moshinsky coefficient, that reads \cite{Berger1985}\footnote{In his article about the separable development \cite{Gogny1975b},  Gogny uses the convention of Baranger and Davies \cite{Baranger1966} for the Moshinsky coefficient. Its Moshinsky coefficient called $M_{r_\mu r_\nu}^{r_{\lambda} r_{\sigma}} \vert_{{G}}$, is linked to the present ones by the relation:
\begin{equation}
M_{r_\mu r_\nu}^{r_{\lambda} r_{\sigma}} \vert_{{G}} = (-)^{l_\lambda + l_\sigma - m_{l \lambda}} M_{r_\mu r_\nu}^{r_{\lambda} r_{\sigma}}.
\end{equation}
}
\begin{multline} \label{defMoshinsky}
M_{r_\mu r_\nu}^{r_{\lambda} r_{\sigma}} = \sum_{r_a r_b} \sum_{r_c r_d} (-)^{X_d} \bigg(\frac{1}{\sqrt{2}}\bigg)^{X_a+X_b+X_c+X_d} \\ C_{r_a r_c}^{r_\mu *} C_{r_b r_d}^{r_\nu *} C_{r_a r_b}^{r_\lambda} C_{r_c r_d}^{r_\sigma}
\end{multline}
with 
\begin{multline}
C_{r_a r_b}^{r_\mu} \equiv (-)^{n_a + n_b - n_\mu} N_{(r_a)(r_b)}^{(r_\mu)} \\
\mel*{l_\mu m_{l\mu}}{Y_{l_a}^{m_{la}}}{l_b m_{lb}}  \delta_{X_{\mu}, X_{a} + X_{b}}
\end{multline}
where $N_{(r_a)(r_b)}^{(r_\mu)}$ is given by \eqref{Ncoeff}.

One can show that the spherical Moshinsky coefficient implies:
\begin{equation} \label{condMoshinsky1spherique}
n_{\sigma} = \frac{X_\mu + X_\nu - X_{\lambda} - l_{\sigma}}{2}
\end{equation}
as well as
\begin{equation} \label{condMoshinsky2spherique}
m_{l\sigma} = m_{l\mu} + m_{l\nu} - m_{l\lambda}
\end{equation}
while the range of values taken by $r_\lambda$ is only constrained by $m_{l \lambda} \le \abs{l_\lambda}$. Thus, the summation over $r_\sigma$ in \eqref{Moshinskytransfospherique} actually corresponds to a simple summation over $l_\sigma$.

In the case $r_{\lambda} = 0$,  the spherical Moshinsky coefficient simplifies according to:
\begin{multline} \label{Mosh0spherique}
M_{r_\mu r_\nu}^{0 \, r_{\sigma}} = (-)^{{l_\mu - l_\nu + l_\sigma}/2}  \sqrt{4 \pi} \bigg( \frac{1}{\sqrt{2}} \bigg)^{X_\mu + X_\nu} \\
\times \left[ \frac{2}{\sqrt{\pi}} \frac{n_\sigma !  \Gamma(n_\sigma + l_\sigma +3/2)}{n_\mu ! \, \Gamma(n_\mu + l_\mu +3/2) n_\nu !  \Gamma(n_\nu + l_\nu +3/2)} \right]^{1/2} \\ \times \mel*{l_\sigma m_{l\sigma}}{Y_{l_\mu}^{m_{l\mu}}}{l_\nu m_{l\nu}} 
\end{multline}
which implies
\begin{equation} \label{condMoshspheriquereduce}
n_{\sigma} = \frac{X_\mu + X_\nu - l_\sigma}{2},
\end{equation}
and
\begin{equation}
m_{l\sigma} = m_{l\mu} + m_{l\nu}.
\end{equation}

In the case $r_\mu = r_\nu =0$,  the spherical Moshinsky coefficient reduces to:
\begin{equation} \label{Moshspherique00}
M_{0 \, 0}^{r_{\lambda} r_{\sigma}} = \delta_{r_{\lambda}, 0} \delta_{r_{\sigma}, 0}.
\end{equation}

\subsection{The Gogny separable expansion} \label{Gognyseparable}

The Gogny separable expansion \cite{Gogny1975b,Berger1985} is a powerful tool that allows to separate the spatial degrees of freedom of a two-body potential, be it central or non-central, in any basis of the harmonic oscillator (HO). Schematically, the separable development reads:
\begin{equation} \label{Gognysim}
v(\vec{r}_1 - \vec{r}_2) \sim \sum_{r_\mu} \hat{\phi}_{r_\mu}(\vec{r}_1) f_{r_\mu}(\vec{r}_2),
\end{equation}
where $v(\vec{r}_1 - \vec{r}_2)$ is the two-body potential, 
$r_\mu$ the set of quantum numbers characterizing the spatial degrees of freedom in some HO basis (namely \eqref{HOstateS} in spherical symmetry), 
$\hat{\phi}_{r_\mu}(\vec{r}_1)$ the HO wave function deprived of its exponential factor (their radial part is given by \eqref{phihatsph} in spherical symmetry), 
and $f_{r_\mu}(\vec{r}_2)$ some function of $\vec{r}_2$. 
Formally, the separable development stipulates that the potential and the development of the right-hand side part are equivalent in the sense that their two-body matrix elements (TBMEs) evaluated in any basis of the harmonic oscillator are equal, i.e.\
\begin{multline}
\mel*{r_a r_b}{v(\vec{r}_1 - \vec{r}_2)}{r_c r_d} = \\ \sum_{r_\mu} \! \mel*{r_a r_b}{\hat{\phi}_{r_\mu}(\vec{r}_1) f_{r_\mu}(\vec{r}_2)}{r_c r_d} \!.
\end{multline}
Thus, the TBMEs of the potential in some HO basis can be written as a product of one-body matrix elements according to:
\begin{multline} \label{Gognysep1}
\mel*{r_a r_b}{v(\vec{r}_1 - \vec{r}_2)}{r_c r_d} = \\ \sum_{r_\mu=1}^{m} \! \mel*{r_a}{\hat{\phi}_{r_\mu}(\vec{r}_1)}{r_c} \mel*{r_b}{f_{r_\mu}(\vec{r}_2)}{r_d}
\end{multline}
or, equivalently, by means of the symmetry under the exchange of the two particles:
\begin{multline} \label{Gognysep2}
\mel*{r_a r_b}{v(\vec{r}_1 - \vec{r}_2)}{r_c r_d} = \\ \sum_{r_\mu=1}^{m'}  \mel*{r_a}{f_{r_\mu}(\vec{r}_1)}{r_c}  \mel*{r_b}{\hat{\phi}_{r_\mu}(\vec{r}_2)}{r_d} 
\end{multline}
where, in general, $m \ne m'$. 
These separations are very useful, both analytically and numerically. 
Analytically, they simplify the calculations by de-correlating the spatial coordinates of the two particles. Numerically, they allow to optimize the loops, especially when implementing fields, by reducing the number of time-demanding operations. Actually, the existence of such a development relies on the following 
property of the HO wave functions \cite{Gogny1975b},
\begin{equation} \label{separableGOGNY}
\phi^*_{r_a}(\vec{r}) \phi_{r_c}(\vec{r}) = \sum_{r_\mu} T_{r_a r_c}^{r_\mu} \phi_0(\vec{r}) \phi_{r_\mu}(\vec{r})
\end{equation}

For the spherical symmetry, the corresponding properties are given by \eqref{Gognysepspherique}, from which one can extract the value of the factor $T_{r_a r_c}^{r_\mu} \equiv \mel*{r_a}{\hat{\phi}_{r_\mu}(\vec{r}_1)}{r_c}$, involving the so-called Talman coefficients. In the next subsections, one gives the path to prove the Gogny separable development for central and non-central potentials.

\subsubsection{Central potentials}

One starts with a central potential which, by definition, only depends on the relative distance between the two nucleons, $v(|\vec{r}_1 - \vec{r}_2|)$. Its TBMEs in some HO representation are:
\begin{multline} \label{centralseparable1}
\mel*{r_a r_b}{v(|\vec{r}_1 - \vec{r}_2|)}{r_c r_d} = \int \dd[3]{r_1} \int \dd[3]{r_2} \\ \times \phi_{r_a}^*(\vec{r}_1) \phi_{r_b}^*(\vec{r}_2) v(|\vec{r}_1 - \vec{r}_2|) \phi_{r_c}(\vec{r}_1) \phi_{r_d}(\vec{r}_2)
\end{multline}
where $\phi$ denotes the HO wave function. Since the potential is central, it commutes with the wave functions. Then, one can apply the relation \eqref{separableGOGNY} to obtain:
\begin{multline}
\mel*{r_a r_b}{v(|\vec{r}_1 - \vec{r}_2|)}{r_c r_d}  = \sum_{r_\mu} T_{r_a r_c}^{r_\mu}
\times \int \dd[3]{r_2} \phi_{r_b}^*(\vec{r}_2) \\ \int \dd[3]{r_1} \phi_0(\vec{r}_1) \phi_{r_\mu}(\vec{r}_1) v(|\vec{r}_1 - \vec{r}_2|) \phi_{r_d}(\vec{r}_2)
\end{multline}
Defining the quantity:
\begin{equation}
f_{r_\mu}(\vec{r}_2) \equiv \int \dd[3]{r_1} \phi_0(\vec{r}_1) \phi_{r_\mu}(\vec{r}_1)v(|\vec{r}_1 - \vec{r}_2|),
\end{equation}
one finally gets the first form of the Gogny separable expansion \eqref{Gognysep1}. If instead of applying the relation \eqref{separableGOGNY} on the wave functions evaluated in $\vec{r}_1$, one applies it on the ones evaluated in $\vec{r}_2$, one equivalently ends up with the second form of the Gogny separable development \eqref{Gognysep2}.

\subsubsection{Non-central potentials}

One continues with the non-central potentials, which, in the case of the generalized Gogny interaction \eqref{gognyDG}, are summarized in the tensor and spin-orbit potentials.

The tensor potential can in fact be written in terms of a central potential as:
\begin{equation}
v^{\text{T}} \equiv v(|\vec{r}_1 - \vec{r}_2|) S_{12}
\end{equation}
where $S_{12}$ is the tensor operator \eqref{tensoroperatormain}. But since the tensor operator commutes with the HO wave functions appearing in Eq.\eqref{centralseparable1}, it barely behaves like a central potential, and its separable expansion is immediately demonstrated, with:
\begin{equation}
f_{r_\mu}(\vec{r}_2, \vec{\sigma}_1, \vec{\sigma}_2) \equiv \int \dd[3]{r_1} \phi_0(\vec{r}_1) \phi_{r_\mu}(\vec{r}_1)v(|\vec{r}_1 - \vec{r}_2|) S_{12}
\end{equation}
Once again, applying the relation \eqref{separableGOGNY} on the wave functions evaluated in $\vec{r}_2$ would lead to the second form of the Gogny separable expansion \eqref{Gognysep2}.

The spin-orbit potential can also be written in terms of a central potential as:
\begin{equation}
v^{\text{SO}} \equiv v(|\vec{r}_1 - \vec{r}_2|) (\vec{r}_{12} \cross \vec{\nabla}_{12}) \cdot \vec{S}
\end{equation}
where one has freed from factors to define the relative position $\vec{r}_{12} \equiv \vec{r}_1 - \vec{r}_2$ and the relative gradient $\vec{\nabla}_{12} \equiv \vec{\nabla}_1 - \vec{\nabla}_2$. Here, the demonstration is a bit trickier as the non-central part of the spin-orbit potential does not commute with the HO wave functions because of the gradient operator. Separating the spatial from the spin degrees of freedom according to:
\begin{multline}
\mel*{r_a r_b}{v^{\text{SO}}}{r_c r_d} = \\ \mel*{r_a r_b}{v(|\vec{r}_1 - \vec{r}_2|) (\vec{r}_{12} \cross \vec{\nabla}_{12})}{r_c r_d} \cdot \vec{S}
\end{multline}
one has,
\begin{multline}
\mel*{r_a r_b}{v(|\vec{r}_1 - \vec{r}_2|) (\vec{r}_{12} \cross \vec{\nabla}_{12})}{r_c r_d} = \\
\int \dd[3]{r_1} \int \dd[3]{r_2} \phi_{r_a}^*(\vec{r}_1) \phi_{r_b}^*(\vec{r}_2) v(|\vec{r}_1 - \vec{r}_2|) \times \\ (\vec{r}_{12} \cross \vec{\nabla}_{12}) \phi_{r_c}(\vec{r}_1) \phi_{r_d}(\vec{r}_2)
\end{multline}
Splitting the integral into two parts, one obtains:
\begin{multline} \label{SOsep}
\mel*{r_a r_b}{v(|\vec{r}_1 - \vec{r}_2|) (\vec{r}_{12} \cross \vec{\nabla}_{12})}{r_c r_d} = \\
\int \dd[3]{r_1} \int \dd[3]{r_2} \phi_{r_a}^*(\vec{r}_1) \phi_{r_b}^*(\vec{r}_2) \phi_{r_d}(\vec{r}_2) \\ \times v(|\vec{r}_1 - \vec{r}_2|) (\vec{r}_{12} \cross \vec{\nabla}_{1}) \phi_{r_c}(\vec{r}_1) \\
- \int \dd[3]{r_1} \int \dd[3]{r_2} \phi_{r_a}^*(\vec{r}_1) \phi_{r_c}(\vec{r}_1) \phi_{r_b}^*(\vec{r}_2) \\ \times v(|\vec{r}_1 - \vec{r}_2|) (\vec{r}_{12} \cross \vec{\nabla}_{2}) \phi_{r_d}(\vec{r}_2)
\end{multline}
By virtue of the formula \eqref{separableGOGNY} applied on both parts, one finds out:
\begin{multline}
\mel*{r_a r_b}{v(|\vec{r}_1 - \vec{r}_2|) (\vec{r}_{12} \cross \vec{\nabla}_{12})}{r_c r_d}  = \\ \sum_{r_\mu} T_{r_b r_d}^{r_\mu} \! \mel*{r_a}{\vec{f}_{r_\mu}(\vec{r}_1)}{r_c} - \sum_{r_\nu} T_{r_a r_c}^{r_\nu} \! \mel*{r_b}{\vec{f}_{r_\nu}(\vec{r}_2)}{r_d}
\end{multline}
where one has defined the vector quantities:
\begin{subequations}
\begin{align}
\vec{f}_{r_\mu}(\vec{r}_1) & \equiv \int \dd[3]{r_2} \phi_0(\vec{r}_2) \phi_{r_\mu}(\vec{r}_2) v(|\vec{r}_1 - \vec{r}_2|) \vec{r}_{12} \cross \vec{\nabla}_{1} \\
\vec{f}_{r_\nu}(\vec{r}_2) & \equiv \int \dd[3]{r_1} \phi_0(\vec{r}_1) \phi_{r_\nu}(\vec{r}_1) v(|\vec{r}_1 - \vec{r}_2|) \vec{r}_{12} \cross \vec{\nabla}_{2}
\end{align}
\end{subequations}
Finally adding the spin part, it comes:
\begin{multline}
\mel*{r_a r_b}{v^{\text{SO}}}{r_c r_d} = \sum_{r_\mu} \! \mel*{r_b}{\hat{\phi}_{r_\mu}(\vec{r}_2)}{r_d} \times \\ \mel*{r_a}{\vec{f}_{r_\mu}(\vec{r}_1)}{r_c} \cdot \vec{S} \\
- \sum_{r_\nu}  \mel*{r_a}{\hat{\phi}_{r_\nu}(\vec{r}_1)}{r_c} \mel*{r_b}{\vec{f}_{r_\nu}(\vec{r}_2)}{r_d} \cdot \vec{S}
\end{multline}
Thus, each of the above parts of the spin-orbit potential can be decomposed into a product of a one-body matrix element in $\vec{r}_1$ and another one in $\vec{r}_2$. This is precisely the definition of the Gogny separable expansion \eqref{Gognysim}, which separately holds for both parts. 
Note that one had no choice but to apply the formula \eqref{separableGOGNY} on the wave functions evaluated in $\vec{r}_2$ in the first integral, and on the ones evaluated in $\vec{r}_1$ in the second integral of \eqref{SOsep}. Therefore, contrary to the central and tensor potentials, one cannot decide to apply this formula on the wave functions depending on the other variables to get the second form of the separable expansion \eqref{Gognysep2}. To do so, one rather notices that the spin-orbit potential is symmetric under the exchange of variables $(\vec{r}_1, \vec{\sigma}_1) \leftrightarrow (\vec{r}_2, \vec{\sigma}_2)$.

\section{Restricted Hartree-Fock fields of the DG Gogny interaction in a spherical one-center harmonic oscillator basis}\label{appenb}

In this appendix, one derives the expressions of the one-body mean field $h_{aa}^{HFR}$ (see Eq.(\ref{hHFR})), the total binding $\mathcal{E}^{\text{HFR}}$
(see Eq.(\ref{EHFR})), as well as the energy difference $\Delta \varepsilon$ (see Eq.(\ref{constraint3})).
All the quantities are evaluated at the restricted Hartree-Fock approximation in the case of the DG Gogny interaction.

\subsection{Expression of the one-body Hamiltonian $h_{aa}^{HFR}$ in the restricted Hartree-Fock approximation}

The one-body Hamiltonian $h_{aa}^{HFR}$ contains a kinetic energy contribution $K_{aa}^{HFR}$, a potential energy contribution $\mathcal{E}_{aa}^{HFR}$ 
and a rearrangement energy contribution $\partial \Gamma_{ab}^{HFR}$ whose expressions are derived in the following.

\subsubsection{Kinetic field}

By virtue of the virial theorem, one can evaluate the contribution of the kinetic field in the spherical HO representation, and obtains:
\begin{equation}
K_{aa}^{HFR} = \frac{\hbar \omega}{2} \Big( 2 n_a + l_a + \frac{3}{2} \Big).
\end{equation}

\subsubsection{Mean field}

For the sake of clarity, one will decompose this field into a direct and an exchange component in the following,  as $\Gamma_{aa}^{HFR} \equiv \Gamma_{aa}\vert_{{D}} + \Gamma_{aa}\vert_{{E}}$, where
\begin{subequations}
\begin{align}
\Gamma_{aa}\vert_{{D}} & \equiv \sum_{b} \! \mel*{ab}{v_{12}}{ab} \! \rho_{bb}, \\
\Gamma_{aa}\vert_{{E}} & \equiv \sum_{b} \! \mel*{ab}{v_{12}}{ba} \! \rho_{bb}.
\end{align}
\end{subequations}
One will evaluate these contributions for the fields of the DG Gogny interaction \eqref{gognyDG}.

\paragraph{Central contribution}

One starts by deriving the direct mean field associated with the central terms. The central terms are given by \eqref{centralintsph}, where the following shorthand notation for the spin-isospin part is used:
\begin{equation} \label{mathP}
\mathcal{P} \equiv W + B P_{\sigma} - H P_{\tau} - M P_{\sigma}P_{\tau}.
\end{equation}
The spatial and spin-isospin degrees of freedom split up in such a way that:
\begin{equation}
\Gamma^{\text{C}}_{aa}\vert_{{D}} = \sum_{r_b u_b} \mel*{r_a r_b}{V(r_{12})}{r_a r_b} \! \! \mel*{u_a u_b}{W_{{D}}}{u_a u_b} \!.
\end{equation}
On the one hand, the spin-isospin TBMEs of the central terms are easy to compute since their spin-isospin part is only a combination of spin- and isospin-exchange operators. One finds out, in general:
\begin{multline} \label{combipara}
\mel*{u_a u_b}{\mathcal{P}}{u_c u_d} = \big( W \delta_{s_a s_c}  \delta_{s_b s_d} + B \delta_{s_a s_d} \delta_{s_b s_c} \big) \\ \times \delta_{t_a t_c} \\ \delta_{t_b t_d}  - \big( H \delta_{s_a s_c} \delta_{s_b s_d} + M \delta_{s_a s_d} \delta_{s_b s_c} \big) \delta_{t_a t_d} \delta_{t_b t_c}
\end{multline}
In the particular case of the HFR approximation, this equation reduces to:
\begin{equation}
\mel*{u_a u_b}{\mathcal{P}}{u_a u_b} = W + B \delta_{s_a s_d} - ( H + M \delta_{s_a s_b} ) \delta_{t_a t_b}
\end{equation}
so that one immediately gets:
\begin{equation} \label{SIcentral}
\sum_{s_b} \! \mel*{s_a t_a \, s_b t_b}{\mathcal{P}}{s_a t_a \, s_b t_b} \! = 2W + B - (2H+M) \delta_{t_a t_b}
\end{equation}
On the other hand, the spatial TBMEs associated with the central terms have been calculated in Eq.\eqref{spatialcentral}. The respective definitions of the integral over three spherical harmonics \eqref{product3SH} and of the particular value of the Moshinsky coefficient \eqref{Mosh0spherique}, coupled with the properties of the Wigner-$3j$ symbols, allow to simplify the spatial part according to:
\begin{multline}
\sum_{m_{lb}}  \mel*{n_a l_a m_{la}  n_b l_b m_{lb}}{V(r_{12})}{n_a l_a m_{la}  n_b l_b m_{lb}} \\  = S^{{C}}_{n_a l_a n_b l_b} \vert_{{D}}
\end{multline}
where one has defined
\begin{multline}
S^{{C}}_{n_a l_a n_b l_b} \vert_{{D}} \equiv \frac{1}{4 \pi} (2l_b+1) \sum_{n_\mu n_\nu} T_{(n_a l_a) (n_a l_a)}^{(n_\mu 0)} \times \\ T_{(n_b l_b) (n_b l_b)}^{(n_\nu 0)} I_{n_\mu n_\nu 0}
\end{multline}
with the quantity
\begin{multline}
I_{n n' l} = \frac{1}{2^{n+n'+l}} \times \\  \frac{(n+n'+l)! \- \Gamma(n+n'+l+3/2)}{[n! n'! \Gamma(n+l+3/2) \Gamma(n'+l+3/2)]^{1/2}} \times \\ \sum_{i=0}^{n+n'+l} \frac{(-)^i G^{-i-3/2}}{i! (n+n'+l-i)!}
\end{multline}
Finally, the direct mean field associated with the central terms writes
\begin{equation} \label{directc}
\Gamma^{{C}}_{aa}\vert_{{D}} = \sum_{n_b l_b t_b} (2W + B - (2H+M) \delta_{t_a t_b}) S^{{C}}_{n_a l_a n_b l_b} \vert_{{D}}
\end{equation}

The calculation of the exchange mean field associated with the central terms is similar. One has
\begin{equation}
\Gamma^{{C}}_{aa}\vert_{{E}} = \sum_{r_b u_b} \mel*{r_a r_b}{V(r_{12})}{r_b r_a} \! \! \mel*{u_a u_b}{W_{{E}}}{u_b u_a}
\end{equation}
Using Eq.\eqref{combipara} in the particular case of the HFR approximation, one directly gets for the spin-isospin part:
\begin{multline} \label{exchWBHM}
\sum_{s_b} \! \mel*{s_a t_a \, s_b t_b}{\mathcal{P}}{s_b t_b \, s_a t_a} \\ = 2M + H - (2B+W) \delta_{t_a t_b}
\end{multline}
The exchange spatial part is obtained in the same way as the direct part, with the last three indices of the spatial TBME reversed. Following the same steps, 
one eventually gets:
\begin{multline}
\sum_{m_{lb}} \mel*{n_a l_a m_{la}  n_b l_b m_{lb}}{V(r_{12})}{n_b l_b m_{lb}  n_a l_a m_{la}} \\ = S^{\text{C}}_{n_a l_a n_b l_b} \vert_{{E}}
\end{multline}
where one has defined
\begin{multline}
S^{\text{C}}_{n_a l_a n_b l_b} \vert_{{E}} = \frac{1}{4 \pi} (2l_b+1) 
\sum_{n_\mu  n_\nu l_\mu} (2l_\mu+1) \times \\ \begin{pmatrix} l_a & l_b & l_\mu \\ 0 & 0 & 0 \end{pmatrix}^2 T_{(n_a l_a) (n_b l_b)}^{(n_\mu l_\mu)} T_{(n_b l_b) (n_a l_a)}^{(n_\nu l_\mu)} I_{n_\mu n_\nu l_\mu}
\end{multline}
where the quantity $I$ is defined in Eq.\eqref{Jsph}. Finally, the exchange mean field associated with the central terms writes:
\begin{equation} \label{exchangec}
\Gamma^{\text{C}}_{aa}\vert_{{E}} = \sum_{n_b l_b t_b} (2M + H - (2B+W) \delta_{q_a q_b}) S^{\text{C}}_{n_a l_a n_b l_b} \vert_{{E}}
\end{equation}
One notes that the ranges of values taken by the quantum numbers $n_\mu$, $n_\nu$ and $l_\mu$ appearing in the equations above can easily be deduced from Eq.\eqref{condY} and Eq.\eqref{condTalmansph}. In the expressions \eqref{directc} and \eqref{exchangec}, one has not performed the summation on the isospin with 
the future purpose of separating the proton and neutron contributions.

\paragraph{Density-dependent contribution}

One continues with the derivation of the direct mean field associated with the density-dependent term. The density-dependent term is given by Eq.\eqref{ddintsph}, where one uses now the shorthand notation for the spin-isospin part \eqref{mathP}, which is precisely the one of the central terms. The spatial and spin-isospin degrees of freedom split up in such a way that:
\begin{multline}
\Gamma^{\text{DD}}_{aa}\vert_{{D}} = \sum_{r_b u_b} \mel*{r_a r_b}{V(r_{12}) D[\rho]}{r_a r_b} \times \\ \mel*{u_a u_b}{\mathcal{P}}{u_a u_b}
\end{multline}
Since the spin-isospin part of the density-dependent term is exactly the same as the one of the central terms \eqref{SIcentral}, one immediately finds:
\begin{multline}
\sum_{s_b}  \mel*{s_a t_a \, s_b t_b}{\mathcal{P}_{{D}}}{s_a t_a \, s_b t_b}  = \\ 2W + B - (2H+M) \delta_{t_a t_b}
\end{multline}
On the other hand, the spatial TBMEs associated with the density-dependent term have been calculated in Eq.\eqref{densitys}. Integrating the angular parts of the integrals appearing in Eq.\eqref{densitys} and using the definition of the integral over three spherical harmonics \eqref{product3SH}, along with the properties of the Wigner-$3j$ symbols, permits to simplify the spatial part according to:
\begin{multline}
\sum_{m_{lb}} \! \mel*{n_a l_a m_{la} \, n_b l_b m_{lb}}{V(r_{12}) D[\rho]}{n_a l_a m_{la} \, n_b l_b m_{lb}} \\ = S^{\text{DD}}_{n_a l_a n_b l_b} \vert_{{D}}
\end{multline}
where one has defined
\begin{multline}
S^{\text{DD}}_{n_a l_a n_b l_b} \vert_{{D}} \equiv \frac{1}{4 \pi} (2l_b+1) \sum_{n_\mu n_\nu} T_{(n_a l_a) (n_a l_a)}^{(n_\mu 0)} \times \\ T_{(n_b l_b) (n_b l_b)}^{(n_\nu 0)} \frac{1}{2} \big[ J_{n_\mu n_\nu 0} + J_{n_\nu n_\mu 0} \big],
\end{multline}
with the integral
\begin{multline}
J_{n n' l} = K \lambda_{(n' l)} \int_0^{\infty} \dd{r} r^2 \phi_{(00)}(r) \phi_{(nl)}(r)  \times \\ \rho^{\alpha}(r) \phi_{(00)}(r, b\sqrt{g}) \phi_{(n'l)}(r, b\sqrt{g})
\end{multline}
which is nothing but the integral appearing in \eqref{intdensitysph}, with $l_\mu = l_\nu$, that has already been evaluated numerically.
Finally, the direct mean field associated with the density-dependent term writes:
\begin{equation} \label{directdd}
\Gamma^{\text{DD}}_{aa}\vert_{{D}} = \sum_{n_b l_b t_b} (2W + B - (2H+M) \delta_{t_a t_b}) S^{\text{DD}}_{n_a l_a n_b l_b} \vert_{{D}}
\end{equation}

The calculation of the exchange mean field associated with the density-dependent term is similar. One obtains:
\begin{equation}
\Gamma^{\text{C}}_{aa}\vert_{{E}} = \sum_{r_b u_b} \mel*{r_a r_b}{V(r_{12}) D[\rho]}{r_b r_a} \! \! \mel*{u_a u_b}{\mathcal{P}}{u_b u_a} \!.
\end{equation}
Since the spin-isospin part of the density-dependent term is exactly the same as the one of the central terms \eqref{exchWBHM}, one immediately finds: 
\begin{equation}
\sum_{s_b} \! \mel*{s_a t_a \, s_b t_b}{\mathcal{P}}{s_b t_b \, s_a t_a} \! = 2M + H - (2B+W) \delta_{t_a t_b}
\end{equation}
The exchange spatial part is obtained in the same way as the direct part, with the last three indices of the spatial TBME reversed. Following the same steps, 
one eventually gets:
\begin{multline}
\sum_{m_{lb}} \! \mel*{n_a l_a m_{la} \, n_b l_b m_{lb}}{V(r_{12}) D[\rho]}{n_b l_b m_{lb} \, n_a l_a m_{la}} \\ = S^{\text{DD}}_{n_a l_a n_b l_b} \vert_{{E}}
\end{multline}
where 
\begin{multline}
S^{\text{DD}}_{n_a l_a n_b l_b} \vert_{{E}} = \frac{1}{4 \pi} (2l_b+1) \sum_{n_\mu  n_\nu l_\mu} (2l_\mu+1) \times \\ \begin{pmatrix} l_a & l_b & l_\mu \\ 0 & 0 & 0 \end{pmatrix}^2 T_{(n_a l_a) (n_b l_b)}^{(n_\mu l_\mu)} T_{(n_b l_b) (n_a l_a)}^{(n_\nu l_\mu)} \frac{1}{2} \big[ J_{n_\mu n_\nu l_\mu} + J_{n_\nu n_\nu l_\mu} \big]
\end{multline}
where the integral $J$ is defined in \eqref{intcentralJ}. Finally, the exchange mean field associated with the density-dependent term writes:
\begin{equation} \label{exchangedd}
\Gamma^{\text{DD}}_{aa}\vert_{{E}} = \sum_{n_b l_b q_b} (2M + H - (2B+W) \delta_{q_a q_b}) S^{\text{DD}}_{n_a l_a n_b l_b} \vert_{{E}}
\end{equation}
Note that the ranges of values taken by the quantum numbers $n_\mu$, $n_\nu$ and $l_\mu$ appearing in the equations above can easily be deduced from Eq.\eqref{condY} and Eq.\eqref{condTalmansph}. As for the central terms, in the expressions \eqref{directdd} and \eqref{exchangedd}, one has not performed the summation on the isospin with the future purpose of separating the proton and neutron contributions.

\paragraph{Tensor contribution} \label{tensorHFR}

Now, one looks at the derivation of the direct mean field associated with the tensor term. The tensor interaction is given by Eq.\eqref{tensorintsph}, 
where one uses the following shorthand notation for the spin-isospin part:
\begin{equation} \label{shorthandTS}
\mathcal{P} \equiv W - H P_{\tau}.
\end{equation}
Separating the spatial and spin-isospin degrees of freedom, one obtains:
\begin{multline}
\Gamma^{\text{T}}_{aa}\vert_{{D}} = \sum_{r_b u_b} \mel*{r_a r_b}{V(r_{12}) [\hat{r}_{12} \otimes \hat{r}_{12}]^{(2)}_{-k}}{r_a r_b} \times \\ \mel*{u_a u_b}{\mathcal{P}[\vec{\sigma}_1 \otimes \vec{\sigma}_2]^{(2)}_k}{u_a u_b}
\end{multline}
Then, one evaluates the spin-isospin TBMEs of the tensor term. The isospin TBMEs are trivial and read, in general:
\begin{equation} \label{tensorisospin}
\mel*{t_a t_b}{\mathcal{P}}{t_c t_d} = W \delta_{t_a t_c} \delta_{t_b t_d}  - H \delta_{t_a t_d} \delta_{t_b t_c}
\end{equation}
As for the spin TBMEs, one has calculated their general expression:
\begin{multline} \label{spinTS}
\mel*{s_a \ s_c}{[\vec{\sigma}_1 \otimes \vec{\sigma}_2]^{(2)}_k}{s_b \ s_d} = 4 s_a s_c \bigg\{ \delta_{s_a s_b} \times \\ \bigg[ \sqrt{\frac{2}{3}} \delta_{s_c s_d} \delta_{k, 0} - \delta_{s_c, -s_d} \delta_{k, 2s_c} \bigg] \\
- \delta_{s_a, -s_b} \bigg[ \delta_{s_c s_d} \delta_{k, 2s_a} - \delta_{s_c,  -s_d} \times \\ 
\bigg( \sqrt{\frac{2}{3}} \delta_{s_a, -s_c} \delta_{k, 0} + 2 \delta_{s_a s_c} \delta_{k, 4s_a} \bigg)  \bigg] \bigg\}
\end{multline}
In the particular case of the HFR approximation, 
this expression reduces to:
\begin{equation}
\mel*{s_a s_b}{[\vec{\sigma}_1 \otimes \vec{\sigma}_2]^{(2)}_k}{s_a s_b} = 4 \sqrt{\frac{2}{3}} s_a s_c \delta_{k,0}
\end{equation}
so that the spin part of the tensor direct mean field vanishes, i.e.:
\begin{equation}
\sum_{s_b} \! \mel*{s_a s_b}{[\vec{\sigma}_1 \otimes \vec{\sigma}_2]^{(2)}_k}{s_a s_b} = 0
\end{equation}
Thus, at the HFR approximation, the direct tensor mean field is zero.

The calculation of the exchange mean field associated with the tensor term is similar. One obtains
\begin{multline}
\Gamma^{\text{T}}_{aa}\vert_{{E}} = \sum_{r_b u_b} \mel*{r_a r_b}{V(r_{12}) [\hat{r}_{12} \otimes \hat{r}_{12}]^{(2)}_{-k}}{r_b r_a} \times \\ \mel*{u_a u_b}{\mathcal{P}[\vec{\sigma}_1 \otimes \vec{\sigma}_2]^{(2)}_k}{u_b u_a}
\end{multline}
Again, from Eq.\eqref{spinTS}, one deduces the following spin TBMEs at the HFR approximation:
\begin{multline}
\mel*{s_a s_b}{[\vec{\sigma}_1 \otimes \vec{\sigma}_2]^{(2)}_k}{s_b s_a} = 4 \sqrt{\frac{2}{3}} s_a s_c \delta_{k,0} \times \\  (\delta_{s_a s_b} + \delta_{s_a,-s_b})
\end{multline}
so that the spin part of the tensor exchange mean field vanishes as well, i.e.:
\begin{equation}
\sum_{s_b} \! \mel*{s_a s_b}{[\vec{\sigma}_1 \otimes \vec{\sigma}_2]^{(2)}_k}{s_b s_a} = 0
\end{equation}
Thus, at the HFR approximation, the exchange tensor mean field is also zero. It follows that, the tensor mean field vanishes at the HFR approximation. \footnote{Here, the HFR approximation is essential to infer that the tensor term gives no contribution. Indeed, at the less restrictive HF approximation, the exchange tensor mean field does not vanish.} It is therefore not necessary to calculate the spatial contribution to the tensor mean field.

\paragraph{Spin-orbit contribution} \label{SOHFR}

In the HFR emulator, the spin-orbit is not introduced to simplify at maximum the model. In spite of this,
one discusses anyway the spin properties of the mean field associated with the spin-orbit term. One starts with the direct mean field. The spin-orbit interaction is given by Eq.\eqref{SOintsph}, where the shorthand notation for the spin-isospin part \eqref{shorthandTS} is used, which is precisely the one of the tensor term. Separating the spatial and spin-isospin degrees of freedom, one obtains:
\begin{multline}
\Gamma^{\text{SO}}_{aa}\vert_{{D}} = \sum_{r_b u_b} \mel*{r_a r_b}{V(r_{12}) [\vec{r}_{12} \otimes \vec{\nabla}_{12}]^{(1)}_{-k}}{r_a r_b} \times \\ \mel*{u_a u_b}{\mathcal{P}[\vec{\sigma}_1 + \vec{\sigma}_2]^{(1)}_k}{u_a u_b}
\end{multline}
One evaluates the spin-isospin TBMEs of the spin-orbit term. The isospin TBMEs are precisely the ones of the tensor term \eqref{tensorisospin}. As for the spin TBME, one has calculated their general expression:
\begin{multline} \label{spinSO}
\mel*{s_a \ s_c}{[\vec{\sigma}_1 + \vec{\sigma}_2]^{(1)}_k}{s_b \ s_d} = 2 (s_a + s_c) \delta_{s_a s_b} \delta_{s_c s_d} \delta_{k,0} \\ - \sqrt{2}  k \big[ \delta_{s_a s_b} \delta_{s_c, -s_d} \delta_{k, 2s_c} + \delta_{s_a, -s_b} \delta_{s_c s_d} \delta_{k, 2s_a} \big]
\end{multline}
In the particular case of the HFR approximation, this expression reduces to:
\begin{equation}
\mel*{s_a s_b}{[\vec{\sigma}_1 + \vec{\sigma}_2]^{(1)}_k}{s_a s_b} = 2 (s_a+s_b) \delta_{k,0}
\end{equation}
so that the spin part of the spin-orbit direct mean field simply reads:
\begin{equation} \label{SOdspin}
\sum_{s_b} \! \mel*{s_a s_b}{[\vec{\sigma}_1 \otimes \vec{\sigma}_2]^{(2)}_k}{s_a s_b} = 4 s_a \delta_{k,0}
\end{equation}

The calculation of the exchange mean field associated with the spin-orbit term is similar. One obtains:
\begin{multline}
\Gamma^{\text{SO}}_{aa}\vert_{{E}} = \sum_{r_b u_b} \mel*{r_a r_b}{V(r_{12}) [\vec{r}_{12} \otimes \vec{\nabla}_{12}]^{(1)}_{-k}}{r_b r_a} \times \\ \mel*{u_a u_b}{\mathcal{P}[\vec{\sigma}_1 + \vec{\sigma}_2]^{(1)}_k}{u_b u_a} .
\end{multline}
Again, from Eq.\eqref{spinSO}, one deduces the following spin TBMEs at the HFR approximation:
\begin{equation}
\mel*{s_a s_b}{[\vec{\sigma}_1 + \vec{\sigma}_2]^{(1)}_k}{s_b s_a} = 2 (s_a+s_b) \delta_{s_a s_b} \delta_{k,0}
\end{equation}
so that the spin part of the spin--orbit exchange mean field vanishes, i.e. \
\begin{equation}
\sum_{s_b} \! \mel*{s_a s_b}{[\vec{\sigma}_1 + \vec{\sigma}_2]^{(1)}_k}{s_b s_a} = 0
\end{equation}
Thus, at the HFR approximation, the spin-orbit exchange mean field is zero, since its spin part is. This is not the case of the direct mean field: its spin part does not vanish and there is no reason for the spatial part to vanish either. 
The only thing that can be infered regarding the direct spatial part is that $k = 0$ since one can deduce from the ranges of values taken by the quantum numbers appearing in Eq.\eqref{SO_1} and Eq.\eqref{SO_2}, that $k = m_{lc} + m_{ld} - m_{la} - m_{lb}$.

\subsubsection{Rearrangement field} \label{rearrangementgeneral}

At the HFR approximation, the rearrangement field is given by \eqref{rfHFB}.  Only the density-dependent term generate such a rearrangement field since it depends on the local nuclear density. By definition, in the spherical HO representation, this density can be written as:
\begin{equation}
\rho(\vec{r}) = \sum_{\sigma \tau} \sum_{ab} \Phi_a^*(\vec{r}, \sigma, \tau) \Phi_b(\vec{r}, \sigma, \tau) \rho_{ba}
\end{equation}
where the wave functions are the ones of the spherical HO, defined by Eq.\eqref{wavefunctionsph}. Thus, the rearrangement field involves the derivative
\begin{multline} \label{rearrangementfield}
\pdv{\rho(\vec{r})}{\rho_{aa}} = \sum_{\sigma \tau} \Phi^*_a(\vec{r}, \sigma, \tau) \Phi_a(\vec{r}, \sigma, \tau) \\
= \phi^*_{r_a}(\vec{r}) \phi_{r_a}(\vec{r})
\end{multline}
where one has used the orthogonality of the spin and isospin wave functions (contained in Eq.\eqref{orthogonalitysph}) when going from the first to the second line. Then, it appears that the rearrangement field is independent of the isospin $t_a$.  Because of this property, one will not need to calculate the rearrangement field for the quantities one is interested in, as one will see.

\subsection{Restricted Hartree-Fock energy}

In this section, one will evaluate both the kinetic and potential contributions to the restricted Hartree-Fock total binding energy.

\subsubsection{Kinetic energy}

At the HFR approximation, the kinetic energy $\mathcal{E}_{{K}}^{\text{HFR}}$ is explicitly given by:
\begin{multline}
\mathcal{E}_{{K}}^{\text{HFR}} = \sum_{a} K_{aa} \rho_{aa} \\ = \sum_{\substack{n_a l_a m_{la} \\ s_a t_a}} \frac{\hbar \omega}{2} \Big( 2 n_a + l_a + \frac{3}{2} \Big) \rho_{aa}
\end{multline}
where one has used the expression of the kinetic field derived in Eq.\eqref{kineticfieldHFR}. One ends up with:
\begin{multline}
\mathcal{E}_{{K}}^{\text{HFR}} = 2 \sum_{(n_a l_a)_\pi} (2l_a+1) \Big( 2 n_a + l_a + \frac{3}{2} \Big) \frac{\hbar \omega}{2} \\ + 2 \sum_{(n_a l_a)_\nu} (2l_a+1) \Big( 2 n_a + l_a + \frac{3}{2} \Big) \frac{\hbar \omega}{2}
\end{multline}
The degeneracy factor $2(2l_a+1)$ appears because a given state has the same kinetic energy no matter the values of its quantum numbers $s_a$ and $m_{la}$, at the origin of factors $2$ and $(2l_a+1)$, respectively. One notes that the summations concern the proton $\pi$ and neutron $\nu$ occupied states $(n_a,l_a)$, in the spherical HO basis. One needs this kinetic energy for the oxygen $\isotope[16]{O}$ and zirconium $\isotope[90]{Zr}$. By adding the successive contributions of each shell up to the proton and neutron Fermi levels according to the above relation, one obtains:
\begin{subequations} \label{EHFRnuclei}
\begin{align}
\mathcal{E}_{{K}}^{\text{HFR}} \big[\isotope[16]{O} \big] /(A=16) & = \frac{9}{8} \hbar \omega \\
\mathcal{E}_{{K}}^{\text{HFR}} \big[\isotope[90]{Zr} \big] /(A=90) & = \frac{71}{36} \hbar \omega 
\end{align}
\end{subequations}
where the kinetic energies have been normalized by the number of nucleons $A$ in the corresponding nuclei.

\subsubsection{Potential energy}

For the sake of clarity, one will decompose the potential energy into a direct and an exchange component in the following,  
as $\mathcal{E}_{{P}}^\text{HFR} \equiv \mathcal{E}_{{P}}\vert_{{D}} + \mathcal{E}_{{P}}\vert_{{E}}$, where:
\begin{subequations}
\begin{align}
\mathcal{E}_{{P}}\vert_{{D}} & \equiv  \frac{1}{2} \sum_{a} \Gamma_{aa}\vert_{{D}} \rho_{aa} \\
\mathcal{E}_{{P}}\vert_{{E}} & \equiv \frac{1}{2} \sum_{a} \Gamma_{aa}\vert_{{E}} \rho_{aa}.
\end{align}
\end{subequations}
In the following, one evaluates these contributions for the fields of the generalized Gogny interaction \eqref{gognyDG}.One will use the notation:
\begin{equation}
\widetilde{S}_{n_a l_a n_b l_b} \equiv (2l_a+1) S_{n_a l_a n_b l_b}
\end{equation}
for the direct and exchange contributions of both the central and density-dependent terms, since they do not explicitly depend on the quantum number $m_{la}$.

\paragraph{Central contribution}

One starts by deriving the direct component of the potential energy associated with the central terms.  One has, using Eq.\eqref{directc}:
\begin{multline}
\mathcal{E}_{{P}}^{{D}} \vert_{{D}} = \frac{1}{2} \sum_{a} \Gamma^{{D}}_{aa}\vert_{{D}} \rho_{aa} \\ = \frac{1}{2} \sum_{s_a t_a t_b} \sum_{n_a l_a} \sum_{n_b l_b} (2M + B -(2H+M) \delta_{t_a t_b}) \times \\ \widetilde{S}_{n_a l_a n_b l_b}^{\text{C}} \vert_{{D}}
\end{multline}
Separating the proton and neutron shells, one obtains in the general case of an asymmetric nucleus:
\begin{multline} \label{Vasym}
\mathcal{E}_{{P}}^{\text{D}} \vert_{{D}} = (2W + B - 2H - M) \times \\ \sum_{(n_a l_a)_{\pi}} \sum_{(n_b l_b)_{\pi}} \widetilde{S}_{n_a l_a n_b l_b}^{\text{C}} \vert_{{D}} \\ + (2W + B - 2H - M) \sum_{(n_a l_a)_{\nu}} \sum_{(n_b l_b)_{\nu}} \widetilde{S}_{n_a l_a n_b l_b}^{\text{C}} \vert_{{D}} \\
+ 2 (2W + B) \sum_{(n_a l_a)_{\pi}} \sum_{(n_b l_b)_{\nu}} \widetilde{S}_{n_a l_a n_b l_b}^{\text{C}} \vert_{{D}}
\end{multline}
where one has taken advantage of the symmetry of the quantity $S_{n_a l_a n_b l_b}^{\text{C}}\vert_{{D}}$ under the exchange of indices $(n_a, l_a) \leftrightarrow (n_b, l_b)$, and where the summations over $(n_a l_a)_{\pi}$ and $(n_b l_b)_{\nu}$ run over the full set of quantum numbers characterizing the states occupied by protons $\pi$ and neutrons $\nu$. When the nucleus is symmetric, this set is the same for both proton and neutron quantum numbers. Then, the potential energy becomes:
\begin{equation} \label{Vsym}
\mathcal{E}_{{P}}^{\text{D}} \vert_{{D}} = (4W + 2B - 2H - M) F^{{D}}[X]
\end{equation}
where one has defined ($X$ denoting the nucleus under study):
\begin{equation}
F^{{D}}[X] \equiv 2 \sum_{(n_a l_a)_{\pi}} \sum_{(n_b l_b)_{\nu}} \widetilde{S}_{n_a l_a n_b l_b}^{\text{C}} \vert_{{D}}.
\end{equation}
This identity holds for the symmetric $\isotope[16]{O}$, but, in principle, not for the asymmetric $\isotope[90]{Zr}$. 
One considers now an asymmetric nucleus that has more neutrons than protons, i.e.\ $\nu_{{max}} = \pi_{{max}} + \nu_+ $, where $\nu_{{max}}$ and $\pi_{{max}}$ are the numbers of neutrons and protons respectively, and where $\nu_+$ is the neutron excess. This choice is conventional, the proton excess is treated in a similar way. From Eq.\eqref{Vasym}, one gets:
\begin{multline}
\mathcal{E}_{{P}}^{\text{D}} \vert_{{D}} = 2 (4W + 2B - 2H - M) \times \\ \sum_{(n_a l_a)_{\pi}} \sum_{(n_b l_b)_{\nu}} \widetilde{S}_{n_a l_a n_b l_b}^{\text{C}} \vert_{{D}} \\ + (2W + B - 2H - M) \times \\ \sum_{(n_a l_a)_{\nu = \pi_{{max}} +1}}^{\nu_{{max}}} \sum_{(n_b l_b)_{\nu = \pi_{{max}} +1}}^{\nu_{{max}}} \widetilde{S}_{n_a l_a n_b l_b}^{\text{C}} \vert_{{D}}
\end{multline}
One observes that a new contribution, proportional to $(2W + B - 2H - M)$, is added in the case of an asymmetric nucleus. However, this contribution is 
negligible compared to that of $(4W + 2B - 2H - M) $ for weakly asymmetric nuclei like the $\isotope[90]{Zr}$, since then $\nu_+ - 1 \ll \nu_{{max}}, \pi_{{max}}$. As a consequence, one will legitimately assume that the identity \eqref{Vsym} also holds for the $\isotope[90]{Zr}$ in the fitting code.

The calculation of the exchange component of the potential energy associated with the central terms is similar. Using Eq.\eqref{exchangec}, one obtains:
\begin{multline} 
\mathcal{E}_{{P}}^{\text{C}} \vert_{{E}} = \frac{1}{2} \sum_{a} \Gamma^{{D}}_{aa}\vert_{{E}} \rho_{aa} \\ = \frac{1}{2} \sum_{s_a t_a t_b} \sum_{n_a l_a} \sum_{n_b l_b} (2M + B -(2H+M) \delta_{t_a t_b}) \times \\ \widetilde{S}_{n_a l_a n_b l_b}^{\text{C}} \vert_{{E}}
\end{multline}
For a symmetric nucleus like the $\isotope[16]{O}$, one ends up with:
\begin{equation} \label{Vsym2}
\mathcal{E}_{{P}}^{\text{C}} \vert_{{E}} = (4M + 2H - 2B - W) F^{{E}}[X]
\end{equation}
where one has defined ($X$ denoting the nucleus under study),
\begin{equation}
F^{{E}}[X] \equiv 2 \sum_{(n_a l_a)_{\pi}} \sum_{(n_b l_b)_{\nu}} \widetilde{S}_{n_a l_a n_b l_b}^{\text{C}} \vert_{{E}}
\end{equation}
while, for an asymmetric nucleus with a neutron excess, one finds:
\begin{multline}
\mathcal{E}_{{P}}^{\text{C}} \vert_{{E}} = 2 (4M + 2H - 2B - W) \times \\ \sum_{(n_a l_a)_{\pi}} \sum_{(n_b l_b)_{\nu}} \widetilde{S}_{n_a l_a n_b l_b}^{\text{C}} \vert_{{E}} + (2M + H - 2B - W) \times \\ \sum_{(n_a l_a)_{\nu = \pi_{{max}} +1}}^{\nu_{{max}}} \sum_{(n_b l_b)_{\nu = \pi_{{max}} +1}}^{\nu_{{max}}} \widetilde{S}_{n_a l_a n_b l_b}^{\text{C}} \vert_{{E}}
\end{multline}
Here again, the contribution proportional to $(2M + H - 2B - W)$ is negligible compared to that of $(4M + 2H - 2B - W)$ for weakly asymmetric nuclei like the $\isotope[90]{Zr}$. One will then assume that the identity \eqref{Vsym2} also holds for the $\isotope[90]{Zr}$ in the fitting code.

\paragraph{Density-dependent contribution}

The procedure introduced above is similar for the density-dependent interaction. Using Eq.\eqref{directdd}, one finds that the direct component of the potential energy associated with the density-dependent term reads:
\begin{multline}
\mathcal{E}_{\text{P}}^{\text{DD}} \vert_{\text{D}} = \frac{1}{2} \sum_{a} \Gamma^{\text{DD}}_{aa}\vert_{\text{D}} \rho_{aa} \\
= \frac{1}{2} \sum_{s_a t_a t_b} \sum_{n_a l_a} \sum_{n_b l_b} (2M + B -(2H+M) \delta_{t_a t_b}) \times \\ \widetilde{S}_{n_a l_a n_b l_b}^{\text{DD}} \vert_{\text{D}}
\end{multline}
Separating the proton and neutron shells, one obtains in the general case of an asymmetric nucleus:
\begin{multline} \label{Vasym2}
\mathcal{E}_{\text{P}}^{\text{DD}} \vert_{\text{D}} = (2W + B - 2H - M) \times \\ \sum_{(n_a l_a)_{\pi}} \sum_{(n_b l_b)_{\pi}} \widetilde{S}_{n_a l_a n_b l_b}^{\text{DD}} \vert_{\text{D}} \\ + (2W + B - 2H - M) \sum_{(n_a l_a)_{\nu}} \sum_{(n_b l_b)_{\nu}} \widetilde{S}_{n_a l_a n_b l_b}^{\text{DD}} \vert_{\text{D}} \\
+ 2 (2W + B) \sum_{(n_a l_a)_{\pi}} \sum_{(n_b l_b)_{\nu}} \widetilde{S}_{n_a l_a n_b l_b}^{\text{DD}} \vert_{\text{D}}
\end{multline}
where one has taken advantage of the symmetry of the quantity $S_{n_a l_a n_b l_b}^{\text{DD}}\vert_{\text{D}}$ under the exchange of indices $(n_a, l_a) \leftrightarrow (n_b, l_b)$. When the nucleus is symmetric, this set is the same for both proton and neutron quantum numbers. Then, the potential energy becomes:
\begin{multline} \label{Vsymdd}
\mathcal{E}_{\text{P}}^{\text{DD}} \vert_{\text{D}} = 2 (4W + 2B - 2H - M) \times \\ \sum_{(n_a l_a)_{\pi}} \sum_{(n_b l_b)_{\nu}} \widetilde{S}_{n_a l_a n_b l_b}^{\text{DD}} \vert_{\text{D}}
\end{multline}
Following the procedure developed for the central terms, one finds, for an asymmetric nucleus with a neutron excess:
\begin{multline}
\mathcal{E}_{\text{P}}^{\text{DD}} \vert_{\text{D}} = 2 (4W + 2B - 2H - M) \times \\ \sum_{(n_a l_a)_{\pi}} \sum_{(n_b l_b)_{\nu}} \widetilde{S}_{n_a l_a n_b l_b}^{\text{DD}} \vert_{\text{D}} \\ + (2W + B - 2H - M) \times \\  \sum_{(n_a l_a)_{\nu = \pi_{\text{max}} +1}}^{\nu_{\text{max}}} \sum_{(n_b l_b)_{\nu = \pi_{\text{max}} +1}}^{\nu_{\text{max}}} \widetilde{S}_{n_a l_a n_b l_b}^{\text{DD}} \vert_{\text{D}}
\end{multline}
As it was the case for the central terms, one observes that a new contribution, proportional to $(2W + B - 2H - M)$, is added in the case of an asymmetric nucleus. Here again, one will neglect this contribution for weakly asymmetric nuclei like the $\isotope[90]{Zr}$ in the fitting code, so that the relation \eqref{Vsymdd} holds for this nucleus as well.

The calculation of the exchange component of the potential energy associated with the density-dependent is similar. Using Eq.\eqref{exchangedd}, one obtains:
\begin{multline} \label{Vsym2dd}
\mathcal{E}_{\text{P}}^{\text{DD}} \vert_{\text{E}} = \frac{1}{2} \sum_{a} \Gamma^{\text{DD}}_{aa}\vert_{\text{E}} \rho_{aa} \\ = \frac{1}{2} \sum_{s_a q_a q_b} \sum_{n_a l_a} \sum_{n_b l_b} (2M + B -(2H+M) \delta_{q_a q_b}) \times \\ \widetilde{S}_{n_a l_a n_b l_b}^{\text{DD}} \vert_{\text{E}}
\end{multline}
For a symmetric nucleus like the $\isotope[16]{O}$, one ends up with:
\begin{multline}
\mathcal{E}_{\text{P}}^{\text{DD}} \vert_{\text{E}} = 2 (4M + 2H - 2B - W) \times \\
\sum_{(n_a l_a)_{\pi}} \sum_{(n_b l_b)_{\nu}} \widetilde{S}_{n_a l_a n_b l_b}^{\text{DD}} \vert_{\text{E}}
\end{multline}
while, for an asymmetric nucleus with a neutron excess, one finds:
\begin{multline}
\mathcal{E}_{\text{P}}^{\text{DD}} \vert_{\text{E}} = 2 (4M + 2H - 2B - W) \times \\\sum_{(n_a l_a)_{\pi}} \sum_{(n_b l_b)_{\nu}} \widetilde{S}_{n_a l_a n_b l_b}^{\text{DD}} \vert_{\text{E}} \\ + (2M + H - 2B - W) \times \\ \sum_{(n_a l_a)_{\nu = \pi_{\text{max}} +1}}^{\nu_{\text{max}}} \sum_{(n_b l_b)_{\nu = \pi_{\text{max}} +1}}^{\nu_{\text{max}}} \widetilde{S}_{n_a l_a n_b l_b}^{\text{DD}} \vert_{\text{E}}
\end{multline}
Here again, the contribution proportional to $(2M + H - 2B - W)$ is negligible compared to that of $(4M + 2H - 2B - W)$ for weakly asymmetric nuclei like the $\isotope[90]{Zr}$. One will then assume that the identity \eqref{Vsym2dd} also holds for the $\isotope[90]{Zr}$ in the fitting code.

\paragraph{Tensor contribution}

We have shown in subsection \ref{tensorHFR} that the mean field associated with the tensor interaction vanishes at the HFR approximation.  Thus, the tensor interaction does not contribute to the HFR potential energy $\mathcal{E}_{\text{P}}^{\text{HFR}}$.

\paragraph{Spin-orbit contribution}

One has shown in subsection \ref{SOHFR} that the exchange mean field associated with the spin-orbit interaction vanishes at the HFR approximation. Thus, it brings no contribution to the potential energy. However, the direct mean field is not zero and may then contribute to the potential energy.
By definition, this direct component of the potential energy associated with the spin-orbit term reads:
\begin{equation}
\mathcal{E}_{\text{P}}^{\text{SO}} \vert_{\text{D}} \equiv \frac{1}{2} \sum_{a} \Gamma^{\text{SO}}_{aa}\vert_{\text{D}} \rho_{aa}
\end{equation}
According to \eqref{SOdspin}, the spin part of the above potential energy reads:
\begin{equation}
\sum_{s_a} 4 s_a \delta_{k,0} = 0
\end{equation}
Thus, the potential energy associated with the spin-orbit interaction vanishes at the HFR approximation, since the spin parts of both its direct and exchange components are zero. Even if the spin-orbit interaction had been considered at the HFR approximation, it would have provided no contribution to the potential energy $\mathcal{E}_{\text{P}}^{\text{HFR}}$.

\subsection{Calculation of the energy difference $\Delta \varepsilon$}

One of the quantities constrained in the fitting code is the energy difference between the neutron and proton $2s_{1/2}$ states in the calcium $\isotope[48]{Ca}$.
At the HFR approximation, it is written as:
\begin{equation} \label{Deltaeps}
\Delta \varepsilon \equiv \varepsilon_{2s_{1/2}}^{\nu} - \varepsilon_{2s_{1/2}}^{\pi}
\end{equation}
where $\varepsilon$ denotes the individual energy of the corresponding state at the HFR approximation, given by Eq.\eqref{epsHFR}. At this approximation, neither the kinetic field given by Eq.\eqref{kineticfieldHFR}, nor the rearrangement field (see Eq.\eqref{rearrangementfield} and the discussion below), depend on the isospin. Thus, only the neutron and proton mean fields contribute to $\Delta \varepsilon$, in such a way that:
\begin{equation}
\Delta \varepsilon = \Gamma_{2s_{1/2}}^{\nu} - \Gamma_{2s_{1/2}}^{\pi}.
\end{equation}
One evaluates now these contributions for the fields of the generalized Gogny interaction \eqref{gognyDG}.

\subsubsection{Central interaction}

One starts by evaluating the contribution to this energy difference $\Delta \varepsilon$ coming from the central terms.  It is:
\begin{equation}
\Delta \varepsilon^{\text{C}} = \Gamma_{2s_{1/2}}^{\text{C} \, \nu} - \Gamma_{2s_{1/2}}^{\text{C} \, \pi}
\end{equation}
where the expressions of the central mean fields are deduced from Eq.\eqref{directc} and Eq.\eqref{exchangec}. These contributions eventually furnish, after filling the shells, for the neutron mean field:
\begin{multline}
\Gamma_{2s_{1/2}}^{\text{C} \, \nu} = \sum_{(n_b l_b)_{\nu}} \big[ \big(2W+B-(2H+M) \big) \widetilde{S}^{\text{C}}_{10 \,  n_b l_b} \vert_{\text{D}}  \\ - \big(2M+H-(2B+W) \big) \widetilde{S}^{\text{C}}_{10 \, n_b l_b} \vert_{\text{E}} \big] \\
  + \sum_{(n_b l_b)_{\pi}} \big[ (2W+B) \widetilde{S}^{\text{C}}_{10 \,  n_b l_b} \vert_{\text{D}} - (2M+H) \widetilde{S}^{\text{C}}_{10 \, n_b l_b} \vert_{\text{E}} \big]
\end{multline}
and, for the proton mean field,
\begin{multline}
\Gamma_{2s_{1/2}}^{\text{C} \, \pi} = \sum_{(n_b l_b)_{\nu}} \big[ (2W+B) \widetilde{S}^{\text{C}}_{10 \,  n_b l_b} \vert_{\text{D}} \\ - (2M+H) \widetilde{S}^{\text{C}}_{10 \, n_b l_b} \vert_{\text{E}} \big] \\ + \sum_{(n_b l_b)_{\pi}} \big[ \big(2W+B-(2H+M) \big) \widetilde{S}^{\text{C}}_{10 \,  n_b l_b} \vert_{\text{D}} \\
 - \big( (2M+H) - (2B+W) \big) \widetilde{S}^{\text{C}}_{10 \, n_b l_b} \vert_{\text{E}} \big]
\end{multline}
Gathering the results, one obtains:
\begin{multline}
\Delta \varepsilon^{\text{C}} = - (2H+M) \bigg[ \sum_{(n_b l_b)_{\nu}} \widetilde{S}^{\text{C}}_{10 \, n_b l_b} \vert_{\text{D}} \\ - \sum_{(n_b l_b)_{\pi}} \widetilde{S}^{\text{C}}_{10 \, n_b l_b} \vert_{\text{D}} \bigg] \\ + (2B+W) \bigg[ \sum_{(n_b l_b)_{\nu}} \widetilde{S}^{\text{C}}_{10 \, n_b l_b} \vert_{\text{E}} - \sum_{(n_b l_b)_{\pi}} \widetilde{S}^{\text{C}}_{10 \, n_b l_b} \vert_{\text{E}} \bigg]
\end{multline}
The $\isotope[48]{Ca}$ isotope has twenty protons and twenty-eight neutrons. The proton Fermi level is therefore $2s$ $(n=1, l=0)$, which is saturated, and the neutron Fermi level $1f$ $(n=0, l=3)$, which contains eight neutrons among the fourteen available slots. Then, the central contribution to the energy difference \eqref{Deltaeps} reads:
\begin{equation}
\Delta \varepsilon^{\text{C}} = (2H+M) f_{\text{D}} + (2B+W) f_{\text{E}}
\end{equation}
where one has defined
\begin{subequations}
\begin{align}
f_{\text{D}} \equiv - \frac{4}{7} \widetilde{S}_{10 \, 03}^{\text{C}} \vert_{\text{D}} \\
f_{\text{E}} \equiv + \frac{4}{7} \widetilde{S}_{10 \, 03}^{\text{C}} \vert_{\text{E}}
\end{align}
\end{subequations}
with the prefactors traducing the occupation of the neutron Fermi level.

\subsubsection{Density-dependent interaction}

The calculation of the contribution from the density-dependent term to $\Delta \varepsilon$ is similar to the central terms. One has:
\begin{equation}
\Delta \varepsilon^{\text{DD}} = \Gamma_{2s_{1/2}}^{\text{DD} \, \nu} - \Gamma_{2s_{1/2}}^{\text{DD} \, \pi},
\end{equation}
where the expressions of the density-dependent mean fields is deduced from Eq.\eqref{directdd} and Eq.\eqref{exchangedd}.  One eventually finds out, 
in a similar way as for the central terms:
\begin{equation}
\Delta \varepsilon^{\text{DD}} = (2H+M) g_{\text{D}} + (2B+W) g_{\text{E}}
\end{equation}
where one has defined
\begin{subequations}
\begin{align}
g_{\text{D}} \equiv - \frac{4}{7} \widetilde{S}_{10 \, 03}^{\text{DD}} \vert_{\text{D}} \\
g_{\text{E}} \equiv + \frac{4}{7} \widetilde{S}_{10 \, 03}^{\text{DD}} \vert_{\text{E}}.
\end{align}
\end{subequations}

\subsubsection{Tensor interaction} \label{DeltaepsTS}

One has shown in subsection \ref{tensorHFR} that the mean field associated with the tensor interaction vanishes at the HFR approximation. Thus, the tensor interaction does not contribute to the energy difference $\Delta \varepsilon$ at this approximation, $\Delta \varepsilon^{\text{T}} = 0$.

\subsubsection{Spin--orbit interaction} \label{DeltaepsSO}

One has shown in subsection \ref{SOHFR} that the exchange mean field associated with the spin-orbit interaction vanishes at the HFR approximation, but the 
direct one does not. Then, the spin-orbit interaction contributes to the energy difference $\Delta \varepsilon$ at the HFR approximation, $\Delta \varepsilon^{\text{SO}} \ne 0$. However, as already mentioned, one neglects this contribution in the fitting code. Otherwise, it should have been taken into account while distinguishing the HO states by their quantum numbers $j$, projections of the total angular momentum.

\section{Two-body matrix elements of the DG Gogny interaction in a spherical one-center HO basis}\label{anexa}

The objective of this Appendix is to give the expressions of the two-body matrix elements (TBMEs) of the DG interaction in spherical symmetry, which are
useful in the building of the emulator used in the fitting \ref{sec2}. The first section is dedicated to the recoupling scheme. Indeed, all the TBMEs have been coupled to the quantum numbers $(J, T)$ as in \cite{Brown1988}, since they are put into relation with the $(J, T)$-coupled TBMEs of \cite{Brown1988,Honma2004} in the inversion procedure of the fitting code.
The second section is devoted to the derivation of the TBMEs for both the common central and density-dependent terms \cite{Gogny1975b,Chappert2007,Berger1985}, and the newly introduced finite-range tensor and spin-orbit terms of DG. Actually those two last terms have already been computed in \cite{Gogny1975b} but in an cryptic way. One proposes here a more explicit derivation. One also details how the radial integral of the density-dependent term is implemented 
in the fitting code.

\subsection{Coupling to $(J,T)$ process} \label{couplingpart}

The TBMEs of the generalized Gogny interaction $v_{12}$, given in Eq.\eqref{gognyDG}, considered in the fitting code are coupled to the quantum numbers $(J,T)$, normalized and antisymmetrized (na), in the same way as in \cite{Brown1988}. They write:
\begin{multline} \label{recoupling}
\mel*{\widetilde{a} \widetilde{b} J T}{v_{12}}{\widetilde{c} \widetilde{d} J T}_{\text{na}} = \frac{1}{\sqrt{(1+\delta_{\widetilde{a} \widetilde{b}})(1+\delta_{\widetilde{c} \widetilde{d}})}} \times \\
\big[ \! \mel*{\widetilde{a} \widetilde{b} J T}{v_{12}}{\widetilde{c} \widetilde{d} J T} \\ + (-)^{j_c + j_d + J + T}  \mel*{\widetilde{a} \widetilde{b} J T}{v_{12}}{\widetilde{d} \widetilde{c} J T}  \big]
\end{multline}
where the states are defined by the set of quantum numbers
\begin{equation}
\ket{\widetilde{a}} = \ket{j_a n_a l_a}
\end{equation}
with $j_i$ the projection of the one-body total angular momentum $\vec{j}_i = \vec{l}_i + \vec{s}_i$, where $ \vec{l}_i$ and $\vec{s}_i$ are respectively the one-body orbital and intrinsic angular momenta. As for $J$ and $T$, they correspond to the quantum numbers associated with the two-body total angular momentum $\vec{J} = \vec{j}_a + \vec{j}_b = \vec{j}_c + \vec{j}_d$ and the two-body isospin operator, $\vec{T} = \vec{t}_a + \vec{t}_b = \vec{t}_c + \vec{t}_d$ (where the $\vec{t}_i$ are the one-body isospin operators), respectively. It is important to differentiate these states from the “pure” spherical HO states \eqref{HOstateS}. Note that the Kronecker delta $\delta_{\widetilde{a} \widetilde{b}} = \delta_{j_a j_b} \delta_{n_a n_b} \delta_{l_a l_b}$ notation has been introduced. 

One notices that those TBMEs are diagonal in $J$ and $T$ and do not depend on their projections $M_J$ and $M_T$, while in general they are written $\! \mel*{\alpha J M_J T M_T}{v_{12}}{\alpha' J' M'_J T' M'_T}$, where $\alpha$ and $\alpha'$ are the set of quantum numbers, here $\tilde{a} \tilde{b}$ and $\tilde{c} \tilde{d}$, characterizing the state. This is a consequence of the invariance of the two-body nuclear interaction $v_{12}$ under rotations in coordinate plus spin and isospin spaces. If the interaction is rotationally invariant, it commutes with $\vec{J}$, and in particular with $\vec{J}^2$, the component $J_z$ and the usual ladder operators $J_+$ and $J_-$. Dropping for a moment the isospin dependence, one can then write down \cite{Berger2008}:
\begin{multline} \label{sphJ1}
\mel*{\alpha J M_J}{v_{12}\vec{J}^2}{\alpha' J' M_J'} \\ - \! \mel*{\alpha J M_J}{\vec{J}^2 v_{12}}{\alpha' J' M_J'} = 0
\end{multline}
so that
\begin{multline}
\mel*{\alpha J M_J}{v_{12}}{\alpha' J' M_J'}  [J(J+1) - J'(J'+1)] = 0
\end{multline}
Therefore, the TBMEs of the interaction are zero unless one has $J=J'$, that is to say:
\begin{equation}
\mel*{\alpha J M_J}{v_{12}}{\alpha' J' M_J'} = \delta_{JJ'} \! \mel*{\alpha J M_J}{v_{12}}{\alpha' J M_J'} 
\end{equation}
Then, the relation \eqref{sphJ1} with $\vec{J}^2$ replaced by $J_z$ similarly provides:
\begin{equation}
\mel*{\alpha J M_J}{v_{12}}{\alpha' J M_J'} \! [M_J - M'_J] = 0
\end{equation}
Therefore, the TBMEs of the interaction are zero unless one also has $M_J=M'_J$, that is to say:
\begin{equation}
\mel*{\alpha J M_J}{v_{12}}{\alpha' J M_J'} = \delta_{M_J M'_J} \! \mel*{\alpha J M_J}{v_{12}}{\alpha' J M_J} 
\end{equation}
Finally, since the action of the ladder operator $J_+$ is:
\begin{equation}
J_+ \! \ket{J M_J-1} \equiv \sqrt{J(J+1) - M_J(M_J-1)} \! \ket{J M_J} 
\end{equation}
one deduces, for $-J+1 \le M_J \le J+1$,
\begin{equation}
\mel*{\alpha J M_J}{v_{12}}{\alpha' J M_J} = \frac{\mel*{\alpha J M_J}{v_{12} J_+}{\alpha' J M_J-1}}{\sqrt{J(J+1) - M_J(M_J-1)}}
\end{equation}
On the other hand, the TBME of the right-hand side can be written
\begin{multline}
\mel*{\alpha J M_J}{v_{12} J_+}{\alpha' J M_J-1} = \\ \sqrt{J(J+1) - M_J(M_J-1)} \times \\
\mel*{\alpha J M_J-1}{v_{12}}{\alpha' J M_J-1} 
\end{multline}
where one has used the fact that $v_{12}$ and $J_+$ commute as well as the identity $J_+ = J_-^{\dagger}$. Combining the last two equations, one obtains, for all allowed values of $M_J$, 
\begin{multline}
\mel*{\alpha J M_J}{v_{12}}{\alpha' J M_J} =  \mel*{\alpha J M_J-1}{v_{12}}{\alpha' J M_J-1} 
\end{multline}
 where $-J < M_J < J$. It is a proof of the $M_J$ independency of the TBMEs. 

As the TBMEs of the interaction are diagonal in $J$ and $M_J$, and do not depend on $M_J$, one can formally express them as:
\begin{multline}
\mel*{\alpha J M_J}{v_{12}}{\alpha' J' M_J'} = \delta_{JJ'} \delta_{M_J M'_J} \! \mel*{\alpha J}{v_{12}}{\alpha' J}
\end{multline}
Actually, the above calculations are a simple application of the Wigner-Eckart theorem to the interaction $v_{12}$, viewed as a scalar (i.e.\ a spherical tensor of rank $0$) \cite{Suhonen2007}, where the reduced TBMEs are conventionally related to the standard TBMEs by
\begin{equation} \label{WEconv}
\langle \alpha J \| v_{12} \| \alpha' J \rangle \equiv \sqrt{2J+1} \! \mel*{\alpha J}{v_{12}}{\alpha' J}
\end{equation}
The same procedure can be undertaken in the isospin space , which leads to the following relation:
\begin{equation}
\mel*{\alpha T M_T}{v_{12}}{\alpha' T' M_T'} = \delta_{TT'} \delta_{M_T M'_T} \! \mel*{\alpha T}{v_{12}}{\alpha' T}
\end{equation}

Thus, it remains to evaluate the contributions of each term of the generalized Gogny interaction to the non-normalized and non-antisymmetrized TBME $\! \mel*{\widetilde{a} \widetilde{b} J T}{v_{12}}{\widetilde{c} \widetilde{d} J T}$ and $\! \mel*{\widetilde{a} \widetilde{b} J T}{v_{12}}{\widetilde{d} \widetilde{c} J T}$. This is done in the next sub-parts.

\subsection{Central two-body matrix elements}

One starts by deriving the TBMEs of the central terms whose expression is recalled below:
\begin{equation} \label{centralintsph}
v_{12}^{\text{C}} = (W + B P_{\sigma} - H P_{\tau} - M P_{\sigma}P_{\tau}) V(r_{12})
\end{equation}
where $V(r_{12})$ is the Gaussian potential given by
\begin{equation} \label{Vr12}
V(r_{12}) \equiv \e^{-(\vec{r}_1 - \vec{r}_2)^2/\mu^2}.
\end{equation}
As explained previously, one needs to express the TBMEs with the conventions chosen in \cite{Brown1988}. The central TBMEs are therefore recoupled to $(J,T)$, normalized and antisymmetrized (na) according to \eqref{recoupling}, i.e.\
\begin{multline} \label{centralTBME}
\mel*{\widetilde{a} \widetilde{b} J T}{v_{12}^{\text{C}}}{\widetilde{c} \widetilde{d} J T}_{\text{na}} = \frac{1}{\sqrt{(1+\delta_{\widetilde{a} \widetilde{b}})(1+\delta_{\widetilde{c} \widetilde{d}})}} \\
\times \big[ \! \mel*{\widetilde{a} \widetilde{b} J T}{v_{12}^{\text{C}}}{\widetilde{c} \widetilde{d} J T} \\ + (-)^{j_c + j_d + J + T} \! \mel*{\widetilde{a} \widetilde{b} J T}{v_{12}^{\text{C}}}{\widetilde{d} \widetilde{c} J T} \! \big]
\end{multline}
One can effortlessly evaluate the isospin dependence of the non-normalized and non-antisymmetrized central TBMEs, and write them as:
\begin{multline}
\mel*{\widetilde{a} \widetilde{b} J T}{v_{12}^{\text{C}}}{\widetilde{c} \widetilde{d} J T} \\ = \mel*{\widetilde{a} \widetilde{b} J}{\underbrace{(W + B P_{\sigma} + (-)^T H + (-)^T M P_{\sigma})V(r_{12})}_{\widetilde{v}_{12}^{\text{C}}}}{\widetilde{c} \widetilde{d} J}
\end{multline}
The transformation from $jj$ coupling to $LS$ coupling provides: \cite{Talmi1993}
\begin{multline}
\mel*{\widetilde{a} \widetilde{b} J}{\widetilde{v}_{12}^{\text{C}}}{\widetilde{c} \widetilde{d} J} = \sum_{LS} \sum_{L' S'} \hat{j}_a \hat{j}_b \hat{j}_c \hat{j}_d \hat{L} \hat{L}' \hat{S} \hat{S}' \times \\ \begin{Bmatrix} l_a & l_b & L \\ 1/2 & 1/2 & S \\ j_a & j_b & J \end{Bmatrix} \begin{Bmatrix} l_c & l_d & L' \\ 1/2 & 1/2 & S' \\ j_c & j_d & J \end{Bmatrix} \\
\times \underbrace{\! \mel*{n_a l_a \frac{1}{2} \, n_b l_b \frac{1}{2} \, L S J}{\widetilde{v}_{12}^{\text{C}}}{n_c l_c \frac{1}{2} \, n_d l_d \frac{1}{2} \, L' S' J} \, }_{B^{\text{C}}}
\end{multline}
where the braces denote the Wigner-$9j$ symbols and where one has defined $\hat{L} \equiv \sqrt{2L+1}$. The spatial and spin parts split up in such a way that:
\begin{multline}
B^{\text{C}} = \sum_{M_L M_S} \sum_{M'_L M'_S}  \braket*{L M_L S M_S}{J M_L+M_S}  \times \\ \braket*{L' M'_L S' M'_S}{J M'_L+M'_S} \\ 
\times \underbrace{\! \mel*{n_a l_a \, n_b l_b \, L M_L}{V(r_{12})}{n_c l_c \, n_d l_d \, L' M'_L}}_{C^{\text{C}}} \\
\times \underbrace{\! \mel*{\frac{1}{2} \, \frac{1}{2} \, S M_S}{(W + B P_{\sigma} + (-)^T H + (-)^T M P_{\sigma})}{\frac{1}{2} \, \frac{1}{2} \, S' M'_S}}_{D^{\text{C}}}
\end{multline}
where the brackets denote the Clebsch-Gordan coefficients.

On the one hand, the action of $\P_{\sigma}$ allow to simplify the above spin TBME as:
\begin{multline} \label{centralspin}
D^{\text{C}} = \big[ (W-B +(-)^T H - (-)^T M) \\ + S(S+1) (B+(-)^TM) \big] \delta_{SS'} \delta_{M_S M'_S}
\end{multline}
One sees that the central TBMEs are divided into two pieces, one proportional to the combination of parameters $(W-B +(-)^T H - (-)^T M)$, the other to $S(S+1) (B+(-)^TM)$. In this way, the isospin and spin degrees of freedom have been decorrelated so that the combination of parameters involved in the central TBMEs \eqref{centralTBME} can be determined for a given total 
isospin $T$.

On the other hand, the above spatial TBMEs read:
\begin{multline}
C^{\text{C}} = \sum_{m_{la} m_{lb}} \sum_{m_{lc} m_{ld}} \! \braket*{l_a m_{la} l_b m_{lb}}{L M_L} \times \\ \braket*{l_c m_{lc} l_c m_{lc}}{L' M'_L} \\
\times \underbrace{\! \mel*{n_a l_a m_{la} \, n_b l_b m_{lb}}{V(r_{12})}{n_c l_c m_{lc} \, n_d l_d m_{ld}}}_{v^{\text{C}}_{r_a r_b r_c r_d}}
\end{multline}
Thus, one needs to evaluate the central TBMEs in the spherical HO basis, whose wave functions are defined in \eqref{swavef}.  By definition, 
\begin{multline}
v_{r_a r_b r_ c r_d}^{\text{C}} \equiv \int \dd[3]{r_1} \int \dd[3]{r_2} \phi_{r_a}^*(\vec{r}_1) \phi_{r_b}^*(\vec{r}_2) \times \\ V(r_{12}) \phi_{r_c}(\vec{r}_1) \phi_{r_d}(\vec{r}_2)
\end{multline}
Since the wave functions and the central potential commute, one can rearrange the terms according to
\begin{multline}
v_{r_a r_b r_ c r_d}^{\text{C}}  = \int \dd[3]{r_1} \int \dd[3]{r_2} \phi_{r_a}^*(\vec{r}_1) \phi_{r_c}(\vec{r}_1) \times \\ V(r_{12}) \phi_{r_b}^*(\vec{r}_2) \phi_{r_d}(\vec{r_2})
\end{multline}
Using twice the Gogny separable development in spherical symmetry \eqref{Gognysepspherique}, one obtains
\begin{multline} \label{centrals}
v_{r_a r_b r_ c r_d}^{\text{C}}  = \sum_{r_\mu r_\nu} T_{(r_a) (r_c)}^{(r_\mu)} T_{(r_b) (r_d)}^{(r_\nu)} \times \\ \mel*{l_a m_{la}}{Y_{l_\mu}^{m_{l\mu}*}}{l_c m_{lc}}  \mel*{l_b m_b}{Y_{l_\nu}^{m_{l\nu}*}}{l_d m_{ld}} \\ 
\times \int \dd[3]{r_1} \int \dd[3]{r_2} \phi_{0}(\vec{r}_1) \phi_{0}(\vec{r}_2) V(r_{12}) \phi_{r_\mu}(\vec{r}_1) \phi_{r_\nu}(\vec{r}_2)
\end{multline}
Then, applying twice the Moshinsky transformation in spherical symmetry \eqref{Moshinskytransfospherique}, one gets
\begin{multline}
v_{r_a r_b r_ c r_d}^{\text{C}}  = \sum_{r_\mu r_\nu} T_{(r_a) (r_c)}^{(r_\mu)} T_{(r_b) (r_d)}^{(r_\nu)} \times \\  \mel*{l_a m_{la}}{Y_{l_\mu}^{m_{l \mu}*}}{l_c m_{lc}} \! \! \mel*{l_b m_{lb}}{Y_{l_\nu}^{m_{l\nu}*}}{l_d m_{ld}} \\
\times \sum_{r_{\lambda} r_{\sigma}} M_{r_{\mu} r_{\nu}}^{r_{\lambda} r_{\sigma}} \int \dd[3]{r} \phi_{0}(\vec{r}) V(\sqrt{2}r) \phi_{r_{\lambda}}(\vec{r}) \times \\ \int \dd[3]{R} \phi^*_{0}(\vec{R}) \phi_{r_{\sigma}}(\vec{R})
\end{multline}
since the Jacobian of the change of variables $(\vec{r}_1, \vec{r}_2) \rightarrow (\vec{r}, \vec{R})$, given by \eqref{rRspherique}, is equal to unity and the spherical Moshinsky coefficient, specified by \eqref{defMoshinsky}, fixes the range of values of $r_\sigma$ according to \eqref{condMoshinsky1spherique} and \eqref{condMoshinsky2spherique}. One notes that in the particular case $r'_\mu = r'_\nu = 0$, the spherical Moshinsky coefficient reduces to \eqref{Moshspherique00}.  One adds also that $\phi_{0}(\vec{R}) = \phi^*_{0}(\vec{R})$ because of \eqref{phi0spherique}. 

The integral over $\vec{R}$ is readily carried out considering the orthogonality relation of the spherical HO wave functions \eqref{orthogonalitysph}. On the other hand, one can simplify the integral over $\vec{r}$ by pulling out its angular part by means of \eqref{phisansSH}, namely:
\begin{multline}
\int \dd[3]{r} \phi_{0}(\vec{r}) V(\sqrt{2}r) \phi_{r_{\sigma}}(\vec{r}) = \\ \int \dd{r} r^2 \phi_{(0)}(r) V(\sqrt{2}r) \phi_{(r_{\sigma})}(r) \times \\ \int \dd[2]{\hat{r}} Y_{l_\lambda}^{m_{l \sigma}}(\hat{r}) Y_0^{0*}(\hat{r})  \\
= \delta_{l_\sigma,0} \delta_{m_{l \sigma},0} \int \dd{r} r^2 \phi_{(0)}(r) V(\sqrt{2}r) \phi_{(r_\sigma)}(r)
\end{multline}
where the orthogonality relation of the spherical harmonics \eqref{orthogonalitySH} has been used and the fact that $Y_0^{0}(\hat{r}) = Y_0^{0*}(\hat{r})$ because of \eqref{Y00}. Thus, the only thing that remains to be done is to evaluate the above radial integral. Considering the particular wave functions: \eqref{phi0spherique} and
\begin{equation}
\phi_{(n_\sigma,0)}(r) = \bigg[ \frac{2}{b^3} \frac{n_\sigma !}{\Gamma(n_\sigma + 3/2)} \bigg]^{1/2} \e^{-r^2/2b^2} L_{n_\sigma}^{1/2} \bigg( \frac{r^2}{b^2} \bigg)
\end{equation}
one can write down
\begin{multline}
\int_0^{\infty} \dd{r} r^2 \phi_{(0, 0)}(r) V(\sqrt{2} r) \phi_{(n_\sigma, 0)}(r) = \\ \bigg[ \frac{8}{b^6 \sqrt{\pi}} \frac{n_\sigma !}{\Gamma(n_\sigma + 3/2)} \bigg]^{1/2} \\
\times \int_0^{\infty} \dd{r} r^2 \e^{-(\mu^2+2b^2)r^2/\mu^2 b^2} L_{n_\sigma}^{1/2} \bigg( \frac{r^2}{b^2} \bigg)
\end{multline}
Taking advantage of the series expansion of the generalized Laguerre polynomial, one finds:
\begin{multline}
\int_0^{\infty} \dd{r} r^2 \phi_{(0, 0)}(r) V(\sqrt{2} r) \phi_{(n_\sigma, 0)}(r) = \\\bigg[ \frac{8}{b^6 \sqrt{\pi}} n_\sigma !  \Gamma(n_\sigma + 3/2) \bigg]^{1/2} \\
\times \sum_{i=0}^{n_\sigma} \frac{(-)^i}{i! (n_\sigma-i)!  \Gamma(i+3/2)} \int_0^{\infty} \dd{r} r^{2i+2} \e^{-r^2/p^2}
\end{multline}
where the change of variable $p \equiv \mu b / \sqrt{\mu^2 + 2b^2}$ has been used. With the formula of the integral representation of the Gamma function, one can determine the above integral, which leads to:
\begin{multline}
\int_0^{\infty} \dd{r} r^2 \phi_{(0, 0)}(r) V(\sqrt{2} r) \phi_{(n_\sigma, 0)}(r) = \\\bigg[ \frac{2}{\sqrt{\pi}} n_\sigma !  \Gamma(n_\sigma + 3/2) \bigg]^{1/2} \sum_{i=0}^{n_\lambda} \frac{(-)^i G^{-i-3/2}}{i! (n_\sigma-i)!}
\end{multline}
where 
\begin{equation} \label{Gspherique}
G \equiv 1 + \frac{2 b^2}{\mu^2}.
\end{equation}
Finally, by gathering all the terms, one gets an expression for the TBMEs of the central terms in the spherical HO basis, that reads:
\begin{multline} \label{spatialcentral}
v_{r_a r_b r_c r_d}^{\text{C}} = \sum_{r_\mu r_\nu} T_{(r_a) (r_c)}^{(r_\mu)} T_{(r_b) (r_d)}^{(r_\nu)} \\ \mel*{l_a m_{la}}{Y_{l_\mu}^{m_{l\mu}*}}{l_c m_{lc}} \! \! \mel*{l_b m_{lb}}{Y_{l_\nu}^{m_{l\nu}*}}{l_d m_{ld}} \\
 \times M_{r_\mu \, r_\nu}^{0 \, r_\sigma} \bigg[ \frac{2}{\sqrt{\pi}} n_\sigma !  \Gamma(n_\sigma + 3/2) \bigg]^{1/2} \sum_{i=0}^{n_\sigma} \frac{(-)^i G^{-i-3/2}}{i! (n_\sigma-i)!}
\end{multline}
where $G$ is given in \eqref{Gspherique}.  In this expression, $r_\sigma = (n_\sigma, 0, 0)$, with $n_\sigma$ given by \eqref{condMoshspheriquereduce},  while the conditions \eqref{condY} and \eqref{condTalmansph} have to be fulfilled at any time.

In the specific cases of $1s$ and $2s$ states, the quantum numbers are $r_a = r_b = r_c = r_d = (0, 0, 0)$ and $r_a = r_b = r_c = r_d = (1, 0, 0)$, respectively. Then, the contributions to the central terms to the $1s$ and $2s$ TBMEs can respectively be written as:
\begin{subequations}
\begin{align}
f_{1s} & = \mel*{000 \, 000}{V(r_{12})}{000 \, 000} \notag \\
& = G^{-3/2} \\
f_{2s} & = \mel*{100 \, 100}{V(r_{12})}{100 \, 100} \notag \\
& = G^{-3/2} \Big( \frac{41}{64} - \frac{79}{48}G^{-1} + \frac{385}{96} G^{-2} \\ 
& ~~~~~~~~~~~~~~~ - \frac{175}{48} G^{-3} + \frac{105}{64} G^{-4} \Big)
\end{align}
\end{subequations}

\subsection{Density-dependent two-body matrix elements}

One continues by deriving the TBME of the finite-range density-dependent interaction whose expression is recalled here:
\begin{equation} \label{ddintsph}
V^{\text{DD}} = (W + B P_{\sigma} - H P_{\tau} - M P_{\sigma} P_{\tau}) V(r_{12}) D[\rho]
\end{equation}
where $V(r_{12})$ is the Gaussian potential \eqref{Vr12}, and 
\begin{equation} \label{Frho}
D[\rho] \equiv \frac{\rho^{\alpha}(\vec{r}_1) + \rho^{\alpha}(\vec{r}_2)}{2}
\end{equation}
the density functional which represents the density of the nucleus under study at its center-of-mass position. For conciseness, one has omitted the factor $1/(\mu \sqrt{\pi})^3$, but it is obviously taken into account in the fitting code.

The coupling procedure to the quantum numbers $(J,T)$ and the spin dependence of the density-dependent term being formally identical to those of the central terms, one refer the reader to the previous subsection for their construction for the density-dependent term. 

Thus, one only needs to evaluate the TBMEs of the density-dependent interaction, expressed as:
\begin{equation}
v_{r_a r_b r_c r_d}^{\text{DD}} \equiv \! \mel*{n_a l_a m_{la} \, n_b l_b m_{lb}}{V(r_{12})D[\rho]}{n_c l_c m_{lc} \, n_d l_d m_{ld}}
\end{equation}
in the spherical HO basis, whose wave functions are defined in \eqref{swavef}. By definition,
\begin{multline}
v_{r_a r_b r_c r_d}^{\text{DD}} = \int \dd[3]{r_1} \int \dd[3]{r_2} \phi_{r_a}^*(\vec{r}_1) \phi_{r_b}^*(\vec{r}_2) \\ V(r_{12}) D[\rho] \phi_{r_c}(\vec{r}_1) \phi_{r_d}(\vec{r}_2)
\end{multline}
Since the density functional commutes with the wave functions, one rearranges the terms and, as for the central terms, consider the Gogny separable expansion \eqref{Gognysepspherique} twice, to express:
\begin{multline} \label{densitys}
v_{r_a r_b r_c r_d}^{\text{DD}} = \sum_{r_\mu r_\nu} T_{(r_a) (r_c)}^{(r_\mu)} T_{(r_b) (r_d)}^{(r_\nu)} \\ \mel*{l_a m_{la}}{Y_{l_\mu}^{m_{l\mu}*}}{l_c m_{lc}} \mel*{l_b m_{lb}}{Y_{l_\nu}^{m_{l\nu}*}}{l_d m_{ld}} \\
\times \frac{1}{2} \bigg[ \int \dd[3]{r_1} \int \dd[3]{r_2} \phi_{0}(\vec{r}_1) \phi_{0}(\vec{r}_2) \rho^{\alpha}(\vec{r}_1) \times \\ V(r_{12}) \phi_{r_\mu}(\vec{r}_1) \phi_{r_\nu}(\vec{r}_2) \\
 + \int \dd[3]{r_1} \int \dd[3]{r_2} \phi_{0}(\vec{r}_1) \phi_{0}(\vec{r}_2)\rho^{\alpha}(\vec{r}_2) \times \\ V(r_{12}) \phi_{r_\mu}(\vec{r}_1) \phi_{r_\nu}(\vec{r}_2) \bigg]
\end{multline}
One notices that the second integral is deduced from the first one by exchanging the quantum numbers $r_\mu$ and $r_\nu$.  Then, one only focuses on the first integral to find an expression for the above TBMEs.  This time, one has no interest in employing the Moshinsky transformation since the density would couple the relative and center-of-mass coordinates. Instead,  one performs 
the integral over $\vec{r}_2$ by means of the following formula \cite{Gogny1975b},\footnote{In fact, this formula is demonstrated from the generating functions of the spherical HO. As the introduction of the generating functions in spherical symmetry would have been long and useful only here, one has chosen to leave it at admitting this relation.}
\begin{multline}
\int \dd[3]{r_2} \e^{-(\vec{r}_1-\vec{r}_2)^2/\mu^2} \phi_0(\vec{r}_2) \phi_{r_\nu}(\vec{r}_2) = \\ K \lambda_{(r_\nu)}  \phi_0(\vec{r}_1, b \sqrt{g}) \phi_{r_\nu}(\vec{r}_1, b \sqrt{g})
\end{multline}
where, in the above wave functions, the oscillator length $b$ is replaced by $b \sqrt{g}$,  with $g \equiv 1+\mu^2/b^2$. The new quantities are expressed in terms of $g$ as $K \equiv \pi [b \sqrt{g-1}]^{3}/2$ and $\lambda_{(r_\nu)} \equiv g^{-n_\nu - l_\nu/2}$. Thus, one will have to evaluate integrals of the form:
\begin{multline} \label{Jdensite}
I_{r_\mu r_\nu} \equiv \int \dd[3]{r} \phi_0(\vec{r}) \phi_{r_\mu}(\vec{r}) \rho^{\alpha}(\vec{r}) \phi_0(\vec{r}, b \sqrt{g}) \times \\ \phi_{r_\nu}(\vec{r}, b \sqrt{g})
\end{multline}
Taking into account that the density is spherically symmetric, i.e.\ $\rho(\vec{r}) = \rho(r)$, as well as the definitions \eqref{phi0spherique} and \eqref{phisansSH}, one gets
\begin{multline}
I_{r_\mu r_\nu} = \frac{1}{4 \pi} \frac{4}{\sqrt{\pi} \- b^3} \int \dd{r} r^2 \rho^{\alpha}(r) \e^{-r^2/2b^2} \e^{-r^2/2gb^2} \times \\  \phi_{(r_\mu)}(r) \phi_{(r_\nu)}(r, b\sqrt{g}) \times  \\
\int \dd[2]{\hat{r}} Y_{l_\mu}^{m_{l \mu}}(\hat{r}) Y_{l_\nu}^{m_{l \nu}}(\hat{r})
\end{multline}
Invoking the orthogonality of the spherical harmonics \eqref{orthogonalitySH} and pulling out the exponentials of the wave functions according to \eqref{phihatsph}, the integral becomes:  
\begin{multline} \label{intdensitysph}
I_{r_\mu r_\nu} = \frac{(-)^{m_{l\mu}}}{(\sqrt{\pi} \- b)^3} \delta_{l_\mu l_\nu} \delta_{m_{l\mu}, -m_{l_\nu}} \times \\ \underbrace{\int_0^{\infty} \dd{r} r^2 \rho^{\alpha}(r) \e^{-(1+g)r^2/gb^2} \hat{\phi}_{(r_\mu)}(r) \hat{\phi}_{(r_\nu)}(r, b\sqrt{g})}_{J_{(r_\mu)(r_\nu)}}
\end{multline}

At this point, there is at least two ways to evaluate the integral $J_{(r_\mu)(r_\nu)}$ numerically. The most straightforward one consists in setting the change of variable $x \equiv (1+g)r^2/gb^2$, to find out:
\begin{equation}
J_{(r_\mu)(r_\nu)} = \frac{1}{2} \bigg[ \frac{gb^2}{1+g} \bigg]^{3/2} \int_0^{\infty} \dd{x} \sqrt{x} \- \e^{-x} f_{(r_\mu)(r_\nu)}(x)
\end{equation}
where the function $f_{(r_\mu)(r_\nu)}(x)$ is defined by:
\begin{multline} \label{fdensite}
f_{(r_\mu)(r_\nu)}(x) = \rho^{\alpha} \bigg( \sqrt{\frac{gb^2}{1+g}} \sqrt{x} \bigg) \times \\ \hat{\phi}_{(r_\mu)}\bigg(\sqrt{\frac{gb^2}{1+g}} \sqrt{x}\bigg) \hat{\phi}_{(r_\nu)}\bigg(\sqrt{\frac{gb^2}{1+g}} \sqrt{x}, b\sqrt{g}\bigg)
\end{multline}
Then, one applies the generalized Gauss--Laguerre quadrature which states that:
\begin{equation}
\int_0^{\infty} \dd{x} \sqrt{x} \- \e^{-x} f_{(r_\mu)(r_\nu)}(x) \simeq \sum_{i=1}^n w_i f_{(r_\mu)(r_\nu)}(x_i)
\end{equation}
where $n$ is the quadrature order, $x_i$ the $i$-th root of the generalized Laguerre polynomial $L_n^{1/2}(x)$ and $w_i$ a weight factor that reads:
\begin{equation} \label{wdensite}
w_i \equiv \frac{\Gamma(n+3/2) x_i}{n! (n+1)^2 \big[L_{n+1}^{1/2}(x_i)\big]^2}
\end{equation}
Gathering the results, the integral \eqref{Jdensite} can be approximated by:
\begin{multline} \label{finaldensity1}
I_{r_\mu r_\nu} \simeq \frac{(-)^{m_{l\mu}}}{2} \bigg[ \frac{1}{\pi} \frac{g}{1+g} \bigg]^{3/2} \delta_{l_\mu l_\nu} \delta_{m_{l\mu}, -m_{l_\nu}} \times \\ \sum_{i=1}^n w_i f_{(r_\mu)(r_\nu)}(x_i)
\end{multline}
with $w_i$ and $f_{(r_\mu)(r_\nu)}(x_i)$ given above. The last summation can then be evaluated numerically, up to a certain order $n$. 

The second method requires more work but is numerically advantageous. This is the one adopted in the fitting procedure. Starting from the general definition of the local nuclear density, one writes:
\begin{equation}
\rho(\vec{r}) = 2 \sum_{r_a} \phi^*_{r_a}(\vec{r}) \phi_{r_a}(\vec{r}) \big( \rho_{r_a,r_a}^\pi + \rho_{r_a,r_a}^\nu \big)
\end{equation}
where the factor $2$ comes from the spin degeneracy and where the proton $\pi$ and neutron $\nu$ density matrices have been separated. Pulling out the angular dependencies of the wave functions and using the addition theorem \eqref{addition}, one obtains:
\begin{multline}
\rho(\vec{r}) = 2 \sum_{nl} \phi_{(n,l)}(r) \phi_{(n,l)}(r) \times \\ 
\sum_{m_l} Y_l^{m_l *}(\hat{r}) Y_l^{m_l}(\hat{r}) \big( \rho_{nl,nl}^\pi + \rho_{nl,nl}^\nu \big) \notag \\
 = \frac{1}{2 \pi} \sum_{nl} (2l+1) \frac{2}{b^3} \frac{n!}{\Gamma(n+l+3/2)} \times \\ 
 \e^{-r^2/b^2} \Big( \frac{r}{b} \Big)^{2l} \Big[L_n^{l+1/2}\bigg( \frac{r^2}{b^2} \bigg) \Big]^2 \big( \rho_{nl,nl}^\pi + \rho_{nl,nl}^\nu \big)
\end{multline}
where one has proved that the nuclear density does not depend on the orientation in space, as expected in spherical symmetry. 
One notes that the density matrix does not depend on $m_l$ as the occupation of the major shells is only specified by $n$ and $l$. The relation between the generalized Laguerre polynomials and the confluent hypergeometric function $_1F_1$ permits to get:
\begin{multline}
\rho(r) = \frac{1}{b^3}\e^{-r^2/b^2} \bigg[ \frac{1}{2 \pi} \sum_{nl} (2l+1) \times \\\frac{2 \Gamma(n+l+3/2)}{n! \Gamma(l+3/2)^2} \Big( \frac{r}{b} \Big)^{2l} \\
\times \! \,_1F_1(-n,l+3/2,r^2/b^2)^2 \big( \rho_{nl,nl}^\pi + \rho_{nl,nl}^\nu \big) \bigg]
\end{multline}
Inserting this expression for the density in the radial integral appearing in \eqref{intdensitysph}, one obtains:
\begin{multline}
J_{(r_\mu)(r_\nu)} = \frac{1}{b^{3(\alpha+1)}} \frac{1}{g^{3/4}} \int_0^{\infty} \dd{r} r^2 \e^{-(\alpha+1+1/g)r^2/gb^2} \times \\ F_{(r_\mu)} \bigg( \frac{r^2}{b^2} \bigg) F_{(r_\nu)} \bigg( \frac{r^2}{gb^2}  \bigg) \widetilde{\rho}^{\alpha}\bigg( \frac{r^2}{b^2} \bigg)
\end{multline}
where the quantities $F_{(r_\mu)}$, $F_{(r_\nu)}$ and $\widetilde{\rho}$ are defined by:
\begin{multline}
F_{(r_\mu)} \bigg( \frac{r^2}{b^2} \bigg) = \bigg[ \frac{2 \Gamma(n_\mu + l_\mu +3/2)}{n_\mu !} \bigg]^{1/2} \times \\ \frac{1}{\Gamma(l_\mu +3/2)} \Big( \frac{r}{b} \Big)^{l_\mu} 
\! \,_1F_1(-n_\mu,l_\mu+3/2,r^2/b^2) \\
F_{(r_\nu)} \bigg( \frac{r^2}{gb^2} \bigg) = \bigg[ \frac{2 \Gamma(n_\nu + l_\mu +3/2)}{n_\nu !} \bigg]^{1/2} \times \\ \frac{1}{\Gamma(l_\mu +3/2)} \bigg( \frac{r}{\sqrt{g} b} \bigg)^{l_\nu}
\ \! \,_1F_1(-n_\nu,l_\mu+3/2,r^2/gb^2)\\
\widetilde{\rho}\bigg( \frac{r^2}{b^2} \bigg) = \frac{1}{2 \pi} \sum_{nl} (2l+1) \frac{2 \Gamma(n+l+3/2)}{n! \Gamma(l+3/2)^2} \Big( \frac{r}{b} \Big)^{2l} \times \\
 \! \,_1F_1(-n,l+3/2,r^2/b^2)^2 \big( \rho_{nl,nl}^\pi + \rho_{nl,nl}^\nu \big)
\end{multline}
The change of variable $x \equiv r/b \sqrt{\alpha+1+1/g}$ eventually furnishes:
\begin{multline} \label{intint}
J_{(r_\mu)(r_\nu)} = \frac{1}{b^{3\alpha}} \frac{1}{g^{3/4}} \frac{1}{\sqrt{\alpha+1+1/g}} \times \\ \int_0^{\infty} \dd{x} \e^{-x^2} f_{(r_\mu)(r_\nu)}(x^2)
\end{multline}
with the function:
\begin{multline} \label{fx2}
f_{(r_\mu)(r_\nu)}(x^2) = \frac{x^2}{\alpha+1+1/g} F_{(r_\mu)} \Big( \frac{x^2}{\alpha+1+1/g} \Big) \times  \\ F_{(r_\nu)} \Big( \frac{x^2}{g(\alpha+1+1/g)} \Big)
\widetilde{\rho}^{\alpha}\Big( \frac{x^2}{\alpha+1+1/g} \Big)
\end{multline}
The Gauss-Hermite quadrature allows to approximate integrals defined on the entire space according to:
\begin{equation} \label{GH}
\int_{-\infty}^{+\infty} \dd{x} \e^{-x^2} f(x) \simeq \sum_{i=0}^n w_i f({x}_i),
\end{equation}
where $n$ is the quadrature order, $x_i$ the $i$-th root of the Hermite polynomial $H_n(x)$ at which the function $f$ is evaluated, and $w_i$ a weight factor that reads:
\begin{equation}
w_i \equiv \frac{2^{n-1} n! \sqrt{\pi}}{n^2 [H_{n-1}(x_i)]^2}.
\end{equation}
Here, the integral \eqref{intint} runs from zero to infinity, but since the roots of the Hermite polynomials are symmetric (to each root $x_i$ corresponds another root $-x_i$) and the integrand is even, one can approximate it with the Gauss-Hermite quadrature according to:
\begin{equation}
\int_{0}^{\infty} \dd{x} \e^{-x^2} f_{(r_\mu)(r_\nu)}(x^2) \simeq \sum_{i=0}^n w_i f_{(r_\mu)(r_\nu)}(x^2),
\end{equation}
where $f_{(r_\mu)(r_\nu)}(x^2)$ is given by \eqref{fx2}. When the series will be truncated at a certain order $n$, it will be necessary to consider $2n$ roots and weight factors associated with the complete quadrature \eqref{GH}. In the fitting code, the quadrature order is $n=10$ as the series converges quite rapidly, so that the first twenty Gauss-Hermite roots and weights are needed. Gathering the results, the integral \eqref{Jdensite} is approximated by:
\begin{multline} \label{finaldensity2}
I_{r_\mu r_\nu} \simeq \frac{(-)^{m_{l \mu}}}{(\sqrt{\pi} g^{1/4} b^{\alpha+1})^3} \frac{1}{\sqrt{\alpha+1+1/g}} \times \\ \delta_{l_\mu l_\nu} \delta_{m_{l\mu}, -m_{l_\nu}} \sum_{i=1}^n w_i f_{(r_\mu)(r_\nu)}(x_i^2)
\end{multline}
with $w_i$ and $f_{(r_\mu)(r_\nu)}(x_i^2)$ given above. 

Finally, in both cases,  the TBMEs of the density-dependent interaction, in the spherical HO basis, read:
\begin{multline}
v_{r_a r_b r_c r_d}^{\text{DD}} = \frac{1}{2} K \sum_{r_\mu r_\nu} T_{(r_a) (r_c)}^{(r_\mu)} T_{(r_b) (r_d)}^{(r_\nu)} \times \\ \mel*{l_a m_{la}}{Y_{l_\mu}^{m_{l\mu}*}}{l_c m_{lc}} \! \! \mel*{l_b m_{lb}}{Y_{l_\nu}^{m_{l\nu}*}}{l_d m_{ld}} \times \\
\big[ \lambda_{(r_\nu)} I_{r_\mu r_\nu} + \lambda_{(r_\mu)} I_{r_\nu r_\mu} \big]
\end{multline}
where the integrals $I_{r_\mu r_\nu}$ and $I_{r_\nu r_\mu}$ are either given by \eqref{finaldensity1} or \eqref{finaldensity2}. In this expression,  one has $l_\mu = l_\nu$ and $m_{l\mu} = - m_{l\nu}$ while the conditions \eqref{condY} and \eqref{condTalmansph} have to be fulfilled at any time. 

\subsection{Tensor two-body matrix elements}

In order to easily compute the TBMEs of the tensor interaction, one shall separate the space and spin-isospin degrees of freedom according to the equivalent form of the tensor operator
$S_{12} = [\hat{r}_{12} \otimes \hat{r}_{12}]^{(2)} \cdot [\vec{\sigma}_1 \otimes \vec{\sigma}_2]^{(2)}$. Then, the tensor interaction reads:
\begin{equation} \label{tensorintsph}
v_{12}^{\text{T}} = (W-H P_{\tau}) V(r_{12}) [\hat{r}_{12} \otimes \hat{r}_{12}]^{(2)} \cdot [\vec{\sigma}_1 \otimes \vec{\sigma}_2]^{(2)},
\end{equation}
where $V(r_{12})$ is the Gaussian potential \eqref{Vr12}. As for the previous terms, the normalized and antisymmetrized (na) TBMEs, recoupled to $(J, T)$, of the tensor interaction can be written by means of \eqref{recoupling}, i.e. \
\begin{multline}
\mel*{\widetilde{a} \widetilde{b} J T}{v_{12}^{\text{T}}}{c d J T}_{\text{na}} = \frac{1}{\sqrt{(1+\delta_{\widetilde{a} \widetilde{b}})(1+\delta_{\widetilde{c} \widetilde{d}})}} \\
\times \big[ \! \mel*{\widetilde{a} \widetilde{b} J T}{v_{12}^{\text{T}}}{\widetilde{c} \widetilde{d} J T} \\ + (-)^{j_c + j_d + J + T} \! \mel*{\widetilde{a} \widetilde{b} J T}{v_{12}^{\text{T}}}{\widetilde{d} \widetilde{c} J T} \! \big]
\end{multline}
One can extract the isospin dependence of the non-normalized and non-antisymmetrized tensor TBMEs according to:
\begin{equation}
\mel*{\widetilde{a} \widetilde{b} J T}{v_{12}^{\text{T}}}{\widetilde{c} \widetilde{d} J T} = (W + (-)^T H) \! \mel*{\widetilde{a} \widetilde{b} J}{\widetilde{v}_{12}^{\text{T}}}{\widetilde{c} \widetilde{d} J}
\end{equation}
where $\widetilde{V}^{\text{T}}$ corresponds to \eqref{tensorintsph} deprived of its combination of parameters. Gathering the results, one finds:
\begin{multline}
\mel*{\widetilde{a} \widetilde{b} J T}{v_{12}^{\text{T}}}{\widetilde{c} \widetilde{d} J T}_{\text{na}} = \frac{W + (-)^T H}{\sqrt{(1+\delta_{\widetilde{a} \widetilde{b}})(1+\delta_{\widetilde{c} \widetilde{d}})}} \times \\
\big[ \! \mel*{\widetilde{a} \widetilde{b} J}{\widetilde{v}_{12}^{\text{T}}}{\widetilde{c} \widetilde{d} J} + (-)^{j_c + j_d + J + T} \! \mel*{\widetilde{a} \widetilde{b} J}{\widetilde{v}_{12}^{\text{T}}}{\widetilde{d} \widetilde{c} J} \! \big]
\end{multline}
In the following, one will evaluate:
\begin{equation}
\mel*{\widetilde{a} \widetilde{b} J}{\widetilde{v}_{12}^{\text{T}}}{\widetilde{c} \widetilde{d} J} = \sum_k (-)^k \! \mel*{\widetilde{a} \widetilde{b} J}{\widetilde{v}_{12}^{\text{T}, \- k}}{\widetilde{c} \widetilde{d} J} \!,
\end{equation}
where the $k$-th component of the tensor interaction is defined as:
\begin{equation}
\widetilde{v}_{12}^{\text{T}, \- k} \equiv V(r_{12}) [\hat{r}_{12} \otimes \hat{r}_{12}]^{(2)}_{-k} [\vec{\sigma}_1 \otimes \vec{\sigma}_2]^{(2)}_k.
\end{equation}
The transformation from $jj$ coupling to $LS$ coupling provides 
\begin{multline}
\mel*{\widetilde{a} \widetilde{b} J}{\widetilde{v}_{12}^{\text{T}, \- k}}{\widetilde{c} \widetilde{d} J} = \sum_{LS} \sum_{L' S'} \hat{j}_a \hat{j}_b \hat{j}_c \hat{j}_d \hat{L} \hat{L}' \hat{S} \hat{S}' \times \\ \begin{Bmatrix} l_a & l_b & L \\ 1/2 & 1/2 & S \\ j_a & j_b & J \end{Bmatrix} \begin{Bmatrix} l_c & l_d & L' \\ 1/2 & 1/2 & S' \\ j_c & j_d & J \end{Bmatrix} \times \\
\underbrace{\! \mel*{n_a l_a \frac{1}{2} \, n_b l_b \frac{1}{2} \, L S J}{\widetilde{v}_{12}^{\text{T}, \- k}}{n_c l_c \frac{1}{2} \, n_d l_d \frac{1}{2} \, L' S' J} \, }_{B^{\text{T}}_k}
\end{multline}
where the space and spin parts split up in such a way that:
\begin{multline}
B^{\text{T}}_k = \sum_{M_L M_S} \sum_{M'_L M'_S}  \braket*{L M_L S M_S}{J M_L+M_S}  \times \\ \braket*{L' M'_L S' M'_S}{J M'_L+M'_S} \\
 \underbrace{\! \mel*{n_a l_a \, n_b l_b \, L M_L}{V(r_{12}) [\hat{r}_{12} \otimes \hat{r}_{12}]^{(2)}_{-k}}{n_c l_c \, n_d l_d \, L' M'_L}}_{C_{-k}^{\text{T}}} \\
\times \underbrace{\! \mel*{\frac{1}{2} \, \frac{1}{2} \, S M_S}{[\vec{\sigma}_1 \otimes \vec{\sigma}_2]^{(2)}_k}{\frac{1}{2} \, \frac{1}{2} \, S' M'_S}}_{D_k^{\text{T}}}
\end{multline}
On the one hand, the Wigner-Eckart theorem relates the above spin TBME to its reduced counterpart according to \cite{Talmi1993}:
\begin{multline}
D_k^{\text{T}} = (-)^{S-M_S} \begin{pmatrix} S & 2 & S' \\ -M_S & k & M'_S \end{pmatrix} \times \\ \langle \frac{1}{2} \, \frac{1}{2} \, S \| [\vec{\sigma}_1 \otimes \vec{\sigma}_2]^{(2)} \| \frac{1}{2} \, \frac{1}{2} \, S' \rangle
\end{multline}
where the double bar denotes the reduced TBME, that can itself be expressed as:
\begin{multline}
\langle \frac{1}{2} \, \frac{1}{2} \, S \| [\vec{\sigma}_1 \otimes \vec{\sigma}_2]^{(2)} \| \frac{1}{2} \, \frac{1}{2} \, S' \rangle = \sqrt{5} \hat{S} \hat{S}' 
\times \\ \begin{Bmatrix} 1/2 & 1/2 & S \\ 1/2 & 1/2 & S' \\ 1 & 1 & 2 \end{Bmatrix} \langle \frac{1}{2} \| \vec{\sigma}_1 \| \frac{1}{2} \rangle \langle \frac{1}{2} \| \vec{\sigma}_2 \| \frac{1}{2} \rangle,
\end{multline}
with the reduced (one-body) matrix elements:
\begin{equation} \label{reducedTBMEspin}
\langle \frac{1}{2} \| \vec{\sigma}_i \| \frac{1}{2} \rangle = \sqrt{6}, \quad \text{for} \ i \in \{1,2\}
\end{equation}
One notes that the Wigner-3$j$ symbol appearing above with parentheses implies $2 \le S + S'$, so that, necessarily, $S = S' = 1$. This result was expected as the tensor term only acts in the $S=1$ channel of the interaction. Gathering the equations, one obtains, for the spin part of the tensor TBMEs:
\begin{multline}
D_k^{\text{T}} = 6 \sqrt{5} (-)^{S-M_S} \hat{S} \hat{S}' \begin{pmatrix} S & 2 & S' \\ -M_S & k & M'_S \end{pmatrix} \times \\ \begin{Bmatrix} 1/2 & 1/2 & S \\ 1/2 & 1/2 & S' \\ 1 & 1 & 2 \end{Bmatrix}
\end{multline}
As for the tensor space part, one finds:
\begin{multline}
C_{-k}^{\text{T}} = \sum_{m_{la} m_{lb}} \sum_{m_{lc} m_{ld}} \braket*{l_a m_{la} l_b m_{lb}}{L M_L} \times \\ \braket*{l_c m_{lc} l_c m_{lc}}{L' M'_L}  \times \\
\underbrace{\! \mel*{n_a l_a m_{la} \, n_b l_b m_{lb}}{V(r_{12}) [\hat{r}_{12} \otimes \hat{r}_{12}]^{(2)}_{-k}}{n_c l_c m_{lc} \, n_d l_d m_{ld}}}_{v_{r_a r_b r_c r_d}^{\text{T} \, -k}}
\end{multline}
Thus, one evaluates the tensor TBMEs in the spherical HO basis, whose wave functions are defined in \eqref{swavef}.  By definition,
\begin{multline}
v_{r_a r_b r_c r_d}^{\text{T} \, -k} \equiv \int \dd[3]{r_1} \int \dd[3]{r_2} \phi_{r_a}^*(\vec{r}_1) \phi_{r_b}^*(\vec{r}_2) \times \\ V(r_{12}) [\hat{r}_{12} \otimes \hat{r}_{12}]^{(2)}_{-k} \phi_{r_c}(\vec{r}_1) \phi_{r_d}(\vec{r}_2)
\end{multline}
Since the sandwiched quantity commutes with the wave functions, one has: 
\begin{multline}
v_{r_a r_b r_c r_d}^{\text{T} \, -k} = \int \dd[3]{r_1} \int \dd[3]{r_2} \phi_{r_a}^*(\vec{r}_1) \phi_{r_c}(\vec{r}_1) \times \\ V(r_{12}) [\hat{r}_{12} \otimes \hat{r}_{12}]^{(2)}_{-k} \phi_{r_b}^*(\vec{r}_2) \phi_{r_d}(\vec{r}_2)
\end{multline}
Applying twice the Gogny separable expansion in spherical symmetry \eqref{Gognysepspherique}, one obtains:
\begin{multline}
v_{r_a r_b r_c r_d}^{\text{T} \, -k} = \sum_{r_\mu r_\nu} T_{(r_a) (r_c)}^{(r_\mu)} T_{(r_b) (r_d)}^{(r_\nu)} \times \\ \mel*{l_a m_{la}}{Y_{l_\mu}^{m_{l \mu}*}}{l_c m_{lc}} \! \! \mel*{l_b m_{lb}}{Y_{l_\nu}^{m_{l\nu}*}}{l_d m_{ld}} \times \\
 \int \dd[3]{r_1} \int \dd[3]{r_2} \phi_{0}(\vec{r}_1) \phi_{0}(\vec{r}_1) \times \\ V(r_{12}) [\hat{r}_{12} \otimes \hat{r}_{12}]^{(2)}_{-k} \phi_{r_{\mu}}(\vec{r}_2) \phi_{r_{\nu}}(\vec{r}_2)
\end{multline}
Now, applying twice the Moshinsky transformation in spherical symmetry \eqref{Moshinskytransfospherique}, one gets:
\begin{multline}
v_{r_a r_b r_c r_d}^{\text{T} \, -k} = \sum_{r_\mu r_\nu} T_{(r_a) (r_c)}^{(r_\mu)} T_{(r_b) (r_d)}^{(r_\nu)} \times \\ \mel*{l_a m_{la}}{Y_{l_\mu}^{m_{l \mu}*}}{l_c m_{lc}} \! \! \mel*{l_b m_{lb}}{Y_{l_\nu}^{m_{l\nu}*}}{l_d m_{ld}} \\
\times \sum_{r_{\lambda} r_{\sigma}} M_{r_{\mu} \, r_{\nu}}^{r_{\lambda} \, r_{\sigma}} \int \dd[3]{r} \phi_{0}(\vec{r}) V(\sqrt{2}r) [\hat{r} \otimes \hat{r}]^{(2)}_{-k} \phi_{r_{\lambda}}(\vec{r}) \times \\ \int \dd[3]{R} \phi^*_{0}(\vec{R}) \phi_{r_{\sigma}}(\vec{R})
\end{multline}
since the Jacobian of the change of variables $(\vec{r}_1, \vec{r}_2) \rightarrow (\vec{r}, \vec{R})$, given by \eqref{rRspherique},  is equal to unity and the spherical Moshinsky coefficient, specified by \eqref{defMoshinsky}, fixes the range of values of $r_\sigma$ according to \eqref{condMoshinsky1spherique} and \eqref{condMoshinsky2spherique}. One notes that in the particular case $r'_\mu = r'_\nu = 0$, the spherical Moshinsky coefficient reduces to \eqref{Moshspherique00} and that $\phi_{0}(\vec{R}) = \phi^*_{0}(\vec{R})$ because of \eqref{phi0spherique}. 

The integral over $\vec{R}$ is readily carried out considering the orthogonality relation of the spherical HO wave functions \eqref{orthogonalitysph}. On the other hand, using the writing \eqref{harmoTS} in spherical symmetry, one can simplify the integral over $\vec{r}$ by pulling out its angular part by means of \eqref{phisansSH}, namely:
\begin{multline}
\int \dd[3]{r} \phi_{0}(\vec{r}) V(\sqrt{2}r) [\hat{r} \otimes \hat{r}]^{(2)}_{-k} \phi_{r_{\sigma}}(\vec{r}) =  \\ 
\sqrt{\frac{8 \pi}{15}} \int \dd[3]{r} \phi_{0}(\vec{r}) V(\sqrt{2}r) \phi_{r_{\sigma}}(\vec{r}) Y_2^{-k}(\hat{r}) \\
 = \sqrt{\frac{8 \pi}{15}} \frac{1}{\sqrt{4 \pi}} \int \dd{r} r^2 \phi_{(0)}(r) V(\sqrt{2}r) \phi_{(r_\sigma)}(r) \times \\ \int \dd[2]{\hat{r}} Y_2^{-k}(\hat{r}) Y_{l_\sigma}^{m_{l \sigma}}(\hat{r}) \\
= (-)^k \sqrt{\frac{2}{15}} \delta_{l_\sigma,2} \delta_{m_{l \sigma},k}  \int \dd{r} r^2 \phi_{(0)}(r) \times \\ V(\sqrt{2}r) \phi_{(r_\sigma)}(r)
\end{multline}
where one has used the orthogonality of the spherical harmonics \eqref{orthogonalitySH}.  Thus, the only thing that remains to be done is to evaluate the above integral. Considering the particular wave functions \eqref{phi0spherique} and
\begin{multline}
\phi_{(n_\sigma,2)} = \bigg[ \frac{2}{b^3} \frac{n_\sigma !}{\Gamma(n_\sigma + 7/2)} \bigg]^{1/2} \times \\ \e^{-r^2/2b^2} \Big( \frac{r}{b} \Big)^2 L_{n_\sigma}^{5/2} \bigg( \frac{r^2}{b^2} \bigg)
\end{multline}
as well as the series expansion of the generalized Laguerre polynomial, one obtains, after a development similar to the one made for the central TBMEs:
\begin{multline}
\int \dd{r} r^2 \phi_{(0,0)}(r) V(\sqrt{2}r) \phi_{(n_\sigma,2)}(r) = \\  \bigg[ \frac{2}{\sqrt{\pi}} n_\sigma ! \, \Gamma(n_\sigma + 7/2) \bigg]^{1/2} \sum_{i=0}^{n_\sigma} \frac{(-)^i G^{-i-5/2}}{i! (n_\sigma-i)! (i+5/2)}
\end{multline}
where $G$ is defined in Eq.\eqref{Gspherique}. Finally, by gathering all the results,  one gets an expression for the TBMEs of the tensor interaction in the spherical HO basis, that reads:
\begin{multline}
v_{r_a r_b r_c r_d}^{\text{T} \, -k} = (-)^k \sqrt{\frac{2}{15}} \sum_{r_\mu r_\nu} T_{(r_a) (r_c)}^{(r_\mu)} T_{(r_b) (r_d)}^{(r_\nu)} \times \\ \mel*{l_a m_{la}}{Y_{l_\mu}^{m_{l \mu}*}}{l_c m_{lc}} \mel*{l_b m_{lb}}{Y_{l_\nu}^{m_{l\nu}*}}{l_d m_{ld}} \times \\
M_{r_{\mu} r_{\nu}}^{0 \, r_{\sigma}} \bigg[ \frac{2}{\sqrt{\pi}} n_\sigma ! \, \Gamma(n_\sigma + 7/2) \bigg]^{1/2} \times \\ \sum_{i=0}^{n_\sigma} \frac{(-)^i G^{-i-5/2}}{i! (n_\sigma-i)! (i+5/2)}
\end{multline}
where $G$ is expressed in \eqref{Gspherique}.  In this expression, $r_\sigma = (n_\sigma, 2, k)$, with $n_\sigma$ given by \eqref{condMoshspheriquereduce},  while the conditions \eqref{condY} and \eqref{condTalmansph} have to be fulfilled at any time.

\subsection{Spin-orbit two-body matrix elements}

In order to easily compute the TBMEs of the spin-orbit interaction, one will separate the space and spin-isospin degrees of freedom according to the equivalent form of the spin-orbit operator $\vec{L} \cdot \vec{S} = - \frac{1}{2\sqrt{2}} [\vec{r}_{12} \otimes \vec{\nabla}_{12}]^{(1)} \cdot [\vec{\sigma}_1 + \vec{\sigma}_2]^{(1)}$. The interaction which is considered in this section then reads:
\begin{multline} \label{SOintsph}
v_{12}^{\text{SO}} = \widetilde{B}(\mu) (W-H P_{\tau}) \times \\ V(r_{12}) [\vec{r}_{12} \otimes \vec{\nabla}_{12}]^{(1)} \cdot [\vec{\sigma}_1 + \vec{\sigma}_2]^{(1)}
\end{multline}
where $V(r_{12})$ is the Gaussian potential \eqref{Vr12} and $\widetilde{B}(\mu)$ the coefficient given by 
\begin{equation}
\widetilde{B}(\mu) \equiv - \frac{1}{2\sqrt{2}} B(\mu) = \frac{\sqrt{2}}{\mu^2} \frac{1}{(\mu \sqrt{\pi})^3}
\end{equation}
with $B(\mu)$ defined in \eqref{coeffB}. As for the previous terms, the normalized and antisymmetrized (na) TBMEs, coupled to $(J, T)$, of the spin-orbit interaction can be written by means of \eqref{recoupling}, i.e.\
\begin{multline}
\mel*{\widetilde{a} \widetilde{b} J T}{v_{12}^{\text{SO}}}{\widetilde{c} \widetilde{d} J T}_{\text{na}} = \frac{1}{\sqrt{(1+\delta_{\widetilde{a} \widetilde{b}})(1+\delta_{\widetilde{c} \widetilde{d}})}} \times \\ \big[ \! \mel*{\widetilde{a} \widetilde{b} J T}{v_{12}^{\text{SO}}}{\widetilde{c} \widetilde{d} J T} \\ + (-)^{j_c + j_d + J + T} \! \mel*{\widetilde{a} \widetilde{b} J T}{v_{12}^{\text{SO}}}{\widetilde{d} \widetilde{c} J T} \! \big]
\end{multline}
One can extract the isospin dependence of the non-normalized and non-antisymmetrized spin-orbit TBMEs, as done for the tensor interaction, according to:
\begin{multline}
\mel*{\widetilde{a} \widetilde{b} J T}{v_{12}^{\text{SO}}}{\widetilde{c} \widetilde{d} J T} = \widetilde{B}(\mu) (W + (-)^T H) \times \\ \mel*{\widetilde{a} \widetilde{b} J}{\widetilde{v}_{12}^{\text{SO}}}{\widetilde{c} \widetilde{d} J}
\end{multline}
where $\widetilde{V}^{\text{SO}}$ corresponds to \eqref{SOintsph} deprived of its combination of parameters. Gathering the results, one finds:
\begin{multline}
\mel*{\widetilde{a} \widetilde{b} J T}{v_{12}^{\text{SO}}}{\widetilde{c} \widetilde{d} J T}_{\text{na}} = \\ \widetilde{B}(\mu) \frac{W + (-)^T H}{\sqrt{(1+\delta_{\widetilde{a} \widetilde{b}})(1+\delta_{\widetilde{c} \widetilde{d}})}} \times \\
\big[ \! \mel*{\widetilde{a} \widetilde{b} J}{\widetilde{v}_{12}^{\text{SO}}}{\widetilde{c} \widetilde{d} J} + (-)^{j_c + j_d + J + T} \! \mel*{\widetilde{a} \widetilde{b} J}{\widetilde{v}_{12}^{\text{SO}}}{\widetilde{d} \widetilde{c} J} \! \big]
\end{multline}
In the following, one will then evaluate:
\begin{equation}
\mel*{\widetilde{a} \widetilde{b} J}{\widetilde{v}_{12}^{\text{SO}}}{\widetilde{c} \widetilde{d} J} = \sum_k (-)^k \! \mel*{\widetilde{a} \widetilde{b} J}{\widetilde{v}_{12}^{\text{SO}, \- k}}{\widetilde{c} \widetilde{d} J}
\end{equation}
where the $k$-th component of the spin--orbit interaction is defined as:
\begin{equation}
\widetilde{v}_{12}^{\text{SO}, \- k} \equiv V(r_{12}) [\vec{r}_{12} \otimes \vec{\nabla}_{12}]^{(1)}_{-k} [\vec{\sigma}_1 + \vec{\sigma}_2]^{(1)}_k
\end{equation}
The transformation from $jj$ coupling to $LS$ coupling provides:
\begin{multline}
\mel*{\widetilde{a} \widetilde{b} J}{\widetilde{v}_{12}^{\text{SO}, \- k}}{\widetilde{c} \widetilde{d} J} = \sum_{LS} \sum_{L' S'} \hat{j}_a \hat{j}_b \hat{j}_c \hat{j}_d \hat{L} \hat{L}' \hat{S} \hat{S}' \times \\ \begin{Bmatrix} l_a & l_b & L \\ 1/2 & 1/2 & S \\ j_a & j_b & J \end{Bmatrix} \begin{Bmatrix} l_c & l_d & L' \\ 1/2 & 1/2 & S' \\ j_c & j_d & J \end{Bmatrix} \times \\
\underbrace{\! \mel*{n_a l_a \frac{1}{2} \, n_b l_b \frac{1}{2} \, L S J}{\widetilde{v}_{12}^{\text{SO}, \- k}}{n_c l_c \frac{1}{2} \, n_d l_d \frac{1}{2} \, L' S' J} \, }_{B^{\text{SO}}_k}
\end{multline}
where the space and spin parts split up in such a way that:
\begin{multline}
B^{\text{SO}}_k = \sum_{M_L M_S} \sum_{M'_L M'_S}   \braket*{L M_L S M_S}{J M_L+M_S} \times \\ \braket*{L' M'_L S' M'_S}{J M'_L+M'_S} \\
 \underbrace{\! \mel*{n_a l_a \, n_b l_b \, L M_L}{V(r_{12}) [\vec{r}_{12} \otimes \hat{\nabla}_{12}]^{(1)}_{-k}}{n_c l_c \, n_d l_d \, L' M'_L}}_{C_{-k}^{\text{SO}}} \times \\ \underbrace{\! \mel*{\frac{1}{2} \, \frac{1}{2} \, S M_S}{[\vec{\sigma}_1 + \vec{\sigma}_2]^{(1)}_k}{\frac{1}{2} \, \frac{1}{2} \, S' M'_S}}_{D_k^{\text{SO}}}
\end{multline}
On the one hand, the Wigner-Eckart theorem relates the above spin TBME to its reduced counterpart according to \cite{Talmi1993}:
\begin{multline} \label{DkSO}
D_k^{\text{SO}} = (-)^{S-M_S} \begin{pmatrix} S & 1 & S' \\ -M_S & k & M'_S \end{pmatrix} \times \\ \langle \frac{1}{2} \, \frac{1}{2} \, S \| \vec{\sigma}_1 + \vec{\sigma}_2 \| \frac{1}{2} \, \frac{1}{2} \, S' \rangle
\end{multline}
Separating the contributions from the first and second particles in the spin reduced TBME, one obtains:
\begin{multline}
\langle \frac{1}{2} \, \frac{1}{2} \, S \| \vec{\sigma}_1 + \vec{\sigma}_2 \| \frac{1}{2} \, \frac{1}{2} \, S' \rangle = \underbrace{\langle \frac{1}{2} \, \frac{1}{2} \, S \| \vec{\sigma}_1 \| \frac{1}{2} \, \frac{1}{2} \, S' \rangle}_{E_1} \\ + \underbrace{\langle \frac{1}{2} \, \frac{1}{2} \, S \| \vec{\sigma}_2 \| \frac{1}{2} \, \frac{1}{2} \, S' \rangle}_{E_2}
\end{multline}
where one has \cite{Talmi1993}:
\begin{subequations}
\begin{align}
E_1 = (-)^{S'} \hat{S} \hat{S}' \begin{Bmatrix} 1/2 & S & 1/2 \\ S' & 1/2 & 1 \end{Bmatrix} \langle \frac{1}{2} \| \vec{\sigma}_1 \| \frac{1}{2} \rangle \\
E_2 = (-)^{S} \hat{S} \hat{S}' \begin{Bmatrix} 1/2 & S & 1/2 \\ S' & 1/2 & 1 \end{Bmatrix} \langle \frac{1}{2} \| \vec{\sigma}_2 \| \frac{1}{2} \rangle
\end{align}
\end{subequations}
with the above reduced matrix elements given by \eqref{reducedTBMEspin}. 
On the one hand, the Wigner-$3j$ symbol appearing in \eqref{DkSO} implies that $S$ and $S'$ cannot be simultaneously equal to $0$. On the other hand, if $S=0$ and $S'=1$ or $S=1$ and $S'=0$, the quantities $E_1$ and $E_2$ expressed above cancel one another so that the spin--orbit TBMEs vanish. Thus, the only possibility is $S=S'=1$,  just like for the tensor interaction (see discussion after Eq.\eqref{reducedTBMEspin}). Once again, this result was expected since the spin-orbit term only acts in the $S=1$ channel of the interaction. Gathering the equations, one obtain, for the spin part of the spin--orbit TBMEs:
\begin{multline}
D_k^{\text{SO}} = 2 \sqrt{6} (-)^{S-M_S} \hat{S} \hat{S}' \begin{pmatrix} S & 1 & S' \\ -M_S & k & M'_S \end{pmatrix} \times \\ \begin{Bmatrix} 1/2 & S & 1/2 \\ S' & 1/2 & 1 \end{Bmatrix}
\end{multline}
As for the spin-orbit spatial part, one finds:
\begin{multline}
C_{-k}^{\text{SO}} = \sum_{m_{la} m_{lb}} \sum_{m_{lc} m_{ld}} \! \braket*{l_a m_{la} l_b m_{lb}}{L M_L} \times \\ \braket*{l_c m_{lc} l_c m_{lc}}{L' M'_L} \times \\ \underbrace{\! \mel*{n_a l_a m_{la} \, n_b l_b m_{lb}}{V(r_{12}) [\vec{r}_{12} \otimes \hat{\nabla}_{12}]^{(1)}_{-k}}{n_c l_c m_{lc} \, n_d l_d m_{ld}}}_{v_{r_a r_b r_c r_d}^{\text{SO} \, -k}}
\end{multline}
Thus, one needs to evaluate the spin-orbit TBME in the spherical HO basis, whose wave functions are defined in \eqref{swavef}.  Decoupling the tensor product, one has:
\begin{multline} \label{debutSO}
v_{r_a r_b r_c r _d}^{\text{SO} \, -k} = \sum_{\alpha \beta} \! \braket*{1 \alpha 1 \beta}{1 -k} \times \\ \underbrace{\! \mel*{n_a l_a m_{la} \, n_b l_b m_{lb}}{r^{\alpha}_{12} V(r_{12}) \nabla^{\beta}_{12}}{n_c l_c m_{lc} \, n_d l_d m_{ld}}}_{v_{r_a r_b r_c r _d}^{\text{SO} \, \alpha \beta}}
\end{multline}
with, by definition,
\begin{multline}
v_{r_a r_b r_c r _d}^{\text{SO} \, \alpha \beta} \equiv \int \dd[3]{r_1} \int \dd[3]{r_2} \phi_{r_a}^*(\vec{r}_1) \phi_{r_b}^*(\vec{r}_2) \times \\ V(r_{12}) r_{12}^{\alpha} \big[ \nabla_{12}^{\beta} \phi_{r_c}(\vec{r}_1) \phi_{r_d}(\vec{r}_2) \big]
\end{multline}
This time, the sandwiched quantity does not commute with the wave functions because of the gradient operator. Then, by separating the gradients acting on the first and second particles, one obtains:
\begin{multline}
v_{r_a r_b r_c r _d}^{\text{SO} \, \alpha \beta} = \int \dd[3]{r_1} \int \dd[3]{r_2} \phi_{r_a}^*(\vec{r}_1) \phi_{r_b}^*(\vec{r}_2) \times \\ V(r_{12}) r_{12}^{\alpha} \big[ \nabla_1^{\beta} \phi_{r_c}(\vec{r}_1) \big] \phi_{r_d}(\vec{r}_2) \\
 - \int \dd[3]{r_1} \int \dd[3]{r_2} \phi_{r_a}^*(\vec{r}_1) \phi_{r_b}^*(\vec{r}_2) \times \\ V(r_{12}) r_{12}^{\alpha} \phi_{r_c}(\vec{r}_1) \big[ \nabla_{2}^{\beta} \phi_{r_d}(\vec{r}_2) \big]
\end{multline}
In the following, $I_1^{\alpha \beta}$ and $I_2^{\alpha \beta}$ 
ar the first and second integrals of the above right-hand side term (the other indices are omitted for conciseness). One focuses on the integral $I_1^{\alpha \beta}$, the integral $I_2^{\alpha \beta}$ being easily deductible from it. One has to evaluate the action of the gradient operator on the wave functions with the help of the gradient formula in spherical symmetry \eqref{grafinal}. One finds:
\begin{multline} \label{int1ab}
I_1^{\alpha \beta} = \frac{1}{b} \sum_{j = \pm 1} (-)^{l_c + m_{lc} + \beta + 1} \begin{pmatrix} l_c & 1 & l_c + j \\ m_{lc} & \beta & -m_{lc} - \beta \end{pmatrix} \\
\times \bigg[ a_j^{(r_c)} \int \dd[3]{r_1} \int \dd[3]{r_2} \phi_{r_a}^*(\vec{r}_1) \times \\ \phi_{n_c-j,  l_c+j, m_{lc} + \beta}(\vec{r}_1) V(r_{12}) r_{12}^{\alpha} \phi_{r_b}^*(\vec{r}_2) \phi_{r_d}(\vec{r}_2) \\
+ c_j^{(r_c)} \int \dd[3]{r_1} \int \dd[3]{r_2} \phi_{r_a}^*(\vec{r}_1) \times \\ \phi_{n_c,  l_c+j,  m_{lc} + \beta}(\vec{r}_1) V(r_{12}) r_{12}^{\alpha} \phi_{r_b}^*(\vec{r}_2) \phi_{r_d}(\vec{r}_2) \bigg]
\end{multline}
where the coefficients $a_j^{(r_c)}$ and $c_j^{(r_c)}$ are defined in \eqref{coefs}. Now,  using the Gogny separable expansion in spherical symmetry \eqref{Gognysepspherique} four times, one ends up with:
\begin{multline}
I_1^{\alpha \beta} = \frac{1}{b} \sum_{j = \pm 1} (-)^{l_c + m_{lc} + \beta + 1} \begin{pmatrix} l_c & 1 & l_c + j \\ m_{lc} & \beta & -m_{lc} - \beta \end{pmatrix} \\
 \times \sum_{l_\mu m_{l\mu}} \sum_{r_\nu} T_{(r_b) (r_d)}^{(r_\nu)} \! \mel*{l_a m_{la}}{Y_{l_\mu}^{m_{l\mu}*}}{l_c+j \, m_{lc} + \beta} \times \\ \mel*{l_b m_{lb}}{Y_{l_\nu}^{m_{l\nu}*}}{l_d m_{ld}} \\
\times \bigg[ a_j^{(r_c)} \sum_{n_\mu} T_{(r_a) (n_c-j, l_c+j)}^{(r_\mu)}  \int \dd[3]{r_1} \int \dd[3]{r_2} \phi_{0}(\vec{r}_1) \times \\ \phi_{0}(\vec{r}_2) V(r_{12}) r_{12}^{\alpha} \phi_{r_\mu}(\vec{r}_1) \phi_{r_\nu}(\vec{r}_2) \\
 + c_j^{(r_c)} \sum_{n'_\mu} T_{(r_a) (n_c, l_c+j)}^{(r'_\mu)} \int \dd[3]{r_1} \int \dd[3]{r_2} \phi_{0}(\vec{r}_1) \times \\ \phi_{0}(\vec{r}_2) V(r_{12}) r_{12}^{\alpha} \phi_{r'_\mu}(\vec{r}_1) \phi_{r_\nu}(\vec{r}_2) \bigg]
\end{multline}
One notes that the notations $r_\mu = (n_\mu, l_\mu,m_{l\mu})$ and $r'_\mu = (n'_\mu, l_\mu,m_{l\mu})$ have been introduced since the quantum numbers $n_\mu$ and $n'_\mu$ run over distinct ranges of values, as they appear in different Talman coefficients (see condition \eqref{condTalmansph}). In the following, one will focus on the first integral, the second one is deduced by the transformation $r_\mu \rightarrow r'_\mu$. Then, applying twice the Moshinsky transformation in spherical symmetry \eqref{Moshinskytransfospherique} on the first integral, one gets:
\begin{multline}
\int \dd[3]{r_1} \int \dd[3]{r_2} \phi_{0}(\vec{r}_1) \phi_{0}(\vec{r}_2) V(r_{12}) \times \\ r_{12}^{\alpha} \phi_{r_\mu}(\vec{r}_1) \phi_{r_\nu}(\vec{r}_2) =  \sqrt{2} \sum_{r_\lambda r_\sigma} M_{r_\mu r_\nu}^{r_\lambda r_\sigma} 
\int \dd[3]{r} \phi_{0}(\vec{r}) \times \\  V(\sqrt{2} r) r^{\alpha} \phi_{r_\sigma}(\vec{r}) \int \dd[3]{R} \phi_{0}^*(\vec{R}) \phi_{r_\lambda}(\vec{R})
\end{multline}
since the Jacobian of the change of variables $(\vec{r}_1, \vec{r}_2) \rightarrow (\vec{r}, \vec{R})$, given by \eqref{rRspherique},  is equal to unity and the spherical Moshinsky coefficient, specified by \eqref{defMoshinsky}, fixes the range of values of $r_\sigma$ according to \eqref{condMoshinsky1spherique} and \eqref{condMoshinsky2spherique}. One note that in the particular case $r'_\mu = r'_\nu = 0$, the spherical Moshinsky coefficient reduces to \eqref{Moshspherique00} and that $\phi_{0}(\vec{R}) = \phi^*_{0}(\vec{R})$ because of \eqref{phi0spherique}. 

The integral over $\vec{R}$ is readily carried out considering the orthogonality relation of the spherical HO wave functions \eqref{orthogonalitysph}.  On the other hand, using the writing \eqref{harmoSO} in spherical symmetry, one simplifies the integral over $\vec{r}$ by pulling out its angular part by means of \eqref{phisansSH}, namely:
\begin{multline}
\int \dd[3]{r} \phi_{0}(\vec{r})V(\sqrt{2} r) r^{\alpha} \phi_{r_\sigma}(\vec{r}) \\ =  
\sqrt{\frac{4 \pi}{3}} \int \dd[3]{r} r \phi_{0}(\vec{r}) V(\sqrt{2}r) \phi_{r_{\sigma}}(\vec{r}) Y_1^{\alpha}(\hat{r}) \\
= \sqrt{\frac{4 \pi}{3}} \frac{1}{\sqrt{4 \pi}} \int \dd{r} r^3 \phi_{(0)}(r) V(\sqrt{2}r) \phi_{(r_\sigma)}(r)  \times \\ \int \dd[2]{\hat{r}} Y_1^{\alpha}(\hat{r}) Y_{l_\sigma}^{m_{l \sigma}}(\hat{r}) \notag \\
 = \frac{(-)^{\alpha}}{\sqrt{3}} \delta_{l_\sigma,1} \delta_{m_{l \sigma},-\alpha} \times \int \dd{r} r^3 \phi_{(0)}(r) V(\sqrt{2}r) \phi_{(r_\sigma)}(r)
\end{multline}
where the orthogonality of the spherical harmonics \eqref{orthogonalitySH} has been used. Finally, using the identity $r \phi_{(0,0)} (r) = \sqrt{3/2} \- b \- \phi_{(0,1)} (r)$ of the spherical HO wave functions \eqref{swavef}, one obtains:
\begin{multline}
\int \dd[3]{r} \phi_{0}(\vec{r})V(\sqrt{2} r) r_{12}^{\alpha} \phi_{r_\sigma}(\vec{r}) = \frac{(-)^{\alpha} b}{\sqrt{2}} \times \\ \delta_{l_\sigma,1} \delta_{m_{l \sigma},-\alpha} \int \dd{r} r^2 \phi_{(0,1)}(r) V(\sqrt{2}r) \phi_{(n_\sigma,1)}(r)
\end{multline}
Thus, the only thing that remains to be done is to evaluate the above integral. Considering the particular wave functions:
\begin{equation}
\phi_{(0,1)}(r) = \bigg[ \frac{8}{3 \sqrt{\pi} \- b^3} \bigg]^{1/2} \frac{r}{b} \, \e^{-r^2/2b^2}
\end{equation}
and
\begin{multline}
\phi_{(n_\sigma,1)}(r) = \bigg[ \frac{2}{b^3} \frac{n_\sigma !}{\Gamma(n_\sigma + 5/2)} \bigg]^{1/2} \frac{r}{b} \, \e^{-r^2/2b^2} \times \\ L_{n_\sigma}^{3/2} \bigg( \frac{r^2}{b^2} \bigg)
\end{multline}
as well as the series expansion of the generalized Laguerre polynomial, one obtains,  after a development similar to the ones made for the central and tensor TBMEs: 
\begin{multline}
\int \dd{r} r^2 \phi_{(0,1)}(r) V(\sqrt{2}r) \phi_{(n_\sigma,1)}(r) = \\ \bigg[ \frac{4}{3 \sqrt{\pi}} n_\sigma! \, \Gamma(n_\sigma + 5/2) \bigg]^{1/2}  \sum_{i=0}^{n_\sigma} \frac{(-)^i G^{-i-5/2}}{i! (n_\sigma-i)!}
\end{multline}
where $G$ is defined in \eqref{Gspherique}. Gathering all the results,  one can write down the expression for the first integral $I_1^{\alpha \beta}$, given by \eqref{int1ab}, namely:
\begin{multline} \label{SO_1}
I_1^{\alpha \beta} = \sum_{j = \pm 1} (-)^{l_c + m_{lc} + \alpha + \beta + 1} \begin{pmatrix} l_c & 1 & l_c + j \\ m_{lc} & \beta & -m_{lc} - \beta \end{pmatrix}  \times \\
\sum_{l'_\mu m'_{l\mu}} \sum_{r_\nu} T_{(r_b) (r_d)}^{(r_\nu)} \! \mel*{l_a m_{la}}{Y_{l'_\mu}^{m'_{l\mu}*}}{l_c+j \, m_{lc} + \beta} \times \\ \mel*{l_b m_{lb}}{Y_{l_\nu}^{m_{l\nu}*}}{l_d m_{ld}} \times \\
\Bigg[ a_j^{(r_c)} T_{(r_a) (n_c-j, l_c+j)}^{(r^{(1)}_\mu)} M_{r^{(1)}_\mu r_\nu}^{0 \, r^{(1)}_\sigma}\times \\ \bigg[ \frac{4}{3 \sqrt{\pi}} n^{(1)}_\sigma! \, \Gamma(n^{(1)}_\sigma + 5/2) \bigg]^{1/2}  \times \\ \sum_{i=0}^{n^{(1)}_\sigma} \frac{(-)^i G^{-i-5/2}}{i! (n^{(1)}_\sigma-i)!} \\
 + c_j^{(r_c)} T_{(r_a) (n_c, l_c+j)}^{(r^{(2)}_\mu)}  M_{r^{(2)}_\mu r_\nu}^{0 \, r^{(2)}_\sigma} \times \\ \bigg[ \frac{4}{3 \sqrt{\pi}} n^{(2)}_\sigma! \, \Gamma(n^{(2)}_\sigma + 5/2) \bigg]^{1/2}  \times \\ \sum_{i=0}^{n^{(2)}_\sigma} \frac{(-)^i G^{-i-5/2}}{i! (n^{(2)}_\sigma-i)!} \Bigg]
\end{multline}
where $G$ is expressed in \eqref{Gspherique}. 
For clarity, one has changed the notations a bit in the above expression. They are $r^{(i)}_\mu = (n^{(i)}_\mu,l'_\mu, m'_{l\mu})$ for $i \in \{1,2\}$ and $r_\nu = (n_\nu,l_\nu, m_{\nu})$. Moreover, one has $r^{(1)}_\sigma = (n^{(1)}_\sigma, 2, k)$ and $r^{(2)}_\sigma = (n^{(2)}_\sigma, 2, k)$ with $n^{(1)}_\sigma = (X^{(1)}_\mu + X_\nu - 2)/2$ and $n^{(2)}_\sigma = (X^{(2)}_\mu + X_\nu - 2)/2$, according to \eqref{condMoshspheriquereduce}. One sees that $n^{(1)}_\sigma$ and $n^{(2)}_\sigma$ respectively depend on the values of $n^{(1)}_\mu$ and $n^{(2)}_\mu$, hence the notations chosen. 
Moreover, the conditions \eqref{condY} and \eqref{condTalmansph} have to be fulfilled at any time. For the sake of completeness, one also writes down the full expression of the integral $I_2^{\alpha \beta}$ whose derivation is analogous to the one of $I_1^{\alpha \beta}$. One has:
\begin{multline} \label{SO_2}
I_2^{\alpha \beta} = \sum_{j = \pm 1} (-)^{l_d + m_{ld} + \alpha + \beta + 1} \begin{pmatrix} l_d & 1 & l_d + j \\ m_{ld} & \beta & -m_{ld} - \beta \end{pmatrix}  \times \\ 
\sum_{l'_\nu m'_{l\nu}} \sum_{r_\mu} T_{(r_a) (r_c)}^{(r_\mu)} \! \mel*{l_a m_{la}}{Y_{l_\mu}^{m_{l\mu}*}}{l_c m_{lc}} \times \\ \mel*{l_b m_{lb}}{Y_{l'_\nu}^{m'_{l\nu}*}}{l_d+j \, m_{ld} + \beta} \times  \\
\bigg[ a_j^{(r_d)} T_{(r_b) (n_d-j, l_d+j)}^{(r^{(1)}_\nu)} M_{r_\mu r^{(1)}_\nu}^{0 \, r^{(3)}_\sigma} \times  \\ \bigg[ \frac{4}{3 \sqrt{\pi}} n^{(3)}_\sigma! \, \Gamma(n^{(3)}_\sigma + 5/2) \bigg]^{1/2}  \times \sum_{i=0}^{n^{(3)}_\sigma} \frac{(-)^i G^{-i-5/2}}{i! (n^{(3)}_\sigma-i)!} \\
+ c_j^{(r_d)} T_{(r_b) (n_d, l_d+j)}^{(r^{(2)}_\nu)} M_{r_\mu r^{(2)}_\nu}^{0 \, r^{(4)}_\sigma} \times \\ \bigg[ \frac{4}{3 \sqrt{\pi}} n^{(4)}_\sigma! \, \Gamma(n^{(4)}_\sigma + 5/2) \bigg]^{1/2}  \times \sum_{i=0}^{n^{(4)}_\sigma} \frac{(-)^i G^{-i-5/2}}{i! (n^{(4)}_\sigma-i)!} \bigg]
\end{multline}
where $G$ is expressed in \eqref{Gspherique}. Once again, the notations have been changed a bit in the above expression. They are $r^{(i)}_\nu = (n^{(i)}_\nu,l'_\nu, m'_{l\nu})$ for $i \in \{1,2\}$ and $r_\mu = (n_\mu,l_\mu, m_{\mu})$. Besides, one has $r^{(3)}_\sigma = (n^{(3)}_\sigma, 2, k)$ and $r^{(4)}_\sigma = (n^{(4)}_\sigma, 2, k)$ with $n^{(3)}_\sigma = (X_\mu + X^{(1)}_\nu - 2)/2$ and $n^{(4)}_\sigma = (X_\mu + X^{(2)}_\nu - 2)/2$, according to \eqref{condMoshspheriquereduce}. One sees that $n^{(3)}_\sigma$ and $n^{(4)}_\sigma$ respectively depend on the values of $n^{(1)}_\nu$ and $n^{(2)}_\nu$, hence the notations chosen (see the previous paragraph).  From equation \eqref{debutSO} and these expressions, one directly deduces the TBME of the spin-orbit interaction in the spherical HO basis.


\section{Derivation of the Landau parameter expressions for the DG Gogny interaction}\label{Landauparam}

In this appendix, a detailed derivation of the Landau parameters $f_l^{ST}(\theta)$ and $h_l^{1T}(\theta)$, associated with the DG Gogny interaction, is given. 
The appropriate limits allow to recover the ones of the D1 and D2-type Gogny interactions. 

\subsection{Central and density-dependent contributions} \label{centralLandauA}

In this subsection,  one will derives the various Landau parameters associated with the finite-range central and density-dependent terms of the generalized Gogny interaction \eqref{gognyDG}. The antisymmetrized central and density-dependent (CDD) interactions are characterized by the set of equations \eqref{CDDpot}-\eqref{FrhoINM}, with their direct and exchange spin-isospin components \eqref{WBHMCDDINM}, written under the equivalent forms, respectively:
\begin{multline} \label{WCDD2}
\mathcal{P}_{\text{D}} = \Big( W + \frac{B}{2} - \frac{H}{2} - \frac{M}{4} \Big) + \Big( \frac{B}{2} - \frac{M}{4} \Big)(\vec{\sigma}_1 \cdot \vec{\sigma}_2) \\ 
- \Big( \frac{H}{2} + \frac{M}{4} \Big)(\vec{\tau}_1 \cdot \vec{\tau}_2) - \frac{M}{4} (\vec{\sigma}_1 \cdot \vec{\sigma}_2) (\vec{\tau}_1 \cdot \vec{\tau}_2)  \\ 
\mathcal{P}_{\text{E}} = \Big( M + \frac{H}{2} - \frac{B}{2} - \frac{W}{4} \Big) + \Big( \frac{H}{2} - \frac{W}{4} \Big)(\vec{\sigma}_1 \cdot \vec{\sigma}_2) \\ 
- \Big( \frac{B}{2} + \frac{W}{4} \Big)(\vec{\tau}_1 \cdot \vec{\tau}_2) - \frac{W}{4} (\vec{\sigma}_1 \cdot \vec{\sigma}_2) (\vec{\tau}_1 \cdot \vec{\tau}_2)
\end{multline}
Onee notices that the direct spin-isospin components of the CDD interactions can be deduced from the exchange ones upon replacing $W, B,H,M$ by $M,H,B,W$. 

In order to get the contributions from the central and density-dependent interactions to the matrix elements of the particle-hole interaction \eqref{MEph}, one 
evaluates the following antisymmetrized matrix elements in the particle-hole representation:
\begin{multline} \label{elmatCDDPW}
\mel*{\vec{k}_p \vec{k}_{h'}}{v_{12}^{\text{CDD,(a)}}}{\vec{k}_h \vec{k}_{p'}} = \\ \! \mel*{\vec{k}_p \vec{k}_{h'}}{V(r_{12}) D[\rho]}{\vec{k}_h \vec{k}_{p'}} \! \mathcal{P}_{\text{D}} \\
+  \mel*{\vec{k}_p \vec{k}_{h'}}{V(r_{12}) D[\rho]}{\vec{k}_{p'} \vec{k}_h}  \mathcal{P}_{\text{E}}
\end{multline}
where one has explicitly separated the direct and exchange as well as the spatial and spin-isospin parts. Since the spin-isospin part is fully specified by \eqref{WCDD2}, one focusses on the spatial parts. In fact, the calculation is pretty similar to the one in Eq.\eqref{elmatPW}. The only difference is that the conservation of the quasiparticle pair momentum is invoked here, i.e.\ $\vec{k}_p + \vec{k}_{h'} = \vec{k}_h + \vec{k}_{p'}$.  One eventually finds out:
\begin{multline}
\mel*{\vec{k}_p \vec{k}_{h'}}{V(r_{12}) D[\rho]}{\vec{k}_h \vec{k}_{p'}} = \\\frac{\rho^{\alpha}}{\mathcal{V}} (\mu \sqrt{\pi})^3 \, \e^{-\mu^2 (\vec{k}_p - \vec{k}_h)^2/4} \\ \\
\mel*{\vec{k}_p \vec{k}_{h'}}{V(r_{12}) D[\rho]}{\vec{k}_{p'} \vec{k}_h} = ~~~~~~~~~~~~~~~~~~~~~~ \\ \frac{\rho^{\alpha}}{\mathcal{V}} (\mu \sqrt{\pi})^3 \, \e^{-\mu^2 (\vec{k}_p - \vec{k}_{p'})^2/4}
\end{multline}
where, from $\alpha = 0$, one deduces the central contributions and from $\alpha \ne 0$ the density-dependent one. Now, at the Landau limit, $|\vec{k}_p| = |\vec{k}_h| = |\vec{k}_{\text{F}}|$ and $|\vec{k}_{p'}| = |\vec{k}_{h'}| = |\vec{k}'_{\text{F}}|$, so that the above relations become:
\begin{equation}
\begin{array}{lcl}
\mel*{\vec{k}_{\text{F}} \vec{k}'_{\text{F}}}{V(r_{12}) D[\rho]}{\vec{k}_{\text{F}} \vec{k}'_{\text{F}}} & =& \frac{\rho^{\alpha}}{\mathcal{V}} (\mu \sqrt{\pi})^3 \\
&=&  G_\mu^\alpha(0) \\ \\
\mel*{\vec{k}_{\text{F}} \vec{k}'_{\text{F}}}{V(r_{12}) D[\rho]}{\vec{k}'_{\text{F}} \vec{k}_{\text{F}}} & = &\frac{\rho^{\alpha}}{\mathcal{V}} (\mu \sqrt{\pi})^3 \, \e^{-\mu^2 \vec{q}_{\text{F}}^{\, 2}/4} \\ 
& =&  G_\mu^\alpha(\vec{q}_{\text{F}})
\end{array}
\end{equation}
where one has defined the function $G_\mu^\alpha(\vec{k}) \equiv \rho^\alpha (\mu \sqrt{\pi})^3/\mathcal{V} \times \, \e^{-\mu^2 \vec{k}^{\, 2}/4}$ and considered the relative momentum at the Fermi surface, $\vec{q}_{\text{F}} = \vec{k}_{\text{F}} - \vec{k}'_{\text{F}}$. In order to connect these expressions to the parameters appearing in Eq.\eqref{QPinteraction}, one also defines: 
\begin{equation} \label{GmuCDD}
\widetilde{G}_\mu^\alpha(\vec{k}) \equiv N_0 G_\mu^\alpha(\vec{k}) = \frac{2 m^* k_{\text{F}} \rho^\alpha}{\pi^2 \hbar^2} (\mu \sqrt{\pi})^3 \, \e^{-\mu^2 \vec{k}^{\, 2}/4}
\end{equation}
Then, with Eqs.\eqref{WCDD2} and \eqref{elmatCDDPW}, one has:
\begin{subequations}
\begin{align}
f^{00}(\theta) & = \Big( W + \frac{B}{2} - \frac{H}{2} - \frac{M}{4} \Big) \widetilde{G}^\alpha_\mu(0) \notag \\ &+ \Big( M + \frac{H}{2} - \frac{B}{2} - \frac{W}{4} \Big) \widetilde{G}^\alpha_\mu(\vec{q}_{\text{F}}) \\
f^{10}(\theta) & = \Big( \frac{B}{2} - \frac{M}{4} \Big) \widetilde{G}^\alpha_\mu(0) + \Big( \frac{H}{2} - \frac{W}{4} \Big) \widetilde{G}^\alpha_\mu(\vec{q}_{\text{F}}) \\
f^{01}(\theta) & = -\Big( \frac{H}{2} + \frac{M}{4} \Big) \widetilde{G}^\alpha_\mu(0) - \Big( \frac{B}{2} + \frac{W}{4} \Big) \widetilde{G}^\alpha_\mu(\vec{q}_{\text{F}}) \\
f^{11}(\theta) & = - \frac{M}{4} \widetilde{G}^\alpha_\mu(0) - \frac{W}{4} \widetilde{G}^\alpha_\mu(\vec{q}_{\text{F}})
\end{align}
\end{subequations}
To get the Landau parameters out of the $f^{ST}$, one expands them in Legendre polynomials according to Eq.\eqref{expansionf}. 
One starts with the function: 
\begin{equation}
\widetilde{G}_\mu^\alpha(\vec{q}_{\text{F}}) = N_0 \frac{\rho^{\alpha}}{\mathcal{V}} (\mu \sqrt{\pi})^3 \, \e^{-\mu^2 k_{\text{F}}^2 (1- \cos \theta)/2}
\end{equation}
where one has used the magnitude of the relative momentum at the Fermi surface $q_{\text{F}} = k_{\text{F}} \sqrt{2 (1 - \cos \theta)}$.
The plane wave expansion of the exponential, $\e^{\ii \vec{k} \cdot \vec{r}} = \sum_{l=0}^{\infty} (2l+1) \- \ii^l j_l(kr) P_l(\cos \theta)$, furnishes:
\begin{multline}
\e^{\mu^2 k^2_{\text{F}} \cos \theta/2} = \e^{\ii (- \ii \mu^2 k^2_{\text{F}}) \cos \theta/2} \\ = \sum_l (2l+1) \- \ii^l j_l \big(- \ii \mu^2 k^2_{\text{F}}/2 \big) P_l(\cos \theta)
\end{multline}
If the argument of the Bessel function of the first kind $j_l$ had been real, onee would have directly considered its analytical expression for a given $l$, the first three examples of which are given by:
\begin{subequations} \label{Besselexplicit}
\begin{align}
j_0(x) & = \frac{\sin x}{x},  \label{Bessel0} \\
j_1(x) & = \frac{\sin x}{x^2} - \frac{\cos x}{x},  \label{Bessel1} \\
j_2(x) & = \bigg( \frac{3}{x^2} - 1 \bigg) \frac{\sin x}{x} - \frac{3 \cos x}{x^2}\label{Bessel2}
\end{align}
\end{subequations}
Here, however, the argument is purely imaginary, so one has no choice but to expand the function in infinite series according to:
\begin{equation} \label{sphBesselimaginary}
j_n(\ii x) = \ii^n \frac{\sqrt{\pi}}{2} \sum_{m=0}^{\infty} \frac{1}{m! \- \Gamma(m+n+3/2)} \Big( \frac{x}{2} \Big)^{2m+n}
\end{equation}
with $n$ $\in \mathbb{Z}$.
In practice this series will be truncated to a finite order, which is not a problem since it converges rather quickly. One therefore finds out:
\begin{multline}
\e^{\mu^2 k^2_{\text{F}} \cos \theta/2} = \frac{\sqrt{\pi}}{2} \sum_l (2l+1) \times \\ 
\sum_m \frac{\big(\mu^2 k_{\text{F}}^2/4\big)^{2m+l}}{m! \- \Gamma(m+l+3/2)} P_l(\cos \theta)
\end{multline}
and finally obtains:
\begin{equation}
\widetilde{G}_\mu^\alpha(\vec{q}_{\text{F}}) = \widetilde{G}_\mu^\alpha(0) \sum_l G_l P_l (\cos \theta)
\end{equation}
where the coefficients of this expansion in Legendre polynomials are defined by:
\begin{equation}
G_l \equiv \frac{\sqrt{\pi}}{2} (2l+1) \, \e^{-\mu^2 k_{\text{F}}^2/2} \sum_m \frac{\big(\mu^2 k_{\text{F}}^2/4\big)^{2m+l}}{m! \- \Gamma(m+l+3/2)}
\end{equation}
Thus, the contributions of the central and density-dependent interactions to the Landau parameters defined in Eq.\eqref{legendreexpansion}, can be evaluated, for $l \in \mathbb{N}$, by means of the relations:
\begin{subequations}
\begin{align}
f_l^{00} & =  \widetilde{G}_\mu^\alpha(0) \bigg[ \Big( W + \frac{B}{2} - \frac{H}{2} - \frac{M}{4} \Big) \delta_{l,0} \notag \\ & + \Big( M + \frac{H}{2} - \frac{B}{2} - \frac{W}{4} \Big) G_l \bigg] \notag \\
f_l^{10} & = \widetilde{G}_\mu^\alpha(0) \bigg[ \Big( \frac{B}{2} - \frac{M}{4} \Big) \delta_{l,0} + \Big( \frac{H}{2} - \frac{W}{4} \Big) G_l \bigg] \notag \\
f_l^{01} & = - \widetilde{G}_\mu^\alpha(0) \bigg[ \Big( \frac{H}{2} + \frac{M}{4} \Big) \delta_{l,0} + \Big( \frac{B}{2} + \frac{W}{4} \Big) G_l \bigg] \notag \\
f_l^{11} & = - \widetilde{G}_\mu^\alpha(0) \bigg[ \frac{M}{4} \delta_{l,0} + \frac{W}{4} G_l \bigg]\notag 
\end{align}
\end{subequations}
One keeps in mind that one must set $\alpha = 0$ in the quantity $\widetilde{G}_\mu^\alpha(0)$ to get the contributions of the central terms, and divide it by the coefficient $(\mu \sqrt{\pi})^3$ to get the one of the density-dependent term.

\subsection{Rearrangement contributions}

In this subsection, one continues by deriving the rearrangement contributions to the Landau parameters associated with the generalized Gogny interaction \eqref{gognyDG}. In DG, only the density-dependent term gives rise to such terms. According to Eq.\eqref{MEph}, three rearrangement terms appear. 

\paragraph{First rearrangement term}

To compute the first one, it will be useful to write the derivative of the antisymmetrized interaction as:
\begin{multline} \label{derivv12}
\pdv{v_{12}^{\text{(\text{a})}}}{\rho_{hp}} = \pdv{v_{12}^{\text{(\text{a})}}}{\rho(\vec{r}_1)}\pdv{\rho(\vec{r}_1)}{\rho_{hp}} + \pdv{v_{12}^{\text{(\text{a})}}}{\rho(\vec{r}_2)}\pdv{\rho(\vec{r}_2)}{\rho_{hp}} \notag \\
= \frac{\alpha}{2} \frac{V(r_{12})}{(\mu \sqrt{\pi})^3} \bigg[ \rho^{\alpha-1}(\vec{r}_1)\pdv{\rho(\vec{r}_1)}{\rho_{hp}} \\ + \rho^{\alpha-1}(\vec{r}_2)\pdv{\rho(\vec{r}_2)}{\rho_{hp}} \bigg] (\mathcal{P}_{\text{D}} + \mathcal{P}_{\text{E}} P_r)
\end{multline}
where the spin-isospin components of the direct $\mathcal{P}_{\text{D}}$ and exchange $\mathcal{P}_{\text{E}}$ density-dependent interaction are given by Eq.\eqref{WCDD2}. Using the definition \eqref{densityorigin} of the local nuclear density, one can rewrite the above derivatives thanks to the formula:
\begin{multline}
\pdv{\rho(\vec{r})}{\rho_{hp}} = \sum_{\sigma \tau} \Phi_p^*(\vec{r}, \sigma, \tau) \Phi_{h}(\vec{r}, \sigma, \tau) \\
= \sum_{\sigma \tau} \int \dd[3]{r'} \Phi_p^*(\vec{r}, \sigma, \tau) \delta(\vec{r} - \vec{r}^{\, \prime}) \Phi_{h}(\vec{r}, \sigma, \tau) \\
= \mel*{p}{ \delta(\vec{r} - \vec{r}^{\, \prime})}{h}
\end{multline}
Then, the derivative \eqref{derivv12} becomes:
\begin{multline} \label{derivativehp}
\displaystyle \pdv{v_{12}^{\text{(\text{a})}}}{\rho_{hp}} =  \displaystyle \frac{\alpha}{2} \frac{V(r_{12})}{(\mu \sqrt{\pi})^3} \langle p \vert \rho^{\alpha-1}(\vec{r}_1) \delta(\vec{r}_1 - \vec{r}_3) \\
  +  \rho^{\alpha-1}(\vec{r}_2) \delta(\vec{r}_2 - \vec{r}_3) \vert h \rangle (\mathcal{P}_{\text{D}} + \mathcal{P}_{\text{E}} P_r)
\end{multline}
and the first rearrangement term appearing in \eqref{MEph} can be written as:
\begin{multline} \label{v12rearr1}
\sum_i  \mel*{h'i}{\pdv{v_{12}^{\text{(a)}}}{\rho_{hp}}}{p'i} = \frac{\alpha}{2} \sum_i \bra*{h'ip} \\ 
\frac{V(r_{12})}{(\mu \sqrt{\pi})^3} \big[ \rho^{\alpha-1}(\vec{r}_1) \delta(\vec{r}_1 - \vec{r}_3) + \rho^{\alpha-1}(\vec{r}_2) \delta(\vec{r}_2 - \vec{r}_3) \big] \\ ( \mathcal{P}^{12}_{\text{D}} +  \mathcal{P}^{12}_{\text{E}} P^{12}_r) \ket*{p'ih}
\end{multline}
since the quantities only acting on the particles $1$ and $2$, spotted by their indices $12$, can be incorporated in the three-body matrix element, having no action on the third particle $p$ and third hole $h$. Defining
\begin{multline} \label{vd1}
v_{12}^{\text{d}_1} = \frac{\alpha}{2} \sum_i \bra*{i} \frac{V(r_{23})}{(\mu \sqrt{\pi})^3} \big[ \rho^{\alpha-1}(\vec{r}_2) \delta(\vec{r}_1 - \vec{r}_2) \\  + \rho^{\alpha-1}(\vec{r}_3) \delta(\vec{r}_1 - \vec{r}_3) \big]  ( \mathcal{P}^{23}_{\text{D}} +  \mathcal{P}^{23}_{\text{E}} P^{23}_r)  \ket*{i}
\end{multline}
one finds out, after a circular permutation of the variables in \eqref{v12rearr1},
\begin{equation} \label{firstrearr}
\sum_i \! \mel*{h'i}{\pdv{v_{12}^{\text{(a)}}}{\rho_{hp}}}{p'i} = \mel*{ph'}{v_{12}^{\text{d}_1}}{hp'} .
\end{equation}
Then, one starts by evaluating the direct component of Eq.\eqref{vd1}, which reads:
\begin{multline}
v_{12}^{\text{d}_1} \vert_{\text{D}} \equiv \frac{\alpha}{2} \sum_i \! \bra*{i}\! \frac{V(r_{23})}{(\mu \sqrt{\pi})^3} \big[ \rho^{\alpha-1}(\vec{r}_2) \delta(\vec{r}_1 - \vec{r}_2) \\ + \rho^{\alpha-1}(\vec{r}_3) \delta(\vec{r}_1 - \vec{r}_3) \big]  \mathcal{P}^{23}_{\text{D}} \! \ket*{i} \!.
\end{multline}
Considering the continuous limit \eqref{transfocontinu}, one obtains (with $i$ renamed $3$ for convenience),
\begin{multline}
v_{12}^{\text{d}_1} \vert_{\text{D}} = \frac{\alpha}{2} \frac{\mathcal{V}}{(2 \pi)^3}  \sum_{u_3} \! \mel*{u_3}{ \mathcal{P}^{23}_{\text{D}}}{u_3} \times \\ 
\int \dd[3]{r_3} \int \dd[3]{k_3} \phi_{\vec{k}_3}^*(\vec{r}_3) \frac{V(r_{23})}{(\mu \sqrt{\pi})^3} \times \\ \big[ \rho^{\alpha-1}(\vec{r}_2) \delta(\vec{r}_1 - \vec{r}_2) + \rho^{\alpha-1}(\vec{r}_3) \delta(\vec{r}_1 - \vec{r}_3) \big] \phi_{\vec{k}_3}(\vec{r}_3)
\end{multline}
On the one hand, the calculation of the nuclear density \eqref{densityINM} at the continuous limit \eqref{transfocontinu} immediately brings:
\begin{equation} \label{densityLandau}
\frac{\mathcal{V}}{(2 \pi)^3} \int \dd[3]{k_3} |\phi_{\vec{k}_3}(\vec{r}_3)|^2 = \frac{\rho(\vec{r}_3)}{4}
\end{equation}
where the factor $4$ comes from the spin and isospin degeneracies. On the other hand, expressing the scalar product of Pauli matrices, one has 
\begin{equation} \label{Pauli1Landau}
\mel*{s_3}{\vec{\sigma}_2 \cdot \vec{\sigma}_3}{s_3} = \sum_k (-)^k \sigma_2^{-k} \! \mel*{s_3}{\sigma_3^k}{s_3} = 2 s_3 \sigma_2^0
\end{equation}
where the unified relation of the matrix elements of the Pauli matrices, $\mel*{s_a}{\sigma_m}{s_b} = 2 s_a \delta_{s_a s_b} \delta_{m,0} - m \sqrt{2} \delta_{s_a, s_b+m}, \text{for } m = 0, \pm 1 $, has been used. One obviously gets an equivalent relation for the Pauli matrices associated with the isospin, so that: 
\begin{equation}
\sum_{s_3} \! \mel*{s_3}{\vec{\sigma}_2 \cdot \vec{\sigma}_3}{s_3} = \sum_{t_3} \! \mel*{t_3}{\vec{\tau}_2 \cdot \vec{\tau}_3}{t_3} = 0
\end{equation}
in such a way that, eventually only the unity operator of $\mathcal{P}^{23}_{\text{D}}$ contributes to the matrix elements, i.e.\
\begin{equation} \label{SIv12d1}
\sum_{s_3 t_3} \! \mel*{s_3 t_3}{\mathcal{P}^{23}_{\text{D}}}{s_3 t_3} = 4 \Big(W + \frac{B}{2} - \frac{H}{2} - \frac{M}{4} \Big)
\end{equation}
Gathering those two results and taking into account the translational invariance of INM, i.e.\ $\rho(\vec{r}_3) = \rho$, it follows that:
\begin{multline} \label{v12d1D}
v_{12}^{\text{d}_1} \vert_{\text{D}}  = \frac{\alpha \rho}{2} \Big(W + \frac{B}{2} - \frac{H}{2} - \frac{M}{4} \Big) \times \\\int \dd[3]{r_3} \frac{V(r_{23})}{(\mu \sqrt{\pi})^3} \big[ \rho^{\alpha-1}(\vec{r}_2) \delta(\vec{r}_1 - \vec{r}_2) \\ + \rho^{\alpha-1}(\vec{r}_3) \delta(\vec{r}_1 - \vec{r}_3) \big]  \\
 = \frac{\alpha \rho}{2} \Big(W + \frac{B}{2} - \frac{H}{2} - \frac{M}{4} \Big) \times  \\ \Big[ \rho^{\alpha-1}(\vec{r}_1) \frac{V(r_{12})}{(\mu \sqrt{\pi})^3} + \rho^{\alpha-1}(\vec{r}_2) \delta(\vec{r}_1 - \vec{r}_2) \Big]
\end{multline}
where the first integral has been evaluated by means of the Gauss integral:
\begin{equation} \label{Gaussintegral}
\int_{-\infty}^{+\infty} \dd[N]{x} \e^{-\alpha x^2} = \Big( \frac{\pi}{\alpha} \Big)^{N/2},\quad \text{with } \alpha \in \mathbb{R}^*
\end{equation}
Now, to get the direct contribution of the first rearrangement term \eqref{firstrearr}, one needs to evaluate the TBMEs of the type:
\begin{multline}
\mel*{p h'}{v_{12}^{\text{d}_1} \vert_{\text{D}}}{h p'} = \mel*{\vec{k}_p \vec{k}_{h'}}{v_{12}^{\text{d}_1} \vert_{\text{D}}}{\vec{k}_h \vec{k}_{p'}} \notag \\
= \frac{\alpha \rho}{2} \Big(W + \frac{B}{2} - \frac{H}{2} - \frac{M}{4} \Big)  \langle \vec{k}_p \vec{k}_{h'} \vert \\ \Big[ \rho^{\alpha-1}(\vec{r}_1) \frac{V(r_{12})}{(\mu \sqrt{\pi})^3} + \rho^{\alpha-1}(\vec{r}_2) \delta(\vec{r}_1 - \vec{r}_2) \Big] \vert \vec{k}_h \vec{k}_{p'} \rangle 
\end{multline}
By definition,
\begin{multline}
\mel*{\vec{k}_p \vec{k}_{h'}}{v_{12}^{\text{d}_1} \vert_{\text{D}}}{\vec{k}_h \vec{k}_{p'}} = \frac{\alpha \rho}{2} \Big(W + \frac{B}{2} - \frac{H}{2} - \frac{M}{4} \Big) \times \\\int \dd[3]{r_1} \int \dd[3]{r_2} \phi^*_{\vec{k}_p}(\vec{r}_1) \phi^*_{\vec{k}_{h'}}(\vec{r}_2) \Big[ \rho^{\alpha-1}(\vec{r}_1) \frac{V(r_{12})}{(\mu \sqrt{\pi})^3} \\ + \rho^{\alpha-1}(\vec{r}_2) \delta(\vec{r}_1 - \vec{r}_2) \Big] \phi_{\vec{k}_h}(\vec{r}_1) \phi_{\vec{k}_{p'}}(\vec{r}_2) \notag \\ \\
= \frac{\alpha \rho}{2 \mathcal{V}^2} \Big(W + \frac{B}{2} - \frac{H}{2} - \frac{M}{4} \Big) \times \\ \int \dd[3]{r_1} \int \dd[3]{r_2} \e^{-\ii (\vec{k}_p - \vec{k}_h) \cdot \vec{r}_1} \e^{\ii (\vec{k}_{p'} - \vec{k}_{h'}) \cdot \vec{r}_2} \times \\ \Big[ \rho^{\alpha-1}(\vec{r}_1) \frac{V(r_{12})}{(\mu \sqrt{\pi})^3} + \rho^{\alpha-1}(\vec{r}_2) \delta(\vec{r}_1 - \vec{r}_2) \Big]
\end{multline}
Using the conservation of the quasiparticle pair momentum $\vec{k}_p + \vec{k}_h = \vec{k}_{p'} + \vec{k}_{h'}$, the translational invariance of INM, i.e.\ $\rho(\vec{r}_1) = \rho(\vec{r}_2) = \rho$, and going from the nucleon to the center-of-mass and relative coordinates defined as
$\vec{r} \equiv \vec{r}_1-\vec{r}_2, \quad \text{and} \quad \vec{R} \equiv \frac{\vec{r}_1+\vec{r}_2}{2}$ and whose Jacobian is equal to unity, 
while splitting the integral into two parts, one successively obtains:
\begin{multline}
\mel*{\vec{k}_p \vec{k}_{h'}}{v_{12}^{\text{d}_1} \vert_{\text{D}}}{\vec{k}_h \vec{k}_{p'}}  = \frac{\alpha \rho^{\alpha}}{2 \mathcal{V}^2} \Big(W + \frac{B}{2} - \frac{H}{2} - \frac{M}{4} \Big) \times \\ \int \dd[3]{r_1} \int \dd[3]{r_2} \e^{-\ii (\vec{k}_p - \vec{k}_h) \cdot (\vec{r}_1 - \vec{r}_2)} \times \\ \Big[ \rho^{\alpha-1}(\vec{r}_1) \frac{V(r_{12})}{(\mu \sqrt{\pi})^3} + \rho^{\alpha-1}(\vec{r}_2) \delta(\vec{r}_1 - \vec{r}_2) \Big]  \notag \\
 ~~~~~~~~~~~~~~~~~~~~~ = \frac{\alpha \rho^\alpha}{2 \mathcal{V}^2} \Big(W + \frac{B}{2} - \frac{H}{2} - \frac{M}{4} \Big) \times \\ \bigg[ \int \dd[3]{R} \int \dd[3]{r} \e^{-\ii (\vec{k}_p - \vec{k}_h) \cdot \vec{r}} \frac{\e^{-r^2/\mu^2}}{(\mu \sqrt{\pi})^3} \\ + \int \dd[3]{R} \int \dd[3]{r} \e^{-\ii (\vec{k}_p - \vec{k}_h) \cdot \vec{r}} \delta(\vec{r}) \bigg]
\end{multline}
Considering an infinite volume, $\mathcal{V} \rightarrow \infty$, and repeating the steps that led to 
\begin{equation} \label{exchangespatialPW}
\mel*{\vec{k}_1 \vec{k}_2}{V(r_{12}) D[\rho]}{\vec{k}_2 \vec{k}_1} = \frac{\rho^{\alpha}}{\mathcal{V}} (\mu \sqrt{\pi})^3 \e^{-(\vec{k}_1 - \vec{k}_2)^2 \mu^2/4}
\end{equation}
one can perform the two integrals with Gaussian integrals \eqref{Gaussintegral}. Indeed,
\begin{multline} \label{finalv12d1}
\mel*{\vec{k}_p \vec{k}_{h'}}{v_{12}^{\text{d}_1} \vert_{\text{D}}}{\vec{k}_h \vec{k}_{p'}} = \frac{\alpha \rho^\alpha}{2 \mathcal{V}} \Big(W + \frac{B}{2} - \frac{H}{2} - \frac{M}{4} \Big) \times \\ \bigg[ \int \dd[3]{r} \e^{-\ii (\vec{k}_p - \vec{k}_h) \cdot \vec{r}} \frac{\e^{-r^2/\mu^2}}{(\mu \sqrt{\pi})^3} + \int \dd[3]{r} \e^{-\ii (\vec{k}_p - \vec{k}_h) \cdot \vec{r}} \delta(\vec{r}) \bigg] \\
 = \frac{\alpha \rho^\alpha}{2 \mathcal{V}} \Big(W + \frac{B}{2} - \frac{H}{2} - \frac{M}{4} \Big) \Big[ \e^{-(\vec{k}_p - \vec{k}_h)^2 \mu^2/4} +1 \Big]
\end{multline}
At the Landau limit, $|\vec{k}_p| = |\vec{k}_h| = |\vec{k}_{\text{F}}|$ and $|\vec{k}_{p'}| = |\vec{k}_{h'}| = |\vec{k}'_{\text{F}}|$, so that the direct component of the first rearrangement term reads:
\begin{equation} \label{Landaurearr1}
\mel*{\vec{k}_{\text{F}} \vec{k}'_{\text{F}}}{v_{12}^{\text{d}_1} \vert_{\text{D}}}{\vec{k}_{\text{F}} \vec{k}'_{\text{F}}} = \frac{\alpha \rho^\alpha}{\mathcal{V}} \Big(W + \frac{B}{2} - \frac{H}{2} - \frac{M}{4} \Big)
\end{equation}
One continues by evaluating the exchange component of Eq.\eqref{vd1}, which is:
\begin{multline}
v_{12}^{\text{d}_1} \vert_{\text{E}} \equiv \frac{\alpha}{2} \sum_i \! \bra*{i}\! \frac{V(r_{23})}{(\mu \sqrt{\pi})^3} \big[ \rho^{\alpha-1}(\vec{r}_2) \delta(\vec{r}_1 - \vec{r}_2) \\ + \rho^{\alpha-1}(\vec{r}_3) \delta(\vec{r}_1 - \vec{r}_3) \big] \mathcal{P}^{23}_{\text{E}} P^{23}_r \! \ket*{i}
\end{multline}
The calculation is trickier than the direct term because of the space exchange operator $P^{23}_r$ that forces to consider directly the two-body matrix element:
\begin{multline} \label{TBMEvd1}
\mel*{p h'}{v_{12}^{\text{d}_1} \vert_{\text{E}}}{h p'} = \frac{\alpha}{2} \sum_i \! \bra*{ph'i}\! \frac{V(r_{23})}{(\mu \sqrt{\pi})^3} \times \\ \big[ \rho^{\alpha-1}(\vec{r}_2) \delta(\vec{r}_1 - \vec{r}_2) + \rho^{\alpha-1}(\vec{r}_3) \delta(\vec{r}_1 - \vec{r}_3) \big] \mathcal{P}_{\text{E}}^{23} \! \ket*{hip'}
\end{multline}
since $P_r^{23} \! \ket*{hp'i} = \ket*{hip'}$. Considering again the continuous limit \eqref{transfocontinu}, one obtains, with $i$ renamed $3$ for convenience:
\begin{multline}
\mel*{p h'}{v_{12}^{\text{d}_1} \vert_{\text{E}}}{h p'} = \frac{\alpha}{2} \frac{\mathcal{V}}{(2 \pi)^3}  \sum_{u_3} \! \mel*{u_3}{\mathcal{P}^{23}_{\text{E}}}{u_3} \times \\ \int \dd[3]{k_3} \int \dd[3]{r_3} \int \dd[3]{r_2} \int \dd[3]{r_1} \phi_{\vec{k}_p}^*(\vec{r}_1) \phi_{\vec{k}_{h'}}^*(\vec{r}_2) \times \\ \phi_{\vec{k}_3}^*(\vec{r}_3) \frac{V(r_{23})}{(\mu \sqrt{\pi})^3} \big[ \rho^{\alpha-1}(\vec{r}_2) \delta(\vec{r}_1 - \vec{r}_2) \\
 + \rho^{\alpha-1}(\vec{r}_3) \delta(\vec{r}_1 - \vec{r}_3) \big]\phi_{\vec{k}_h}(\vec{r}_1) \phi_{\vec{k}_3}(\vec{r}_2) \phi_{\vec{k}_{p'}}(\vec{r}_3)
\end{multline}
To get the spin-isospin contribution, one needs to change $W, B, H, M$ for $M, H, B, W$ in Eq.\eqref{SIv12d1}, so that:
\begin{equation} \label{SIv12d1E}
\sum_{s_3 t_3} \! \mel*{s_3 t_3}{\mathcal{P}^{23}_{\text{E}}}{s_3 t_3} = 4 \Big(M + \frac{H}{2} - \frac{B}{2} - \frac{W}{4} \Big)
\end{equation}
Then,
\begin{multline}
\mel*{p h'}{v_{12}^{\text{d}_1} \vert_{\text{E}}}{h p'} = \mel*{\vec{k}_p \vec{k}_{h'}}{v_{12}^{\text{d}_1} \vert_{\text{E}}}{\vec{k}_h \vec{k}_{p'}} \\
= \frac{2}{(2\pi)^3} \frac{\alpha}{\mathcal{V}^2} \Big(M + \frac{H}{2} - \frac{B}{2} - \frac{W}{4} \Big) \times \\ \int \dd[3]{k_3} \int \dd[3]{r_3} \int \dd[3]{r_2} \int \dd[3]{r_1} \times \\ \e^{-\ii (\vec{k}_p - \vec{k}_h) \cdot \vec{r}_1} \e^{-\ii (\vec{k}_3 - \vec{k}_{p'}) \cdot \vec{r}_2} \e^{-\ii (\vec{k}_{h'} - \vec{k}_3) \cdot \vec{r}_3} \times \\
\frac{V(r_{23})}{(\mu \sqrt{\pi})^3} \big[ \rho^{\alpha-1}(\vec{r}_2) \delta(\vec{r}_1 - \vec{r}_2) + \rho^{\alpha-1}(\vec{r}_3) \delta(\vec{r}_1 - \vec{r}_3) \big]
\end{multline}
Splitting the integral into two parts and evaluating the integrals over $\vec{r}_1$, one gets:
\begin{multline}
\mel*{\vec{k}_p \vec{k}_{h'}}{v_{12}^{\text{d}_1} \vert_{\text{E}}}{\vec{k}_h \vec{k}_{p'}} = \\
~~~~~~~~~~~~~~~~~~~ \frac{2}{(2\pi)^3} \frac{\alpha}{\mathcal{V}^2} \Big(M + \frac{H}{2} - \frac{B}{2} - \frac{W}{4} \Big) \times \\
\bigg[ \int \dd[3]{k_3} \int \dd[3]{r_3} \int \dd[3]{r_2} \int \dd[3]{r_1} \rho^{\alpha-1}(\vec{r}_2) \delta(\vec{r}_1 - \vec{r}_2) \times \\ \frac{V(r_{23})}{(\mu \sqrt{\pi})^3} 
 \e^{-\ii (\vec{k}_p - \vec{k}_h) \cdot \vec{r}_1} \e^{-\ii (\vec{k}_3 - \vec{k}_{p'}) \cdot \vec{r}_2} \e^{-\ii (\vec{k}_{h'} - \vec{k}_3) \cdot \vec{r}_3}  \\
+ \int \dd[3]{k_3} \int \dd[3]{r_3} \int \dd[3]{r_2} \int \dd[3]{r_1} \rho^{\alpha-1}(\vec{r}_3) \delta(\vec{r}_1 - \vec{r}_3) \times \\ \frac{V(r_{23})}{(\mu \sqrt{\pi})^3} 
\e^{-\ii (\vec{k}_p - \vec{k}_h) \cdot \vec{r}_1} \e^{-\ii (\vec{k}_3 - \vec{k}_{p'}) \cdot \vec{r}_2} \e^{-\ii (\vec{k}_{h'} - \vec{k}_3) \cdot \vec{r}_3} \bigg] \\
~~~~~~~~~~~~~~~~~~~ = \frac{2}{(2\pi)^3} \frac{\alpha}{\mathcal{V}^2} \Big(M + \frac{H}{2} - \frac{B}{2} - \frac{W}{4} \Big) \times \\
\bigg[ \int \dd[3]{k_3} \int \dd[3]{r_3} \int \dd[3]{r_2} \rho^{\alpha-1}(\vec{r}_2) \times \\
\frac{V(r_{23})}{(\mu \sqrt{\pi})^3} \e^{-\ii (\vec{k}_p + \vec{k}_3 - \vec{k}_h - \vec{k}_{p'}) \cdot \vec{r}_2} \e^{-\ii (\vec{k}_{h'} - \vec{k}_3) \cdot \vec{r}_3} \\
 + \int \dd[3]{k_3} \int \dd[3]{r_3} \int \dd[3]{r_2} \rho^{\alpha-1}(\vec{r}_3) \times \\
 \frac{V(r_{23})}{(\mu \sqrt{\pi})^3}  \e^{-\ii (\vec{k}_3 - \vec{k}_{p'}) \cdot \vec{r}_2} \e^{-\ii (\vec{k}_p + \vec{k}_{h'} - \vec{k}_h - \vec{k}_3) \cdot \vec{r}_3} \bigg]
\end{multline}
The conservation of the quasiparticle pair momentum $\vec{k}_p + \vec{k}_{h'} = \vec{k}_h + \vec{k}_{p'}$ provides:
\begin{multline}
\mel*{\vec{k}_p \vec{k}_{h'}}{v_{12}^{\text{d}_1} \vert_{\text{E}}}{\vec{k}_h \vec{k}_{p'}}  = \\ 
~~~~~~~~~~~~~~~~~~~ \frac{2}{(2\pi)^3} \frac{\alpha}{\mathcal{V}^2} \Big(M + \frac{H}{2} - \frac{B}{2} - \frac{W}{4} \Big) \notag \\
\times \bigg[ \int \dd[3]{k_3} \int \dd[3]{r_3} \int \dd[3]{r_2} \rho^{\alpha-1}(\vec{r}_2) \times \\
\frac{V(r_{23})}{(\mu \sqrt{\pi})^3} \e^{-\ii (\vec{k}_{h'} - \vec{k}_3) \cdot (\vec{r}_3-\vec{r}_2)} \\
+ \int \dd[3]{k_3} \int \dd[3]{r_3} \int \dd[3]{r_2} \rho^{\alpha-1}(\vec{r}_3) \times \\ \frac{V(r_{23})}{(\mu \sqrt{\pi})^3} \e^{-\ii (\vec{k}_{p'} - \vec{k}_3) \cdot (\vec{r}_3-\vec{r}_2)} \bigg] \\
~~~~~~~~~~~~~~~~~~~ = \frac{2}{(2\pi)^3} \frac{\alpha \rho^{\alpha-1}}{\mathcal{V}^2} \Big(M + \frac{H}{2} - \frac{B}{2} - \frac{W}{4} \Big) \\
\times \bigg[ \int \dd[3]{r_1} \int \dd[3]{r_2} \frac{V(r_{12})}{(\mu \sqrt{\pi})^3} \e^{-\ii \vec{k}_{h'} \cdot (\vec{r}_1-\vec{r}_2)} \times \\ 
\int \dd[3]{k_3} \e^{-\ii \vec{k}_{3} \cdot (\vec{r}_1-\vec{r}_2)} \\
 + \int \dd[3]{r_1} \int \dd[3]{r_2}  \frac{V(r_{12})}{(\mu \sqrt{\pi})^3} \e^{-\ii \vec{k}_{p'} \cdot (\vec{r}_1-\vec{r}_2)} \times \\ \int \dd[3]{k_3} \e^{-\ii \vec{k}_{3} \cdot (\vec{r}_1-\vec{r}_2)} \bigg]
\end{multline}
where one has renamed the integration variable $\vec{r}_3$ into $\vec{r}_1$ and used the translational invariance of INM, i.e.\ $\rho(\vec{r}_1) = \rho(\vec{r}_2) = \rho$. The integral over $\vec{k}_3$ reads:
\begin{multline}
\int \dd[3]{k_3} \e^{-\ii \vec{k}_{3} \cdot (\vec{r}_1-\vec{r}_2)} = \\ 2 \pi \int_0^{k_{\text{F}}} \dd{k_3} k_3^2 
\int_0^{\pi} \dd{\theta} \sin \theta \, \e^{-\ii k_3 r \cos \theta}  \\ = 2 \pi \int_0^{k_{\text{F}}} \dd{k_3} k_3 \frac{1}{\ii r} \big[ \e^{-\ii k_3 r \cos \theta} \big]^{\pi}_0 \notag \\
= \frac{4 \pi}{r} \int_0^{k_{\text{F}}} \dd{k_3} k_3 \sin(k_3 r)  \\
= \frac{4 \pi}{r} \frac{1}{r^2} \int_0^{k_{\text{F}} r} \dd{(k_3 r)} k_3 r \sin(k_3 r) \notag \\
 = \frac{4 \pi}{r} \frac{\sin k_{\text{F}}r - k_{\text{F}} r \cos k_{\text{F}}r}{r^2} \\ = \frac{4 \pi k_{\text{F}}^2}{r} j_1(k_{\text{F}} r) ~~~~~~~~~~~~~~~~~~~
\end{multline}
with $\vec{r} \equiv \vec{r}_1 - \vec{r}_2$ and the spherical Bessel function of the first kind given by \eqref{Bessel2}. Finally, the expression of the constant nuclear density in INM \eqref{densityINM} gives
\begin{equation} \label{intj1}
\frac{1}{(2 \pi)^3} \int \dd[3]{k_3} \e^{-\ii \vec{k}_{3} \cdot (\vec{r}_1-\vec{r}_2)} = \frac{3 \rho}{4} \frac{j_1(k_{\text{F}}r)}{k_{\text{F}}r}
\end{equation}
Moving from the nucleon to the center-of-mass and relative coordinates defined by 
\begin{equation} \label{cdmrelativePW}
\vec{r} \equiv \vec{r}_1-\vec{r}_2, \quad \text{and} \quad \vec{R} \equiv \frac{\vec{r}_1+\vec{r}_2}{2}
\end{equation}
and considering an infinite volume, $\mathcal{V} \rightarrow \infty$, one finds out:
\begin{multline} \label{finalechv12d1}
\mel*{\vec{k}_p \vec{k}_{h'}}{v_{12}^{\text{d}_1} \vert_{\text{E}}}{\vec{k}_h \vec{k}_{p'}}  = \frac{3 \alpha \rho^\alpha}{2 \mathcal{V}^2} \Big(M + \frac{H}{2} - \frac{B}{2} - \frac{W}{4} \Big) \times \\
 \bigg[ \int \dd[3]{R} \int \dd[3]{r} \e^{-\ii \vec{k}_{h'} \cdot \vec{r}} \frac{\e^{-r^2/\mu^2}}{(\mu \sqrt{\pi})^3} \frac{j_1(k_{\text{F}}r)}{k_{\text{F}}r} \\ + \int \dd[3]{R} \int \dd[3]{r} \e^{-\ii \vec{k}_{p'} \cdot \vec{r}} \frac{\e^{-r^2/\mu^2}}{(\mu \sqrt{\pi})^3} \frac{j_1(k_{\text{F}}r)}{k_{\text{F}}r} \bigg] \\ \\
= \frac{3 \alpha \rho^\alpha}{2 \mathcal{V}} \Big(M + \frac{H}{2} - \frac{B}{2} - \frac{W}{4} \Big) \\
\times \bigg[ \int \dd[3]{r} \e^{-\ii \vec{k}_{h'} \cdot \vec{r}} \frac{\e^{-r^2/\mu^2}}{(\mu \sqrt{\pi})^3} \frac{j_1(k_{\text{F}}r)}{k_{\text{F}}r} \\ + \int \dd[3]{r} \e^{-\ii \vec{k}_{p'} \cdot \vec{r}} \frac{\e^{-r^2/\mu^2}}{(\mu \sqrt{\pi})^3} \frac{j_1(k_{\text{F}}r)}{k_{\text{F}}r} \bigg]
\end{multline}
At the Landau limit,  $|\vec{k}_p| = |\vec{k}_h| = |\vec{k}_{\text{F}}|$ and $|\vec{k}_{p'}| = |\vec{k}_{h'}| = |\vec{k}'_{\text{F}}|$, so that the two integrals coincide and:
\begin{multline}
\mel*{\vec{k}_{\text{F}} \vec{k}'_{\text{F}}}{v_{12}^{\text{d}_1} \vert_{\text{E}}}{\vec{k}_{\text{F}} \vec{k}'_{\text{F}}} = \frac{3 \alpha \rho^\alpha}{\mathcal{V}} \Big(M + \frac{H}{2} - \frac{B}{2} - \frac{W}{4} \Big) \times \\ \int \dd[3]{r} \e^{-\ii k_{\text{F}} r \cos \theta} \frac{\e^{-r^2/\mu^2}}{(\mu \sqrt{\pi})^3} \frac{j_1(k_{\text{F}}r)}{k_{\text{F}}r} \\
~~~~~~~~~~~~~~~~= \frac{6 \pi \alpha \rho^\alpha}{\mathcal{V}} \Big(M + \frac{H}{2} - \frac{B}{2} - \frac{W}{4} \Big) \times \\ \int \dd{r} r^2 \frac{\e^{-r^2/\mu^2}}{(\mu \sqrt{\pi})^3} \frac{j_1(k_{\text{F}}r)}{k_{\text{F}}r} \int_0^{\pi} \dd{\theta} \sin \theta \, \e^{-\ii k_{\text{F}} r \cos \theta}
\end{multline}
The angular integral is simply:
\begin{multline}
\int_0^{\pi} \dd{\theta} \sin \theta \, \e^{-\ii k_{\text{F}} r \cos \theta} \\ = \frac{1}{\ii k_{\text{F}}r} \int_0^{\pi} \dd{\theta} \ii k_{\text{F}}r \sin \theta \, \e^{-\ii k_{\text{F}} r \cos \theta} \\ = \frac{2 \sin k_{\text{F}} r}{k_{\text{F}} r} = 2 j_0(k_{\text{F}} r)
\end{multline}
with the spherical Bessel function of the first kind given by \eqref{Bessel0}. Finally, the exchange component of the first rearrangement term reads:
\begin{multline}
\mel*{\vec{k}_{\text{F}} \vec{k}'_{\text{F}}}{v_{12}^{\text{d}_1} \vert_{\text{E}}}{\vec{k}_{\text{F}} \vec{k}'_{\text{F}}}  = \frac{12 \pi \alpha \rho^\alpha}{\mathcal{V}} \Big(M + \frac{H}{2} - \frac{B}{2} - \frac{W}{4} \Big) \times \\
\int_0^{\infty} \dd{r} r^2 \frac{\e^{-r^2/\mu^2}}{(\mu \sqrt{\pi})^3} \frac{j_1(k_{\text{F}}r)}{k_{\text{F}}r} j_0(k_{\text{F}}r)
\end{multline}
This is consistently the expression found in \cite{Chappert2015}. One evaluate now the above radial integral, hereafter called $I^{\text{d}_1}$, analytically. 
Using the expressions of the spherical Bessel functions of the first kind \eqref{Bessel0} and \eqref{Bessel1}, and linearizing the sine and cosine functions, 
one has to evaluate:
\begin{multline} \label{exchv12d1_0}
I^{\text{d}_1} \equiv \frac{1}{2 k_{\text{F}}^2} \frac{1}{(\mu \sqrt{\pi})^3} \int_0^{\infty} \dd{r} \e^{-r^2/\mu^2} \times \\ \bigg[ \frac{1 - \cos 2 k_{\text{F}}r}{(k_{\text{F}}r)^2} - \frac{\sin 2 k_{\text{F}} r}{k_{\text{F}} r} \bigg]
\end{multline}
The series expansions of the sine and cosine functions, and the parity of the integrands provide:
\begin{multline}
I^{\text{d}_1}  = \frac{1}{k_{\text{F}}^2} \frac{1}{(\mu \sqrt{\pi})^3} \int_{-\infty}^{+\infty} \dd{r} \e^{-r^2/\mu^2} \times \\ \bigg[ \sum_n \frac{(-)^n}{(2n+2)!} (2 k_{\text{F}} r)^{2n} - \frac{1}{2} \sum_n \frac{(-)^n}{(2n+1)!} (2 k_{\text{F}} r)^{2n} \bigg] \notag \\
= \frac{\sqrt{\pi}}{k_{\text{F}}^3} \frac{1}{(\mu \sqrt{\pi})^3} \bigg[ \sum_n \frac{(-)^n (\mu k_{\text{F}})^{2n+1}}{(2n+1)(2n+2)n!} \\ - \frac{1}{2} \sum_n \frac{(-)^n (\mu k_{\text{F}})^{2n+1}}{(2n+1) n!} \bigg]
\end{multline}
where the integrals and sums have been inverted, and the expression:
\begin{multline} \label{Gammaintegralinfinity}
\int_{-\infty}^{+\infty} \dd{x} x^n \e^{-\alpha x^2} = \\
\begin{cases} 
 \displaystyle \frac{\Gamma[(n+1)/2]}{\alpha^{(n+1)/2}} = \frac{\sqrt{\pi}}{\alpha^{(n+1)/2}} \frac{(2n)!}{4^n n!}          \quad & \text{if } n \text{ is even} \\
  0 \quad & \text{if } n \text{ is odd}
\end{cases}
\end{multline}
has been used. One recognizes the series expansion of the error function in the second right-hand side, so that:
\begin{multline}
I^{\text{d}_1}  = \frac{\sqrt{\pi}}{2k_{\text{F}}^3} \frac{1}{(\mu \sqrt{\pi})^3} \times \\ \bigg[ 2 \sum_n \frac{(-)^n (\mu k_{\text{F}})^{2n+1}}{(2n+1)n!} \bigg( 1 - \frac{2n+1}{2n+2} \bigg) \\ - \frac{\sqrt{\pi}}{2} \erf(\mu k_{\text{F}}) \bigg] \notag \\
 = \frac{\sqrt{\pi}}{2k_{\text{F}}^3} \frac{1}{(\mu \sqrt{\pi})^3} \bigg[ \sqrt{\pi} \erf(\mu k_{\text{F}}) \\ -  \sum_n \frac{(-)^n (\mu k_{\text{F}})^{2n+1}}{(n+1)!} - \frac{\sqrt{\pi}}{2} \erf(\mu k_{\text{F}}) \bigg]
\end{multline}
Setting $X \equiv \mu k_{\text{F}}$ and recognizing the series expansion of the function $(\e^{-X^2}-1)/X$ in the second right-hand side term, one finally gets:
\begin{equation} \label{Int1rearr}
I^{\text{d}_1} = \frac{1}{2 \pi X^3} \bigg[\frac{1}{X} \big( \e^{-X^2} - 1 \big) + \frac{\sqrt{\pi}}{2} \erf(X) \bigg]
\end{equation}
Thus, the exchange first rearrangement term becomes
\begin{multline} \label{exchv12d1}
\mel*{\vec{k}_{\text{F}} \vec{k}'_{\text{F}}}{v_{12}^{\text{d}_1} \vert_{\text{E}}}{\vec{k}_{\text{F}} \vec{k}'_{\text{F}}} = \frac{6 \alpha \rho^\alpha}{\mathcal{V} X^3} \Big(M + \frac{H}{2} - \frac{B}{2} - \frac{W}{4} \Big) \times \\ \bigg[\frac{1}{X} \big( \e^{-X^2} - 1 \big) + \frac{\sqrt{\pi}}{2} \erf(X) \bigg]
\end{multline}
where the values of the error function are tabulated in the fitting code.

\paragraph{Second rearrangement term}

In order to compute the second rearrangement term of the quasiparticle interaction \eqref{MEph}, it is useful to give an expression for the derivative appearing therein. According to \eqref{derivativehp}, it directly reads:
\begin{multline} 
\pdv{v_{12}^{\text{(\text{a})}}}{\rho_{p'h'}} = \frac{\alpha}{2} \frac{V(r_{12})}{(\mu \sqrt{\pi})^3} \langle h' \vert \rho^{\alpha-1}(\vec{r}_1) \delta(\vec{r}_1 - \vec{r}_3) \\ + \rho^{\alpha-1}(\vec{r}_2) \delta(\vec{r}_2 - \vec{r}_3) \vert p' \rangle (\mathcal{P}_{\text{D}} + \mathcal{P}_{\text{E}} P_r)
\end{multline}
In the same way as for the first rearrangement term, the second rearrangement one can be written as: 
\begin{equation} 
\sum_i  \mel*{pi}{\pdv{v_{12}^{\text{(a)}}}{\rho_{p'h'}}}{hi} =  \mel*{ph'}{v_{12}^{\text{d}_2}}{hp'}
\end{equation}
where
\begin{multline} \label{vd2}
v_{12}^{\text{d}_2} = \frac{\alpha}{2} \sum_i \bra*{i}\! \frac{V(r_{13})}{(\mu \sqrt{\pi})^3} \big[ \rho^{\alpha-1}(\vec{r}_1) \delta(\vec{r}_1 - \vec{r}_2) \\ + \rho^{\alpha-1}(\vec{r}_3) \delta(\vec{r}_2 - \vec{r}_3) \big] (\mathcal{P}^{13}_{\text{D}} + \mathcal{P}^{13}_{\text{E}} P^{13}_r) \! \ket*{i}
\end{multline}
Then, one starts by evaluating the direct component of the above one-body matrix element, which reads:
\begin{multline}
v_{12}^{\text{d}_2} \vert_{\text{D}} = \frac{\alpha}{2} \sum_i \! \bra*{i}\! \frac{V(r_{13})}{(\mu \sqrt{\pi})^3} \big[ \rho^{\alpha-1}(\vec{r}_1) \delta(\vec{r}_1 - \vec{r}_2) \\ + \rho^{\alpha-1}(\vec{r}_3) \delta(\vec{r}_2 - \vec{r}_3) \big] \mathcal{P}^{13}_{\text{D}} \! \ket*{i}
\end{multline}
The procedure is similar to the one undertaken for the direct component of the first rearrangement term, leading to the result Eq.\eqref{Landaurearr1}. Here, it comes:
\begin{multline} \label{v12d2D}
v_{12}^{\text{d}_2} \vert_{\text{D}} = \frac{\alpha \rho}{2} \Big(W + \frac{B}{2} - \frac{H}{2} - \frac{M}{4} \Big) \times \\ \Big[ \rho^{\alpha-1}(\vec{r}_1) \frac{V(r_{12})}{(\mu \sqrt{\pi})^3} + \rho^{\alpha-1}(\vec{r}_1) \delta(\vec{r}_1 - \vec{r}_2) \Big]
\end{multline}
Now, one evaluates the two-body matrix elements:
\begin{multline}
\mel*{p h'}{v_{12}^{\text{d}_2} \vert_{\text{D}}}{h p'}  = \mel*{\vec{k}_p \vec{k}_{h'}}{v_{12}^{\text{d}_2} \vert_{\text{D}}}{\vec{k}_h \vec{k}_{p'}} \notag \\
 = \frac{\alpha \rho}{2} \Big(W + \frac{B}{2} - \frac{H}{2} - \frac{M}{4} \Big) \langle \vec{k}_p \vec{k}_{h'} \vert \Big[ \rho^{\alpha-1}(\vec{r}_1) \frac{V(r_{12})}{(\mu \sqrt{\pi})^3} \\+ \rho^{\alpha-1}(\vec{r}_1) \delta(\vec{r}_1 - \vec{r}_2) \Big] \vert \vec{k}_h \vec{k}_{p'} \rangle
\end{multline}
By following the steps leading to \eqref{finalv12d1}, one finally gets:
\begin{multline}
\mel*{\vec{k}_p \vec{k}_{h'}}{v_{12}^{\text{d}_2} \vert_{\text{D}}}{\vec{k}_h \vec{k}_{p'}} = \frac{\alpha \rho^\alpha}{2 \mathcal{V}} \Big(W + \frac{B}{2} - \frac{H}{2} - \frac{M}{4} \Big) \times \\  \Big[ \e^{-(\vec{k}_p - \vec{k}_h)^2 \mu^2/4} +1 \Big]
\end{multline}
which is exactly the expression found for the first rearrangement term \eqref{finalv12d1}. At the Landau limit, $|\vec{k}_p| = |\vec{k}_h| = |\vec{k}_{\text{F}}|$ and $|\vec{k}_{p'}| = |\vec{k}_{h'}| = |\vec{k}'_{\text{F}}|$, so that the direct component of the second rearrangement term reads:
\begin{equation}
\mel*{\vec{k}_{\text{F}} \vec{k}'_{\text{F}}}{v_{12}^{\text{d}_2} \vert_{\text{D}}}{\vec{k}_{\text{F}} \vec{k}'_{\text{F}}} = \frac{\alpha \rho^\alpha}{\mathcal{V}} \Big(W + \frac{B}{2} - \frac{H}{2} - \frac{M}{4} \Big)
\end{equation}
This expression is the same as the direct component of the first rearrangement term \eqref{Landaurearr1}.\\

\noindent One continues by evaluating the exchange component of Eq.\eqref{vd2}, which is:
\begin{multline}
v_{12}^{\text{d}_2} \vert_{\text{E}} \equiv \frac{\alpha}{2} \sum_i \! \bra*{i}\! \frac{V(r_{13})}{(\mu \sqrt{\pi})^3} \big[ \rho^{\alpha-1}(\vec{r}_1) \delta(\vec{r}_1 - \vec{r}_2) \\ + \rho^{\alpha-1}(\vec{r}_3) \delta(\vec{r}_2 - \vec{r}_3) \big] W^{13}_{\text{E}} \mathcal{P}^{13}_r \! \ket*{i}
\end{multline}
Once again, onee has to deal with the space exchange operator $P^{13}_r$ that forces to consider directly the two-body matrix element:
\begin{multline} \label{TBMEvd2}
\mel*{p h'}{v_{12}^{\text{d}_2} \vert_{\text{E}}}{h p'}  = \frac{\alpha}{2} \sum_i  \bra*{ph'i}\! \frac{V(r_{13})}{(\mu \sqrt{\pi})^3} \times \\ \big[ \rho^{\alpha-1}(\vec{r}_1) \delta(\vec{r}_1 - \vec{r}_2) + \rho^{\alpha-1}(\vec{r}_3) \delta(\vec{r}_2 - \vec{r}_3) \big] \mathcal{P}_{\text{E}}^{13} \! \ket*{ip'h}
\end{multline}
since $P_r^{13} \! \ket*{hp'i} = \ket*{ip'h}$. The derivation is analogous to the first rearrangement term leading to Eq.\eqref{finalechv12d1}. One obtains:
\begin{multline}
\mel*{p h'}{v_{12}^{\text{d}_2} \vert_{\text{E}}}{h p'}  = \mel*{\vec{k}_p \vec{k}_{h'}}{v_{12}^{\text{d}_2} \vert_{\text{E}}}{\vec{k}_h \vec{k}_{p'}} \\
 = \frac{3 \alpha \rho^\alpha}{2 \mathcal{V}} \Big(M + \frac{H}{2} - \frac{B}{2} - \frac{W}{4} \Big) \times \\ \bigg[ \int \dd[3]{r} \e^{-\ii \vec{k}_{h} \cdot \vec{r}} \frac{\e^{-r^2/\mu^2}}{(\mu \sqrt{\pi})^3} \frac{j_1(k_{\text{F}}r)}{k_{\text{F}}r} \\ + \int \dd[3]{r} \e^{-\ii \vec{k}_{p} \cdot \vec{r}} \frac{\e^{-r^2/\mu^2}}{(\mu \sqrt{\pi})^3} \frac{j_1(k_{\text{F}}r)}{k_{\text{F}}r} \bigg]
\end{multline}
At the Landau limit, $|\vec{k}_p| = |\vec{k}_h| = |\vec{k}_{\text{F}}|$. One ends up with the same expression obtained for the first rearrangement term, namely:
\begin{multline}
\mel*{\vec{k}_{\text{F}} \vec{k}'_{\text{F}}}{v_{12}^{\text{d}_2} \vert_{\text{E}}}{\vec{k}_{\text{F}} \vec{k}'_{\text{F}}} = \frac{3 \alpha \rho^\alpha}{\mathcal{V}} \Big(M + \frac{H}{2} - \frac{B}{2} - \frac{W}{4} \Big) \times \\  \int \dd[3]{r} \e^{-\ii k_{\text{F}} r \cos \theta} \frac{\e^{-r^2/\mu^2}}{(\mu \sqrt{\pi})^3} \frac{j_1(k_{\text{F}}r)}{k_{\text{F}}r}
\end{multline}
so that the exchange component of the second rearrangement term eventually reads:
\begin{multline}
\mel*{\vec{k}_{\text{F}} \vec{k}'_{\text{F}}}{v_{12}^{\text{d}_2} \vert_{\text{E}}}{\vec{k}_{\text{F}} \vec{k}'_{\text{F}}}  = \frac{12 \pi \alpha \rho^\alpha}{\mathcal{V}} \Big(M + \frac{H}{2} - \frac{B}{2} - \frac{W}{4} \Big) \times \\ \int_0^{\infty} \dd{r} r^2 \frac{\e^{-r^2/\mu^2}}{(\mu \sqrt{\pi})^3} \frac{j_1(k_{\text{F}}r)}{k_{\text{F}}r} j_0(k_{\text{F}}r)
\end{multline}
This is the same exchange term as the first rearrangement term \eqref{exchv12d1}, where the radial integral has been calculated in Eq.\eqref{Int1rearr}. 
Then, one gets:
\begin{multline} \label{exchv12d2}
\mel*{\vec{k}_{\text{F}} \vec{k}'_{\text{F}}}{v_{12}^{\text{d}_2} \vert_{\text{E}}}{\vec{k}_{\text{F}} \vec{k}'_{\text{F}}}  = \frac{6 \alpha \rho^\alpha}{\mathcal{V} X^3} \Big(M + \frac{H}{2} - \frac{B}{2} - \frac{W}{4} \Big) \times \\  \bigg[\frac{1}{X} \big( \e^{-X^2} - 1 \big) + \frac{\sqrt{\pi}}{2} \erf(X) \bigg]
\end{multline}
which is identical to the first rearrangement term.

\paragraph{Third rearrangement term}

Finally, to derive the third rearrangement term of the quasiparticle interaction \eqref{MEph}, one proceeds in the way as before, with a second derivative. 
Taking advantage of the first derivative \eqref{derivativehp}, this latter reads:
\begin{multline} 
\pdv{v_{12}^{\text{(\text{a})}}}{\rho_{hp}}{\rho_{p'h'}} = \pdv{\rho_{p'h'}} \bigg( \pdv{v_{12}^{\text{(\text{a})}}}{\rho_{hp}} \bigg) \notag \\
= \frac{\alpha(\alpha-1)}{2} \frac{V(r_{12})}{(\mu \sqrt{\pi})^3}  \bra*{ph'}  \rho^{\alpha-2}(\vec{r}_1) \delta(\vec{r}_1 - \vec{r}_3) \delta(\vec{r}_1 - \vec{r}_4) \\ + \rho^{\alpha-2}(\vec{r}_2) \delta(\vec{r}_2 - \vec{r}_3) \delta(\vec{r}_2 - \vec{r}_4)  \ket*{hp'}  (\mathcal{P}_{\text{D}} + \mathcal{P}_{\text{E}} P_r)
\end{multline}
Thus, the third rearrangement term appearing in \eqref{MEph} can be written:
\begin{multline} \label{v12rearr3}
\frac{1}{2} \sum_{ij}  \mel*{ij}{\pdv{v_{12}^{\text{(a)}}}{\rho_{hp}}{\rho_{p'h'}}}{ij} = \frac{\alpha(\alpha-1)}{4} \times \\ \sum_{ij}  \bra*{ijph'}  \frac{V(r_{12})}{(\mu \sqrt{\pi})^3} \big[ \rho^{\alpha-2}(\vec{r}_1) \delta(\vec{r}_1 - \vec{r}_3) \delta(\vec{r}_1 - \vec{r}_4) \\ + \rho^{\alpha-2}(\vec{r}_2) \delta(\vec{r}_2 - \vec{r}_3) \delta(\vec{r}_2 - \vec{r}_4) \big] (\mathcal{P}^{12}_{\text{D}} + \mathcal{P}^{12}_{\text{E}} P^{12}_r)  \ket*{ijhp'}
\end{multline}
since the quantities only acting on the particles $1$ and $2$, spotted by their indices $12$, can be incorporated in the four-body matrix element, having no action on the third particle $p$ and third hole $h$, nor on the fourth particle $p'$ and fourth hole $h'$. Defining:
\begin{multline} \label{vd3}
v_{12}^{\text{d}_3} = \frac{\alpha(\alpha-1)}{4} \sum_{ij}  \bra*{ij} \frac{V(r_{34})}{(\mu \sqrt{\pi})^3} \times \\
\big[ \rho^{\alpha-2}(\vec{r}_3) \delta(\vec{r}_1 - \vec{r}_3) \delta(\vec{r}_2 - \vec{r}_3) \\ + \rho^{\alpha-1}(\vec{r}_4) \delta(\vec{r}_1 - \vec{r}_4) \delta(\vec{r}_2 - \vec{r}_4) \big] (\mathcal{P}^{34}_{\text{D}} + \mathcal{P}^{34}_{\text{E}} P^{34}_r) \ket*{ij}
\end{multline}
one finds out, after two circular permutations of the variables in \eqref{v12rearr3},
\begin{equation} \label{thirdrearr}
\frac{1}{2} \sum_{ij}  \mel*{ij}{\pdv{v_{12}^{\text{(a)}}}{\rho_{hp}}{\rho_{p'h'}}}{ij} =  \mel*{ph'}{v_{12}^{\text{d}_3}}{hp'}
\end{equation}
Then, one starts by evaluating the direct component of \eqref{vd3}, which reads:
\begin{multline}
v_{12}^{\text{d}_3} \vert_{\text{D}} = \frac{\alpha(\alpha-1)}{4} \sum_{ij}  \bra*{ij} \frac{V(r_{34})}{(\mu \sqrt{\pi})^3} \times \\ \big[ \rho^{\alpha-2}(\vec{r}_3) \delta(\vec{r}_1 - \vec{r}_3) \delta(\vec{r}_2 - \vec{r}_3) \\+ \rho^{\alpha-1}(\vec{r}_4) \delta(\vec{r}_1 - \vec{r}_4) \delta(\vec{r}_2 - \vec{r}_4) \big] \mathcal{P}^{34}_{\text{D}} \! \ket*{ij}
\end{multline}
Considering the continuous limit \eqref{transfocontinu}, one obtains, with $i$ and $j$ respectively renamed $3$ and $4$ for convenience:
\begin{multline}
v_{12}^{\text{d}_3} \vert_{\text{D}}  = \frac{\alpha(\alpha-1)}{4} \frac{\mathcal{V}^2}{(2 \pi)^6}  \sum_{u_3 u_4} \! \mel*{u_3 u_4}{\mathcal{P}^{34}_{\text{D}}}{u_3 u_4} \times \\
\int \dd[3]{r_3} \int \dd[3]{r_4} \int \dd[3]{k_3} \int \dd[3]{k_4} \\
\times \phi_{\vec{k}_3}^*(\vec{r}_3) \phi_{\vec{k}_4}^*(\vec{r}_4) \frac{V(r_{34})}{(\mu \sqrt{\pi})^3} \times \\ \big[ \rho^{\alpha-2}(\vec{r}_3) \delta(\vec{r}_1 - \vec{r}_3) \delta(\vec{r}_2 - \vec{r}_3) \\ + \rho^{\alpha-2}(\vec{r}_4) \delta(\vec{r}_1 - \vec{r}_4) \delta(\vec{r}_2 - \vec{r}_4) \big] \phi_{\vec{k}_3}(\vec{r}_3) \phi_{\vec{k}_4}(\vec{r}_4)
\end{multline}
On the one hand, expressing the scalar product of Pauli matrices asdone in \eqref{Pauli1Landau}, one finds:
\begin{equation}
\begin{array}{lcl}
\displaystyle \mel*{s_3 s_4}{\vec{\sigma}_3 \cdot \vec{\sigma}_4}{s_3 s_4} &= & \displaystyle \sum_k (-)^k \! \mel*{s_3}{\sigma_3^k}{s_3}  \mel*{s_4}{\sigma_4^{-k}}{s_4} \\
&= & \displaystyle 4 s_3 s_4
\end{array}
\end{equation}
where the unified relation of the matrix elements of the Pauli matrices, $\mel*{s_a}{\sigma_m}{s_b} = 2 s_a \delta_{s_a s_b} \delta_{m,0} - m \sqrt{2} \delta_{s_a, s_b+m}, \text{for } m = 0, \pm 1 $, has been used. One obviously gets an equivalent relation for the Pauli matrices associated with the isospin, so that: 
\begin{equation}
\sum_{s_3 s_4} \! \mel*{s_3 s_4}{\vec{\sigma}_3 \cdot \vec{\sigma}_4}{s_3 s_4} = \sum_{t_3 t_4} \! \mel*{t_3 t_4}{\vec{\tau}_3 \cdot \vec{\tau}_4}{t_3 t_4} = 0
\end{equation}
in such a way that, eventually only the unity operator of $\mathcal{P}^{34}_{\text{D}}$ contributes to the matrix elements, i.e.\
\begin{multline} \label{SIv12d3}
\sum_{s_3 t_3} \sum_{s_4 t_4} \! \mel*{s_3 t_3 \, s_4 t_4}{\mathcal{P}^{34}_{\text{D}}}{s_3 t_3 \, s_4 t_4} = \\ 16 \Big(W + \frac{B}{2} - \frac{H}{2} - \frac{M}{4} \Big).
\end{multline}
Combining this result with the nuclear density \eqref{densityLandau}, and invoking the translational invariance of INM, i.e.\ $\rho(\vec{r}_1) = \rho(\vec{r}_2) = \rho$, it follows that:
\begin{multline} 
v_{12}^{\text{d}_3} \vert_{\text{D}} = \frac{\alpha (\alpha-1) \rho^2}{4} \Big(W + \frac{B}{2} - \frac{H}{2} - \frac{M}{4} \Big) \times \\ \int \dd[3]{r_3} \int \dd[3]{r_4} \frac{V(r_{34})}{(\mu \sqrt{\pi})^3} \\
\times \big[ \rho^{\alpha-2}(\vec{r}_3) \delta(\vec{r}_1 - \vec{r}_3) \delta(\vec{r}_2 - \vec{r}_3) \\ 
+ \rho^{\alpha-2}(\vec{r}_4) \delta(\vec{r}_1 - \vec{r}_4) \delta(\vec{r}_2 - \vec{r}_4) \big]
\end{multline}
Now, to get the direct contribution of the third rearrangement term \eqref{thirdrearr}, one evaluates the two-body matrix elements:
\begin{multline}
\mel*{p h'}{v_{12}^{\text{d}_3} \vert_{\text{D}}}{h p'} = \mel*{\vec{k}_p \vec{k}_{h'}}{v_{12}^{\text{d}_3} \vert_{\text{D}}}{\vec{k}_h \vec{k}_{p'}} \notag \\
= \frac{\alpha (\alpha-1) \rho^2}{4} \Big(W + \frac{B}{2} - \frac{H}{2} - \frac{M}{4} \Big) \int \dd[3]{r_3} \int \dd[3]{r_4} \times \\
\bra*{\vec{k}_p \vec{k}_{h'}} \! \frac{V(r_{34})}{(\mu \sqrt{\pi})^3} \big[ \rho^{\alpha-2}(\vec{r}_3) \delta(\vec{r}_1 - \vec{r}_3) \delta(\vec{r}_2 - \vec{r}_3) \\ 
+ \rho^{\alpha-2}(\vec{r}_4) \delta(\vec{r}_1 - \vec{r}_4) \delta(\vec{r}_2 - \vec{r}_4) \big] \ket*{\vec{k}_h \vec{k}_{p'}}
\end{multline}
By definition,
\begingroup
\allowdisplaybreaks
\begin{multline}
\mel*{\vec{k}_p \vec{k}_{h'}}{v_{12}^{\text{d}_3} \vert_{\text{D}}}{\vec{k}_h \vec{k}_{p'}}  = \\ \frac{\alpha (\alpha-1) \rho^2}{4} \Big(W + \frac{B}{2} - \frac{H}{2} - \frac{M}{4} \Big) \times \\ 
\int \dd[3]{r_1} \int \dd[3]{r_2} \int \dd[3]{r_3} \int \dd[3]{r_4} 
  \phi^*_{\vec{k}_p}(\vec{r}_1) \phi^*_{\vec{k}_{h'}}(\vec{r}_2)  \times  \\  \frac{V(r_{34})}{(\mu \sqrt{\pi})^3} 
 \Big[ \rho^{\alpha-2}(\vec{r}_3) \delta(\vec{r}_1 - \vec{r}_3) \delta(\vec{r}_2 - \vec{r}_3) \\ + \rho^{\alpha-2}(\vec{r}_4) \delta(\vec{r}_1 - \vec{r}_4) \delta(\vec{r}_2 - \vec{r}_4) \Big] \phi_{\vec{k}_h}(\vec{r}_1) \phi_{\vec{k}_{p'}}(\vec{r}_2) \notag \\
 = \frac{\alpha (\alpha-1) \rho^2}{4 \mathcal{V}^2} \Big(W + \frac{B}{2} - \frac{H}{2} - \frac{M}{4} \Big) \times \\
\bigg[ \int \dd[3]{r_3} \int \dd[3]{r_4} \e^{-\ii (\vec{k}_p + \vec{k}_{h'} - \vec{k}_h - \vec{k}_{p^{\prime}}) \cdot \vec{r}_3} \times \\ \frac{V(r_{34})}{(\mu \sqrt{\pi})^3} \rho^{\alpha-2}(\vec{r}_3) + \int \dd[3]{r_3} \int \dd[3]{r_4} \times \\ \e^{-\ii (\vec{k}_p + \vec{k}_{h'} - \vec{k}_h - \vec{k}_{p^{\prime}}) \cdot \vec{r}_4} \frac{V(r_{34})}{(\mu \sqrt{\pi})^3} \rho^{\alpha-2}(\vec{r}_4) \bigg]
\end{multline}
\endgroup
where one has performed the straight integrals over $\vec{r}_1$ and $\vec{r}_2$, and separated the expression into two parts. 
Then, using the conservation of the quasiparticle pair momentum $\vec{k}_p + \vec{k}_h = \vec{k}_{p'} + \vec{k}_{h'}$, one obtains:
\begin{multline}
\mel*{\vec{k}_p \vec{k}_{h'}}{v_{12}^{\text{d}_3} \vert_{\text{D}}}{\vec{k}_h \vec{k}_{p'}} = \frac{\alpha (\alpha-1) \rho^2}{2 \mathcal{V}^2} \times \\ \Big(W + \frac{B}{2} - \frac{H}{2} - \frac{M}{4} \Big) \int \dd[3]{r_1} \int \dd[3]{r_2} \frac{V(r_{12})}{(\mu \sqrt{\pi})^3} \rho^{\alpha-2}(\vec{r}_1)
\end{multline}
since the two integrals are equal, when exchanging the integration variables $\vec{r}_3$ and $\vec{r}_4$ in the second integral. Note that one has also relabelled $\vec{r}_3$ as $\vec{r}_1$ and $\vec{r}_4$ as $\vec{r}_2$. Using the translational invariance of INM, i.e.\ $\rho(\vec{r}_1) = \rho$, and going from the nucleon to the center-of-mass and relative coordinates defined in Eq.\eqref{cdmrelativePW}, it comes:
\begin{multline}
\mel*{\vec{k}_p \vec{k}_{h'}}{v_{12}^{\text{d}_3} \vert_{\text{D}}}{\vec{k}_h \vec{k}_{p'}} = \\ \frac{\alpha (\alpha-1) \rho^\alpha}{2 \mathcal{V}^2} \Big(W + \frac{B}{2} - \frac{H}{2} - \frac{M}{4} \Big) \int \dd[3]{r} \int \dd[3]{R} \frac{\e^{-r^2/\mu^2}}{(\mu \sqrt{\pi})^3} \\
 = \frac{\alpha (\alpha-1) \rho^\alpha}{2 \mathcal{V}} \Big(W + \frac{B}{2} - \frac{H}{2} - \frac{M}{4} \Big) \int \dd[3]{r} \frac{\e^{-r^2/\mu^2}}{(\mu \sqrt{\pi})^3} \\
 = \frac{\alpha (\alpha-1) \rho^\alpha}{2 \mathcal{V}} \Big(W + \frac{B}{2} - \frac{H}{2} - \frac{M}{4} \Big)
\end{multline}
where an infinite volume $\mathcal{V} \rightarrow \infty$ has been considered and the integral has been performed by means of the Gauss integral \eqref{Gaussintegral}. At the Landau limit, $|\vec{k}_p| = |\vec{k}_h| = |\vec{k}_{\text{F}}|$ and $|\vec{k}_{p'}| = |\vec{k}_{h'}| = |\vec{k}'_{\text{F}}|$, so that the direct contribution of the third rearrangement term reads:
\begin{multline} \label{Landaurearr3}
\mel*{\vec{k}_{\text{F}} \vec{k}'_{\text{F}}}{v_{12}^{\text{d}_3} \vert_{\text{D}}}{\vec{k}_{\text{F}} \vec{k}'_{\text{F}}} = \frac{\alpha (\alpha-1) \rho^\alpha}{2\mathcal{V}} \times \\ \Big(W + \frac{B}{2} - \frac{H}{2} - \frac{M}{4} \Big)
\end{multline}
One continues by evaluating the exchange component of \eqref{vd3}, which is:
\begin{multline}
v_{12}^{\text{d}_3} \vert_{\text{E}} = \frac{\alpha(\alpha-1)}{4} \sum_{ij} \! \bra*{ij}\! \frac{V(r_{34})}{(\mu \sqrt{\pi})^3} \times \\ \big[ \rho^{\alpha-2}(\vec{r}_3) \delta(\vec{r}_1 - \vec{r}_3) \delta(\vec{r}_2 - \vec{r}_3) \\ 
+ \rho^{\alpha-1}(\vec{r}_4) \delta(\vec{r}_1 - \vec{r}_4) \delta(\vec{r}_2 - \vec{r}_4) \big] \mathcal{P}^{34}_{\text{D}} P_r^{34} \! \ket*{ij}
\end{multline}
Here, contrary to the first two rearrangement terms \eqref{TBMEvd1} and \eqref{TBMEvd2}, one can evaluate the above quantity without having to consider the full two-body matrix element $  \mel*{p h'}{v_{12}^{\text{d}_2} \vert_{\text{E}}}{h p'} $ since $P_r^{34}  \ket*{ij} = \! \ket*{ji}$. However, it turns out that it is not possible to simplify the result as for the direct terms in Eq.\eqref{v12d1D} and Eq.\eqref{v12d2D}. Therefore, it will be more efficient to simply follow the procedure adopted for the first two exchange terms. Then, one has:
\begin{multline} 
\mel*{p h'}{v_{12}^{\text{d}_3} \vert_{\text{E}}}{h p'} = \frac{V(r_{34})}{(\mu \sqrt{\pi})^3} \times \\ \big[ \rho^{\alpha-2}(\vec{r}_3) \delta(\vec{r}_1 - \vec{r}_3) \delta(\vec{r}_2 - \vec{r}_3) \\ + \rho^{\alpha-1}(\vec{r}_4) \delta(\vec{r}_1 - \vec{r}_4) \delta(\vec{r}_2 - \vec{r}_4) \big] \mathcal{P}^{34}_{\text{E}} \ket*{hp'ji} 
\end{multline}
Considering again the continuous limit \eqref{transfocontinu}, one obtains, with $i$ and $j$ respectively renamed $3$ and $4$ for convenience:
\begin{multline}
\mel*{p h'}{v_{12}^{\text{d}_3} \vert_{\text{E}}}{h p'} = \frac{\alpha (\alpha-1)}{4} \frac{\mathcal{V}^2}{(2 \pi)^6}  \sum_{u_3 u_4} \times \\ \mel*{u_3 u_4}{\mathcal{P}^{34}_{\text{E}}}{u_3 u_4} \times \\
 \int \dd[3]{k_3} \int \dd[3]{k_4} \int \dd[3]{r_4} \int \dd[3]{r_3} \int \dd[3]{r_2} \int \dd[3]{r_1} \times \\ \phi_{\vec{k}_p}^*(\vec{r}_1) \phi_{\vec{k}_{h'}}^*(\vec{r}_2) \phi_{\vec{k}_3}^*(\vec{r}_3) \phi_{\vec{k}_4}^*(\vec{r}_4) \times \\
\frac{V(r_{34})}{(\mu \sqrt{\pi})^3} \big[ \rho^{\alpha-2}(\vec{r}_3) \delta(\vec{r}_1 - \vec{r}_3) \delta(\vec{r}_2 - \vec{r}_3) \\ + \rho^{\alpha-2}(\vec{r}_4) \delta(\vec{r}_1 - \vec{r}_4) \delta(\vec{r}_2 - \vec{r}_4) \big] \phi_{\vec{k}_h}(\vec{r}_1) \times \\ \phi_{\vec{k}_{p'}}(\vec{r}_2) \phi_{\vec{k}_4}(\vec{r}_3) \phi_{\vec{k}_3}(\vec{r}_4)
\end{multline}
To get the spin-isospin contribution, one needs to change $W, B, H, M$ for $M, H, B, W$ in \eqref{SIv12d3}, so that:
\begin{equation} \label{SIv12d3E}
\sum_{s_3 t_3} \sum_{s_4 t_4} \! \mel*{s_3 t_3 \, s_4 t_4}{\mathcal{P}^{34}_{\text{E}}}{s_3 t_3 \, s_4 t_4} = 16 \Big(M + \frac{H}{2} - \frac{B}{2} - \frac{W}{4} \Big)
\end{equation}
Thus,
\begin{multline}
\mel*{p h'}{v_{12}^{\text{d}_3} \vert_{\text{E}}}{h p'}  = \mel*{\vec{k}_p \vec{k}_{h'}}{v_{12}^{\text{d}_3} \vert_{\text{E}}}{\vec{k}_h \vec{k}_{p'}} \\
 = \frac{4 \alpha (\alpha-1)}{(2 \pi)^6 \mathcal{V}^2} \Big(M + \frac{H}{2} - \frac{B}{2} - \frac{W}{4} \Big)  \times \\
 \int \dd[3]{k_3} \int \dd[3]{k_4} \int \dd[3]{r_4} \int \dd[3]{r_3} \int \dd[3]{r_2} \int \dd[3]{r_1} \times \\ \e^{-\ii (\vec{k}_p - \vec{k}_h) \cdot \vec{r}_1} \e^{-\ii (\vec{k}_{h'} - \vec{k}_{p'}) \cdot \vec{r}_2} \frac{V(r_{34})}{(\mu \sqrt{\pi})^3} \times  \\
 \big[ \rho^{\alpha-2}(\vec{r}_3) \delta(\vec{r}_1 - \vec{r}_3) \delta(\vec{r}_2 - \vec{r}_3) \\ + \rho^{\alpha-2}(\vec{r}_4) \delta(\vec{r}_1 - \vec{r}_4) \delta(\vec{r}_2 - \vec{r}_4) \big] \times \\ \e^{-\ii (\vec{k}_3 - \vec{k}_4) \cdot \vec{r}_3} \e^{-\ii (\vec{k}_4 - \vec{k}_3) \cdot \vec{r}_4}
\end{multline}
Splitting the integral into two parts and evaluating the integrals over $\vec{r}_1$ and $\vec{r}_2$ taking into account the conservation of the quasiparticle pair momentum $\vec{k}_p + \vec{k}_h = \vec{k}_{p'} + \vec{k}_{h'}$, one obtains:
\begin{multline}
\mel*{\vec{k}_p \vec{k}_{h'}}{v_{12}^{\text{d}_3} \vert_{\text{E}}}{\vec{k}_h \vec{k}_{p'}} = \\ \frac{4 \alpha (\alpha-1)}{(2 \pi)^6 \mathcal{V}^2} \Big(M + \frac{H}{2} - \frac{B}{2} - \frac{W}{4} \Big) \times \\
\bigg[ \int \dd[3]{k_3} \int \dd[3]{k_4} \int \dd[3]{r_4} \int \dd[3]{r_3} \int \dd[3]{r_2} \int \dd[3]{r_1} \times \\ \frac{V(r_{34})}{(\mu \sqrt{\pi})^3}
\rho^{\alpha-2}(\vec{r}_3) \delta(\vec{r}_1 - \vec{r}_3) \delta(\vec{r}_2 - \vec{r}_3) \times \\ \e^{-\ii (\vec{k}_p - \vec{k}_h) \cdot (\vec{r}_1 - \vec{r}_2)} \e^{-\ii (\vec{k}_3 - \vec{k}_4) \cdot (\vec{r}_3 - \vec{r}_4)} \\
+ \int \dd[3]{k_3} \int \dd[3]{k_4} \int \dd[3]{r_4} \int \dd[3]{r_3} \int \dd[3]{r_2} \int \dd[3]{r_1} \times \\ \frac{V(r_{34})}{(\mu \sqrt{\pi})^3} 
 \rho^{\alpha-2}(\vec{r}_4) \delta(\vec{r}_1 - \vec{r}_4) \delta(\vec{r}_2 - \vec{r}_4) \times \\ \e^{-\ii (\vec{k}_p - \vec{k}_h) \cdot (\vec{r}_1 - \vec{r}_2)} \e^{-\ii (\vec{k}_3 - \vec{k}_4) \cdot (\vec{r}_3 - \vec{r}_4)} \bigg] \\
 = \frac{4 \alpha (\alpha-1)}{(2 \pi)^6 \mathcal{V}^2} \Big(M + \frac{H}{2} - \frac{B}{2} - \frac{W}{4} \Big) \times \\
\bigg[ \int \dd[3]{k_3} \int \dd[3]{k_4} \int \dd[3]{r_4} \int \dd[3]{r_3} \rho^{\alpha-2}(\vec{r}_3) \times \\ \frac{V(r_{34})}{(\mu \sqrt{\pi})^3} \e^{-\ii (\vec{k}_3 - \vec{k}_4) \cdot (\vec{r}_3 - \vec{r}_4)} \times \\
\int \dd[3]{k_3} \int \dd[3]{k_4} \int \dd[3]{r_4} \int \dd[3]{r_3} \rho^{\alpha-2}(\vec{r}_4) \times \\ \frac{V(r_{34})}{(\mu \sqrt{\pi})^3}  \e^{-\ii (\vec{k}_3 - \vec{k}_4) \cdot (\vec{r}_3 - \vec{r}_4)}
\end{multline}
Now, one relabels the integration variables $\vec{r}_3$ as $\vec{r}_1$ and $\vec{r}_4$ as $\vec{r}_2$, and use the translational invariance of INM, i.e.\ $\rho(\vec{r}_1) = \rho(\vec{r}_2) = \rho$ to find out that the above integrals are the same and read:
\begin{multline}
\mel*{\vec{k}_p \vec{k}_{h'}}{v_{12}^{\text{d}_3} \vert_{\text{E}}}{\vec{k}_h \vec{k}_{p'}} = \frac{8 \alpha (\alpha-1) \rho^{\alpha-2}}{(2 \pi)^6 \mathcal{V}^2} \times \\ \Big(M + \frac{H}{2} - \frac{B}{2} - \frac{W}{4} \Big) \times \\ \int \dd[3]{r_1} \int \dd[3]{r_2} \frac{V(r_{12})}{(\mu \sqrt{\pi})^3} \int \dd[3]{k_3}  \e^{-\ii \vec{k}_3 \cdot (\vec{r}_1 - \vec{r}_2)} \times \\ \int \dd[3]{k_4}  \e^{\ii \vec{k}_4 \cdot (\vec{r}_1 - \vec{r}_2)}
\end{multline}
It is not difficult to see that:
\begin{equation}
\int \dd[3]{k_3}  \e^{-\ii \vec{k}_3 \cdot (\vec{r}_1 - \vec{r}_2)} = \int \dd[3]{k_4}  \e^{\ii \vec{k}_4 \cdot (\vec{r}_1 - \vec{r}_2)}
\end{equation}
already evaluated in Eq.\eqref{intj1}. Now, moving from the nucleon to the center-of-mass and relative coordinates defined in \eqref{cdmrelativePW} and considering an infinite volume, $\mathcal{V} \rightarrow \infty$, one obtains:
\begin{multline}
\mel*{\vec{k}_p \vec{k}_{h'}}{v_{12}^{\text{d}_3} \vert_{\text{E}}}{\vec{k}_h \vec{k}_{p'}} = \frac{9 \alpha (\alpha-1) \rho^\alpha}{2 \mathcal{V}^2} \times \\ \Big(M + \frac{H}{2} - \frac{B}{2} - \frac{W}{4} \Big) \int \dd[3]{r} \int \dd[3]{R} \frac{\e^{-r^2/\mu^2}}{(\mu \sqrt{\pi})^3} \bigg[ \frac{j_1(k_{\text{F}}r)}{k_{\text{F}}r} \bigg]^2 \\
= \frac{18 \pi \alpha (\alpha-1) \rho^\alpha}{\mathcal{V}} \Big(M + \frac{H}{2} - \frac{B}{2} - \frac{W}{4} \Big) \times \\ \int_0^{\infty} \dd{r} r^2 \frac{\e^{-r^2/\mu^2}}{(\mu \sqrt{\pi})^3} \bigg[ \frac{j_1(k_{\text{F}}r)}{k_{\text{F}}r} \bigg]^2
\end{multline}
At the Landau limit,  $|\vec{k}_p| = |\vec{k}_h| = |\vec{k}_{\text{F}}|$ and $|\vec{k}_{p'}| = |\vec{k}_{h'}| = |\vec{k}'_{\text{F}}|$, so that the exchange component of the third rearrangement term reads:
\begin{multline}
\mel*{\vec{k}_{\text{F}} \vec{k}'_{\text{F}}}{v_{12}^{\text{d}_3} \vert_{\text{E}}}{\vec{k}_{\text{F}} \vec{k}'_{\text{F}}}  = \frac{18 \pi \alpha (\alpha-1) \rho^\alpha}{\mathcal{V}} \times \\ \Big(M + \frac{H}{2} - \frac{B}{2} - \frac{W}{4} \Big) \int_0^{\infty} \dd{r} r^2 \frac{\e^{-r^2/\mu^2}}{(\mu \sqrt{\pi})^3} \bigg[ \frac{j_1(k_{\text{F}}r)}{k_{\text{F}}r} \bigg]^2
\end{multline}
This is consistently the expression found in \cite{Chappert2015}. One, one evaluates the above radial integral, hereafter called $I^{\text{d}_3}$, analytically. The procedure is the same as the one that led to \eqref{Int1rearr}, although more complex. One eventually obtains:
\begin{equation}
I^{\text{d}_3} = \frac{1}{6 \pi X^6} \big[ \e^{-X^2}(X^2-2) - (3X^2-2) + \sqrt{\pi} X^3 \erf(X) \big]
\end{equation}
The exchange contribution of the third rearrangement terms then becomes:
\begin{multline}
\mel*{\vec{k}_{\text{F}} \vec{k}'_{\text{F}}}{v_{12}^{\text{d}_3} \vert_{\text{E}}}{\vec{k}_{\text{F}} \vec{k}'_{\text{F}}}  = \frac{3 \alpha (\alpha-1) \rho^\alpha}{\mathcal{V} X^6} \times \\ \Big(M + \frac{H}{2} - \frac{B}{2} - \frac{W}{4} \Big) \times \\  \big[ \e^{-X^2}(X^2-2) - (3X^2-2) + \sqrt{\pi} X^3 \erf(X) \big]
\end{multline}
where $X \equiv \mu k_{\text{F}}$, and the values of the error function are tabulated in the fitting code.

\vspace{\baselineskip}

Now that all the rearrangement terms of the quasiparticle interaction have been calculated, one transforms them in terms of the Landau parameters. 
It has been shown that these terms contribute only to the channel $(S=0,T=0)$ of the quasiparticle interaction (see Eqs. \eqref{SIv12d1} and \eqref{SIv12d1E} for the first two terms, as well as Eqs. \eqref{SIv12d3} and \eqref{SIv12d3E} for the third term). Putting together the direct contributions on one side and the exchange contributions on the other side of the three rearrangement terms, one writes:
\begin{multline} \label{f00rearr}
f^{00}(\theta) = \widetilde{G}_\mu^\alpha(0) \bigg[ \frac{(3+\alpha)\alpha}{2} \Big(W + \frac{B}{2} - \frac{H}{2} - \frac{M}{4} \Big) \\  + \frac{3\alpha}{X^3} \big( 4 K_1(X) + \frac{\alpha-1}{X^3} K_2(X) \big) \times \\  \Big(M + \frac{H}{2} - \frac{B}{2} - \frac{W}{4} \Big) \bigg]
\end{multline}
where the functions $K_i$ are defined by
\begin{multline}
K_1(X) = \frac{1}{X} \big( \e^{-X^2} - 1 \big) + \frac{\sqrt{\pi}}{2} \erf(X) \\
K_2(X) = \e^{-X^2}(X^2-2) - (3X^2-2) \\ + \sqrt{\pi} X^3 \erf(X)
\end{multline}
as well as 
\begin{equation}\label{tructruc}
\widetilde{G}_\mu^\alpha(0) \equiv N_0 G_\mu^\alpha(0) = N_0 \frac{\rho^\alpha}{\mathcal{V}}
\end{equation}
In Eq.\ref{tructruc}, $N_0$ is the density of quasiparticle states at the Fermi surface $N_0 \equiv \frac{2 m^{*} \mathcal{V} k_{\text{F}}}{\pi^2 \hbar^2}$, ensuring the connection of the previous matrix elements with the parameters appearing in \eqref{QPinteraction}. Moreover, one has defined $G_\mu^\alpha(0)$ without the factor $(\mu \sqrt{\pi})^3$, contrary to what has been done in Eq.\eqref{GmuCDD}. The reason is that the rearrangement terms are exclusively derived from the density-dependent term, for which the potential \eqref{gaussianpotential} must then be divided by this factor. Now, since the right-hand side of \eqref{f00rearr} does not depend on the angle $\theta$ between the quasiparticles, the rearrangement terms only contribute to the component $l=0$ of the Landau parameter $f^{00}_l$. Indeed, the expansion \eqref{expansionf} is independent of $\theta$ only for $l=0$ as $P_0(\cos \theta) = 1$. Thus, the contribution of the rearrangement terms to the Landau parameters reads:
\begin{multline}
f^{00}_l = \widetilde{G}_\mu^\alpha(0) \delta_{l,0} \bigg[ \frac{(3+\alpha)\alpha}{2} \Big(W + \frac{B}{2} - \frac{H}{2} - \frac{M}{4} \Big) \\ + \frac{3 \alpha}{X^3} \big( 4 K_1(X) + \frac{\alpha-1}{X^3} K_2(X) \big) \times \\ \Big(M + \frac{H}{2} - \frac{B}{2} - \frac{W}{4} \Big) \bigg]
\end{multline}

\subsection{Tensor contribution}

In this subsection,  one derives the Landau parameters associated with the finite-range tensor term of the generalized Gogny interaction \eqref{gognyDG}. The antisymmetrized tensor interaction is characterized by Eq.(\eqref{Tpot}), with its direct and exchange spin-isospin components \eqref{WHtensorINM}, respectively written under the equivalent forms:
\begin{subequations} \label{coeffT}
\begin{align}
\mathcal{P}_{\text{D}} \equiv \Big( W - \frac{H}{2} \Big) - \frac{H}{2} \vec{\tau}_1 \cdot \vec{\tau}_2 \\
\mathcal{P}_{\text{E}} \equiv \Big( H - \frac{W}{2} \Big) - \frac{W}{2} \vec{\tau}_1 \cdot \vec{\tau}_2 \label{coeffTE}
\end{align}
\end{subequations}
One notices that the direct isospin components of the tensor interaction can be deduced from the exchange ones by switching $W$ and $H$.

In order to get the contribution from the tensor interaction to the matrix elements of the particle-hole interaction \eqref{MEph}, one evaluates 
the following antisymmetrized matrix elements in the particle-hole representation:
\begin{multline} \label{elmatTPW}
\mel*{\vec{k}_p \vec{k}_{h'}}{v_{12}^{\text{T,(a)}}}{\vec{k}_h \vec{k}_{p'}} = \\\mel*{\vec{k}_p \vec{k}_{h'}}{V(r_{12}) S_{12}(\hat{r}_{12})}{\vec{k}_h \vec{k}_{p'}} \! \mathcal{P}_{\text{D}} 
 \\ + \! \mel*{\vec{k}_p \vec{k}_{h'}}{V(r_{12}) S_{12}(\hat{r}_{12})}{\vec{k}_{p'} \vec{k}_h} \! \mathcal{P}_{\text{E}}
\end{multline}
where one has explicitly separated the direct and exchange parts as well as the spatial-spin and isospin ones. One has also specified that the above tensor operator acts in the coordinate space by adding $\hat{r}_{12}$ as argument. Since the isospin part is fully specified by \eqref{coeffT}, one focusses 
on the spatial-spin parts. For the direct spatial-spin part, one has, by definition:
\begin{multline}
\mel*{\vec{k}_p \vec{k}_{h'}}{V(r_{12}) S_{12}(\hat{r}_{12})}{\vec{k}_h \vec{k}_{p'}} = \\ \int \dd[3]{r_1} \int \dd[3]{r_2} \phi^*_{\vec{k}_p}(\vec{r}_1) \phi^*_{\vec{k}_{h'}}(\vec{r}_2) V(r_{12}) \times \\ S_{12}(\hat{r}_{12}) \phi_{\vec{k}_h}(\vec{r}_1) \phi_{\vec{k}_{p'}}(\vec{r}_2) \\
= \sum_k (-)^k [\vec{\sigma}_1 \otimes \vec{\sigma}_2]^{(2)}_{-k} \frac{1}{\mathcal{V}^2} \int \dd[3]{r_1} \int \dd[3]{r_2} \e^{\ii (\vec{k}_h - \vec{k}_p) \cdot (\vec{r}_1 - \vec{r}_2)} \\ V(r_{12}) [\hat{r}_{12} \otimes \hat{r}_{12}]^{(2)}_{k}  \\
= \sum_k (-)^k [\vec{\sigma}_1 \otimes \vec{\sigma}_2]^{(2)}_{-k} \frac{1}{\mathcal{V}}  \int \dd[3]{r}  \e^{\ii \vec{q}_{12} \cdot \vec{r}} V(r) [\hat{r} \otimes \hat{r}]^{(2)}_{k}
\end{multline}
From the first to the second line, one has used the equivalent form of the tensor operator 
\begin{equation} \label{tensor1}
S_{12} = [\hat{r}_{12} \otimes \hat{r}_{12}]^{(2)} \cdot [\vec{\sigma}_1 \otimes \vec{\sigma}_2]^{(2)}
\end{equation}
and invoked the conservation the quasiparticle pair momentum, that is to say $\vec{k}_p + \vec{k}_{h'} = \vec{k}_h + \vec{k}_{p'}$. From the second to the third line, one has defined the relative momentum of the quasiparticle pair $\vec{q}_{12} \equiv \vec{k}_h - \vec{k}_p = \vec{k}_{h'} - \vec{k}_{p'}$ and moved from the nucleon to the relative and center-of-mass coordinates defined in \eqref{cdmrelativePW}. Now, using the writing \eqref{harmoTS} and the plane wave expansion of the exponential in terms of the spherical harmonics $
\e^{\ii \vec{k} \cdot \vec{r}} = 4 \pi \sum_{l=0}^{\infty} \sum_{m=-l}^l \ii^l j_l(kr) Y_l^{m *}(\hat{r}) Y_l^m(\hat{k})$, the above integral becomes:
\begin{multline}
\frac{1}{\mathcal{V}}  \int \dd[3]{r}  \e^{\ii \vec{q}_{12} \cdot \vec{r}} V(r) [\hat{r} \otimes \hat{r}]^{(2)}_{k}  \\ = \frac{4 \pi}{\mathcal{V}} \sqrt{\frac{8 \pi}{15}} \int \dd{r} r^2 V(r) \sum_{lm} \ii^l j_l(q_{12}r) Y_l^{m*}(\hat{q}_{12}) \\ \times \int \dd{\hat{r}} Y_l^m(\hat{r}) Y_2^k(\hat{r}) \\
 = \frac{4 \pi}{\mathcal{V}} (-)^{k+1} Y_2^{-k *}(\hat{q}_{12}) \int \dd{r} r^2 V(r) j_2(q_{12}r) \\
 = -\frac{4 \pi}{\mathcal{V}} [\hat{q}_{12} \otimes \hat{q}_{12}]^{(2)}_{k} \int \dd{r} r^2 V(r) j_2(q_{12}r)
\end{multline}
From the first to the second line, one has used the orthogonality relation of the spherical harmonics \eqref{orthogonalitySH} and from the second to the third line the writing \eqref{harmoTS} in momentum space. Considering the tensor operator in momentum space \eqref{S12q}, one finally obtains:
\begin{multline}
\mel*{\vec{k}_p \vec{k}_{h'}}{V(r_{12}) S_{12}(\hat{r}_{12})}{\vec{k}_h \vec{k}_{p'}} = - \frac{4 \pi}{\mathcal{V}} S_{12}(\hat{q}_{12}) \times \\ \int \dd{r} r^2 V(r) j_2(q_{12}r)
\end{multline}
The calculation of the exchange spatial-spin part is done in the same manner. One eventually finds out:
\begin{multline}
\mel*{\vec{k}_p \vec{k}_{h'}}{V(r_{12}) S_{12}(\hat{r}_{12})}{\vec{k}_{p'} \vec{k}_h} = - \frac{4 \pi}{\mathcal{V}} S_{12}(\hat{q}^{\- \prime}_{12}) \times \\ \int \dd{r} r^2 V(r) j_2(q'_{12}r)
\end{multline}
with a relative momentum between the quasiparticle pair given by $\vec{q}^{\, \prime}_{12} \equiv \vec{k}_{p'} - \vec{k}_{p} = \vec{k}_{h'} - \vec{k}_{h}$. 
At the Landau limit,  one has $|\vec{k}_p| = |\vec{k}_h| = |\vec{k}_{\text{F}}|$ and $|\vec{k}_{p'}| = |\vec{k}_{h'}| = |\vec{k}'_{\text{F}}|$, so that $|\vec{q}_{12}| = 0$ and $|\vec{q}^{\, \prime}_{12}| = q_{\text{F}}$, the relative momentum at the Fermi surface, which is given by:
\begin{equation} \label{relativesurface}
q_{\text{F}} = k_{\text{F}} \sqrt{2 (1 - \cos \theta)}.
\end{equation}
This implies:
\begin{multline}
\mel*{\vec{k}_{\text{F}} \vec{k}'_{\text{F}}}{V(r_{12}) S_{12}(\hat{r}_{12})}{\vec{k}_{\text{F}} \vec{k}'_{\text{F}}} = 0  \label{nulTS} \\
\mel*{\vec{k}_{\text{F}} \vec{k}'_{\text{F}}}{V(r_{12}) S_{12}(\hat{r}_{12})}{\vec{k}'_{\text{F}} \vec{k}_{\text{F}}} = - \frac{4 \pi}{\mathcal{V}} S_{12}(\hat{q}^{\- \prime}_{12}) \times \\ \int \dd{r} r^2 V(r) j_2(q_{\text{F}}r)
\end{multline}
Therefore, the direct spatial-spin part of the tensor interaction vanishes at the Landau limit. The direct component of the finite-range tensor force does not contribute to the Landau parameters. Moreover, from Eqs.\eqref{MEph}, \eqref{coeffTE} and \eqref{nulTS}, one gets:
\begin{multline}
h^{10}(\theta) \frac{q_{\text{F}}^2}{k_{\text{F}}^2} S_{12}(\hat{q}_{\text{F}}) = \Big( H - \frac{W}{2} \Big) \mel*{\vec{k}_{\text{F}} \vec{k}'_{\text{F}}}{V(r_{12}) S_{12}(\hat{r}_{12})}{\vec{k}'_{\text{F}} \vec{k}_{\text{F}}}  \\
h^{11}(\theta) \frac{q_{\text{F}}^2}{k_{\text{F}}^2} S_{12}(\hat{q}_{\text{F}})  = - \frac{W}{2} \! \mel*{\vec{k}_{\text{F}} \vec{k}'_{\text{F}}}{V(r_{12}) S_{12}(\hat{r}_{12})}{\vec{k}'_{\text{F}} \vec{k}_{\text{F}}} 
\end{multline}
since $\hat{q}_{12} = \hat{q}_{\text{F}}$ as this quantity only indicates the direction of the relative momentum between the quasiparticles. Expanding the parameters $h^{1T}$ in series of Legendre polynomials according to \eqref{expansionh}, the resulting coefficients, the Landau parameters associated with the tensor interaction, are given by:
\begin{equation} \label{tensorcoeff}
a_l \equiv \frac{2l+1}{2} \int_{-1}^1 \dd{x} f(x) P_l(x)
\end{equation}
In the present case:
\begin{subequations}
\begin{align}
h_l^{10} & = \Big( H - \frac{W}{2} \Big) H_l(q_{\text{F}}) \\
h_l^{11} & = - \frac{W}{2} H_l(q_{\text{F}})
\end{align}
\end{subequations}
where, after the change of variable $u \equiv \cos \theta$, one has defined the function:
\begin{multline}
H_l(k) \equiv - \frac{2 \pi}{\mathcal{V}} (2l+1) \int_{-1}^1 \dd{u} \frac{P_l(u)}{2(1-u)} \times \\ \int_0^{\infty} \dd{r} r^2 V(r) j_2(kr)
\end{multline}
This final expression is also present in the review paper on linear response function in homogeneous INM \cite{Pastore2015a}.
In order to simplify this expression, one first sets:
\begin{equation}
A_l(r) \equiv \int_{-1}^1 \dd{u} \frac{P_l(u)}{2(1-u)} j_2(q_{\text{F}} r)
\end{equation}
to evaluate the first integral. The simple change of variable $x \equiv q_{\text{F}} r = x_{\text{F}} \sqrt{2(1-u)}$ with $x_{\text{F}} = k_{\text{F}} r$ provides:
\begin{multline}
A_l(r) = \int_0^{2 x_{\text{F}}} \dd{x} \frac{x}{x_{\text{F}}^2} \frac{x_{\text{F}}^2}{x^2} P_l \Big( 1 - \frac{x^2}{2 x_{\text{F}}^2} \Big) j_2(x) \\
= \int_0^{2 x_{\text{F}}} \frac{\dd{x}}{x} \bigg[ \Big( \frac{3}{x^2} - \frac{1}{x} \Big) \sin x - \frac{3}{x^2} \cos x \bigg] \times \\ P_l \Big( 1 - \frac{x^2}{2 x_{\text{F}}^2} \Big)
\end{multline}
where one has plugged the expression of the second spherical Bessel function of the first kind \eqref{Bessel2}. By virtue of the series expansion of the Legendre polynomial, one obtains:
\begin{equation} \label{lastAl}
A_l(r) = \sum_{i=0}^{l} \frac{(-)^i (l+i)!}{(i!)^2 (l-i)!} \frac{1}{(2 x_{\text{F}})^{2i}} I_{2(i-2)}(r)
\end{equation}
with the integral
\begin{equation}
I_{2(i-2)}(r) \equiv \int_0^{2 x_{\text{F}}} \dd{x} \big[ (3-x^2) \sin x - 3x \cos x \big] x^{2(i-2)}
\end{equation}
Setting $n \equiv i-2$,  one will evaluate the integral:
\begin{multline}
I_{2n}(r) = \int_0^{2 x_{\text{F}}} \dd{x} \big( 3 x^{2n} \sin x \\ - x^{2n+2} \sin x - 3x^{2n+1} \cos x \big)
\end{multline}
for $n \ge -2$. For $n \ge 0$, one derives the following results using the integration by parts,
\begin{multline}
\int_0^{2 x_{\text{F}}} \dd{x} x^{2n+1} \cos x  = (2 x_{\text{F}})^{2n+1} \sin 2 x_{\text{F}} \\ - (2n+1) J_{2n}(r)  \\
\int_0^{2 x_{\text{F}}} \dd{x} x^{2n+2} \sin x  = -(2 x_{\text{F}})^{2n+2} \cos 2 x_{\text{F}}
\\ + (2n+2) \big[ (2 x_{\text{F}})^{2n+1} \sin 2 x_{\text{F}} - (2n+1) J_{2n}(r) \Big]
\end{multline}
where the integral $J_{2n}(r)$ as been defined by:
\begin{equation}
J_{2n}(r) \equiv \int_0^{2 x_{\text{F}}} \dd{x} x^{2n} \sin x.
\end{equation}
Gathering all the contributions, one obtains:
\begin{multline}
I_{2n}(r) = (2n+2)(2n+4)J_{2n}(r) \\ - (2n+5)(2 x_{\text{F}})^{2n+1} \sin 2 x_{\text{F}} + (2 x_{\text{F}})^{2n+2} \cos 2 x_{\text{F}}
\end{multline}
It only remains to evaluate the integral $J_{2n}(r)$. From the formula 
\begin{multline} \label{xnsin}
\int \dd{x} x^n \sin x  = \cos x \sum_{k=0}^{\lfloor n/2 \rfloor} (-)^{k+1} \frac{n!}{(n-2k)!} x^{n-2k} \\
 + \sin x \sum_{k=0}^{\lfloor (n-1)/2 \rfloor} (-)^k \frac{n!}{(n-2k-1)!} x^{n-2k-1} 
\end{multline}
with $n$ $\in \mathbb{N}$, one finds:
\begin{multline}
J_{2n} = (2n)! \Bigg[ \sum_{t=0}^n (-)^{t+1} \frac{x^{2n-2t}}{(2n-2t)!} \cos x \\ 
+ \sum_{t=0}^{n-1} (-)^t \frac{x^{2n-2t-1}}{(2n-2t-1)!} \sin x \Bigg]^{2 x_{\text{F}}}_0 \\
= (2n)!  \Bigg[ (-)^{n+1} \cos x \\ + \sum_{t=0}^{n-1} (-)^{t} \frac{x^{2n-2t}}{(2n-2t)!} \bigg( (2n-2t) \frac{\sin x}{x} - \cos x \bigg) \Bigg]^{2 x_{\text{F}}}_0 \\
= (2n)!  \Bigg[ (-)^n (1 - \cos 2 x_{\text{F}})
\\ + \sum_{t=0}^{n-1} (-)^{t} \frac{(2x_{\text{F}})^{2n-2t}}{(2n-2t)!} \bigg( (2n-2t) \frac{\sin 2x_{\text{F}}}{2x_{\text{F}}} - \cos 2x_{\text{F}} \bigg) \Bigg]
\end{multline}
For $n=-1$, one has to evaluate:
\begin{multline}
I_{-2}(r) = \int_0^{2 x_{\text{F}}} \dd{x} \bigg( \frac{3 \sin x}{x^2} - \sin x - \frac{3 \cos x}{x} \bigg)
\end{multline}
Integration by parts provides
\begin{equation}
\int_0^{2 x_{\text{F}}} \dd{x} \frac{\cos x}{x} = \frac{\sin 2 x_{\text{F}}}{2 x_{\text{F}}} - 1 + \int_0^{2 x_{\text{F}}} \dd{x} \frac{\sin x}{x^2}
\end{equation}
so that one ends up with
\begin{equation}
I_{-2}(r) = \cos 2 x_{\text{F}} - \frac{3 \sin 2 x_{\text{F}}}{2 x_{\text{F}}} + 2
\end{equation}
For $n=-2$,  one has to evaluate:
\begin{equation}
I_{-4}(r) = \int_0^{2 x_{\text{F}}} \dd{x} \bigg( \frac{3 \sin x}{x^4} - \frac{\sin x}{x^2} - \frac{3 \cos x}{x^3} \bigg)
\end{equation}
An integration by parts provides:
\begin{subequations}
\begin{align}
\int_0^{2 x_{\text{F}}} \dd{x} \frac{\sin x}{x^2} & = - \bigg[ \frac{\cos x}{x^2} \bigg]^{2 x_{\text{F}}}_0 - 2 \int_0^{2 x_{\text{F}}} \dd{x} \frac{\cos x}{x^3} \\
\int_0^{2 x_{\text{F}}} \dd{x} \frac{\cos x}{x^3} & = \bigg[ \frac{\sin x}{x^3} \bigg]^{2 x_{\text{F}}}_0 +3 \int_0^{2 x_{\text{F}}} \dd{x} \frac{\sin x}{x^4} 
\end{align}
\end{subequations}
so that one ends up with:
\begin{multline}
I_{-4}(r) = \bigg[ \frac{x \cos x - \sin x}{x^3} \bigg]^{2 x_{\text{F}}}_0 = \\  \frac{2 x_{\text{F}} \cos 2 x_{\text{F}} - \sin 2 x_{\text{F}}}{(2 x_{\text{F}})^3} + \frac{1}{3}
\end{multline}
Thus, the function $A_l(r)$ defined in \eqref{lastAl} is fully specified. Injecting it in \eqref{Hl}, one finally obtains:
\begin{multline}
H_l(q_{\text{F}}) = - \frac{2 \pi}{\mathcal{V}} (2l+1) \sum_{i=0}^{l} \frac{(-)^i (l+i)!}{(i!)^2 (l-i)!} \times \\ \int_0^{\infty} \dd{r} r^{2(1-i)} \, \e^{-r^2/\mu^2} I_{2(i-2)}(r).
\end{multline}
In the following, one gives the expressions of the above quantity for the first values of $l$, which have been implemented in the fitting code. 
The corresponding integrals are evaluated in a similar manner as done for the previous terms. One has:
\begingroup
\allowdisplaybreaks
\begin{multline}
H_0  = -\frac{\pi^{3/2}}{4 \mathcal{V} k_{\text{F}}^3} \bigg[ \frac{2}{3} X^3 + X \e^{-X^2} - \frac{\sqrt{\pi}}{2} \erf(X) \bigg] \\ \\
H_1  = -\frac{3 \pi^{3/2}}{4 \mathcal{V} k_{\text{F}}^3} \bigg[ \frac{2}{3} X^3 - X \big( \e^{-X^2} + 4 \big) + \frac{5}{2} \sqrt{\pi} \erf(X) \bigg] \\ \\
H_2  = -\frac{5 \pi^{3/2}}{4 \mathcal{V} k_{\text{F}}^3} \bigg[ \frac{2}{3} X^3 + X \big( \e^{-X^2} - 12 \big) \\ + \frac{24}{X} \big( \e^{-X^2} - 1 \big) + \frac{35}{2} \sqrt{\pi} \erf(X) \bigg] \\ \\
H_3  = -\frac{7 \pi^{3/2}}{4 \mathcal{V} k_{\text{F}}^3} \bigg[ \frac{2}{3} X^3 - X \big( \e^{-X^2} + 24 \big) \\ + \frac{40}{X} \big( \e^{-X^2} - 3 \big) - \frac{80}{X^3} \big( \e^{-X^2} - 1 \big)  + \frac{105}{2} \sqrt{\pi} \erf(X) \bigg] \\ \\
H_4  = -\frac{9 \pi^{3/2}}{4 \mathcal{V} k_{\text{F}}^3} \bigg[ \frac{2}{3} X^3 + X \big( \e^{-X^2} - 40 \big) \\ + \frac{8}{X} \big(17 \- \e^{-X^2} - 45 \big) + \frac{112}{X^3}\big( \e^{-X^2} + 5 \big) \\ 
+ \frac{672}{X^5} \big( \e^{-X^2} - 1 \big) + \frac{231}{2} \sqrt{\pi} \erf(X) \bigg] \\ \\
H_5  = -\frac{11 \pi^{3/2}}{4 \mathcal{V} k_{\text{F}}^3} \bigg[ \frac{2}{3} X^3 - X \big( \e^{-X^2} + 60 \big) \\ + \frac{8}{X} \big(23 \- \e^{-X^2} - 105 \big)  
- \frac{64}{X^3}\big( 8 \- \e^{-X^2} - 35 \big) \\ - \frac{864}{X^5} \big( 3 \- \e^{-X^2} + 7 \big) - \frac{8 \, 640}{X^7} \big( \e^{-X^2} + 1 \big)  + \frac{429}{2} \sqrt{\pi} \erf(X) \bigg] \\ \\
H_6  = -\frac{13 \pi^{3/2}}{4 \mathcal{V} k_{\text{F}}^3} \bigg[ \frac{2}{3} X^3 + X \big( \e^{-X^2} - 84 \big) \\+ \frac{80}{X} \big(5 \- \e^{-X^2} - 21 \big) 
- \frac{320}{X^3}\big( 2 \- \e^{-X^2} + 21 \big) \\ + \frac{480}{X^5} \big( 19 \- \e^{-X^2} - 63 \big) \\ + \frac{10 \, 560}{X^7} \big(5 \- \e^{-X^2} + 9 \big) \\ 
 + \frac{147 \, 840}{X^9} \big( \e^{-X^2} - 1 \big) + \frac{715}{2} \sqrt{\pi} \erf(X) \bigg] \\ \\
H_7  = -\frac{15 \pi^{3/2}}{4 \mathcal{V} k_{\text{F}}^3} \bigg[ \frac{2}{3} X^3 - X \big( \e^{-X^2} + 112 \big) \\ + \frac{16}{X} \big(31 \- \e^{-X^2} - 189 \big) 
 - \frac{160}{X^3}\big( 11 \- \e^{-X^2} - 105 \big) \\- \frac{480}{X^5} \big( 47 \- \e^{-X^2} + 231 \big) - \frac{1 \, 920}{X^7} \big(11 \- \e^{-X^2} - 297 \big) \\
 - \frac{174 \, 720}{X^9} \big(7 \- \e^{-X^2} + 11 \big) \\ - \frac{3 \, 144 \, 960}{X^{11}} \big( \e^{-X^2} - 1 \big) + \frac{1 \, 105}{2} \sqrt{\pi} \erf(X) \bigg]
\end{multline}
\endgroup
where w$X \equiv \mu k_{\text{F}}$ and the error function $erf$ is defined by
\begin{equation} \label{errorfunction}
\erf(x) \equiv \frac{2}{\sqrt{\pi}} \int_0^x \dd{t} \, \e^{-t^2}
\end{equation}
Thanks to the relations \eqref{Landautensor}, these expressions allow to fully determine the Landau parameters associated with the tensor force.

\subsection{Spin-orbit contribution} \label{spinorbitLandauA}

In this subsection, one proves that the spin-orbit interaction characterized by \eqref{SOpot} does not contribute to the Landau parameters in INM.
Following the steps that led to Eq.\eqref{directSOelmat} and Eq.\eqref{exchangeSOelmat} along with the conservation of the quasiparticle pair momentum, $\vec{k}_p + \vec{k}_{h'} = \vec{k}_h + \vec{k}_{p'}$, the direct and exchange spatial matrix elements of the spin-orbit interaction in the particle-hole representation 
read, respectively:
\begin{multline}
\mel*{\vec{k}_p \vec{k}_{h'}}{\big[ \cev{\nabla}_{12} G(r_{12}) \cross \vec{\nabla}_{12} \big]}{\vec{k}_h \vec{k}_{p'}}
= \\ \frac{1}{\mathcal{V}^2} \int \dd[3]{r_1} \int \dd[3]{r_2} \e^{-\ii (\vec{k}_p - \vec{k}_h) \cdot (\vec{r}_1 - \vec{r}_2)} G(r_{12}) \times \\ \big[ (\vec{k}_p - \vec{k}_{h'}) \cross (\vec{k}_h - \vec{k}_{p'}) \big] \\ \\
\mel*{\vec{k}_p \vec{k}_{h'}}{\big[ \cev{\nabla}_{12} G(r_{12}) \cross \vec{\nabla}_{12} \big]}{\vec{k}_{p'} \vec{k}_h} 
= ~~~~~~~~~~~~~~~~ \\
-\frac{1}{\mathcal{V}^2} \int \dd[3]{r_1} \int \dd[3]{r_2} \e^{-\ii (\vec{k}_p - \vec{k}_h) \cdot (\vec{r}_1 - \vec{r}_2)}  G(r_{12}) \times \\ \big[ (\vec{k}_p - \vec{k}_{h'}) \cross (\vec{k}_h - \vec{k}_{p'}) \big]
\end{multline}
At the Landau limit,  one has $|\vec{k}_p| = |\vec{k}_h| = |\vec{k}_{\text{F}}|$ and $|\vec{k}_{p'}| = |\vec{k}_{h'}| = |\vec{k}'_{\text{F}}|$, so that:
\begin{multline}
\mel*{\vec{k}_{\text{F}} \vec{k}'_{\text{F}}}{\big[ \cev{\nabla}_{12} G(r_{12}) \cross \vec{\nabla}_{12} \big]}{\vec{k}_{\text{F}} \vec{k}'_{\text{F}}} = \\ 
\frac{1}{\mathcal{V}^2} \int \dd[3]{r_1} \int \dd[3]{r_2} G(r_{12}) \big[ (\vec{k}_{\text{F}} - \vec{k}'_{\text{F}}) \cross (\vec{k}_{\text{F}} - \vec{k}'_{\text{F}}) \big] \\ = 0\\ \\
\mel*{\vec{k}_{\text{F}} \vec{k}'_{\text{F}}}{\big[ \cev{\nabla}_{12} G(r_{12}) \cross \vec{\nabla}_{12} \big]}{\vec{k}'_{\text{F}} \vec{k}_{\text{F}}} 
= ~~~~~~~~~~~~~~~ \\
-\frac{1}{\mathcal{V}^2} \int \dd[3]{r_1} \int \dd[3]{r_2} G(r_{12}) \big[ (\vec{k}_{\text{F}} - \vec{k}'_{\text{F}}) \cross (\vec{k}_{\text{F}} - \vec{k}'_{\text{F}}) \big] \\ = 0
\end{multline}
Therefore, at the Landau limit, the spatial matrix elements associated with the spin-orbit interaction vanish. Thus, the spin-orbit interaction does not contribute to the Landau parameters in INM. This is the reason why there are no such parameters in the expression of the quasiparticle interaction \eqref{QPinteraction}.

\section{Brief reminder on reaction mechanisms}
\label{app:reac}
Nuclear reactions induced by nucleons or light ions at energies ranging from a few keV up to several hundred MeV per nucleon generally proceed via three main mechanisms: direct, pre-equilibrium, and compound reactions. Each mechanism is characterized by distinct time scales, dominant dynamics, and theoretical frameworks. Central to modeling these processes is the optical potential. The following sections summarize these key concepts and then explore each reaction mechanism.

\subsection{Optical Model}
\label{sec:intro:omp}

The optical model provides a complex potential, the optical potential, that encodes the average nucleon-nucleus interaction, including absorption into non-elastic channels. It can be constructed either phenomenologically, typically with local Woods-Saxon forms, or microscopically from fundamental interactions. Following Feshbach's projection-operator formalism, the full many-body scattering problem is effectively replaced by a one-body, energy-dependent complex and generally non-local potential, defined as:
\begin{equation}
U(r,r';E)=V(r,r';E)+i\,W(r,r';E).
\end{equation}
Here, the real part \(V\) describes elastic dynamics within the elastic space, while the imaginary component \(W\) quantifies absorption, accounting for the loss of flux into inelastic excitations, transfer, and compound-nucleus formation.

Elastic scattering observables are derived from this potential. The scattering amplitude,
\begin{equation}
    f(\theta)=\frac{1}{2ik}\sum_{\ell}(2\ell+1)\left(e^{2i\delta_\ell}-1\right)P_\ell(\cos\theta),
\end{equation}
directly yields the differential cross-section,
\begin{equation}
    \frac{d\sigma}{d\Omega}=|f(\theta)|^2,
\end{equation}
and via the optical theorem, the total cross-section,
\begin{equation}
    \sigma_{\mathrm{tot}}=\frac{4\pi}{k}\Im f(0),
\end{equation}
with the reaction cross-section defined as \(\sigma_{\mathrm{r}}=\sigma_{\mathrm{tot}}-\sigma_{\mathrm{el}}\).

Phenomenological potentials commonly incorporate volume and surface terms for both \(V\) and \(W\), accompanied by a spin-orbit coupling term, frequently expressed using Woods-Saxon profiles. Advanced treatments include non-locality and dispersion relations linking energy dependencies of \(V\) and \(W\) to maintain causality.

Microscopic approaches explicitly incorporate nuclear structure. Brueckner-Hartree-Fock theory utilizes the in-medium \(G\)-matrix derived from bare nucleon-nucleon interactions in nuclear matter. By applying a local-density approximation, the resulting optical potential for finite nuclei emerges through folding procedures involving the effective nuclear interaction and the target ground-state density. Alternatively, in the microscopic RPA-Green's function framework, the optical potential arises from dynamic polarization of the target nucleus: particle-hole excitations handled through RPA methods produce both real and imaginary components naturally, yielding a dispersive, non-local, and fully microscopic optical potential.

\subsection{Direct reactions}
\label{sec:intro:dir}
Direct nuclear reactions involve a projectile interacting with only one or a few nucleons of the target (or equivalently, involving one or a few degrees of freedom) in one or a few fast steps (on the order of $10^{-21}$–$10^{-22}$ seconds, the average time for the projectile to cross the target diameter). They become dominant over compound processes when the incident energy increases, and include processes such as elastic scattering, inelastic scattering (excitation of nuclear states), and one- or few-nucleon transfer reactions. These reactions are distinguished by prompt emission of reaction products and strongly forward-peaked, anisotropic angular distributions in the center-of-mass frame. The angular distributions in direct reactions are very sensitive to the momentum transfer and any parity change during the interaction. Consequently, the pattern of angles at which the outgoing particles are observed (for example, the number and position of diffraction maxima or minima in the distribution) reflects the transferred orbital angular momentum. Using the selection rules for angular momentum and parity, one can often deduce the spin-parity ($J^{\pi}$) of the excited states populated in the reaction from the characteristic shape of the angular distribution. The theoretical framework for describing direct reactions uses the optical model to account for initial- and final-state interactions. In practice, one treats the projectile–target system with complex optical potentials that produce incoming and outgoing $\emph{distorted waves}$ (i.e. solutions to the Schrödinger equation that include absorption and phase shifts from the average nuclear potential). To calculate the cross section for populating a specific final state, one can apply the Distorted Wave Born Approximation (DWBA) or, for strongly coupled channels, a Coupled-Channels approach. These models take nuclear structure inputs such as transition densities or deformation parameters to define the coupling interaction that links the initial state to the chosen excited state. As an example, consider an inelastic scattering reaction $A(a,a')A'$, in which a projectile $a$ scatters from nucleus $A$ and leaves it in an excited state $A'$. In DWBA the transition amplitude for this process can be written as
\begin{equation}
T_{if} \;=\; \Big\langle\, \chi_f^{(-)}\,\Phi_f \;\Big|\; V_{if} \;\Big|\; \chi_i^{(+)}\,\Phi_i \,\Big\rangle \,
\end{equation}
where $\chi_i^{(+)}$ and $\chi_f^{(-)}$ are the incoming and outgoing distorted-wave functions, $\Phi_i$ and $\Phi_f$ are the internal wavefunctions of the target nucleus in the initial (ground) and final (excited) states, and $V_{if}$ is the effective coupling potential responsible for the transition. From this amplitude, the differential cross section for the reaction (in the center-of-mass frame) is obtained as
\begin{equation}
\label{eq:dwba}
\frac{d\sigma}{d\Omega} \;=\; \frac{\mu_i,\mu_f}{(2\pi\hbar^2)^2}\;\frac{k_f}{k_i}\;\Big|\,T_{if}\,\Big|^2 \,
\end{equation}
with $\mu_{i,f}$ the reduced masses and $k_{i,f}$ the wave numbers for the initial and final channels, respectively. This formalism allows one to calculate angular distributions and absolute cross sections for direct reactions, which can then be compared with experimental data to extract nuclear structure information or to challenge nuclear structure models.

\subsection{Pre–equilibrium (multi–step) reactions}
\label{sec:intro:preeq}
Pre-equilibrium or multi-step reactions (with characteristic time scales of \(\tau\sim10^{-21}\! -\!10^{-18}\,\mathrm{s}\)) describe processes in which the nuclear system undergoes several intermediate excitations (such as particle–hole creations and annihilations) before reaching a fully equilibrated compound nucleus state. Unlike direct reactions, which involve only one or very few nucleons in a single step, pre-equilibrium reactions involve multiple internal degrees of freedom over intermediate timescales.

The  exciton model provides a widely-used semi-classical description, representing the intermediate nuclear configurations in terms of particle–hole states (excitons). This model tracks the stepwise evolution of particle–hole configurations as the nucleus progresses towards equilibrium, using transition probabilities to characterize the successive excitations and de-excitations.

In contrast, microscopic pre-equilibrium theories, such as the Multi-Step Direct (MSD) and Multi-Step Compound (MSC) models, provide a fully quantum-mechanical treatment. These theories expand the full reaction amplitude as a series of terms in powers of the residual nuclear interaction \(V\):
\begin{equation}
\begin{array}{lcl}
T_{fi} &=& \langle \chi_f^{(-)}\,\Phi_f|\;V + V\,G_0\,V + \\ 
& & ~~~~V\,G_0\,V\,G_0\,V + \dots\;|\Phi_i\,\chi_i^{(+)}\rangle
\end{array}
\end{equation}
where \(G_0\) is the free Green's function, and \(\chi^{(\pm)}\) represents distorted waves determined by optical potentials.

Microscopic theories resemble direct reaction treatments in that they use optical potentials for distorted waves and require detailed nuclear structure inputs, such as transition densities. However, they differ fundamentally by explicitly accounting for multiple intermediate interactions rather than just a single, fast collision. Additionally, intermediate nuclear states and transitions may be treated statistically, with approximations such as sudden or adiabatic dynamics applied depending on the energy regime and reaction type.

In practical applications, these quantum-mechanical approaches can better capture complex reaction dynamics, providing insights into excitation mechanisms and better predictive capabilities for cross-sections and spectra.

\subsection{Compound nucleus reactions}
\label{sec:intro:cn}

Compound–nucleus (CN) reactions occur when the projectile is fully absorbed, forming an excited intermediate system that loses memory of the entrance channel. After a time scale of \(\tau\sim10^{-18}\!-\!10^{-16}\,\mathrm{s}\), the system reaches statistical equilibrium. At this stage, only conserved quantities like total energy, angular momentum \(J\), and parity \(\pi\) govern the decay. All accessible quantum states are considered equally probable, and the CN decays through statistical emission of nucleons, light ions, \(\gamma\)-rays and fission.

The Hauser–Feshbach (HF) model \cite{Hauser1952} describes these decays in terms of transmission coefficients and level densities. The decay probability from an entrance channel \(\alpha\) to an exit channel \(\beta\) (e.g., neutron, proton, \(\gamma\)) is given by:

\begin{equation}
\begin{array}{lcl}
\displaystyle \sigma_{\alpha \rightarrow \beta}(E) &=& \displaystyle \sum_{J, \pi} \frac{(2J+1)}{(2s_P+1)(2s_T+1)} \times \\
 & & ~~~~ \displaystyle \frac{T_\alpha^{J\pi}(E)\, T_\beta^{J\pi}(E)}{\sum_\gamma T_\gamma^{J\pi}(E)}\,W_{\alpha\beta}^{J\pi}\,
\end{array}
\end{equation}

where \(s_P\) and \(s_T\) are the projectile and target spins, \(T_c^{J\pi}\) is the transmission coefficient for channel \(c\) with spin-parity \(J^\pi\), and \(W_{\alpha\beta}^{J\pi}\) is a width-fluctuation correction factor.

Decays can populate either known discrete levels or the continuum. For low excitation energies, transitions to individually known final states are calculated explicitly. Beyond a cutoff energy, a statistical treatment using level densities \(\rho(E,J,\pi)\) is applied, and the total transmission to the continuum is integrated over final excitation energy. A unified expression for the partial decay width \(\Gamma_c(E)\) for channel \(c\) is:

\begin{equation}
\begin{array}{lcl}
\Gamma_c(E) &=& \displaystyle \sum_f T_c(E_c^f) + \int_{E_c^{\mathrm{cut}}}^{E_c^{\mathrm{max}}} T_c(E_c)\, \times \\ 
 & & \displaystyle ~~~~~~~~~~~~~~~~~~~ \rho(E_c, J', \pi')\, dE_c
\end{array}
\end{equation}
where \(E_c^f\) are the energies of discrete final states, and \(\rho(E_c, J', \pi')\) is the level density in the residual nucleus.

Hauser–Feshbach calculations require the following ingredients:
\begin{itemize}[label=$-$]
  \item Level densities \(\rho(E, J, \pi)\) for final nuclei and/or $E, J, \pi$ values of discrete levels.
  \item Transmission coefficients \(T_c(E,\ell)\) from optical model potentials for nucleon and light-ion channels.
  \item Gamma-ray strength functions \(f_\gamma(E_\gamma)\), used to derive \(T_\gamma(E_\gamma)\) for electromagnetic decays,
 as well as branching ratios and electron conversion coefficients for decay between discrete levels.
 \item Transmission coefficient for the tunneling  through the deformation energy barrier that separates the initial (ground-state) well from the scission configuration.
\end{itemize}




\end{appendices}


\end{document}